%% file: _main.tex
\documentclass[english]{lipics-v2016}

\renewcommand{\copyrightline}{}
\newcommand{\texInputDir}{.}
\usepackage[utf8]{inputenc}

\usepackage{multicol}

\usepackage{graphicx}
\usepackage{epstopdf}
\usepackage{color}
\usepackage[dvipsnames]{xcolor}
\usepackage{xspace}

\usepackage{wrapfig}
\usepackage{comment}
\usepackage{enumitem}
\usepackage{array, booktabs, afterpage}

\usepackage{caption,subcaption,graphicx}

\usepackage{amsmath, amssymb,bbm,mathrsfs}
\renewcommand{\leq}{\leqslant}
\renewcommand{\geq}{\geqslant}

\usepackage{amsthm}
\theoremstyle{plain}
\newtheorem{proposition}[theorem]{Proposition}
\newtheorem{fact}[theorem]{Fact}

\usepackage{adjustbox}
\usepackage{varioref}

\usepackage{caption,subcaption,graphicx}

\input{\texInputDir/macros.tex}

\title{Proving the Turing Universality of Oritatami Co-\!Transcriptional Folding (Full Text)}

\author[1]{Cody Geary}
\author[2]{Pierre-\'Etienne Meunier}
\author[3]{Nicolas Schabanel}
\author[4]{Shinnosuke Seki}
\affil[1]{California Institute of Technology, Pasadena, CA, USA. \texttt{\href{mailto:codyge@gmail.com}{codyge@gmail.com}}.}
\affil[2]{Tapdance, Inria Paris, France. \texttt{\href{mailto:pe@pijul.org}{pe@pijul.org}}}
\affil[3]{CNRS, U. Lyon (LIP), France and IXXI, U. Lyon, France. \texttt{\href{http://perso.ens-lyon.fr/nicolas.schabanel/}{perso.ens-lyon.fr/nicolas.schabanel/}}}
\affil[4]{oritatami Lab, University of Electro-Communications, Tokyo, Japan. \texttt{\href{http://www.sseki.lab.uec.ac.jp/}{www.sseki.lab.uec.ac.jp/}}}
\authorrunning{C. Geary, P.-É. Meunier, N. Schabanel, S. Seki}
\ArticleNo{1}

\date{\today}                                           

\begin{document}
\maketitle

\begin{abstract}
We study the oritatami model for molecular co-transcriptional folding. 
In oritatami systems, the transcript (the ``molecule'') folds as it is synthesized (transcribed), according to a local energy optimisation process, which is similar to how actual biomolecules such as RNA fold into complex shapes and functions as they are transcribed. 
We prove that there is an oritatami system embedding universal computation in the folding process itself. 

Our result relies on the development of a generic toolbox, which is easily reusable for future work to design complex functions in oritatami systems. 
%
We develop  ``low-level'' tools that allow to easily spread apart the encoding of different ``functions'' in the transcript, even if they are required to be applied at the same geometrical location in the folding.
We build upon these low-level tools, a programming framework with increasing levels of abstraction, from encoding of instructions into the transcript to logical analysis. This framework is similar to the hardware-to-algorithm levels of abstractions in standard algorithm theory. These various levels of abstractions allow to separate the proof of correctness of the global behavior of our system, from the proof of correctness of its implementation. Thanks to this framework, we were able to computerise the proof of correctness of its implementation and produce certificates, in the form of a relatively small number of proof trees, compact and easily readable/checkable by human, while encapsulating huge case enumerations. We believe this particular type of certificates can be generalised to other discrete dynamical systems, where proofs involve large case enumerations as well.
\end{abstract}




\input{\texInputDir/intro.tex}
\section{Definitions and Main results}

\subsection{Oritatami Systems}

Let $B$ be a finite set of \emph{bead types}.
A \emph{configuration} $c$ of a bead type sequence $p \in B^* \cup B^{\mathbb{N}}$ is a directed self-avoiding path in the triangular lattice $\Tlat$,\footnote{The triangular lattice is defined as $\mathbb{T} = (\mathbb{Z}^2, \sim)$, where $(x, y) \sim (u, v)$ if and only if $(u, v) \in \cup_{\epsilon = \pm1}\{{(x+\epsilon, y)},  {(x, y+\epsilon)}, {(x+\epsilon, y+\epsilon)}\}$. Every position $(x,y)$ in $\Tlat$ is mapped in the euclidean plane to $x\cdot \vec E + y \cdot \vec{SW}$ using the vector basis $\vec E = (1,0)$ and $\vec{SW} = \rotateClockwiseAA{\vec E} = (-\frac12, -\frac{\sqrt3}2)$.} where for all integer $i$, vertex $c_i$ of~$c$ is labelled by $p_i$. $c_i$ is the \emph{position} in $\Tlat$ of the $(i+1)$th bead, of type $p_i$, in configuration~$c$. A \emph{partial configuration} of a sequence $p$ is a configuration of a prefix of $p$. 

For any partial configuration $c$ of some sequence $p$, an \emph{elongation} of $c$ by $k$ beads (or \emph{$k$-elongation}) is a partial configuration of $p$ of length $|c|+k$ extending by $k$ positions the self-avoiding path $c$. We denote by $\Ccal_p$ the set of all partial configurations of $p$ (the index $p$ will be omitted when the context is clear). We denote by $\elong{c}{k}$ the set of all  $k$-elongations of a partial configuration~$c$ of sequence~$p$.

\subparagraph{Oritatami systems.} An \emph{oritatami system} $\TMO=(p,\heart,\delay)$ is composed of (1) a (possibly infinite) bead type sequence $p$, called the \emph{transcript}, (2) an \emph{attraction rule}, which is a symmetric relation $\heart\subseteq B^2$, (3) a parameter $\delay$ called the \emph{delay}. $\TMO$ is said \emph{periodic} if $p$ is infinite and periodic. Periodicity ensures that the ``program'' $p$ embedded in the oritatami system is finite (does not hardcode any specific behavior) and at the same time allows arbitrary long computation. 

We say that two bead types $a$ and $b$ \emph{attract} each other when $a\heart b$. Furthermore, given a (partial) configuration $c$ of a bead type sequence $q$, we say that there is a \emph{bond} between two adjacent positions $c_i$ and $c_j$ of $c$ in $\Tlat$ if  $q_i\heart q_j$ and $|i-j|>1$. The \emph{number of bonds} of configuration $c$ of~$q$ is denoted by ${H(c)=  |\{(i,j)\,:\,c_i\sim c_j,\, j>i+1,\text{ and }q_i\heart q_j\}|}$.

\subparagraph{Oritatami dynamics.}
The folding of an oritatami system is controlled by the delay $\delta$. Informally, the configuration grows from a \emph{seed configuration} (the input), one bead at a time. This new bead adopts the position(s) that maximise the potential number of bonds the configuration can make when elongated by $\delta$ beads in total.  This dynamics is \emph{oblivious} as it keeps no memory of the previously preferred positions; it differs thus slightly from the hasty dynamics studied in~\cite{GeMeScSe2016}; it might also be considered as closer to experimental conditions such as in \cite{GearyRothemundAndersen2014}.

Formally,  given an oritatami system $\TMO = (p, \heart, \delay)$ and a \emph{seed configuration} $\sigma$ of a \emph{seed bead type sequence} $s$, we denote by $\Ccal_{\sigma,p}$ the set of all partial  configurations of the sequence $s\cdot p$ elongating the seed configuration $\sigma$. The considered \emph{dynamics} ${\Dynamics:2^{\Ccal_{\sigma,p}}\rightarrow2^{\Ccal_{\sigma,p}}}$ maps every subset $S$ of partial configurations of length~$\ell$ elongating $\sigma$ of the sequence $s\cdot p$ to the subset $\Dynamics(S)$ of partial configurations of length~$\ell+1$ of $s\cdot p$ as follows:
$$
\Dynamics(S) = \bigcup_{\mbox{\footnotesize $c\in S$}}\, \underset{\mbox{\footnotesize$\gamma\in\elong{c}{1}$}}{\arg\max} \left(\, 
\max_{\mbox{\footnotesize$\eta\in\elong{\gamma}{(\delta-1)}$}} H(\eta)\,\right)
$$
The possible configurations at time~$t$ of the oritatami system $\TMO$ are the elongations of the seed configuration~$\sigma$ by $t$ beads in the set $\Dynamics^t(\{\sigma\})$.

We say that the oritatami system is \emph{deterministic} if at all time~$t$, $\Dynamics^t(\{\sigma\})$ is either a singleton or the empty set. In this case, we denote by $c^t$ the configuration at time~$t$, such that: $c^0 = \sigma$ and $\Dynamics^t(\{\sigma\}) = \{c^t\}$ for all $t>0$; we say that the partial configuration $c^t$ \emph{folds (co-transcriptionally) into} the partial configuration $c^{t+1}$ deterministically. In this case, at time $t$, the $(t+1)$-th bead of $p$ is placed in $c^{t+1}$ at the position that maximises the number of bonds that can be made in a $\delta$-elongation of $c^t$.

We say that the oritatami system \emph{halts} at time $t$ if $t$ is the first time for which ${\Dynamics^t(\{\sigma\}) = \varnothing}$. The folding process may only stop because of a geometric obstruction (no more elongation is possible because the configuration is trapped in a closed area).
\smallskip

Please refer to Fig.~\ref{fig:glider:folding} and~\ref{fig:switchback:folding} for examples of the dynamical folding of a transcript. Observe that a given transcript may fold (deterministically) into different paths because of its interactions with its local environment (see section~\ref{sec:glider:switchback} for more details). 
\medskip

\subsection{Main result}

Our main result consists in proving the following theorem that demonstrates that oritatami systems are able to complete arbitrary Turing computation. It shows in particular that deciding whether a given oritatami system folds into a finite size shape for a given seed is undecidable.

\begin{theorem}[Main result]
\label{thm:main}
There is a fixed set $B$ of 542 bead types with a fixed attraction rule $\heart$ on $B$, together with two encodings:
\begin{itemize}[topsep=2pt]
\item $\pi$ that maps in polynomial time, any single tape Turing machine $\TMM$ to a bead type sequence $\pi_\TMM \in B^*$;
\item 
$(s,\sigma)$ that maps in polynomial-time, any single-tape Turing machine $\TMM$ and any input $x$ of $\TMM$ to a seed configuration $\sigma_\TMM(x)$ of a bead type sequence $s_\TMM(x)$ of length $O_\TMM(|x|)$, linear in the size of the input $x$ (and polynomial in $|\TMM|$);    
\end{itemize}
such that: For any single tape Turing machine $\TMM$ and every input $x$ of \TMM, the deterministic and periodic oritatami system $\TMO_\TMM = ((\pi_\TMM)^\infty, \heart, 3)$ whose transcript has period $\pi_\TMM$ and whose delay is $\delta = 3$,  halts its folding from the seed configuration $\sigma_\TMM(x)$ if and only if $\TMM$ halts on input $x$. Furthermore, for all $t$ and all input~$x$ of $\TMM$, if $\TMM$ halts on $x$ after $t$ steps, then the folding of $\TMO_\TMM$ from seed configuration $\sigma_\TMM(x)$ halts after folding $O_\TMM(t^4\log^2 t)$ beads. %
%
%
\end{theorem}

\subparagraph{There is one Turing-universal periodic transcript.} Note that if we apply this theorem to an intrinsically universal single tape Turing machine $\TMU$ (see \cite{Ollinger2002ICALP}), then we obtain one single \emph{absolutely fixed} transcript $\pi_\TMU$ such that the deterministic and periodic oritatami system $\TMO_\TMU = ((\pi_\TMU)^\infty,\heart,3)$ with 542 bead types can simulate efficiently the halting of any Turing machine $\TMM$ on any input $x$ using a suitable seed configuration obtained via the encoding of $\TMM$ and $x$ in $\TMU$. It follows that this absolutely fixed oritatami system consisting of one single periodic transcript is able of arbitrary Turing computation. 
\medskip

From now on, we only consider deterministic periodic oritatami systems with delay $\delta = 3$. 

\subsection{Basic design tool: Glider/Switchback}
\label{sec:glider:switchback}

As a warm-up, let us introduce a special type of bead sequence (see Fig.~\ref{fig:glider:switchback}) that, depending on the initial context of its folding, either folds as a \emph{glider} (a long and thin self-supported shape heading in a fixed direction) or as \emph{switchbacks} (a narrow and high shape allowing compact storage). This only requires a small number of distinct beads types (12 per switchbacks, that can be repeated every 4 switchbacks).  This is achieved by designing a rule with minimum interactions ensuring minimum interferences between both folding patterns. Compatibility between the glider and the turns in switchbacks is ensured by aligning the switchback turns with the turns of the glider, exploiting thus the similarity of their finger-like shape there.   

This glider/switchback sequence will be used to store (as switchbacks) and expose (as glider) specific information encoded in the transcript when needed. 

{
\newcommand{\sbgscale}{.3}
\newcommand{\gscale}{1}
\begin{figure}[t]
 \center
	\begin{subfigure}{.25\textwidth}
	\includegraphics[scale=\sbgscale,trim={6cm 0cm 6cm 0cm},clip]{\texInputDir/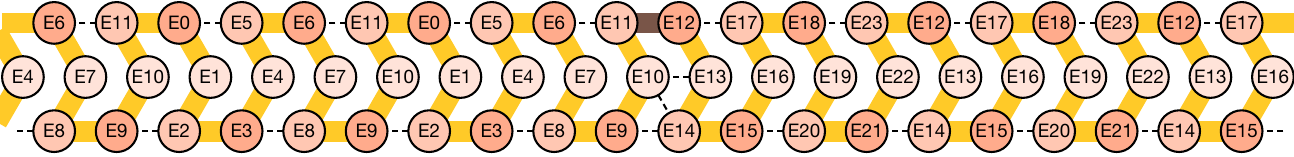}
	\caption{Glider}
	\end{subfigure}
	\begin{subfigure}{.25\textwidth}
	 \includegraphics[scale=\sbgscale,trim={2mm 0cm 2mm 0cm},clip]{\texInputDir/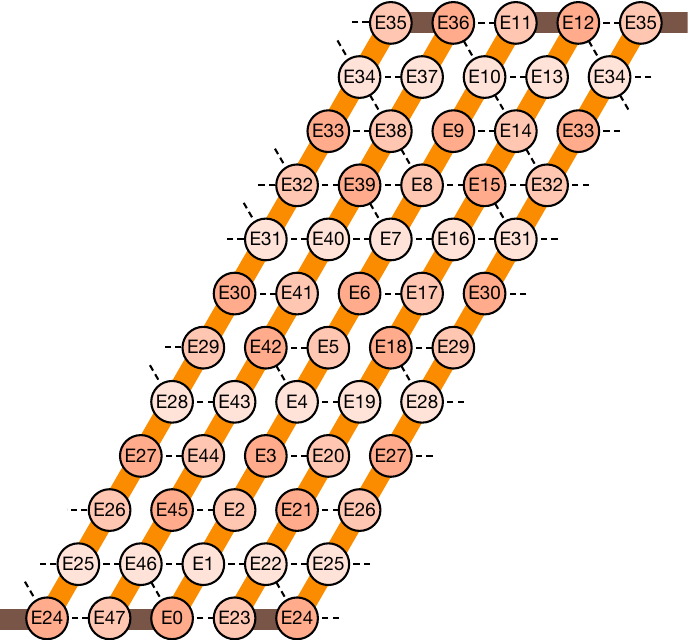}
	 \caption{Switchback}
	\end{subfigure}
	\begin{subfigure}{.45\textwidth}
	 \includegraphics[scale=\sbgscale]{\texInputDir/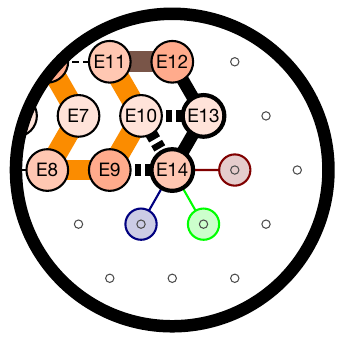}
	\hspace*{1cm}
	\includegraphics[scale=\sbgscale]{\texInputDir/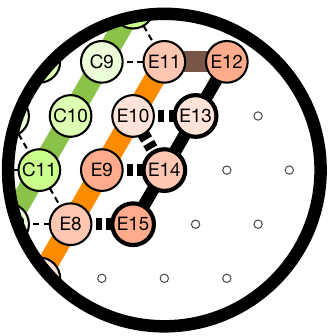}
	 \caption{Glider/Switchback turn folding compatibility.}
	\end{subfigure}
	\\[-6mm]
	\begin{subfigure}{\textwidth}
 	\adjustbox{scale=2,max width=\textwidth}{\includegraphics[scale=\gscale]{\texInputDir/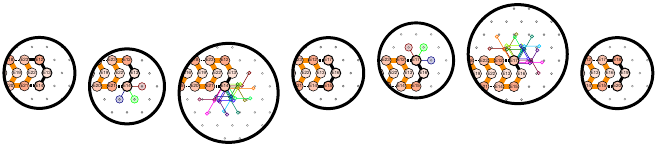}}
	\caption{\captionpar{from left to right:} the folding of the subsequence as a glider.}
	\label{fig:glider:folding}
	\end{subfigure}
	\\[2mm]
	\begin{subfigure}{\textwidth}		
\adjustbox{scale=2,max width=\textwidth}{ \includegraphics[scale=\sbgscale]{\texInputDir/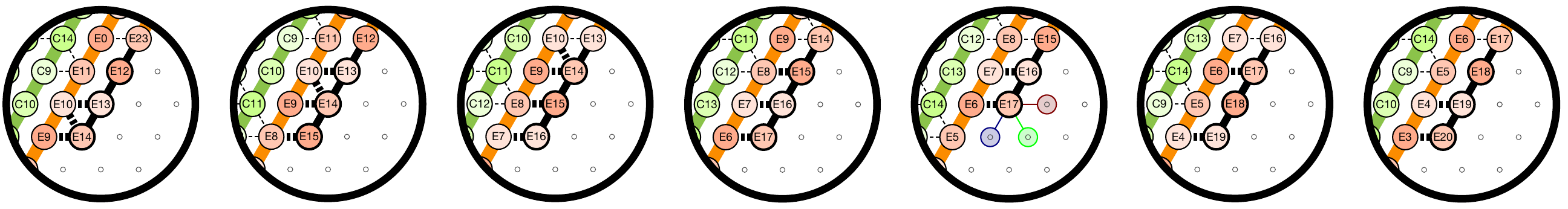}}
	\caption{\captionpar{from left to right:} the folding of the subsequence as a switchback.}
	\label{fig:switchback:folding}
	\end{subfigure}
 \caption{\captionpar{Glider/switchback subsequence.} The folding path of the  transcript is represented as the thick colorful line and the $\heart$-bonds between beads are represented as dashed lines. The bond-maximizing path for the $\delta=3$ lastly produced beads is represented  by a thick black line, possibly terminated by several colorful paths if several paths realize the maximum of number of bonds.}
 \label{fig:glider:switchback}
 \end{figure}

\subsection{Skipping Cyclic Tag Systems and Turing-Universality}

Our proof of  the Turing-universality of oritatami systems consists in simulating a special kind of cyclic tag systems (CTS), called skipping cyclic tag system. Cook introduced CTS in \cite{Cook2004} and proved that they combined the tremendous advantage of simulating efficiently any Turing machines, while not requiring a random access lookup table, which makes simulation a lot easier.

\subparagraph{A skipping cyclic tag system (SCTS)}
consists of a cyclic list of $n$ words ${\alpha = \tuple{\alpha^0,\ldots,\alpha^{n-1}}\in\{\word0,\word1\}^*}$, called  \emph{appendants}, and an initial \emph{dataword} $u^0\in\{\word0,\word1\}^*$. Intuitively, $\alpha$ encodes the program and $u^0$ encodes the input. Its configuration at time $t$ consists of a \emph{marker} $m^t$, recording the index of the current appendant at time~$t$, and a dataword $u^t$.  Initially, $m^0 = 0$ and the dataword is $u^0$. At each time step~$t$, the SCTS acts deterministically on configuration $(m^t,u^t)$ in one of three ways:
\begin{description}[topsep=2pt,itemsep=-1pt]
\item[(Halt step) If $u^t$ is the empty word $\epsilon$,] then the SCTS halts;%
\footnote{Note that SCTS halting condition requires the dataword to be empty as opposed to \cite{Cook2004,WoodsNeary2006} where the computation of a cyclic tag system is said to end also if it repeats a configuration.}
%
%
\item[(Nop step) If the first letter $u^t_0$ of $u^t$ is $\word 0$,] then $u^t_0$ is deleted and the marker moves to the next appendant cyclically: i.e., $m^{t+1} = (m^t+1) \mod n$ and ${u^{t+1} = u^t_1\cdots u^t_{|u^t|-1}}$; 
\item[(Skip-append step) If $u^t_0 = \word1$,] then $u^t_0$ is deleted, the next appendant  $\alpha^{(m^t+1~\mod~n)}$ is appended onto the right end of $u^t$, and the marker moves to the second next appendant: i.e., $u^{t+1} = u^t_1\cdots u^t_{|u^t|-1}\cdot \alpha^{(m^t+1~\mod~n)}$ and ${m^{t+1} = (m^t+2)\mod n}$. 
\end{description}
%

\noindent 
For example, consider the SCTS $\TME=(\tuple{\word{110}, \epsilon, \word{11}, \word{0}}; u^0 = \word{010})$. Its execution $(\pQ{m^t}u^t)_t$ is: \\
\centerline{$
\pQ0\word{010} 
\rightarrow \pQ1\word{10}
\tagTransition{Append}{[2:\word{11}]} 
\pQ3\word{011}
\rightarrow \pQ0\word{11}
\tagTransition{Append}{[1:\epsilon]} 
\pQ2\word{1} 
\tagTransition{Append}{[3:\word{0}]} 
\pQ0\word{0}
\rightarrow  \pQ1\,\texttt{Halt}
$}
%

\subparagraph{Turing universality.}
A sequence of articles and thesis by Cook~\cite{Cook2004}, and Neary and Woods~\cite{WoodsNeary2006,NearyPhD}, allows to show that SCTS are able to simulate any Turing machine efficiently in the following sense: (proof deferred to appendix \vpageref{proof:prop:TM>SCTS})

\begin{proposition}[\cite{WoodsNeary2006,NearyPhD}]
\label{prop:TM>SCTS}
Let $\TMM$ be a deterministic Turing machine using a single tape. There is a polynomial algorithm that computes a skipping cyclic tag system $\TMS_\TMM$, together with a linear-time encoding $u_\TMM(x)$ of the input $x$ of $\TMM$ as an input dataword for $\TMS_\TMM$, such that for all input $x$: $\TMS_\TMM$ halts on input dataword $u_\TMM(x)$ if and only if $\TMM$ halts on input $x$. Furthermore, for all $t$, if $\TMM$ halts after $t$ steps, then $\TMS$ halts after $O_\TMM(t^2\log t)$ steps.
%
%
Moreover, the number of appendants of $\TMS$ is a multiple of~4.
\end{proposition}

In order to prove Theorem~\ref{thm:main}, we are thus left with proving that there is an oritatami system that simulates in quadratic time any SCTS system (see Theorem~\ref{thm:key} in appendix} for a precise statement). 

%

\input{\texInputDir/block-simulation}





%


\section{Advanced Design Tool box}
\label{sec:toolbox}

\input{\texInputDir/design-toolbox}

\section{Correctness of local folding: Proof tree certificates}
\label{sec:proof:local:folding}

The goal of this section is to conclude the proof of our design by proving Key Lemma~\ref{lem:key}. The proof works by induction, assuming that the preceding beads of the transcript fold at the locations claimed by the lemma. We proceed in 3 steps:
\begin{itemize} 
\item We first enumerate all the possible environments for every part of the transcript. As, we carefully aligned our design, most of the beads only see a small number of different environments.
\item For the few cases (three in total) where the number of environments is unbounded, we give an explicit proof of correctness of their folding (Lemmas~\ref{lem:env:below:write}, \ref{lem:env:top:read1}, and~\ref{lem:exp:col} in section~\ref{sec:enum:env}). This is where the concealing feature of socks and the exponential bead type coloring play a crucial role. 
\item For all the other cases, we designed human-checkable computer-generated certificates, called \emph{proof trees}. It consists in listing in a compact but readable manner all the possible paths for the transcript in every possible environment.  In order to match human readability, paths with identical bonding patterns are grouped into one single ball. Balls containing the paths maximizing the number of bonds are highlighted in bold and organized in a tree. This reduces the number of cases to less than 5 balls in most of the levels of the tree, achieving human-checkability of the computed certificate (see Fig.~\ref{fig:proof:tree:example} in appendix). Proof trees are available at \url{https://www.irif.fr/~nschaban/oritatami/}
\end{itemize} 

This ensures the highest level of certification of the correctness of our design.

\bibliographystyle{plain}
\bibliography{biblio.bib}

\newpage

\begin{figure}[p]
\vspace*{-5mm}
\begin{subfigure}{\textwidth}
\adjustbox{max width = \textwidth}{\includegraphics{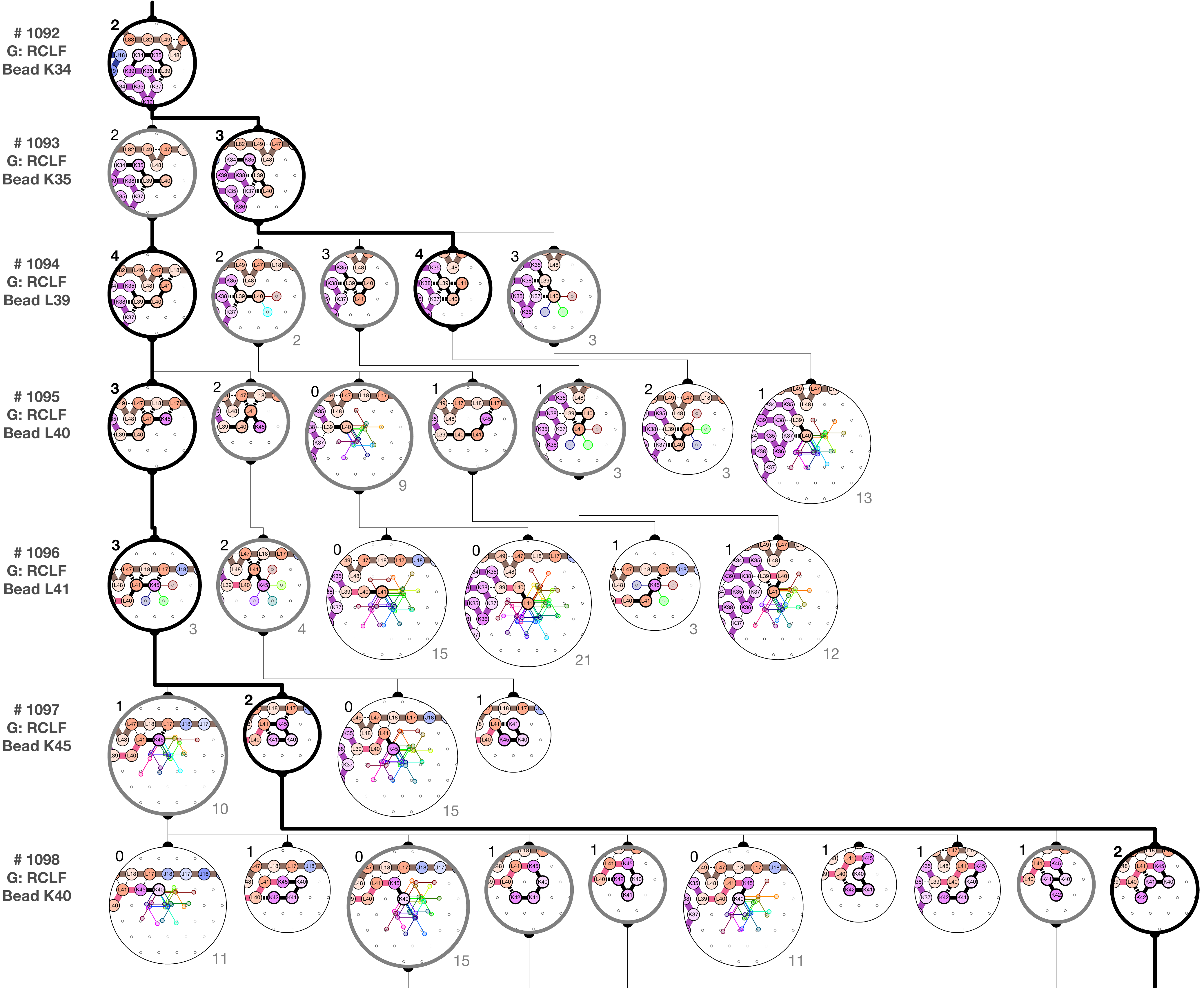}}
\caption{Proof tree for the glider turn in $\GReadZ$.}
\end{subfigure}
\begin{subfigure}{\textwidth}
\adjustbox{max width = \textwidth}{\includegraphics{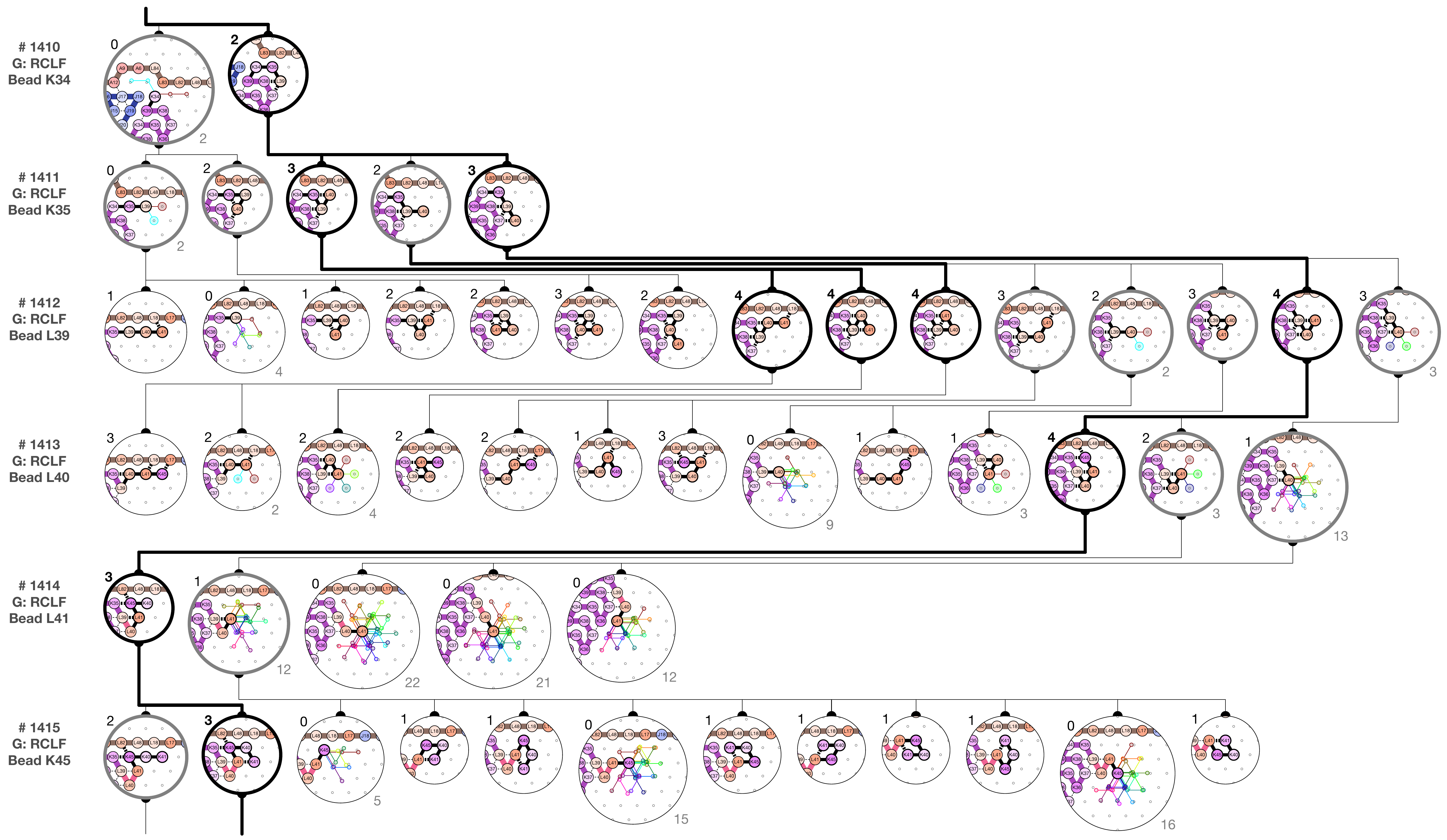}}
\caption{Proof tree for the glider turn in $\GReadU$.}
\end{subfigure}
\caption{\captionpar{Two examples of proof trees for the same subsequence in two different environements.} The number at the upper-left corner of every ball stands for the number of bonds for the path inside the ball. The number at the lower right corner of each ball stands for the number of paths grouped in the ball, allowing to check that no path was omitted. Balls highlighted in black bold contain the bonds-maximizing paths. Balls highlighted in grey bold contain the paths that places the bead at the same location as the bonds-maximizing paths, and which must thus be considered in the next level as well.}
\label{fig:proof:tree:example}
\end{figure}

\afterpage{\clearpage}
\newpage


\appendix

\input{\texInputDir/omitted}

\newpage

\input{\texInputDir/all-bricks}

\input{\texInputDir/blocks}

\input{\texInputDir/geometry}

\afterpage{\clearpage} 
\newpage

\input{\texInputDir/modules+bricks}

\afterpage{\clearpage} 
\newpage

\section{Computerized proof of correctness of the STCS oritatami simulation}

\subsection{Enumerating all possible environments for each module}
\input{\texInputDir/proof-environments}

\afterpage{\clearpage}
\newpage

\subsection{Computer-generated proof trees for each possible environment}
\input{\texInputDir/proof-trees}


\afterpage{\clearpage} 
\newpage


\input{\texInputDir/rule}






\end{document}

%% file: macros.tex
\usepackage{pifont,bbding}
\usepackage{ifthen}

\newenvironment{omittedproof}[2]{
	\begin{proof}[Proof of {#1}~\ref{#2}]
	\label{proof:#2}
}{
	\end{proof}
}
\newcommand{\bZig}{\ensuremath{\blacktriangleright}}
\newcommand{\bZag}{\ensuremath{\blacktriangleleft}}
\newcommand{\empile}[2]{\makebox[.6em]{\scalebox{.8}{\makebox[0em]{\raisebox{.4em}[0em]{#1}}\makebox[0em]{\raisebox{0em}[0em]{#2}}}}}
\newcommand{\bZigZag}{{\empile{\bZig}\bZag}} 
\newcommand{\bZagZig}{{\empile{\bZag}\bZig}} 

\newcommand{\retraitAlign}{\hspace*{8mm}}

\newcommand{\rot}{\ensuremath{\operatorname{Rotate_{180^\circ}}}}
\newcommand{\mirror}{\ensuremath{\operatorname{HorizontalMirror}}}
\newcommand{\vmirror}{\ensuremath{\operatorname{VerticalMirror}}}
\newcommand{\rotateClockwise}{\ensuremath{\operatorname{RotateClockwise}}}
\newcommand{\rotateClockwiseAA}[1]{\ensuremath{\rotateClockwise\left(#1,120^\circ\right)}}

\newcommand{\newBrickName}[4]{%
	\ensuremath{\Module{#1}{\ensuremath{\text{#2}\operatorname{#3}\text{#4}}}}
	}

\newcommand{\newBrickNameTwo}[5]{%
	\ensuremath{\ModuleTwo{#1}{#2}{\ensuremath{\text{#3}\operatorname{#4}\text{#5}}}}
	}

\newcommand{\newBrickNameThree}[6]{%
	\ensuremath{\ModuleThree{#1}{#2}{#3}{\ensuremath{\text{#4}\operatorname{#5}\text{#6}}}}
	}

\newcommand{\ZigUp}{\ensuremath{\raisebox{.17em}{\ensuremath{\underline{\bZig}}}}}
\newcommand{\ZigDown}{\ensuremath{\overline{\bZig}}}
\newcommand{\Zag}{\ensuremath{\overline{\bZag}}}
\newcommand{\Halt}{\ensuremath{\blacksquare}}

\newcommand{\AZigUp}{\newBrickName{couleurA}{A}{\ZigUp}{}}
\newcommand{\AZigDown}{\newBrickName{couleurA}{A}{\ZigDown}{}}
\newcommand{\AZag}{\newBrickName{couleurA}{A}{\Zag}{}}

\newcommand{\BZigUp}{\newBrickName{couleurB}{B}{\ZigUp}{}}
\newcommand{\BZigDown}{\newBrickName{couleurB}{B}{\ZigDown}{}}
\newcommand{\BZag}{\newBrickName{couleurB}{B}{\Zag}{}}
\newcommand{\BHalt}{\newBrickName{couleurB}{B}{\Halt}{}}

\newcommand{\CZigUp}{\newBrickName{couleurC}{C}{\ZigUp}{}}
\newcommand{\CZigDown}{\newBrickName{couleurC}{C}{\ZigDown}{}}
\newcommand{\CZag}{\newBrickName{couleurC}{C}{\Zag}{}}
\newcommand{\CEnd}{\newBrickName{couleurC}{C}{\ZigDown}{End}}

\newcommand{\DZigUp}{\newBrickNameTwo{couleurDZ}{couleurDU}{D}{\ZigUp}{}}
\newcommand{\DZigDown}{\newBrickNameTwo{couleurDZ}{couleurDU}{D}{\ZigDown}{}}
\newcommand{\DZag}{\newBrickNameTwo{couleurDZ}{couleurDU}{D}{\Zag}{}}
\newcommand{\Append}{\raisebox{-.2em}{\text{\!\PencilRightDown}}}
\newcommand{\DAppend}{\newBrickNameTwo{couleurDZ}{couleurDU}{D}{\Append}{}}
\newcommand{\DWrite}{\DAppend}

\newcommand{\DxrtZigUp}[2]{\newBrickNameTwo{couleurDZ}{couleurDU}{D\word{#1}\ensuremath{_{#2}}}{\ZigUp}{}}
\newcommand{\DxrtZigDown}[2]{\newBrickNameTwo{couleurDZ}{couleurDU}{D\word{#1}\ensuremath{_{#2}}}{\ZigDown}{}}
\newcommand{\DxrtZag}[2]{\newBrickNameTwo{couleurDZ}{couleurDU}{D\word{#1}\ensuremath{_{#2}}}{\Zag}{}}

\newcommand{\DZZigUp}[1]{\newBrickName{couleurDZ}{D\word0\ensuremath{_{#1}}}{\ZigUp}{}}
\newcommand{\DZZigDown}[1]{\newBrickName{couleurDZ}{D\word0\ensuremath{_{#1}}}{\ZigDown}{}}
\newcommand{\DZZag}[1]{\newBrickName{couleurDZ}{D\word0\ensuremath{_{#1}}}{\Zag}{}}

\newcommand{\DUZigUp}[1]{\newBrickName{couleurDU}{D\word1\ensuremath{_{#1}}}{\ZigUp}{}}
\newcommand{\DUZigDown}[1]{\newBrickName{couleurDU}{D\word1\ensuremath{_{#1}}}{\ZigDown}{}}
\newcommand{\DUZag}[1]{\newBrickName{couleurDU}{D\word1\ensuremath{_{#1}}}{\Zag}{}}

\newcommand{\DZAppend}[1]{\newBrickName{couleurDZ}{~\Append D\word0\ensuremath{_{#1}}}{\!}{}}
\newcommand{\DUAppend}[1]{\newBrickName{couleurDU}{~\Append D\word1\ensuremath{_{#1}}}{\!}{}}

\newcommand{\DxrtAppend}[2]{\newBrickNameTwo{couleurDZ}{couleurDU}{~\Append D\word{#1}\ensuremath{_{#2}}}{\!}{}}

\newcommand{\SegDZigUp}[1]{\newBrickName{couleurSEG}{SegD\ensuremath{_{#1}}}{\ZigUp}{}}
\newcommand{\SegDAppend}[1]{\newBrickName{couleurSEG}{SegD\ensuremath{_{#1}}}{\Append}{}}

\newcommand{\EZigUp}{\newBrickName{couleurE}{E}{\ZigUp}{}}
\newcommand{\EZigDown}{\newBrickName{couleurE}{E}{\ZigDown}{}}
\newcommand{\EZag}{\newBrickName{couleurE}{E}{\Zag}{}}
\newcommand{\Return}{\text{Return}}
\newcommand{\ECR}{\newBrickName{couleurE}{E}{\ZigDown}{\Return}}

\newcommand{\EaZigUp}[1]{\newBrickName{couleurE}{E\ensuremath{_{#1}}}{\ZigUp}{}}
\newcommand{\EaZigDown}[1]{\newBrickName{couleurE}{E\ensuremath{_{#1}}}{\ZigDown}{}}
\newcommand{\EaZag}[1]{\newBrickName{couleurE}{E\ensuremath{_{#1}}}{\Zag}{}}
\newcommand{\EaCR}[1]{\newBrickName{couleurE}{E\ensuremath{_{#1}}}{\bZigZag}{}}

\newcommand{\WordName}{Appendant} 
\newcommand{\WordZigUp}[1]{\newBrickNameThree{couleurE}{couleurDZ}{couleurDU}{\WordName\ensuremath{(#1)}}{\ZigUp}{}}
\newcommand{\WordZigDown}[1]{\newBrickNameThree{couleurE}{couleurDZ}{couleurDU}{\WordName\ensuremath{(#1)}}{\ZigDown}{}}
\newcommand{\WordZag}[1]{\newBrickNameThree{couleurE}{couleurDZ}{couleurDU}{\WordName\ensuremath{(#1)}}{\Zag}{}}
\newcommand{\WordAppend}[1]{\newBrickNameThree{couleurE}{couleurDZ}{couleurDU}{Append\WordName\ensuremath{(#1)}}{\bZigZag}{}}

\newcommand{\FZigUp}{\newBrickName{couleurF}{F}{\ZigUp}{}}
\newcommand{\FZigDown}{\newBrickName{couleurF}{F}{\ZigDown}{}}
\newcommand{\FZag}{\newBrickName{couleurF}{F}{\Zag}{}}

\newcommand{\GRead}{\newBrickName{couleurG}{G}{\ZigUp}{Read}}
\newcommand{\GReadZ}{\newBrickName{couleurG}{G}{\ZigUp}{Read\word{0}}}
\newcommand{\GReadU}{\newBrickName{couleurG}{G}{\ZigUp}{Read\word{1}}}
\newcommand{\GZigCopy}{\newBrickName{couleurG}{G}{\ZigDown}{Copy}}
\newcommand{\GZigCopyZ}{\newBrickName{couleurG}{G}{\ZigDown}{Copy\word0}}
\newcommand{\GZigCopyU}{\newBrickName{couleurG}{G}{\ZigDown}{Copy\word1}}

\newcommand{\GZagCopyZ}{\newBrickName{couleurG}{G}{\Zag}{Copy\word0}}
\newcommand{\GZagCopyU}{\newBrickName{couleurG}{G}{\Zag}{Copy\word1}}
\newcommand{\GLF}{\newBrickName{couleurG}{G}{\bZagZig}{LineFeed}}

\DeclareSymbolFont{extraup}{U}{zavm}{m}{n}
\DeclareMathSymbol{\varheart}{\mathalpha}{extraup}{86}
\newcommand{\heart}{\ensuremath{\mathbin{\textcolor{red}{\ensuremath\varheart}}}}

\newcommand{\bkHALT}{\blockName{Halt\,\Halt}}
\newcommand{\bkREAD}[1]{\blockName{Read\ensuremath{#1}\bZig}}
\newcommand{\bkZIGCOPY}[1]{\blockName{Copy\ensuremath{#1}\bZig}}
\newcommand{\bkZAGCOPY}[1]{\blockName{\bZag Copy\ensuremath{#1}}}
\newcommand{\bkUNIT}[1]{\blockName{Unit}}

\newcommand{\bkAPPENDRETURN}[1]{\blockName{Append\bZigZag\,Return\ensuremath{#1}}}

\newcommand{\bkRETURNLINEFEEDHALT}{\blockName{CarriageReturn\bZigZag\,LineFeed\bZagZig\,Halt\,\Halt}}
\newcommand{\bkZAGCOPYLF}[1]{\blockName{Copy\ensuremath{#1}\bZagZig \,LineFeed}}

\newcommand{\word}[1]{\texttt{\upshape #1}}
\newcommand{\blockName}[1]{\textsf{\upshape #1}}
\newcommand{\markedApp}[1]{\ensuremath{#1}}

\newcommand{\bnSEED}{\blockName{Seed}}
\newcommand{\bnHALT}{\blockName{Halt}}
\newcommand{\bnREAD}[2]{\blockName{Read{\word{#1\markedApp{#2}}}\bZig}}
\newcommand{\bnZIGCOPY}[2]{\blockName{Copy{\word{#1\markedApp{#2}}}\bZig}}
\newcommand{\bnZIGCOPYUNIT}[2]{\blockName{Copy{\word{#1\markedApp{#2}}\bZig Unit}}}
\newcommand{\bnZAGCOPYUNIT}[2]{\blockName{Copy{\word{#1\markedApp{#2}}}\bZag  Unit}}
\newcommand{\bnZAGCOPY}[2]{\blockName{Copy{\word{#1\markedApp{#2}}}\bZag}}
\newcommand{\bnAPPENDRETURN}[1]{\blockName{Append\markedApp{#1}\bZigZag Return}}
\newcommand{\bnRETURNLINEFEED}{\blockName{CarriageReturn\bZigZag LineFeed\bZagZig}}
\newcommand{\bnRETURNLINEFEEDHALT}{\blockName{CarriageReturn\bZagZig LineFeed\bZag Halt\Halt}}
\newcommand{\bnCOPYLINEFEEDUNIT}[1]{\blockName{LineFeed\ensuremath{#1}\bZagZig Unit}}
\newcommand{\bnCOPYLINEFEED}[2]{\blockName{Copy{\word{#1\markedApp{#2}}}\bZagZig  LineFeed}}

\newcommand{\bSEED}[1]{\ensuremath{\bnSEED(\word{#1})}}

\newcommand{\blockNameState}[2]{\ensuremath{{#1}\markedApp{#2}}}
\newcommand{\bHALT}[1]{\blockNameState{\bnHALT}{#1}}

\newcommand{\tagTransition}[2]{\begin{array}{c}{\xrightarrow{\texttt{#1}}}\\[-.8em]\scriptstyle #2\end{array}} 
\newcommand{\tuple}[1]{\ensuremath{\langle{#1}\rangle}}
\newcommand{\TM}[1]{\ensuremath{\mathcal{#1}}}

\newcommand{\TMM}{{\TM M}}
\newcommand{\TMC}{{\TM C}}
\newcommand{\TMS}{{\TM S}}
\newcommand{\TMT}{{\TM T}}
\newcommand{\TMU}{{\TM U}}

\newcommand{\TMO}{{\TM O}}

\newcommand{\pQ}[1]{\ensuremath{{}^{[#1]}}}
\newcommand{\kase}[1]{\multicolumn{1}{|@{\,}r@{\,}}{#1}}

\makeatletter
\newcommand\xleftrightarrow[2]{%
  \ext@arrow 9999{\longleftrightarrowfill@}{#1}{#2}}
\newcommand\longleftrightarrowfill@{%
  \arrowfill@\leftarrow\relbar\rightarrow}
\makeatother

\newcommand{\pointHere}[1]{%
	\multicolumn{1}{c}{\ensuremath{\overset{#1}{\scriptstyle\downarrow}}}%
}

\newcommand{\Ebond}{\ensuremath{\mathbin{\overset{\rightarrow}{_\texttt{E}}}}}
\newcommand{\Wbond}{\ensuremath{\mathbin{\overset{\leftarrow}{_\texttt{W}}}}}
\newcommand{\NEbond}{\ensuremath{\mathbin{\overset{\texttt{NE}}{_\nearrow}}}}
\newcommand{\NWbond}{\ensuremath{\mathbin{\overset{\texttt{NW}}{_\nwarrow}}}}
\newcommand{\SEbond}{\ensuremath{\mathbin{\overset{\searrow}{_{\texttt{SE}}}}}}
\newcommand{\SWbond}{\ensuremath{\mathbin{\overset{\swarrow}{_\texttt{SW}}}}}

\newcommand{\Eb}{\Ebond}
\newcommand{\Wb}{\Wbond}
\newcommand{\NEb}{\NEbond}
\newcommand{\NWb}{\NWbond}
\newcommand{\SEb}{\SEbond}
\newcommand{\SWb}{\SWbond}

\usepackage{soul}

\newcommand{\bead}[2]{{\textsf{\upshape\bfseries \ul{#1#2}}}}
\newcommand{\bA}[1]{{\setulcolor{couleurA}\bead A{#1}}}
\newcommand{\bB}[1]{{\setulcolor{couleurB}\bead B{#1}}}
\newcommand{\bC}[1]{{\setulcolor{couleurC}\bead C{#1}}}
\newcommand{\bD}[1]{{\setulcolor{couleurDchaussetteBump}\bead D{#1}}}
\newcommand{\bE}[1]{{\setulcolor{couleurDZ}\bead E{#1}}}
\newcommand{\bF}[1]{{\setulcolor{couleurE}\bead F{#1}}}
\newcommand{\bG}[1]{{\setulcolor{couleurEpaddingG}\bead G{#1}}}
\newcommand{\bH}[1]{{\setulcolor{couleurEpaddingCliff}\bead H{#1}}}
\newcommand{\bI}[1]{{\setulcolor{couleurEpaddingSB3}\bead I{#1}}}
\newcommand{\bJ}[1]{{\setulcolor{couleurF}\bead J{#1}}}
\newcommand{\bK}[1]{{\setulcolor{couleurG}\bead K{#1}}}
\newcommand{\bL}[1]{{\setulcolor{couleurGspecial}\bead L{#1}}}
\newcommand{\bM}[1]{{\setulcolor{couleurGspecial}\bead M{#1}}}
\newcommand{\bN}[1]{{\setulcolor{gray}\bead N{#1}}}

\newcommand{\bbA}[2]{{\setulcolor{couleurA}\bead A{{#1}..{#2}}}}
\newcommand{\bbB}[2]{{\setulcolor{couleurB}\bead B{{#1}..{#2}}}}
\newcommand{\bbC}[2]{{\setulcolor{couleurC}\bead C{{#1}..{#2}}}}
\newcommand{\bbD}[2]{{\setulcolor{couleurDchaussetteBump}\bead D{{#1}..{#2}}}}
\newcommand{\bbE}[2]{{\setulcolor{couleurDZ}\bead E{{#1}..{#2}}}}
\newcommand{\bbF}[2]{{\setulcolor{couleurE}\bead F{{#1}..{#2}}}}
\newcommand{\bbG}[2]{{\setulcolor{couleurEpaddingG}\bead G{{#1}..{#2}}}}
\newcommand{\bbH}[2]{{\setulcolor{couleurEpaddingCliff}\bead H{{#1}..{#2}}}}
\newcommand{\bbI}[2]{{\setulcolor{couleurEpaddingSB3}\bead I{{#1}..{#2}}}}
\newcommand{\bbJ}[2]{{\setulcolor{couleurF}\bead J{{#1}..{#2}}}}
\newcommand{\bbK}[2]{{\setulcolor{couleurG}\bead K{{#1}..{#2}}}}
\newcommand{\bbL}[2]{{\setulcolor{couleurGspecial}\bead L{{#1}..{#2}}}}
\newcommand{\bbM}[2]{{\setulcolor{couleurGspecial}\bead M{{#1}..{#2}}}}

\definecolor{couleurSeed}{RGB}{132,86,0} 
\definecolor{couleurA}{RGB}{211,47,47} 
\definecolor{couleurB}{RGB}{63, 81, 181} 
\definecolor{couleurC}{RGB}{139, 195, 74} 
\definecolor{couleurDZ}{RGB}{255,202,40} 
\definecolor{couleurDU}{RGB}{251,140,0} 
\definecolor{couleurSEG}{RGB}{180,180,180} 
\definecolor{couleurSEGdarker}{RGB}{80,80,80} 
\definecolor{couleurHead}{RGB}{120,120,120} 
\definecolor{couleurE}{RGB}{3, 169, 244} 
\definecolor{couleurF}{RGB}{48,63,159} 
\definecolor{couleurG}{RGB}{171,71,188} 
\definecolor{couleurAppendant}{RGB}{75,75,75}

\definecolor{couleurDchaussetteBump}{HTML}{795548} 
\definecolor{couleurESB1}{HTML}{03A9F4} 
\definecolor{couleurESB1bottes}{HTML}{00BCD4} 
\definecolor{couleurESB2}{HTML}{1976D2} 
\definecolor{couleurEpaddingC}{HTML}{455A64} 
\definecolor{couleurEpaddingG}{HTML}{009688} 
\definecolor{couleurEpaddingSB3}{HTML}{00ACC1} 
\definecolor{couleurEpaddingCliff}{HTML}{455A64} 
\definecolor{couleurFspecial}{HTML}{5E35B1} 
\definecolor{couleurGspecial}{HTML}{F06292} 
\definecolor{couleurEexp}{HTML}{FF80AB} 
\definecolor{couleurEpaddingCliff}{HTML}{455A64} 

\setul{0.2ex}{2pt}
\newcommand{\couleurBead}[1]{
	\ifthenelse{\equal{#1}{A}}{\setulcolor{couleurA}}{} 
	\ifthenelse{\equal{#1}{B}}{\setulcolor{couleurB}}{} 
	\ifthenelse{\equal{#1}{C}}{\setulcolor{couleurC}}{} 
	\ifthenelse{\equal{#1}{D}}{\setulcolor{couleurDchaussetteBump}}{} 
	\ifthenelse{\equal{#1}{E}}{\setulcolor{couleurDZ}}{} 
	\ifthenelse{\equal{#1}{F}}{\setulcolor{couleurE}}{} 
	\ifthenelse{\equal{#1}{G}}{\setulcolor{couleurEpaddingG}}{} 
	\ifthenelse{\equal{#1}{H}}{\setulcolor{couleurEpaddingCliff}}{} 
	\ifthenelse{\equal{#1}{I}}{\setulcolor{couleurEpaddingSB3}}{} 
	\ifthenelse{\equal{#1}{J}}{\setulcolor{couleurF}}{} 
	\ifthenelse{\equal{#1}{K}}{\setulcolor{couleurG}}{} 
	\ifthenelse{\equal{#1}{L}}{\setulcolor{couleurGspecial}}{} 
	\ifthenelse{\equal{#1}{M}}{\setulcolor{couleurGspecial}}{} 
	\ifthenelse{\equal{#1}{N}}{\setulcolor{gray}}{} 
}

\definecolor{couleurSpeciale}{RGB}{233, 30, 99,} 
\definecolor{couleurCopyWrite}{RGB}{170,109,109}
\definecolor{couleurRead}{RGB}{86,252,193}
\definecolor{couleurChaussettes}{RGB}{121, 85, 72} 
\definecolor{couleurBottes}{RGB}{253,33,2}
\definecolor{couleurPointe}{RGB}{9,107,45}
\definecolor{couleurNeutre}{RGB}{75,75,75}
\definecolor{couleurExponential}{RGB}{150,150,25}

\newcommand{\bScolor}[2]{#2} 
\newcommand{\bS}[1]{\bScolor{couleurSpeciale}{#1}}
\newcommand{\bSCW}[1]{\bScolor{couleurCopyWrite}{#1}}
\newcommand{\bSR}[1]{\bScolor{couleurRead}{#1}}
\newcommand{\bSC}[1]{\bScolor{couleurChaussettes}{#1}}
\newcommand{\bSB}[1]{\bScolor{couleurBottes}{#1}}
\newcommand{\bSP}[1]{\bScolor{couleurPointe}{#1}}
\newcommand{\bSN}[1]{\bScolor{couleurNeutre}{#1}}
\newcommand{\bSE}[1]{\bScolor{couleurExponential}{#1}}

\renewcommand{\mod}{\mathbin{\operatorname{mod}}}

\newcommand{\Module}[2]{{\scalebox{0.8}{\adjustbox{cfbox=#1 1.5pt 2pt}{\sffamily\upshape\bfseries{#2}}}}}

\newcommand{\ModuleTwo}[3]{{\scalebox{0.8}{\adjustbox{cframe=#1}{\adjustbox{cfbox=#2 1.5pt 1pt}{\sffamily\upshape\bfseries{#3}}}}}}

\newcommand{\ModuleThree}[4]{{\scalebox{0.8}{\adjustbox{cframe=#1}{\adjustbox{cframe=#2}{\adjustbox{cfbox=#3 1.5pt 1pt}{\sffamily\upshape\bfseries{#4}}}}}}}

\newcommand{\ModuleSeed}[1]{\Module{couleurSeed}{Seed\ensuremath{#1}}}
\newcommand{\ModuleA}{\ensuremath{\Module{couleurA}{A}}}
\newcommand{\ModuleB}{\ensuremath{\Module{couleurB}{B}}}
\newcommand{\ModuleC}{\ensuremath{\Module{couleurC}{C}}}
\newcommand{\ModuleDZ}[1]{\Module{couleurDZ}{D\word0\ensuremath{_{#1}}}}
\newcommand{\ModuleDU}[1]{\Module{couleurDU}{D\word1\ensuremath{_{#1}}}}
\newcommand{\ModuleDx}[1]{\ModuleTwo{couleurDZ}{couleurDU}{D\ensuremath{_{#1}}}}
\newcommand{\ModuleDxrt}[2]{\ModuleTwo{couleurDZ}{couleurDU}{D\word{#1}\ensuremath{_{#2}}}}
\newcommand{\ModuleE}{\ensuremath{\Module{couleurE}{E}}}
\newcommand{\ModuleEa}[1]{\Module{couleurE}{E\ensuremath{_{#1}}}}
\newcommand{\ModuleF}{\ensuremath{\Module{couleurF}{F}}}
\newcommand{\ModuleG}{\ensuremath{\Module{couleurG}{G}}}

\newcommand{\modA}{module $\ModuleA$}
\newcommand{\modB}{module $\ModuleB$}
\newcommand{\modC}{module $\ModuleC$}
\newcommand{\modD}{module $\ModuleDx{}$}

\newcommand{\modE}{module $\ModuleEa{}$}
\newcommand{\modEa}[1]{module $\ModuleEa{#1}$}
\newcommand{\modF}{module $\ModuleF$}
\newcommand{\modG}{module $\ModuleG$}

\newcommand{\ModA}{Module $\ModuleA$}
\newcommand{\ModB}{Module $\ModuleB$}
\newcommand{\ModC}{Module $\ModuleC$}
\newcommand{\ModD}{Module $\ModuleDx{}$}
\newcommand{\ModDxrt}[2]{Module $\ModuleDxrt{#1}{#2}$}

\newcommand{\ModEa}[1]{Module $\ModuleEa{#1}$}
\newcommand{\ModF}{Module $\ModuleF$}
\newcommand{\ModG}{Module $\ModuleG$}

\newcommand{\BigSEG}[2]{\Module{couleurSEGdarker}{Seg{#1}{#2}}}
\newcommand{\SEG}[2]{\Module{couleurSEG}{Seg{#1}{#2}}}
\newcommand{\SEGD}[1]{\Module{couleurSEG}{SegD\ensuremath{_{#1}}}}
\newcommand{\SEGE}[1]{\Module{couleurSEG}{SegE{#1}}}

\newcommand{\HEAD}[1]{\Module{couleurHead}{Head{#1}}}
\newcommand{\TAIL}[1]{\Module{couleurHead}{Tail{#1}}}

\newcommand{\APPENDANTa}[1]{\Module{couleurAppendant}{Appendant \ensuremath{\alpha^{#1}}}}
\newcommand{\WORD}[1]{\Module{couleurAppendant}{word\ensuremath{(#1)}}}

\newcommand{\subst}[1]{\ensuremath{{\substLeft{#1}\substRight}}}
\newcommand{\substLeft}{\ensuremath{\langle\!\langle}}
\newcommand{\substRight}{\ensuremath{\rangle\!\rangle}}
\newcommand{\substOpen}[1]{\ensuremath{{\substLeft{#1}}}}
\newcommand{\substClose}[1]{\ensuremath{{{#1}\substRight}}}

\newcommand{\conc}{\cdot}
\newcommand{\concatop}{\bigodot}
\newcommand{\concat}[1]{\underset{#1}{\concatop}\,}
\newcommand{\concatup}[2]{\overset{#2}{\underset{#1}{\concatop}}\,}

\newcommand{\pathstyle}[1]{\ensuremath{\operatorname{\textsf{#1}}}}
\newcommand{\conformation}{\pathstyle{Conformation}}
\newcommand{\dirPath}[1]{\pathstyle{{#1}-path}}
\newcommand{\NEpath}{\dirPath{NE}}
\newcommand{\Epath}{\dirPath{E}}
\newcommand{\SEpath}{\dirPath{SE}}
\newcommand{\SWpath}{\dirPath{SW}}
\newcommand{\Wpath}{\dirPath{W}}
\newcommand{\NWpath}{\dirPath{NW}}

\newcommand{\dirGlider}[1]{\ensuremath{\operatorname{\textsf{{#1}-glider}}}}
\newcommand{\dirRGlider}[1]{\ensuremath{\operatorname{\textsf{{#1}-rev-glider}}}}
\newcommand{\NEglider}{\dirGlider{NE}}
\newcommand{\Eglider}{\dirGlider{E}}

\newcommand{\SERglider}{\dirRGlider{SE}}

\usepackage{tikz}
\usetikzlibrary{
  arrows,%
  decorations.pathmorphing,%
  decorations.pathreplacing,%
}

\newcommand{\elong}[2]{{#1}^{\hspace*{.05em}\triangleright{#2}}}

\usepackage{mathrsfs}

\newcommand{\Dynamics}{\operatorname{\mathscr D}}

\newcommand{\Ccal}{{\mathcal C}}
\newcommand{\Tlat}{{\mathbb T}}

\newcommand{\delay}{\delta}

\newcommand{\letterSet}{\ensuremath{\{\word0,\word1\}}}

\newcommand{\defaultBrickScale}{2.5}
\makeatletter
\def\maxwidth#1{\ifdim\defaultBrickScale\Gin@nat@width>#1 #1\else\defaultBrickScale\Gin@nat@width\fi}
\def\maxheight#1{\ifdim\defaultBrickScale\Gin@nat@height>#1 #1\else\defaultBrickScale\Gin@nat@height\fi}
\makeatother

\newcommand{\fitFigSize}[3]{%
	\includegraphics[width=\maxwidth{#2},height=\maxheight{#3},keepaspectratio]{ModulesDescriptions/{#1}}
}

\newcommand{\fitFiggSize}[3]{%
	\includegraphics[width=\maxwidth{#2},height=\maxheight{#3},keepaspectratio]{#1}
}

\newcommand{\figBrickHere}[3]{%
	\figBrickHereSize{#1}{#2}{#3}{\textwidth}{\textheight}
}

\newcommand{\figBrickHereSize}[5]{%
	\begin{figure}[h]
	\centerline{\fitFigSize{#1.pdf}{#4}{#5}}
	\caption{\protect{#2}}
	\label{#3}
	\end{figure}
}

\newcommand{\figBrickHereLandscapeSize}[5]{%
	\begin{sidewaysfigure}[h]
	\centerline{\fitFigSize{#1.pdf}{#4}{#5}}
	\caption{\protect{#2}}
	\label{#3}
	\end{sidewaysfigure}
}

\newcommand{\subfigBrickWidth}[4]{%
	\begin{subfigure}{#4}
	\centering\includegraphics[width=#4]{ModulesDescriptions/{#1}}
	\vspace*{-1.25em}
	\caption{\protect{#2}}
	\label{#3}
	\end{subfigure}
}

\newcommand{\subfigBrickLandscape}[3]{%
	\subfigBrickWidth{#1}{#2}{#3}{\textheight}
}

\newcommand{\subfigBrickPortrait}[3]{%
	\subfigBrickWidth{#1}{#2}{#3}{\textwidth}
}

\newcommand{\nodeLabel}[4]{%
	\node () at (#1 cm, #2 cm) {\small\begin{tabular}{#3} #4\end{tabular}} ;
}

\newcommand{\newSubFigBrick}[6]{%
	\begin{subfigure}{#4}
    		\centering
    		\adjustbox{max size= {#4}{\textheight}}{
    		\begin{tikzpicture}
    			\node () at (0,0){\includegraphics[width=15cm,keepaspectratio]{ModulesDescriptions/{#1}}};
			\ifthenelse{\equal{#6}{grid}}{
				\foreach \x in {-10,...,10}
					\draw (\x cm, -10cm)--(\x cm, 10cm) (-10 cm, \x cm)--(10 cm, \x cm) ;
				\draw [draw=red] (0, -10cm)--(0 cm, 10cm) (-10 cm, 0 cm)--(10 cm, 0 cm) ;
			}{}
			#5
		\end{tikzpicture}}
  		\caption{#2}%
  		\label{#3}
	\end{subfigure}
}

\newcommand{\scaleFig}{1.5}

\newcommand{\blockFigs}[2]{\centerline{\adjustbox{scale=\scaleFig, max width=\textwidth, max height=.7\textheight}{#2}}}
\newlength{\figSkip}
\newlength{\figHeight}

\pgfmathsetmacro{\s}{0.075}
\pgfmathsetmacro{\st}{\s}
\newcommand{\TikzBlock}[5]{ 
\node[draw,rounded corners=6pt,scale=0.75,fill=#4,fill opacity=0.5,text opacity=1,draw opacity = 0,anchor=#5] () at (#2*\st pt, #3*\st pt) {#1};  
}
\newcommand{\appendant}[2]{\TMEapp{#1}}
\newcommand{\TIKZBLOCKREAD}[5]{ %
	\ifthenelse{\equal{#1}{0}}%
	{\TikzBlock{\bnREAD{#1}{\appendant{#2}{#3}}}{#4}{#5}{blockReadZeroOutlineColor}{center}}%
	{\TikzBlock{\bnREAD{#1}{\appendant{#2}{#3}}}{#4}{#5}{blockReadOneOutlineColor}{center}}
}
\newcommand{\TIKZBLOCKZIGCOPY}[5]{ %
	\ifthenelse{\equal{#1}{0}}%
	{\TikzBlock{\bnZIGCOPY{#1}{\appendant{#2}{#3}}}{#4}{#5}{blockZigCopyZeroOutlineColor}{center}}%
	{\TikzBlock{\bnZIGCOPY{#1}{\appendant{#2}{#3}}}{#4}{#5}{blockZigCopyOneOutlineColor}{center}}
	}
\newcommand{\TIKZBLOCKZAGCOPY}[5]{ %
	\ifthenelse{\equal{#1}{0}}%
	{\TikzBlock{\bnZAGCOPY{#1}{\appendant{#2}{#3}}}{#4}{#5}{blockZagCopyZeroOutlineColor}{center}}%
	{\TikzBlock{\bnZAGCOPY{#1}{\appendant{#2}{#3}}}{#4}{#5}{blockZagCopyOneOutlineColor}{center}}
 }
\newcommand{\TIKZBLOCKZAGCOPYLINEFEED}[5]{  %
	\ifthenelse{\equal{#1}{0}}%
	{\TikzBlock{\bnCOPYLINEFEED{#1}{\appendant{#2}{#3}}}{#4}{#5}{blockZagCopyZeroOutlineColor}{center}}%
	{\TikzBlock{\bnCOPYLINEFEED{#1}{\appendant{#2}{#3}}}{#4}{#5}{blockZagCopyOneOutlineColor}{center}}
}
\newcommand{\TIKZBLOCKSEED}[3]{
	\TikzBlock{\bnSEED\ensuremath{[\word{#1}]}}{#2}{#3}{blockSeedBeginOutlineColor}{south}
}
\newcommand{\TIKZBLOCKSEEDLETTER}[3]{ 
	\ifthenelse{\equal{#1}{0}}%
	{\TikzBlock{\bnSEED\word{#1}}{#2}{#3}{blockSeedZeroOutlineColor}{north} }%
	{\TikzBlock{\bnSEED\word{#1}}{#2}{#3}{blockSeedOneOutlineColor}{north} }%
}
\newcommand{\TIKZBLOCKAPPENDLETTER}[4]{%
	\ifthenelse{\equal{#1}{0}}%
	{ \TikzBlock{\blockName{APPEND\word{#1}\ensuremath{_{#2}}}}{#3}{#4}{blockAppendZeroOutlineColor}{north} }%
	{ \TikzBlock{\blockName{APPEND\word{#1}\ensuremath{_{#2}}}}{#3}{#4}{blockAppendOneOutlineColor}{north} }%
}
\newcommand{\TIKZBLOCKAPPENDCARRIAGERETURN}[3]{
	\TikzBlock{\bnAPPENDRETURN{\appendant{#1}{}}}{#2}{#3}{blockAppendOutlineColor} {south}
}
\newcommand{\TIKZBLOCKHALT}[3]{%
	\TikzBlock{\bHALT{\TMEapp{#1}}}{#2}{#3}{blockHaltOutlineColor}{west} 
}
\newcommand{\TIKZBLOCKRETURNLINEFEEDHALT}[3]{%
	\TikzBlock{\bnRETURNLINEFEEDHALT{#1}}{#2}{#3}{blockHaltOutlineColor}{north west} 
}

\newlength{\TIKZtextwidth}
\newlength{\TIKZtextheight}

\newcommand{\TIKZPointOfInterest}[9]{%
	\pgfmathsetmacro{\spoi}{#8*\s}
	\pgfmathsetmacro{\lwpoi}{0.2}
	\pgfmathsetmacro{\poifontscale}{0.5}
	\node [anchor=center] (poi#2zoom) at #7 {\includegraphics[scale =\spoi]{#9}};
	\node[fill,circle,scale = \s,minimum size=0.3 pt] (poi#2) at #1 {};
	\ifthenelse{\equal{#4}{E}}{%
		\node[shift={(#5,0)},scale=\poifontscale,anchor=west] (poi#2label) at (poi#2) {\textsf{\bfseries (#2)} #3};
		\draw[scale=\s, line width = \lwpoi pt] (poi#2.center) -- (poi#2label.west);
		}{}
	\ifthenelse{\equal{#4}{W}}{%
		\node[shift={(-#5,0)},scale=\poifontscale,anchor=east] (poi#2label) at (poi#2) {\textsf{\bfseries (#2)} #3};
		\draw[scale=\s, line width = \lwpoi pt] (poi#2.center) -- (poi#2label.east);
		}{}
	\ifthenelse{\equal{#4}{NW}}{%
		\node[shift={(-#5,#5)},scale=\poifontscale,anchor=east] (poi#2label) at (poi#2) {\textsf{\bfseries (#2)} #3};
		\draw[scale=\s, line width = \lwpoi pt] (poi#2.center) -- (poi#2label.east);
		}{}
	\ifthenelse{\equal{#4}{NE}}{%
		\node[shift={(#5,#5)},scale=\poifontscale,anchor=west] (poi#2label) at (poi#2) {\textsf{\bfseries (#2)} #3};
		\draw[scale=\s, line width = \lwpoi pt] (poi#2.center) -- (poi#2label.west);
		}{}
	\ifthenelse{\equal{#4}{SE}}{%
		\node[shift={(#5,-#5)},scale=\poifontscale,anchor=west] (poi#2label) at (poi#2) {\textsf{\bfseries (#2)} #3};
		\draw[scale=\s, line width = 0.2 pt] (poi#2.center) -- (poi#2label.west);
		}{}
	\ifthenelse{\equal{#4}{SW}}{%
		\node[shift={(-#5,-#5)},scale=\poifontscale,anchor=east] (poi#2label) at (poi#2) {\textsf{\bfseries (#2)} #3};
		\draw[scale=\s, line width = \lwpoi pt] (poi#2.center) -- (poi#2label.east);
		}{}
	\node[scale=\poifontscale,anchor=center] at (poi#2zoom.south){\textsf{\bfseries (#2)} #6};
}

\newcommand{\TIKZlabel}[4]{%
	\pgfmathsetmacro{\lwpoi}{0.2}
	\pgfmathsetmacro{\poifontscale}{0.5}
	\node[fill,circle,scale = \s,minimum size=0.3 pt] (disc) at #1 {};
	\ifthenelse{\equal{#3}{E}}{%
		\node[shift={(#4,0)},scale=\poifontscale,anchor=west] (label) at (disc) {\textsf{\bfseries #2}};
		\draw[scale=\s, line width = \lwpoi pt] (disc.center) -- (label.west);
		}{}
	\ifthenelse{\equal{#3}{W}}{%
		\node[shift={(-#4,0)},scale=\poifontscale,anchor=east] (label) at (disc) {\textsf{\bfseries #2}};
		\draw[scale=\s, line width = \lwpoi pt] (disc.center) -- (label.east);
		}{}
	\ifthenelse{\equal{#3}{NW}}{%
		\node[shift={(-#4,#4)},scale=\poifontscale,anchor=east] (label) at (disc) {\textsf{\bfseries #2}};
		\draw[scale=\s, line width = \lwpoi pt] (disc.center) -- (label.east);
		}{}
	\ifthenelse{\equal{#3}{NE}}{%
		\node[shift={(#4,#4)},scale=\poifontscale,anchor=west] (label) at (disc) {\textsf{\bfseries #2}};
		\draw[scale=\s, line width = \lwpoi pt] (disc.center) -- (label.west);
		}{}
	\ifthenelse{\equal{#3}{SE}}{%
		\node[shift={(#4,-#4)},scale=\poifontscale,anchor=west] (label) at (disc) {\textsf{\bfseries #2}};
		\draw[scale=\s, line width = 0.2 pt] (disc.center) -- (label.west);
		}{}
	\ifthenelse{\equal{#3}{SW}}{%
		\node[shift={(-#4,-#4)},scale=\poifontscale,anchor=east] (label) at (disc) {\textsf{\bfseries #2}};
		\draw[scale=\s, line width = \lwpoi pt] (disc.center) -- (label.east);
		}{}
}

\newcommand{\explodedZigZagSize}[1]{\scalebox{.75}{\textcolor{gray}{#1}}}

\usepackage{float}

\newcommand{\figSubRuleHereSize}[5]{
	\begin{figure}[H]
	\centerline{\fitFiggSize{rule/#1.pdf}{#4}{#5}}
	\caption{\protect{#2}}
	\label{#3}
	\end{figure}
}
\newcommand{\figSubRuleHere}[5]{
	\figSubRuleHereSize{#1}{#2}{#3}{\textwidth}{\textheight}
}

\graphicspath{{./OritatamiTuringPatterns/n=4-L=3-P=1/}{.}{./OritatamiTuringPatterns/n=8-L=3-P=1/}}

\newcommand{\homo}{{\ensuremath{\chi}}}

\input{\texInputDir/tikz-blocks}

\input{\texInputDir/tikz-exploded}

\usepackage[section]{placeins}

\newcommand{\TME}{{\TM E}}

\newcommand{\TMEapp}[1]{
	\ifthenelse{\equal{#1}{0}}{{\ensuremath{[0{:}\word{110}]}}}{}
	\ifthenelse{\equal{#1}{1}}{{\ensuremath{[1{:}\epsilon]}}}{}
	\ifthenelse{\equal{#1}{2}}{{\ensuremath{[2{:}\word{11}]}}}{}
	\ifthenelse{\equal{#1}{3}}{{\ensuremath{[3{:}\word{0}]}}}{}
}

\newcommand{\captionpar}[1]{\textcolor{darkgray}{\sffamily\bfseries #1}}

\input{./OritatamiTuringPatterns/n=4-L=3-P=1/blockColors.tex}

%% file: tikz-blocks.tex
\input{\tikzBlockDir/AppendLineFeedHalt_0__n=4-L=3-P=1.tex}
\input{\tikzBlockDir/AppendLineFeedHalt_110__n=4-L=3-P=1.tex}
\input{\tikzBlockDir/AppendLineFeedHalt_11__n=4-L=3-P=1.tex}
\input{\tikzBlockDir/AppendLineFeedHalt_____n=4-L=3-P=1.tex}
\input{\tikzBlockDir/Append_0__n=4-L=3-P=1.tex}
\input{\tikzBlockDir/Append_110__n=4-L=3-P=1.tex}
\input{\tikzBlockDir/Append_11__n=4-L=3-P=1.tex}
\input{\tikzBlockDir/Append_____n=4-L=3-P=1.tex}
\input{\tikzBlockDir/Halt_0__n=4-L=3-P=1.tex}
\input{\tikzBlockDir/Halt_110__n=4-L=3-P=1.tex}
\input{\tikzBlockDir/Halt_11__n=4-L=3-P=1.tex}
\input{\tikzBlockDir/Halt_____n=4-L=3-P=1.tex}
\input{\tikzBlockDir/Read0_0__n=4-L=3-P=1.tex}
\input{\tikzBlockDir/Read0_110__n=4-L=3-P=1.tex}
\input{\tikzBlockDir/Read0_11__n=4-L=3-P=1.tex}
\input{\tikzBlockDir/Read0_____n=4-L=3-P=1.tex}
\input{\tikzBlockDir/Read1_0__n=4-L=3-P=1.tex}
\input{\tikzBlockDir/Read1_110__n=4-L=3-P=1.tex}
\input{\tikzBlockDir/Read1_11__n=4-L=3-P=1.tex}
\input{\tikzBlockDir/Read1_____n=4-L=3-P=1.tex}
\input{\tikzBlockDir/Seed010_n=4-L=3-P=1.tex}
\input{\tikzBlockDir/ZagCopy0LineFeed_0__n=4-L=3-P=1.tex}
\input{\tikzBlockDir/ZagCopy0LineFeed_110__n=4-L=3-P=1.tex}
\input{\tikzBlockDir/ZagCopy0LineFeed_11__n=4-L=3-P=1.tex}
\input{\tikzBlockDir/ZagCopy0LineFeed_____n=4-L=3-P=1.tex}
\input{\tikzBlockDir/ZagCopy0_0__n=4-L=3-P=1.tex}
\input{\tikzBlockDir/ZagCopy0_110__n=4-L=3-P=1.tex}
\input{\tikzBlockDir/ZagCopy0_11__n=4-L=3-P=1.tex}
\input{\tikzBlockDir/ZagCopy0_____n=4-L=3-P=1.tex}
\input{\tikzBlockDir/ZagCopy1LineFeed_0__n=4-L=3-P=1.tex}
\input{\tikzBlockDir/ZagCopy1LineFeed_110__n=4-L=3-P=1.tex}
\input{\tikzBlockDir/ZagCopy1LineFeed_11__n=4-L=3-P=1.tex}
\input{\tikzBlockDir/ZagCopy1LineFeed_____n=4-L=3-P=1.tex}
\input{\tikzBlockDir/ZagCopy1_0__n=4-L=3-P=1.tex}
\input{\tikzBlockDir/ZagCopy1_110__n=4-L=3-P=1.tex}
\input{\tikzBlockDir/ZagCopy1_11__n=4-L=3-P=1.tex}
\input{\tikzBlockDir/ZagCopy1_____n=4-L=3-P=1.tex}
\input{\tikzBlockDir/ZigCopy0_0__n=4-L=3-P=1.tex}
\input{\tikzBlockDir/ZigCopy0_110__n=4-L=3-P=1.tex}
\input{\tikzBlockDir/ZigCopy0_11__n=4-L=3-P=1.tex}
\input{\tikzBlockDir/ZigCopy0_____n=4-L=3-P=1.tex}
\input{\tikzBlockDir/ZigCopy1_0__n=4-L=3-P=1.tex}
\input{\tikzBlockDir/ZigCopy1_110__n=4-L=3-P=1.tex}
\input{\tikzBlockDir/ZigCopy1_11__n=4-L=3-P=1.tex}
\input{\tikzBlockDir/ZigCopy1_____n=4-L=3-P=1.tex}

\input{\tikzBlockDir/execution_n=4-L=3-P=1-outlined.tex}

%% file: OritatamiTuringPatterns/n=4-L=3-P=1/AppendLineFeedHalt_0__n=4-L=3-P=1.tex
\newcommand{\AppendLineFeedHaltBlockZ}[1]{
    \begin{tikzpicture}
        \node [anchor=center] () at (0,0) {\includegraphics[scale =\s]{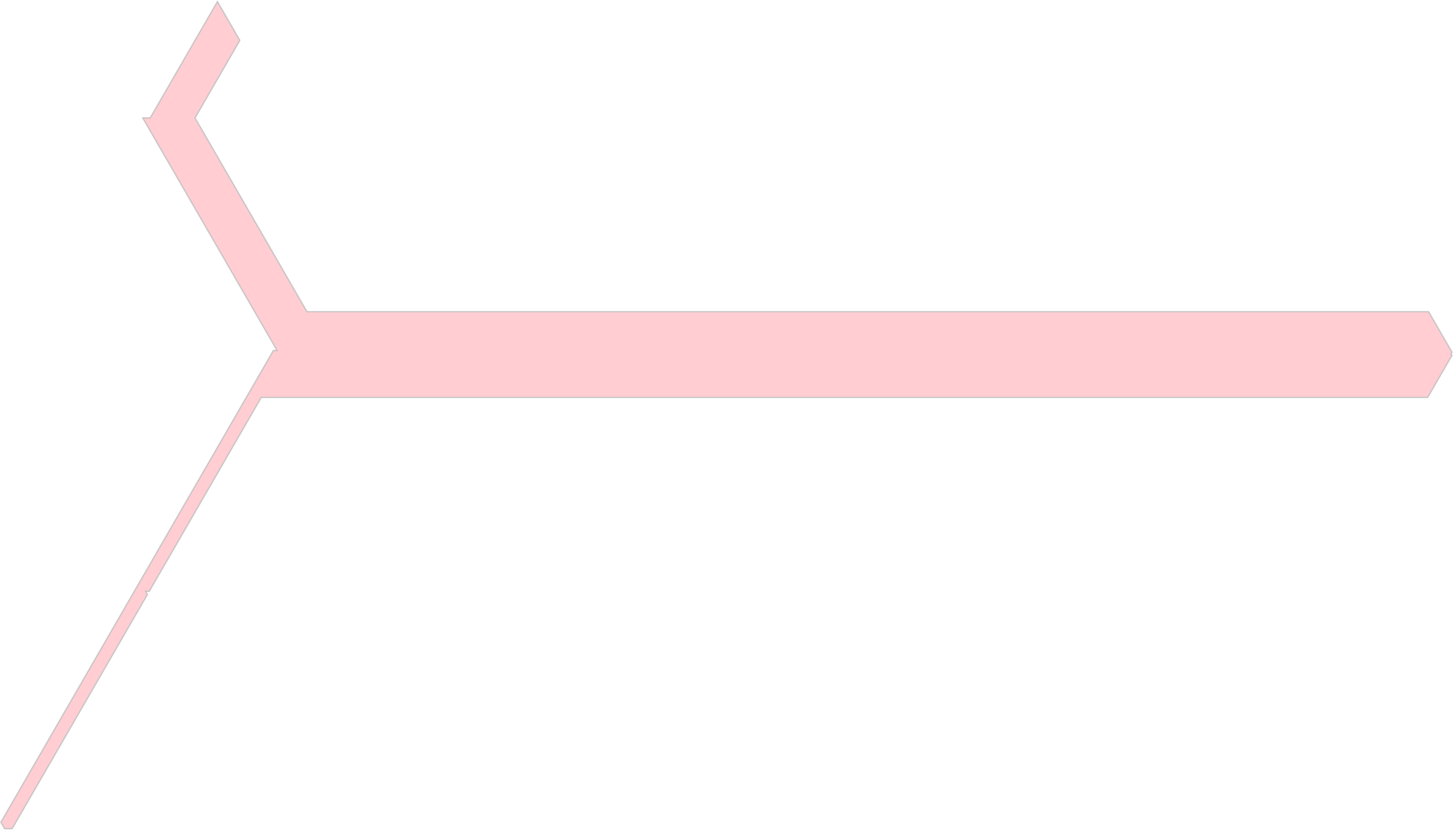}};
        \node [anchor=center] () at (0,0) {\includegraphics[scale =\s]{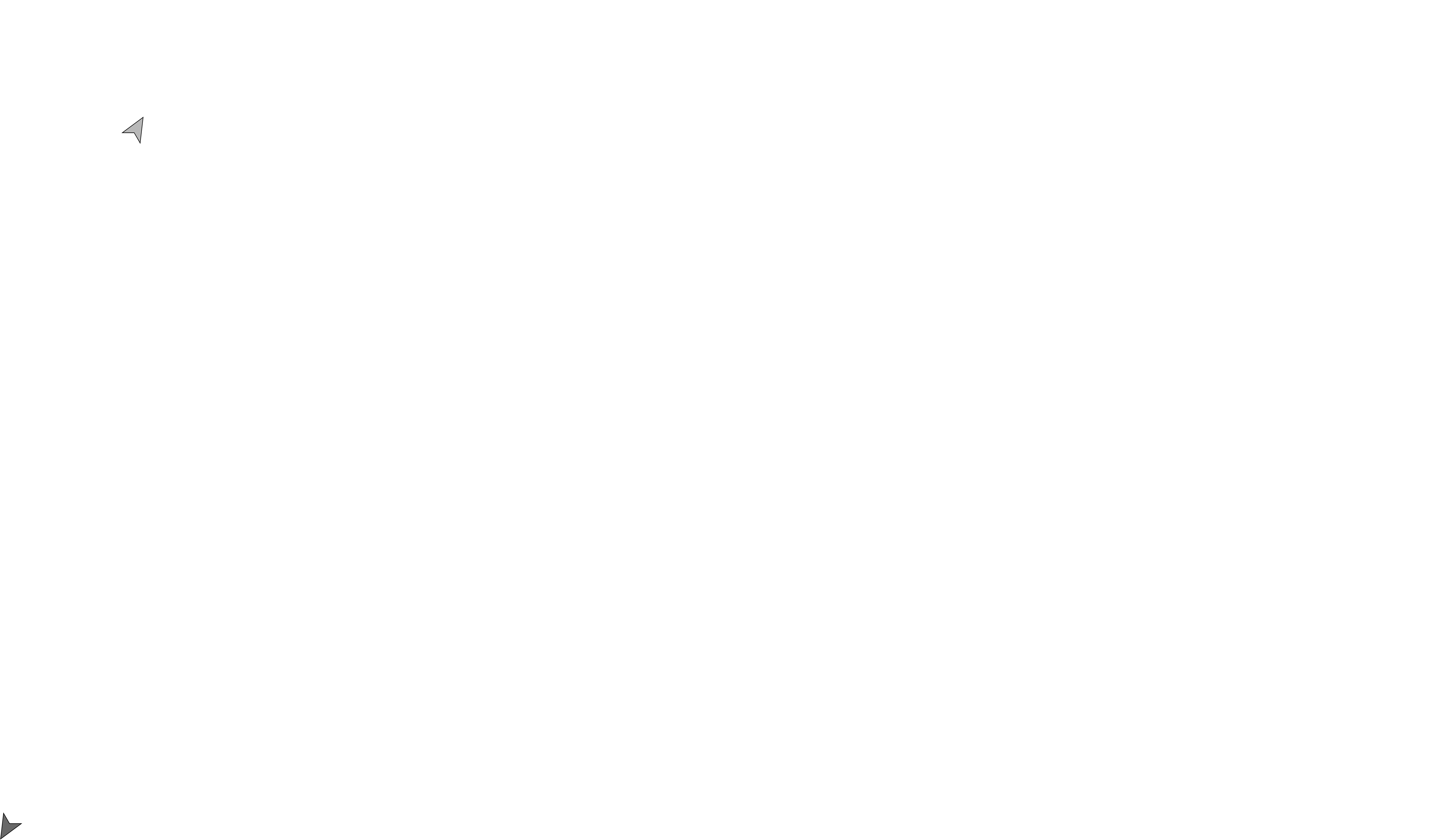}};
        \ifthenelse{\equal{#1}{}}{
\TIKZPointOfInterest{(-1563.75000000*\st pt, 809.73375254*\st pt)}{a}{$(-1,1-h)$}{W}{100.0*\st pt}{Start}{(-1800.00000000*\st pt, -1621.50289790*\st pt)}{8.0}{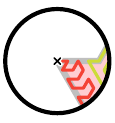}
\TIKZPointOfInterest{(-1363.75000000*\st pt, 1121.50289790*\st pt)}{b}{$\left(3,-\cfrac{3h-5}2\right)$}{E}{100.0*\st pt}{}{(-1200.00000000*\st pt, -1621.50289790*\st pt)}{8.0}{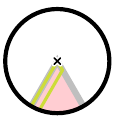}
\TIKZPointOfInterest{(1941.25000000*\st pt, 177.53520778*\st pt)}{c}{$(h+c(0),0)$}{E}{100.0*\st pt}{}{(-600.00000000*\st pt, -1621.50289790*\st pt)}{8.0}{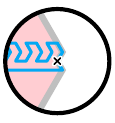}
\TIKZPointOfInterest{(-1546.25000000*\st pt, -458.99346401*\st pt)}{d}{$(h+2,h)$}{E}{100.0*\st pt}{}{(0.00000000*\st pt, -1621.50289790*\st pt)}{8.0}{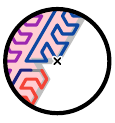}
\TIKZPointOfInterest{(-1196.25000000*\st pt, 173.20508076*\st pt)}{e}{$(h-1,1)$}{SE}{75.0*\st pt}{}{(600.00000000*\st pt, -1621.50289790*\st pt)}{8.0}{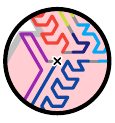}
\TIKZPointOfInterest{(-1928.75000000*\st pt, -1095.52213579*\st pt)}{f}{$(h-1,2h)$}{SW}{100.0*\st pt}{}{(1200.00000000*\st pt, -1621.50289790*\st pt)}{8.0}{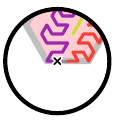}
\TIKZPointOfInterest{(-1913.75000000*\st pt, -1095.52213579*\st pt)}{g}{$(h+2,2h)$}{NE}{100.0*\st pt}{}{(1800.00000000*\st pt, -1621.50289790*\st pt)}{8.0}{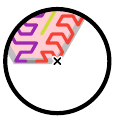}
\TIKZPointOfInterest{(-1558.75000000*\st pt, -463.32359102*\st pt)}{h}{$(h,h+1)$}{NW}{100.0*\st pt}{Trapped}{(-1800.00000000*\st pt, -2221.50289790*\st pt)}{8.0}{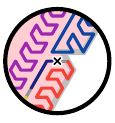}
\TIKZPointOfInterest{(-1921.25000000*\st pt, -1082.53175473*\st pt)}{i}{$(h-1,2h-3)$}{NW}{100.0*\st pt}{Stop}{(-1200.00000000*\st pt, -2221.50289790*\st pt)}{8.0}{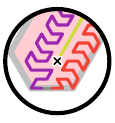}
            }{}
\node [anchor=center] () at (0,0) {\includegraphics[scale =\s]{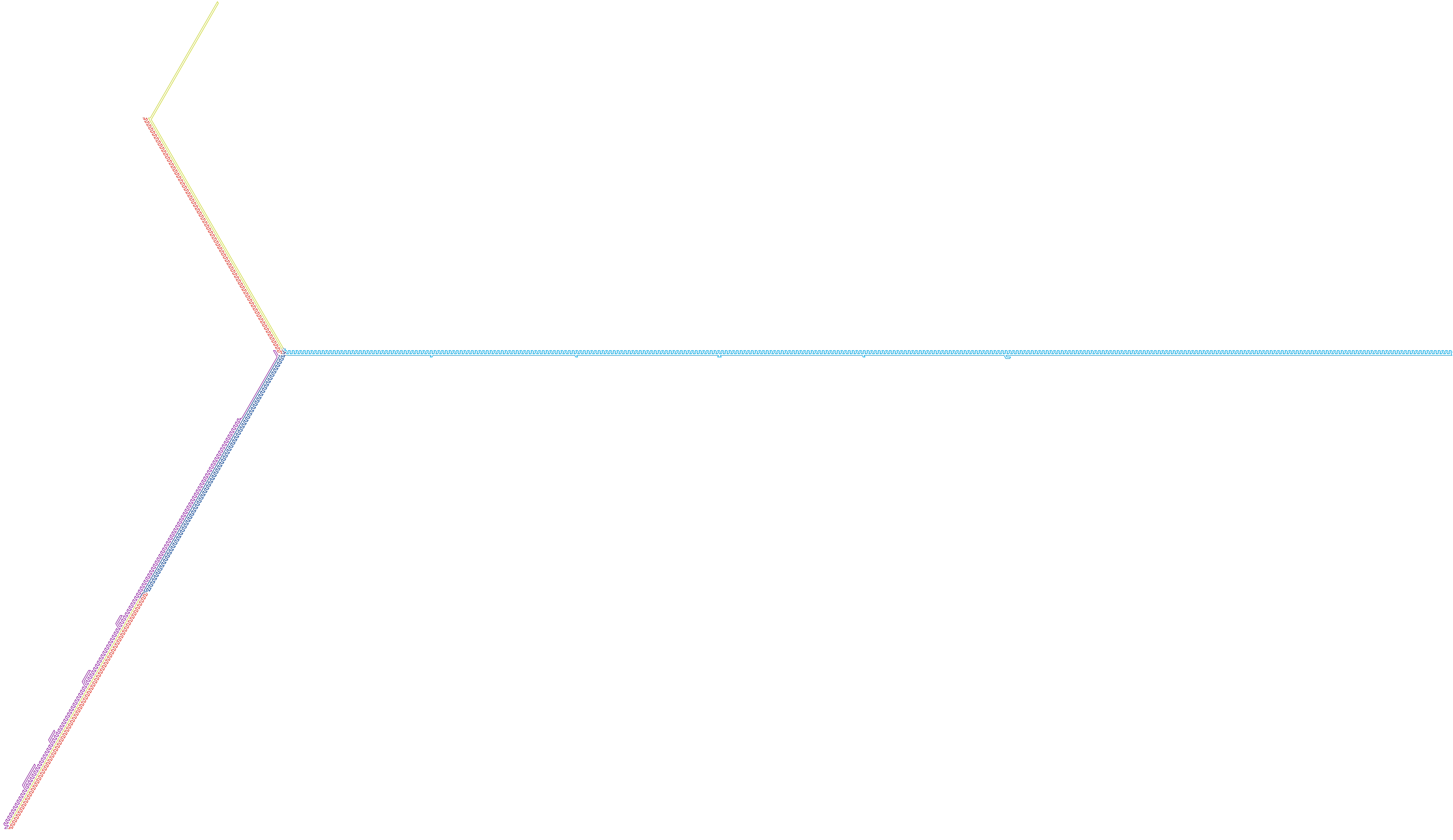}};
            \ifthenelse{\equal{#1}{}}{
\TIKZBLOCKRETURNLINEFEEDHALT{3}{-406.125}{77.726}
            }{}
    \end{tikzpicture}
}

%% file: OritatamiTuringPatterns/n=4-L=3-P=1/AppendLineFeedHalt_110__n=4-L=3-P=1.tex
\newcommand{\AppendLineFeedHaltBlockUUZ}[1]{
    \begin{tikzpicture}
        \node [anchor=center] () at (0,0) {\includegraphics[scale =\s]{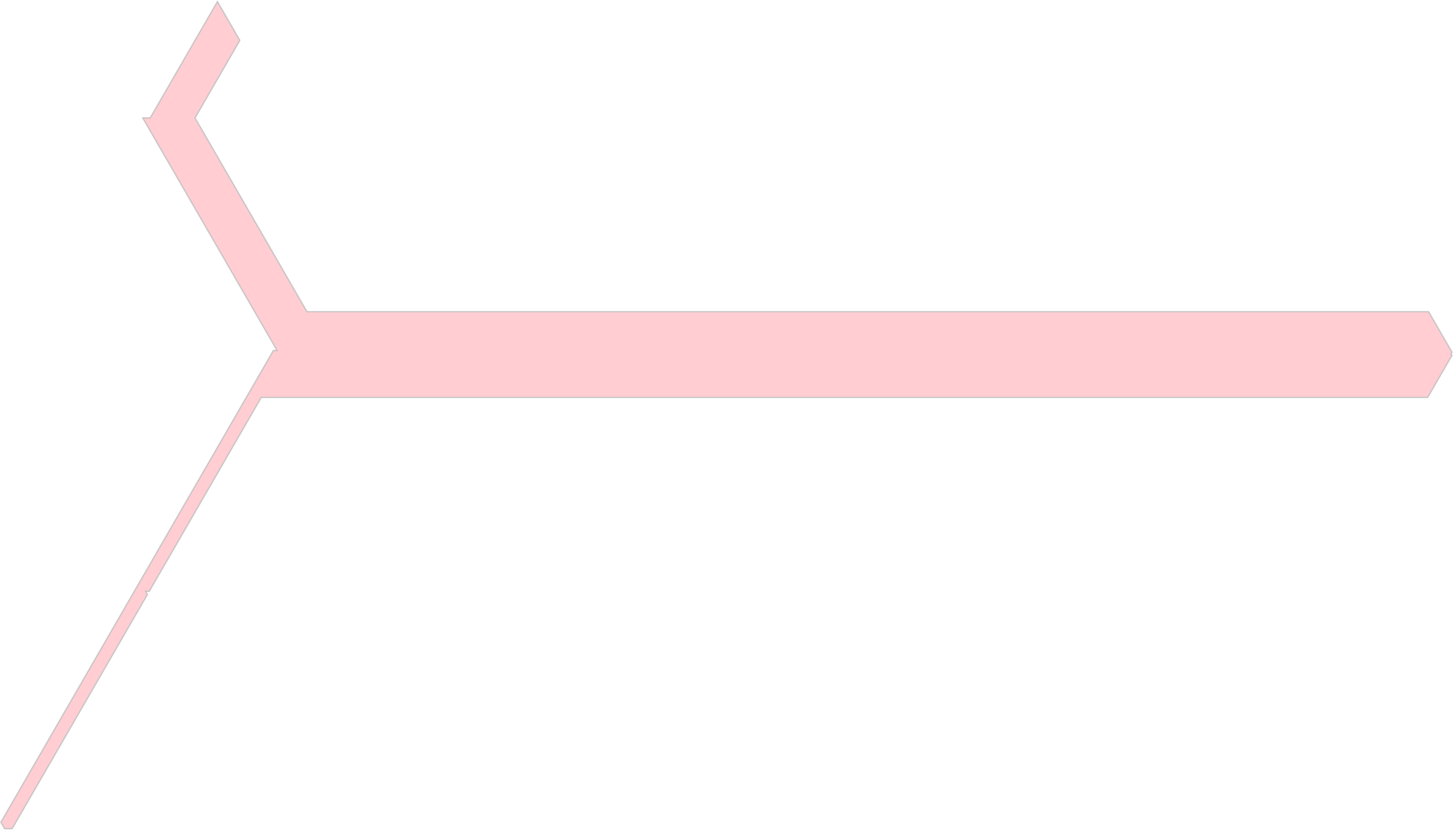}};
        \node [anchor=center] () at (0,0) {\includegraphics[scale =\s]{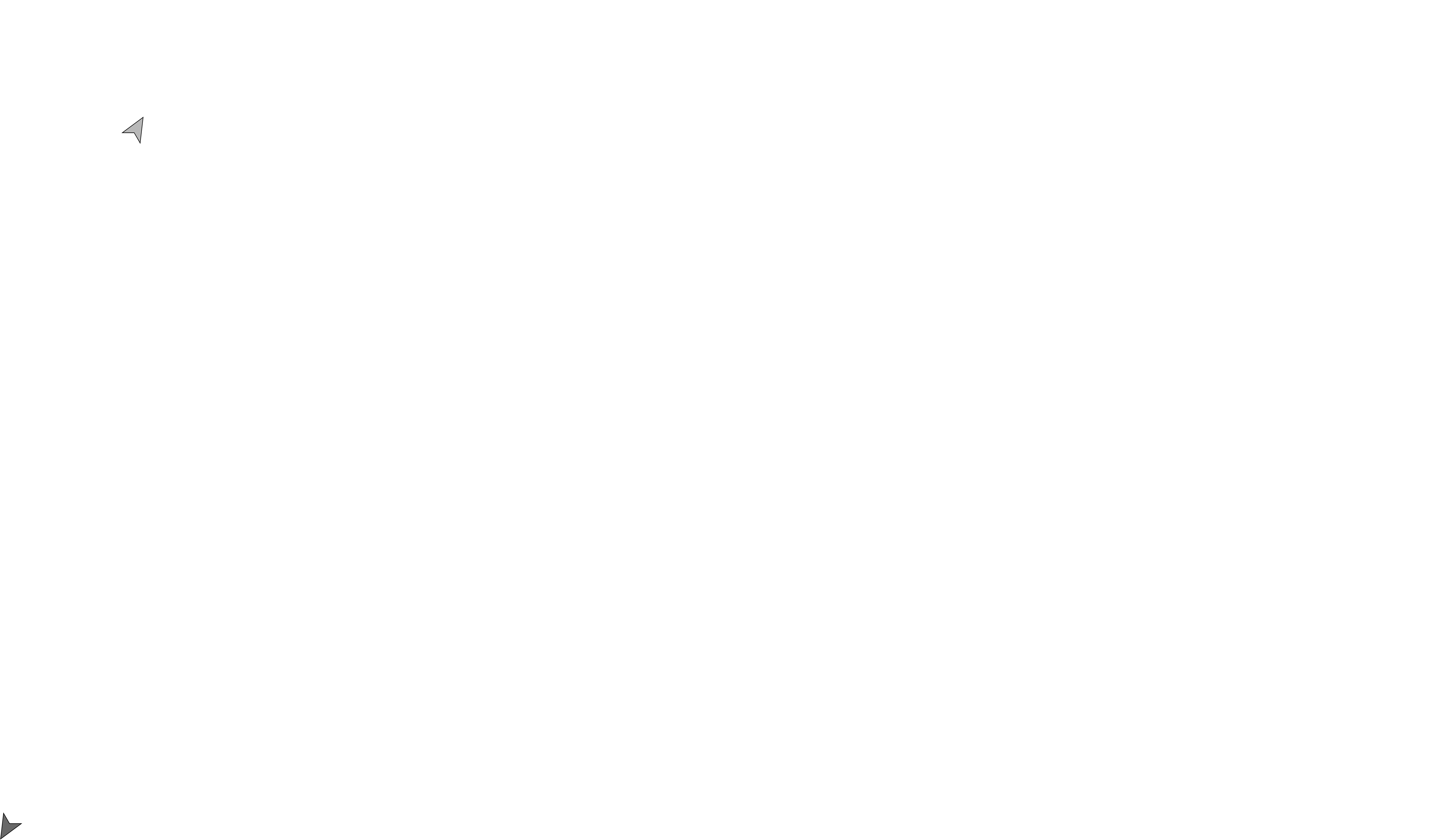}};
        \ifthenelse{\equal{#1}{}}{
\TIKZPointOfInterest{(-1563.75000000*\st pt, 809.73375254*\st pt)}{a}{$(-1,1-h)$}{W}{100.0*\st pt}{Start}{(-1800.00000000*\st pt, -1621.50289790*\st pt)}{8.0}{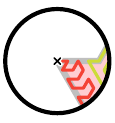}
\TIKZPointOfInterest{(-1363.75000000*\st pt, 1121.50289790*\st pt)}{b}{$\left(3,-\cfrac{3h-5}2\right)$}{E}{100.0*\st pt}{}{(-1200.00000000*\st pt, -1621.50289790*\st pt)}{8.0}{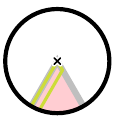}
\TIKZPointOfInterest{(1941.25000000*\st pt, 177.53520778*\st pt)}{c}{$(h+c(0),0)$}{E}{100.0*\st pt}{}{(-600.00000000*\st pt, -1621.50289790*\st pt)}{8.0}{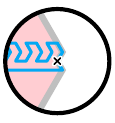}
\TIKZPointOfInterest{(-1546.25000000*\st pt, -458.99346401*\st pt)}{d}{$(h+2,h)$}{E}{100.0*\st pt}{}{(0.00000000*\st pt, -1621.50289790*\st pt)}{8.0}{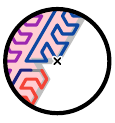}
\TIKZPointOfInterest{(-1196.25000000*\st pt, 173.20508076*\st pt)}{e}{$(h-1,1)$}{SE}{75.0*\st pt}{}{(600.00000000*\st pt, -1621.50289790*\st pt)}{8.0}{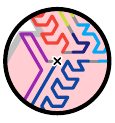}
\TIKZPointOfInterest{(-1928.75000000*\st pt, -1095.52213579*\st pt)}{f}{$(h-1,2h)$}{SW}{100.0*\st pt}{}{(1200.00000000*\st pt, -1621.50289790*\st pt)}{8.0}{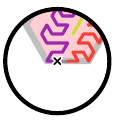}
\TIKZPointOfInterest{(-1913.75000000*\st pt, -1095.52213579*\st pt)}{g}{$(h+2,2h)$}{NE}{100.0*\st pt}{}{(1800.00000000*\st pt, -1621.50289790*\st pt)}{8.0}{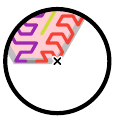}
\TIKZPointOfInterest{(-1558.75000000*\st pt, -463.32359102*\st pt)}{h}{$(h,h+1)$}{NW}{100.0*\st pt}{Trapped}{(-1800.00000000*\st pt, -2221.50289790*\st pt)}{8.0}{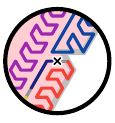}
\TIKZPointOfInterest{(-1921.25000000*\st pt, -1082.53175473*\st pt)}{i}{$(h-1,2h-3)$}{NW}{100.0*\st pt}{Stop}{(-1200.00000000*\st pt, -2221.50289790*\st pt)}{8.0}{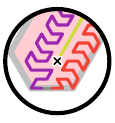}
            }{}
\node [anchor=center] () at (0,0) {\includegraphics[scale =\s]{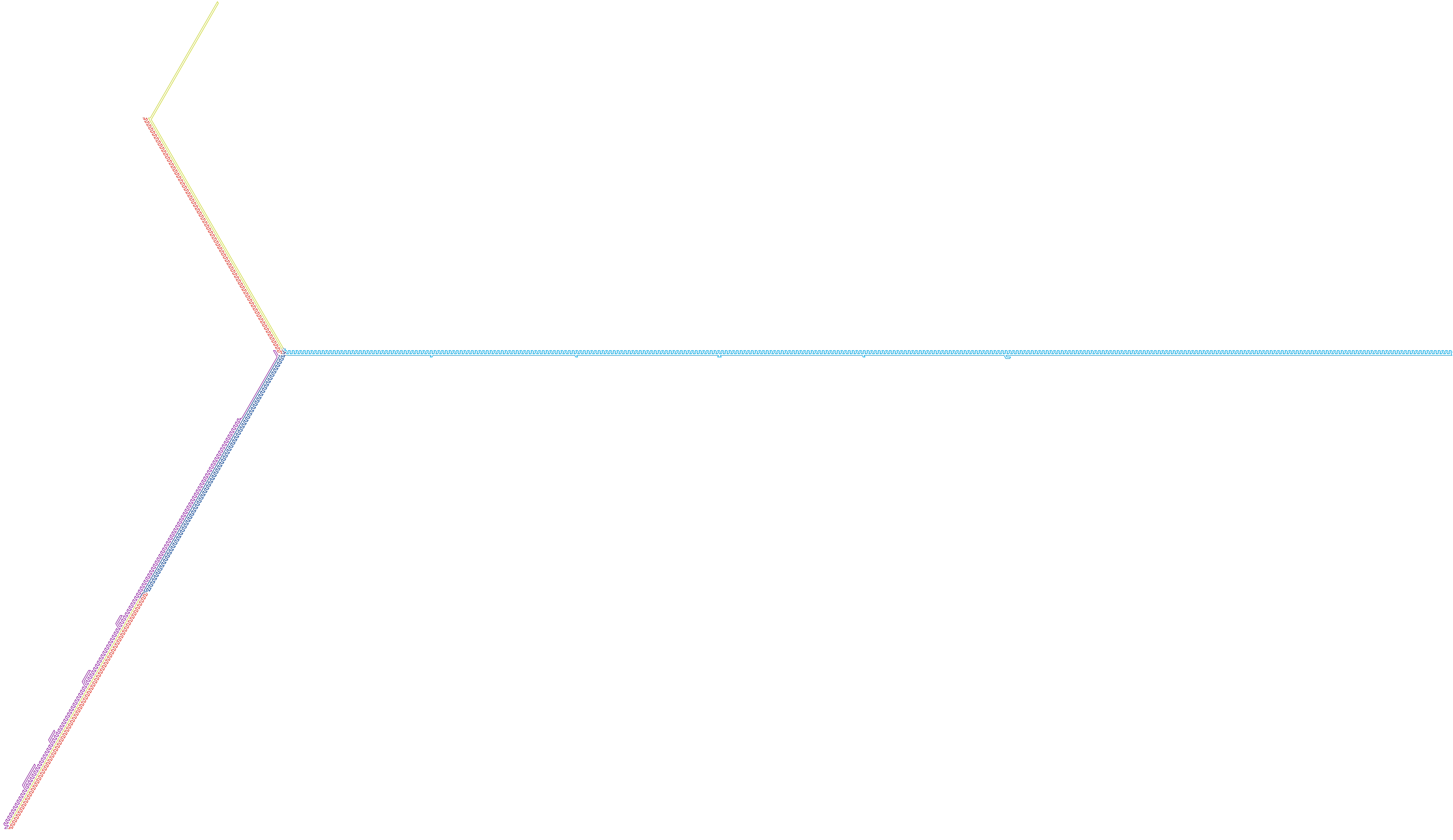}};
            \ifthenelse{\equal{#1}{}}{
\TIKZBLOCKRETURNLINEFEEDHALT{0}{-406.125}{77.726}
            }{}
    \end{tikzpicture}
}

%% file: OritatamiTuringPatterns/n=4-L=3-P=1/AppendLineFeedHalt_11__n=4-L=3-P=1.tex
\newcommand{\AppendLineFeedHaltBlockUU}[1]{
    \begin{tikzpicture}
        \node [anchor=center] () at (0,0) {\includegraphics[scale =\s]{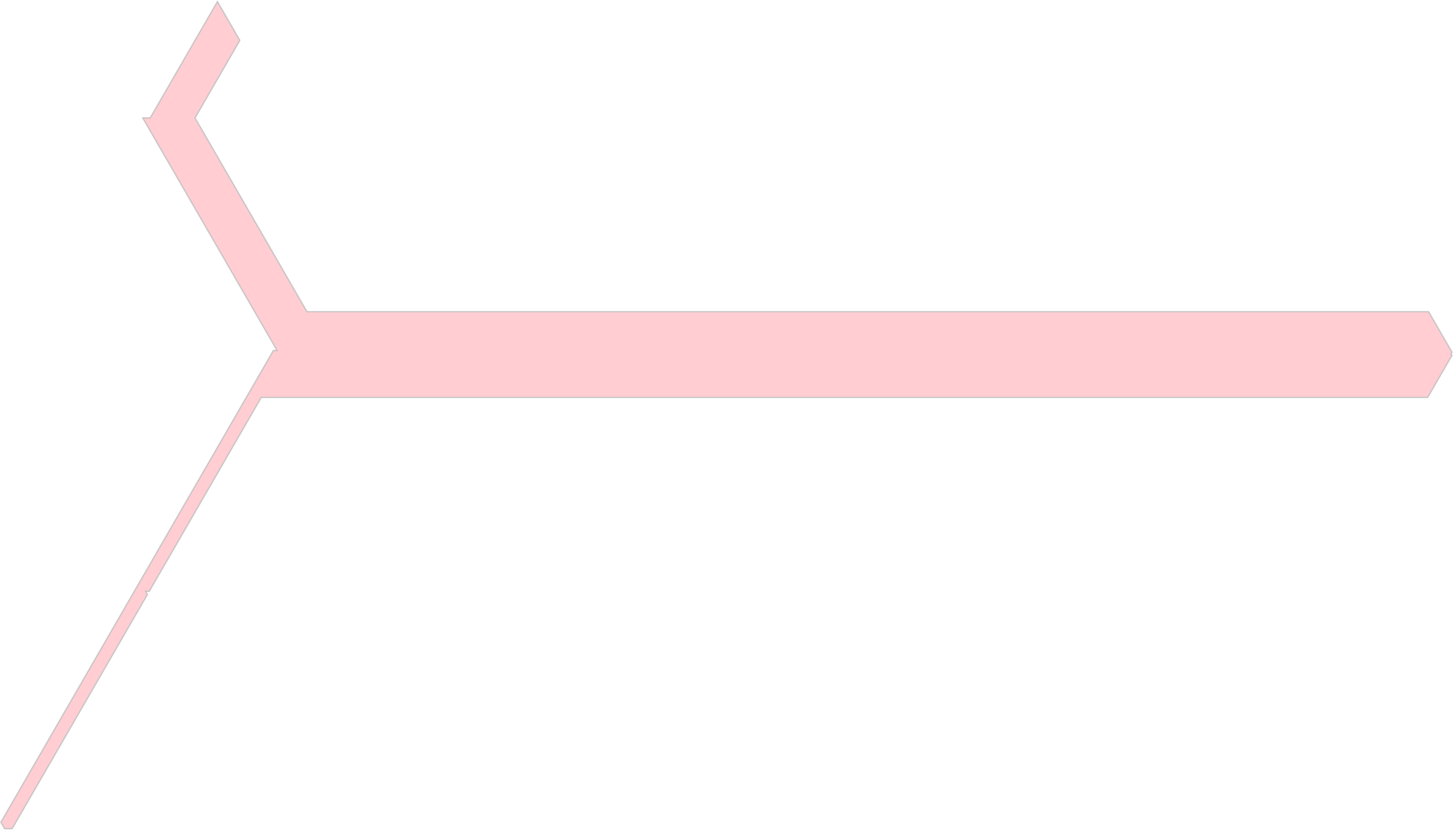}};
        \node [anchor=center] () at (0,0) {\includegraphics[scale =\s]{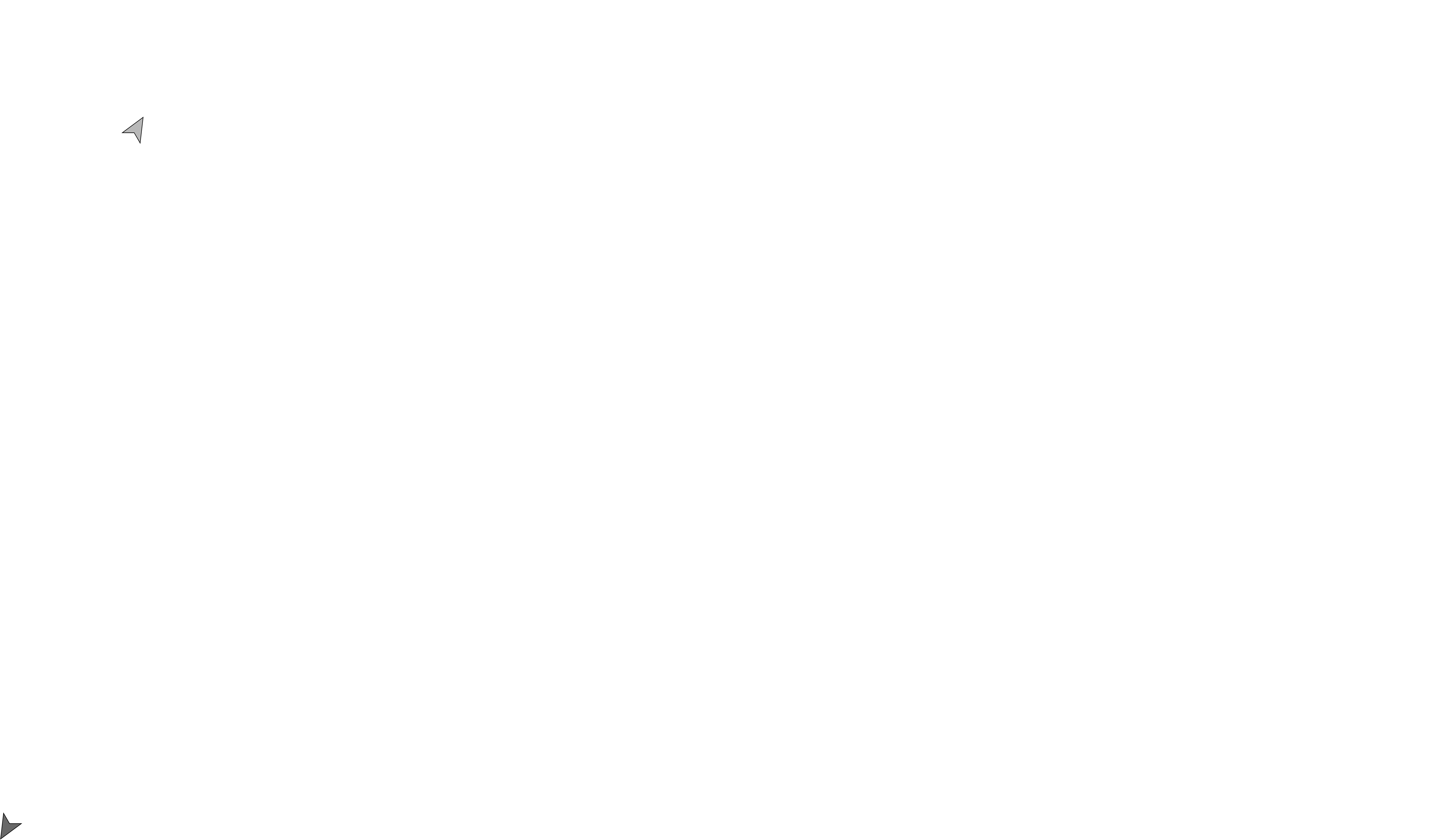}};
        \ifthenelse{\equal{#1}{}}{
\TIKZPointOfInterest{(-1563.75000000*\st pt, 809.73375254*\st pt)}{a}{$(-1,1-h)$}{W}{100.0*\st pt}{Start}{(-1800.00000000*\st pt, -1621.50289790*\st pt)}{8.0}{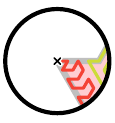}
\TIKZPointOfInterest{(-1363.75000000*\st pt, 1121.50289790*\st pt)}{b}{$\left(3,-\cfrac{3h-5}2\right)$}{E}{100.0*\st pt}{}{(-1200.00000000*\st pt, -1621.50289790*\st pt)}{8.0}{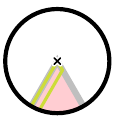}
\TIKZPointOfInterest{(1941.25000000*\st pt, 177.53520778*\st pt)}{c}{$(h+c(0),0)$}{E}{100.0*\st pt}{}{(-600.00000000*\st pt, -1621.50289790*\st pt)}{8.0}{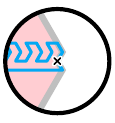}
\TIKZPointOfInterest{(-1546.25000000*\st pt, -458.99346401*\st pt)}{d}{$(h+2,h)$}{E}{100.0*\st pt}{}{(0.00000000*\st pt, -1621.50289790*\st pt)}{8.0}{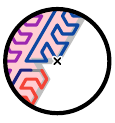}
\TIKZPointOfInterest{(-1196.25000000*\st pt, 173.20508076*\st pt)}{e}{$(h-1,1)$}{SE}{75.0*\st pt}{}{(600.00000000*\st pt, -1621.50289790*\st pt)}{8.0}{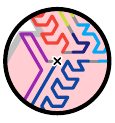}
\TIKZPointOfInterest{(-1928.75000000*\st pt, -1095.52213579*\st pt)}{f}{$(h-1,2h)$}{SW}{100.0*\st pt}{}{(1200.00000000*\st pt, -1621.50289790*\st pt)}{8.0}{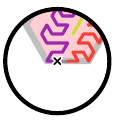}
\TIKZPointOfInterest{(-1913.75000000*\st pt, -1095.52213579*\st pt)}{g}{$(h+2,2h)$}{NE}{100.0*\st pt}{}{(1800.00000000*\st pt, -1621.50289790*\st pt)}{8.0}{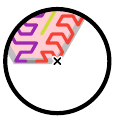}
\TIKZPointOfInterest{(-1558.75000000*\st pt, -463.32359102*\st pt)}{h}{$(h,h+1)$}{NW}{100.0*\st pt}{Trapped}{(-1800.00000000*\st pt, -2221.50289790*\st pt)}{8.0}{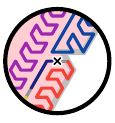}
\TIKZPointOfInterest{(-1921.25000000*\st pt, -1082.53175473*\st pt)}{i}{$(h-1,2h-3)$}{NW}{100.0*\st pt}{Stop}{(-1200.00000000*\st pt, -2221.50289790*\st pt)}{8.0}{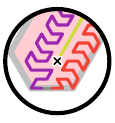}
            }{}
\node [anchor=center] () at (0,0) {\includegraphics[scale =\s]{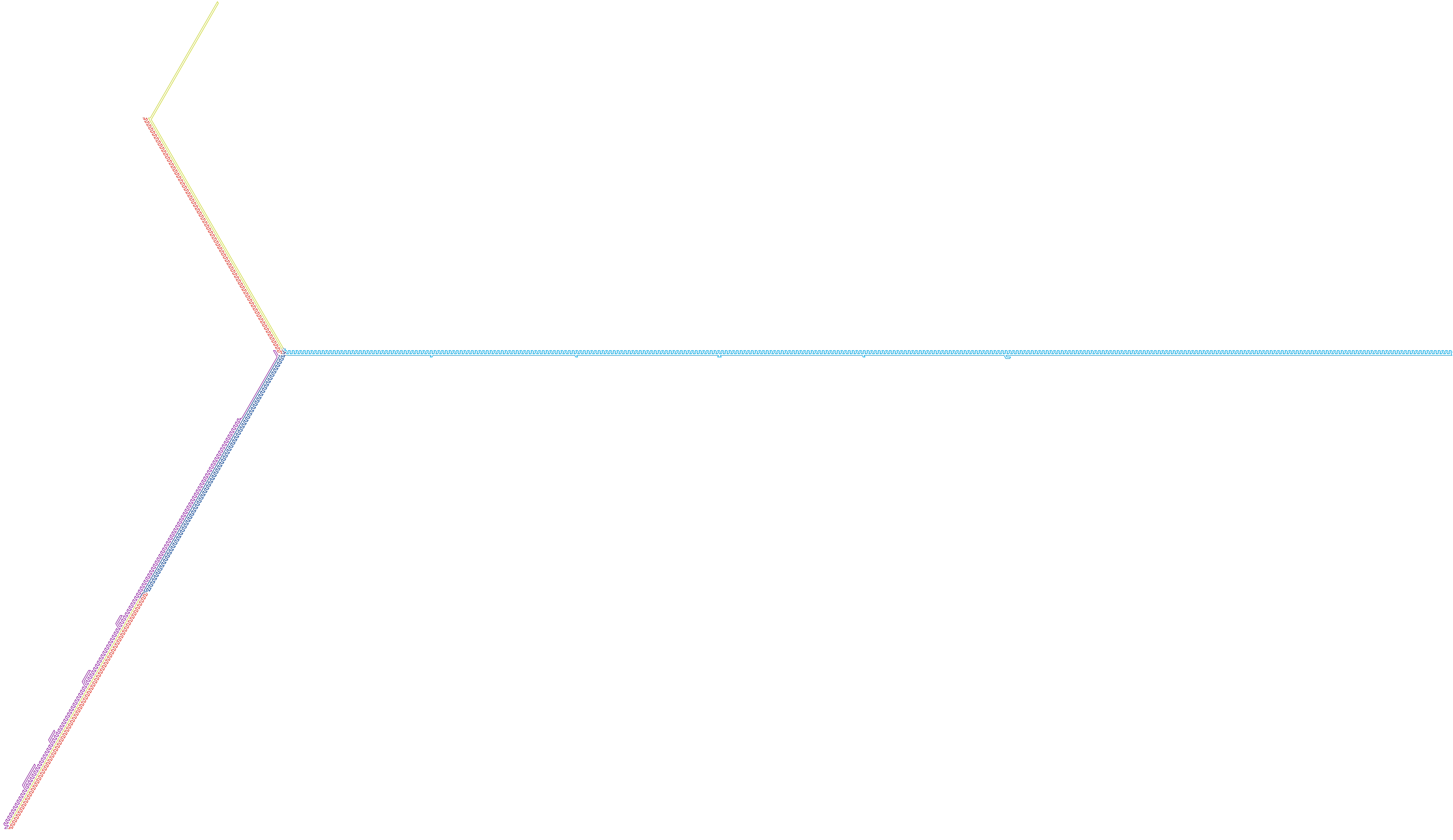}};
            \ifthenelse{\equal{#1}{}}{
\TIKZBLOCKRETURNLINEFEEDHALT{2}{-406.125}{77.726}
            }{}
    \end{tikzpicture}
}

%% file: OritatamiTuringPatterns/n=4-L=3-P=1/AppendLineFeedHalt_____n=4-L=3-P=1.tex
\newcommand{\AppendLineFeedHaltBlockEpsilon}[1]{
    \begin{tikzpicture}
        \node [anchor=center] () at (0,0) {\includegraphics[scale =\s]{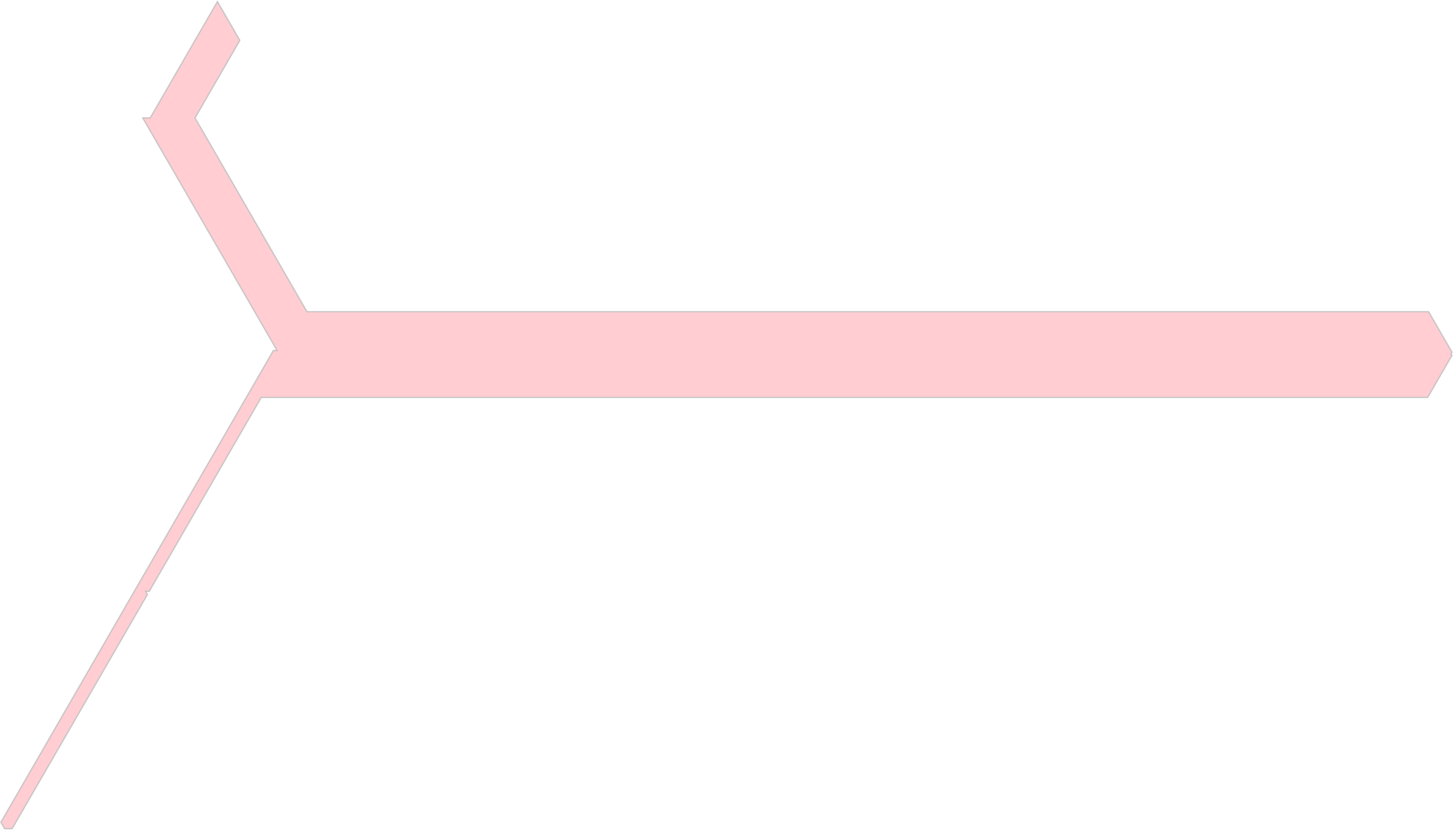}};
        \node [anchor=center] () at (0,0) {\includegraphics[scale =\s]{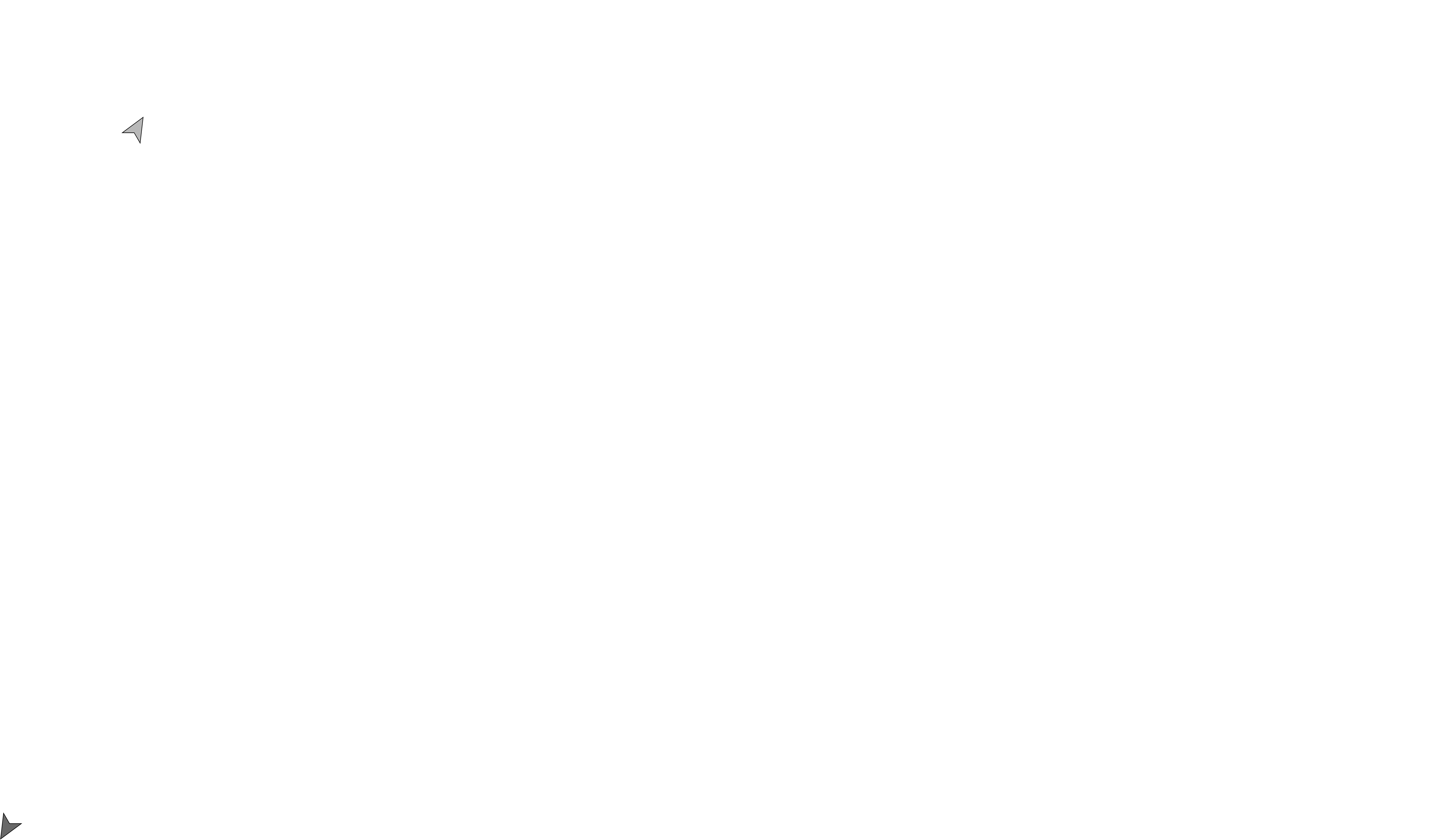}};
        \ifthenelse{\equal{#1}{}}{
\TIKZPointOfInterest{(-1563.75000000*\st pt, 809.73375254*\st pt)}{a}{$(-1,1-h)$}{W}{100.0*\st pt}{Start}{(-1800.00000000*\st pt, -1621.50289790*\st pt)}{8.0}{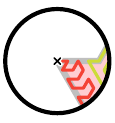}
\TIKZPointOfInterest{(-1363.75000000*\st pt, 1121.50289790*\st pt)}{b}{$\left(3,-\cfrac{3h-5}2\right)$}{E}{100.0*\st pt}{}{(-1200.00000000*\st pt, -1621.50289790*\st pt)}{8.0}{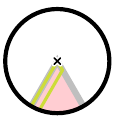}
\TIKZPointOfInterest{(1941.25000000*\st pt, 177.53520778*\st pt)}{c}{$(h+c(0),0)$}{E}{100.0*\st pt}{}{(-600.00000000*\st pt, -1621.50289790*\st pt)}{8.0}{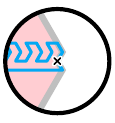}
\TIKZPointOfInterest{(-1546.25000000*\st pt, -458.99346401*\st pt)}{d}{$(h+2,h)$}{E}{100.0*\st pt}{}{(0.00000000*\st pt, -1621.50289790*\st pt)}{8.0}{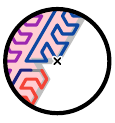}
\TIKZPointOfInterest{(-1196.25000000*\st pt, 173.20508076*\st pt)}{e}{$(h-1,1)$}{SE}{75.0*\st pt}{}{(600.00000000*\st pt, -1621.50289790*\st pt)}{8.0}{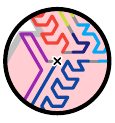}
\TIKZPointOfInterest{(-1928.75000000*\st pt, -1095.52213579*\st pt)}{f}{$(h-1,2h)$}{SW}{100.0*\st pt}{}{(1200.00000000*\st pt, -1621.50289790*\st pt)}{8.0}{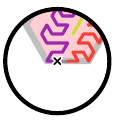}
\TIKZPointOfInterest{(-1913.75000000*\st pt, -1095.52213579*\st pt)}{g}{$(h+2,2h)$}{NE}{100.0*\st pt}{}{(1800.00000000*\st pt, -1621.50289790*\st pt)}{8.0}{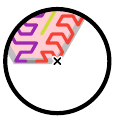}
\TIKZPointOfInterest{(-1558.75000000*\st pt, -463.32359102*\st pt)}{h}{$(h,h+1)$}{NW}{100.0*\st pt}{Trapped}{(-1800.00000000*\st pt, -2221.50289790*\st pt)}{8.0}{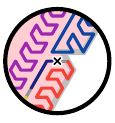}
\TIKZPointOfInterest{(-1921.25000000*\st pt, -1082.53175473*\st pt)}{i}{$(h-1,2h-3)$}{NW}{100.0*\st pt}{Stop}{(-1200.00000000*\st pt, -2221.50289790*\st pt)}{8.0}{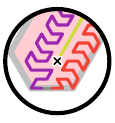}
            }{}
\node [anchor=center] () at (0,0) {\includegraphics[scale =\s]{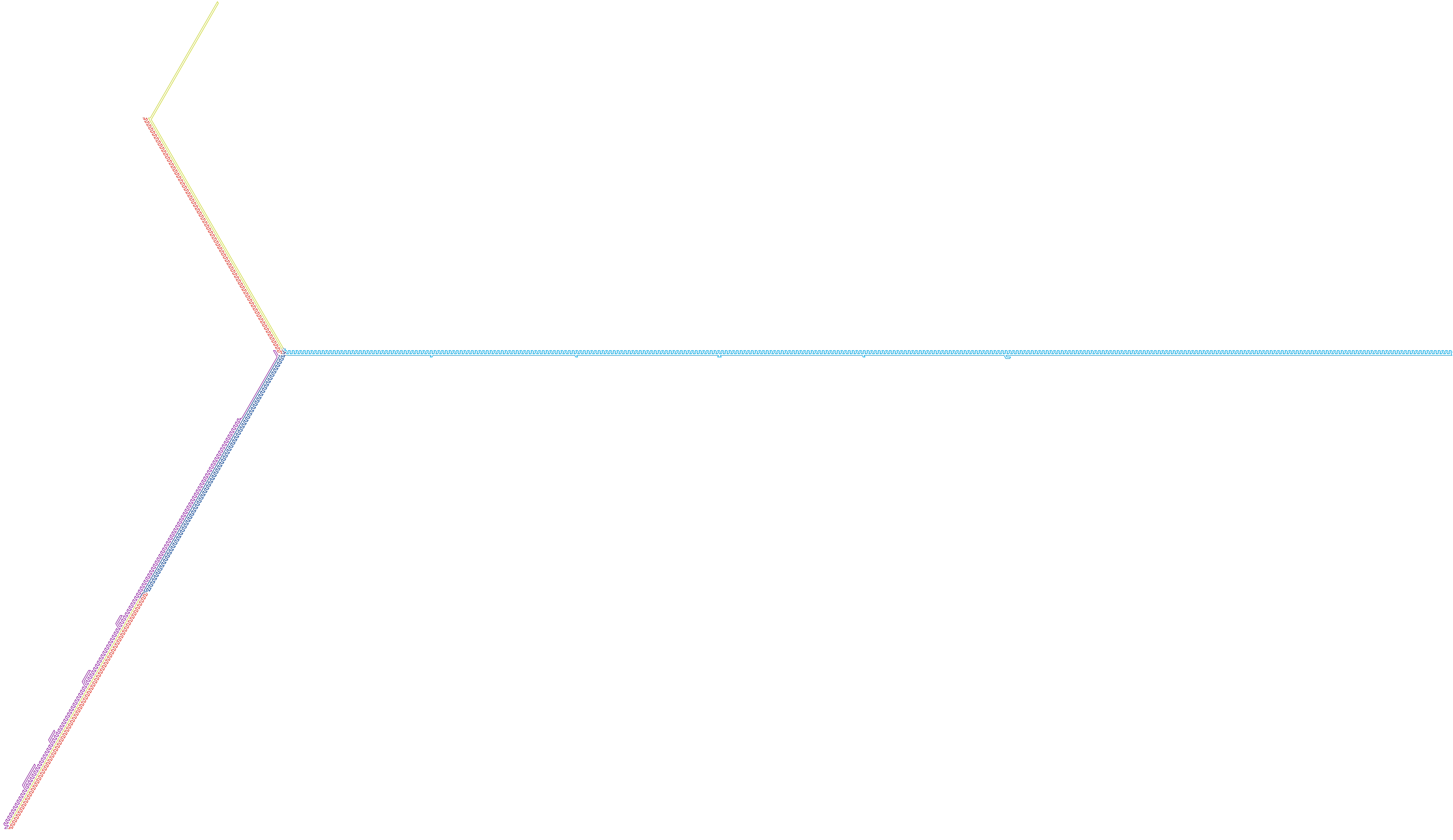}};
            \ifthenelse{\equal{#1}{}}{
\TIKZBLOCKRETURNLINEFEEDHALT{1}{-406.125}{77.726}
            }{}
    \end{tikzpicture}
}

%% file: OritatamiTuringPatterns/n=4-L=3-P=1/Append_0__n=4-L=3-P=1.tex
\newcommand{\AppendBlockZ}[1]{
    \begin{tikzpicture}
        \node [anchor=center] () at (0,0) {\includegraphics[scale =\s]{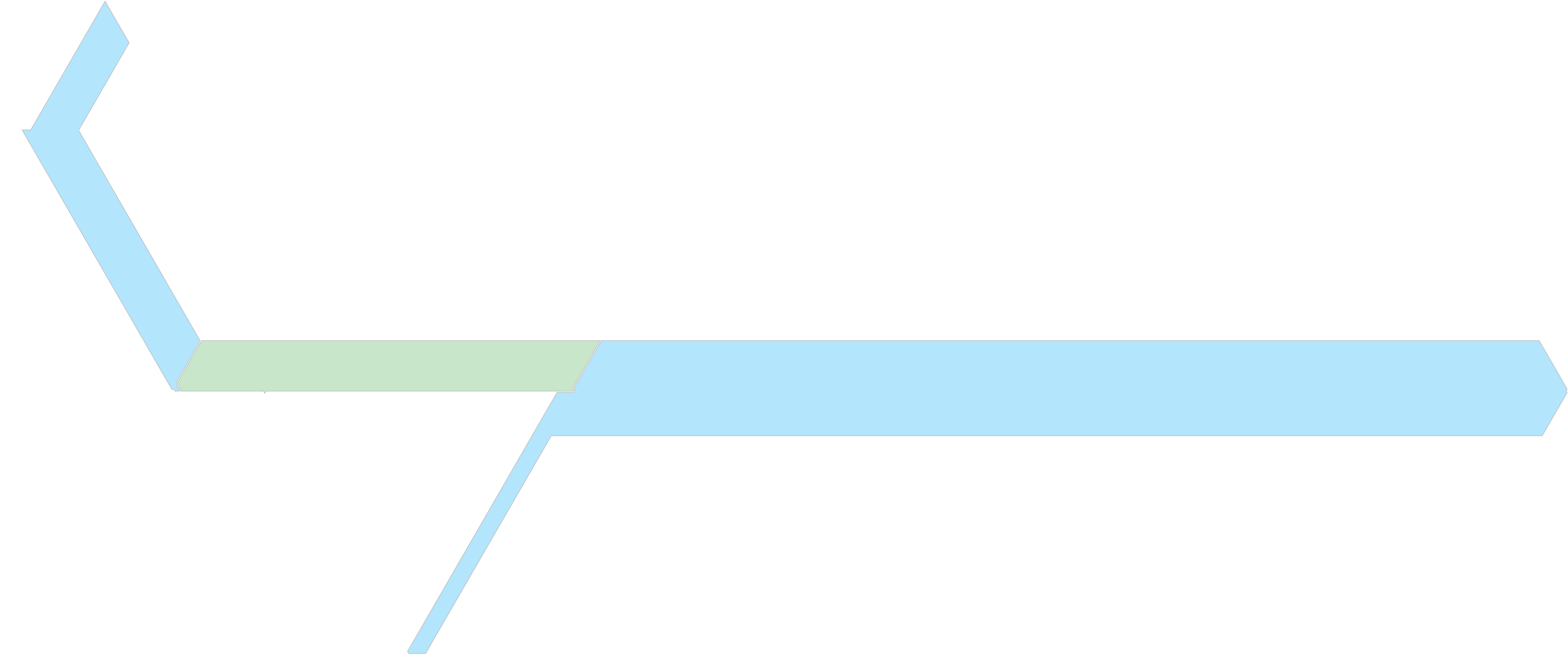}};
        \node [anchor=center] () at (0,0) {\includegraphics[scale =\s]{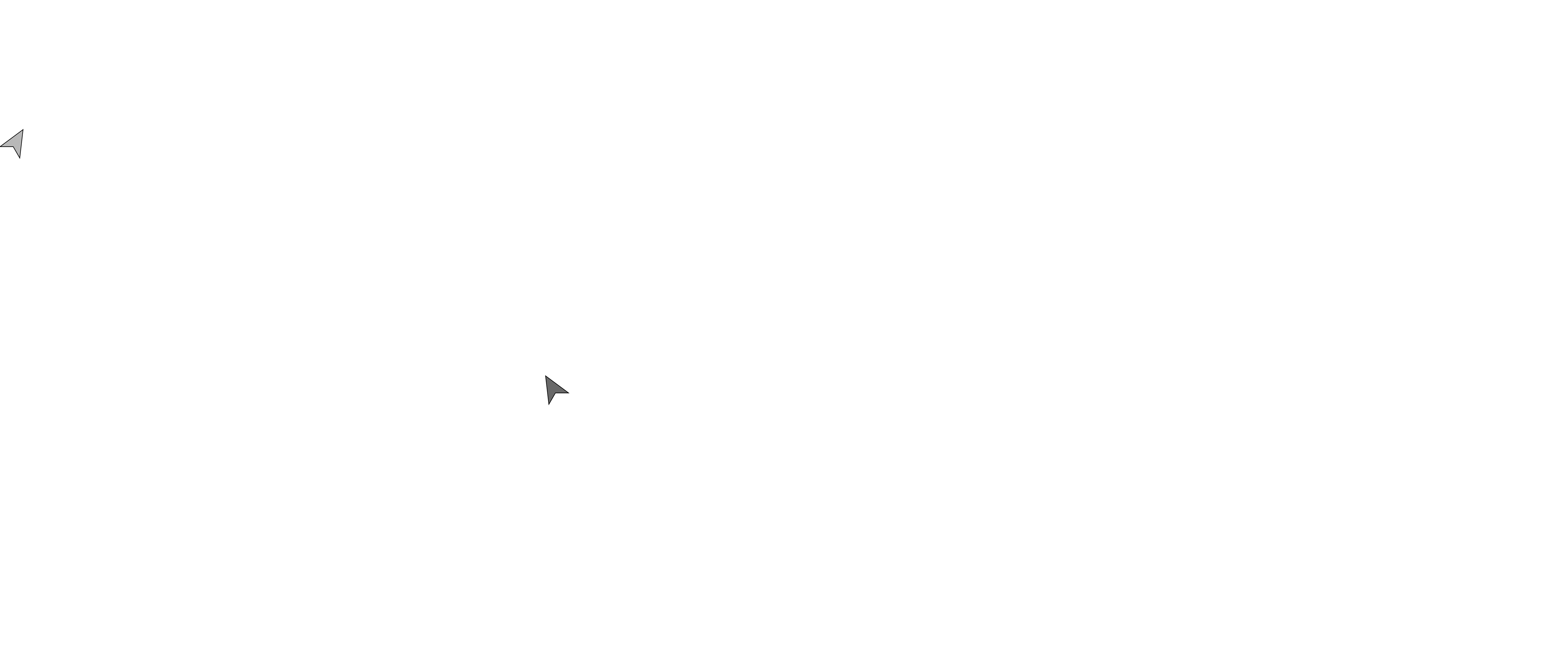}};
        \ifthenelse{\equal{#1}{}}{
\TIKZPointOfInterest{(-1847.50000000*\st pt, 478.47903559*\st pt)}{a}{$(-1,1-h)$}{W}{100.0*\st pt}{Start}{(-1800.00000000*\st pt, -1090.24818095*\st pt)}{8.0}{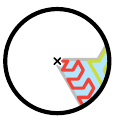}
\TIKZPointOfInterest{(-1647.50000000*\st pt, 790.24818095*\st pt)}{b}{$\left(3,-\cfrac{3h-5}2\right)$}{E}{100.0*\st pt}{}{(-1200.00000000*\st pt, -1090.24818095*\st pt)}{8.0}{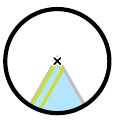}
\TIKZPointOfInterest{(-1470.00000000*\st pt, -140.72912811*\st pt)}{c}{$\left(h-1,-3\right)$}{W}{100.0*\st pt}{}{(-600.00000000*\st pt, -1090.24818095*\st pt)}{8.0}{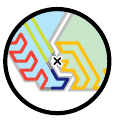}
\TIKZPointOfInterest{(-1260.00000000*\st pt, -158.04963619*\st pt)}{d}{$(w+h+1+iW,1)$}{SW}{100.0*\st pt}{Write \word0}{(0.00000000*\st pt, -1090.24818095*\st pt)}{8.0}{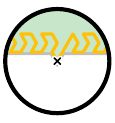}
\TIKZPointOfInterest{(-507.50000000*\st pt, -153.71950917*\st pt)}{e}{$(|u|W+h+1,0)$}{SE}{75.0*\st pt}{Start return}{(600.00000000*\st pt, -1090.24818095*\st pt)}{8.0}{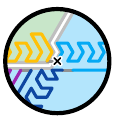}
\TIKZPointOfInterest{(1897.50000000*\st pt, -153.71950917*\st pt)}{f}{$(|u|W+h+c(|u|),0)$}{E}{100.0*\st pt}{}{(1200.00000000*\st pt, -1090.24818095*\st pt)}{8.0}{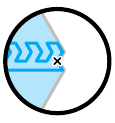}
\TIKZPointOfInterest{(-870.00000000*\st pt, -790.24818095*\st pt)}{g}{$(|u|W+h+2,h)$}{E}{100.0*\st pt}{}{(1800.00000000*\st pt, -1090.24818095*\st pt)}{8.0}{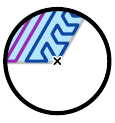}
\TIKZPointOfInterest{(-915.00000000*\st pt, -790.24818095*\st pt)}{h}{$(|u|W+h-7,h)$}{W}{100.0*\st pt}{}{(-1800.00000000*\st pt, -1690.24818095*\st pt)}{8.0}{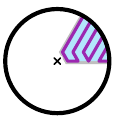}
\TIKZPointOfInterest{(-555.00000000*\st pt, -158.04963619*\st pt)}{i}{$(|u|W+h-8,1)$}{SW}{75.0*\st pt}{Next}{(-1200.00000000*\st pt, -1690.24818095*\st pt)}{8.0}{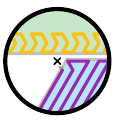}
            }{}
\node [anchor=center] () at (0,0) {\includegraphics[scale =\s]{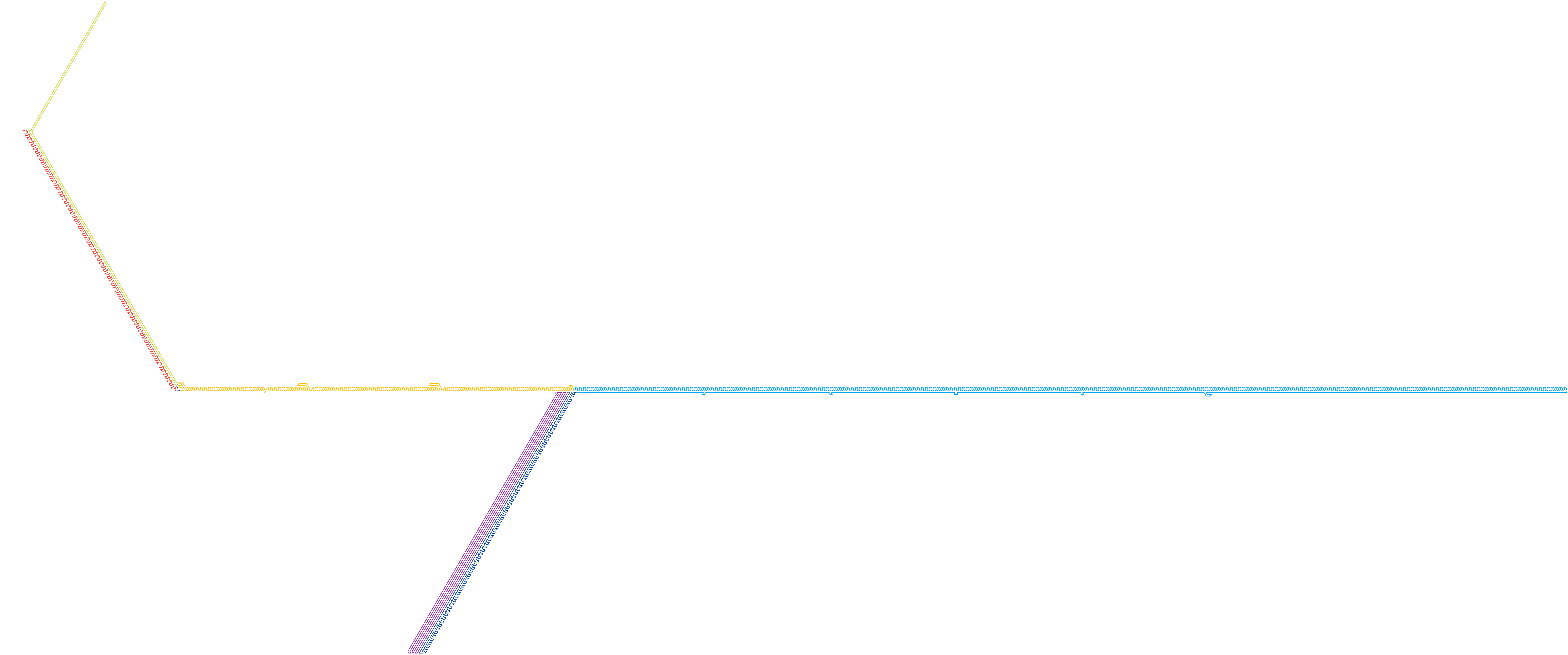}};
            \ifthenelse{\equal{#1}{}}{
\TIKZBLOCKAPPENDCARRIAGERETURN{3}{219.500}{223.868}
\TIKZBLOCKAPPENDLETTER{0}{0,1}{-905.250}{177.535}
            }{}
    \end{tikzpicture}
}

%% file: OritatamiTuringPatterns/n=4-L=3-P=1/Append_110__n=4-L=3-P=1.tex
\newcommand{\AppendBlockUUZ}[1]{
    \begin{tikzpicture}
        \node [anchor=center] () at (0,0) {\includegraphics[scale =\s]{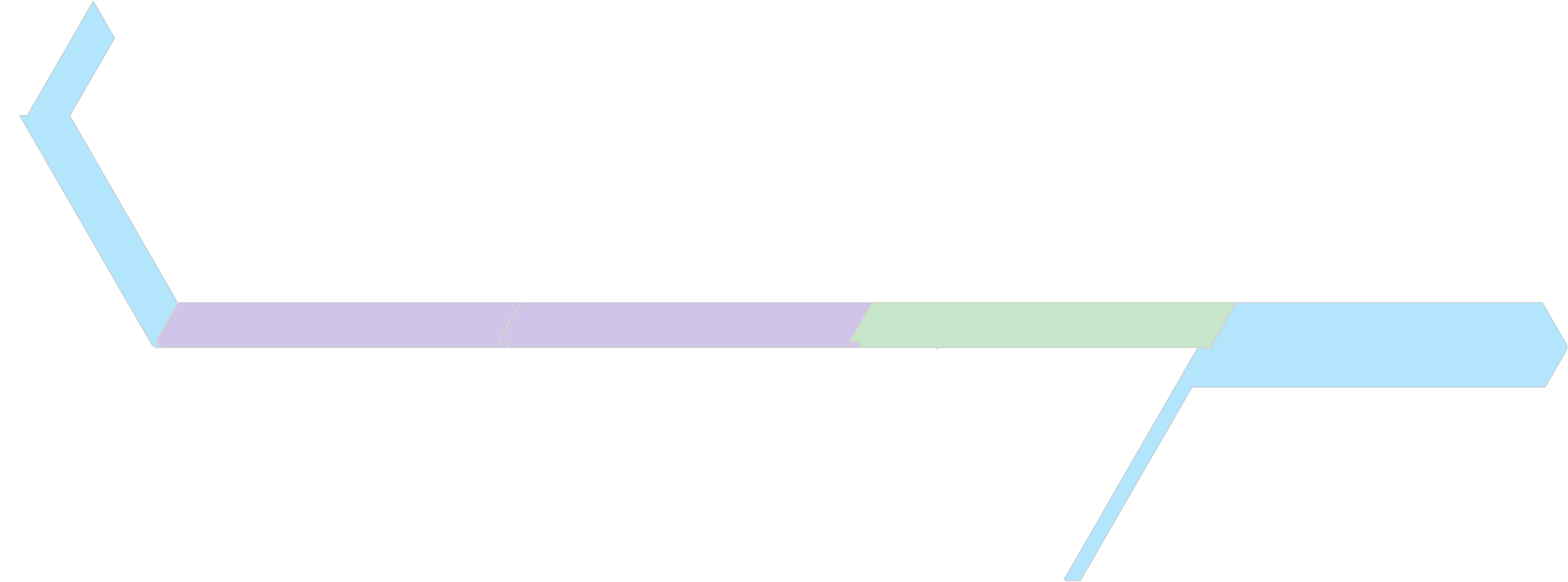}};
        \node [anchor=center] () at (0,0) {\includegraphics[scale =\s]{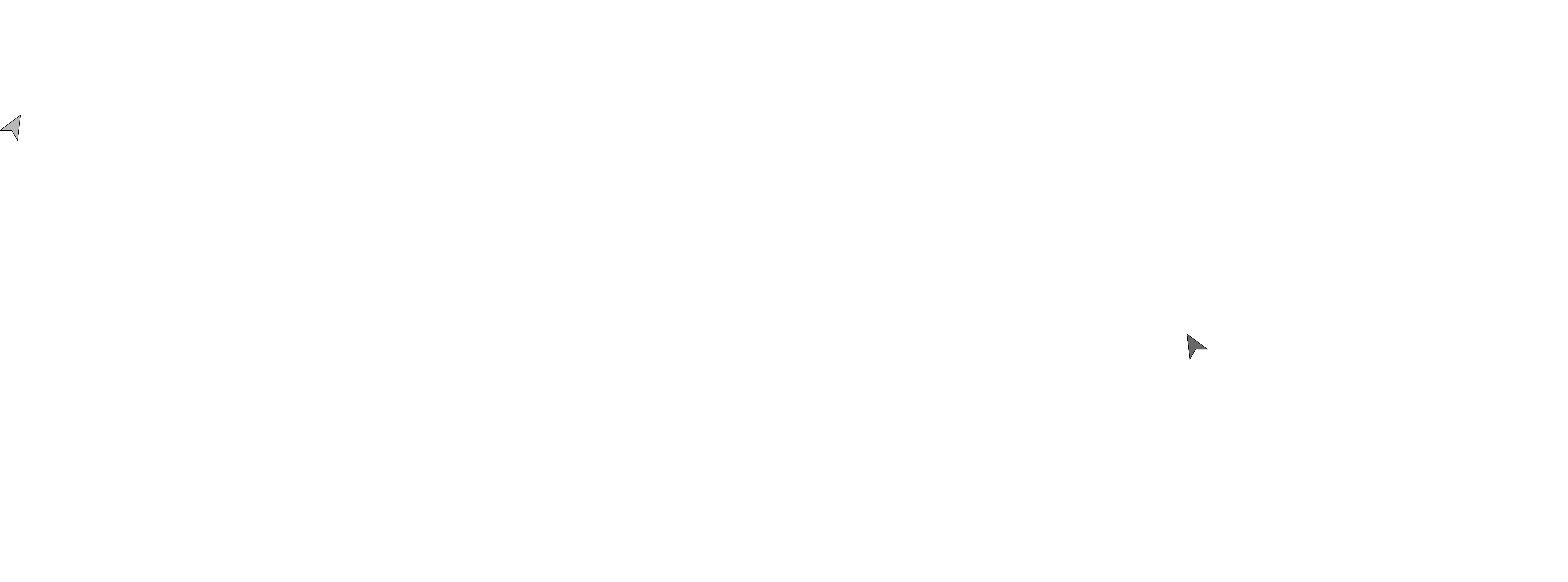}};
        \ifthenelse{\equal{#1}{}}{
\TIKZPointOfInterest{(-2087.50000000*\st pt, 478.47903559*\st pt)}{a}{$(-1,1-h)$}{W}{100.0*\st pt}{Start}{(-2100.00000000*\st pt, -1090.24818095*\st pt)}{8.0}{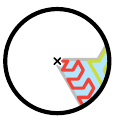}
\TIKZPointOfInterest{(-1887.50000000*\st pt, 790.24818095*\st pt)}{b}{$\left(3,-\cfrac{3h-5}2\right)$}{E}{100.0*\st pt}{}{(-1500.00000000*\st pt, -1090.24818095*\st pt)}{8.0}{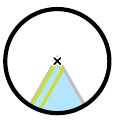}
\TIKZPointOfInterest{(-1710.00000000*\st pt, -140.72912811*\st pt)}{c}{$\left(h-1,-3\right)$}{W}{100.0*\st pt}{}{(-900.00000000*\st pt, -1090.24818095*\st pt)}{8.0}{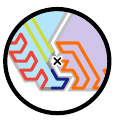}
\TIKZPointOfInterest{(-1497.50000000*\st pt, -153.71950917*\st pt)}{d}{$(w+h+1+iW,0)$}{SW}{100.0*\st pt}{Write \word1}{(-300.00000000*\st pt, -1090.24818095*\st pt)}{8.0}{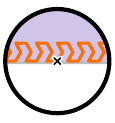}
\TIKZPointOfInterest{(-777.50000000*\st pt, -136.39900110*\st pt)}{e}{$((i+1)W+h-7,-4)$}{NW}{50.0*\st pt}{}{(300.00000000*\st pt, -1090.24818095*\st pt)}{8.0}{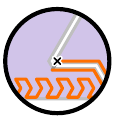}
\TIKZPointOfInterest{(-537.50000000*\st pt, -153.71950917*\st pt)}{f}{$(w+h+1+iW,0)$}{SW}{100.0*\st pt}{Write \word1}{(900.00000000*\st pt, -1090.24818095*\st pt)}{8.0}{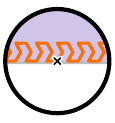}
\TIKZPointOfInterest{(182.50000000*\st pt, -136.39900110*\st pt)}{g}{$((i+1)W+h-7,-4)$}{NW}{50.0*\st pt}{}{(1500.00000000*\st pt, -1090.24818095*\st pt)}{8.0}{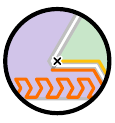}
\TIKZPointOfInterest{(420.00000000*\st pt, -158.04963619*\st pt)}{h}{$(w+h+1+iW,1)$}{SW}{100.0*\st pt}{Write \word0}{(2100.00000000*\st pt, -1090.24818095*\st pt)}{8.0}{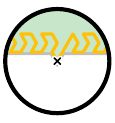}
\TIKZPointOfInterest{(1172.50000000*\st pt, -153.71950917*\st pt)}{i}{$(|u|W+h+1,0)$}{SE}{75.0*\st pt}{Start return}{(-2100.00000000*\st pt, -1690.24818095*\st pt)}{8.0}{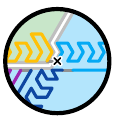}
\TIKZPointOfInterest{(2137.50000000*\st pt, -153.71950917*\st pt)}{j}{$(|u|W+h+c(|u|),0)$}{E}{100.0*\st pt}{}{(-1500.00000000*\st pt, -1690.24818095*\st pt)}{8.0}{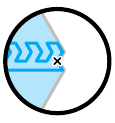}
\TIKZPointOfInterest{(810.00000000*\st pt, -790.24818095*\st pt)}{k}{$(|u|W+h+2,h)$}{E}{100.0*\st pt}{}{(-900.00000000*\st pt, -1690.24818095*\st pt)}{8.0}{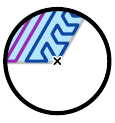}
\TIKZPointOfInterest{(765.00000000*\st pt, -790.24818095*\st pt)}{l}{$(|u|W+h-7,h)$}{W}{100.0*\st pt}{}{(-300.00000000*\st pt, -1690.24818095*\st pt)}{8.0}{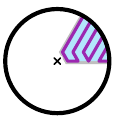}
\TIKZPointOfInterest{(1125.00000000*\st pt, -158.04963619*\st pt)}{m}{$(|u|W+h-8,1)$}{SW}{75.0*\st pt}{Next}{(300.00000000*\st pt, -1690.24818095*\st pt)}{8.0}{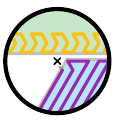}
            }{}
\node [anchor=center] () at (0,0) {\includegraphics[scale =\s]{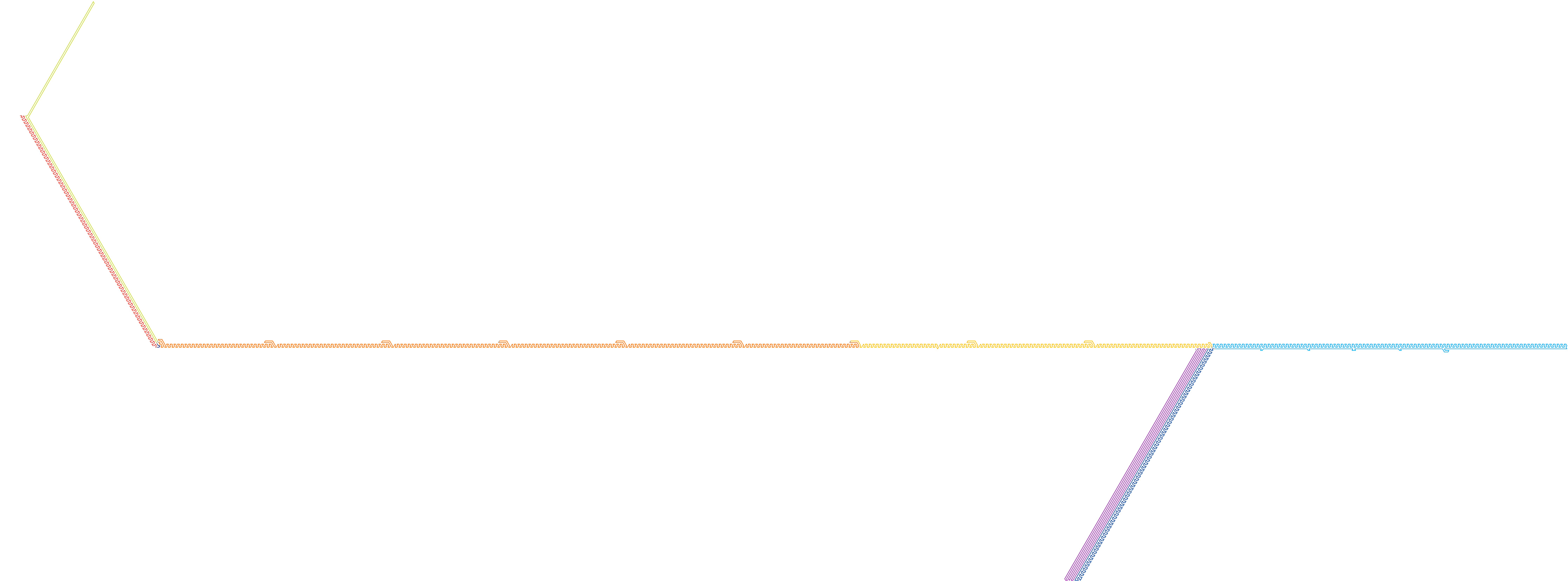}};
            \ifthenelse{\equal{#1}{}}{
\TIKZBLOCKAPPENDCARRIAGERETURN{0}{-20.500}{223.868}
\TIKZBLOCKAPPENDLETTER{1}{0,0}{-1145.250}{177.535}
\TIKZBLOCKAPPENDLETTER{1}{1,0}{-212.750}{181.865}
\TIKZBLOCKAPPENDLETTER{0}{2,1}{747.250}{181.865}
            }{}
    \end{tikzpicture}
}

%% file: OritatamiTuringPatterns/n=4-L=3-P=1/Append_11__n=4-L=3-P=1.tex
\newcommand{\AppendBlockUU}[1]{
    \begin{tikzpicture}
        \node [anchor=center] () at (0,0) {\includegraphics[scale =\s]{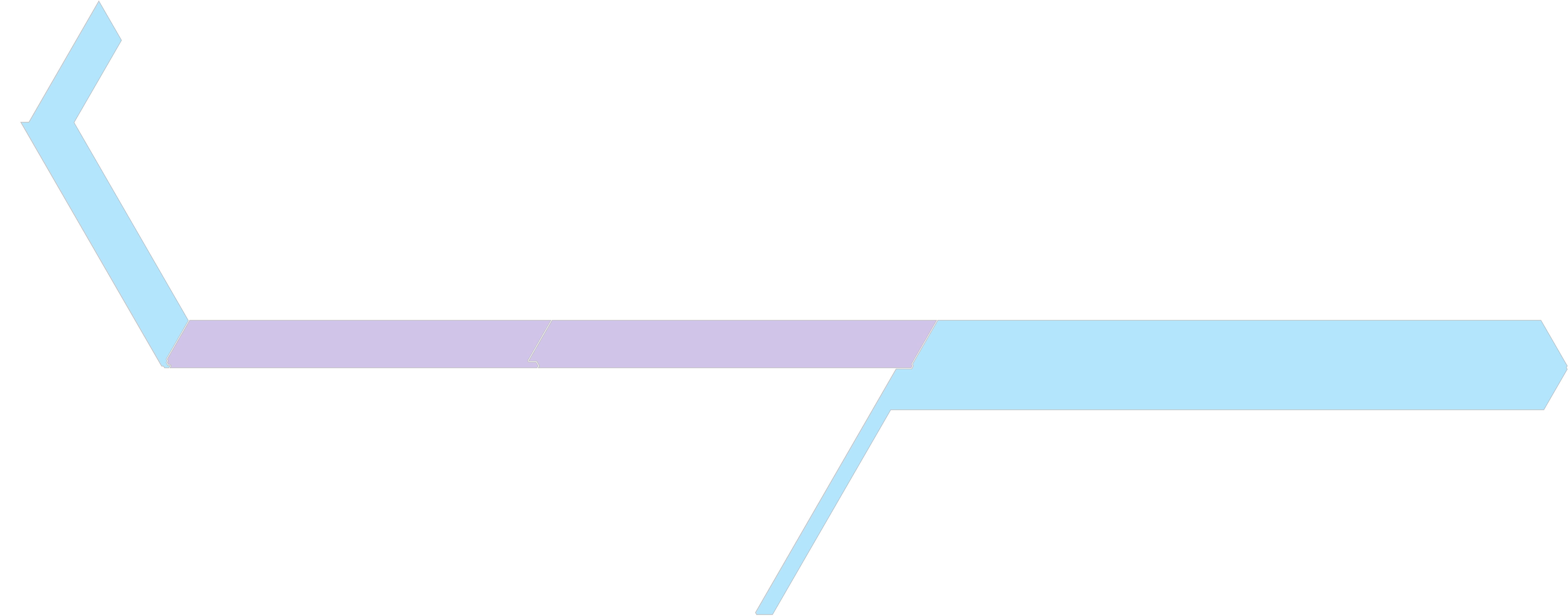}};
        \node [anchor=center] () at (0,0) {\includegraphics[scale =\s]{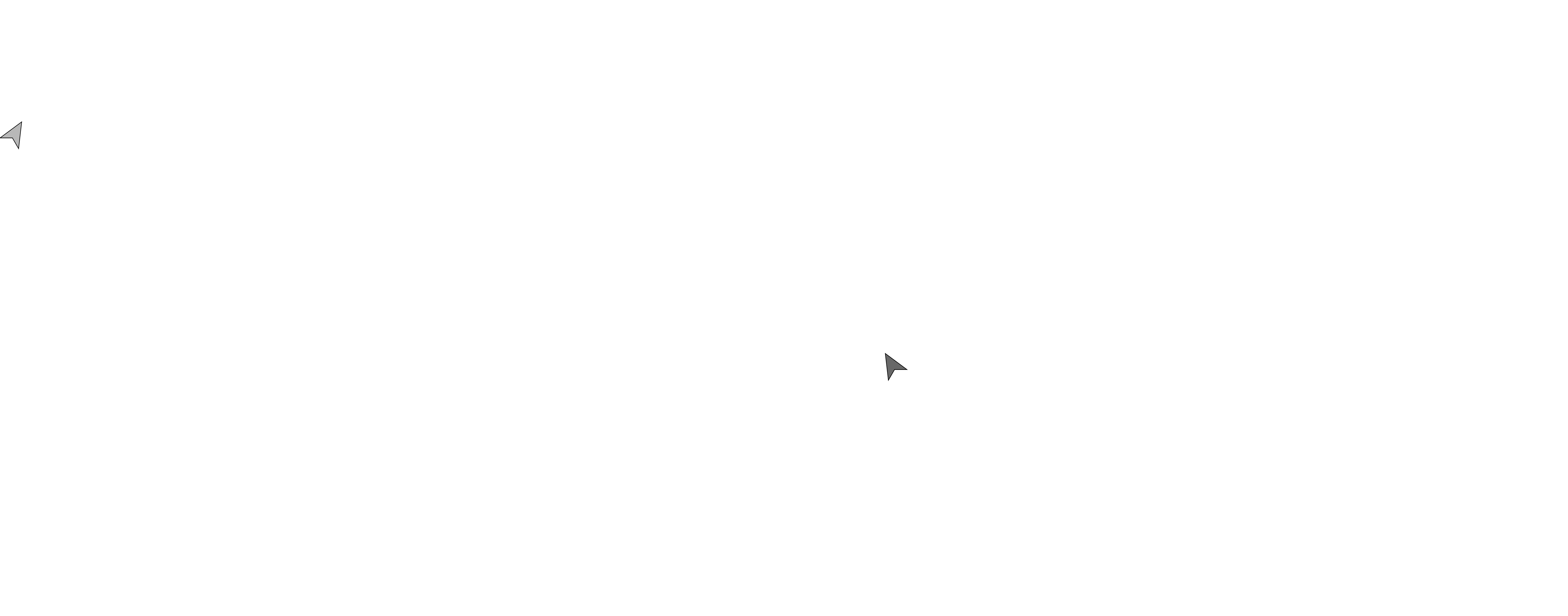}};
        \ifthenelse{\equal{#1}{}}{
\TIKZPointOfInterest{(-1967.50000000*\st pt, 478.47903559*\st pt)}{a}{$(-1,1-h)$}{W}{100.0*\st pt}{Start}{(-2100.00000000*\st pt, -1090.24818095*\st pt)}{8.0}{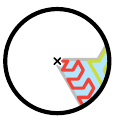}
\TIKZPointOfInterest{(-1767.50000000*\st pt, 790.24818095*\st pt)}{b}{$\left(3,-\cfrac{3h-5}2\right)$}{E}{100.0*\st pt}{}{(-1500.00000000*\st pt, -1090.24818095*\st pt)}{8.0}{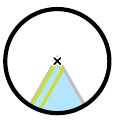}
\TIKZPointOfInterest{(-1590.00000000*\st pt, -140.72912811*\st pt)}{c}{$\left(h-1,-3\right)$}{W}{100.0*\st pt}{}{(-900.00000000*\st pt, -1090.24818095*\st pt)}{8.0}{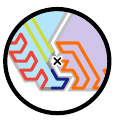}
\TIKZPointOfInterest{(-1377.50000000*\st pt, -153.71950917*\st pt)}{d}{$(w+h+1+iW,0)$}{SW}{100.0*\st pt}{Write \word1}{(-300.00000000*\st pt, -1090.24818095*\st pt)}{8.0}{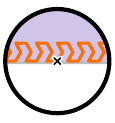}
\TIKZPointOfInterest{(-657.50000000*\st pt, -136.39900110*\st pt)}{e}{$((i+1)W+h-7,-4)$}{NW}{50.0*\st pt}{}{(300.00000000*\st pt, -1090.24818095*\st pt)}{8.0}{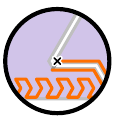}
\TIKZPointOfInterest{(-417.50000000*\st pt, -153.71950917*\st pt)}{f}{$(w+h+1+iW,0)$}{SW}{100.0*\st pt}{Write \word1}{(900.00000000*\st pt, -1090.24818095*\st pt)}{8.0}{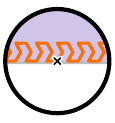}
\TIKZPointOfInterest{(332.50000000*\st pt, -153.71950917*\st pt)}{g}{$(|u|W+h+1,0)$}{SE}{75.0*\st pt}{Start return}{(1500.00000000*\st pt, -1090.24818095*\st pt)}{8.0}{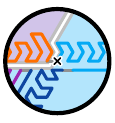}
\TIKZPointOfInterest{(2017.50000000*\st pt, -153.71950917*\st pt)}{h}{$(|u|W+h+c(|u|),0)$}{E}{100.0*\st pt}{}{(2100.00000000*\st pt, -1090.24818095*\st pt)}{8.0}{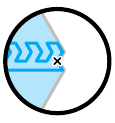}
\TIKZPointOfInterest{(-30.00000000*\st pt, -790.24818095*\st pt)}{i}{$(|u|W+h+2,h)$}{E}{100.0*\st pt}{}{(-2100.00000000*\st pt, -1690.24818095*\st pt)}{8.0}{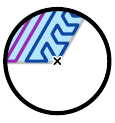}
\TIKZPointOfInterest{(-75.00000000*\st pt, -790.24818095*\st pt)}{j}{$(|u|W+h-7,h)$}{W}{100.0*\st pt}{}{(-1500.00000000*\st pt, -1690.24818095*\st pt)}{8.0}{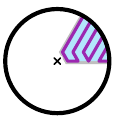}
\TIKZPointOfInterest{(285.00000000*\st pt, -158.04963619*\st pt)}{k}{$(|u|W+h-8,1)$}{SW}{75.0*\st pt}{Next}{(-900.00000000*\st pt, -1690.24818095*\st pt)}{8.0}{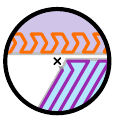}
            }{}
\node [anchor=center] () at (0,0) {\includegraphics[scale =\s]{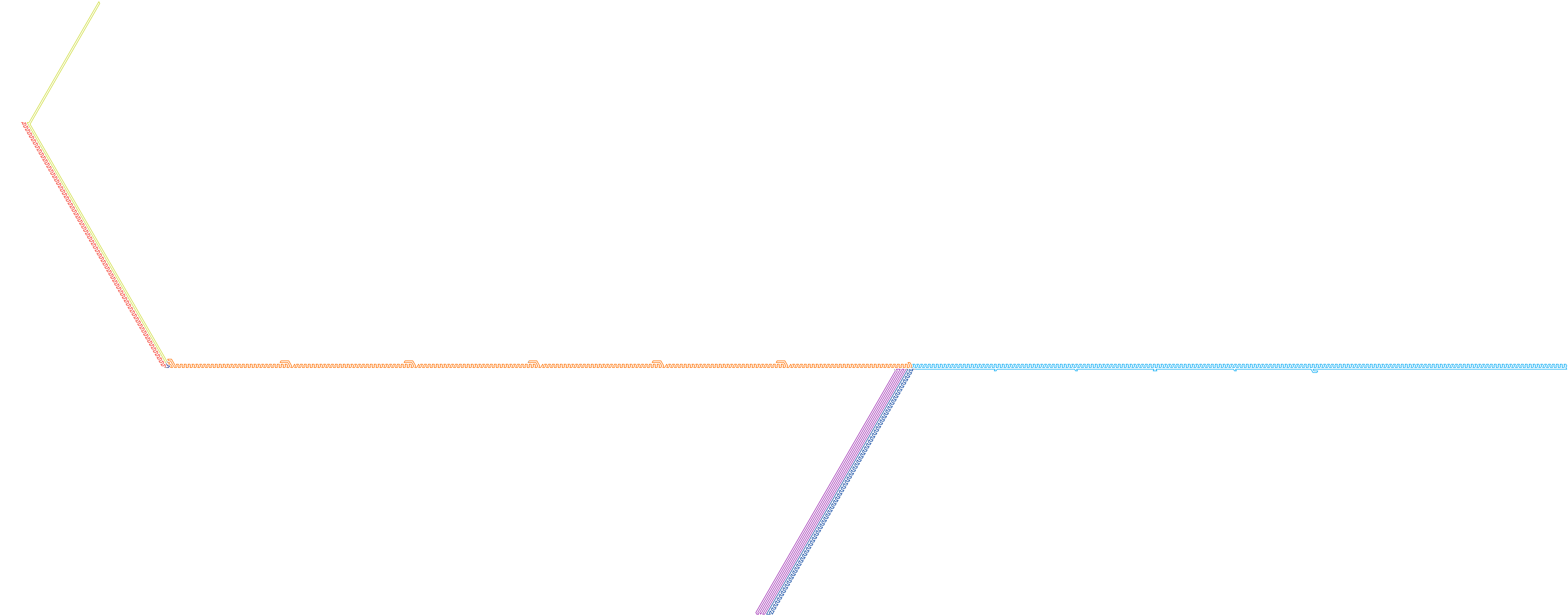}};
            \ifthenelse{\equal{#1}{}}{
\TIKZBLOCKAPPENDCARRIAGERETURN{2}{99.500}{223.868}
\TIKZBLOCKAPPENDLETTER{1}{0,0}{-1025.250}{177.535}
\TIKZBLOCKAPPENDLETTER{1}{1,1}{-92.750}{181.865}
            }{}
    \end{tikzpicture}
}

%% file: OritatamiTuringPatterns/n=4-L=3-P=1/Append_____n=4-L=3-P=1.tex
\newcommand{\AppendBlockEpsilon}[1]{
    \begin{tikzpicture}
        \node [anchor=center] () at (0,0) {\includegraphics[scale =\s]{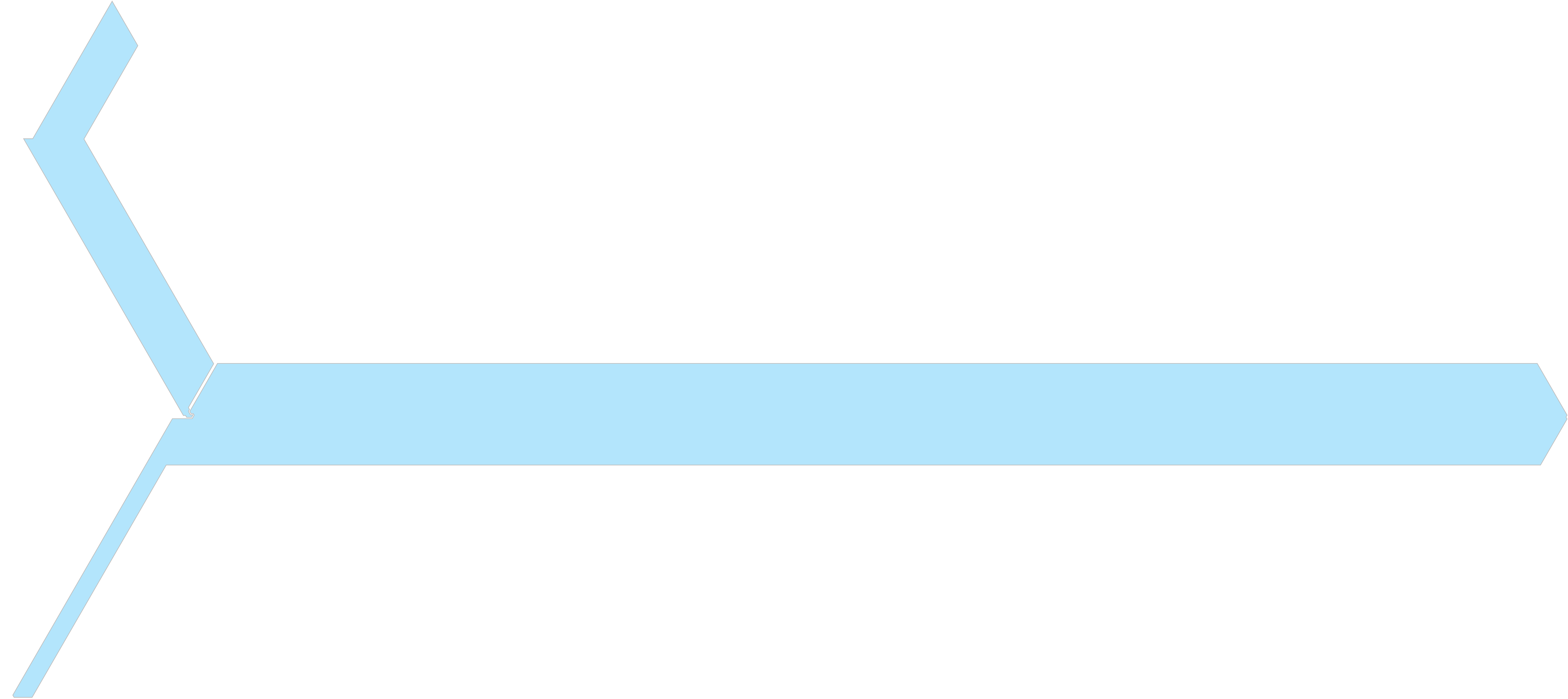}};
        \node [anchor=center] () at (0,0) {\includegraphics[scale =\s]{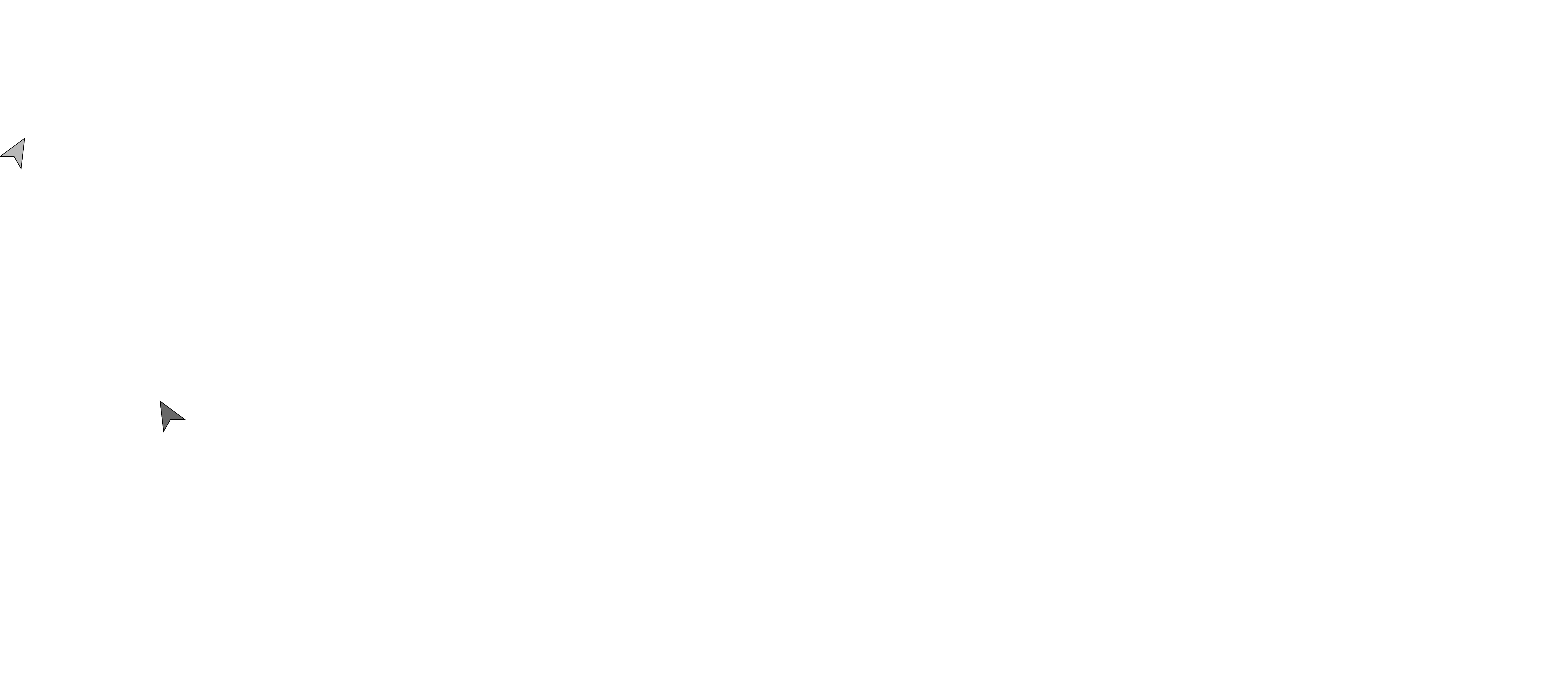}};
        \ifthenelse{\equal{#1}{}}{
\TIKZPointOfInterest{(-1727.50000000*\st pt, 478.47903559*\st pt)}{a}{$(-1,1-h)$}{W}{100.0*\st pt}{Start}{(-1800.00000000*\st pt, -1090.24818095*\st pt)}{8.0}{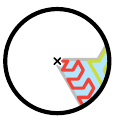}
\TIKZPointOfInterest{(-1527.50000000*\st pt, 790.24818095*\st pt)}{b}{$\left(3,-\cfrac{3h-5}2\right)$}{E}{100.0*\st pt}{}{(-1200.00000000*\st pt, -1090.24818095*\st pt)}{8.0}{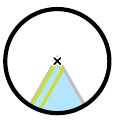}
\TIKZPointOfInterest{(-1350.00000000*\st pt, -140.72912811*\st pt)}{c}{$\left(h-1,-3\right)$}{W}{100.0*\st pt}{}{(-600.00000000*\st pt, -1090.24818095*\st pt)}{8.0}{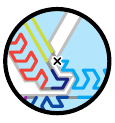}
\TIKZPointOfInterest{(1777.50000000*\st pt, -153.71950917*\st pt)}{d}{$(h+c(0),0)$}{E}{100.0*\st pt}{}{(0.00000000*\st pt, -1090.24818095*\st pt)}{8.0}{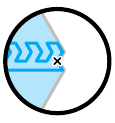}
\TIKZPointOfInterest{(-1710.00000000*\st pt, -790.24818095*\st pt)}{e}{$(h+2,h)$}{E}{100.0*\st pt}{}{(600.00000000*\st pt, -1090.24818095*\st pt)}{8.0}{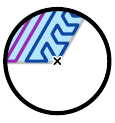}
\TIKZPointOfInterest{(-1755.00000000*\st pt, -790.24818095*\st pt)}{f}{$(h-7,h)$}{W}{100.0*\st pt}{}{(1200.00000000*\st pt, -1090.24818095*\st pt)}{8.0}{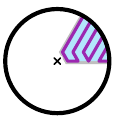}
\TIKZPointOfInterest{(-1395.00000000*\st pt, -158.04963619*\st pt)}{g}{$(h-8,1)$}{SW}{75.0*\st pt}{Next}{(1800.00000000*\st pt, -1090.24818095*\st pt)}{8.0}{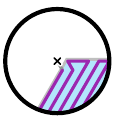}
            }{}
\node [anchor=center] () at (0,0) {\includegraphics[scale =\s]{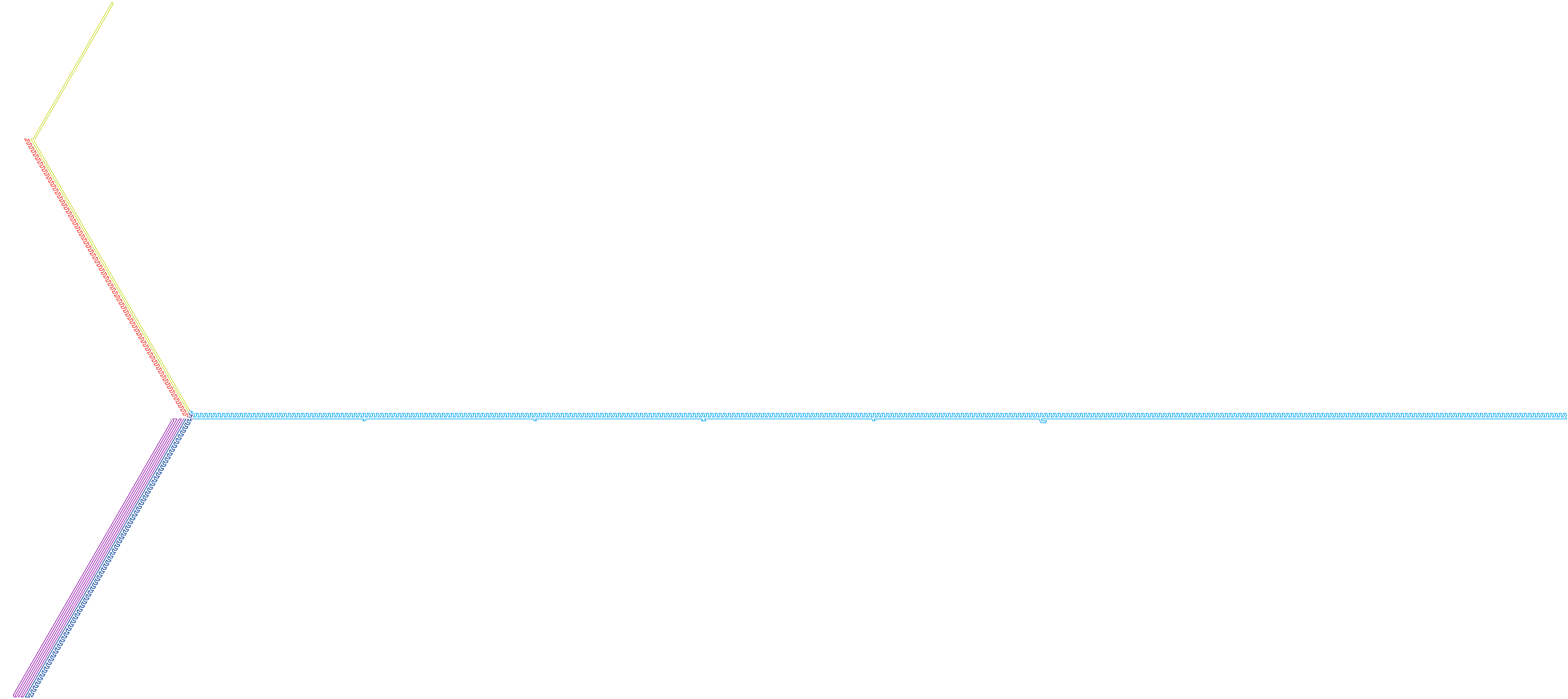}};
            \ifthenelse{\equal{#1}{}}{
\TIKZBLOCKAPPENDCARRIAGERETURN{1}{486.500}{-30.744}
            }{}
    \end{tikzpicture}
}

%% file: OritatamiTuringPatterns/n=4-L=3-P=1/Halt_0__n=4-L=3-P=1.tex
\newcommand{\HaltBlockZ}[1]{
    \begin{tikzpicture}
        \node [anchor=center] () at (0,0) {\includegraphics[scale =\s]{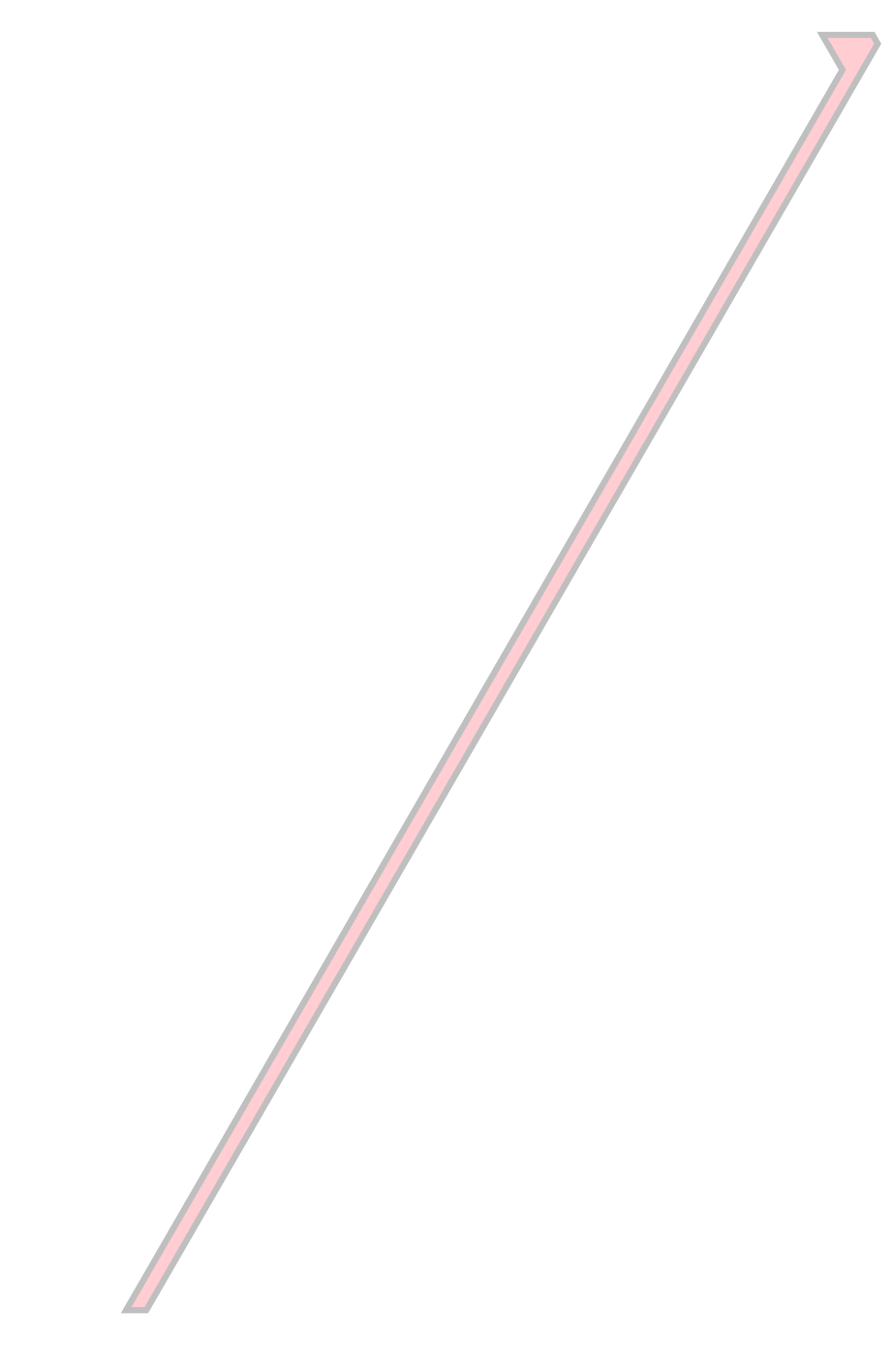}};
        \node [anchor=center] () at (0,0) {\includegraphics[scale =\s]{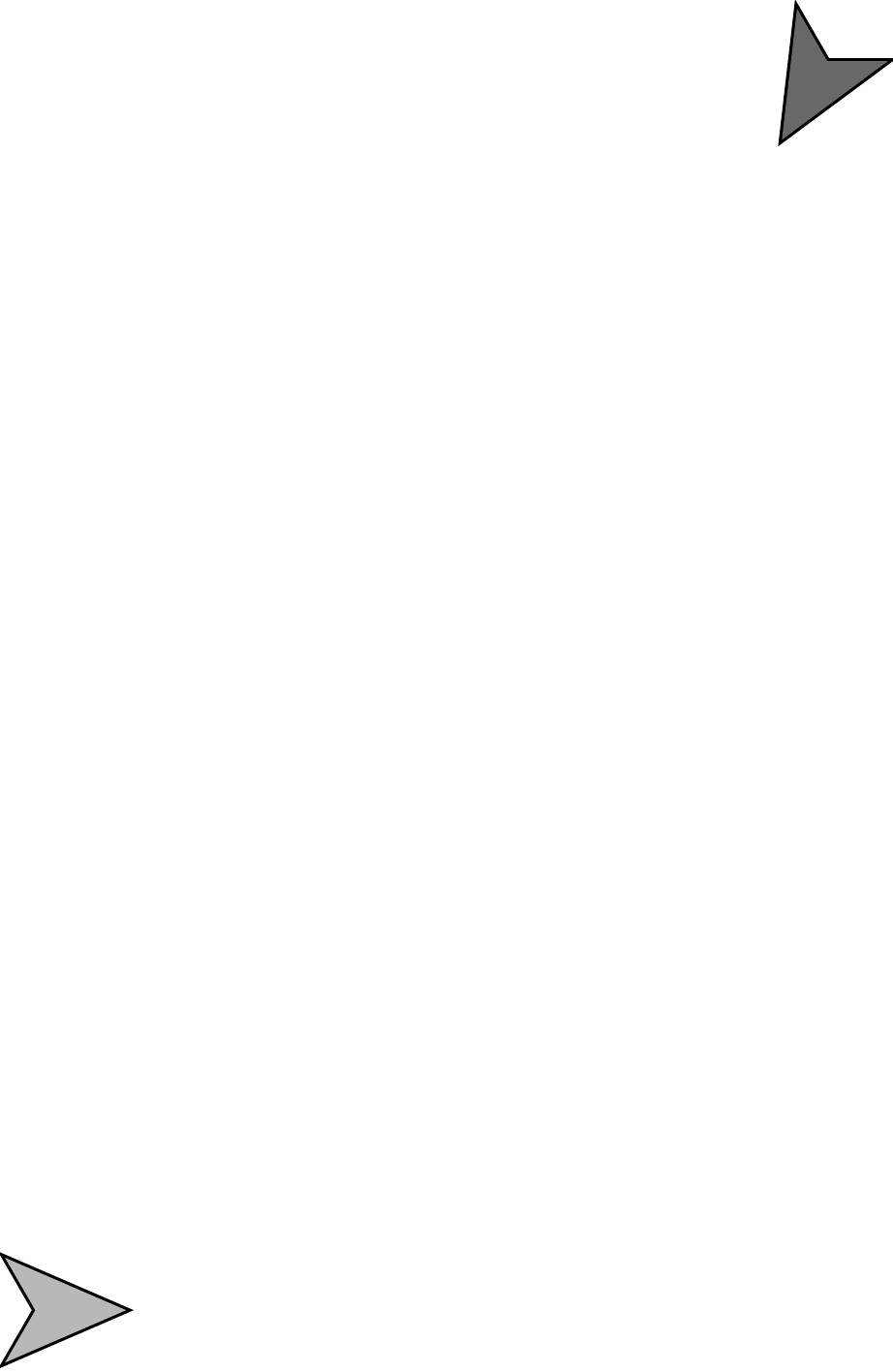}};
        \ifthenelse{\equal{#1}{}}{
\TIKZPointOfInterest{(-158.75000000*\st pt, -309.60408185*\st pt)}{a}{$(0,0)$}{NW}{100.0*\st pt}{Start}{(-600.00000000*\st pt, -635.58484397*\st pt)}{8.0}{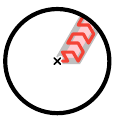}
\TIKZPointOfInterest{(216.25000000*\st pt, 322.59446291*\st pt)}{b}{$(2,1-h)$}{E}{100.0*\st pt}{}{(0.00000000*\st pt, -635.58484397*\st pt)}{8.0}{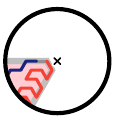}
\TIKZPointOfInterest{(201.25000000*\st pt, 322.59446291*\st pt)}{c}{$(-1,1-h)$}{NW}{50.0*\st pt}{Trapped}{(600.00000000*\st pt, -635.58484397*\st pt)}{8.0}{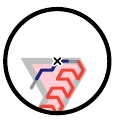}
            }{}
\node [anchor=center] () at (0,0) {\includegraphics[scale =\s]{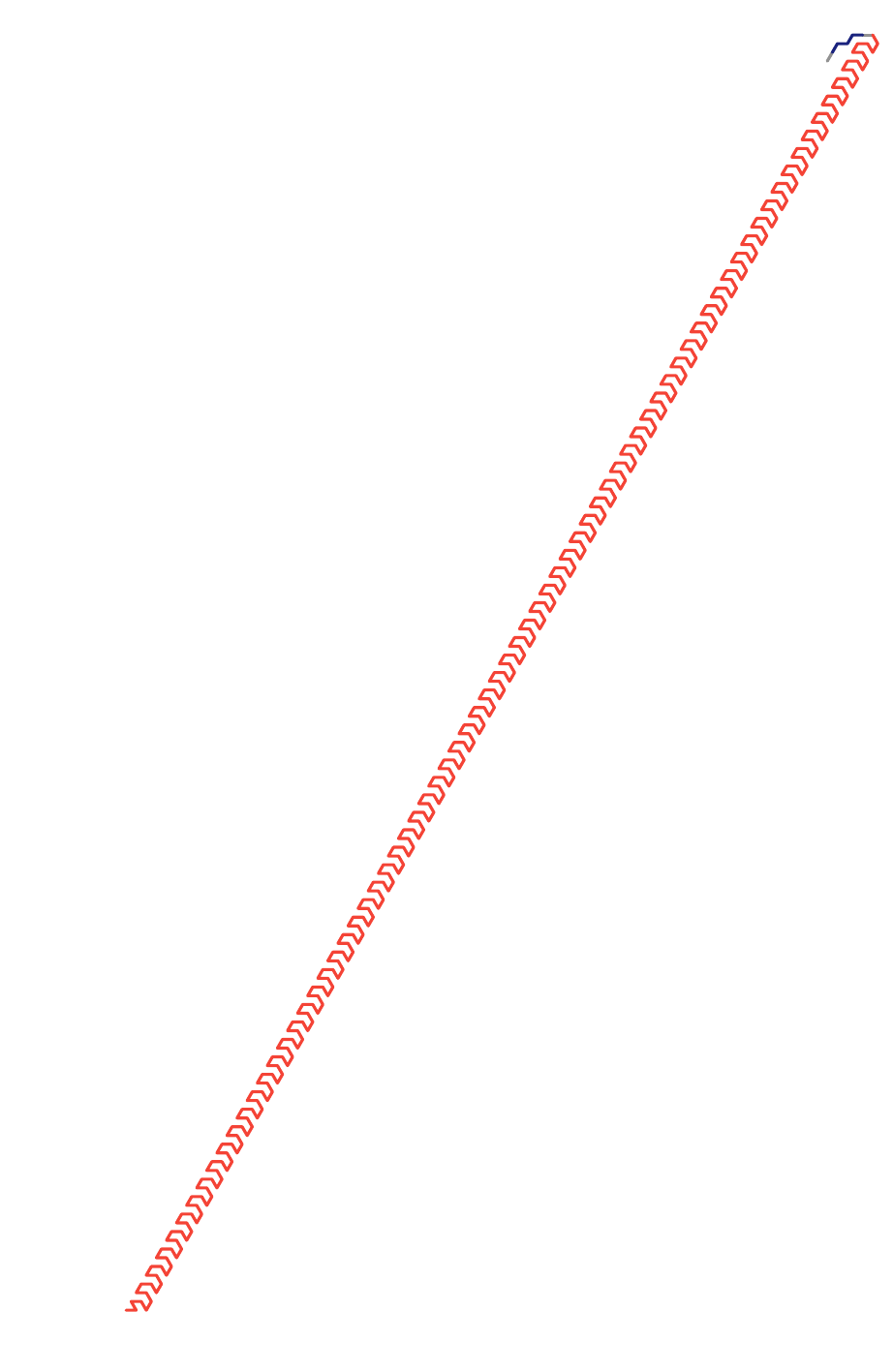}};
            \ifthenelse{\equal{#1}{}}{
\TIKZBLOCKHALT{3}{121.000}{8.660}
            }{}
    \end{tikzpicture}
}

%% file: OritatamiTuringPatterns/n=4-L=3-P=1/Halt_110__n=4-L=3-P=1.tex
\newcommand{\HaltBlockUUZ}[1]{
    \begin{tikzpicture}
        \node [anchor=center] () at (0,0) {\includegraphics[scale =\s]{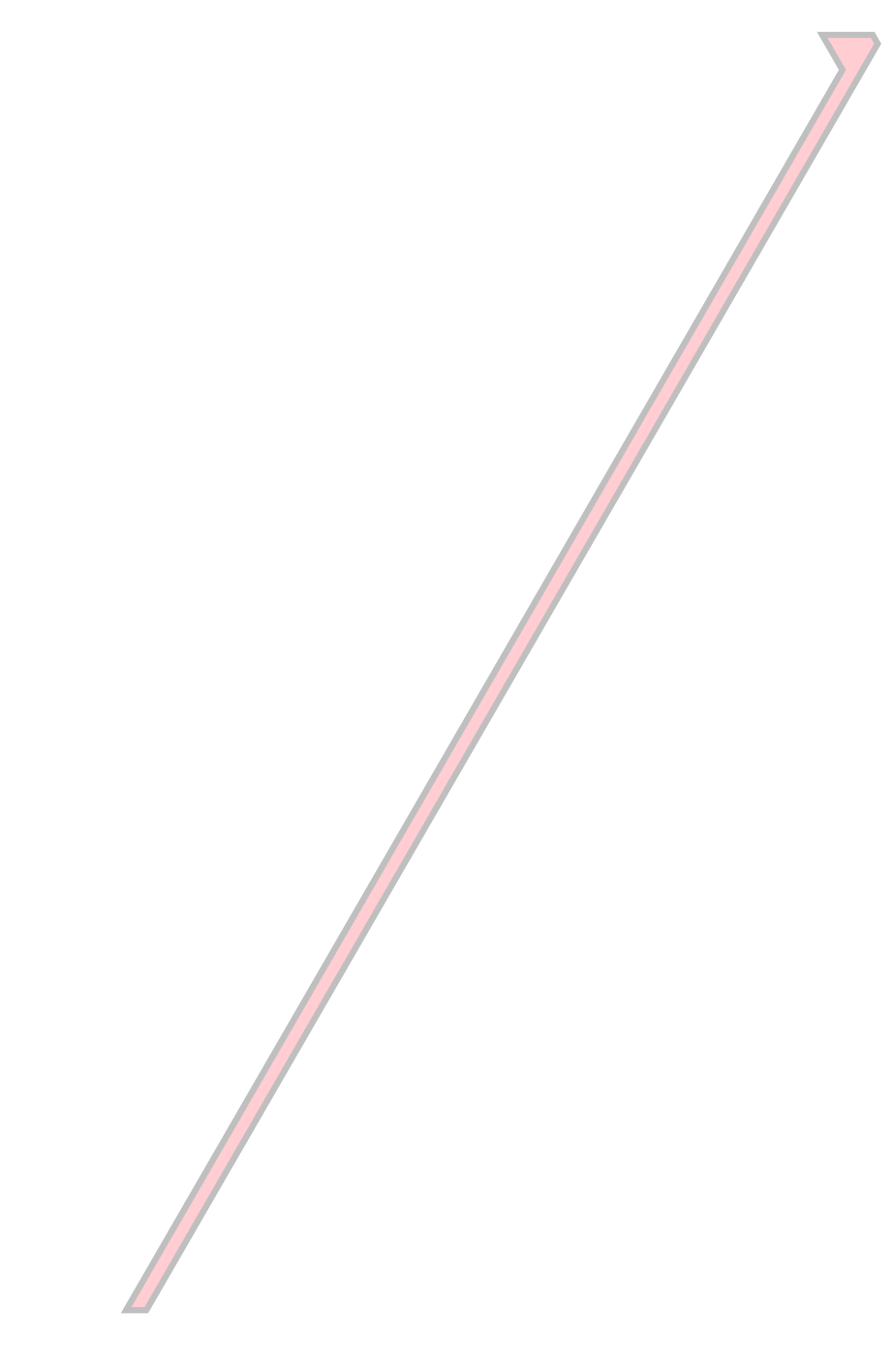}};
        \node [anchor=center] () at (0,0) {\includegraphics[scale =\s]{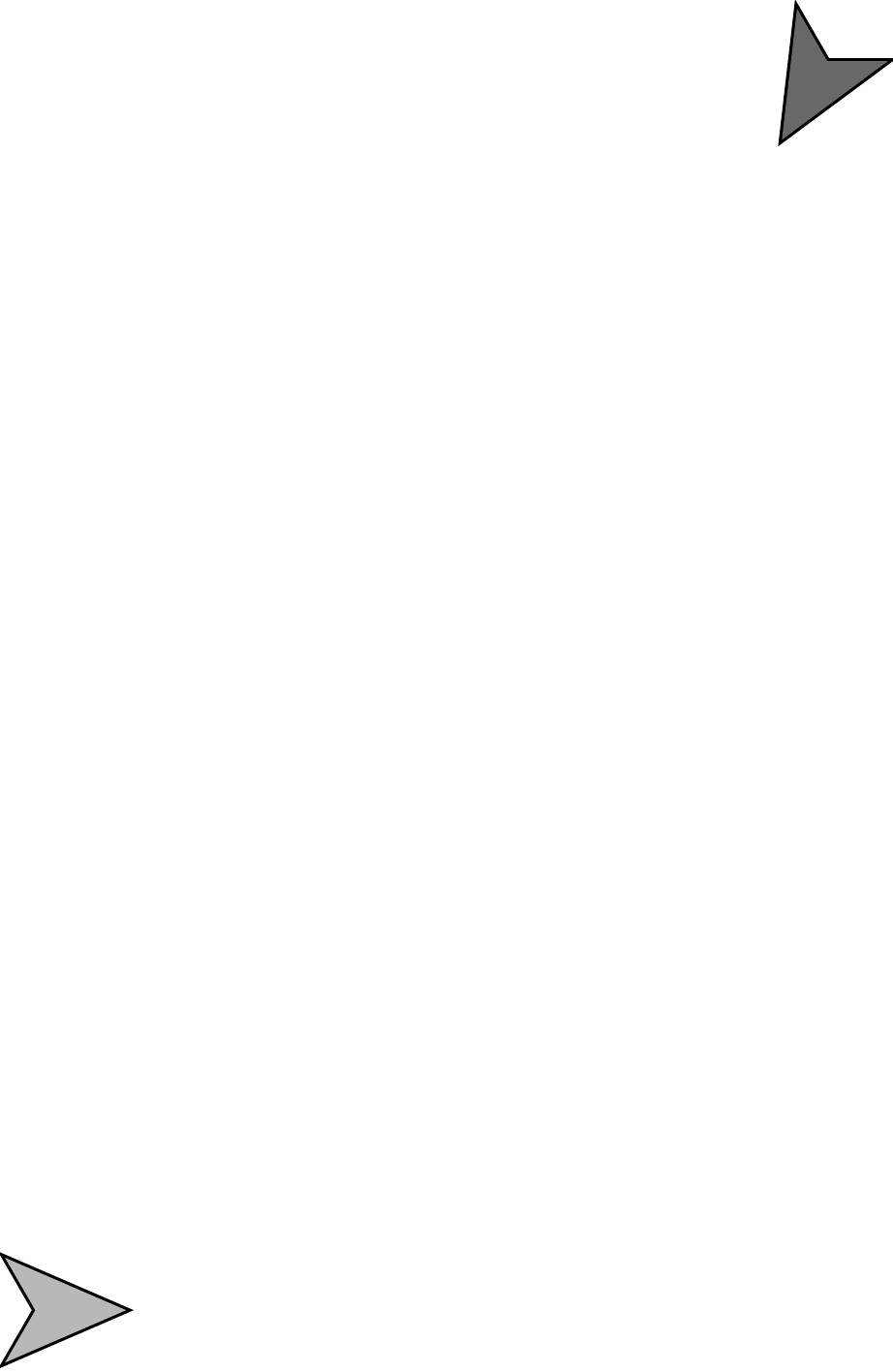}};
        \ifthenelse{\equal{#1}{}}{
\TIKZPointOfInterest{(-158.75000000*\st pt, -309.60408185*\st pt)}{a}{$(0,0)$}{NW}{100.0*\st pt}{Start}{(-600.00000000*\st pt, -635.58484397*\st pt)}{8.0}{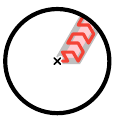}
\TIKZPointOfInterest{(216.25000000*\st pt, 322.59446291*\st pt)}{b}{$(2,1-h)$}{E}{100.0*\st pt}{}{(0.00000000*\st pt, -635.58484397*\st pt)}{8.0}{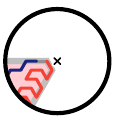}
\TIKZPointOfInterest{(201.25000000*\st pt, 322.59446291*\st pt)}{c}{$(-1,1-h)$}{NW}{50.0*\st pt}{Trapped}{(600.00000000*\st pt, -635.58484397*\st pt)}{8.0}{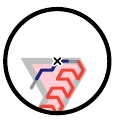}
            }{}
\node [anchor=center] () at (0,0) {\includegraphics[scale =\s]{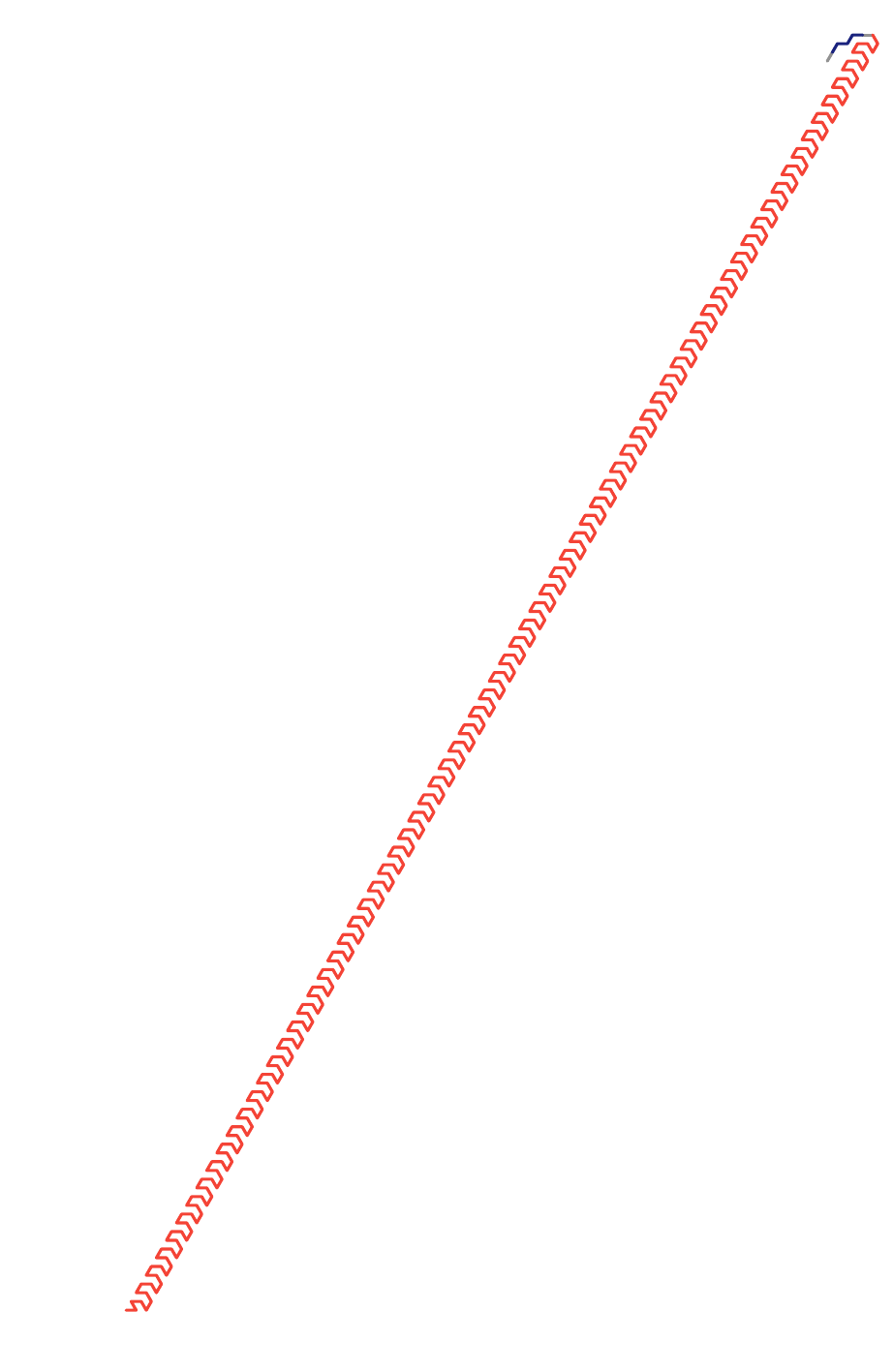}};
            \ifthenelse{\equal{#1}{}}{
\TIKZBLOCKHALT{0}{121.000}{8.660}
            }{}
    \end{tikzpicture}
}

%% file: OritatamiTuringPatterns/n=4-L=3-P=1/Halt_11__n=4-L=3-P=1.tex
\newcommand{\HaltBlockUU}[1]{
    \begin{tikzpicture}
        \node [anchor=center] () at (0,0) {\includegraphics[scale =\s]{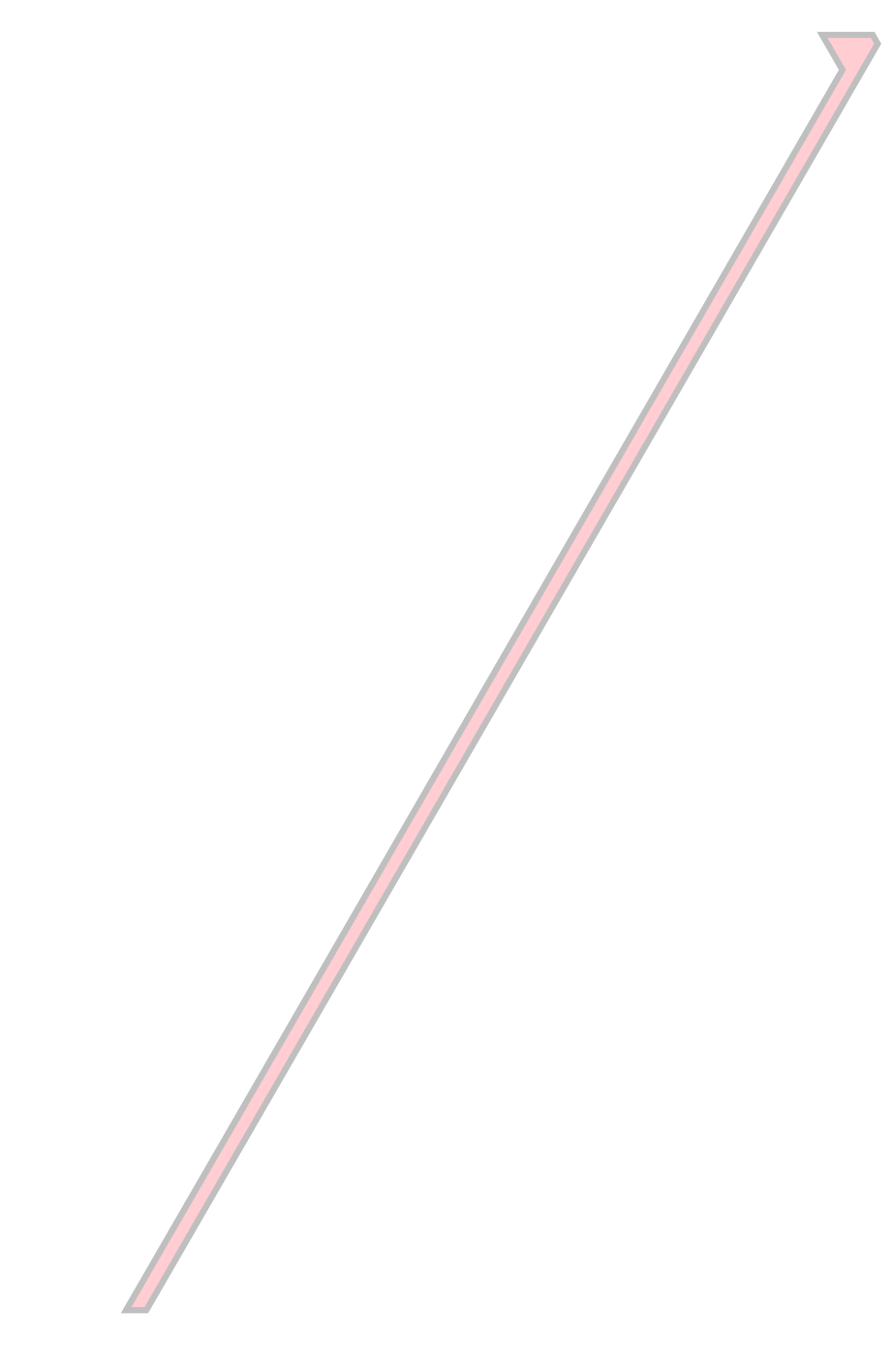}};
        \node [anchor=center] () at (0,0) {\includegraphics[scale =\s]{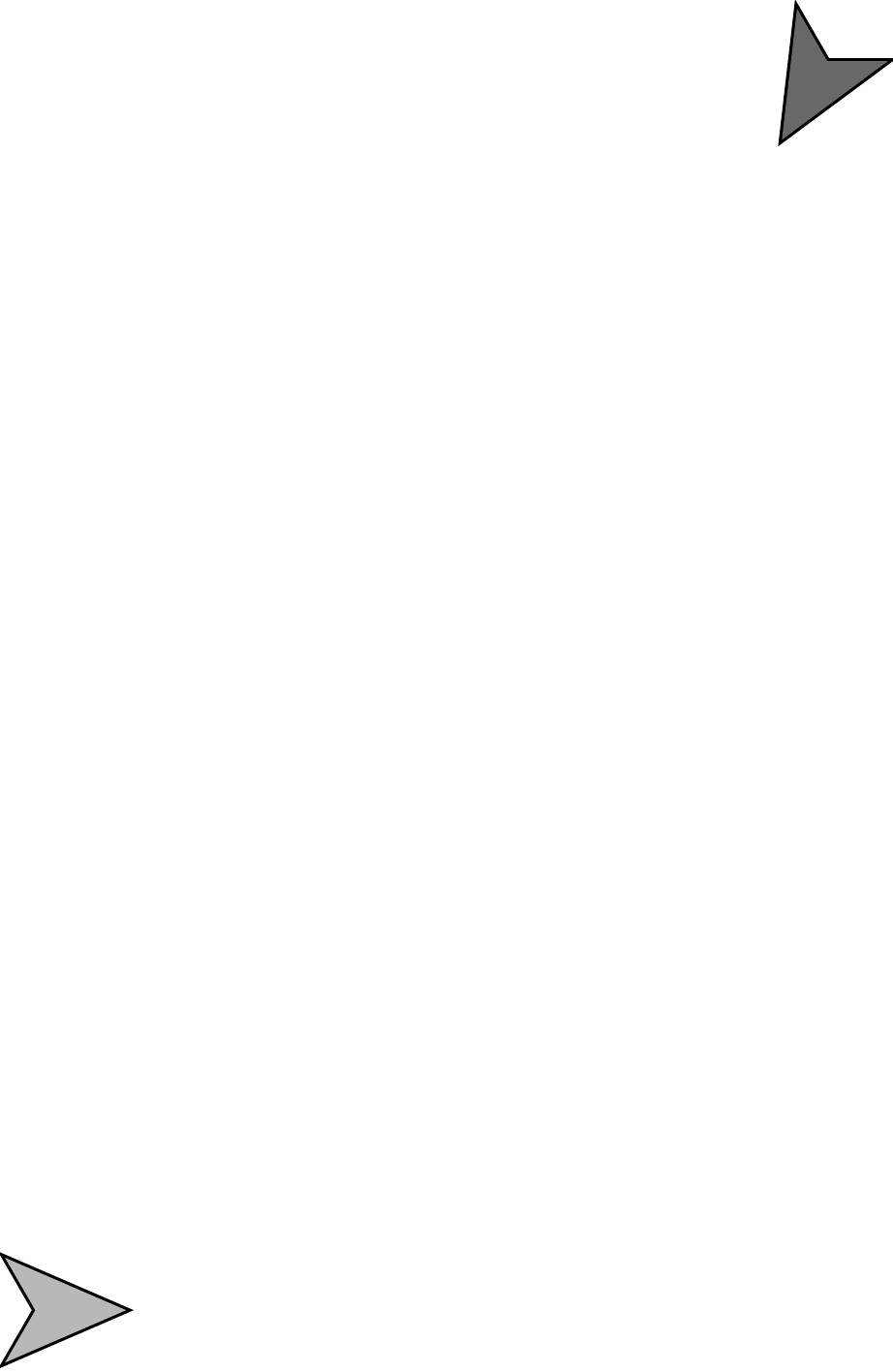}};
        \ifthenelse{\equal{#1}{}}{
\TIKZPointOfInterest{(-158.75000000*\st pt, -309.60408185*\st pt)}{a}{$(0,0)$}{NW}{100.0*\st pt}{Start}{(-600.00000000*\st pt, -635.58484397*\st pt)}{8.0}{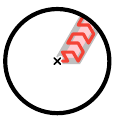}
\TIKZPointOfInterest{(216.25000000*\st pt, 322.59446291*\st pt)}{b}{$(2,1-h)$}{E}{100.0*\st pt}{}{(0.00000000*\st pt, -635.58484397*\st pt)}{8.0}{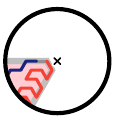}
\TIKZPointOfInterest{(201.25000000*\st pt, 322.59446291*\st pt)}{c}{$(-1,1-h)$}{NW}{50.0*\st pt}{Trapped}{(600.00000000*\st pt, -635.58484397*\st pt)}{8.0}{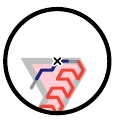}
            }{}
\node [anchor=center] () at (0,0) {\includegraphics[scale =\s]{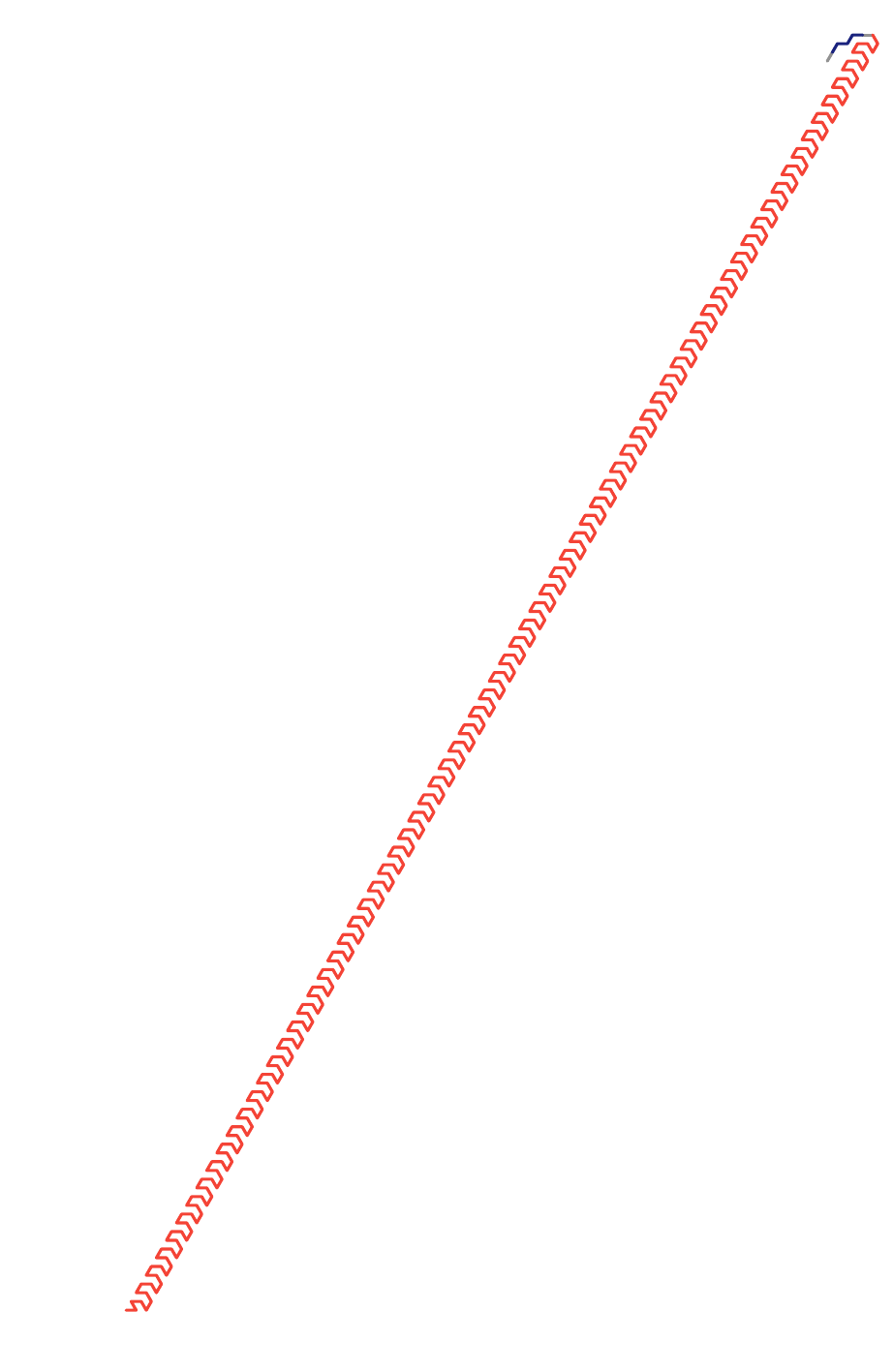}};
            \ifthenelse{\equal{#1}{}}{
\TIKZBLOCKHALT{2}{121.000}{8.660}
            }{}
    \end{tikzpicture}
}

%% file: OritatamiTuringPatterns/n=4-L=3-P=1/Halt_____n=4-L=3-P=1.tex
\newcommand{\HaltBlockEpsilon}[1]{
    \begin{tikzpicture}
        \node [anchor=center] () at (0,0) {\includegraphics[scale =\s]{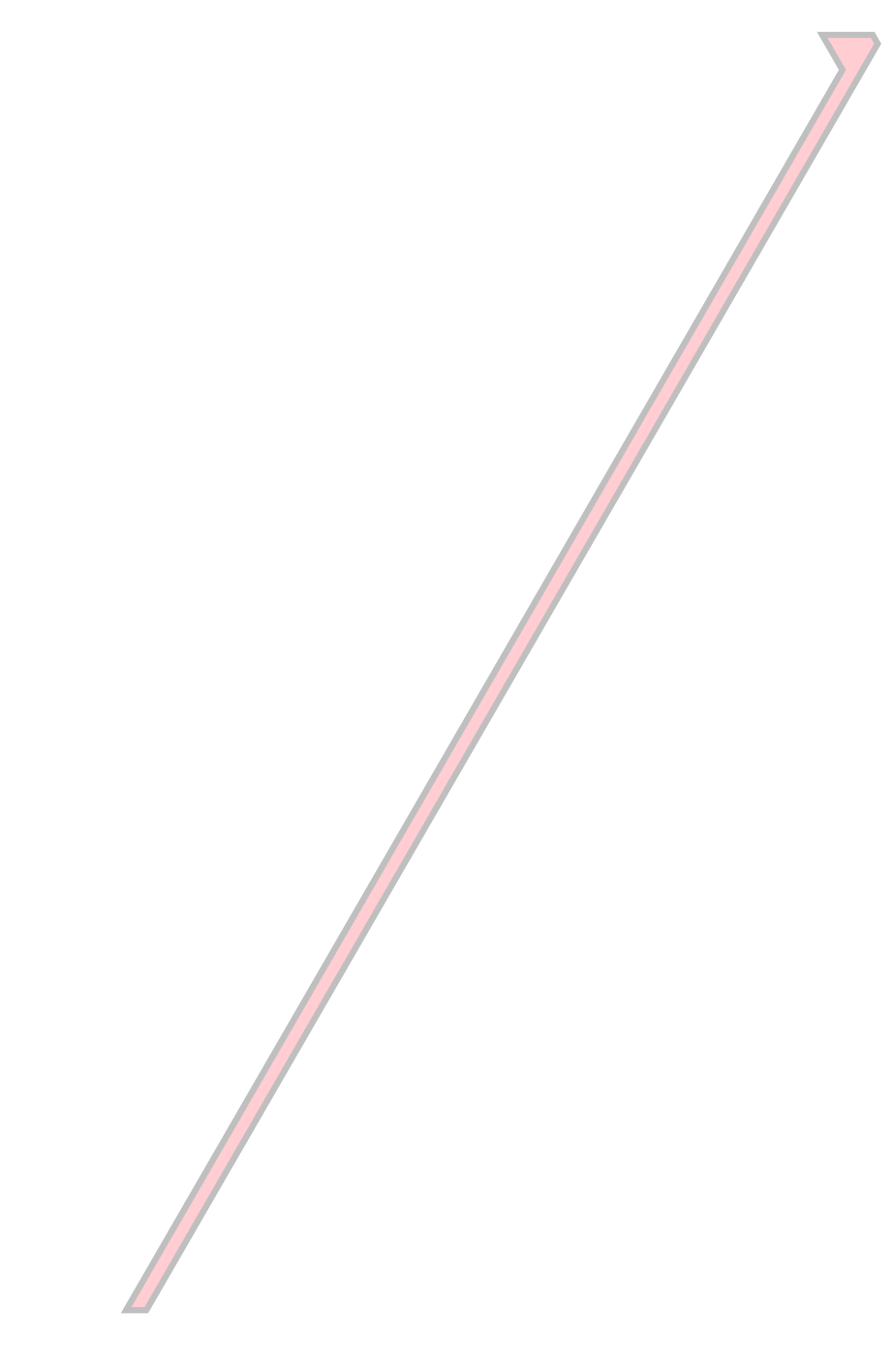}};
        \node [anchor=center] () at (0,0) {\includegraphics[scale =\s]{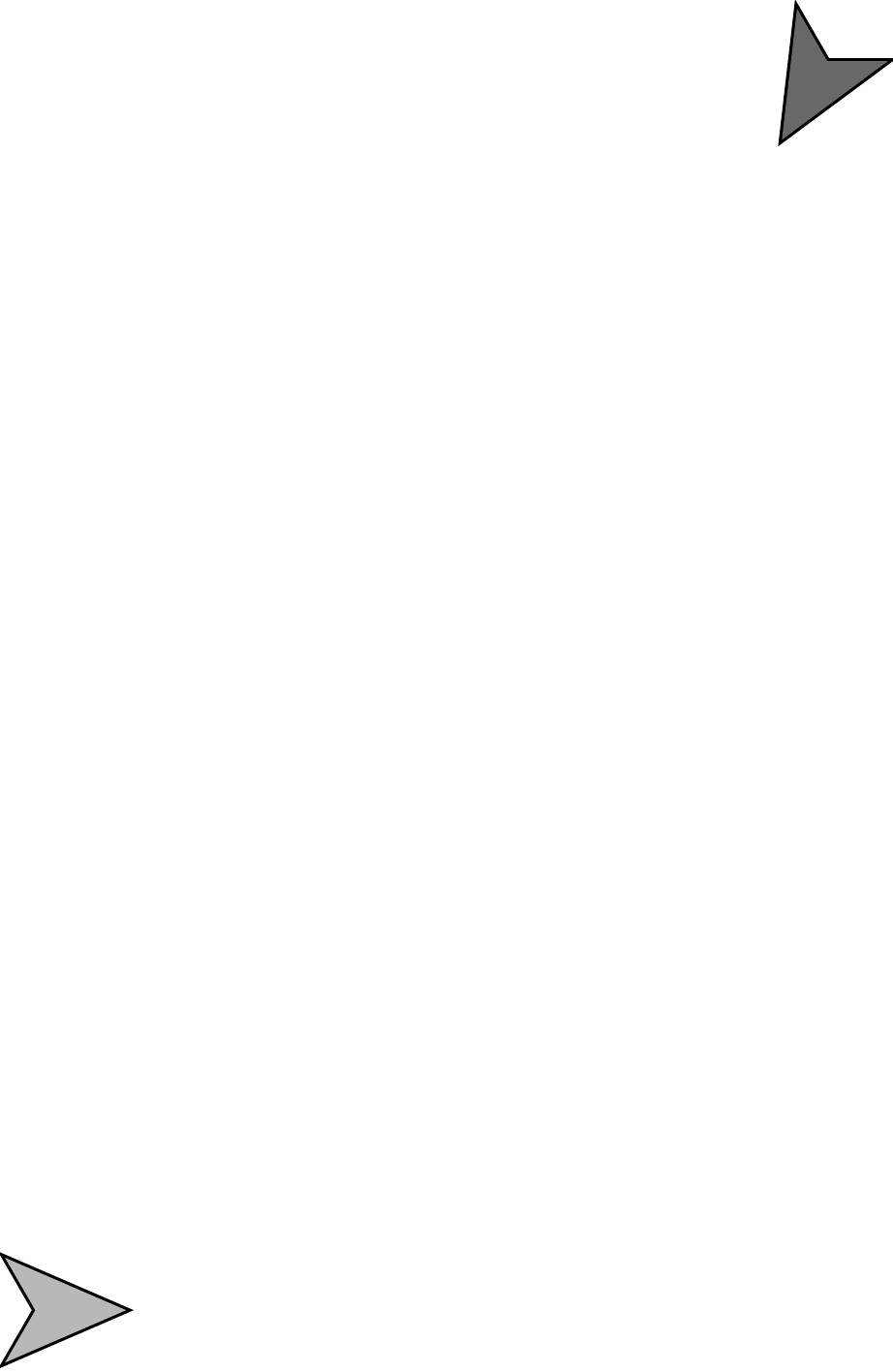}};
        \ifthenelse{\equal{#1}{}}{
\TIKZPointOfInterest{(-158.75000000*\st pt, -309.60408185*\st pt)}{a}{$(0,0)$}{NW}{100.0*\st pt}{Start}{(-600.00000000*\st pt, -635.58484397*\st pt)}{8.0}{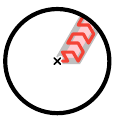}
\TIKZPointOfInterest{(216.25000000*\st pt, 322.59446291*\st pt)}{b}{$(2,1-h)$}{E}{100.0*\st pt}{}{(0.00000000*\st pt, -635.58484397*\st pt)}{8.0}{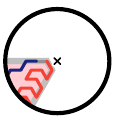}
\TIKZPointOfInterest{(201.25000000*\st pt, 322.59446291*\st pt)}{c}{$(-1,1-h)$}{NW}{50.0*\st pt}{Trapped}{(600.00000000*\st pt, -635.58484397*\st pt)}{8.0}{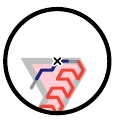}
            }{}
\node [anchor=center] () at (0,0) {\includegraphics[scale =\s]{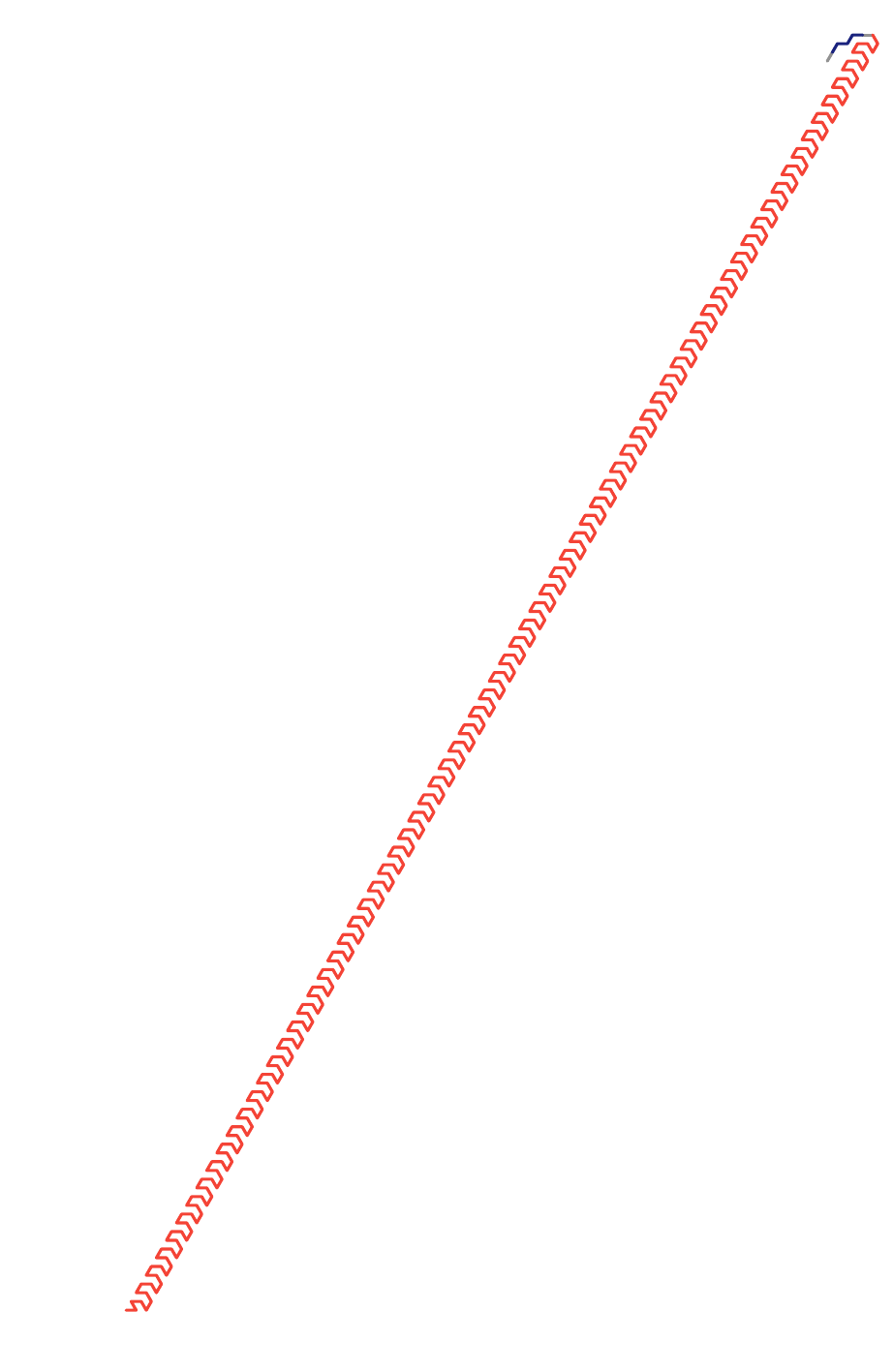}};
            \ifthenelse{\equal{#1}{}}{
\TIKZBLOCKHALT{1}{121.000}{8.660}
            }{}
    \end{tikzpicture}
}

%% file: OritatamiTuringPatterns/n=4-L=3-P=1/Read0_0__n=4-L=3-P=1.tex
\newcommand{\ReadZBlockZ}[1]{
    \begin{tikzpicture}
        \node [anchor=center] () at (0,0) {\includegraphics[scale =\s]{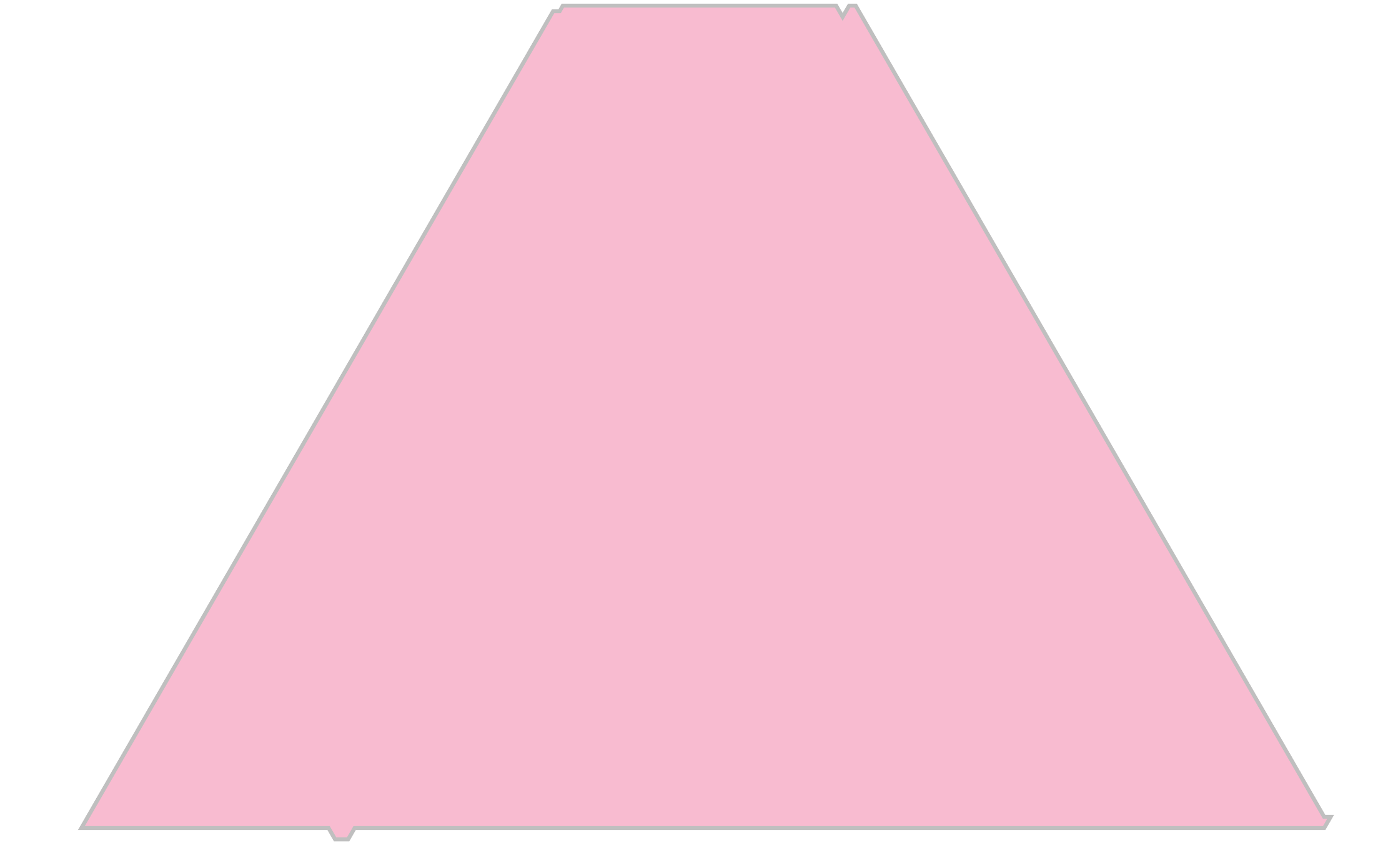}};
        \node [anchor=center] () at (0,0) {\includegraphics[scale =\s]{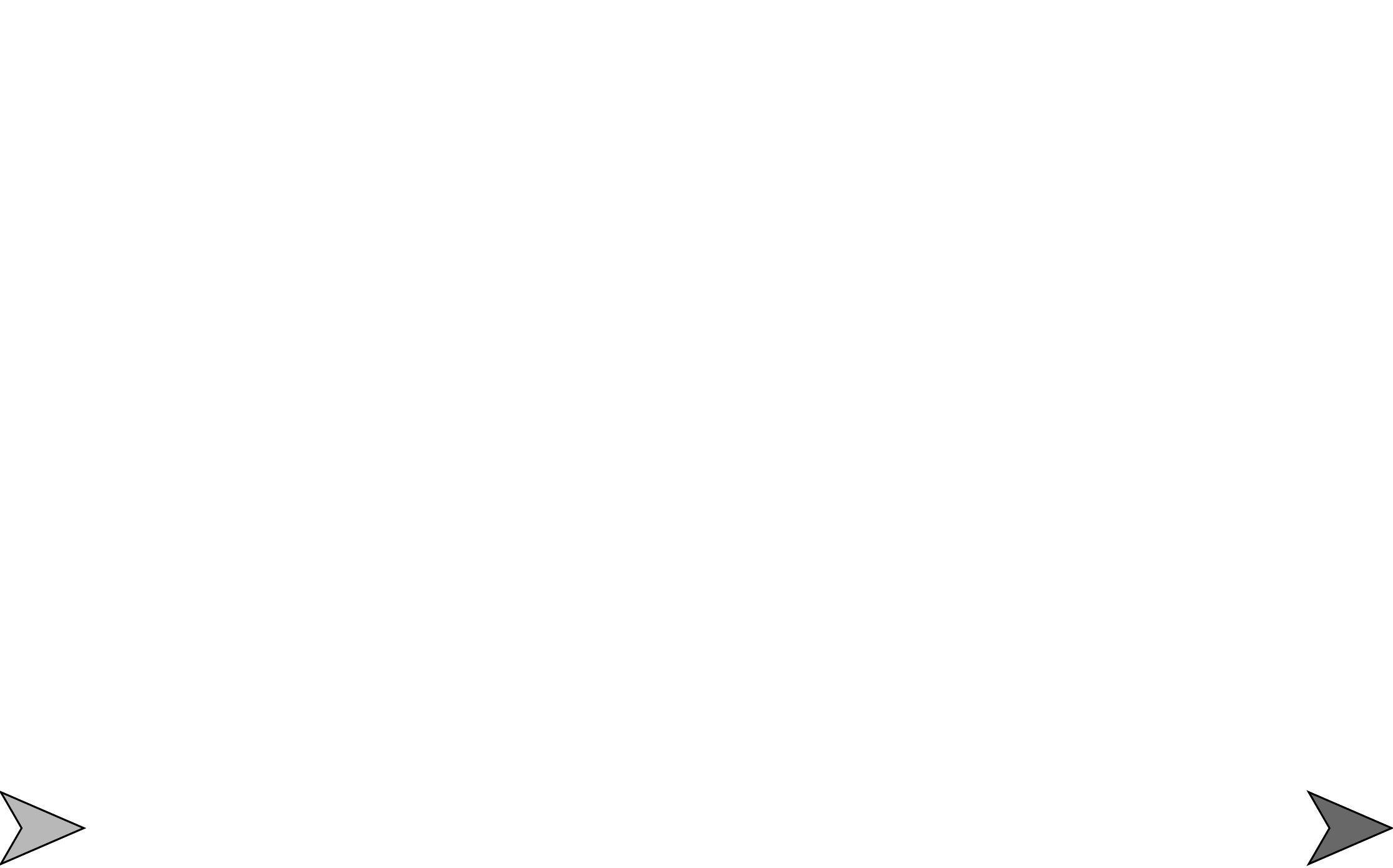}};
        \ifthenelse{\equal{#1}{}}{
\TIKZPointOfInterest{(-472.50000000*\st pt, -303.10889132*\st pt)}{a}{$(0,0)$}{NW}{100.0*\st pt}{Start}{(-1200.00000000*\st pt, -629.08965344*\st pt)}{8.0}{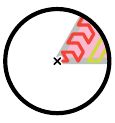}
\TIKZPointOfInterest{(-107.50000000*\st pt, 329.08965344*\st pt)}{b}{$(0,1-h)$}{NW}{50.0*\st pt}{}{(-600.00000000*\st pt, -629.08965344*\st pt)}{8.0}{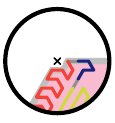}
\TIKZPointOfInterest{(-267.50000000*\st pt, -303.10889132*\st pt)}{c}{$(w-1,0)$}{SE}{50.0*\st pt}{}{(0.00000000*\st pt, -629.08965344*\st pt)}{8.0}{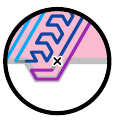}
\TIKZPointOfInterest{(112.50000000*\st pt, 329.08965344*\st pt)}{d}{$(w+2,1-h)$}{NE}{50.0*\st pt}{Read \word0}{(600.00000000*\st pt, -629.08965344*\st pt)}{8.0}{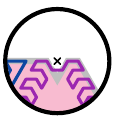}
\TIKZPointOfInterest{(487.50000000*\st pt, -303.10889132*\st pt)}{e}{$(W,0)$}{NE}{100.0*\st pt}{Next}{(1200.00000000*\st pt, -629.08965344*\st pt)}{8.0}{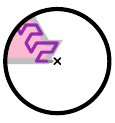}
            }{}
\node [anchor=center] () at (0,0) {\includegraphics[scale =\s]{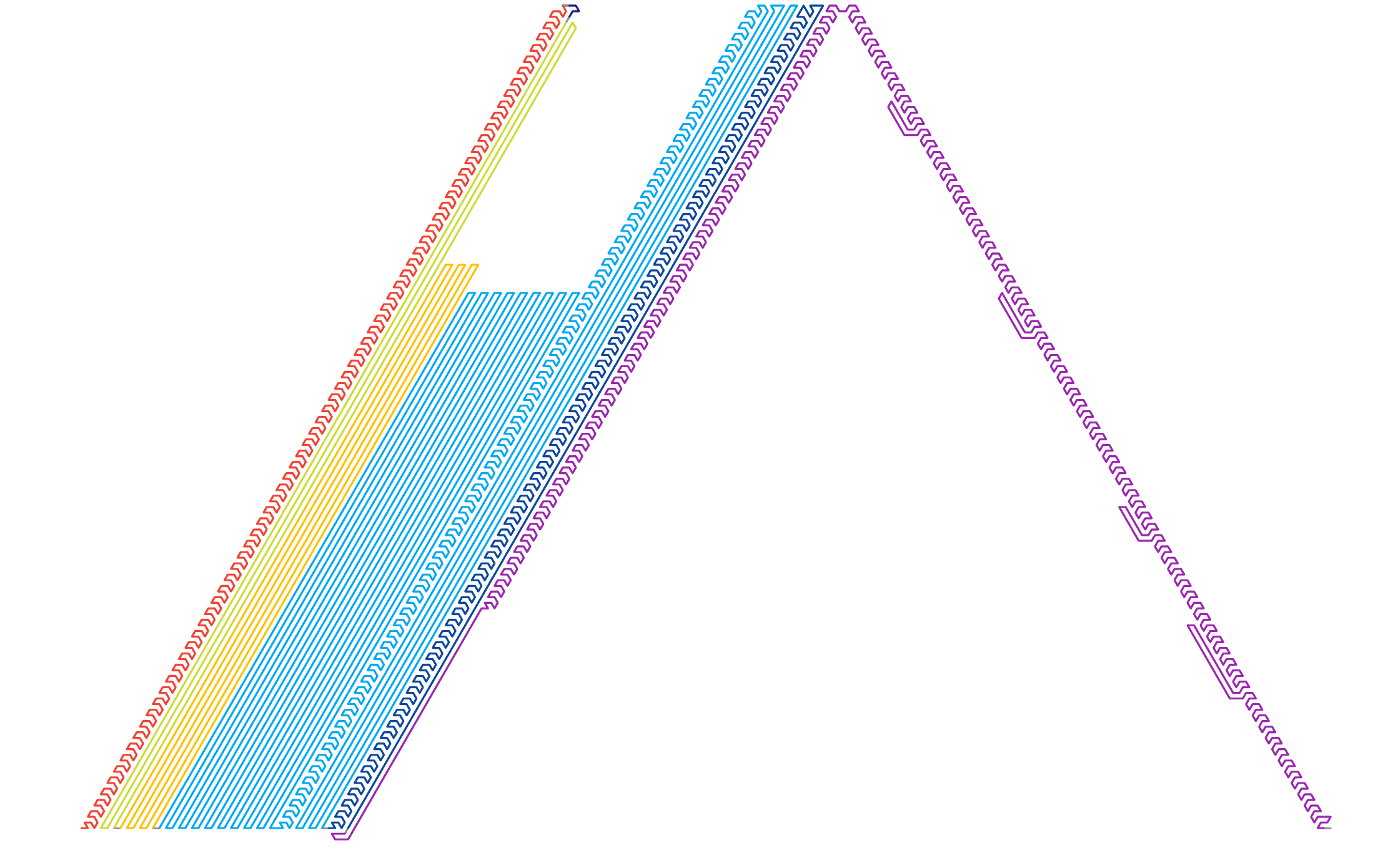}};
            \ifthenelse{\equal{#1}{}}{
\TIKZBLOCKREAD{0}{3}{0}{190.000}{12.990}
            }{}
    \end{tikzpicture}
}

%% file: OritatamiTuringPatterns/n=4-L=3-P=1/Read0_110__n=4-L=3-P=1.tex
\newcommand{\ReadZBlockUUZ}[1]{
    \begin{tikzpicture}
        \node [anchor=center] () at (0,0) {\includegraphics[scale =\s]{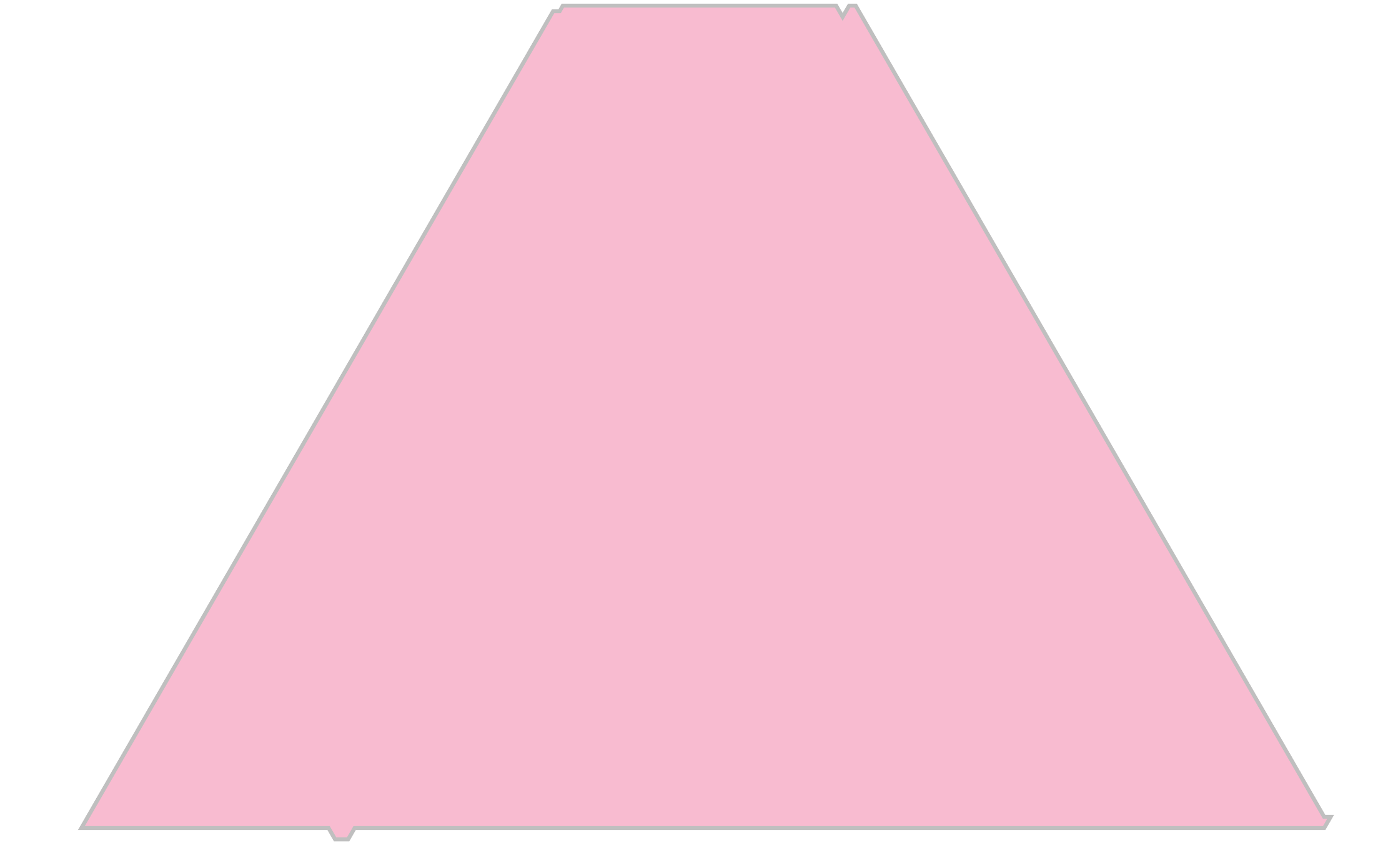}};
        \node [anchor=center] () at (0,0) {\includegraphics[scale =\s]{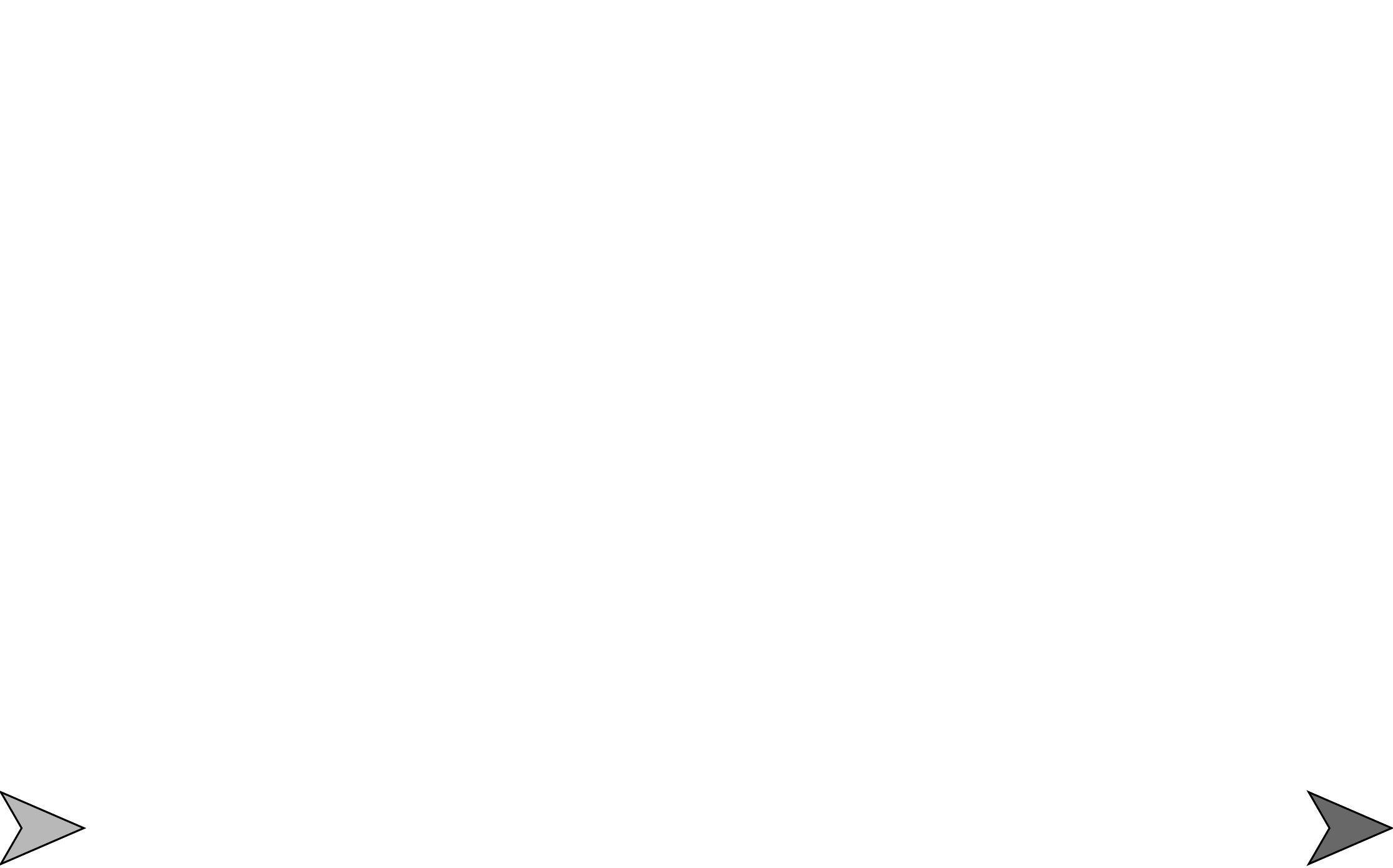}};
        \ifthenelse{\equal{#1}{}}{
\TIKZPointOfInterest{(-472.50000000*\st pt, -303.10889132*\st pt)}{a}{$(0,0)$}{NW}{100.0*\st pt}{Start}{(-1200.00000000*\st pt, -629.08965344*\st pt)}{8.0}{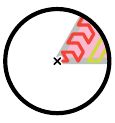}
\TIKZPointOfInterest{(-107.50000000*\st pt, 329.08965344*\st pt)}{b}{$(0,1-h)$}{NW}{50.0*\st pt}{}{(-600.00000000*\st pt, -629.08965344*\st pt)}{8.0}{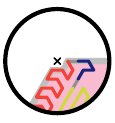}
\TIKZPointOfInterest{(-267.50000000*\st pt, -303.10889132*\st pt)}{c}{$(w-1,0)$}{SE}{50.0*\st pt}{}{(0.00000000*\st pt, -629.08965344*\st pt)}{8.0}{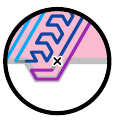}
\TIKZPointOfInterest{(112.50000000*\st pt, 329.08965344*\st pt)}{d}{$(w+2,1-h)$}{NE}{50.0*\st pt}{Read \word0}{(600.00000000*\st pt, -629.08965344*\st pt)}{8.0}{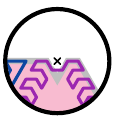}
\TIKZPointOfInterest{(487.50000000*\st pt, -303.10889132*\st pt)}{e}{$(W,0)$}{NE}{100.0*\st pt}{Next}{(1200.00000000*\st pt, -629.08965344*\st pt)}{8.0}{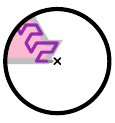}
            }{}
\node [anchor=center] () at (0,0) {\includegraphics[scale =\s]{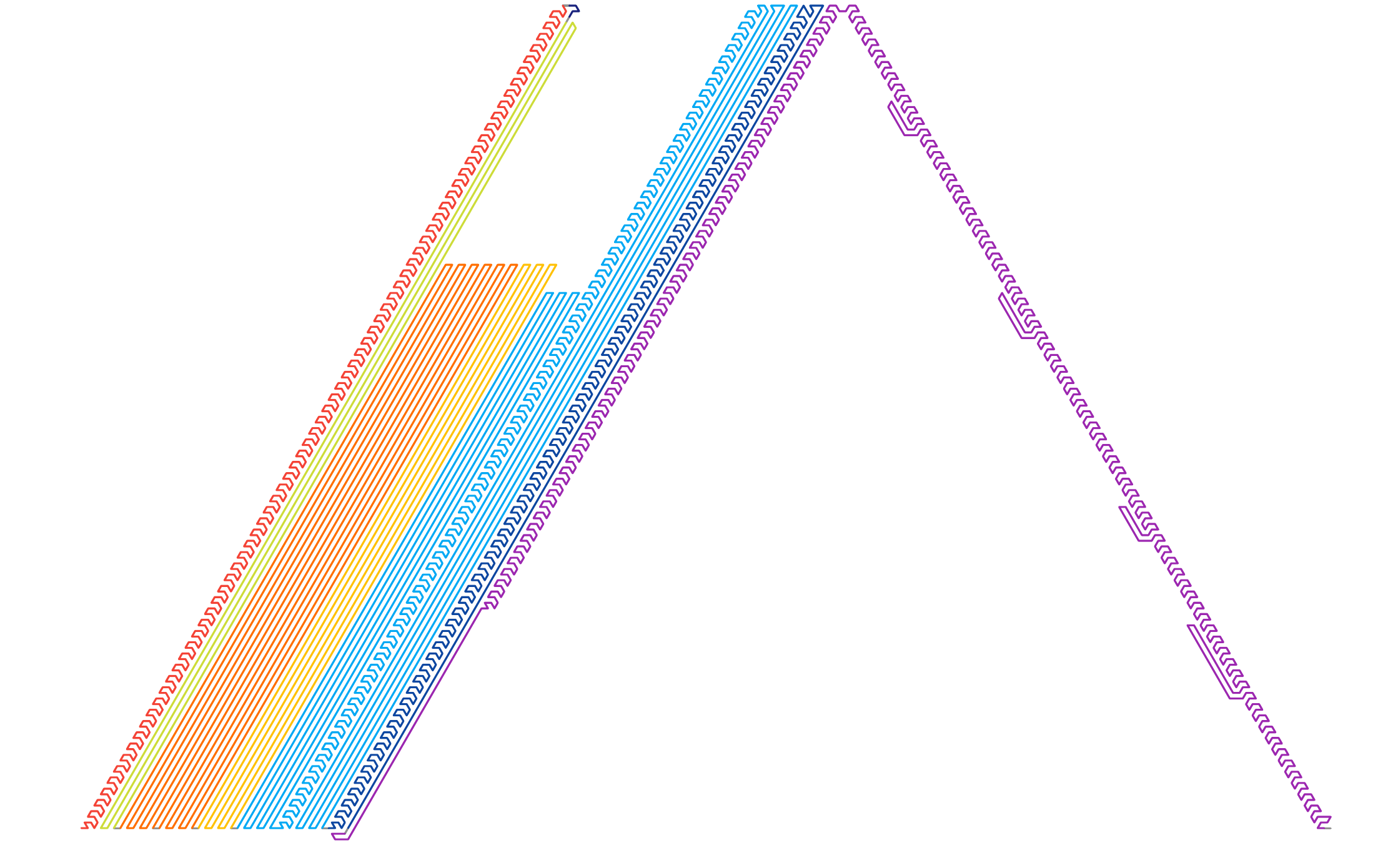}};
            \ifthenelse{\equal{#1}{}}{
\TIKZBLOCKREAD{0}{0}{110}{190.000}{12.990}
            }{}
    \end{tikzpicture}
}

%% file: OritatamiTuringPatterns/n=4-L=3-P=1/Read0_11__n=4-L=3-P=1.tex
\newcommand{\ReadZBlockUU}[1]{
    \begin{tikzpicture}
        \node [anchor=center] () at (0,0) {\includegraphics[scale =\s]{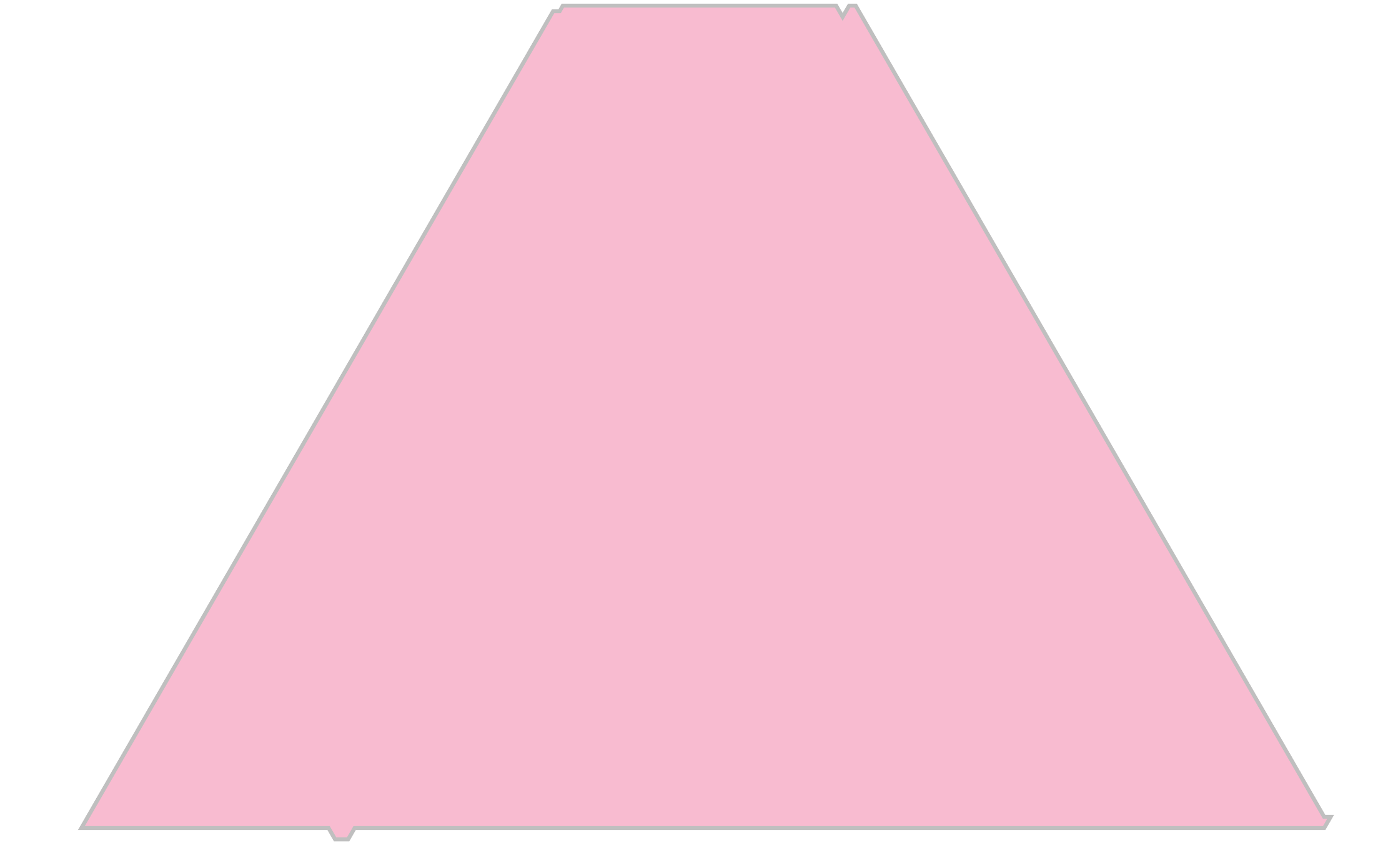}};
        \node [anchor=center] () at (0,0) {\includegraphics[scale =\s]{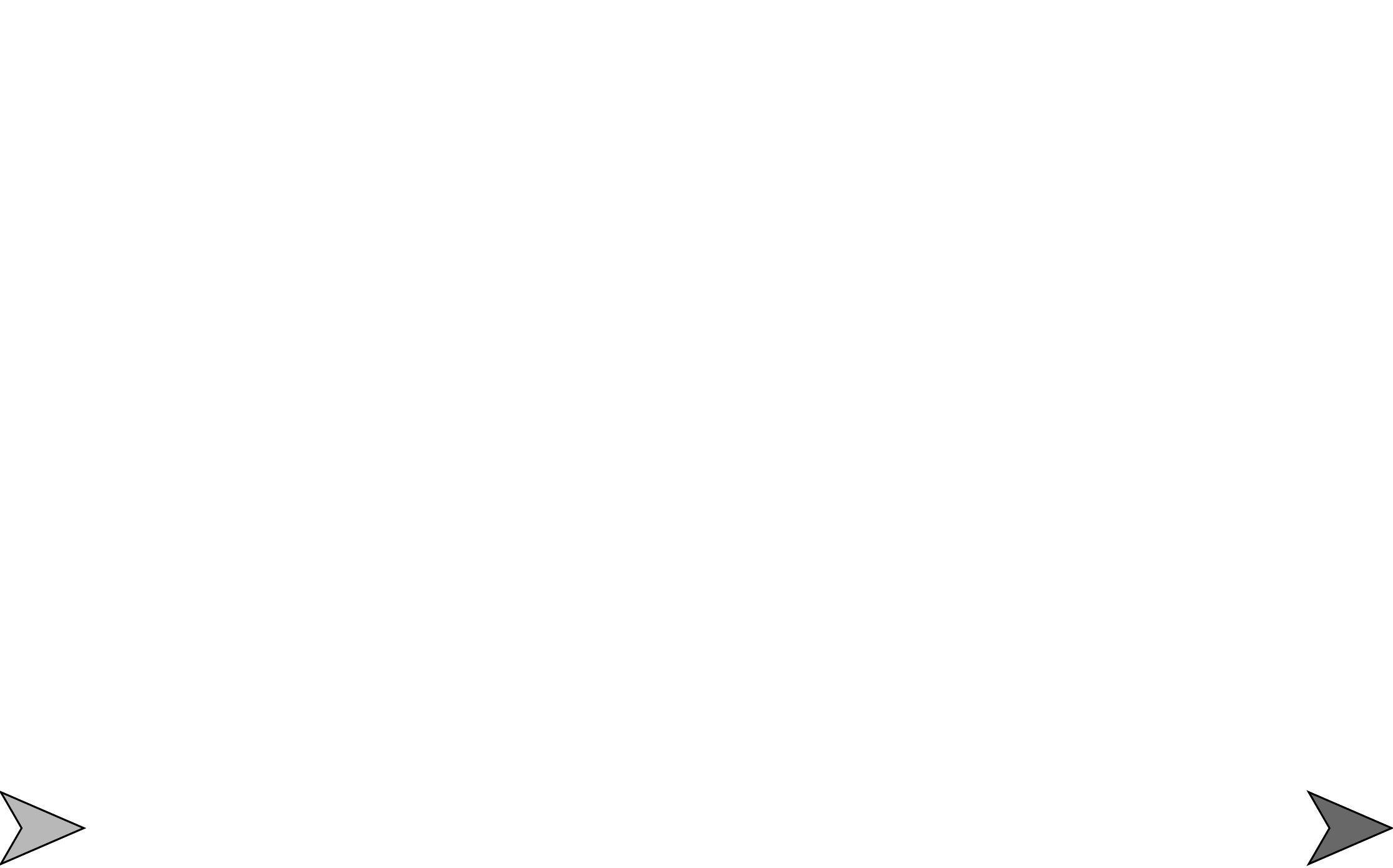}};
        \ifthenelse{\equal{#1}{}}{
\TIKZPointOfInterest{(-472.50000000*\st pt, -303.10889132*\st pt)}{a}{$(0,0)$}{NW}{100.0*\st pt}{Start}{(-1200.00000000*\st pt, -629.08965344*\st pt)}{8.0}{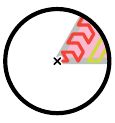}
\TIKZPointOfInterest{(-107.50000000*\st pt, 329.08965344*\st pt)}{b}{$(0,1-h)$}{NW}{50.0*\st pt}{}{(-600.00000000*\st pt, -629.08965344*\st pt)}{8.0}{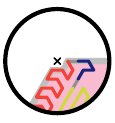}
\TIKZPointOfInterest{(-267.50000000*\st pt, -303.10889132*\st pt)}{c}{$(w-1,0)$}{SE}{50.0*\st pt}{}{(0.00000000*\st pt, -629.08965344*\st pt)}{8.0}{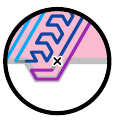}
\TIKZPointOfInterest{(112.50000000*\st pt, 329.08965344*\st pt)}{d}{$(w+2,1-h)$}{NE}{50.0*\st pt}{Read \word0}{(600.00000000*\st pt, -629.08965344*\st pt)}{8.0}{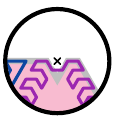}
\TIKZPointOfInterest{(487.50000000*\st pt, -303.10889132*\st pt)}{e}{$(W,0)$}{NE}{100.0*\st pt}{Next}{(1200.00000000*\st pt, -629.08965344*\st pt)}{8.0}{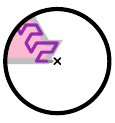}
            }{}
\node [anchor=center] () at (0,0) {\includegraphics[scale =\s]{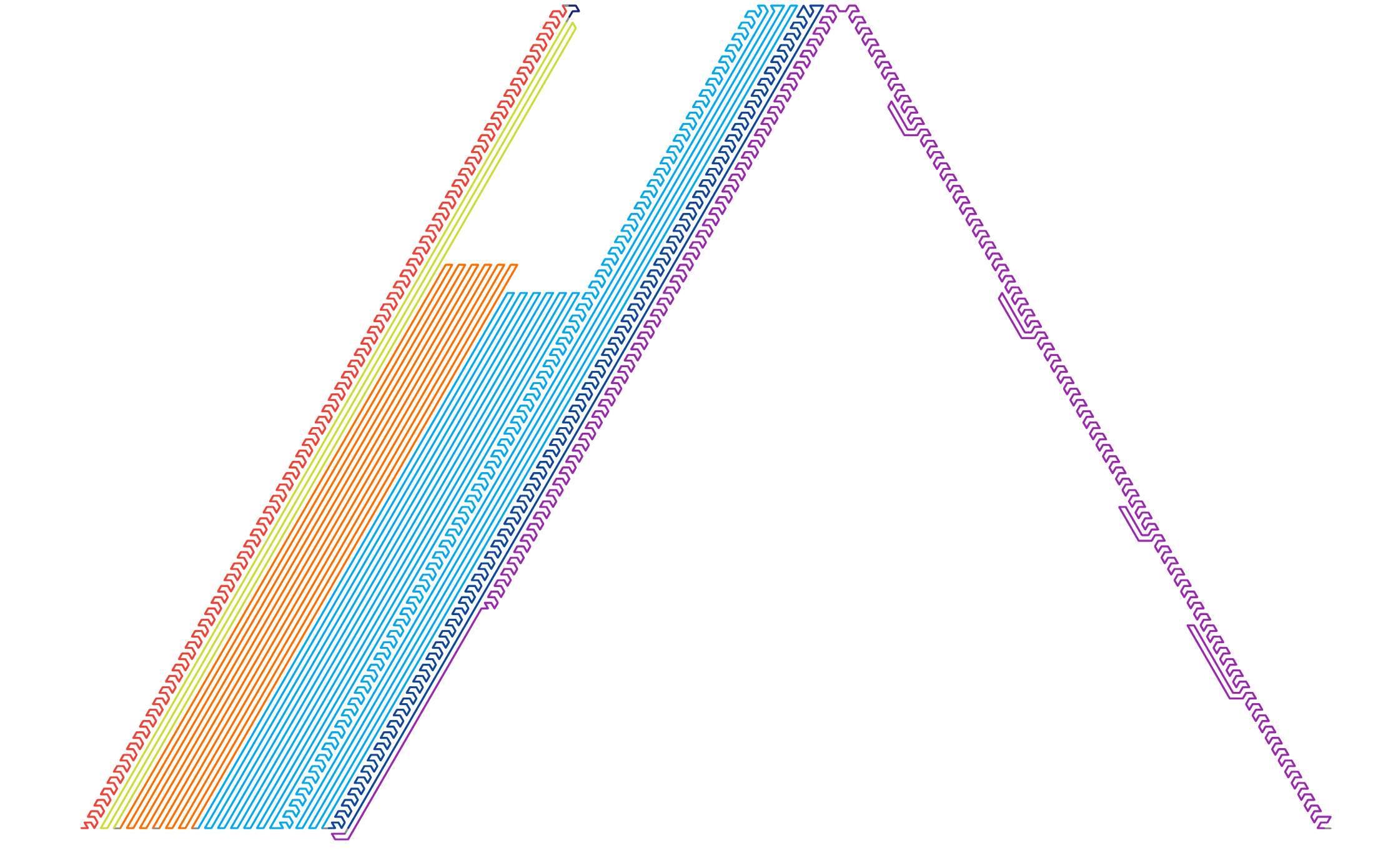}};
            \ifthenelse{\equal{#1}{}}{
\TIKZBLOCKREAD{0}{2}{11}{190.000}{12.990}
            }{}
    \end{tikzpicture}
}

%% file: OritatamiTuringPatterns/n=4-L=3-P=1/Read0_____n=4-L=3-P=1.tex
\newcommand{\ReadZBlockEpsilon}[1]{
    \begin{tikzpicture}
        \node [anchor=center] () at (0,0) {\includegraphics[scale =\s]{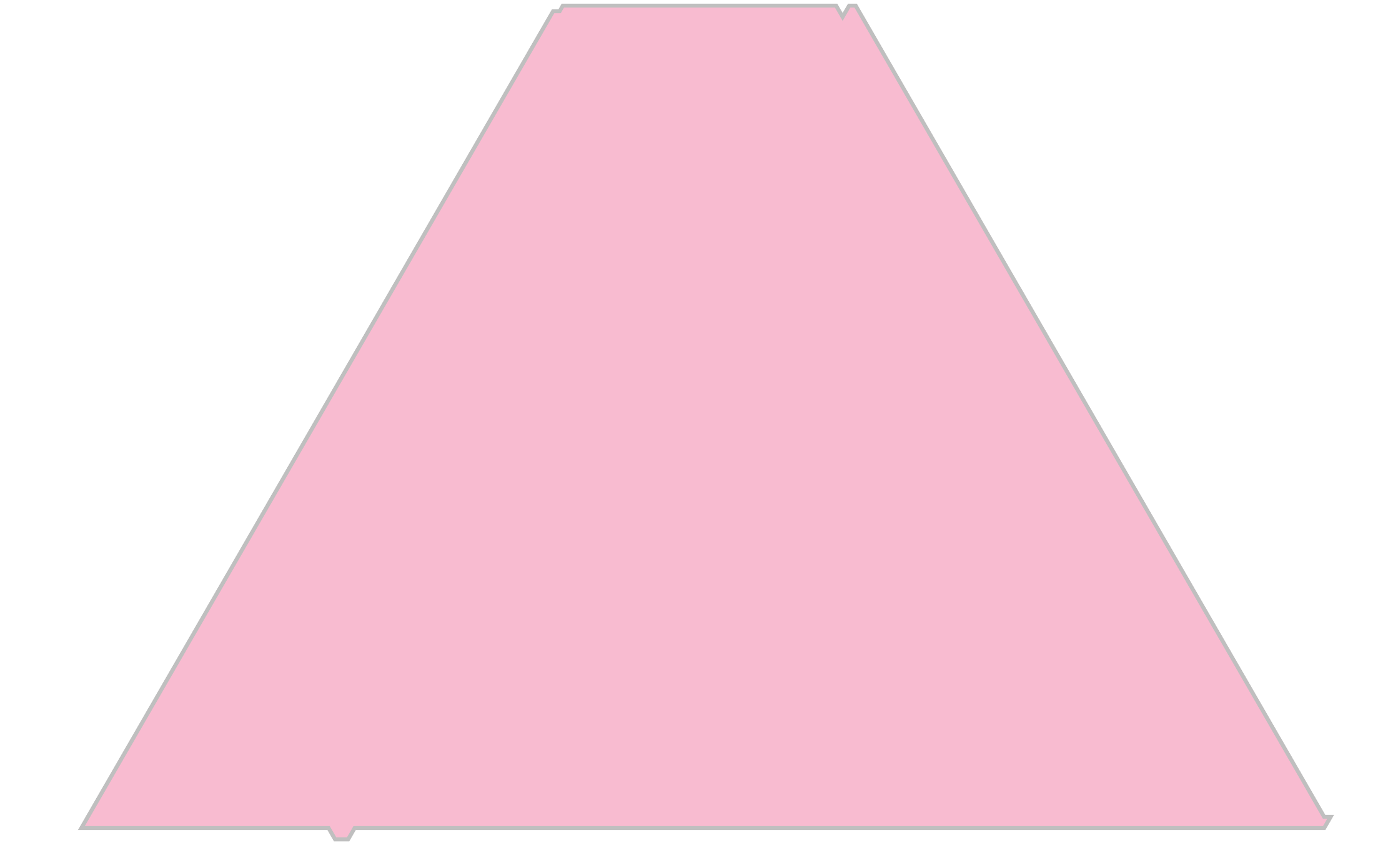}};
        \node [anchor=center] () at (0,0) {\includegraphics[scale =\s]{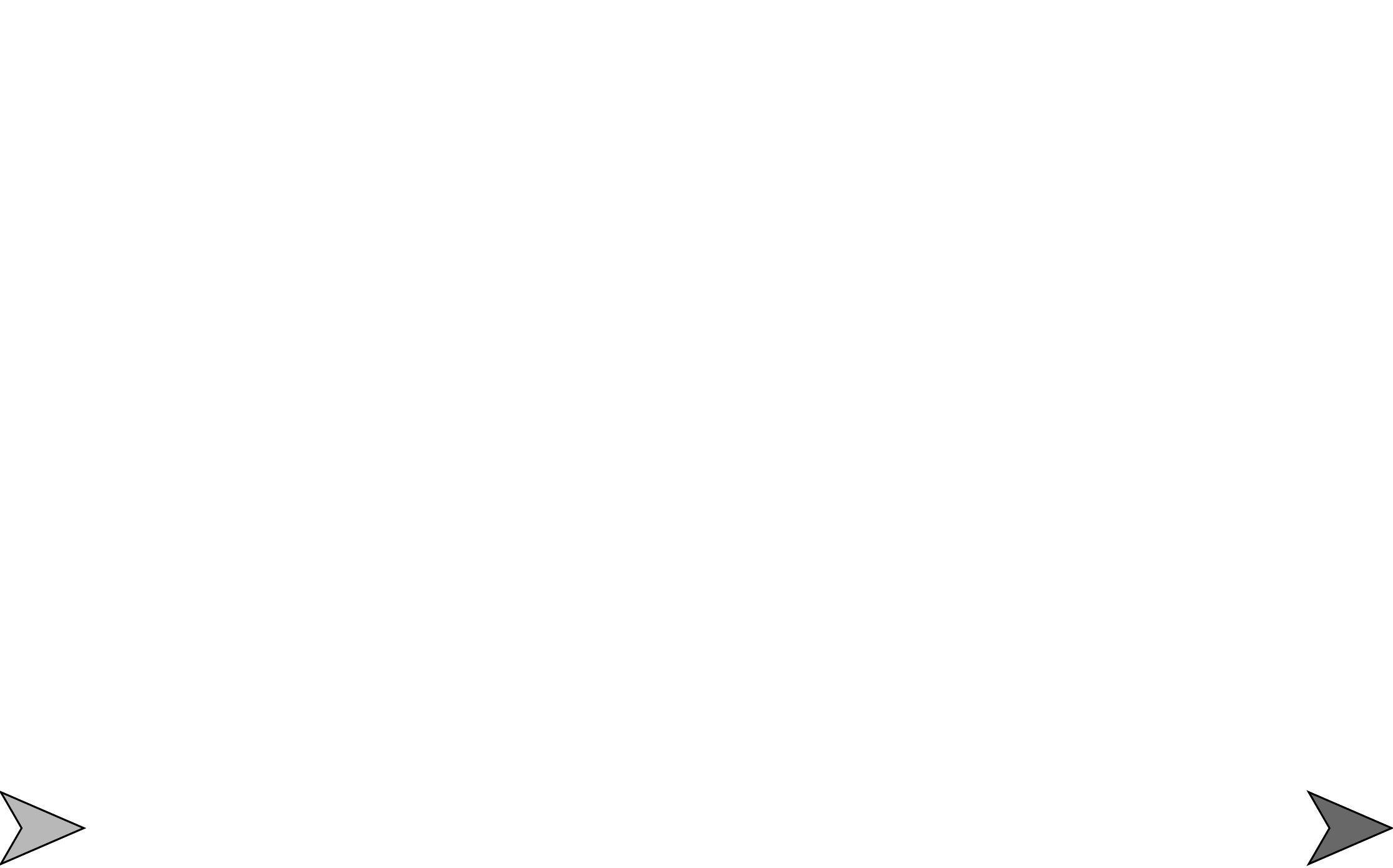}};
        \ifthenelse{\equal{#1}{}}{
\TIKZPointOfInterest{(-472.50000000*\st pt, -303.10889132*\st pt)}{a}{$(0,0)$}{NW}{100.0*\st pt}{Start}{(-1200.00000000*\st pt, -629.08965344*\st pt)}{8.0}{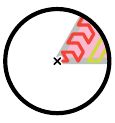}
\TIKZPointOfInterest{(-107.50000000*\st pt, 329.08965344*\st pt)}{b}{$(0,1-h)$}{NW}{50.0*\st pt}{}{(-600.00000000*\st pt, -629.08965344*\st pt)}{8.0}{Read0____n=4-L=3-P=1-PI-B-_0,-146_-top-left.pdf}
\TIKZPointOfInterest{(-267.50000000*\st pt, -303.10889132*\st pt)}{c}{$(w-1,0)$}{SE}{50.0*\st pt}{}{(0.00000000*\st pt, -629.08965344*\st pt)}{8.0}{Read0____n=4-L=3-P=1-PI-C-_41,0_-w0.pdf}
\TIKZPointOfInterest{(112.50000000*\st pt, 329.08965344*\st pt)}{d}{$(w+2,1-h)$}{NE}{50.0*\st pt}{Read \word0}{(600.00000000*\st pt, -629.08965344*\st pt)}{8.0}{Read0____n=4-L=3-P=1-PI-D-_44,-146_-read0.pdf}
\TIKZPointOfInterest{(487.50000000*\st pt, -303.10889132*\st pt)}{e}{$(W,0)$}{NE}{100.0*\st pt}{Next}{(1200.00000000*\st pt, -629.08965344*\st pt)}{8.0}{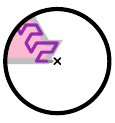}
            }{}
\node [anchor=center] () at (0,0) {\includegraphics[scale =\s]{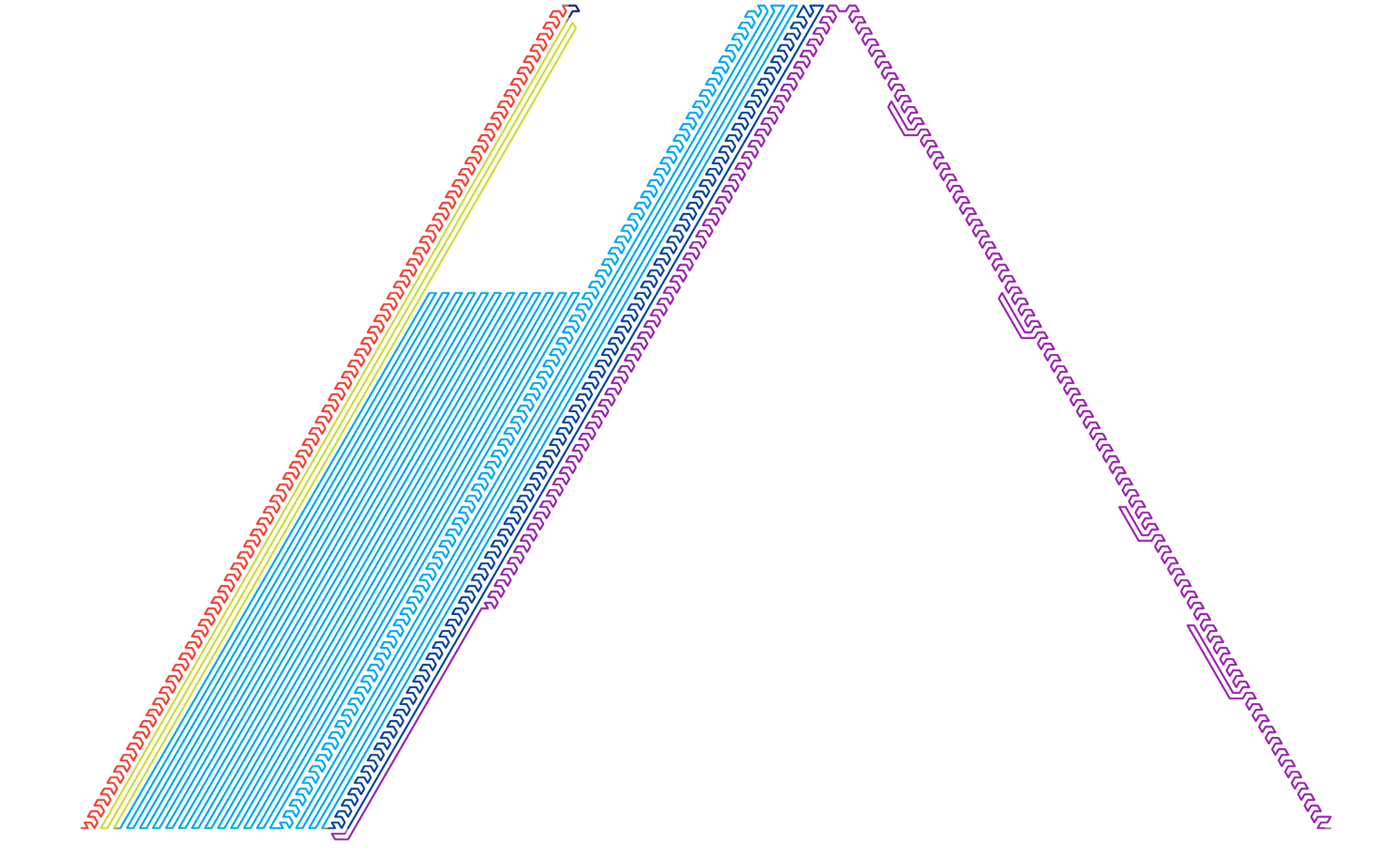}};
            \ifthenelse{\equal{#1}{}}{
\TIKZBLOCKREAD{0}{1}{\ensuremath{\epsilon}}{190.000}{12.990}
            }{}
    \end{tikzpicture}
}

%% file: OritatamiTuringPatterns/n=4-L=3-P=1/Read1_0__n=4-L=3-P=1.tex
\newcommand{\ReadUBlockZ}[1]{
    \begin{tikzpicture}
        \node [anchor=center] () at (0,0) {\includegraphics[scale =\s]{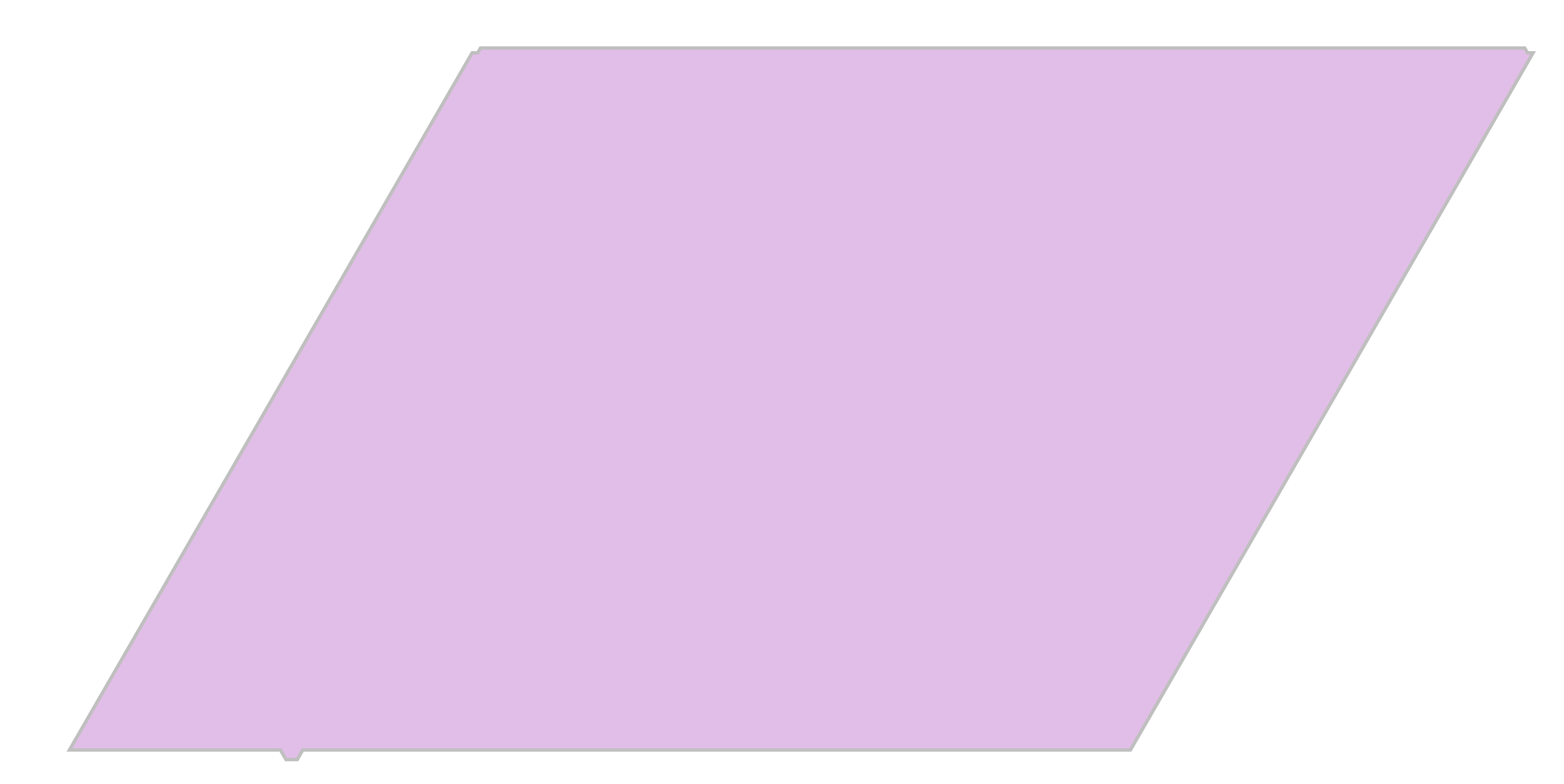}};
        \node [anchor=center] () at (0,0) {\includegraphics[scale =\s]{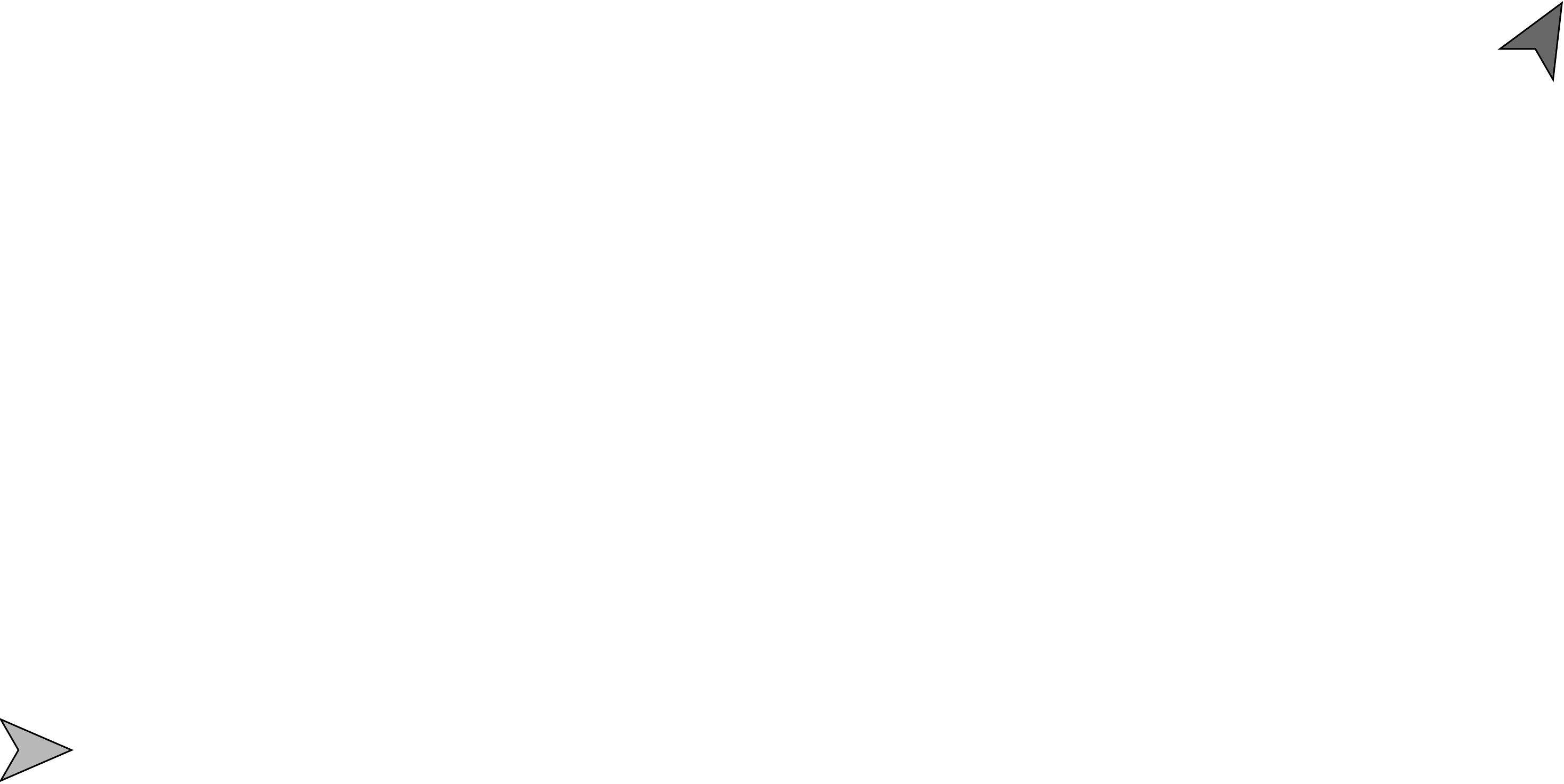}};
        \ifthenelse{\equal{#1}{}}{
\TIKZPointOfInterest{(-641.25000000*\st pt, -322.59446291*\st pt)}{a}{$(0,0)$}{NW}{100.0*\st pt}{Start}{(-1200.00000000*\st pt, -648.57522502*\st pt)}{8.0}{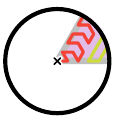}
\TIKZPointOfInterest{(-276.25000000*\st pt, 309.60408185*\st pt)}{b}{$(0,1-h)$}{NW}{50.0*\st pt}{}{(-600.00000000*\st pt, -648.57522502*\st pt)}{8.0}{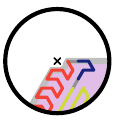}
\TIKZPointOfInterest{(-436.25000000*\st pt, -322.59446291*\st pt)}{c}{$(w-1,0)$}{SE}{50.0*\st pt}{}{(0.00000000*\st pt, -648.57522502*\st pt)}{8.0}{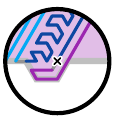}
\TIKZPointOfInterest{(-56.25000000*\st pt, 309.60408185*\st pt)}{d}{$(w+2,1-h)$}{NE}{50.0*\st pt}{Read \word1}{(600.00000000*\st pt, -648.57522502*\st pt)}{8.0}{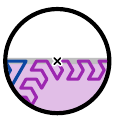}
\TIKZPointOfInterest{(678.75000000*\st pt, 309.60408185*\st pt)}{e}{$(W-1,1-h)$}{SE}{100.0*\st pt}{Next}{(1200.00000000*\st pt, -648.57522502*\st pt)}{8.0}{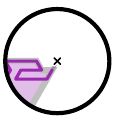}
            }{}
\node [anchor=center] () at (0,0) {\includegraphics[scale =\s]{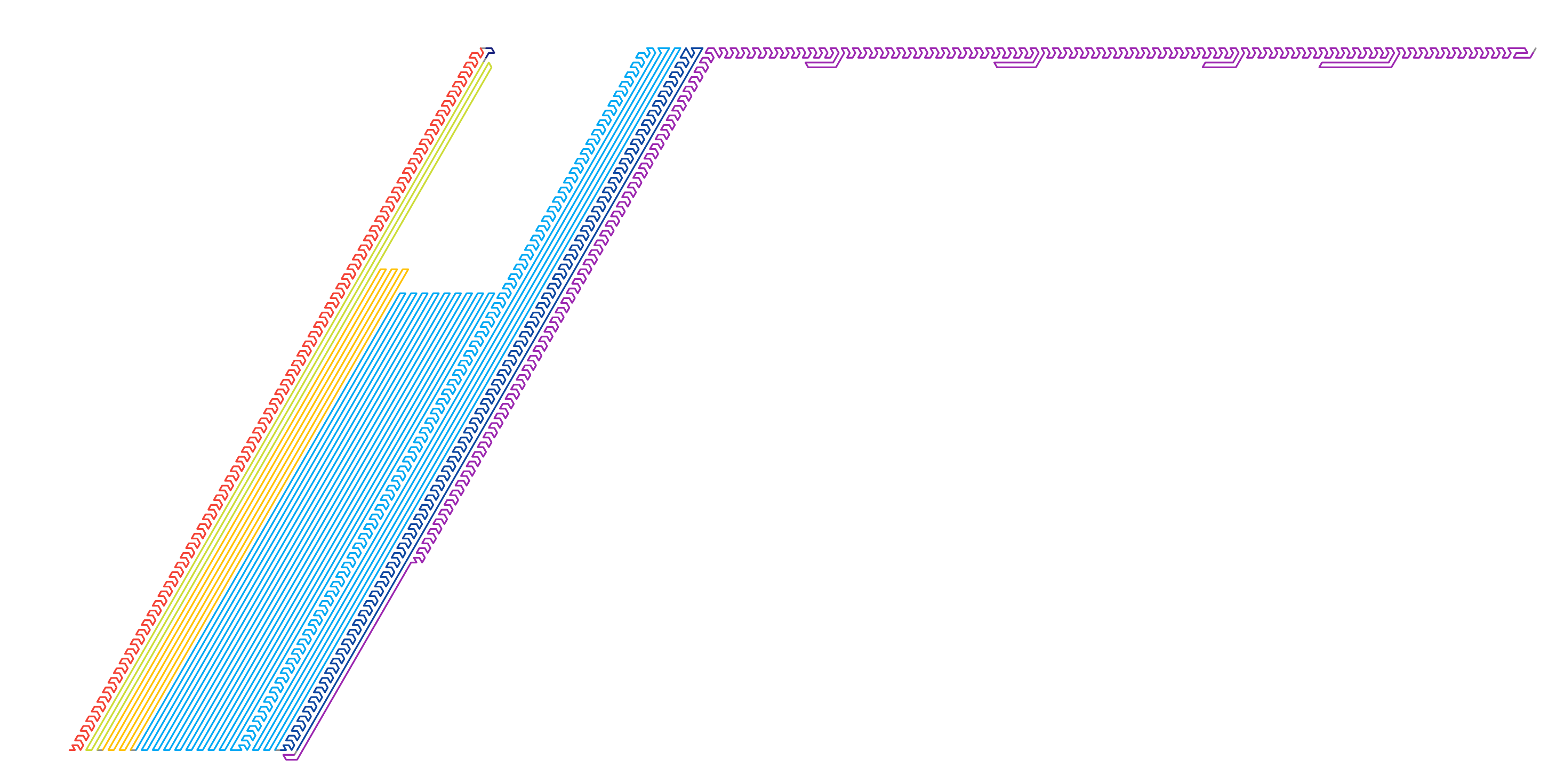}};
            \ifthenelse{\equal{#1}{}}{
\TIKZBLOCKREAD{1}{3}{0}{21.250}{-6.495}
            }{}
    \end{tikzpicture}
}

%% file: OritatamiTuringPatterns/n=4-L=3-P=1/Read1_110__n=4-L=3-P=1.tex
\newcommand{\ReadUBlockUUZ}[1]{
    \begin{tikzpicture}
        \node [anchor=center] () at (0,0) {\includegraphics[scale =\s]{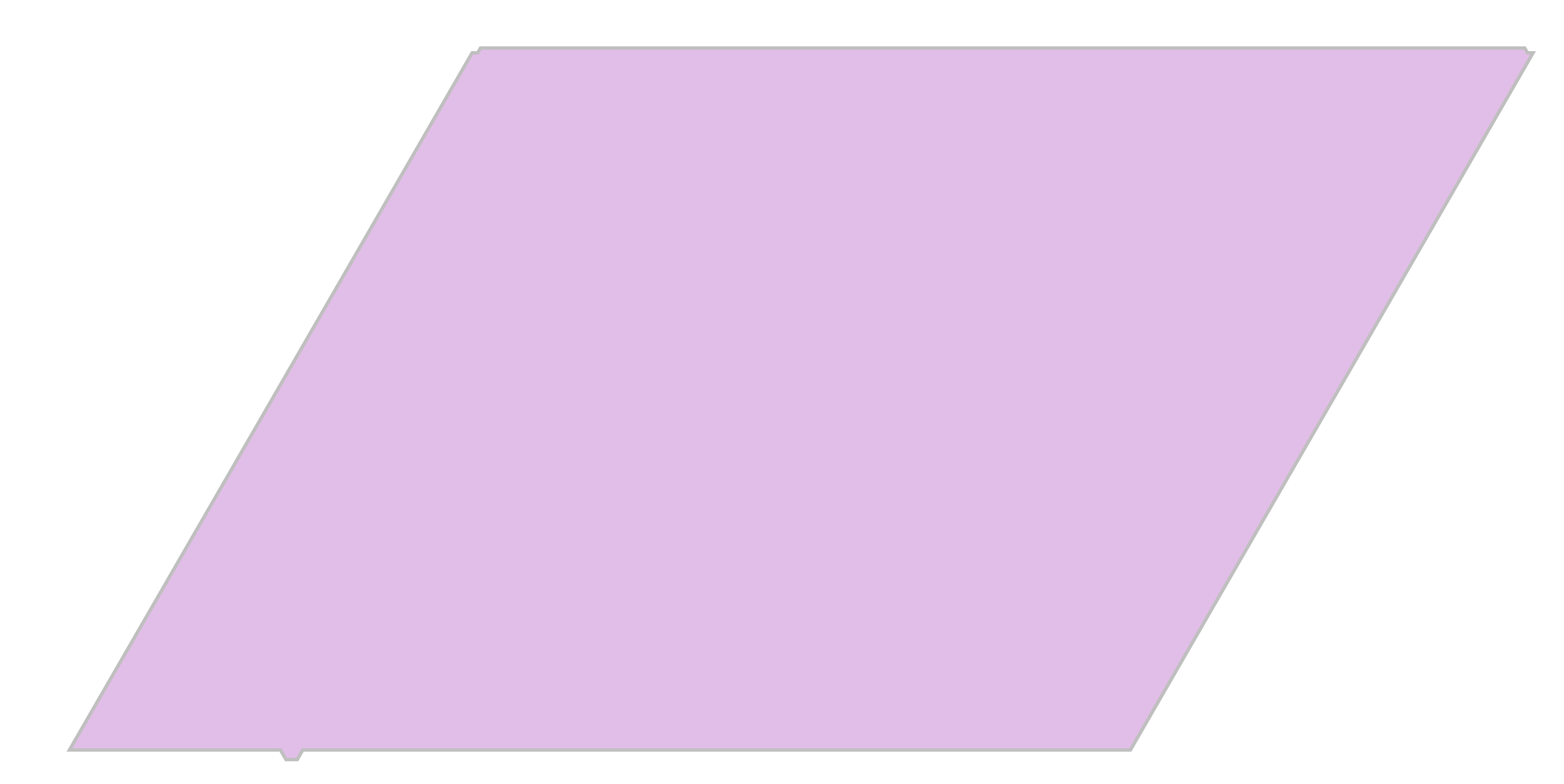}};
        \node [anchor=center] () at (0,0) {\includegraphics[scale =\s]{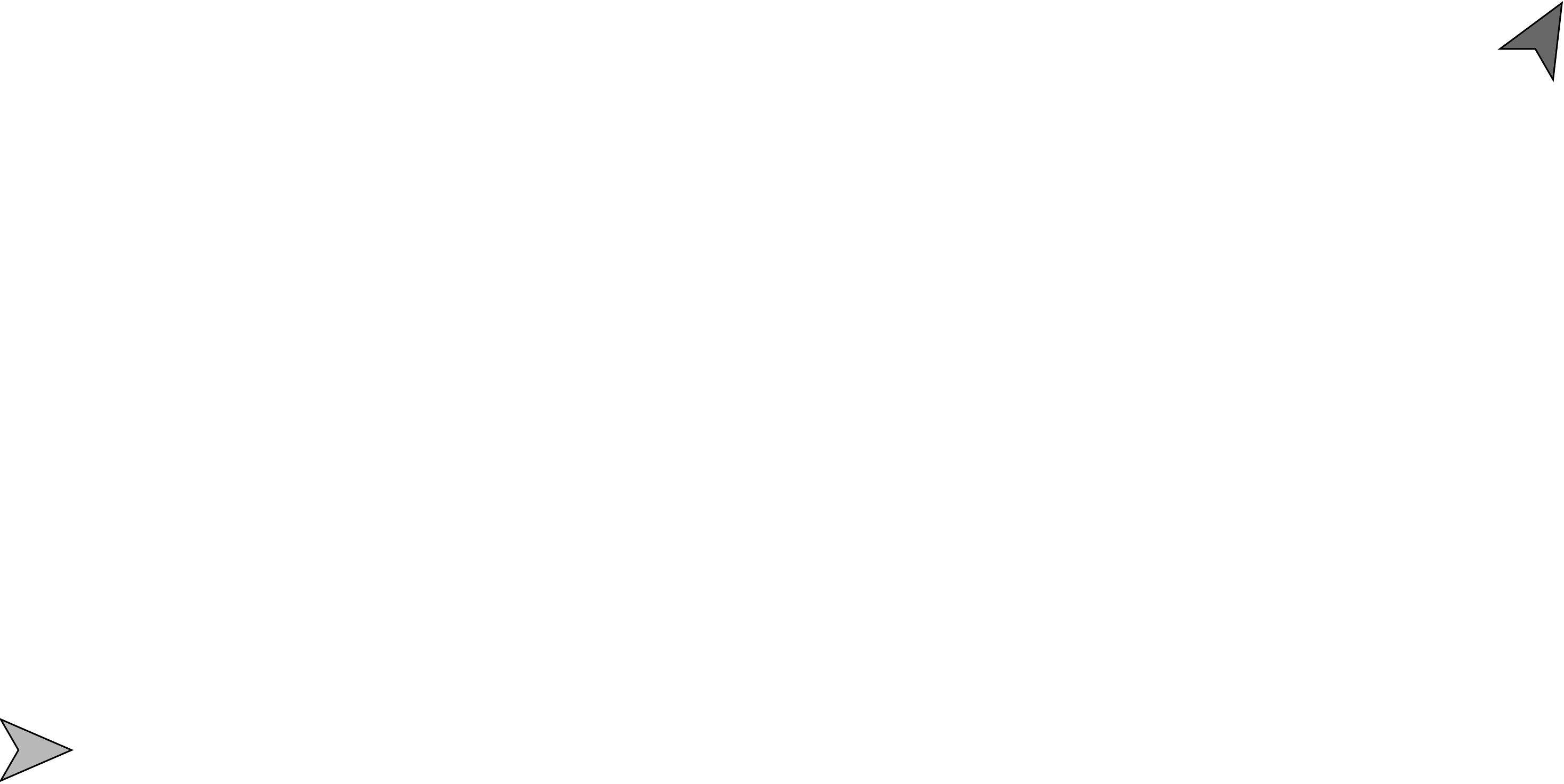}};
        \ifthenelse{\equal{#1}{}}{
\TIKZPointOfInterest{(-641.25000000*\st pt, -322.59446291*\st pt)}{a}{$(0,0)$}{NW}{100.0*\st pt}{Start}{(-1200.00000000*\st pt, -648.57522502*\st pt)}{8.0}{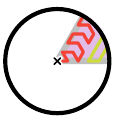}
\TIKZPointOfInterest{(-276.25000000*\st pt, 309.60408185*\st pt)}{b}{$(0,1-h)$}{NW}{50.0*\st pt}{}{(-600.00000000*\st pt, -648.57522502*\st pt)}{8.0}{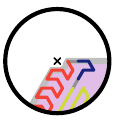}
\TIKZPointOfInterest{(-436.25000000*\st pt, -322.59446291*\st pt)}{c}{$(w-1,0)$}{SE}{50.0*\st pt}{}{(0.00000000*\st pt, -648.57522502*\st pt)}{8.0}{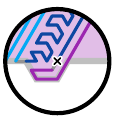}
\TIKZPointOfInterest{(-56.25000000*\st pt, 309.60408185*\st pt)}{d}{$(w+2,1-h)$}{NE}{50.0*\st pt}{Read \word1}{(600.00000000*\st pt, -648.57522502*\st pt)}{8.0}{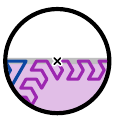}
\TIKZPointOfInterest{(678.75000000*\st pt, 309.60408185*\st pt)}{e}{$(W-1,1-h)$}{SE}{100.0*\st pt}{Next}{(1200.00000000*\st pt, -648.57522502*\st pt)}{8.0}{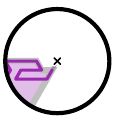}
            }{}
\node [anchor=center] () at (0,0) {\includegraphics[scale =\s]{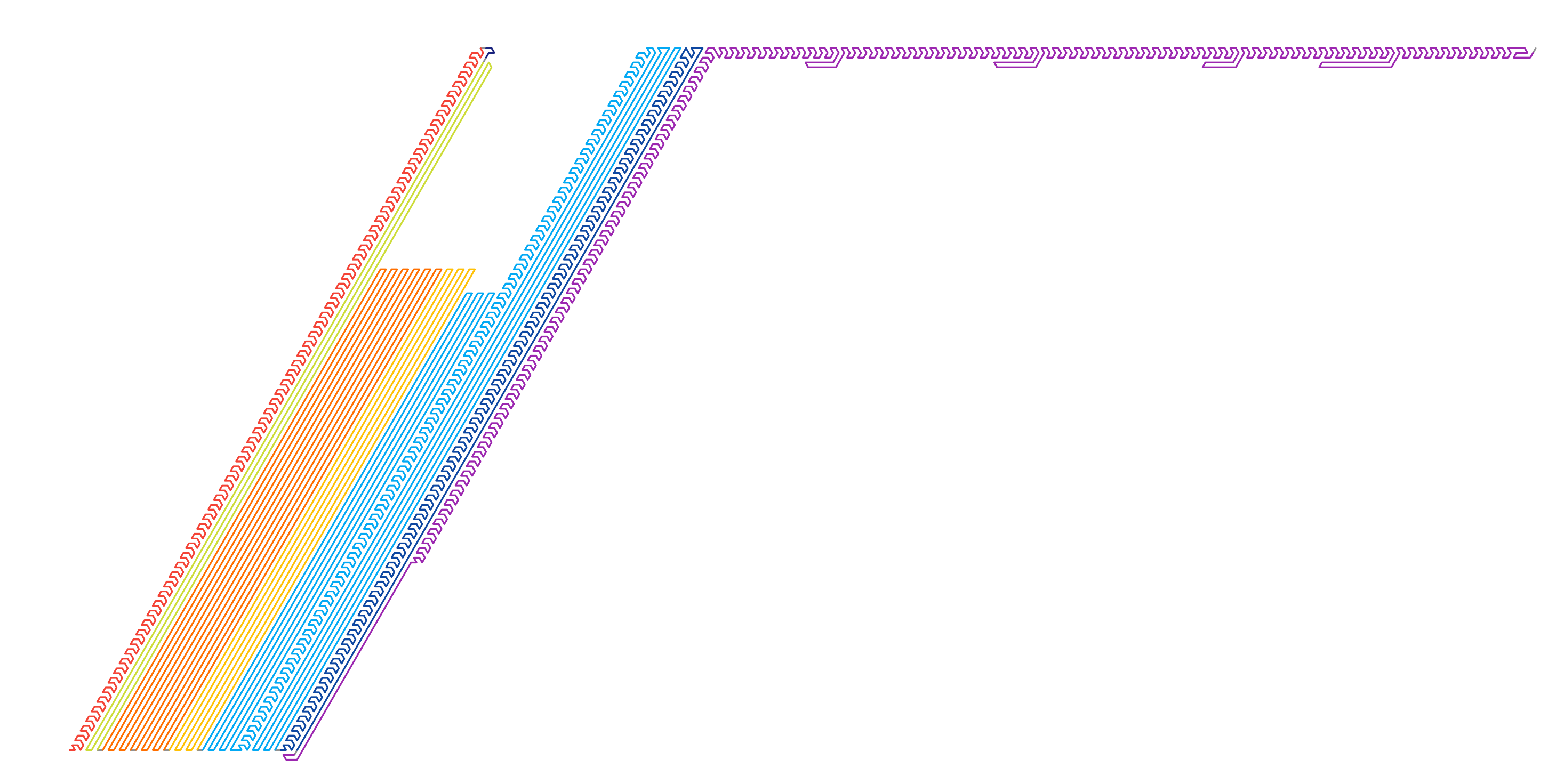}};
            \ifthenelse{\equal{#1}{}}{
\TIKZBLOCKREAD{1}{0}{110}{21.250}{-6.495}
            }{}
    \end{tikzpicture}
}

%% file: OritatamiTuringPatterns/n=4-L=3-P=1/Read1_11__n=4-L=3-P=1.tex
\newcommand{\ReadUBlockUU}[1]{
    \begin{tikzpicture}
        \node [anchor=center] () at (0,0) {\includegraphics[scale =\s]{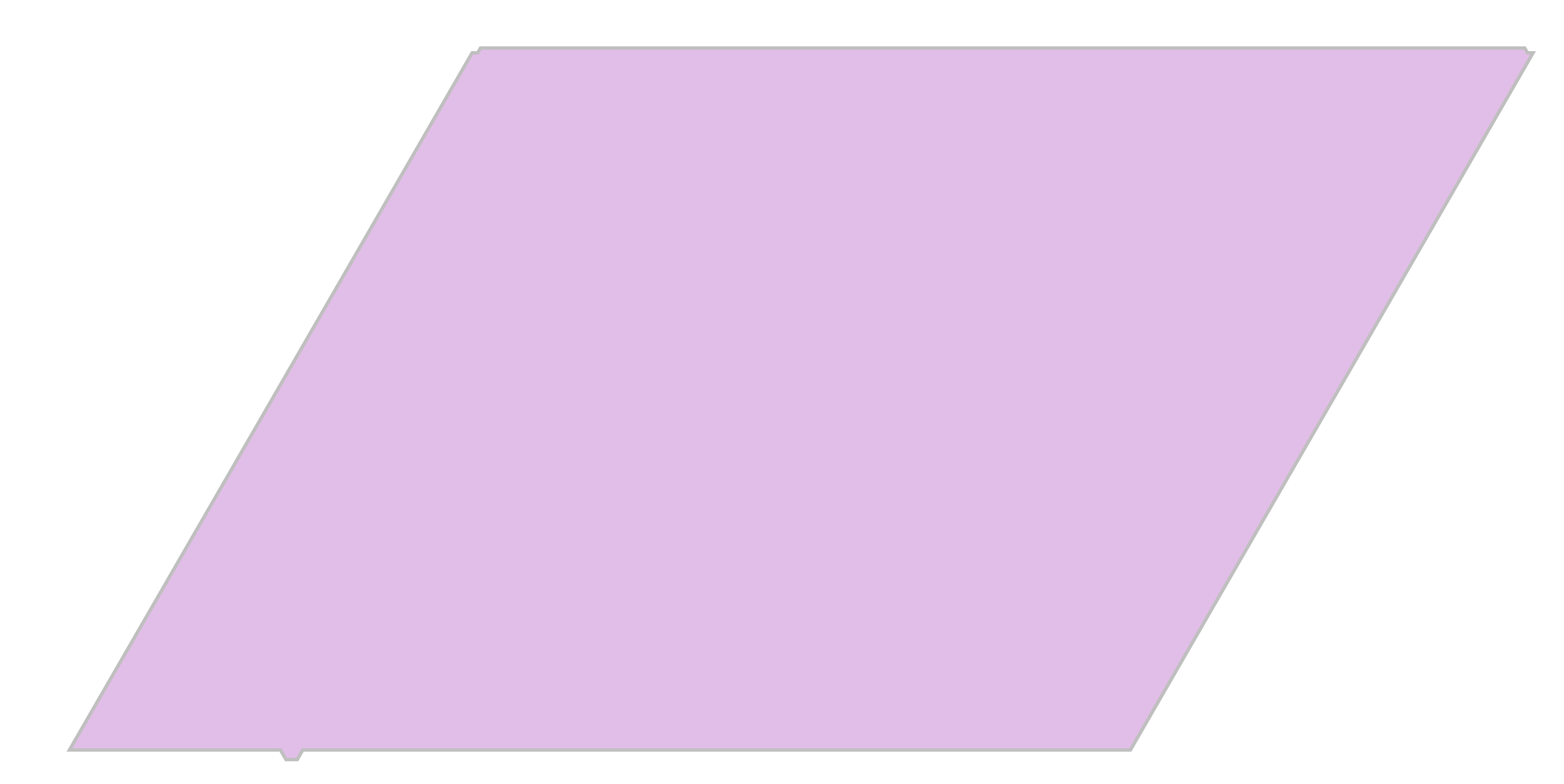}};
        \node [anchor=center] () at (0,0) {\includegraphics[scale =\s]{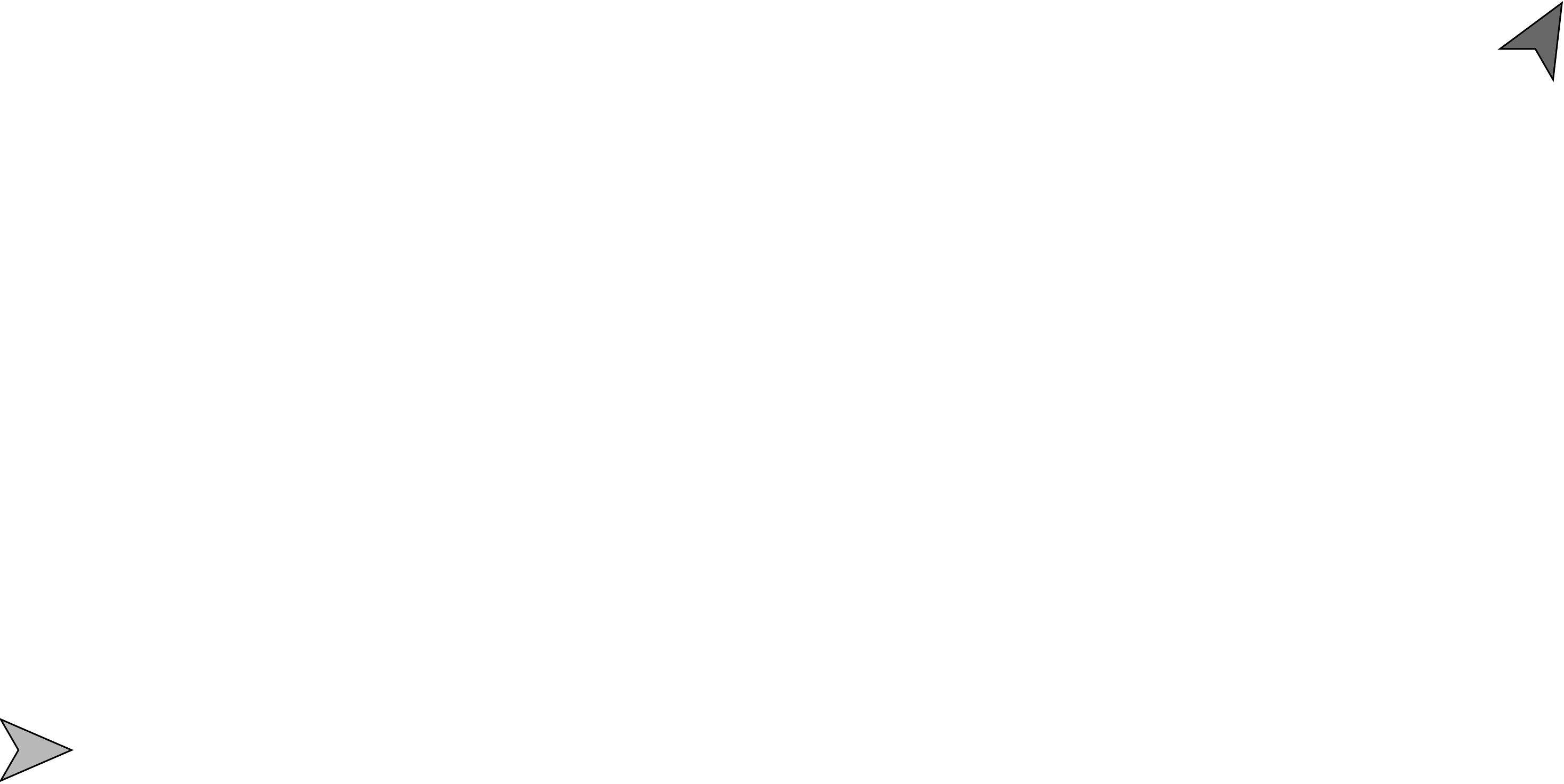}};
        \ifthenelse{\equal{#1}{}}{
\TIKZPointOfInterest{(-641.25000000*\st pt, -322.59446291*\st pt)}{a}{$(0,0)$}{NW}{100.0*\st pt}{Start}{(-1200.00000000*\st pt, -648.57522502*\st pt)}{8.0}{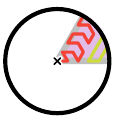}
\TIKZPointOfInterest{(-276.25000000*\st pt, 309.60408185*\st pt)}{b}{$(0,1-h)$}{NW}{50.0*\st pt}{}{(-600.00000000*\st pt, -648.57522502*\st pt)}{8.0}{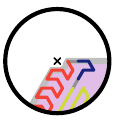}
\TIKZPointOfInterest{(-436.25000000*\st pt, -322.59446291*\st pt)}{c}{$(w-1,0)$}{SE}{50.0*\st pt}{}{(0.00000000*\st pt, -648.57522502*\st pt)}{8.0}{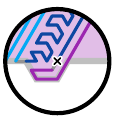}
\TIKZPointOfInterest{(-56.25000000*\st pt, 309.60408185*\st pt)}{d}{$(w+2,1-h)$}{NE}{50.0*\st pt}{Read \word1}{(600.00000000*\st pt, -648.57522502*\st pt)}{8.0}{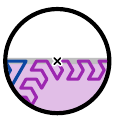}
\TIKZPointOfInterest{(678.75000000*\st pt, 309.60408185*\st pt)}{e}{$(W-1,1-h)$}{SE}{100.0*\st pt}{Next}{(1200.00000000*\st pt, -648.57522502*\st pt)}{8.0}{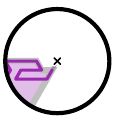}
            }{}
\node [anchor=center] () at (0,0) {\includegraphics[scale =\s]{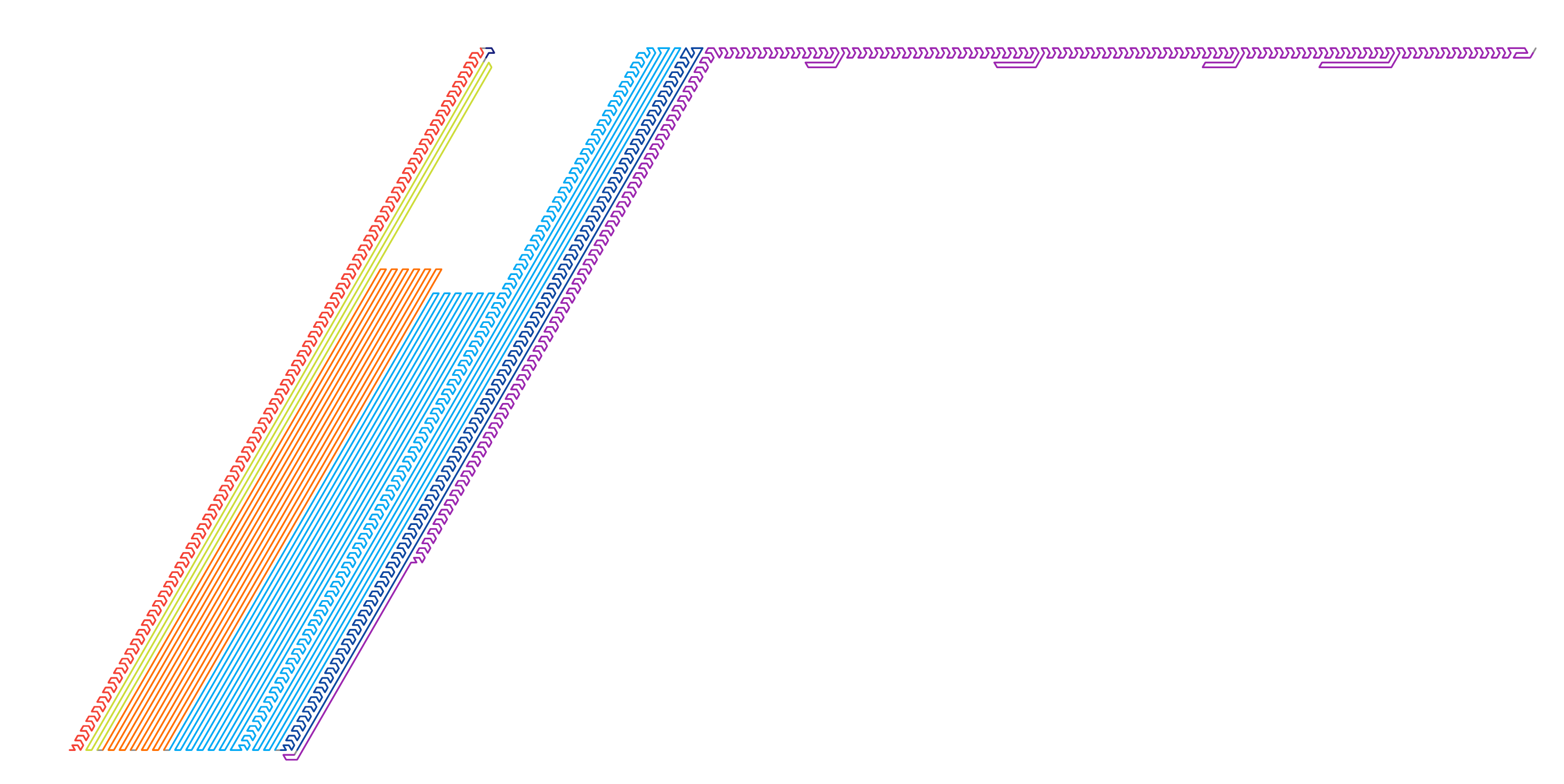}};
            \ifthenelse{\equal{#1}{}}{
\TIKZBLOCKREAD{1}{2}{11}{21.250}{-6.495}
            }{}
    \end{tikzpicture}
}

%% file: OritatamiTuringPatterns/n=4-L=3-P=1/Read1_____n=4-L=3-P=1.tex
\newcommand{\ReadUBlockEpsilon}[1]{
    \begin{tikzpicture}
        \node [anchor=center] () at (0,0) {\includegraphics[scale =\s]{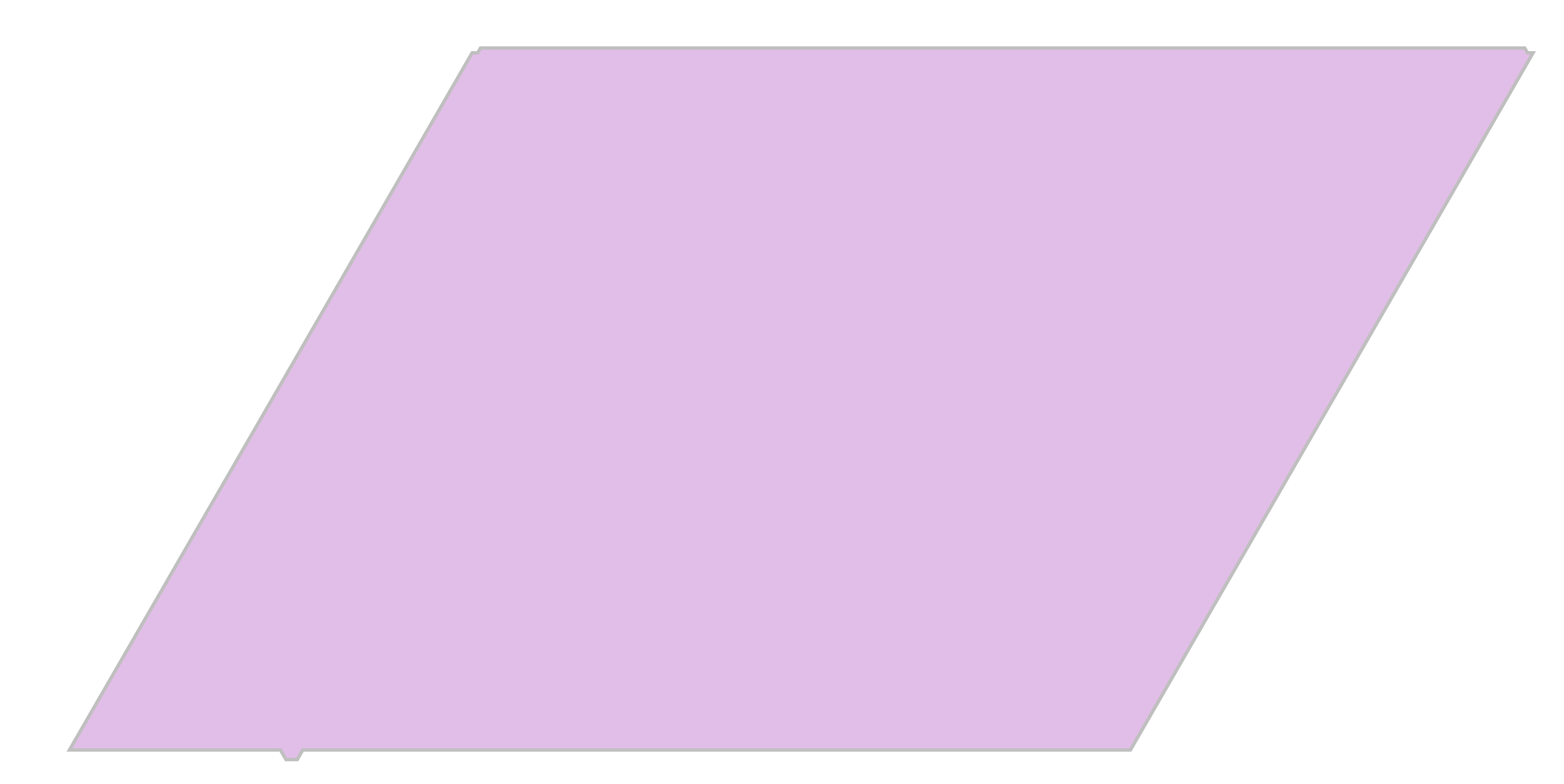}};
        \node [anchor=center] () at (0,0) {\includegraphics[scale =\s]{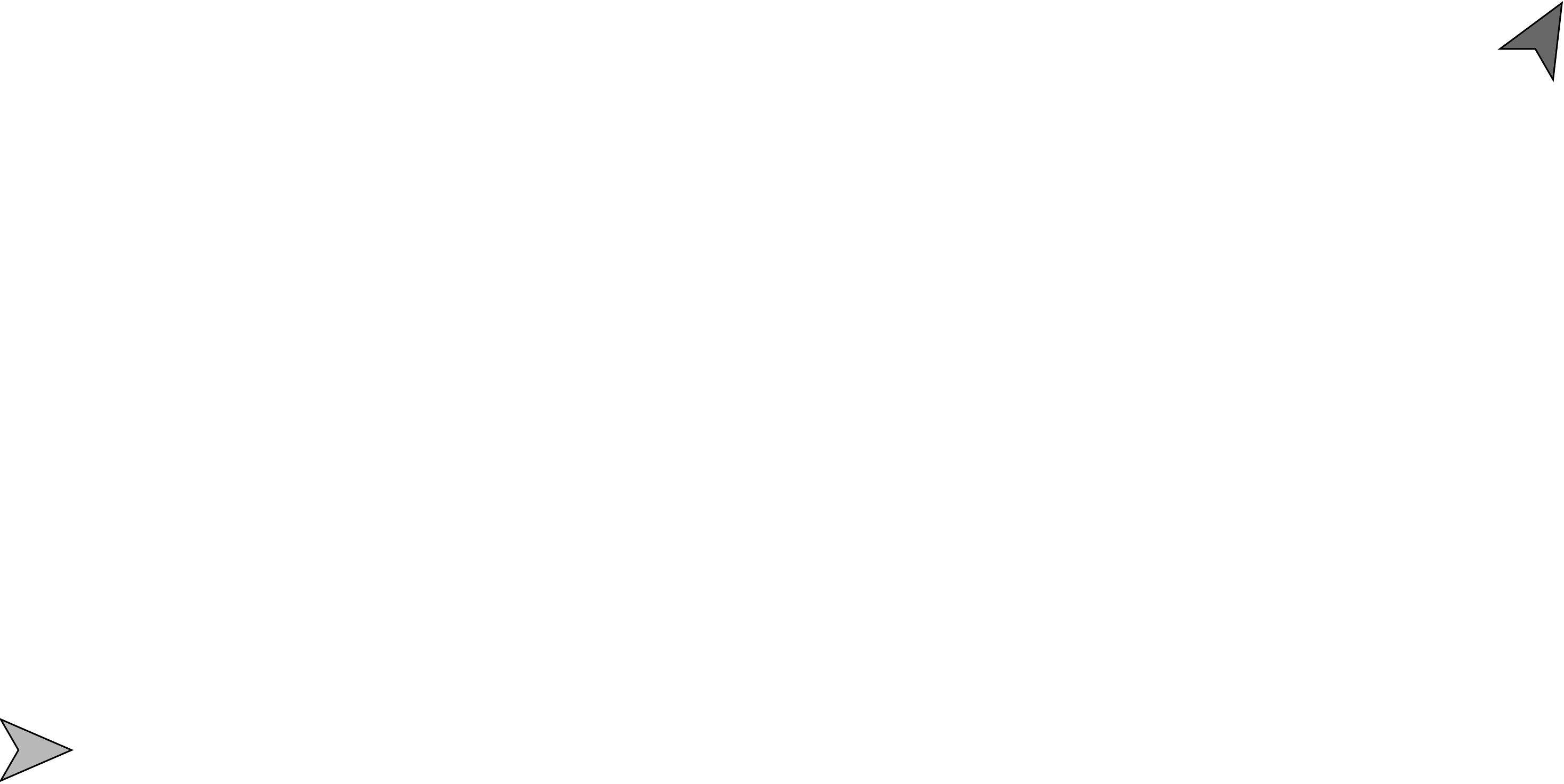}};
        \ifthenelse{\equal{#1}{}}{
\TIKZPointOfInterest{(-641.25000000*\st pt, -322.59446291*\st pt)}{a}{$(0,0)$}{NW}{100.0*\st pt}{Start}{(-1200.00000000*\st pt, -648.57522502*\st pt)}{8.0}{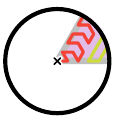}
\TIKZPointOfInterest{(-276.25000000*\st pt, 309.60408185*\st pt)}{b}{$(0,1-h)$}{NW}{50.0*\st pt}{}{(-600.00000000*\st pt, -648.57522502*\st pt)}{8.0}{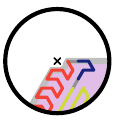}
\TIKZPointOfInterest{(-436.25000000*\st pt, -322.59446291*\st pt)}{c}{$(w-1,0)$}{SE}{50.0*\st pt}{}{(0.00000000*\st pt, -648.57522502*\st pt)}{8.0}{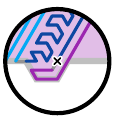}
\TIKZPointOfInterest{(-56.25000000*\st pt, 309.60408185*\st pt)}{d}{$(w+2,1-h)$}{NE}{50.0*\st pt}{Read \word1}{(600.00000000*\st pt, -648.57522502*\st pt)}{8.0}{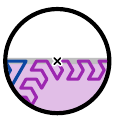}
\TIKZPointOfInterest{(678.75000000*\st pt, 309.60408185*\st pt)}{e}{$(W-1,1-h)$}{SE}{100.0*\st pt}{Next}{(1200.00000000*\st pt, -648.57522502*\st pt)}{8.0}{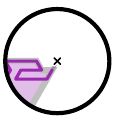}
            }{}
\node [anchor=center] () at (0,0) {\includegraphics[scale =\s]{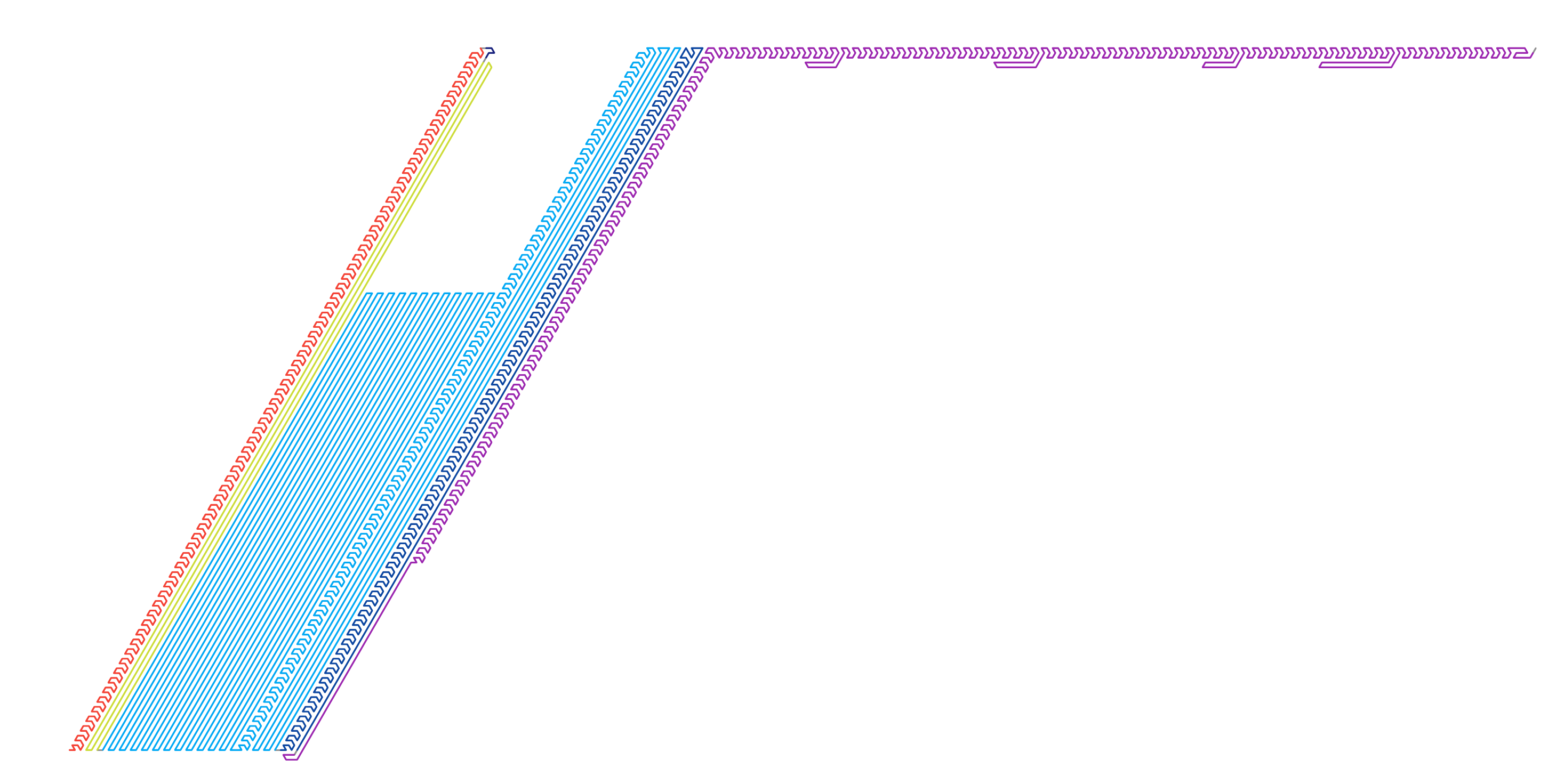}};
            \ifthenelse{\equal{#1}{}}{
\TIKZBLOCKREAD{1}{1}{\ensuremath{\epsilon}}{21.250}{-6.495}
            }{}
    \end{tikzpicture}
}

%% file: OritatamiTuringPatterns/n=4-L=3-P=1/Seed010_n=4-L=3-P=1.tex
\newcommand{\SeedZUZ}[1]{
    \begin{tikzpicture}
        \node [anchor=center] () at (0,0) {\includegraphics[scale =\s]{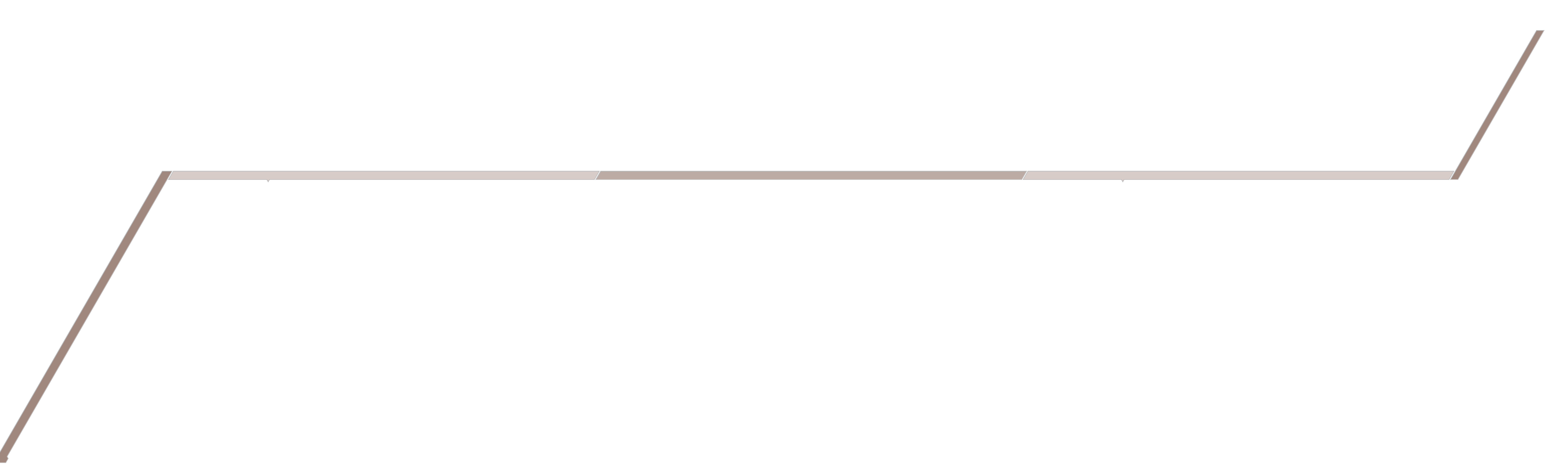}};
        \node [anchor=center] () at (0,0) {\includegraphics[scale =\s]{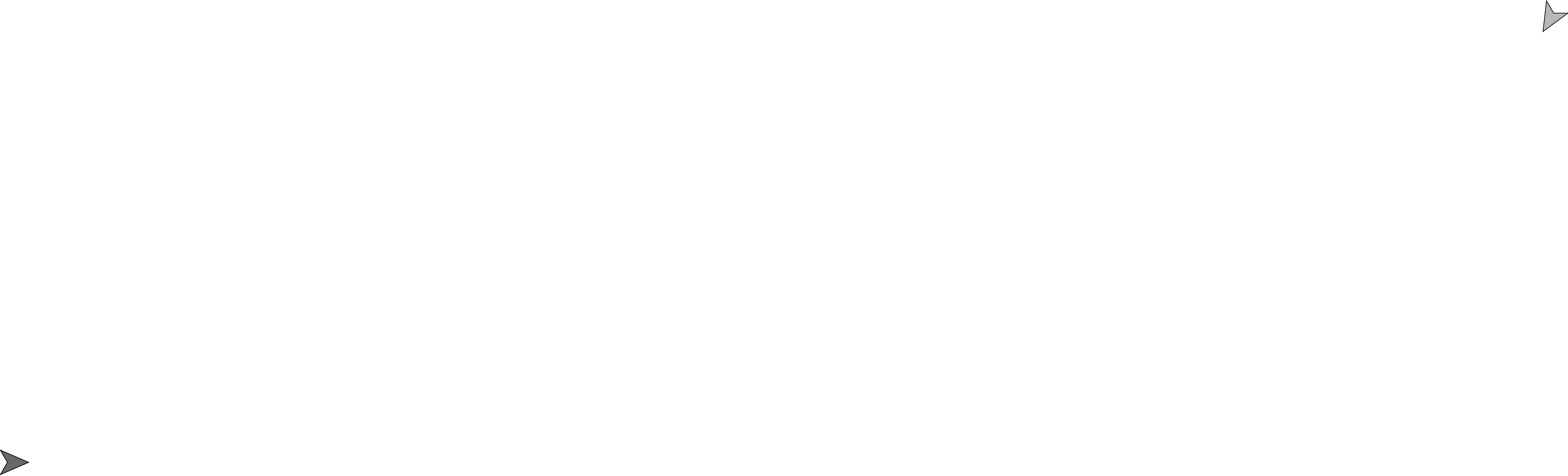}};
        \ifthenelse{\equal{#1}{}}{
\TIKZPointOfInterest{(1706.25000000*\st pt, 465.48865453*\st pt)}{a}{$\left(2+|u|W,-\cfrac{3h+7}2\right)$}{NE}{100.0*\st pt}{Start}{(-1800.00000000*\st pt, -830.44055982*\st pt)}{8.0}{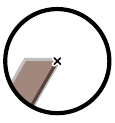}
\TIKZPointOfInterest{(1513.75000000*\st pt, 132.06887408*\st pt)}{b}{$(2+|u|W,-h)$}{SE}{100.0*\st pt}{}{(-1200.00000000*\st pt, -830.44055982*\st pt)}{8.0}{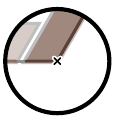}
\TIKZPointOfInterest{(761.25000000*\st pt, 127.73874706*\st pt)}{c}{$(w+2+iW,1-h)$}{SE}{100.0*\st pt}{Letter \word0}{(-600.00000000*\st pt, -830.44055982*\st pt)}{8.0}{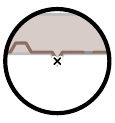}
\TIKZPointOfInterest{(-196.25000000*\st pt, 132.06887408*\st pt)}{d}{$(w+2+iW,-h)$}{SE}{100.0*\st pt}{Letter \word1}{(0.00000000*\st pt, -830.44055982*\st pt)}{8.0}{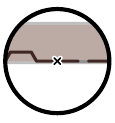}
\TIKZPointOfInterest{(-1158.75000000*\st pt, 127.73874706*\st pt)}{e}{$(w+2+iW,1-h)$}{SE}{100.0*\st pt}{Letter \word0}{(600.00000000*\st pt, -830.44055982*\st pt)}{8.0}{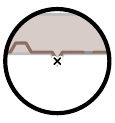}
\TIKZPointOfInterest{(-1386.25000000*\st pt, 132.06887408*\st pt)}{f}{$(-2,-h)$}{NW}{100.0*\st pt}{}{(1200.00000000*\st pt, -830.44055982*\st pt)}{8.0}{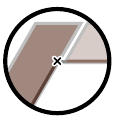}
\TIKZPointOfInterest{(-1743.75000000*\st pt, -504.45979770*\st pt)}{g}{$(0,0)$}{NE}{50.0*\st pt}{Next}{(1800.00000000*\st pt, -830.44055982*\st pt)}{8.0}{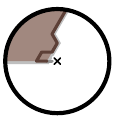}
            }{}
\node [anchor=center] () at (0,0) {\includegraphics[scale =\s]{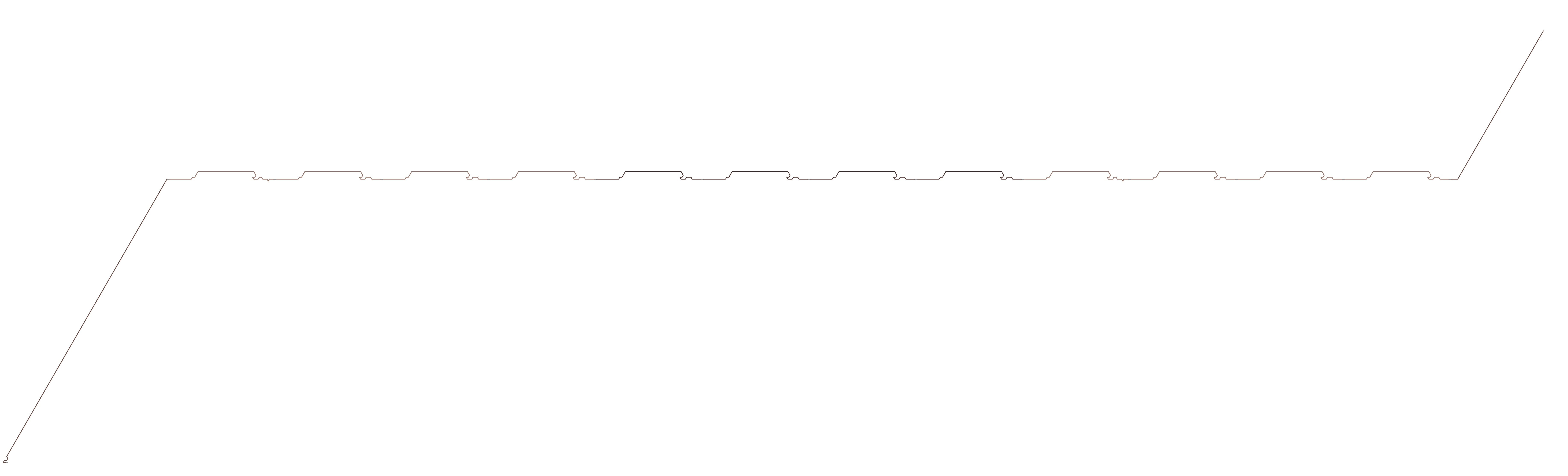}};
            \ifthenelse{\equal{#1}{}}{
\TIKZBLOCKSEED{010}{229.500}{401.836}
\TIKZBLOCKSEEDLETTER{0}{1135.025}{342.123}
\TIKZBLOCKSEEDLETTER{1}{175.025}{342.123}
\TIKZBLOCKSEEDLETTER{0}{-784.975}{342.123}
            }{}
    \end{tikzpicture}
}

%% file: OritatamiTuringPatterns/n=4-L=3-P=1/ZagCopy0LineFeed_0__n=4-L=3-P=1.tex
\newcommand{\ZagCopyZLineFeedBlockZ}[1]{
    \begin{tikzpicture}
        \node [anchor=center] () at (0,0) {\includegraphics[scale =\s]{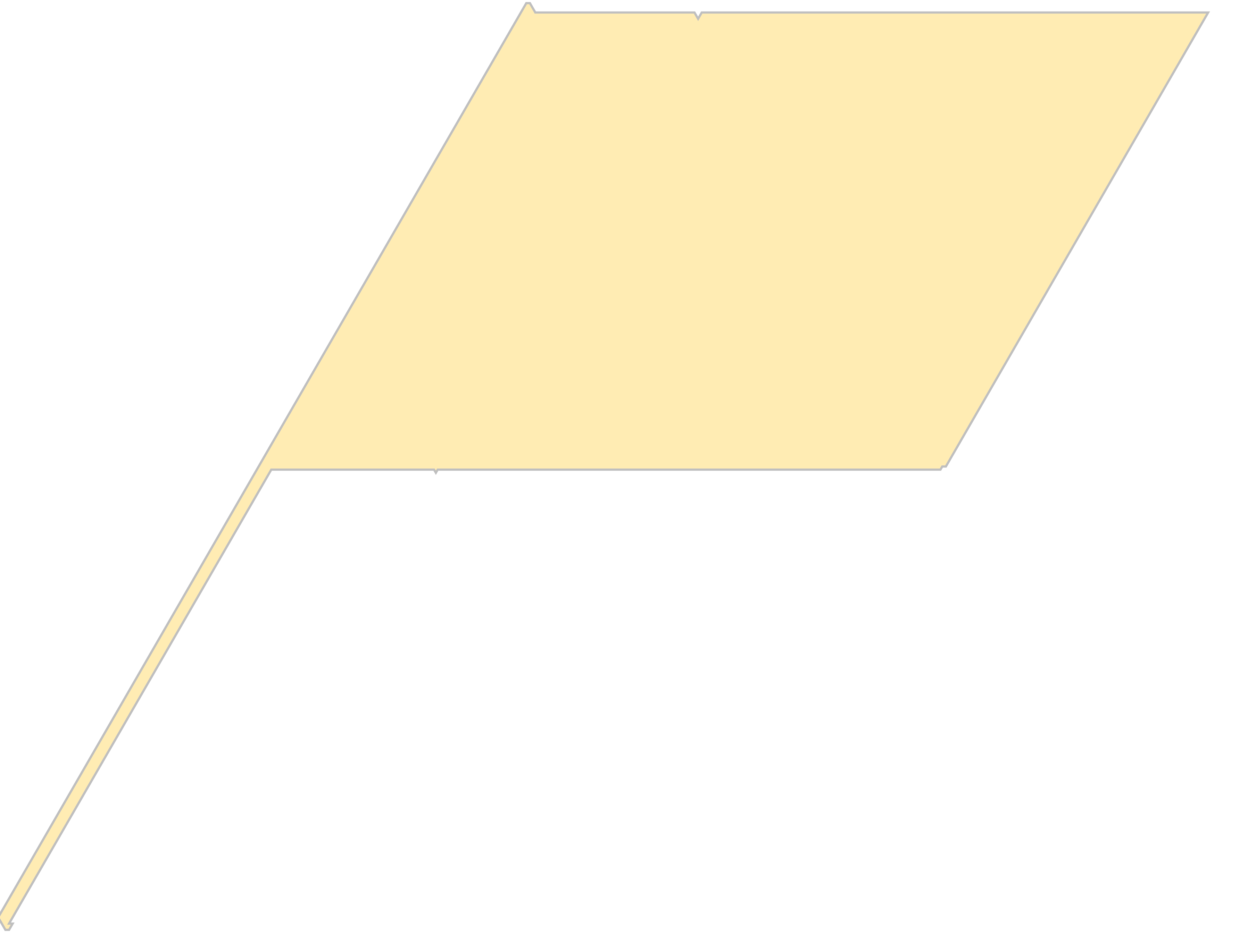}};
        \node [anchor=center] () at (0,0) {\includegraphics[scale =\s]{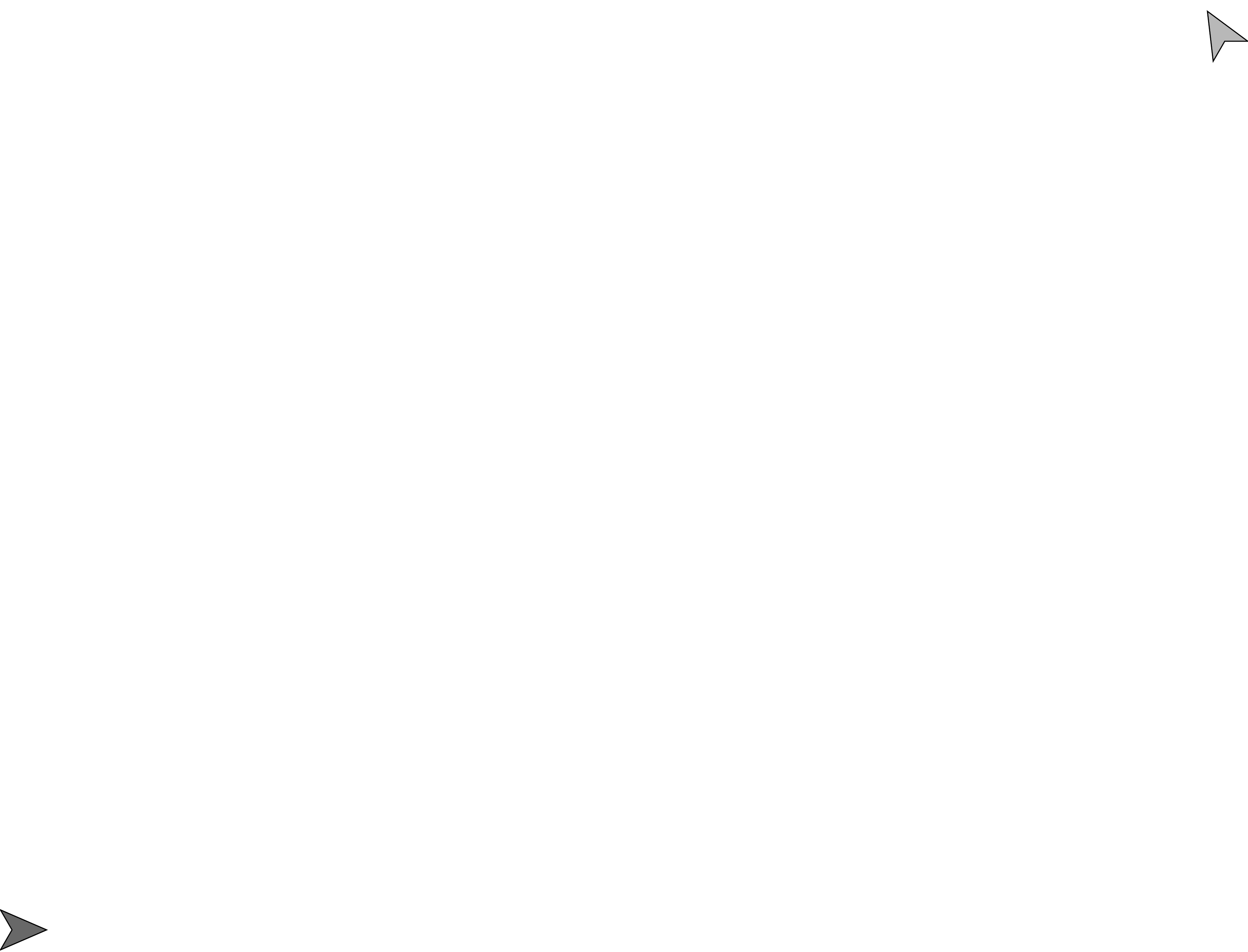}};
        \ifthenelse{\equal{#1}{}}{
\TIKZPointOfInterest{(807.50000000*\st pt, 640.85879880*\st pt)}{a}{$(h-8,1)$}{E}{100.0*\st pt}{start}{(-1500.00000000*\st pt, -953.84917986*\st pt)}{8.0}{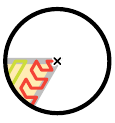}
\TIKZPointOfInterest{(442.50000000*\st pt, 8.66025404*\st pt)}{b}{$(h-8,h)$}{E}{150.0*\st pt}{}{(-900.00000000*\st pt, -953.84917986*\st pt)}{8.0}{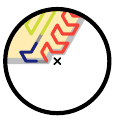}
\TIKZPointOfInterest{(102.50000000*\st pt, 640.85879880*\st pt)}{c}{$(h-W+w+1,1)$}{NE}{75.0*\st pt}{Read \word0}{(-300.00000000*\st pt, -953.84917986*\st pt)}{8.0}{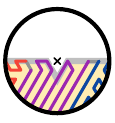}
\TIKZPointOfInterest{(-260.00000000*\st pt, 4.33012702*\st pt)}{d}{$(h-W+w+2,1+h)$}{SE}{100.0*\st pt}{Copy \word0}{(300.00000000*\st pt, -953.84917986*\st pt)}{8.0}{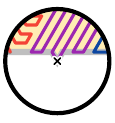}
\TIKZPointOfInterest{(-122.50000000*\st pt, 640.85879880*\st pt)}{e}{$(h-W-2,1)$}{NE}{175.0*\st pt}{}{(900.00000000*\st pt, -953.84917986*\st pt)}{8.0}{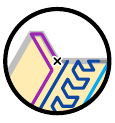}
\TIKZPointOfInterest{(-135.00000000*\st pt, 653.84917986*\st pt)}{f}{$(h-W-6,-2)$}{W}{100.0*\st pt}{}{(1500.00000000*\st pt, -953.84917986*\st pt)}{8.0}{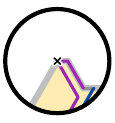}
\TIKZPointOfInterest{(-497.50000000*\st pt, 8.66025404*\st pt)}{g}{$(h-W-4,h)$}{NW}{100.0*\st pt}{}{(-1500.00000000*\st pt, -1553.84917986*\st pt)}{8.0}{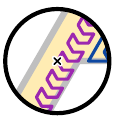}
\TIKZPointOfInterest{(-845.00000000*\st pt, -627.86841774*\st pt)}{h}{$(h-W,2h)$}{NE}{50.0*\st pt}{Next}{(-900.00000000*\st pt, -1553.84917986*\st pt)}{8.0}{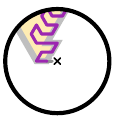}
            }{}
\node [anchor=center] () at (0,0) {\includegraphics[scale =\s]{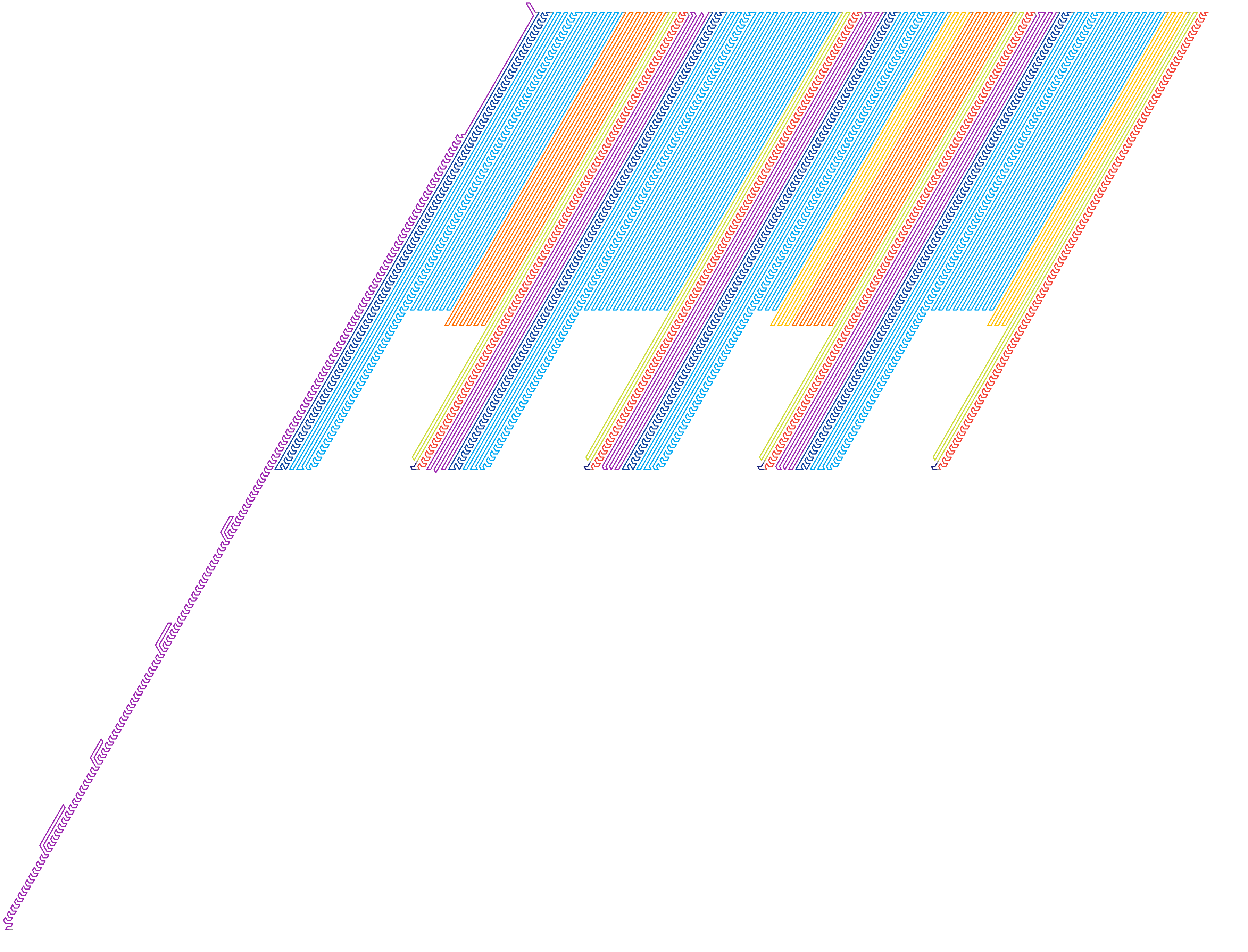}};
            \ifthenelse{\equal{#1}{}}{
\TIKZBLOCKZAGCOPYLINEFEED{0}{3}{0}{145.000}{324.760}
            }{}
    \end{tikzpicture}
}

%% file: OritatamiTuringPatterns/n=4-L=3-P=1/ZagCopy0LineFeed_110__n=4-L=3-P=1.tex
\newcommand{\ZagCopyZLineFeedBlockUUZ}[1]{
    \begin{tikzpicture}
        \node [anchor=center] () at (0,0) {\includegraphics[scale =\s]{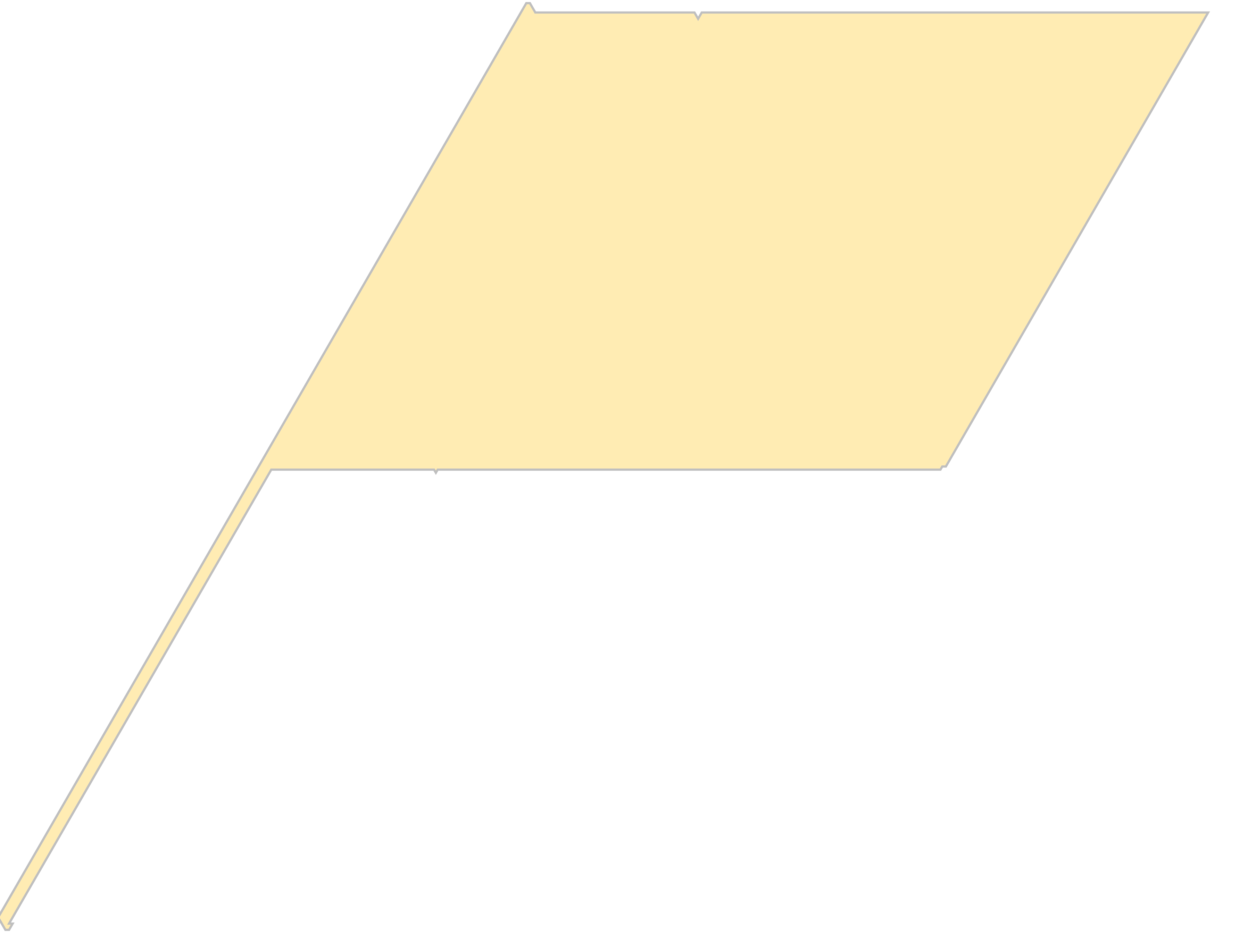}};
        \node [anchor=center] () at (0,0) {\includegraphics[scale =\s]{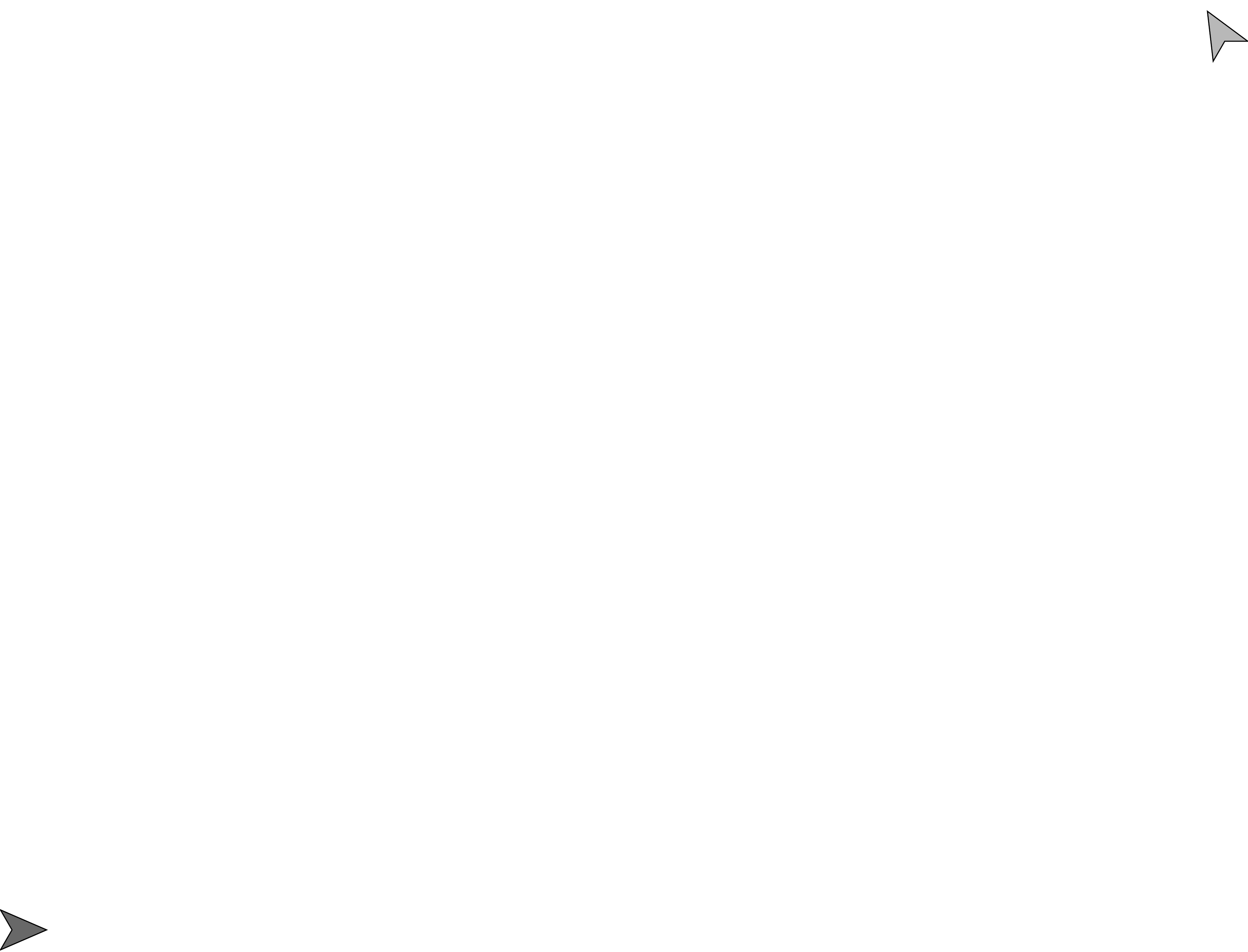}};
        \ifthenelse{\equal{#1}{}}{
\TIKZPointOfInterest{(807.50000000*\st pt, 640.85879880*\st pt)}{a}{$(h-8,1)$}{E}{100.0*\st pt}{start}{(-1500.00000000*\st pt, -953.84917986*\st pt)}{8.0}{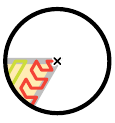}
\TIKZPointOfInterest{(442.50000000*\st pt, 8.66025404*\st pt)}{b}{$(h-8,h)$}{E}{150.0*\st pt}{}{(-900.00000000*\st pt, -953.84917986*\st pt)}{8.0}{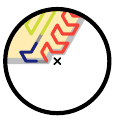}
\TIKZPointOfInterest{(102.50000000*\st pt, 640.85879880*\st pt)}{c}{$(h-W+w+1,1)$}{NE}{75.0*\st pt}{Read \word0}{(-300.00000000*\st pt, -953.84917986*\st pt)}{8.0}{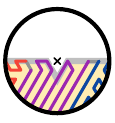}
\TIKZPointOfInterest{(-260.00000000*\st pt, 4.33012702*\st pt)}{d}{$(h-W+w+2,1+h)$}{SE}{100.0*\st pt}{Copy \word0}{(300.00000000*\st pt, -953.84917986*\st pt)}{8.0}{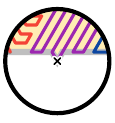}
\TIKZPointOfInterest{(-122.50000000*\st pt, 640.85879880*\st pt)}{e}{$(h-W-2,1)$}{NE}{175.0*\st pt}{}{(900.00000000*\st pt, -953.84917986*\st pt)}{8.0}{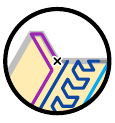}
\TIKZPointOfInterest{(-135.00000000*\st pt, 653.84917986*\st pt)}{f}{$(h-W-6,-2)$}{W}{100.0*\st pt}{}{(1500.00000000*\st pt, -953.84917986*\st pt)}{8.0}{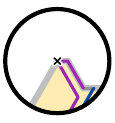}
\TIKZPointOfInterest{(-497.50000000*\st pt, 8.66025404*\st pt)}{g}{$(h-W-4,h)$}{NW}{100.0*\st pt}{}{(-1500.00000000*\st pt, -1553.84917986*\st pt)}{8.0}{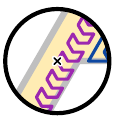}
\TIKZPointOfInterest{(-845.00000000*\st pt, -627.86841774*\st pt)}{h}{$(h-W,2h)$}{NE}{50.0*\st pt}{Next}{(-900.00000000*\st pt, -1553.84917986*\st pt)}{8.0}{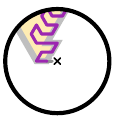}
            }{}
\node [anchor=center] () at (0,0) {\includegraphics[scale =\s]{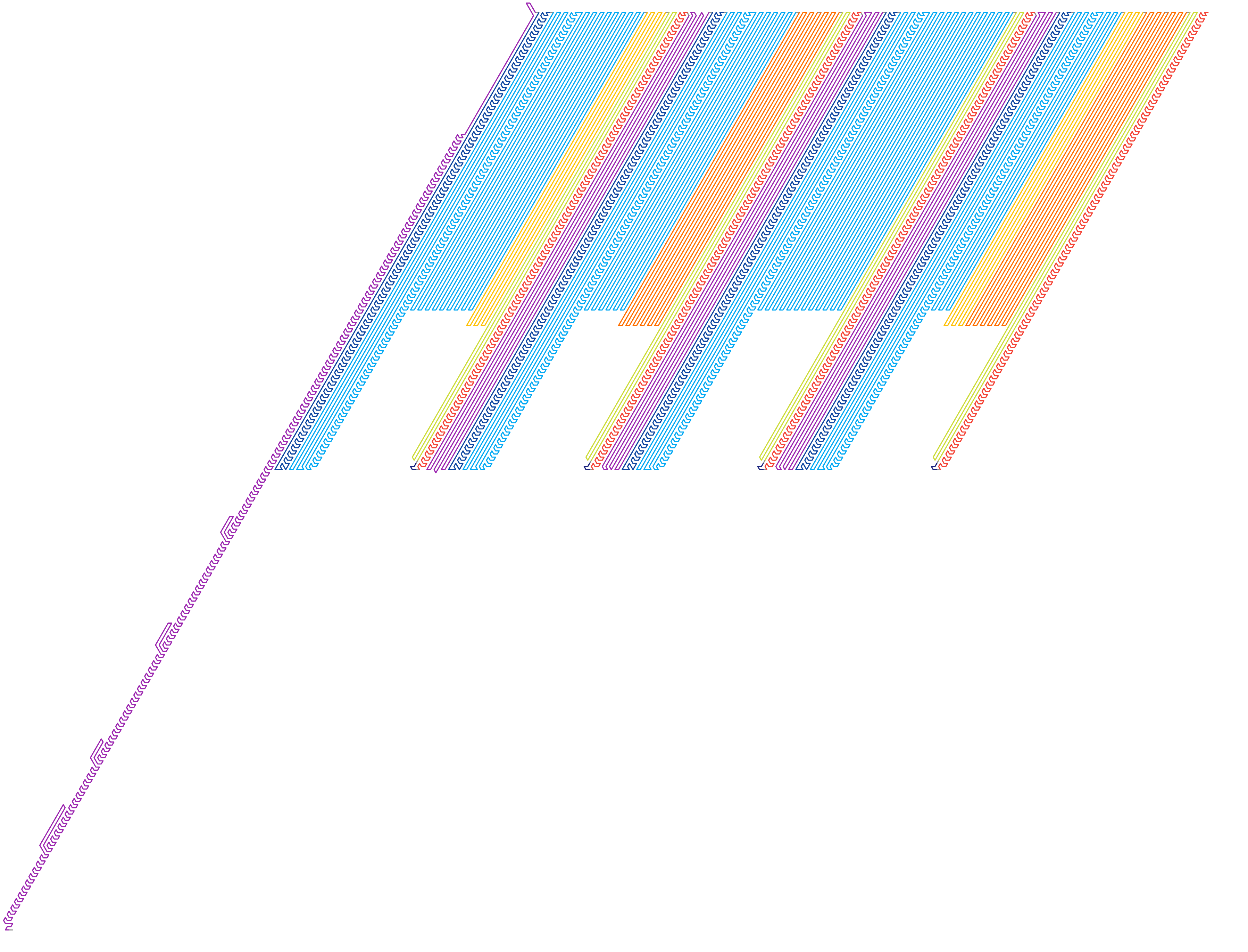}};
            \ifthenelse{\equal{#1}{}}{
\TIKZBLOCKZAGCOPYLINEFEED{0}{0}{110}{145.000}{324.760}
            }{}
    \end{tikzpicture}
}

%% file: OritatamiTuringPatterns/n=4-L=3-P=1/ZagCopy0LineFeed_11__n=4-L=3-P=1.tex
\newcommand{\ZagCopyZLineFeedBlockUU}[1]{
    \begin{tikzpicture}
        \node [anchor=center] () at (0,0) {\includegraphics[scale =\s]{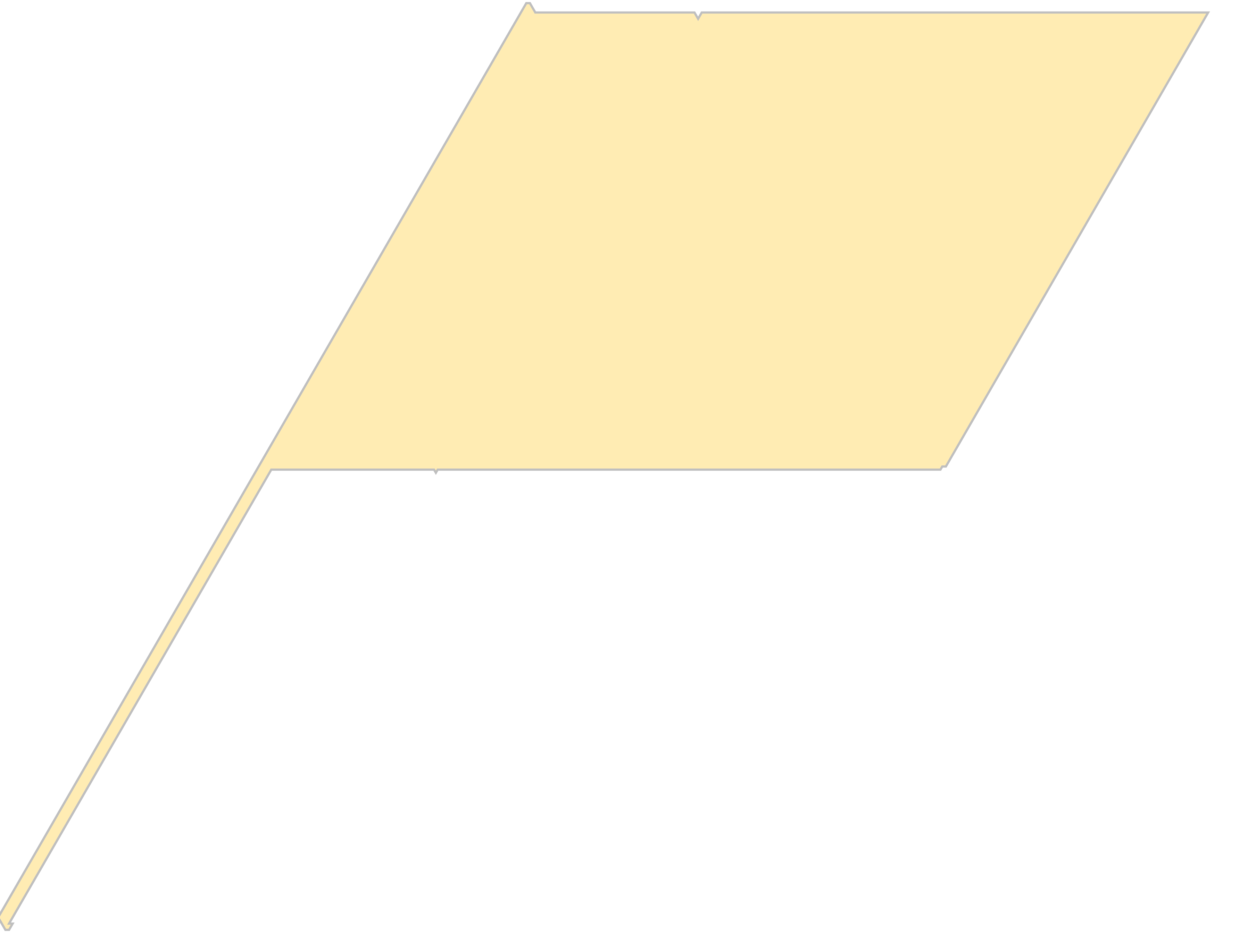}};
        \node [anchor=center] () at (0,0) {\includegraphics[scale =\s]{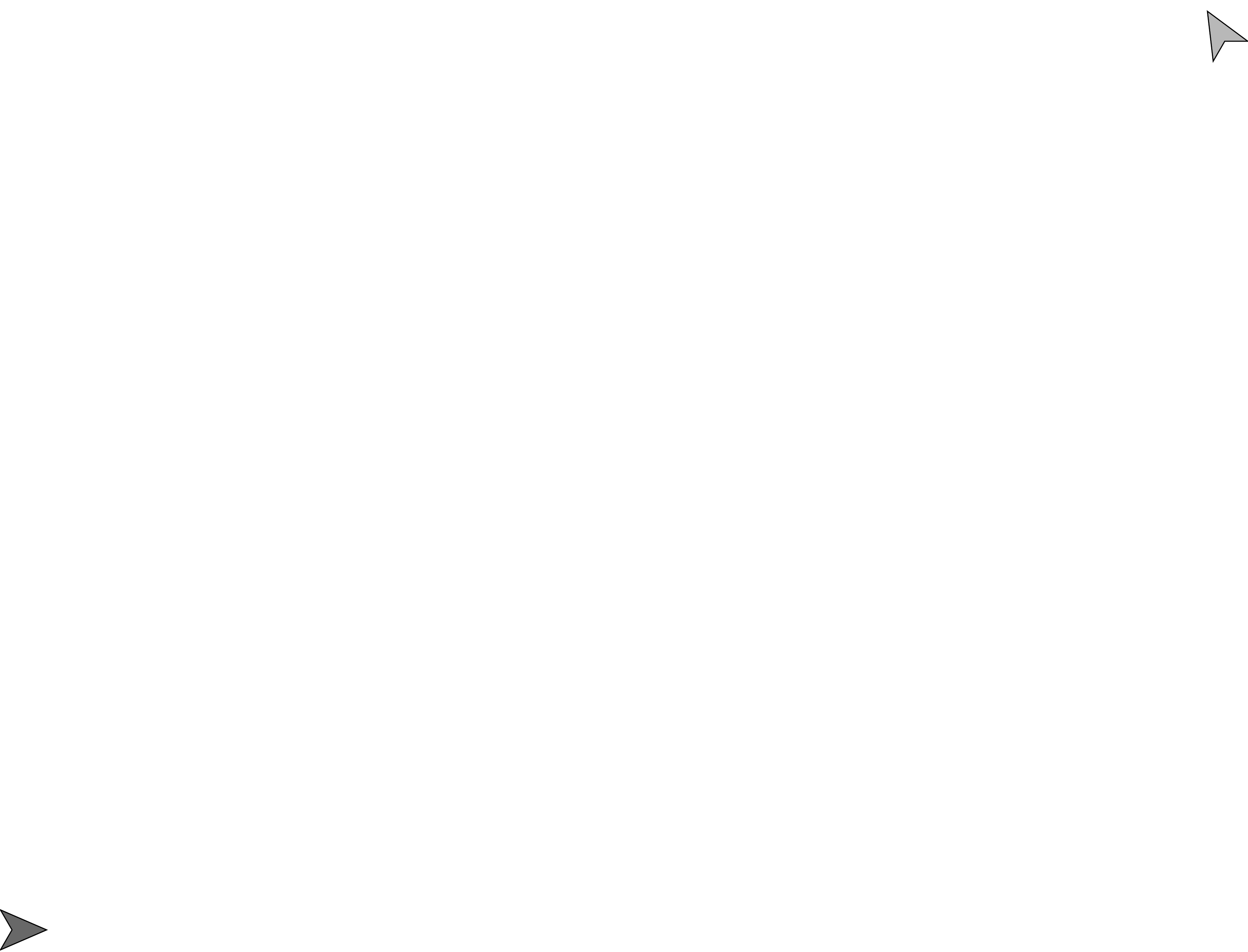}};
        \ifthenelse{\equal{#1}{}}{
\TIKZPointOfInterest{(807.50000000*\st pt, 640.85879880*\st pt)}{a}{$(h-8,1)$}{E}{100.0*\st pt}{start}{(-1500.00000000*\st pt, -953.84917986*\st pt)}{8.0}{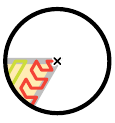}
\TIKZPointOfInterest{(442.50000000*\st pt, 8.66025404*\st pt)}{b}{$(h-8,h)$}{E}{150.0*\st pt}{}{(-900.00000000*\st pt, -953.84917986*\st pt)}{8.0}{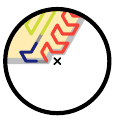}
\TIKZPointOfInterest{(102.50000000*\st pt, 640.85879880*\st pt)}{c}{$(h-W+w+1,1)$}{NE}{75.0*\st pt}{Read \word0}{(-300.00000000*\st pt, -953.84917986*\st pt)}{8.0}{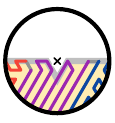}
\TIKZPointOfInterest{(-260.00000000*\st pt, 4.33012702*\st pt)}{d}{$(h-W+w+2,1+h)$}{SE}{100.0*\st pt}{Copy \word0}{(300.00000000*\st pt, -953.84917986*\st pt)}{8.0}{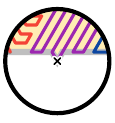}
\TIKZPointOfInterest{(-122.50000000*\st pt, 640.85879880*\st pt)}{e}{$(h-W-2,1)$}{NE}{175.0*\st pt}{}{(900.00000000*\st pt, -953.84917986*\st pt)}{8.0}{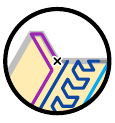}
\TIKZPointOfInterest{(-135.00000000*\st pt, 653.84917986*\st pt)}{f}{$(h-W-6,-2)$}{W}{100.0*\st pt}{}{(1500.00000000*\st pt, -953.84917986*\st pt)}{8.0}{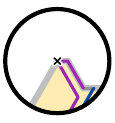}
\TIKZPointOfInterest{(-497.50000000*\st pt, 8.66025404*\st pt)}{g}{$(h-W-4,h)$}{NW}{100.0*\st pt}{}{(-1500.00000000*\st pt, -1553.84917986*\st pt)}{8.0}{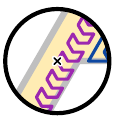}
\TIKZPointOfInterest{(-845.00000000*\st pt, -627.86841774*\st pt)}{h}{$(h-W,2h)$}{NE}{50.0*\st pt}{Next}{(-900.00000000*\st pt, -1553.84917986*\st pt)}{8.0}{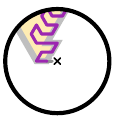}
            }{}
\node [anchor=center] () at (0,0) {\includegraphics[scale =\s]{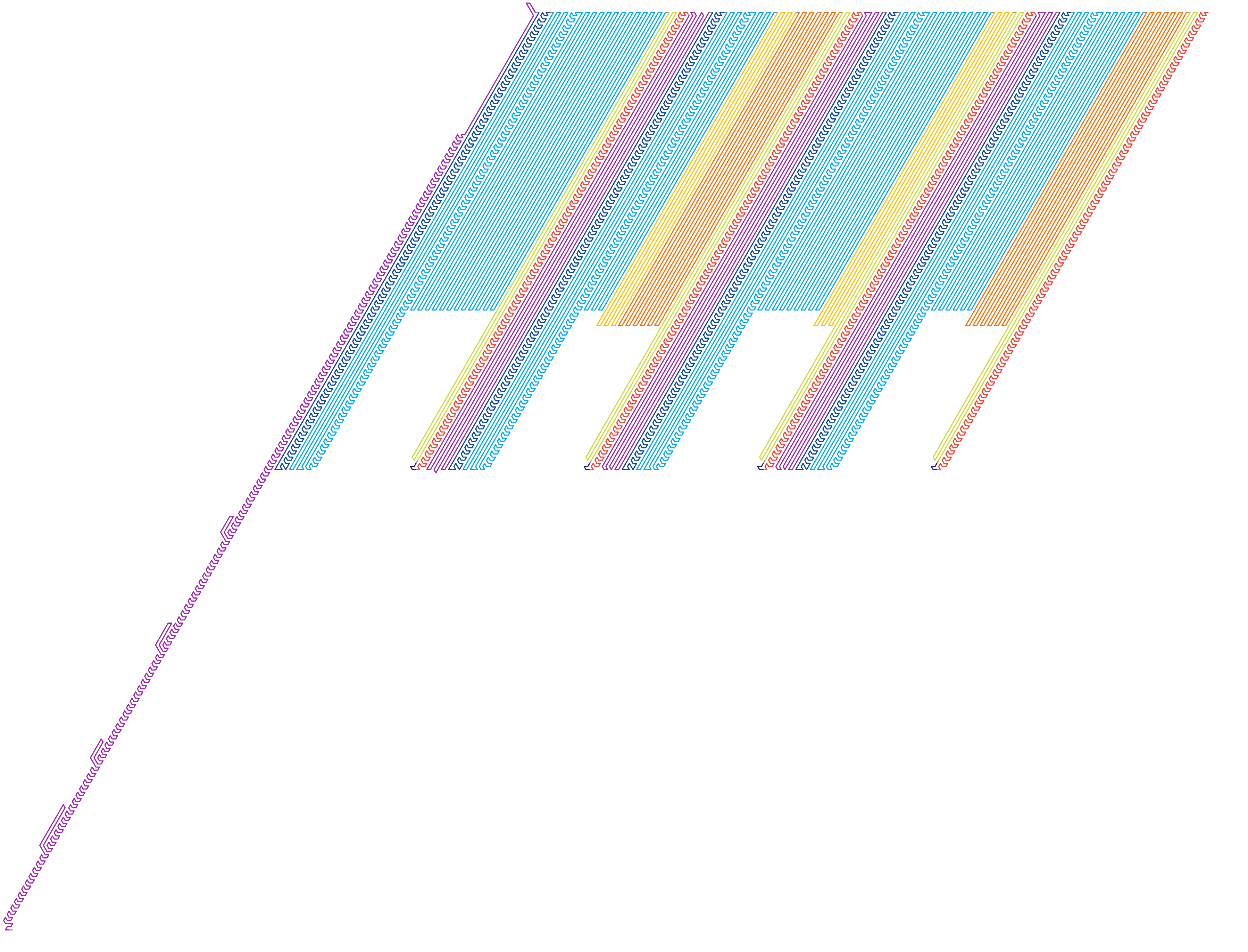}};
            \ifthenelse{\equal{#1}{}}{
\TIKZBLOCKZAGCOPYLINEFEED{0}{2}{11}{145.000}{324.760}
            }{}
    \end{tikzpicture}
}

%% file: OritatamiTuringPatterns/n=4-L=3-P=1/ZagCopy0LineFeed_____n=4-L=3-P=1.tex
\newcommand{\ZagCopyZLineFeedBlockEpsilon}[1]{
    \begin{tikzpicture}
        \node [anchor=center] () at (0,0) {\includegraphics[scale =\s]{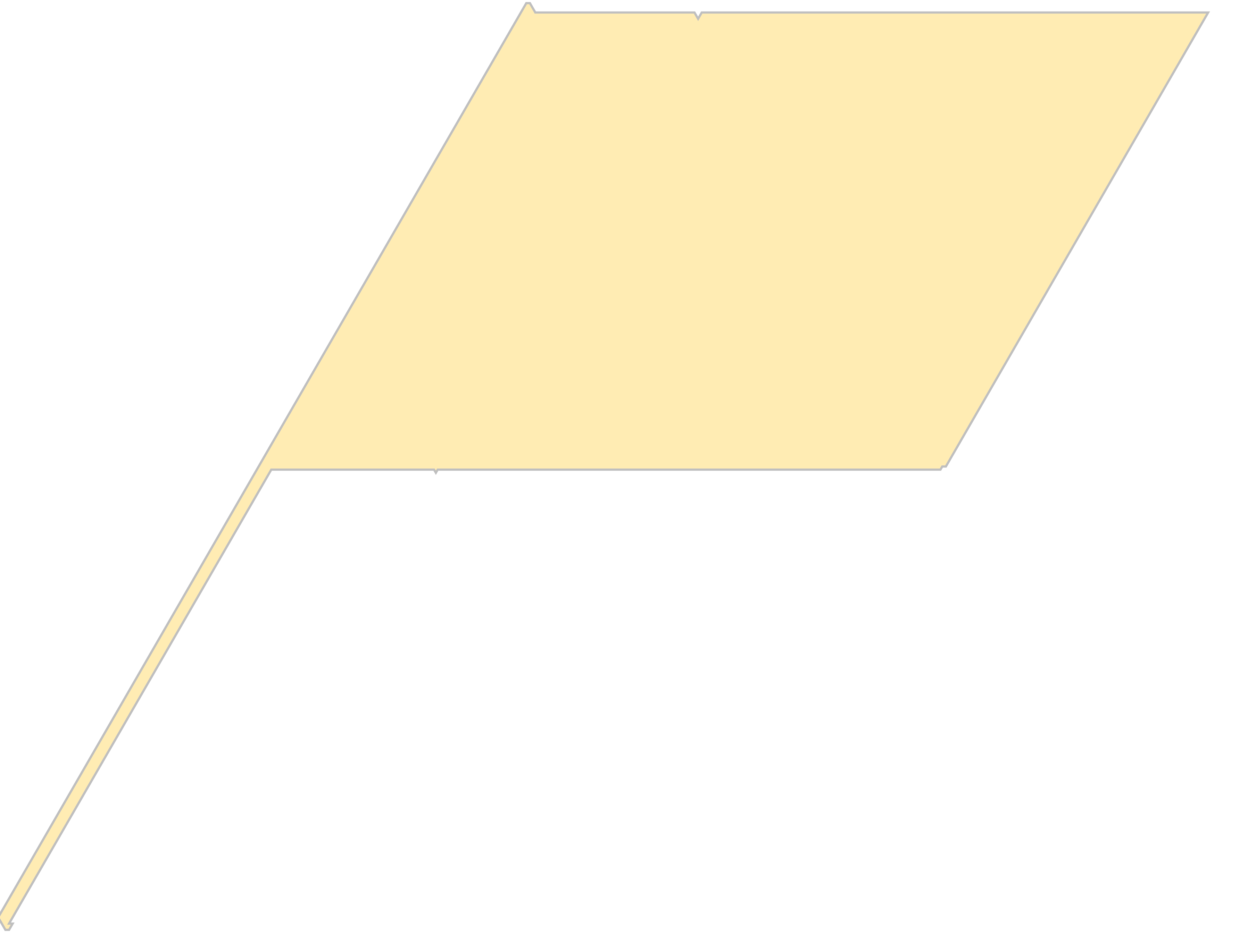}};
        \node [anchor=center] () at (0,0) {\includegraphics[scale =\s]{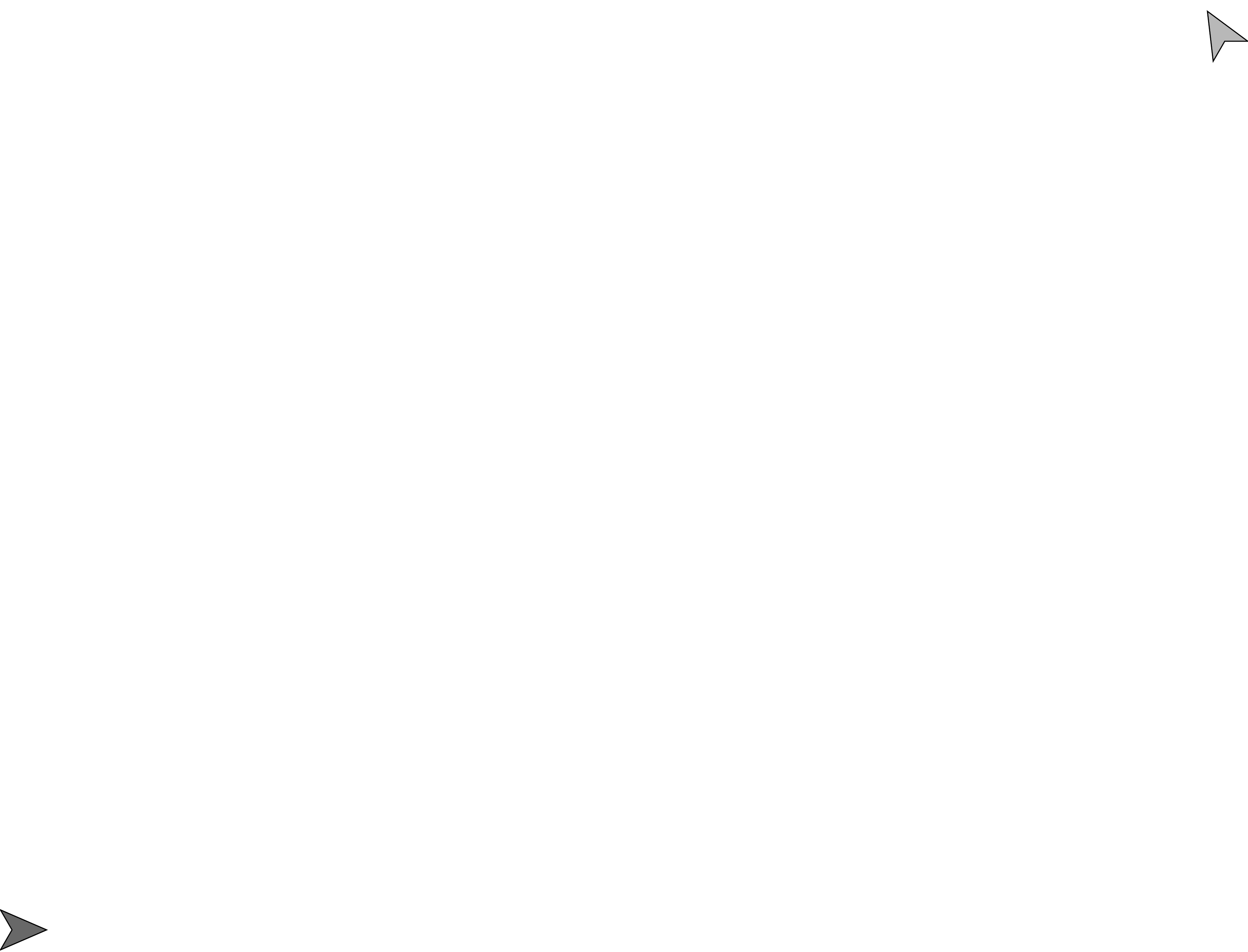}};
        \ifthenelse{\equal{#1}{}}{
\TIKZPointOfInterest{(807.50000000*\st pt, 640.85879880*\st pt)}{a}{$(h-8,1)$}{E}{100.0*\st pt}{start}{(-1500.00000000*\st pt, -953.84917986*\st pt)}{8.0}{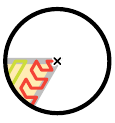}
\TIKZPointOfInterest{(442.50000000*\st pt, 8.66025404*\st pt)}{b}{$(h-8,h)$}{E}{150.0*\st pt}{}{(-900.00000000*\st pt, -953.84917986*\st pt)}{8.0}{ZagCopy0LineFeed____n=4-L=3-P=1-PI-B-_0,146_-bottom-right.pdf}
\TIKZPointOfInterest{(102.50000000*\st pt, 640.85879880*\st pt)}{c}{$(h-W+w+1,1)$}{NE}{75.0*\st pt}{Read \word0}{(-300.00000000*\st pt, -953.84917986*\st pt)}{8.0}{ZagCopy0LineFeed____n=4-L=3-P=1-PI-C-_-141,0_-read0.pdf}
\TIKZPointOfInterest{(-260.00000000*\st pt, 4.33012702*\st pt)}{d}{$(h-W+w+2,1+h)$}{SE}{100.0*\st pt}{Copy \word0}{(300.00000000*\st pt, -953.84917986*\st pt)}{8.0}{ZagCopy0LineFeed____n=4-L=3-P=1-PI-D-_-140,147_-copy0.pdf}
\TIKZPointOfInterest{(-122.50000000*\st pt, 640.85879880*\st pt)}{e}{$(h-W-2,1)$}{NE}{175.0*\st pt}{}{(900.00000000*\st pt, -953.84917986*\st pt)}{8.0}{ZagCopy0LineFeed____n=4-L=3-P=1-PI-E-_-186,0_-top-left.pdf}
\TIKZPointOfInterest{(-135.00000000*\st pt, 653.84917986*\st pt)}{f}{$(h-W-6,-2)$}{W}{100.0*\st pt}{}{(1500.00000000*\st pt, -953.84917986*\st pt)}{8.0}{ZagCopy0LineFeed____n=4-L=3-P=1-PI-F-_-190,-3_-top-left-corner.pdf}
\TIKZPointOfInterest{(-497.50000000*\st pt, 8.66025404*\st pt)}{g}{$(h-W-4,h)$}{NW}{100.0*\st pt}{}{(-1500.00000000*\st pt, -1553.84917986*\st pt)}{8.0}{ZagCopy0LineFeed____n=4-L=3-P=1-PI-G-_-188,146_-mid-left.pdf}
\TIKZPointOfInterest{(-845.00000000*\st pt, -627.86841774*\st pt)}{h}{$(h-W,2h)$}{NE}{50.0*\st pt}{Next}{(-900.00000000*\st pt, -1553.84917986*\st pt)}{8.0}{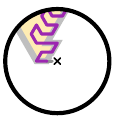}
            }{}
\node [anchor=center] () at (0,0) {\includegraphics[scale =\s]{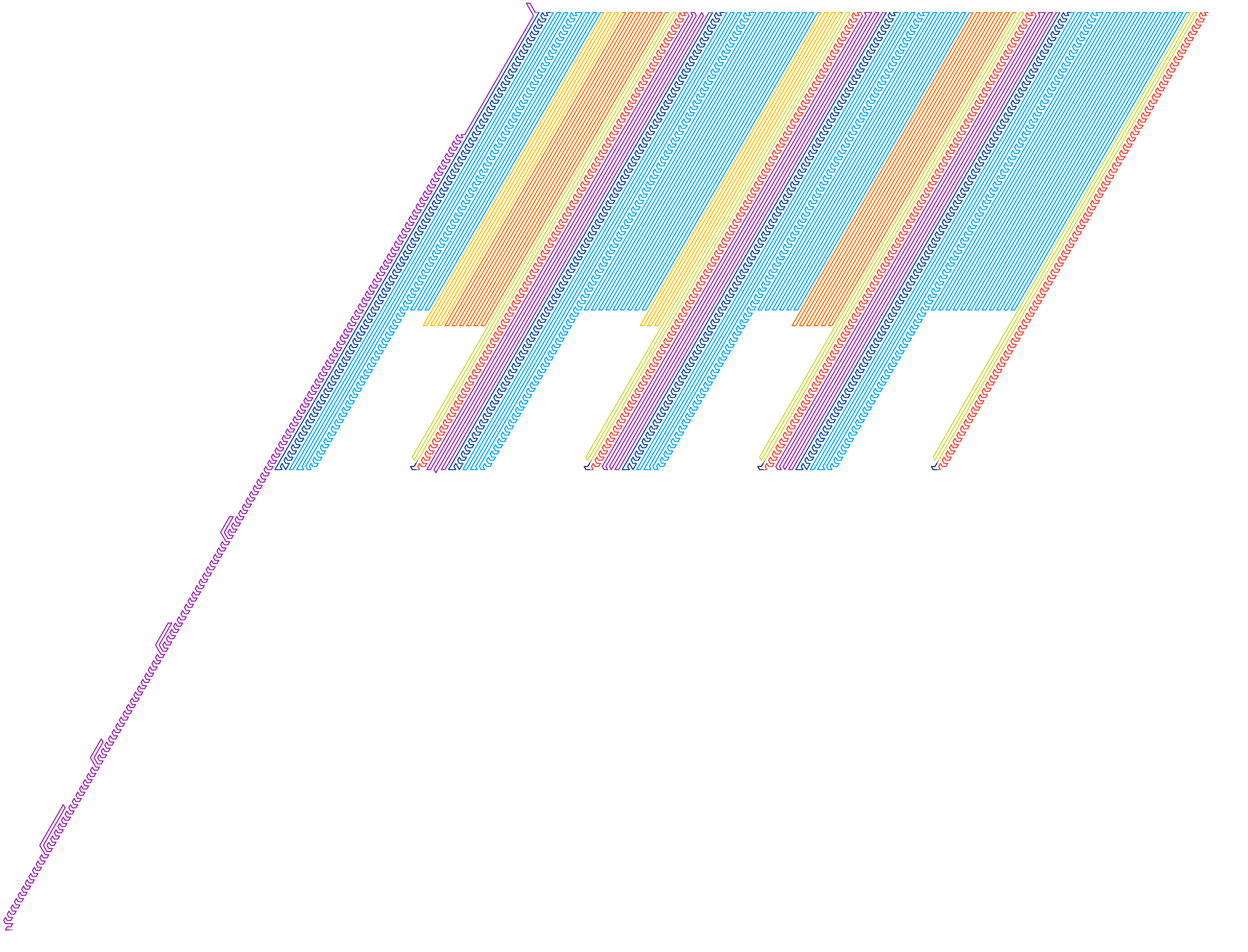}};
            \ifthenelse{\equal{#1}{}}{
\TIKZBLOCKZAGCOPYLINEFEED{0}{1}{\ensuremath{\epsilon}}{145.000}{324.760}
            }{}
    \end{tikzpicture}
}

%% file: OritatamiTuringPatterns/n=4-L=3-P=1/ZagCopy0_0__n=4-L=3-P=1.tex
\newcommand{\ZagCopyZBlockZ}[1]{
    \begin{tikzpicture}
        \node [anchor=center] () at (0,0) {\includegraphics[scale =\s]{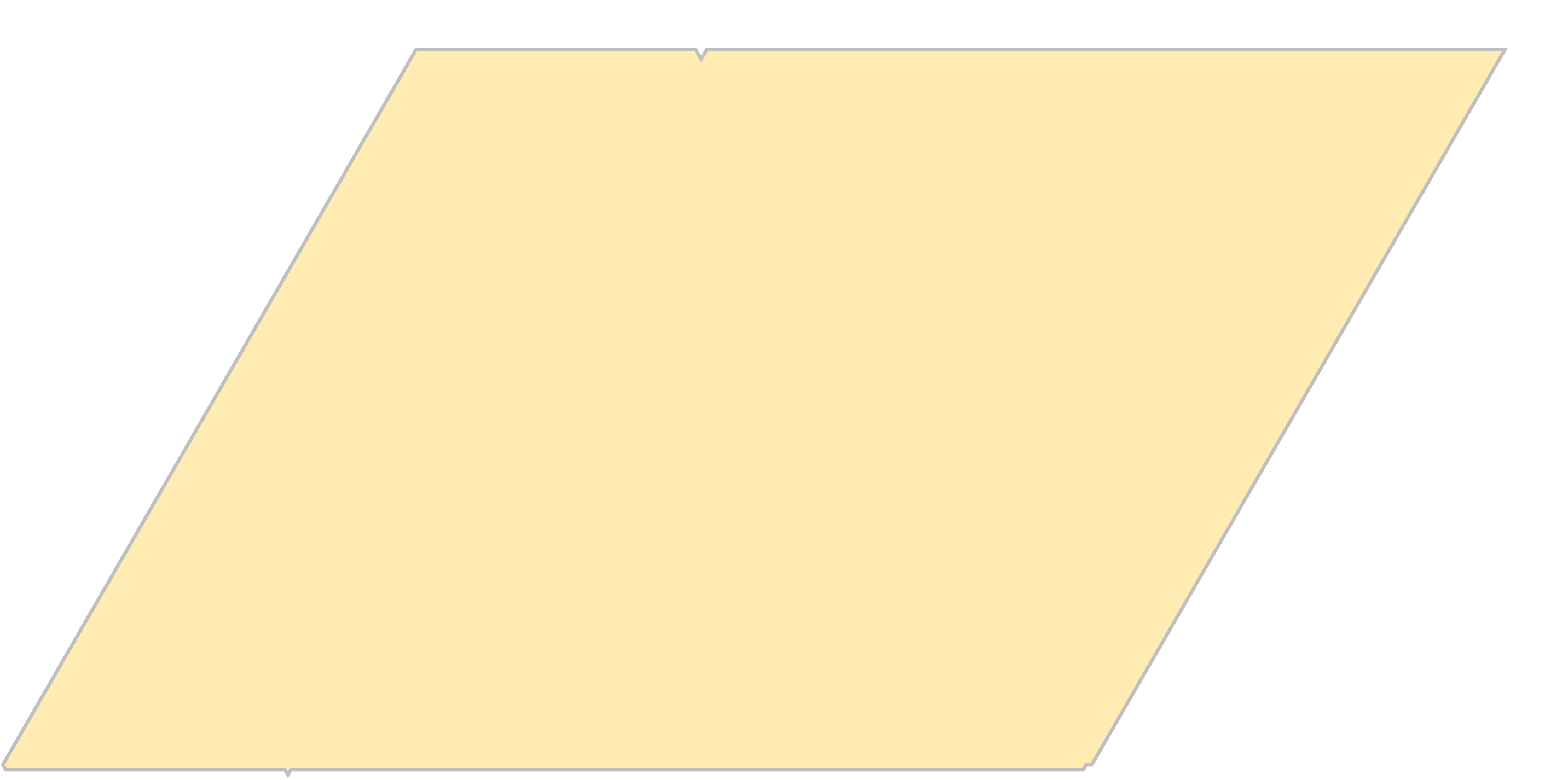}};
        \node [anchor=center] () at (0,0) {\includegraphics[scale =\s]{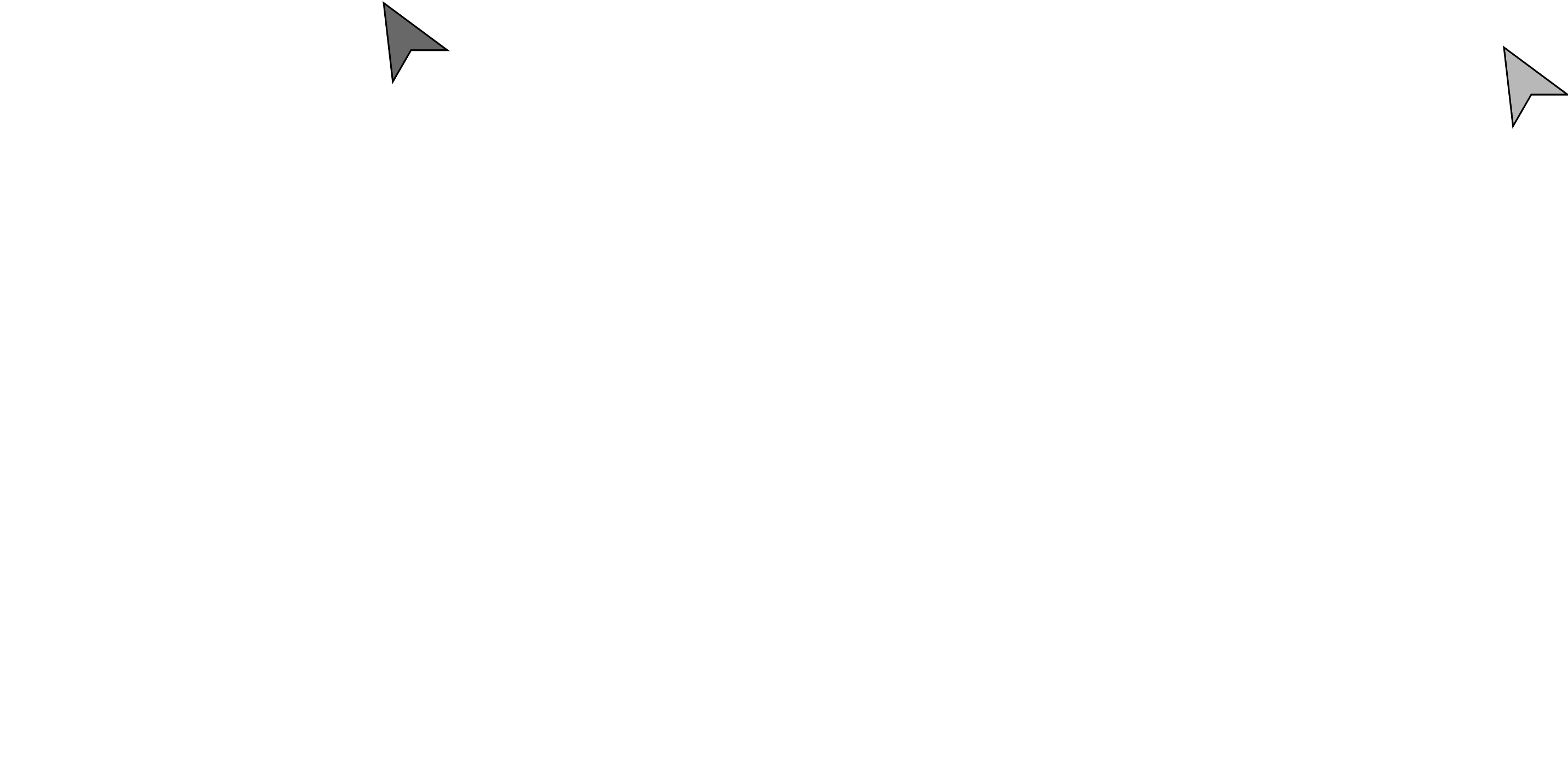}};
        \ifthenelse{\equal{#1}{}}{
\TIKZPointOfInterest{(632.50000000*\st pt, 298.77876431*\st pt)}{a}{$(h-8,1)$}{E}{100.0*\st pt}{Start}{(-1500.00000000*\st pt, -837.74990748*\st pt)}{8.0}{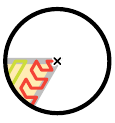}
\TIKZPointOfInterest{(267.50000000*\st pt, -333.41978046*\st pt)}{b}{$(h-8,h)$}{E}{150.0*\st pt}{}{(-900.00000000*\st pt, -837.74990748*\st pt)}{8.0}{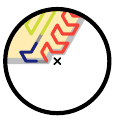}
\TIKZPointOfInterest{(-72.50000000*\st pt, 298.77876431*\st pt)}{c}{$(h-W+w+1,1)$}{NE}{100.0*\st pt}{Read \word0}{(-300.00000000*\st pt, -837.74990748*\st pt)}{8.0}{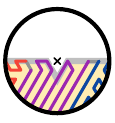}
\TIKZPointOfInterest{(-435.00000000*\st pt, -337.74990748*\st pt)}{d}{$(h-W+w+2,1+h)$}{SE}{100.0*\st pt}{Copy \word0}{(300.00000000*\st pt, -837.74990748*\st pt)}{8.0}{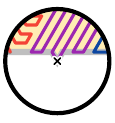}
\TIKZPointOfInterest{(-687.50000000*\st pt, -333.41978046*\st pt)}{e}{$(h-W-7,h)$}{NW}{100.0*\st pt}{}{(900.00000000*\st pt, -837.74990748*\st pt)}{8.0}{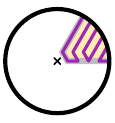}
\TIKZPointOfInterest{(-327.50000000*\st pt, 298.77876431*\st pt)}{f}{$(h-W-8,1)$}{W}{100.0*\st pt}{Next}{(1500.00000000*\st pt, -837.74990748*\st pt)}{8.0}{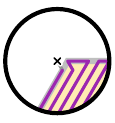}
            }{}
\node [anchor=center] () at (0,0) {\includegraphics[scale =\s]{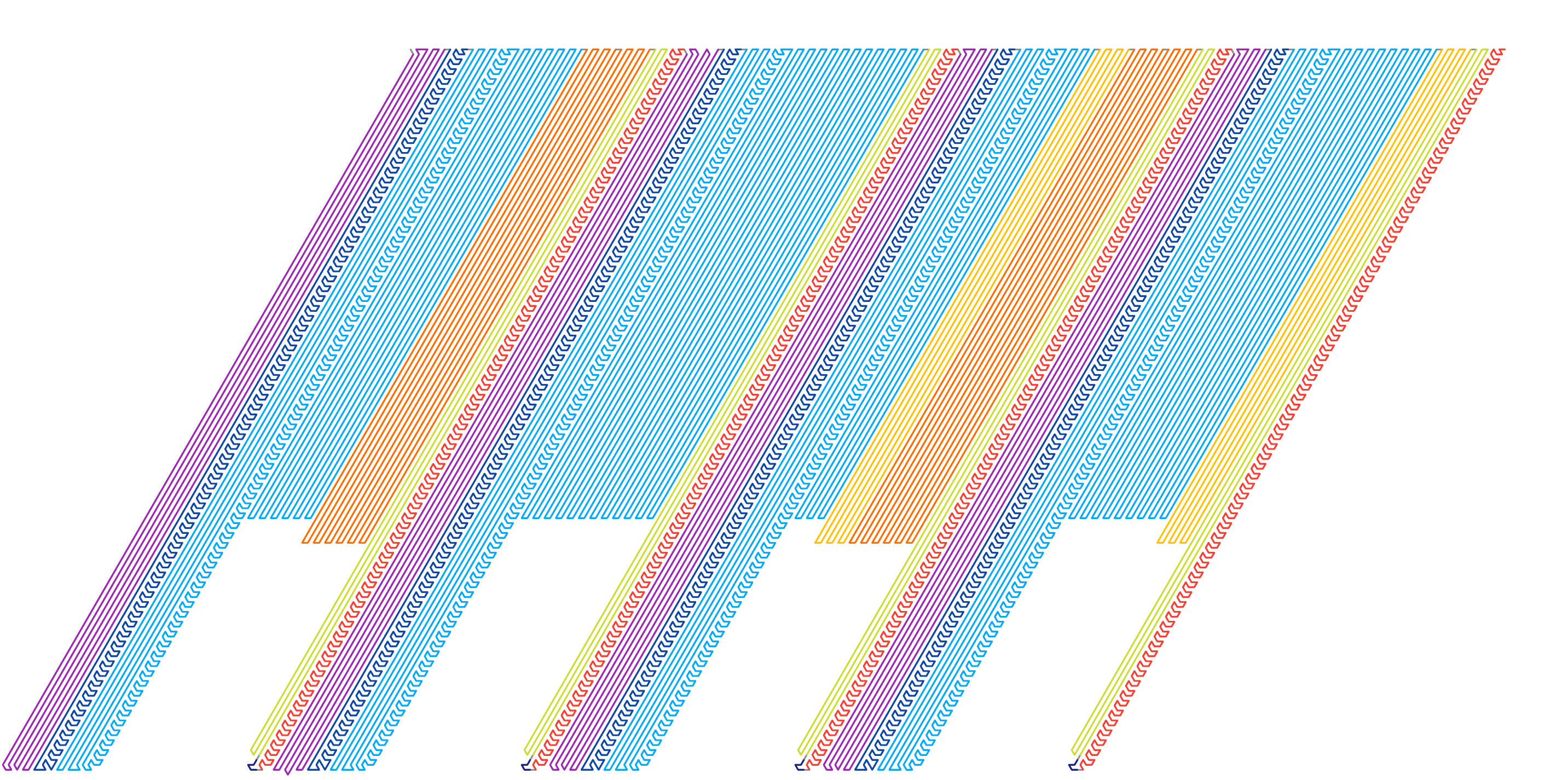}};
            \ifthenelse{\equal{#1}{}}{
\TIKZBLOCKZAGCOPY{0}{3}{0}{-30.000}{-17.321}
            }{}
    \end{tikzpicture}
}

%% file: OritatamiTuringPatterns/n=4-L=3-P=1/ZagCopy0_110__n=4-L=3-P=1.tex
\newcommand{\ZagCopyZBlockUUZ}[1]{
    \begin{tikzpicture}
        \node [anchor=center] () at (0,0) {\includegraphics[scale =\s]{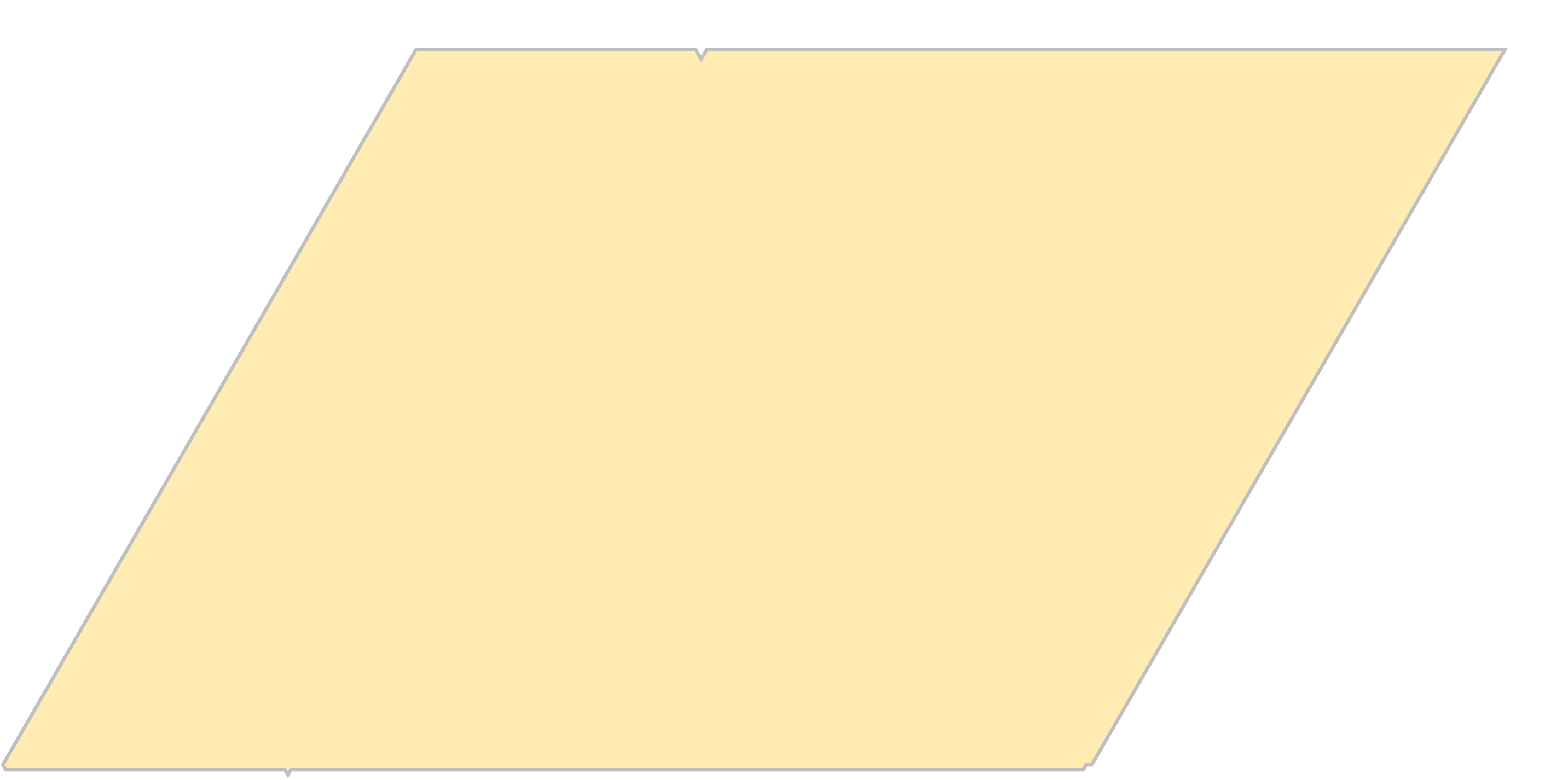}};
        \node [anchor=center] () at (0,0) {\includegraphics[scale =\s]{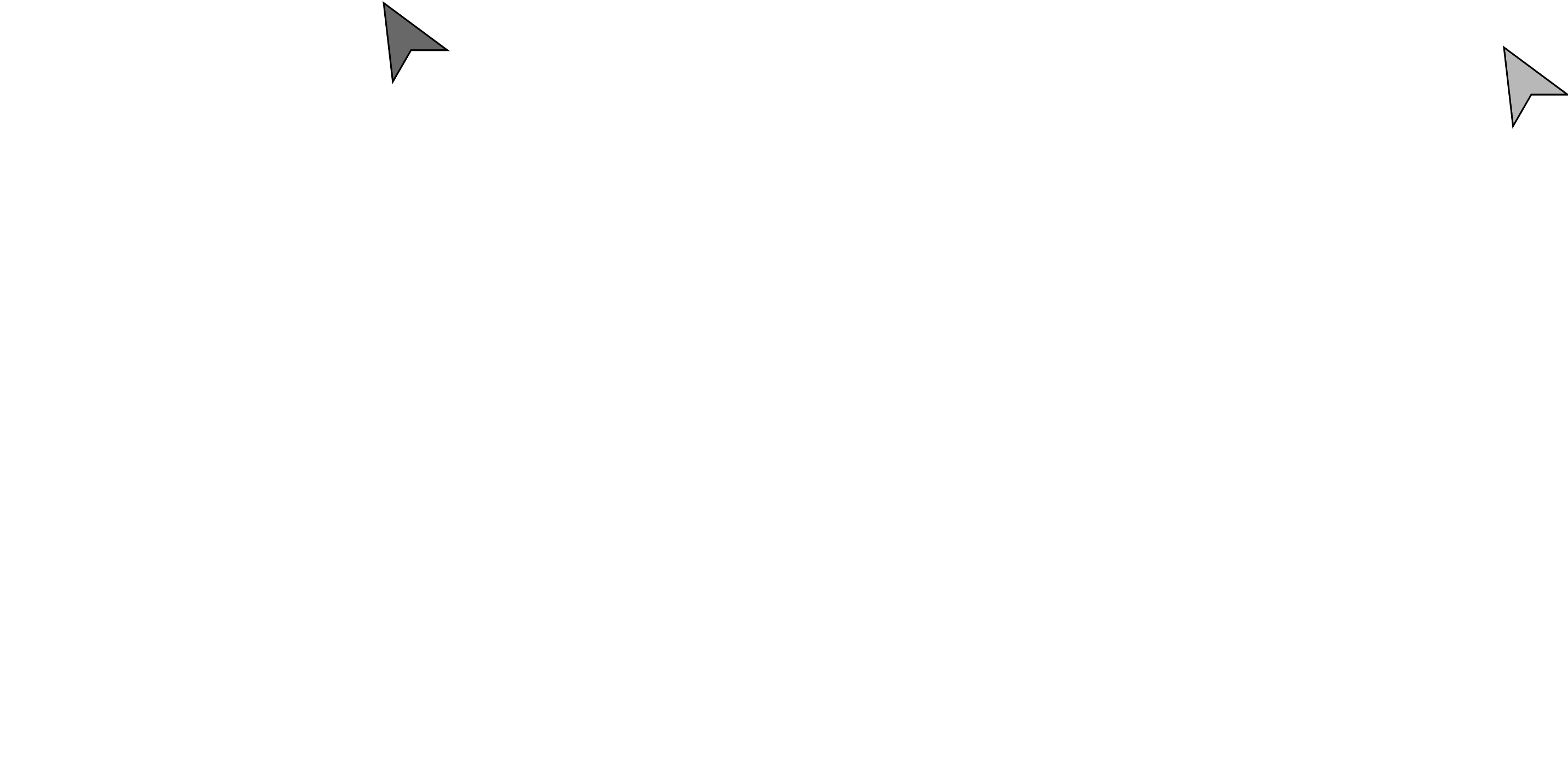}};
        \ifthenelse{\equal{#1}{}}{
\TIKZPointOfInterest{(632.50000000*\st pt, 298.77876431*\st pt)}{a}{$(h-8,1)$}{E}{100.0*\st pt}{Start}{(-1500.00000000*\st pt, -837.74990748*\st pt)}{8.0}{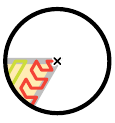}
\TIKZPointOfInterest{(267.50000000*\st pt, -333.41978046*\st pt)}{b}{$(h-8,h)$}{E}{150.0*\st pt}{}{(-900.00000000*\st pt, -837.74990748*\st pt)}{8.0}{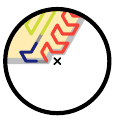}
\TIKZPointOfInterest{(-72.50000000*\st pt, 298.77876431*\st pt)}{c}{$(h-W+w+1,1)$}{NE}{100.0*\st pt}{Read \word0}{(-300.00000000*\st pt, -837.74990748*\st pt)}{8.0}{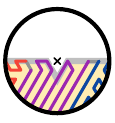}
\TIKZPointOfInterest{(-435.00000000*\st pt, -337.74990748*\st pt)}{d}{$(h-W+w+2,1+h)$}{SE}{100.0*\st pt}{Copy \word0}{(300.00000000*\st pt, -837.74990748*\st pt)}{8.0}{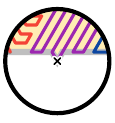}
\TIKZPointOfInterest{(-687.50000000*\st pt, -333.41978046*\st pt)}{e}{$(h-W-7,h)$}{NW}{100.0*\st pt}{}{(900.00000000*\st pt, -837.74990748*\st pt)}{8.0}{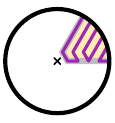}
\TIKZPointOfInterest{(-327.50000000*\st pt, 298.77876431*\st pt)}{f}{$(h-W-8,1)$}{W}{100.0*\st pt}{Next}{(1500.00000000*\st pt, -837.74990748*\st pt)}{8.0}{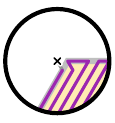}
            }{}
\node [anchor=center] () at (0,0) {\includegraphics[scale =\s]{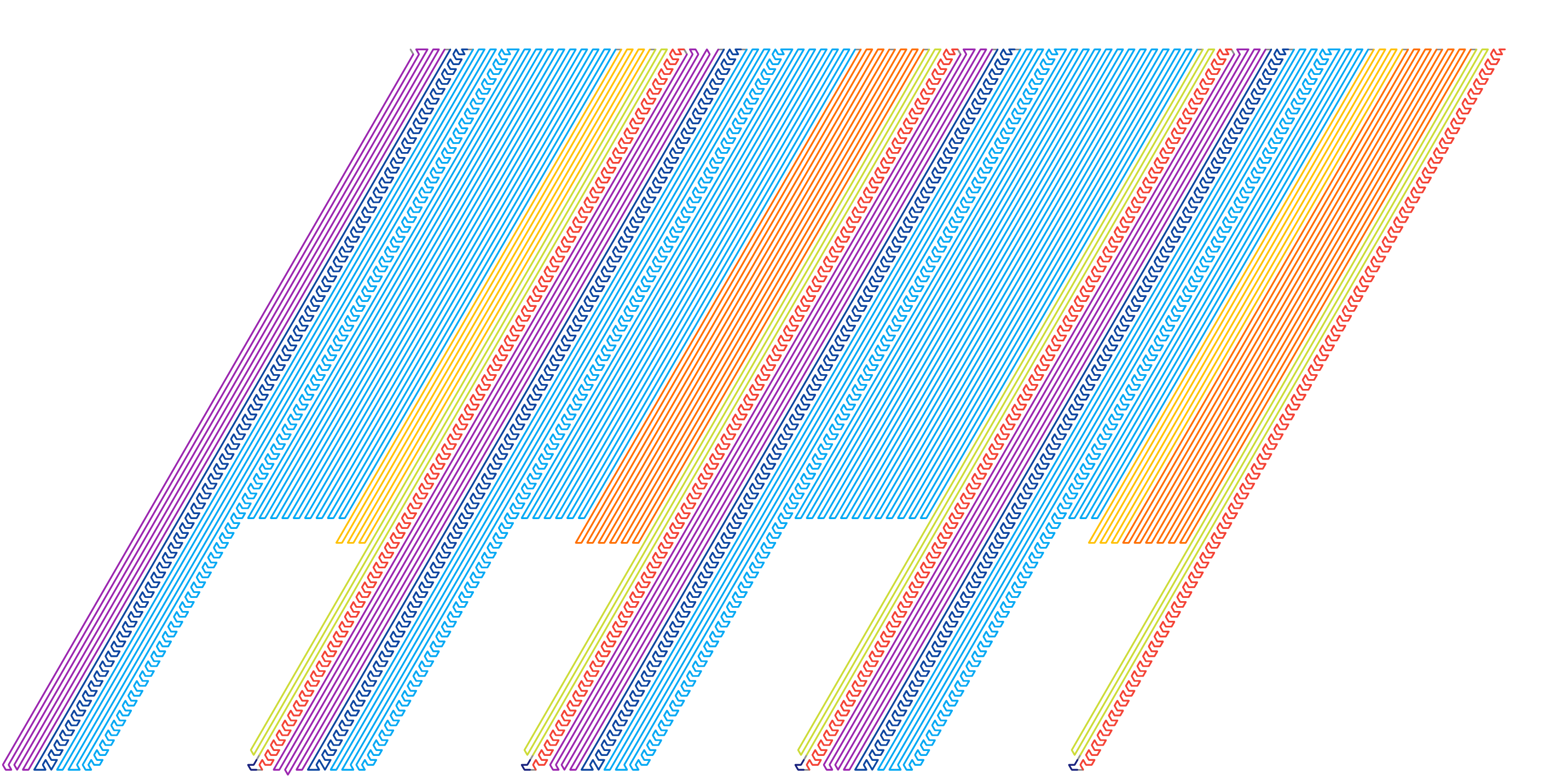}};
            \ifthenelse{\equal{#1}{}}{
\TIKZBLOCKZAGCOPY{0}{0}{110}{-30.000}{-17.321}
            }{}
    \end{tikzpicture}
}

%% file: OritatamiTuringPatterns/n=4-L=3-P=1/ZagCopy0_11__n=4-L=3-P=1.tex
\newcommand{\ZagCopyZBlockUU}[1]{
    \begin{tikzpicture}
        \node [anchor=center] () at (0,0) {\includegraphics[scale =\s]{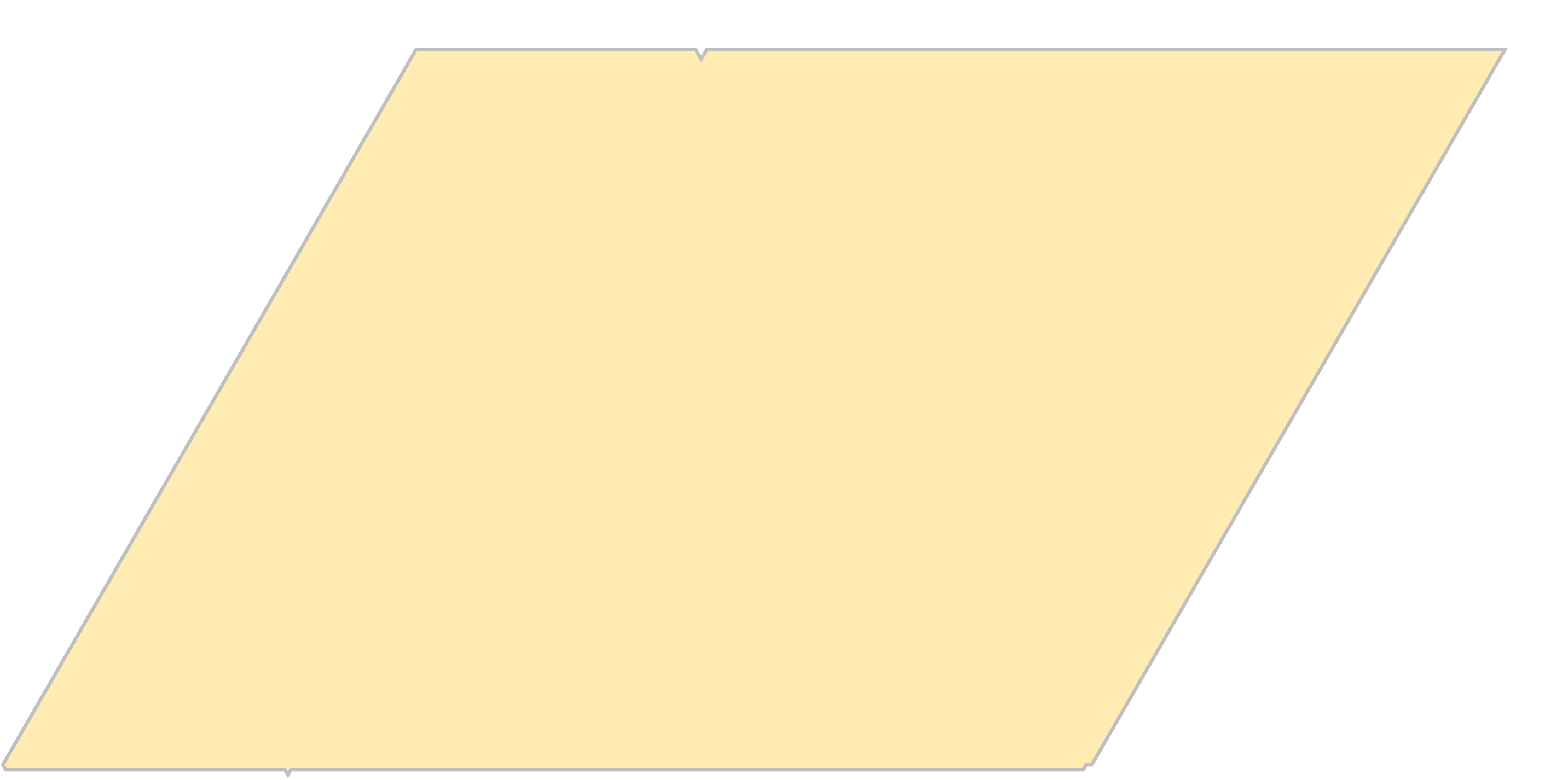}};
        \node [anchor=center] () at (0,0) {\includegraphics[scale =\s]{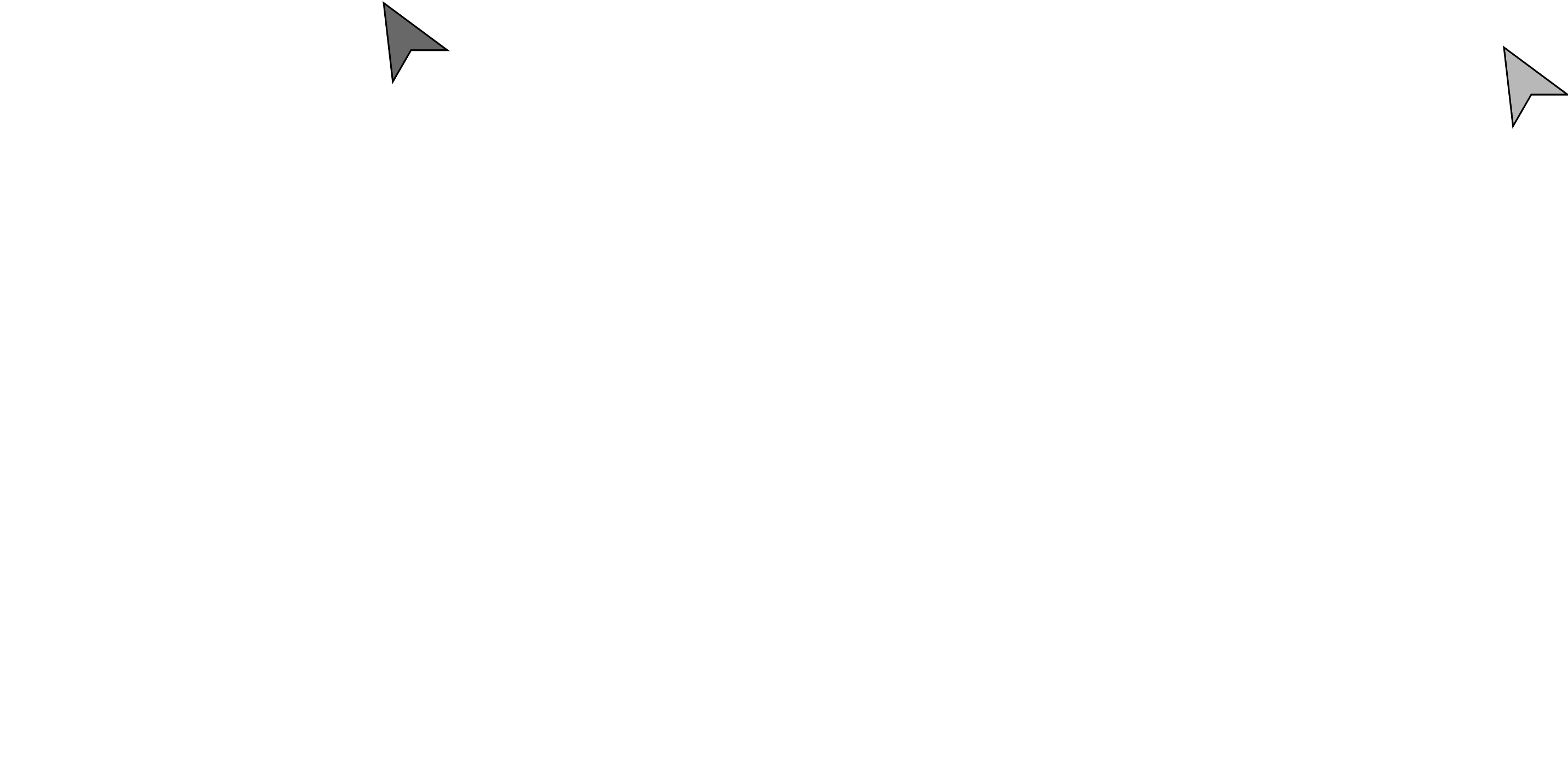}};
        \ifthenelse{\equal{#1}{}}{
\TIKZPointOfInterest{(632.50000000*\st pt, 298.77876431*\st pt)}{a}{$(h-8,1)$}{E}{100.0*\st pt}{Start}{(-1500.00000000*\st pt, -837.74990748*\st pt)}{8.0}{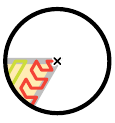}
\TIKZPointOfInterest{(267.50000000*\st pt, -333.41978046*\st pt)}{b}{$(h-8,h)$}{E}{150.0*\st pt}{}{(-900.00000000*\st pt, -837.74990748*\st pt)}{8.0}{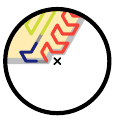}
\TIKZPointOfInterest{(-72.50000000*\st pt, 298.77876431*\st pt)}{c}{$(h-W+w+1,1)$}{NE}{100.0*\st pt}{Read \word0}{(-300.00000000*\st pt, -837.74990748*\st pt)}{8.0}{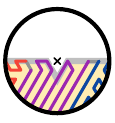}
\TIKZPointOfInterest{(-435.00000000*\st pt, -337.74990748*\st pt)}{d}{$(h-W+w+2,1+h)$}{SE}{100.0*\st pt}{Copy \word0}{(300.00000000*\st pt, -837.74990748*\st pt)}{8.0}{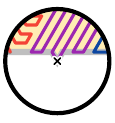}
\TIKZPointOfInterest{(-687.50000000*\st pt, -333.41978046*\st pt)}{e}{$(h-W-7,h)$}{NW}{100.0*\st pt}{}{(900.00000000*\st pt, -837.74990748*\st pt)}{8.0}{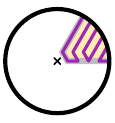}
\TIKZPointOfInterest{(-327.50000000*\st pt, 298.77876431*\st pt)}{f}{$(h-W-8,1)$}{W}{100.0*\st pt}{Next}{(1500.00000000*\st pt, -837.74990748*\st pt)}{8.0}{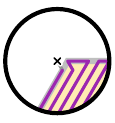}
            }{}
\node [anchor=center] () at (0,0) {\includegraphics[scale =\s]{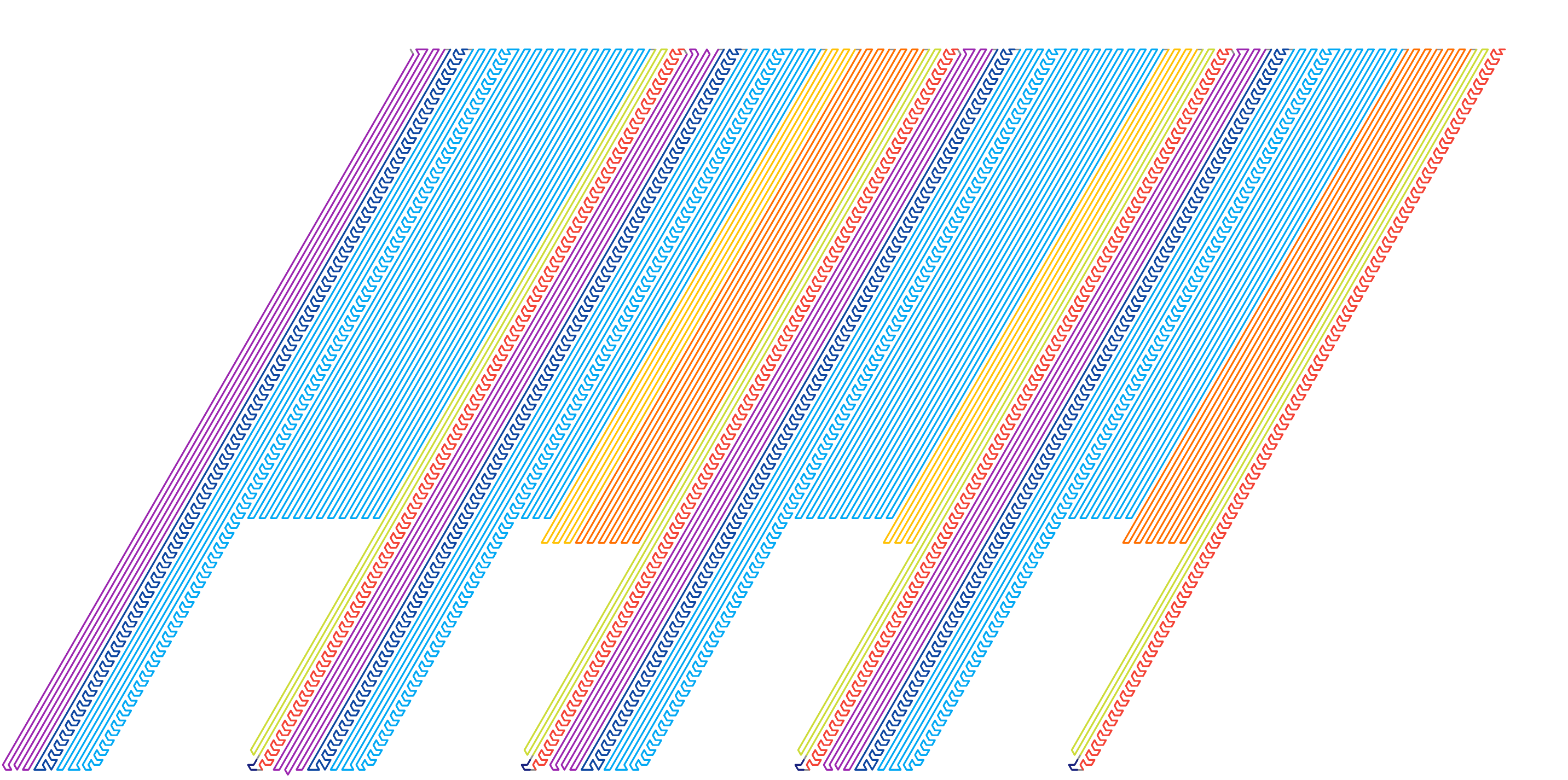}};
            \ifthenelse{\equal{#1}{}}{
\TIKZBLOCKZAGCOPY{0}{2}{11}{-30.000}{-17.321}
            }{}
    \end{tikzpicture}
}

%% file: OritatamiTuringPatterns/n=4-L=3-P=1/ZagCopy0_____n=4-L=3-P=1.tex
\newcommand{\ZagCopyZBlockEpsilon}[1]{
    \begin{tikzpicture}
        \node [anchor=center] () at (0,0) {\includegraphics[scale =\s]{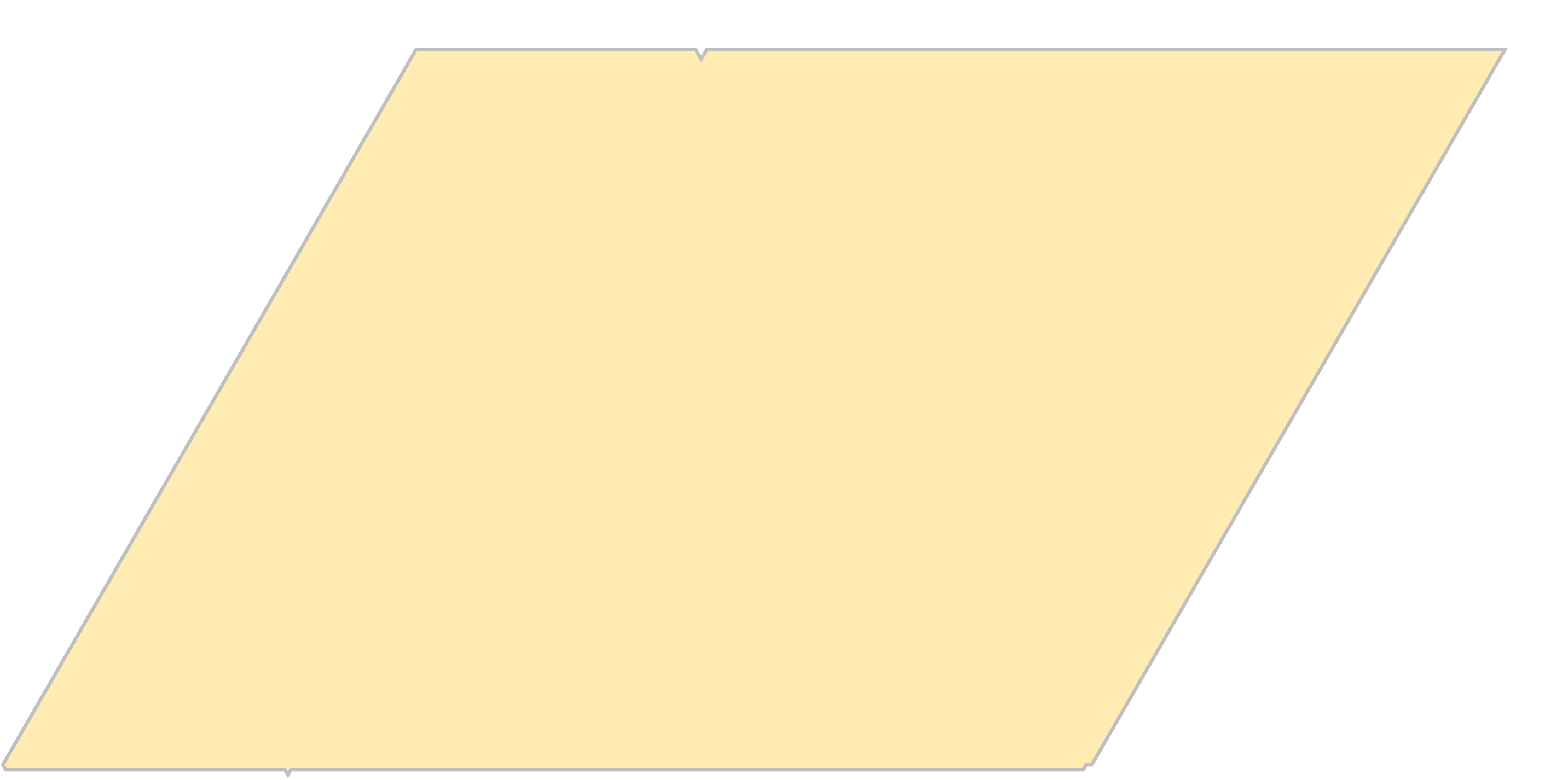}};
        \node [anchor=center] () at (0,0) {\includegraphics[scale =\s]{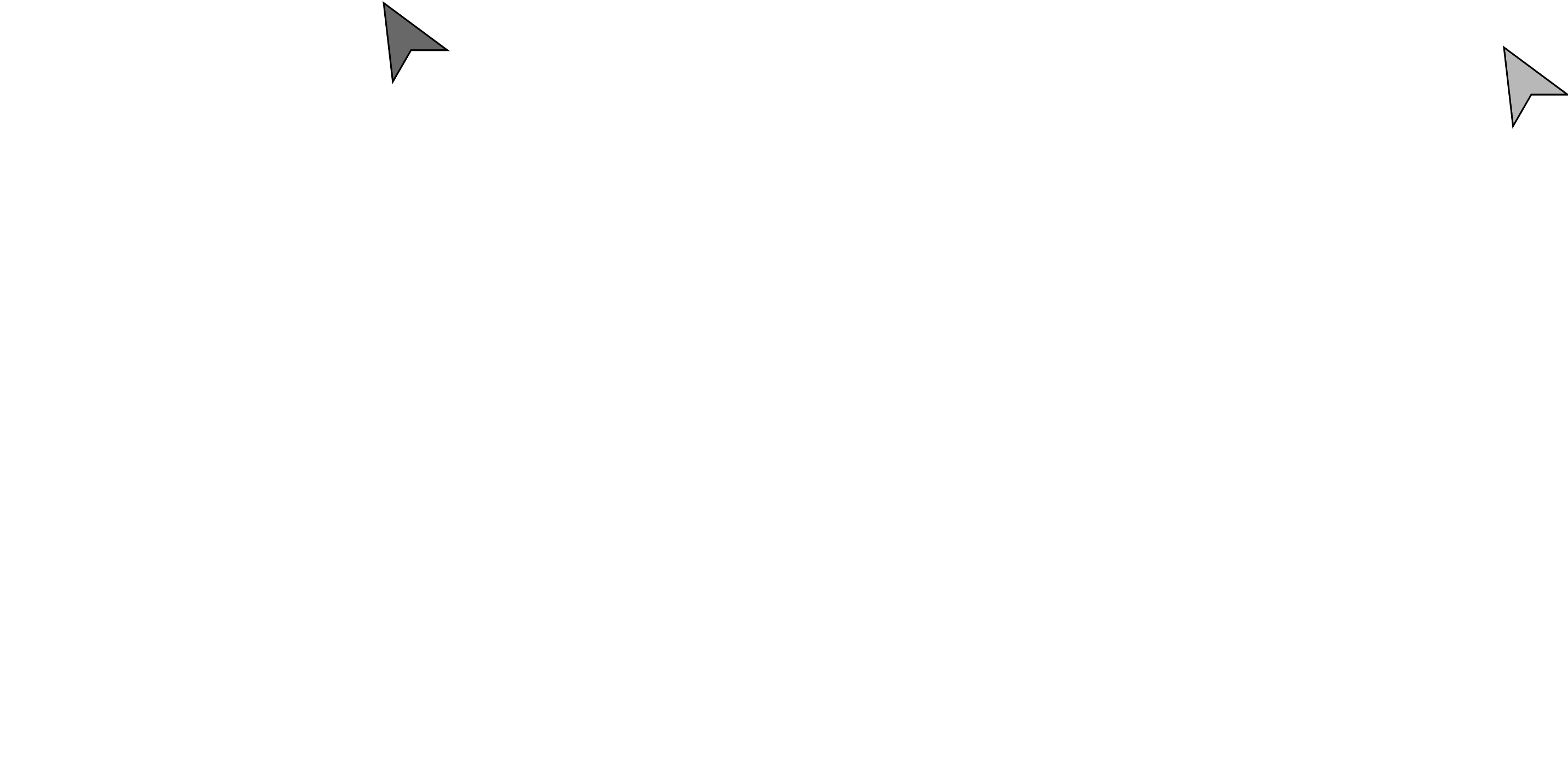}};
        \ifthenelse{\equal{#1}{}}{
\TIKZPointOfInterest{(632.50000000*\st pt, 298.77876431*\st pt)}{a}{$(h-8,1)$}{E}{100.0*\st pt}{Start}{(-1500.00000000*\st pt, -837.74990748*\st pt)}{8.0}{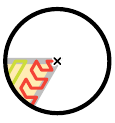}
\TIKZPointOfInterest{(267.50000000*\st pt, -333.41978046*\st pt)}{b}{$(h-8,h)$}{E}{150.0*\st pt}{}{(-900.00000000*\st pt, -837.74990748*\st pt)}{8.0}{ZagCopy0____n=4-L=3-P=1-PI-B-_0,146_-bottom-right.pdf}
\TIKZPointOfInterest{(-72.50000000*\st pt, 298.77876431*\st pt)}{c}{$(h-W+w+1,1)$}{NE}{100.0*\st pt}{Read \word0}{(-300.00000000*\st pt, -837.74990748*\st pt)}{8.0}{ZagCopy0____n=4-L=3-P=1-PI-C-_-141,0_-read0.pdf}
\TIKZPointOfInterest{(-435.00000000*\st pt, -337.74990748*\st pt)}{d}{$(h-W+w+2,1+h)$}{SE}{100.0*\st pt}{Copy \word0}{(300.00000000*\st pt, -837.74990748*\st pt)}{8.0}{ZagCopy0____n=4-L=3-P=1-PI-D-_-140,147_-copy0.pdf}
\TIKZPointOfInterest{(-687.50000000*\st pt, -333.41978046*\st pt)}{e}{$(h-W-7,h)$}{NW}{100.0*\st pt}{}{(900.00000000*\st pt, -837.74990748*\st pt)}{8.0}{ZagCopy0____n=4-L=3-P=1-PI-E-_-191,146_-bottom-left.pdf}
\TIKZPointOfInterest{(-327.50000000*\st pt, 298.77876431*\st pt)}{f}{$(h-W-8,1)$}{W}{100.0*\st pt}{Next}{(1500.00000000*\st pt, -837.74990748*\st pt)}{8.0}{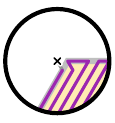}
            }{}
\node [anchor=center] () at (0,0) {\includegraphics[scale =\s]{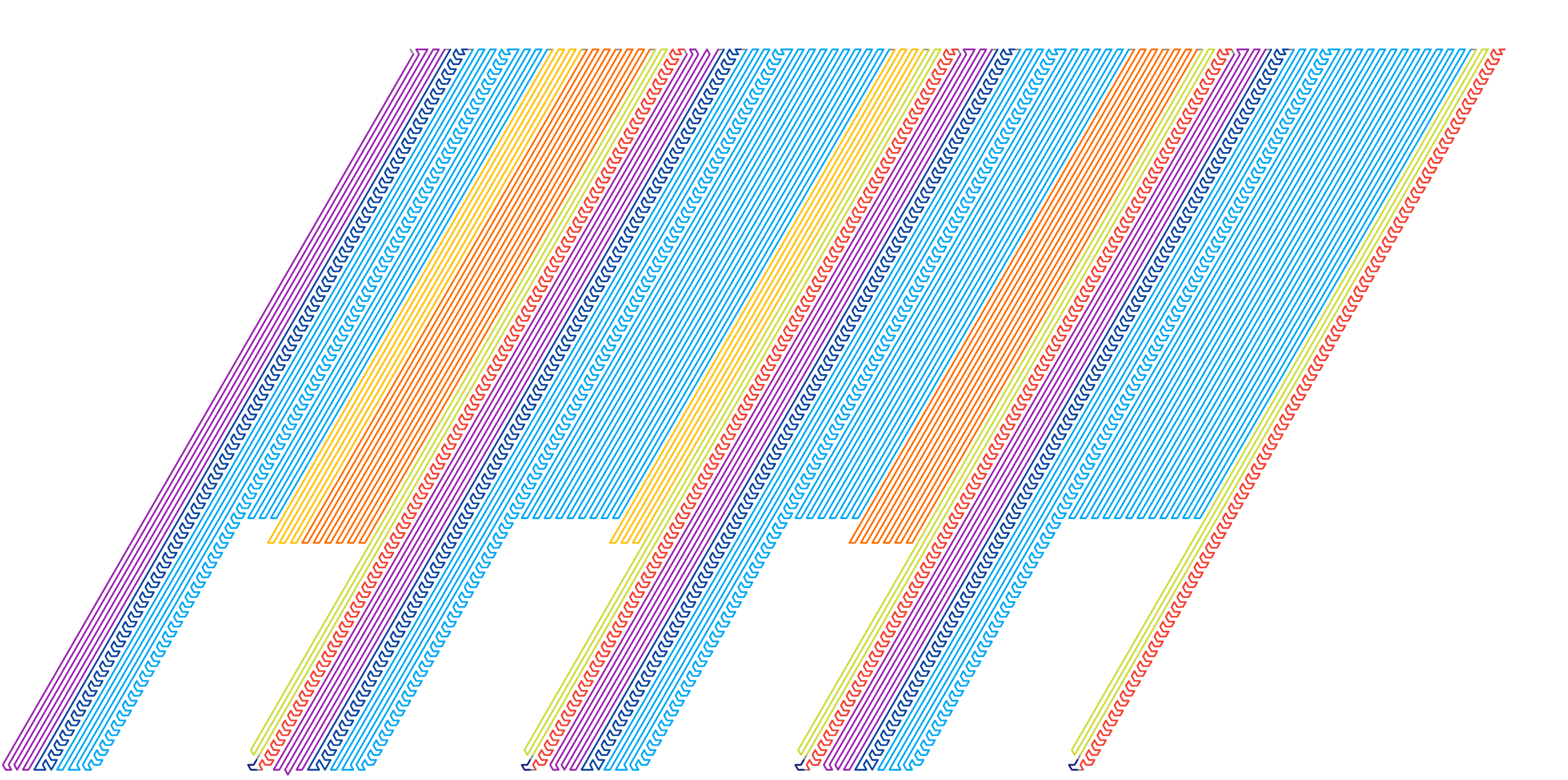}};
            \ifthenelse{\equal{#1}{}}{
\TIKZBLOCKZAGCOPY{0}{1}{\ensuremath{\epsilon}}{-30.000}{-17.321}
            }{}
    \end{tikzpicture}
}

%% file: OritatamiTuringPatterns/n=4-L=3-P=1/ZagCopy1LineFeed_0__n=4-L=3-P=1.tex
\newcommand{\ZagCopyULineFeedBlockZ}[1]{
    \begin{tikzpicture}
        \node [anchor=center] () at (0,0) {\includegraphics[scale =\s]{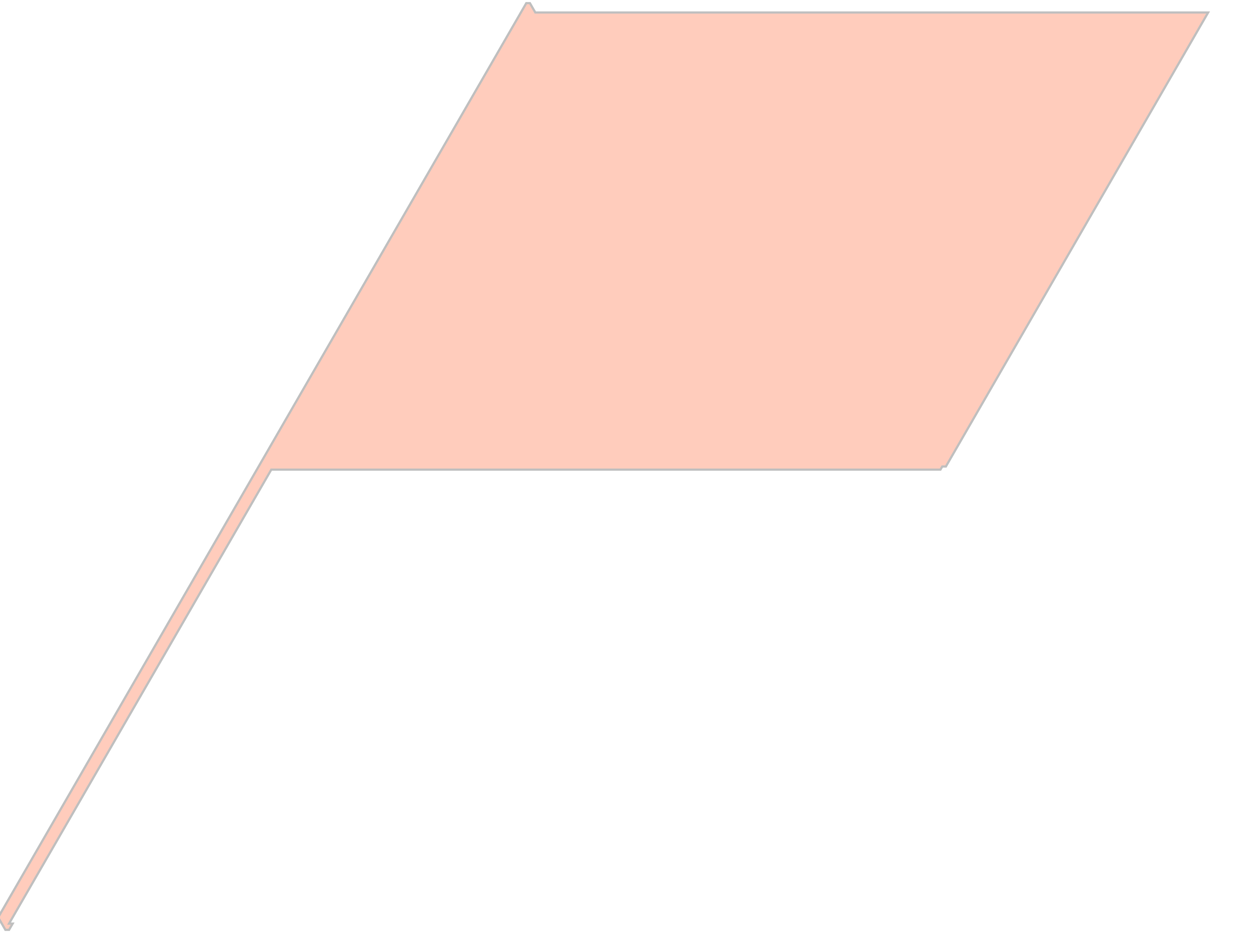}};
        \node [anchor=center] () at (0,0) {\includegraphics[scale =\s]{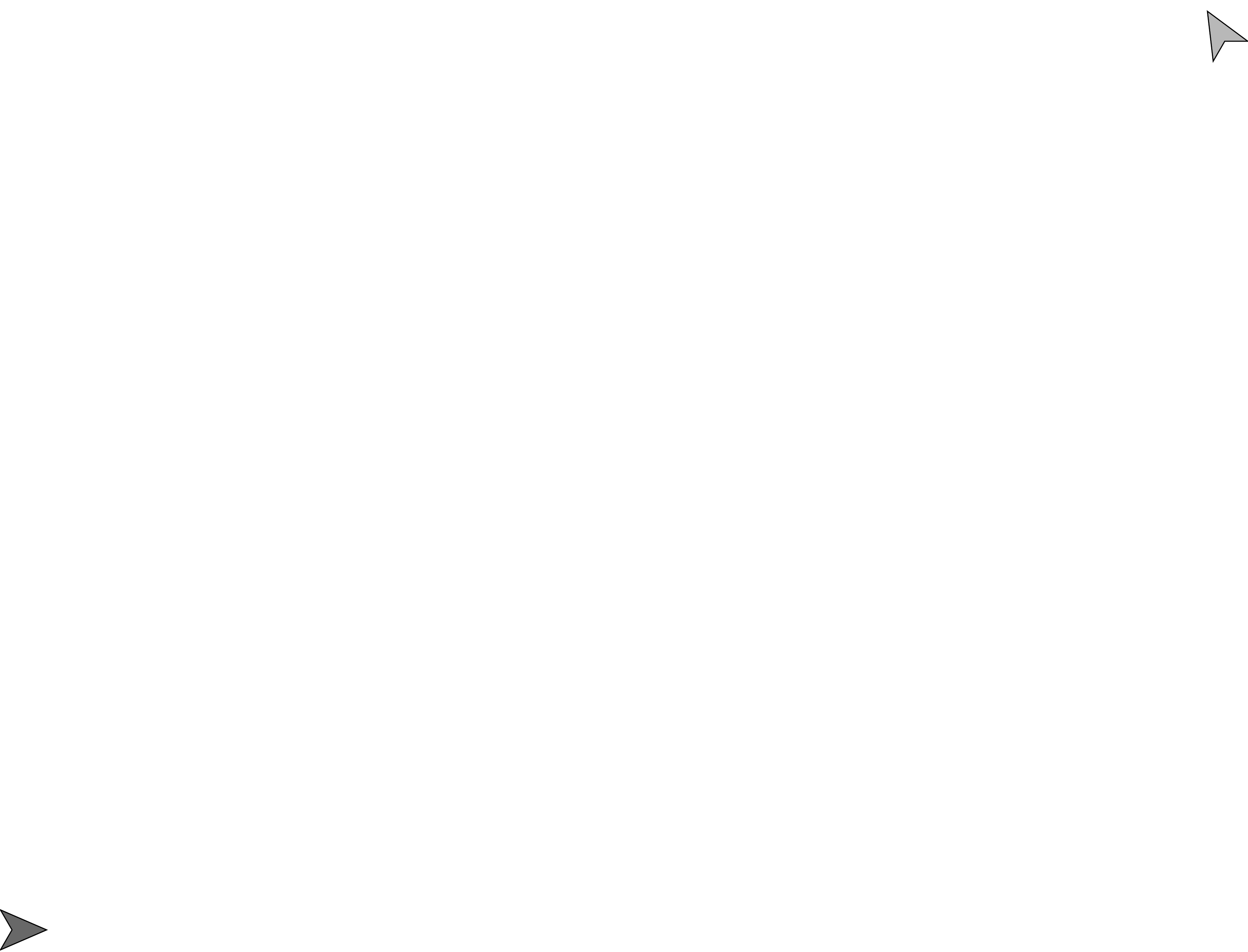}};
        \ifthenelse{\equal{#1}{}}{
\TIKZPointOfInterest{(807.50000000*\st pt, 640.85879880*\st pt)}{a}{$(h-8,1)$}{E}{100.0*\st pt}{start}{(-1500.00000000*\st pt, -953.84917986*\st pt)}{8.0}{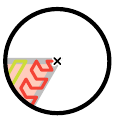}
\TIKZPointOfInterest{(442.50000000*\st pt, 8.66025404*\st pt)}{b}{$(h-8,h)$}{E}{150.0*\st pt}{}{(-900.00000000*\st pt, -953.84917986*\st pt)}{8.0}{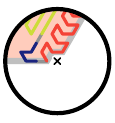}
\TIKZPointOfInterest{(102.50000000*\st pt, 640.85879880*\st pt)}{c}{$(h-W+w+1,1)$}{NE}{75.0*\st pt}{Read \word1}{(-300.00000000*\st pt, -953.84917986*\st pt)}{8.0}{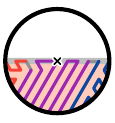}
\TIKZPointOfInterest{(-257.50000000*\st pt, 8.66025404*\st pt)}{d}{$(h-W+w+2,h)$}{SE}{100.0*\st pt}{Copy \word1}{(300.00000000*\st pt, -953.84917986*\st pt)}{8.0}{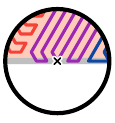}
\TIKZPointOfInterest{(-122.50000000*\st pt, 640.85879880*\st pt)}{e}{$(h-W-2,1)$}{NE}{175.0*\st pt}{}{(900.00000000*\st pt, -953.84917986*\st pt)}{8.0}{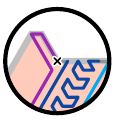}
\TIKZPointOfInterest{(-135.00000000*\st pt, 653.84917986*\st pt)}{f}{$(h-W-6,-2)$}{W}{100.0*\st pt}{}{(1500.00000000*\st pt, -953.84917986*\st pt)}{8.0}{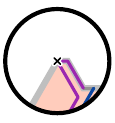}
\TIKZPointOfInterest{(-497.50000000*\st pt, 8.66025404*\st pt)}{g}{$(h-W-4,h)$}{NW}{100.0*\st pt}{}{(-1500.00000000*\st pt, -1553.84917986*\st pt)}{8.0}{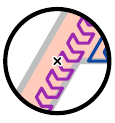}
\TIKZPointOfInterest{(-845.00000000*\st pt, -627.86841774*\st pt)}{h}{$(h-W,2h)$}{NE}{50.0*\st pt}{Next}{(-900.00000000*\st pt, -1553.84917986*\st pt)}{8.0}{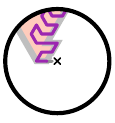}
            }{}
\node [anchor=center] () at (0,0) {\includegraphics[scale =\s]{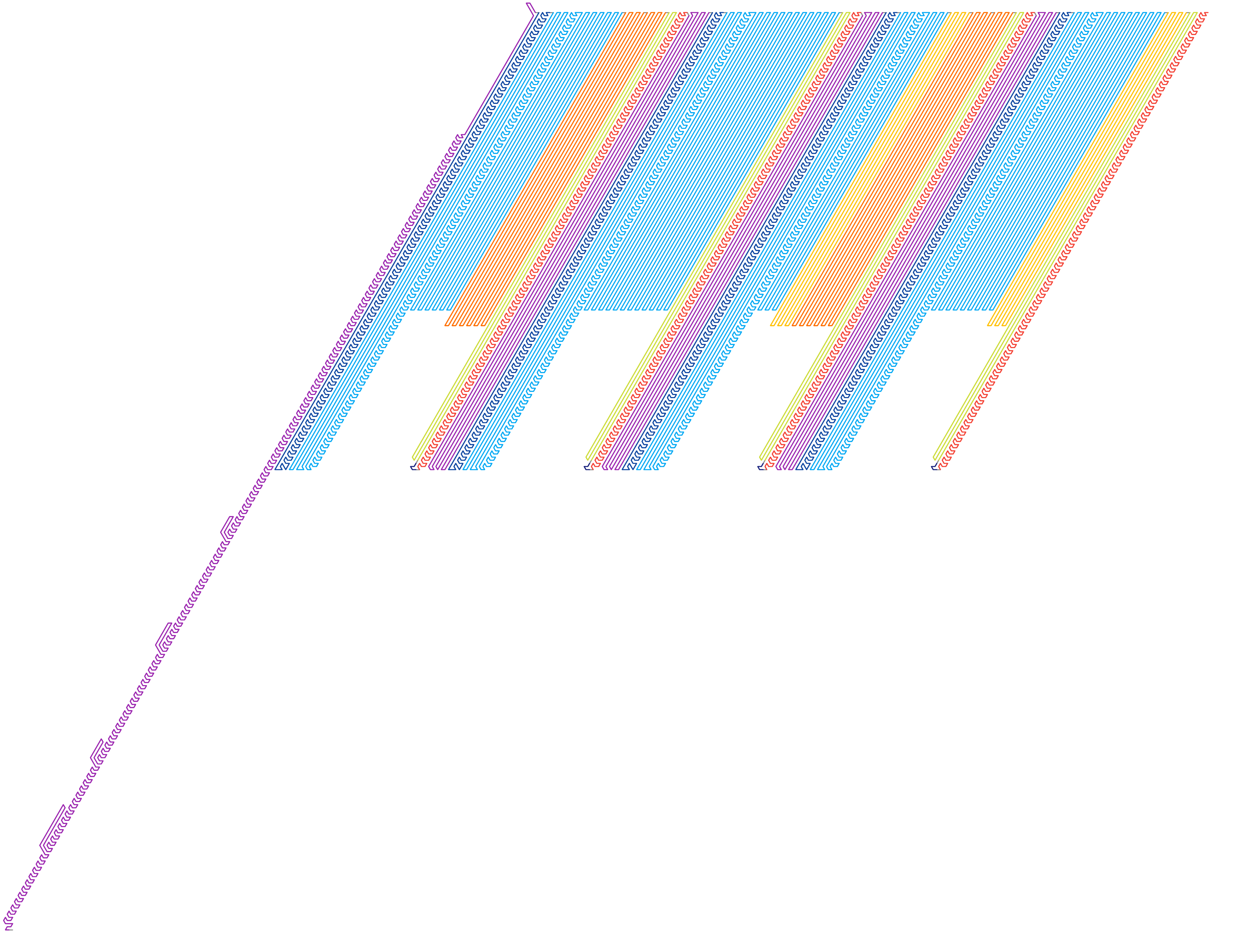}};
            \ifthenelse{\equal{#1}{}}{
\TIKZBLOCKZAGCOPYLINEFEED{1}{3}{0}{145.000}{324.760}
            }{}
    \end{tikzpicture}
}

%% file: OritatamiTuringPatterns/n=4-L=3-P=1/ZagCopy1LineFeed_110__n=4-L=3-P=1.tex
\newcommand{\ZagCopyULineFeedBlockUUZ}[1]{
    \begin{tikzpicture}
        \node [anchor=center] () at (0,0) {\includegraphics[scale =\s]{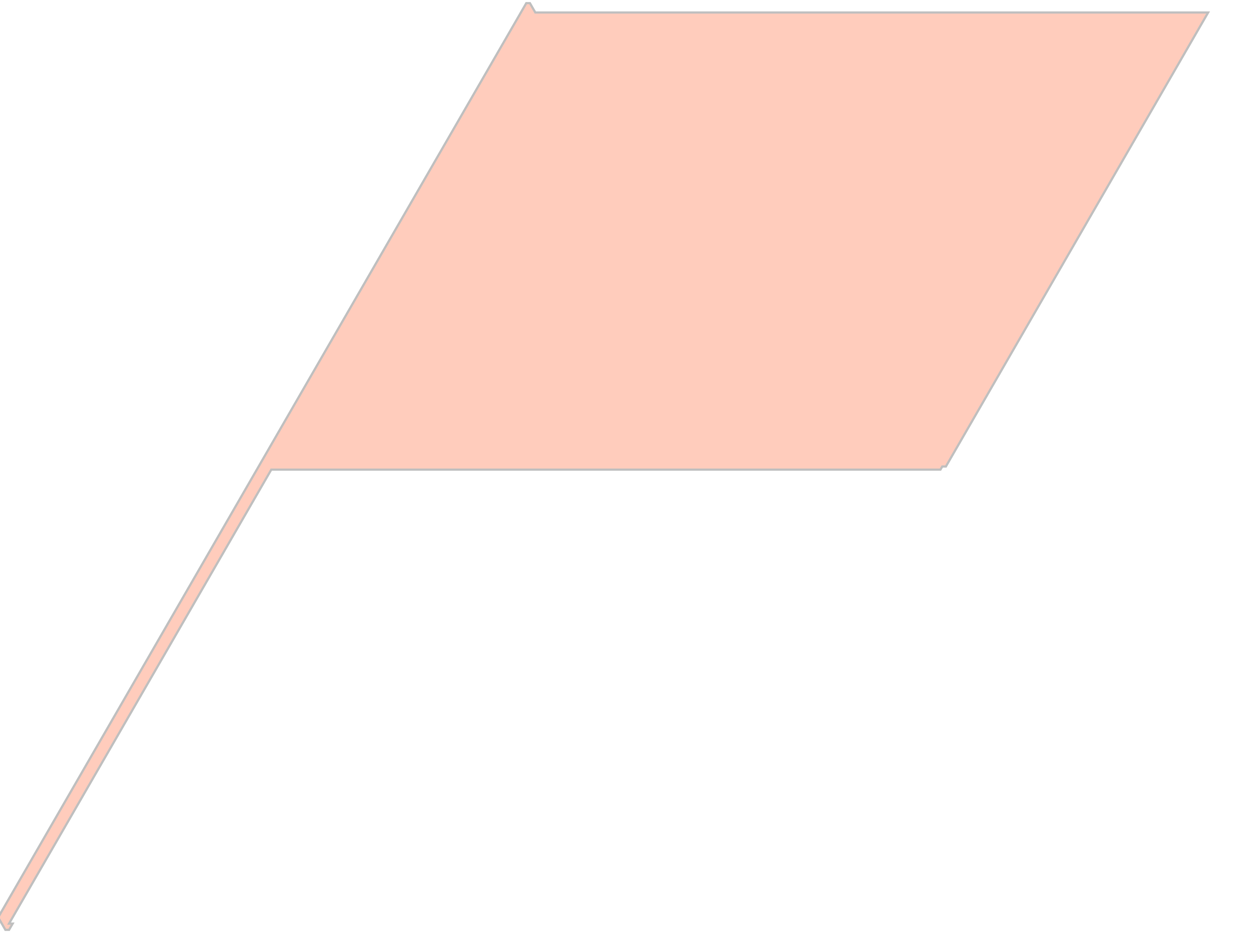}};
        \node [anchor=center] () at (0,0) {\includegraphics[scale =\s]{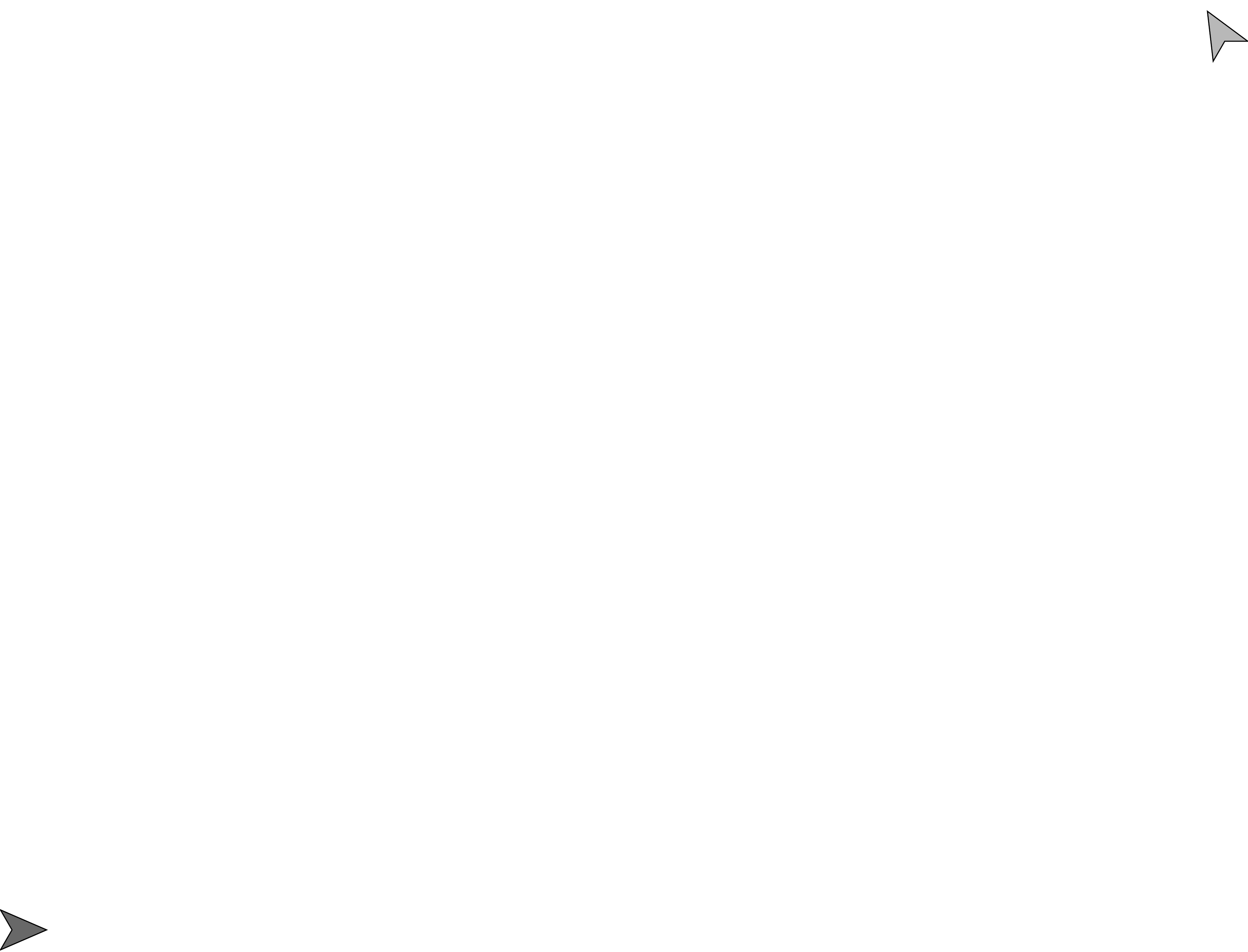}};
        \ifthenelse{\equal{#1}{}}{
\TIKZPointOfInterest{(807.50000000*\st pt, 640.85879880*\st pt)}{a}{$(h-8,1)$}{E}{100.0*\st pt}{start}{(-1500.00000000*\st pt, -953.84917986*\st pt)}{8.0}{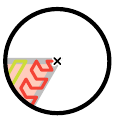}
\TIKZPointOfInterest{(442.50000000*\st pt, 8.66025404*\st pt)}{b}{$(h-8,h)$}{E}{150.0*\st pt}{}{(-900.00000000*\st pt, -953.84917986*\st pt)}{8.0}{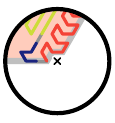}
\TIKZPointOfInterest{(102.50000000*\st pt, 640.85879880*\st pt)}{c}{$(h-W+w+1,1)$}{NE}{75.0*\st pt}{Read \word1}{(-300.00000000*\st pt, -953.84917986*\st pt)}{8.0}{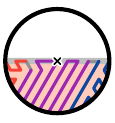}
\TIKZPointOfInterest{(-257.50000000*\st pt, 8.66025404*\st pt)}{d}{$(h-W+w+2,h)$}{SE}{100.0*\st pt}{Copy \word1}{(300.00000000*\st pt, -953.84917986*\st pt)}{8.0}{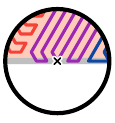}
\TIKZPointOfInterest{(-122.50000000*\st pt, 640.85879880*\st pt)}{e}{$(h-W-2,1)$}{NE}{175.0*\st pt}{}{(900.00000000*\st pt, -953.84917986*\st pt)}{8.0}{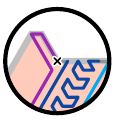}
\TIKZPointOfInterest{(-135.00000000*\st pt, 653.84917986*\st pt)}{f}{$(h-W-6,-2)$}{W}{100.0*\st pt}{}{(1500.00000000*\st pt, -953.84917986*\st pt)}{8.0}{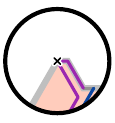}
\TIKZPointOfInterest{(-497.50000000*\st pt, 8.66025404*\st pt)}{g}{$(h-W-4,h)$}{NW}{100.0*\st pt}{}{(-1500.00000000*\st pt, -1553.84917986*\st pt)}{8.0}{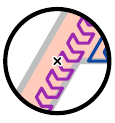}
\TIKZPointOfInterest{(-845.00000000*\st pt, -627.86841774*\st pt)}{h}{$(h-W,2h)$}{NE}{50.0*\st pt}{Next}{(-900.00000000*\st pt, -1553.84917986*\st pt)}{8.0}{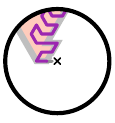}
            }{}
\node [anchor=center] () at (0,0) {\includegraphics[scale =\s]{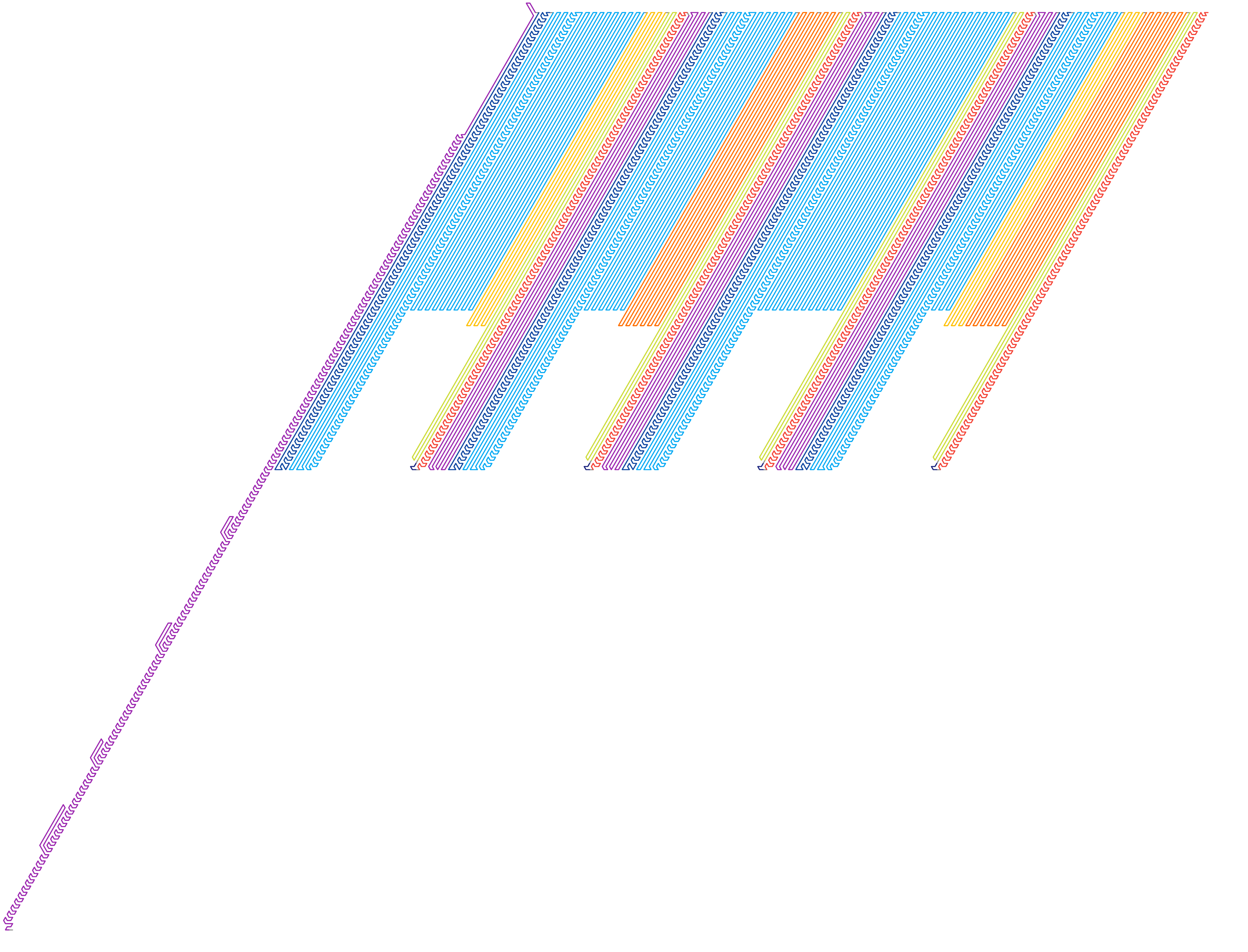}};
            \ifthenelse{\equal{#1}{}}{
\TIKZBLOCKZAGCOPYLINEFEED{1}{0}{110}{145.000}{324.760}
            }{}
    \end{tikzpicture}
}

%% file: OritatamiTuringPatterns/n=4-L=3-P=1/ZagCopy1LineFeed_11__n=4-L=3-P=1.tex
\newcommand{\ZagCopyULineFeedBlockUU}[1]{
    \begin{tikzpicture}
        \node [anchor=center] () at (0,0) {\includegraphics[scale =\s]{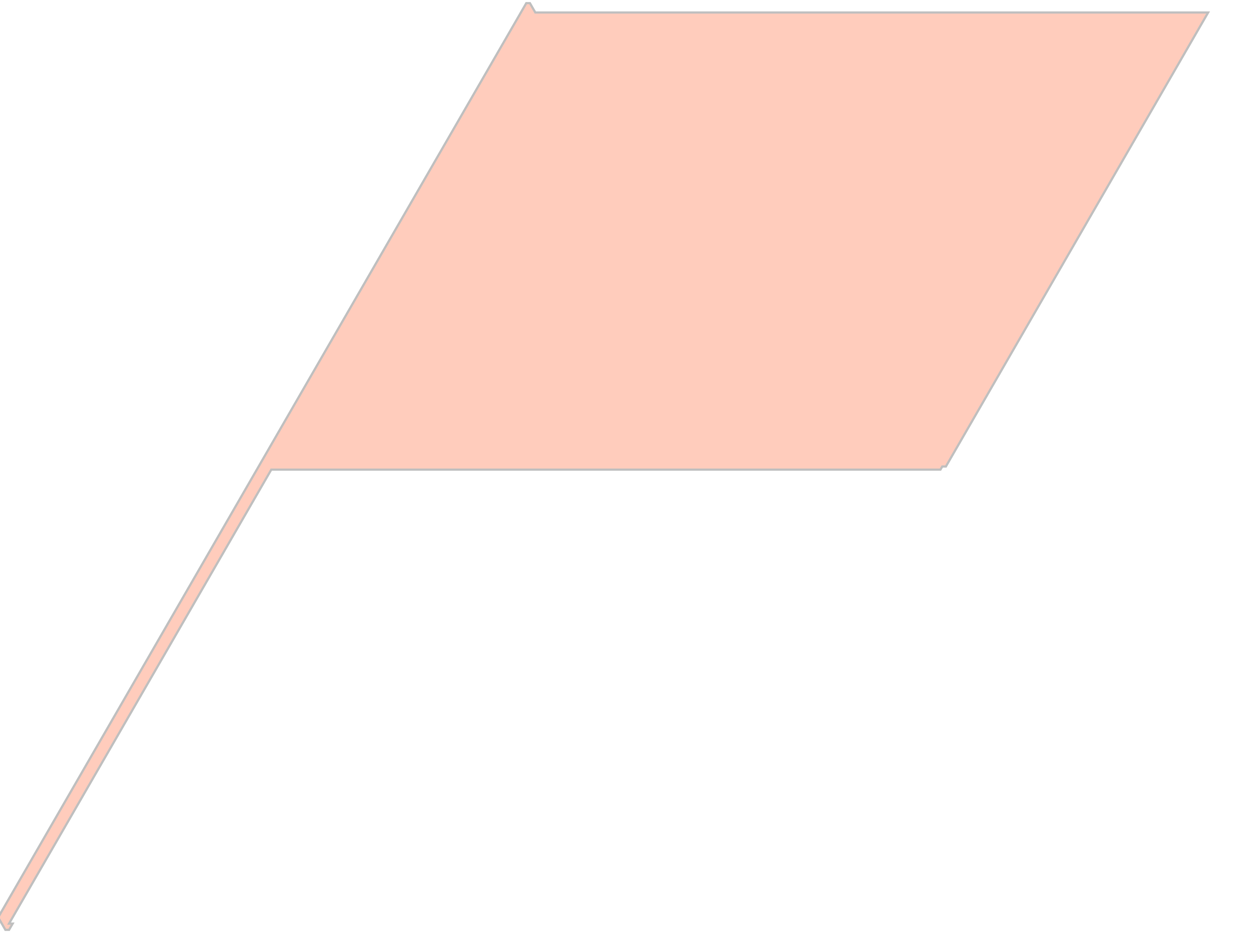}};
        \node [anchor=center] () at (0,0) {\includegraphics[scale =\s]{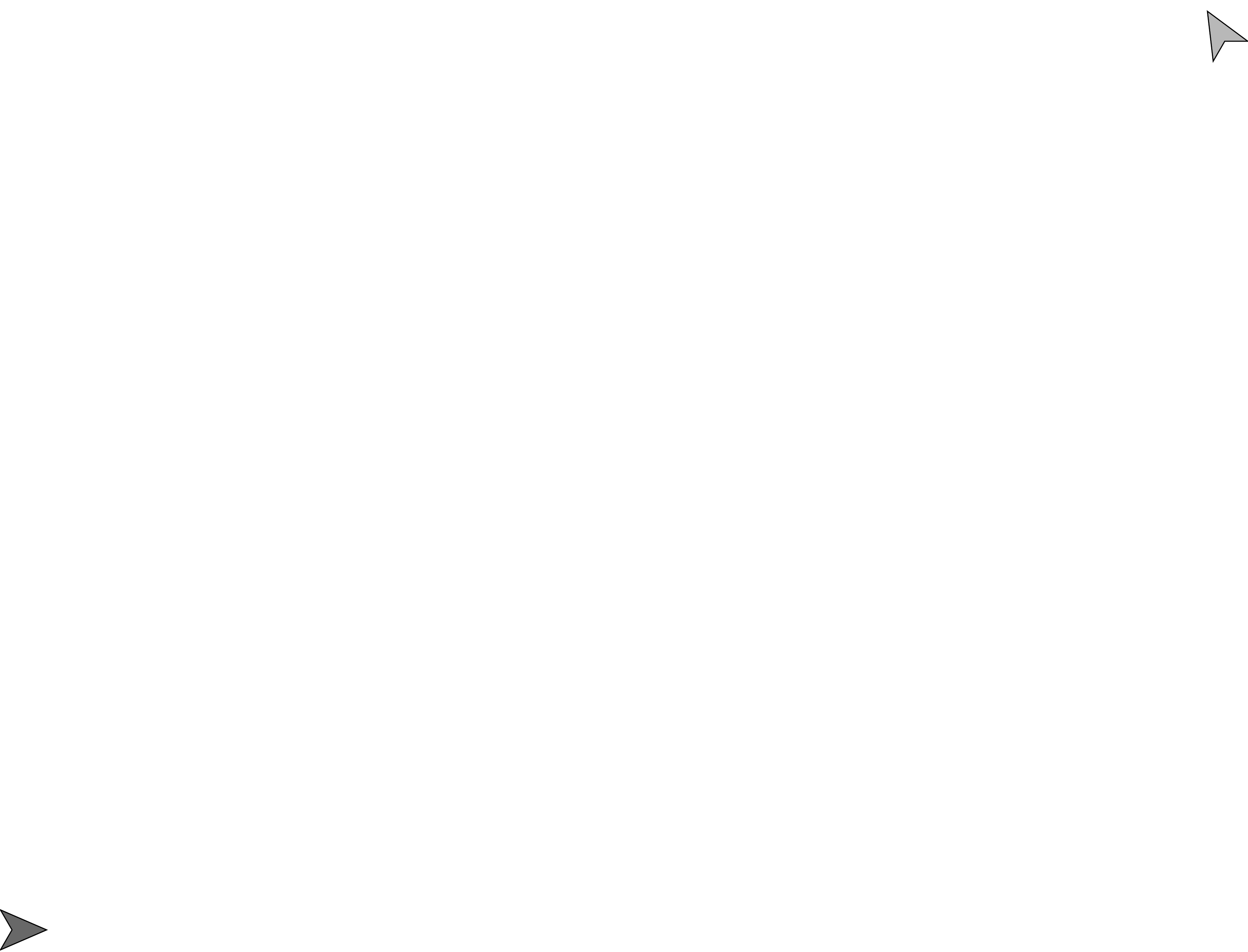}};
        \ifthenelse{\equal{#1}{}}{
\TIKZPointOfInterest{(807.50000000*\st pt, 640.85879880*\st pt)}{a}{$(h-8,1)$}{E}{100.0*\st pt}{start}{(-1500.00000000*\st pt, -953.84917986*\st pt)}{8.0}{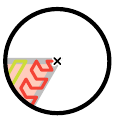}
\TIKZPointOfInterest{(442.50000000*\st pt, 8.66025404*\st pt)}{b}{$(h-8,h)$}{E}{150.0*\st pt}{}{(-900.00000000*\st pt, -953.84917986*\st pt)}{8.0}{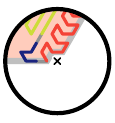}
\TIKZPointOfInterest{(102.50000000*\st pt, 640.85879880*\st pt)}{c}{$(h-W+w+1,1)$}{NE}{75.0*\st pt}{Read \word1}{(-300.00000000*\st pt, -953.84917986*\st pt)}{8.0}{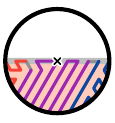}
\TIKZPointOfInterest{(-257.50000000*\st pt, 8.66025404*\st pt)}{d}{$(h-W+w+2,h)$}{SE}{100.0*\st pt}{Copy \word1}{(300.00000000*\st pt, -953.84917986*\st pt)}{8.0}{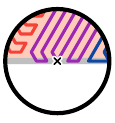}
\TIKZPointOfInterest{(-122.50000000*\st pt, 640.85879880*\st pt)}{e}{$(h-W-2,1)$}{NE}{175.0*\st pt}{}{(900.00000000*\st pt, -953.84917986*\st pt)}{8.0}{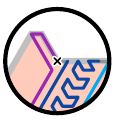}
\TIKZPointOfInterest{(-135.00000000*\st pt, 653.84917986*\st pt)}{f}{$(h-W-6,-2)$}{W}{100.0*\st pt}{}{(1500.00000000*\st pt, -953.84917986*\st pt)}{8.0}{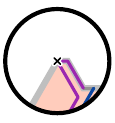}
\TIKZPointOfInterest{(-497.50000000*\st pt, 8.66025404*\st pt)}{g}{$(h-W-4,h)$}{NW}{100.0*\st pt}{}{(-1500.00000000*\st pt, -1553.84917986*\st pt)}{8.0}{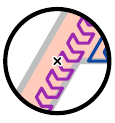}
\TIKZPointOfInterest{(-845.00000000*\st pt, -627.86841774*\st pt)}{h}{$(h-W,2h)$}{NE}{50.0*\st pt}{Next}{(-900.00000000*\st pt, -1553.84917986*\st pt)}{8.0}{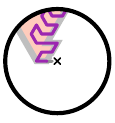}
            }{}
\node [anchor=center] () at (0,0) {\includegraphics[scale =\s]{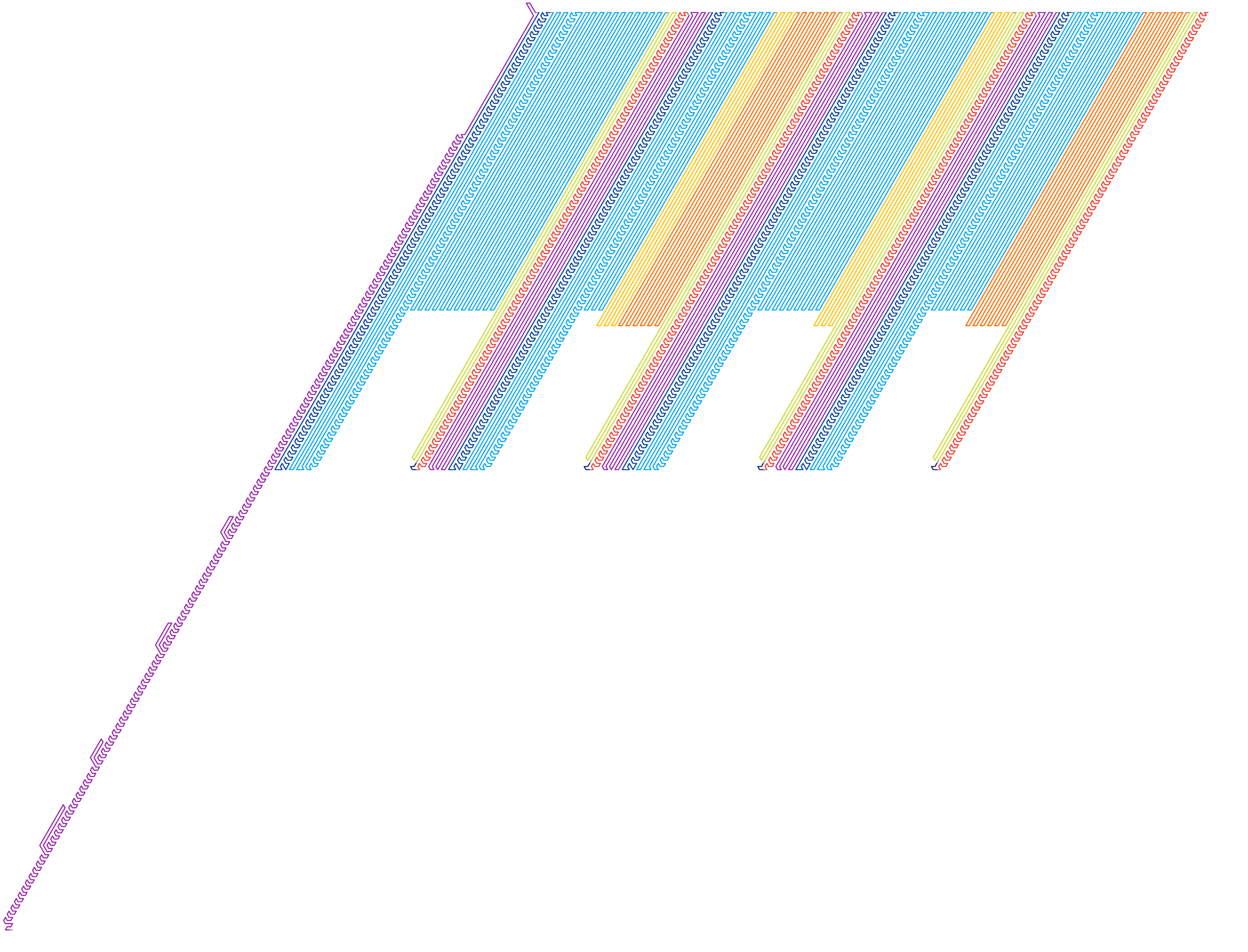}};
            \ifthenelse{\equal{#1}{}}{
\TIKZBLOCKZAGCOPYLINEFEED{1}{2}{11}{145.000}{324.760}
            }{}
    \end{tikzpicture}
}

%% file: OritatamiTuringPatterns/n=4-L=3-P=1/ZagCopy1LineFeed_____n=4-L=3-P=1.tex
\newcommand{\ZagCopyULineFeedBlockEpsilon}[1]{
    \begin{tikzpicture}
        \node [anchor=center] () at (0,0) {\includegraphics[scale =\s]{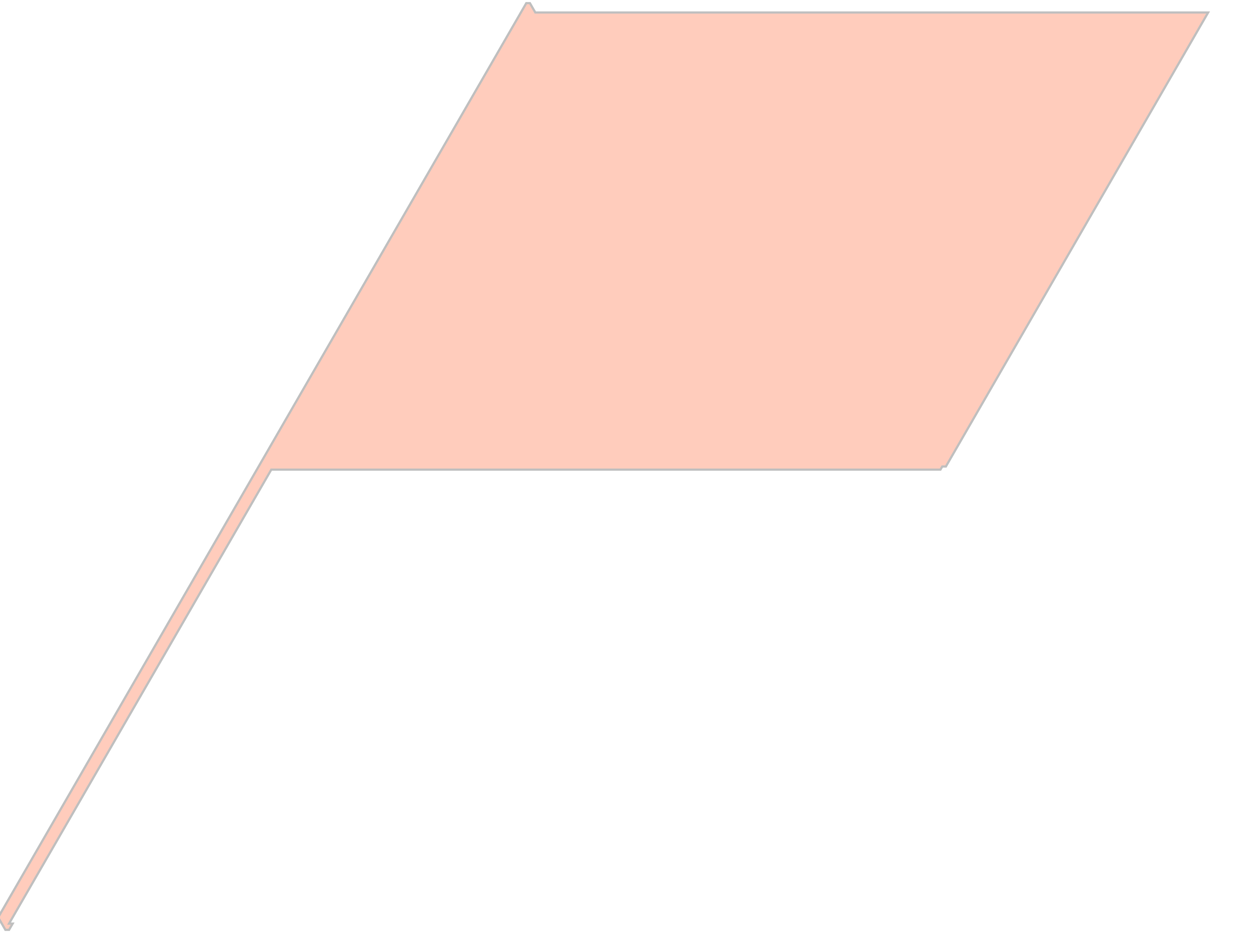}};
        \node [anchor=center] () at (0,0) {\includegraphics[scale =\s]{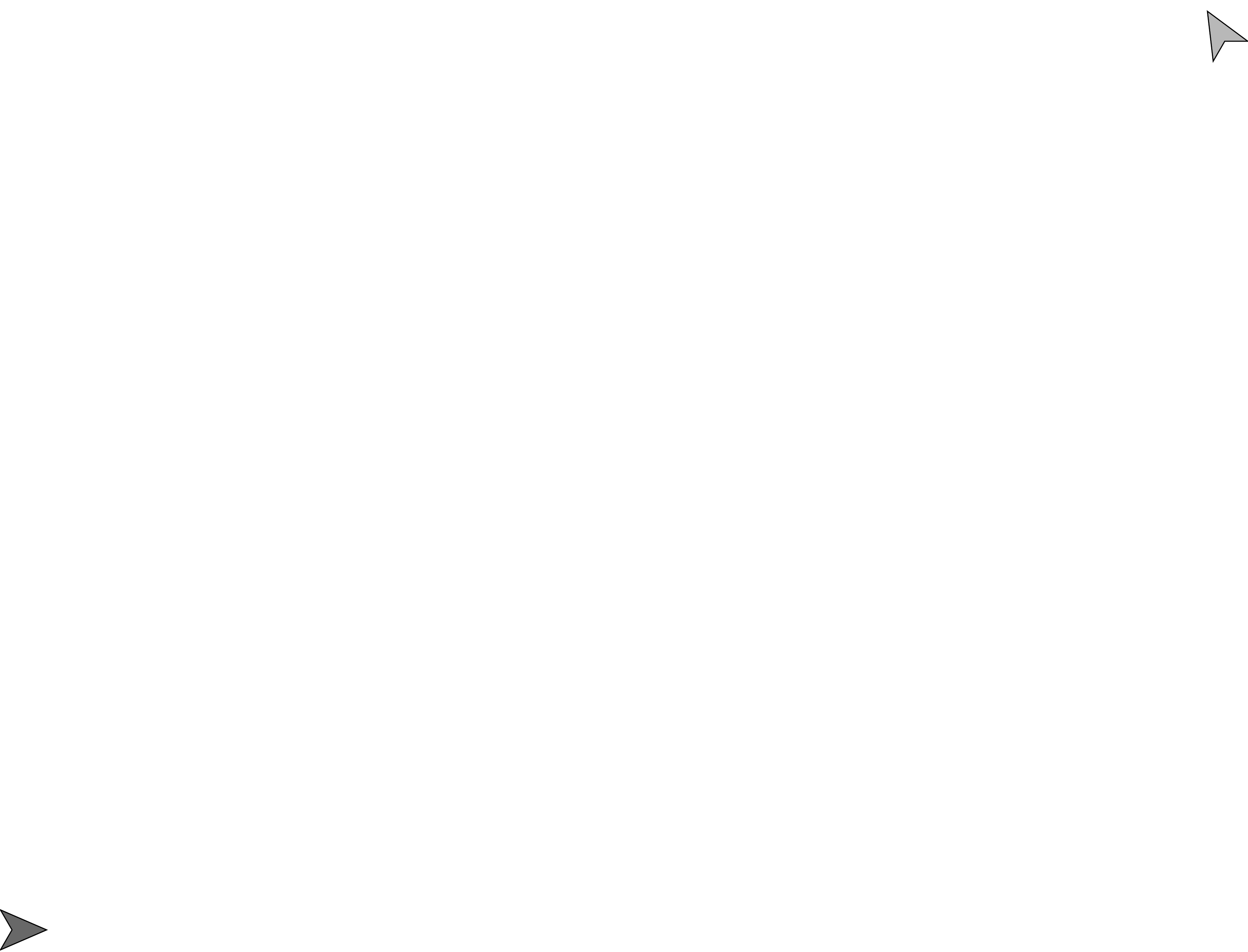}};
        \ifthenelse{\equal{#1}{}}{
\TIKZPointOfInterest{(807.50000000*\st pt, 640.85879880*\st pt)}{a}{$(h-8,1)$}{E}{100.0*\st pt}{start}{(-1500.00000000*\st pt, -953.84917986*\st pt)}{8.0}{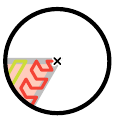}
\TIKZPointOfInterest{(442.50000000*\st pt, 8.66025404*\st pt)}{b}{$(h-8,h)$}{E}{150.0*\st pt}{}{(-900.00000000*\st pt, -953.84917986*\st pt)}{8.0}{ZagCopy1LineFeed____n=4-L=3-P=1-PI-B-_0,146_-bottom-right.pdf}
\TIKZPointOfInterest{(102.50000000*\st pt, 640.85879880*\st pt)}{c}{$(h-W+w+1,1)$}{NE}{75.0*\st pt}{Read \word1}{(-300.00000000*\st pt, -953.84917986*\st pt)}{8.0}{ZagCopy1LineFeed____n=4-L=3-P=1-PI-C-_-141,0_-read1.pdf}
\TIKZPointOfInterest{(-257.50000000*\st pt, 8.66025404*\st pt)}{d}{$(h-W+w+2,h)$}{SE}{100.0*\st pt}{Copy \word1}{(300.00000000*\st pt, -953.84917986*\st pt)}{8.0}{ZagCopy1LineFeed____n=4-L=3-P=1-PI-D-_-140,146_-copy1.pdf}
\TIKZPointOfInterest{(-122.50000000*\st pt, 640.85879880*\st pt)}{e}{$(h-W-2,1)$}{NE}{175.0*\st pt}{}{(900.00000000*\st pt, -953.84917986*\st pt)}{8.0}{ZagCopy1LineFeed____n=4-L=3-P=1-PI-E-_-186,0_-top-left.pdf}
\TIKZPointOfInterest{(-135.00000000*\st pt, 653.84917986*\st pt)}{f}{$(h-W-6,-2)$}{W}{100.0*\st pt}{}{(1500.00000000*\st pt, -953.84917986*\st pt)}{8.0}{ZagCopy1LineFeed____n=4-L=3-P=1-PI-F-_-190,-3_-top-left-corner.pdf}
\TIKZPointOfInterest{(-497.50000000*\st pt, 8.66025404*\st pt)}{g}{$(h-W-4,h)$}{NW}{100.0*\st pt}{}{(-1500.00000000*\st pt, -1553.84917986*\st pt)}{8.0}{ZagCopy1LineFeed____n=4-L=3-P=1-PI-G-_-188,146_-mid-left.pdf}
\TIKZPointOfInterest{(-845.00000000*\st pt, -627.86841774*\st pt)}{h}{$(h-W,2h)$}{NE}{50.0*\st pt}{Next}{(-900.00000000*\st pt, -1553.84917986*\st pt)}{8.0}{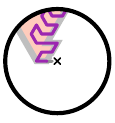}
            }{}
\node [anchor=center] () at (0,0) {\includegraphics[scale =\s]{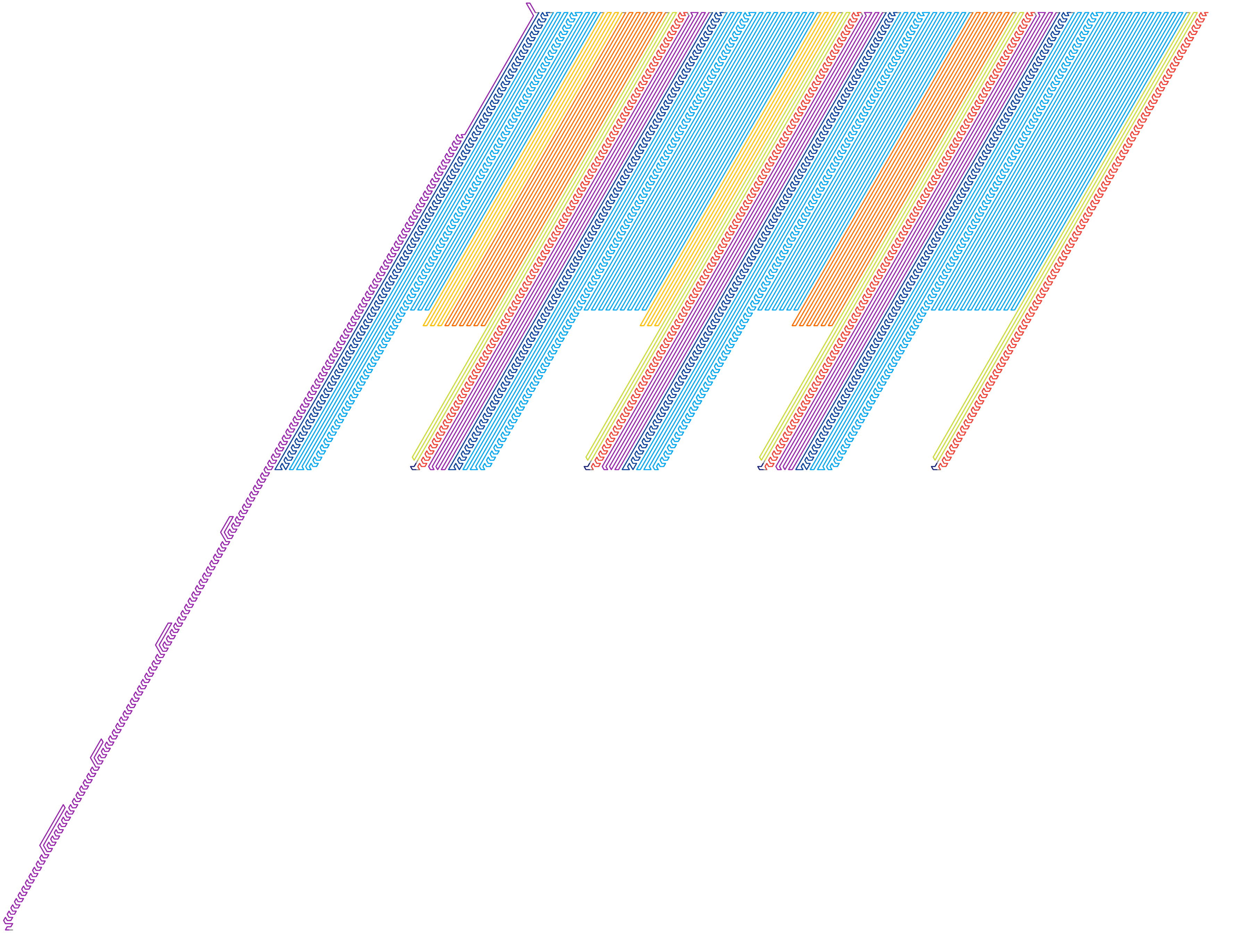}};
            \ifthenelse{\equal{#1}{}}{
\TIKZBLOCKZAGCOPYLINEFEED{1}{1}{\ensuremath{\epsilon}}{145.000}{324.760}
            }{}
    \end{tikzpicture}
}

%% file: OritatamiTuringPatterns/n=4-L=3-P=1/ZagCopy1_0__n=4-L=3-P=1.tex
\newcommand{\ZagCopyUBlockZ}[1]{
    \begin{tikzpicture}
        \node [anchor=center] () at (0,0) {\includegraphics[scale =\s]{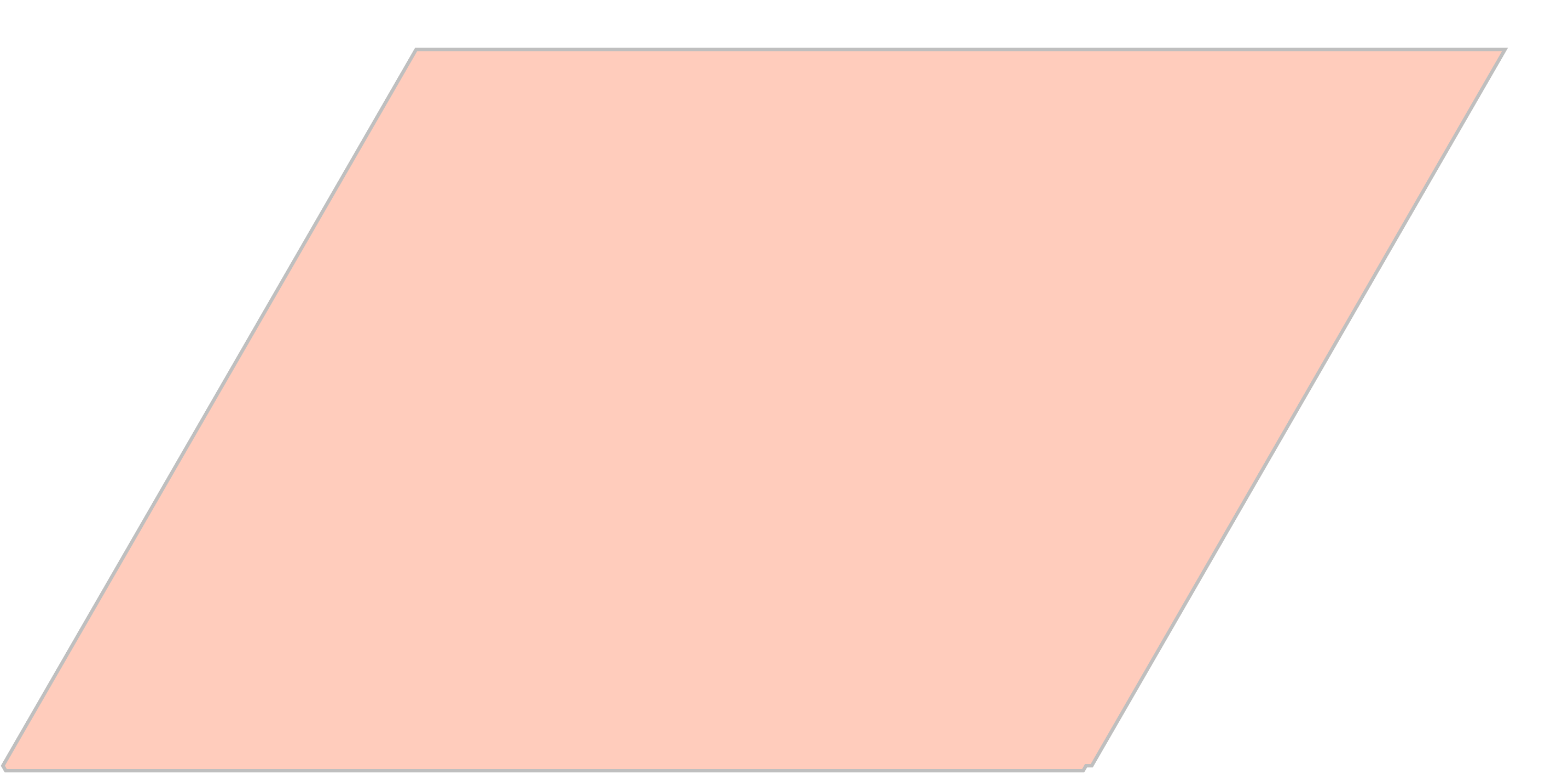}};
        \node [anchor=center] () at (0,0) {\includegraphics[scale =\s]{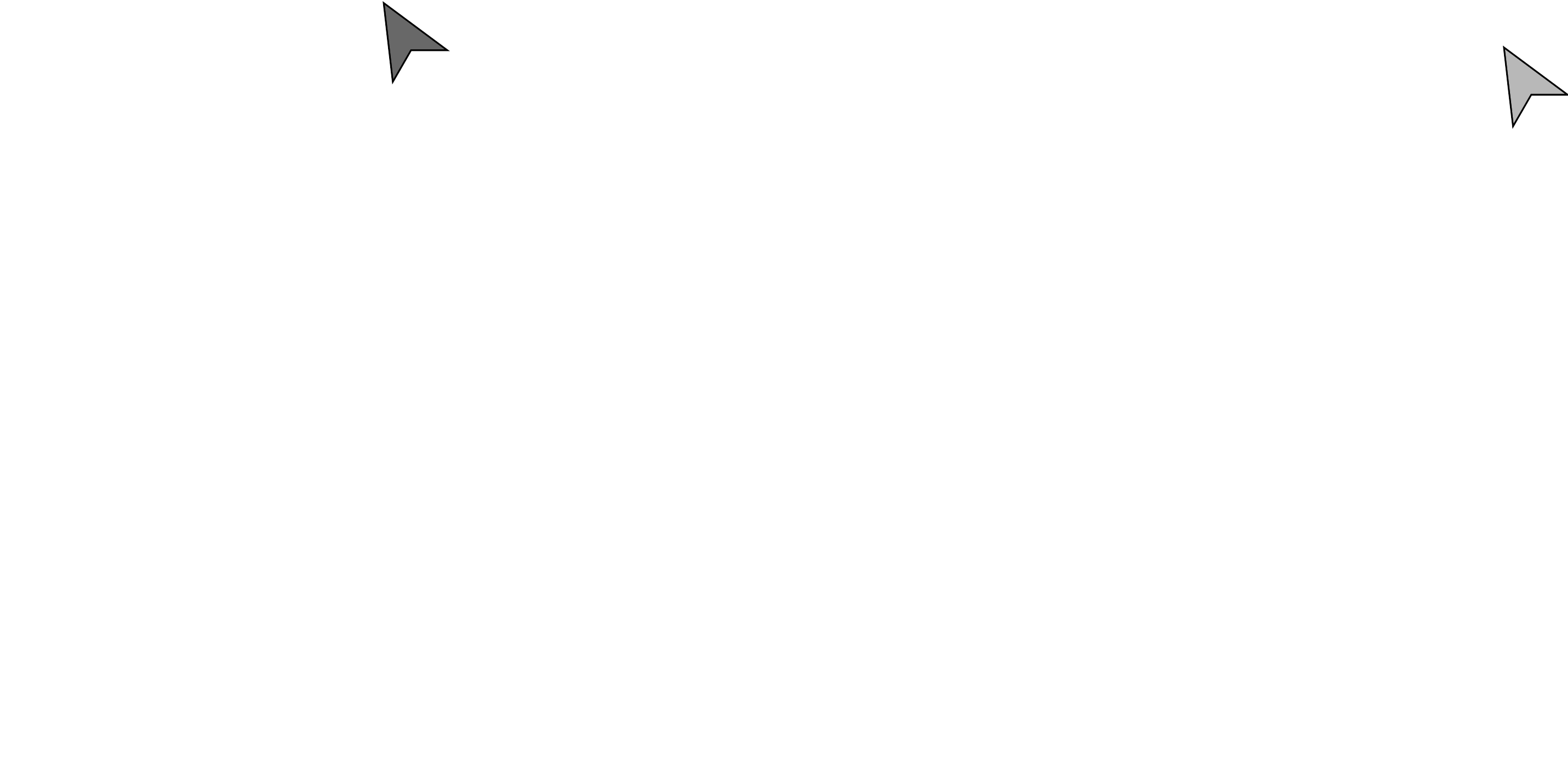}};
        \ifthenelse{\equal{#1}{}}{
\TIKZPointOfInterest{(632.50000000*\st pt, 296.61370080*\st pt)}{a}{$(h-8,1)$}{E}{100.0*\st pt}{Start}{(-1500.00000000*\st pt, -835.58484397*\st pt)}{8.0}{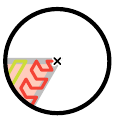}
\TIKZPointOfInterest{(267.50000000*\st pt, -335.58484397*\st pt)}{b}{$(h-8,h)$}{E}{150.0*\st pt}{}{(-900.00000000*\st pt, -835.58484397*\st pt)}{8.0}{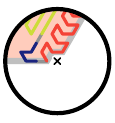}
\TIKZPointOfInterest{(-72.50000000*\st pt, 296.61370080*\st pt)}{c}{$(h-W+w+1,1)$}{NE}{100.0*\st pt}{Read \word1}{(-300.00000000*\st pt, -835.58484397*\st pt)}{8.0}{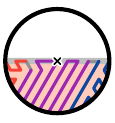}
\TIKZPointOfInterest{(-432.50000000*\st pt, -335.58484397*\st pt)}{d}{$(h-W+w+2,h)$}{SE}{100.0*\st pt}{Copy \word1}{(300.00000000*\st pt, -835.58484397*\st pt)}{8.0}{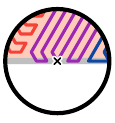}
\TIKZPointOfInterest{(-687.50000000*\st pt, -335.58484397*\st pt)}{e}{$(h-W-7,h)$}{NW}{100.0*\st pt}{}{(900.00000000*\st pt, -835.58484397*\st pt)}{8.0}{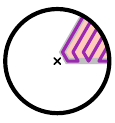}
\TIKZPointOfInterest{(-327.50000000*\st pt, 296.61370080*\st pt)}{f}{$(h-W-8,1)$}{W}{100.0*\st pt}{Next}{(1500.00000000*\st pt, -835.58484397*\st pt)}{8.0}{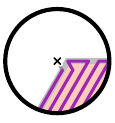}
            }{}
\node [anchor=center] () at (0,0) {\includegraphics[scale =\s]{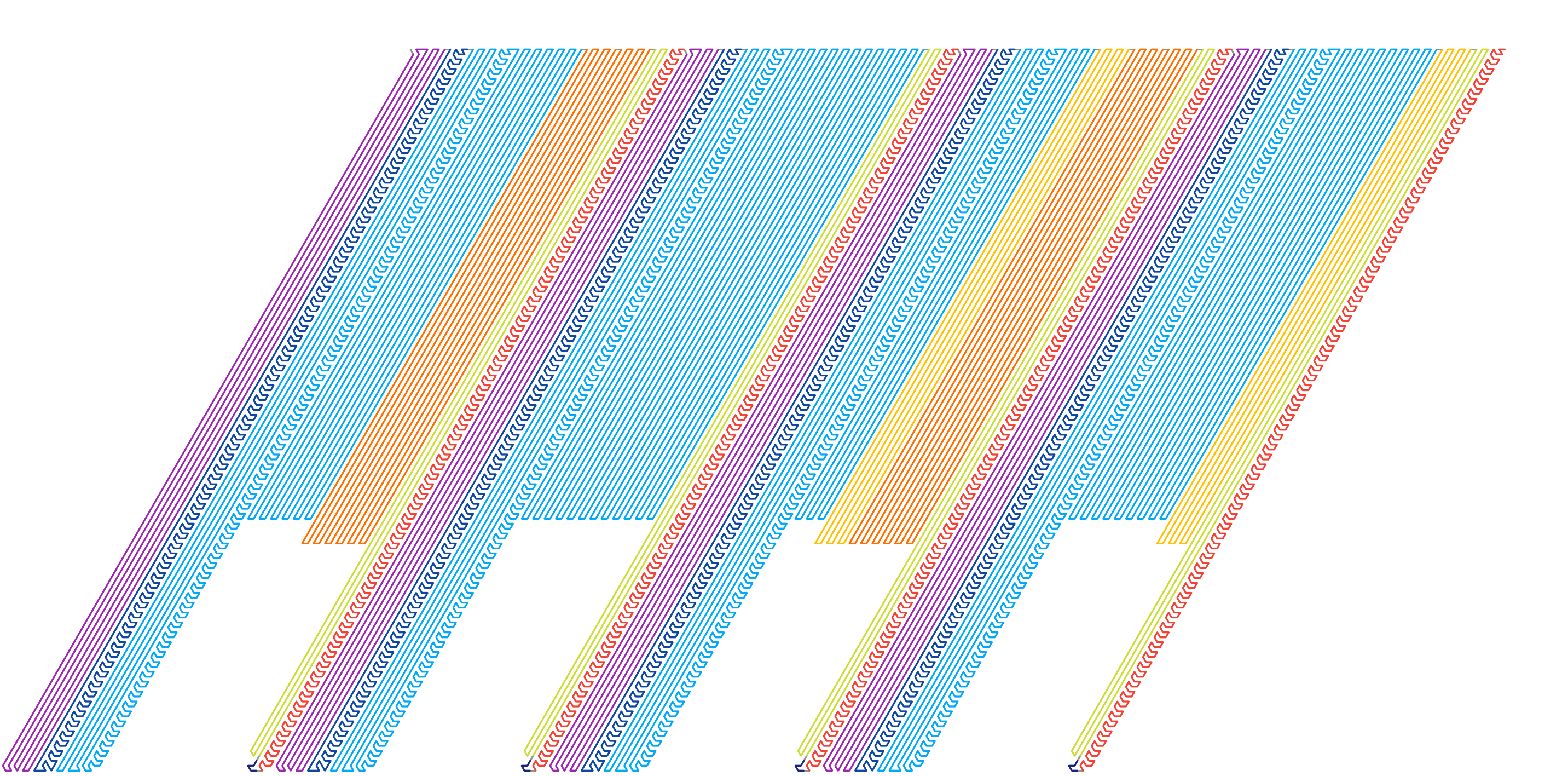}};
            \ifthenelse{\equal{#1}{}}{
\TIKZBLOCKZAGCOPY{1}{3}{0}{-30.000}{-19.486}
            }{}
    \end{tikzpicture}
}

%% file: OritatamiTuringPatterns/n=4-L=3-P=1/ZagCopy1_110__n=4-L=3-P=1.tex
\newcommand{\ZagCopyUBlockUUZ}[1]{
    \begin{tikzpicture}
        \node [anchor=center] () at (0,0) {\includegraphics[scale =\s]{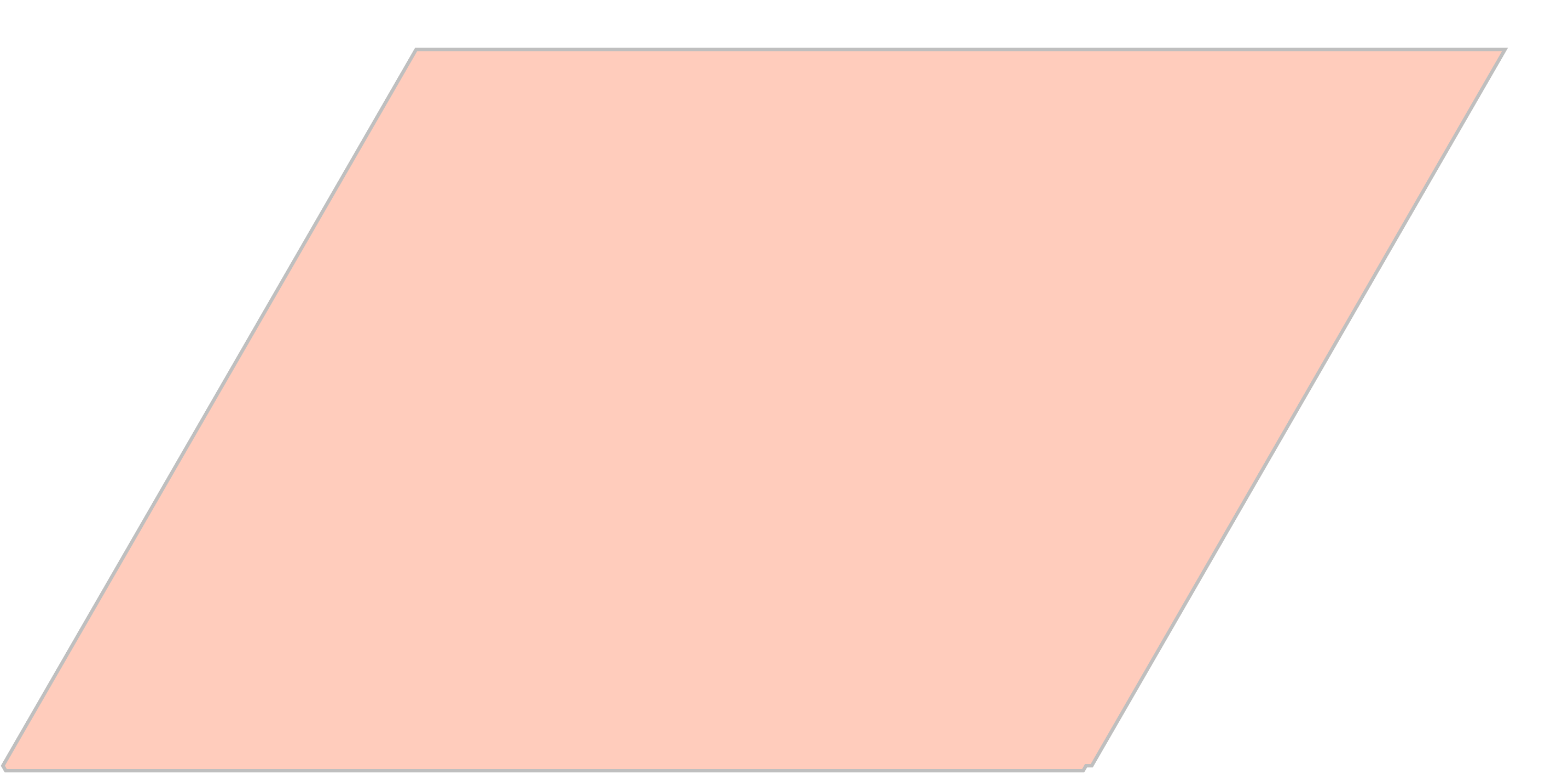}};
        \node [anchor=center] () at (0,0) {\includegraphics[scale =\s]{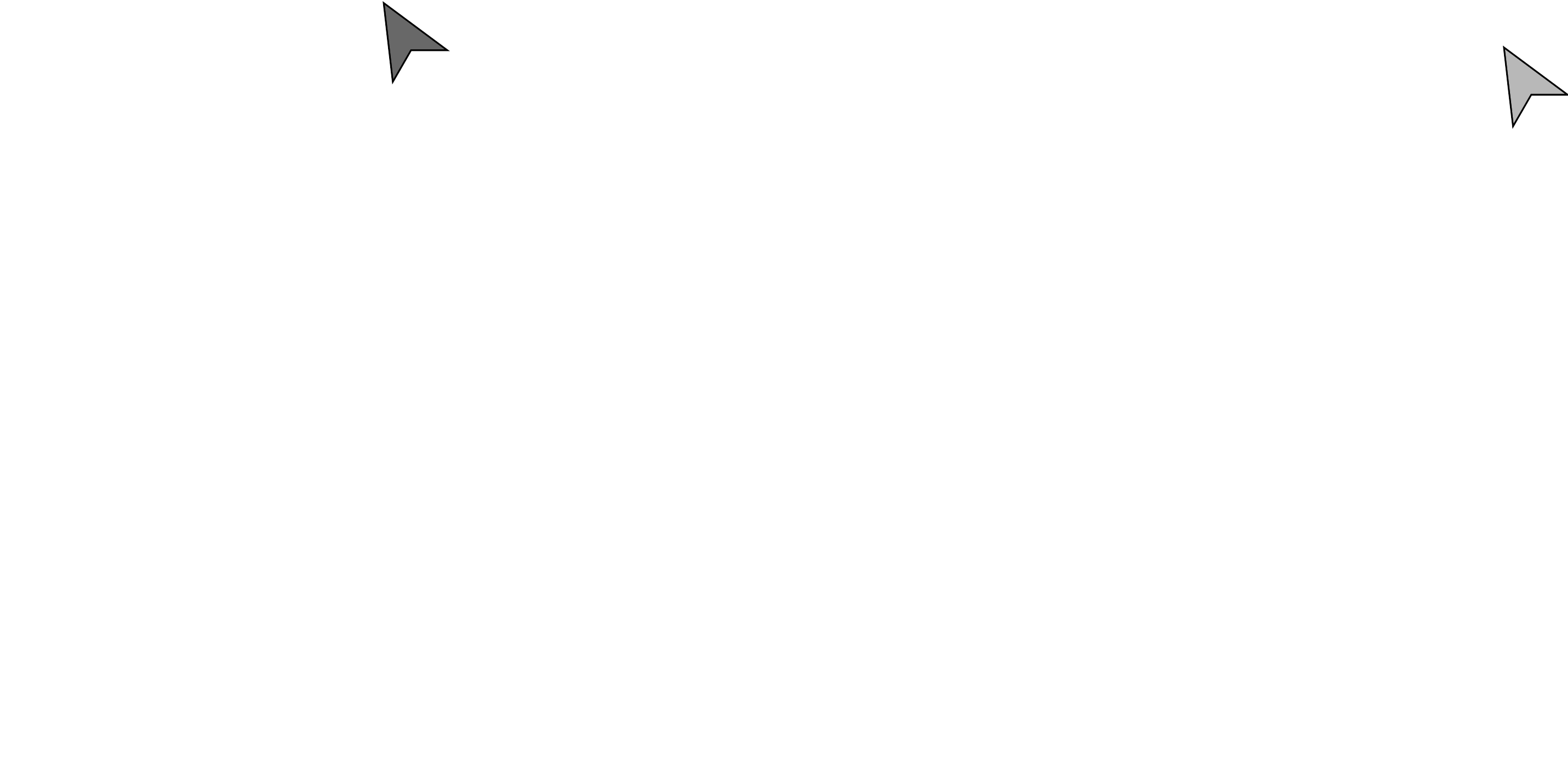}};
        \ifthenelse{\equal{#1}{}}{
\TIKZPointOfInterest{(632.50000000*\st pt, 296.61370080*\st pt)}{a}{$(h-8,1)$}{E}{100.0*\st pt}{Start}{(-1500.00000000*\st pt, -835.58484397*\st pt)}{8.0}{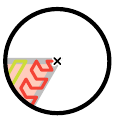}
\TIKZPointOfInterest{(267.50000000*\st pt, -335.58484397*\st pt)}{b}{$(h-8,h)$}{E}{150.0*\st pt}{}{(-900.00000000*\st pt, -835.58484397*\st pt)}{8.0}{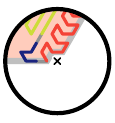}
\TIKZPointOfInterest{(-72.50000000*\st pt, 296.61370080*\st pt)}{c}{$(h-W+w+1,1)$}{NE}{100.0*\st pt}{Read \word1}{(-300.00000000*\st pt, -835.58484397*\st pt)}{8.0}{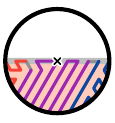}
\TIKZPointOfInterest{(-432.50000000*\st pt, -335.58484397*\st pt)}{d}{$(h-W+w+2,h)$}{SE}{100.0*\st pt}{Copy \word1}{(300.00000000*\st pt, -835.58484397*\st pt)}{8.0}{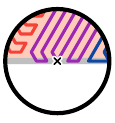}
\TIKZPointOfInterest{(-687.50000000*\st pt, -335.58484397*\st pt)}{e}{$(h-W-7,h)$}{NW}{100.0*\st pt}{}{(900.00000000*\st pt, -835.58484397*\st pt)}{8.0}{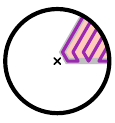}
\TIKZPointOfInterest{(-327.50000000*\st pt, 296.61370080*\st pt)}{f}{$(h-W-8,1)$}{W}{100.0*\st pt}{Next}{(1500.00000000*\st pt, -835.58484397*\st pt)}{8.0}{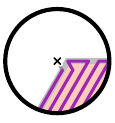}
            }{}
\node [anchor=center] () at (0,0) {\includegraphics[scale =\s]{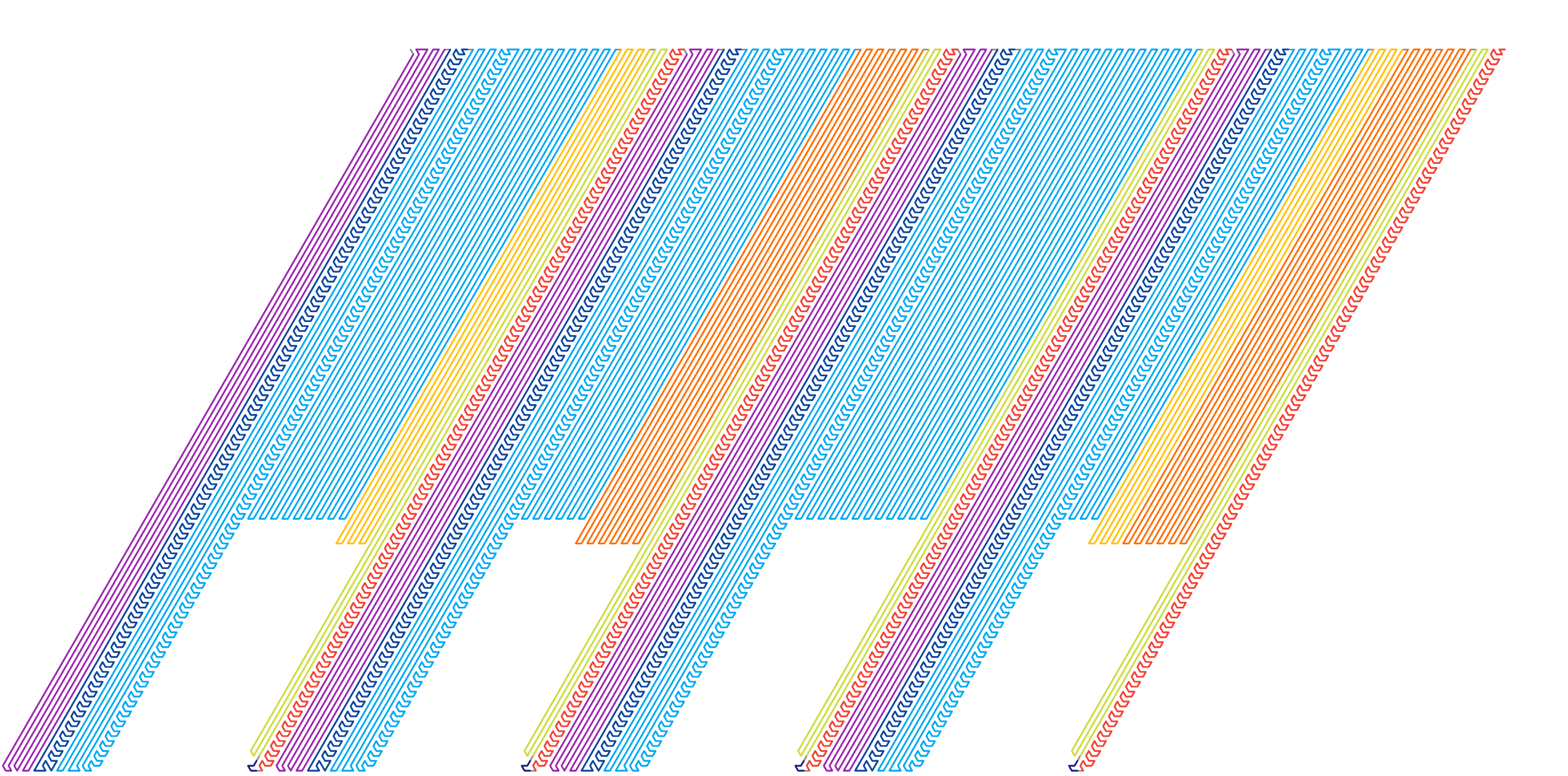}};
            \ifthenelse{\equal{#1}{}}{
\TIKZBLOCKZAGCOPY{1}{0}{110}{-30.000}{-19.486}
            }{}
    \end{tikzpicture}
}

%% file: OritatamiTuringPatterns/n=4-L=3-P=1/ZagCopy1_11__n=4-L=3-P=1.tex
\newcommand{\ZagCopyUBlockUU}[1]{
    \begin{tikzpicture}
        \node [anchor=center] () at (0,0) {\includegraphics[scale =\s]{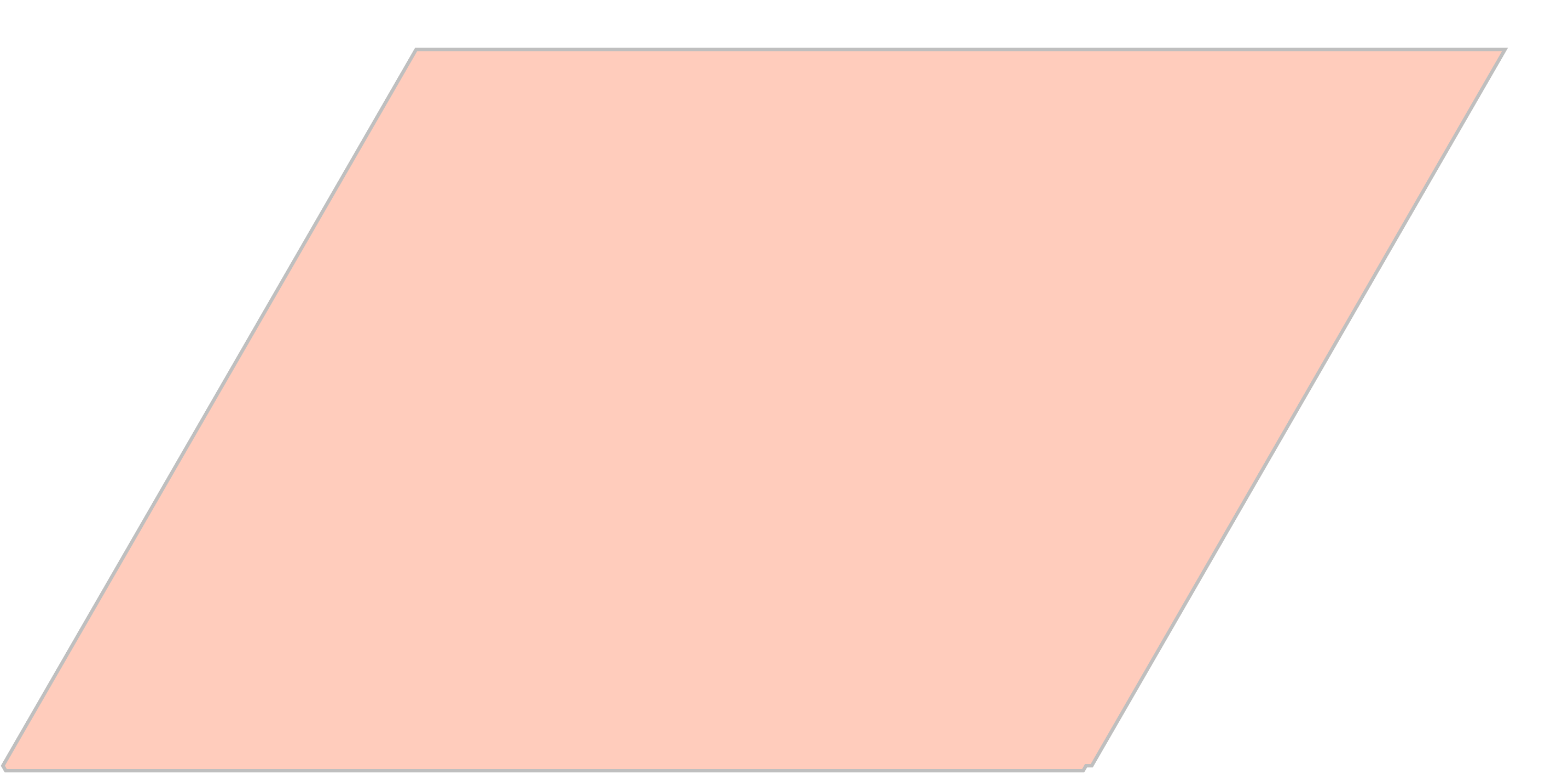}};
        \node [anchor=center] () at (0,0) {\includegraphics[scale =\s]{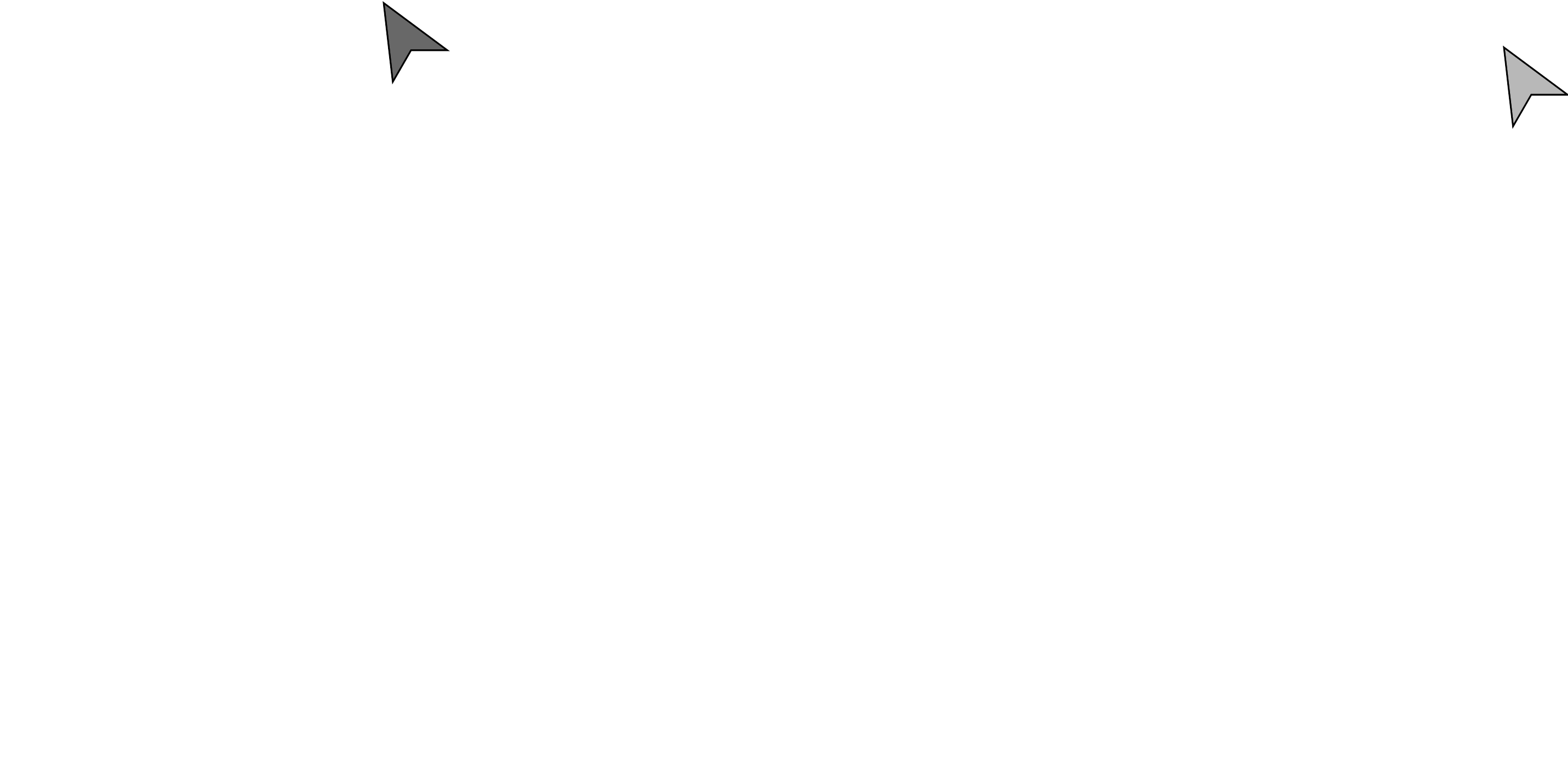}};
        \ifthenelse{\equal{#1}{}}{
\TIKZPointOfInterest{(632.50000000*\st pt, 296.61370080*\st pt)}{a}{$(h-8,1)$}{E}{100.0*\st pt}{Start}{(-1500.00000000*\st pt, -835.58484397*\st pt)}{8.0}{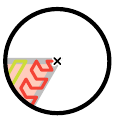}
\TIKZPointOfInterest{(267.50000000*\st pt, -335.58484397*\st pt)}{b}{$(h-8,h)$}{E}{150.0*\st pt}{}{(-900.00000000*\st pt, -835.58484397*\st pt)}{8.0}{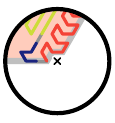}
\TIKZPointOfInterest{(-72.50000000*\st pt, 296.61370080*\st pt)}{c}{$(h-W+w+1,1)$}{NE}{100.0*\st pt}{Read \word1}{(-300.00000000*\st pt, -835.58484397*\st pt)}{8.0}{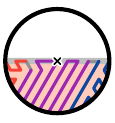}
\TIKZPointOfInterest{(-432.50000000*\st pt, -335.58484397*\st pt)}{d}{$(h-W+w+2,h)$}{SE}{100.0*\st pt}{Copy \word1}{(300.00000000*\st pt, -835.58484397*\st pt)}{8.0}{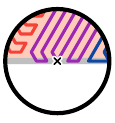}
\TIKZPointOfInterest{(-687.50000000*\st pt, -335.58484397*\st pt)}{e}{$(h-W-7,h)$}{NW}{100.0*\st pt}{}{(900.00000000*\st pt, -835.58484397*\st pt)}{8.0}{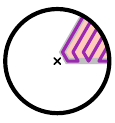}
\TIKZPointOfInterest{(-327.50000000*\st pt, 296.61370080*\st pt)}{f}{$(h-W-8,1)$}{W}{100.0*\st pt}{Next}{(1500.00000000*\st pt, -835.58484397*\st pt)}{8.0}{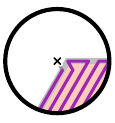}
            }{}
\node [anchor=center] () at (0,0) {\includegraphics[scale =\s]{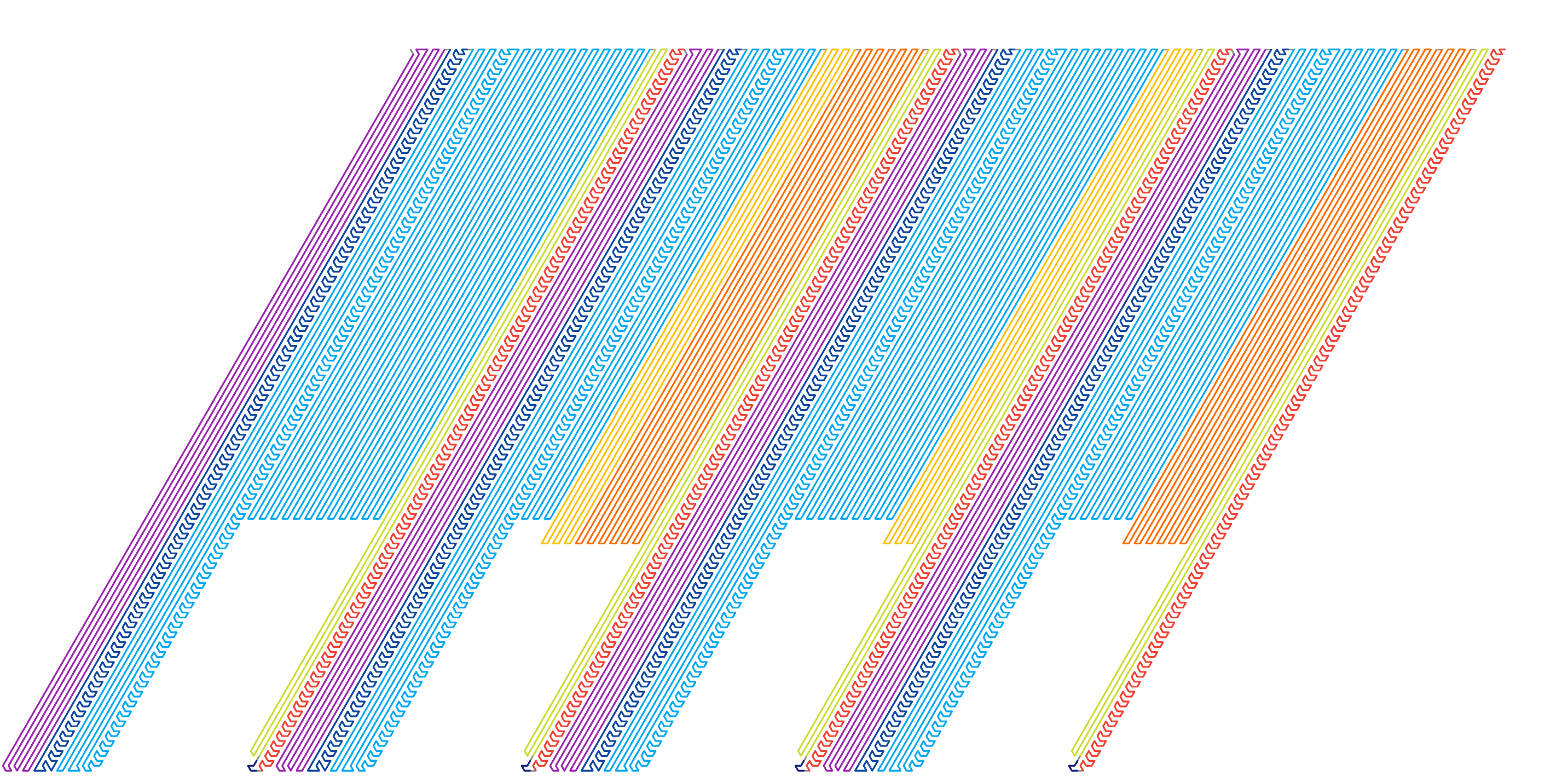}};
            \ifthenelse{\equal{#1}{}}{
\TIKZBLOCKZAGCOPY{1}{2}{11}{-30.000}{-19.486}
            }{}
    \end{tikzpicture}
}

%% file: OritatamiTuringPatterns/n=4-L=3-P=1/ZagCopy1_____n=4-L=3-P=1.tex
\newcommand{\ZagCopyUBlockEpsilon}[1]{
    \begin{tikzpicture}
        \node [anchor=center] () at (0,0) {\includegraphics[scale =\s]{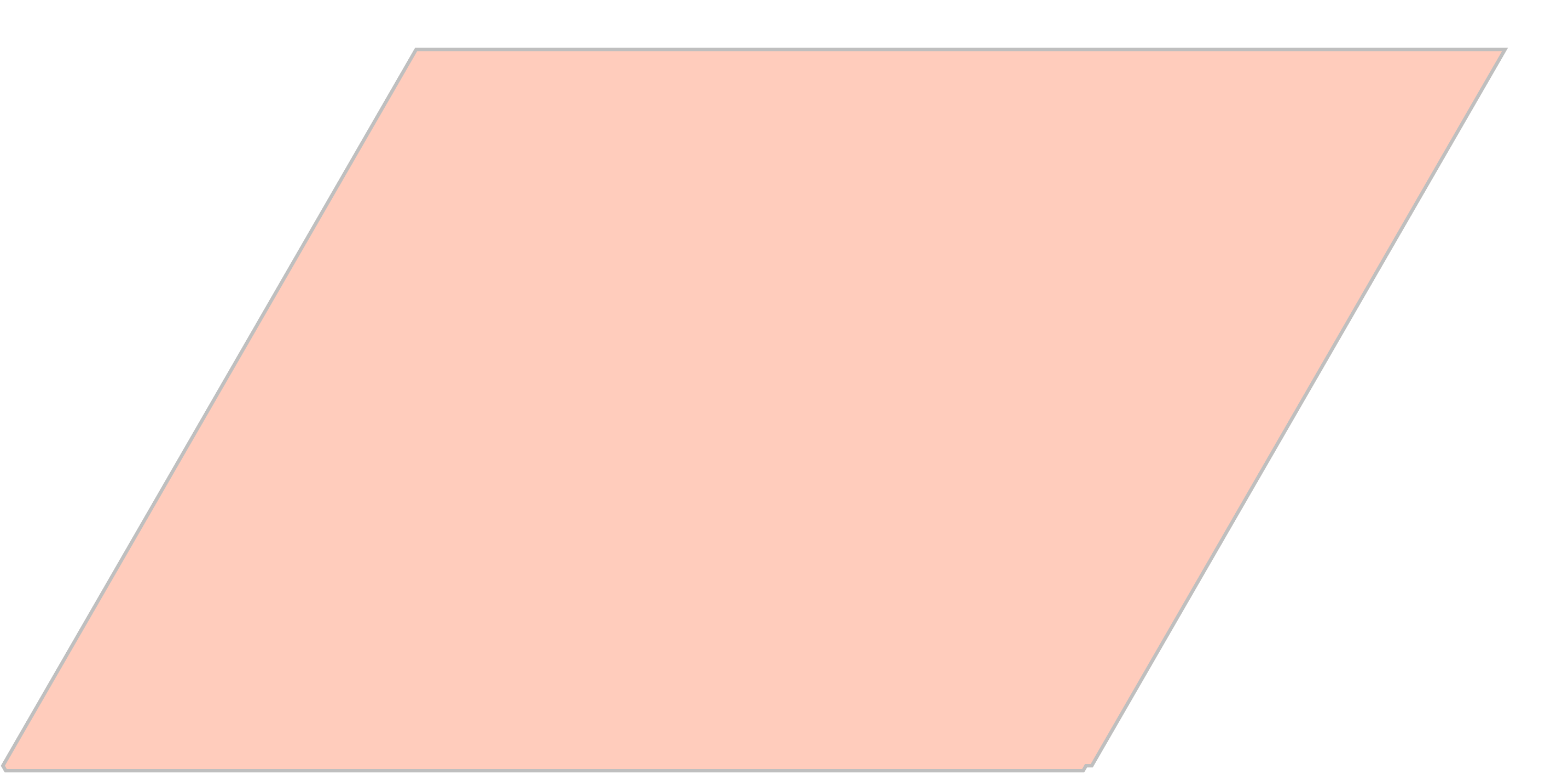}};
        \node [anchor=center] () at (0,0) {\includegraphics[scale =\s]{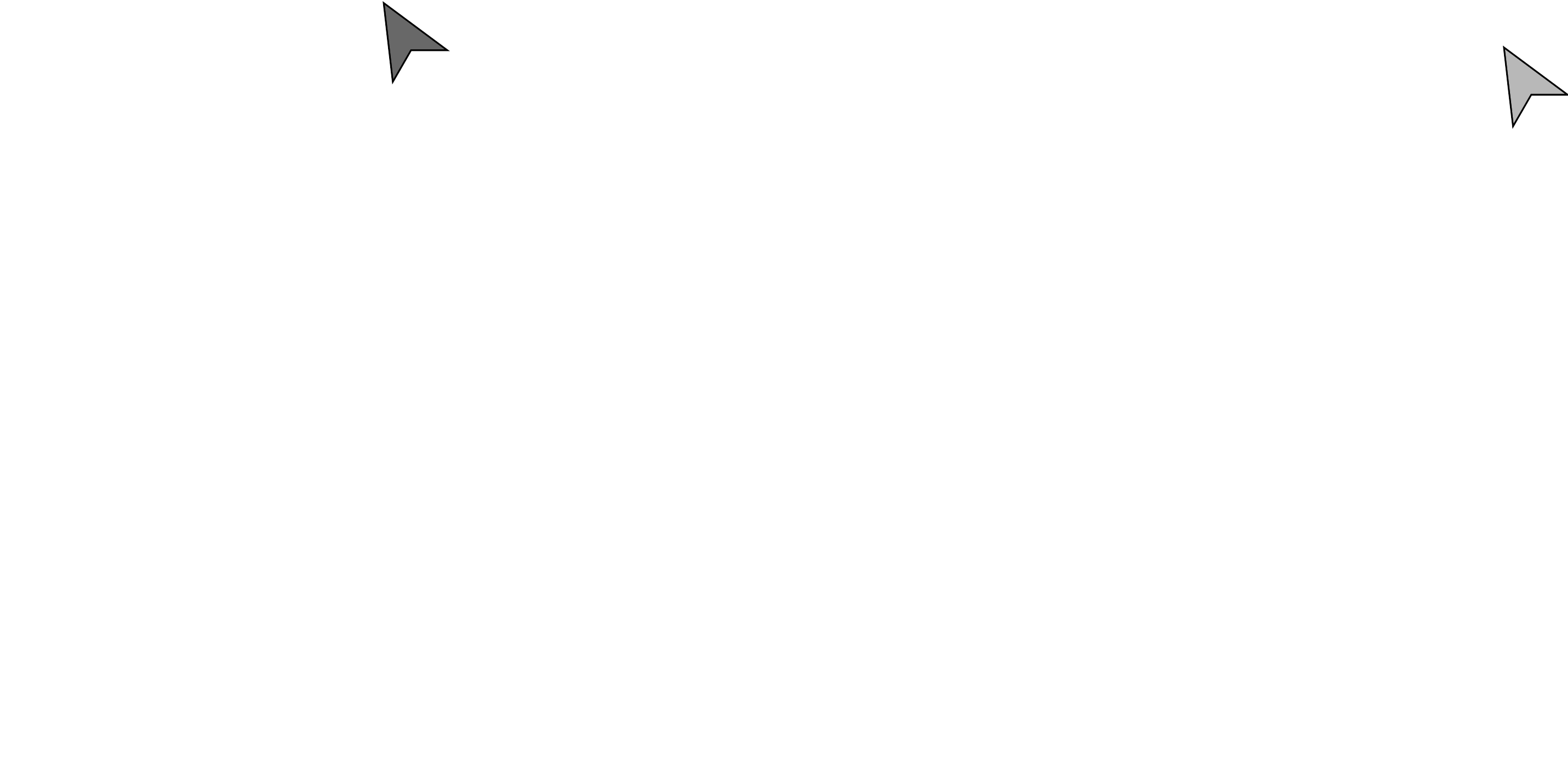}};
        \ifthenelse{\equal{#1}{}}{
\TIKZPointOfInterest{(632.50000000*\st pt, 296.61370080*\st pt)}{a}{$(h-8,1)$}{E}{100.0*\st pt}{Start}{(-1500.00000000*\st pt, -835.58484397*\st pt)}{8.0}{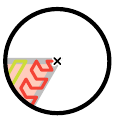}
\TIKZPointOfInterest{(267.50000000*\st pt, -335.58484397*\st pt)}{b}{$(h-8,h)$}{E}{150.0*\st pt}{}{(-900.00000000*\st pt, -835.58484397*\st pt)}{8.0}{ZagCopy1____n=4-L=3-P=1-PI-B-_0,146_-bottom-right.pdf}
\TIKZPointOfInterest{(-72.50000000*\st pt, 296.61370080*\st pt)}{c}{$(h-W+w+1,1)$}{NE}{100.0*\st pt}{Read \word1}{(-300.00000000*\st pt, -835.58484397*\st pt)}{8.0}{ZagCopy1____n=4-L=3-P=1-PI-C-_-141,0_-read1.pdf}
\TIKZPointOfInterest{(-432.50000000*\st pt, -335.58484397*\st pt)}{d}{$(h-W+w+2,h)$}{SE}{100.0*\st pt}{Copy \word1}{(300.00000000*\st pt, -835.58484397*\st pt)}{8.0}{ZagCopy1____n=4-L=3-P=1-PI-D-_-140,146_-copy1.pdf}
\TIKZPointOfInterest{(-687.50000000*\st pt, -335.58484397*\st pt)}{e}{$(h-W-7,h)$}{NW}{100.0*\st pt}{}{(900.00000000*\st pt, -835.58484397*\st pt)}{8.0}{ZagCopy1____n=4-L=3-P=1-PI-E-_-191,146_-bottom-left.pdf}
\TIKZPointOfInterest{(-327.50000000*\st pt, 296.61370080*\st pt)}{f}{$(h-W-8,1)$}{W}{100.0*\st pt}{Next}{(1500.00000000*\st pt, -835.58484397*\st pt)}{8.0}{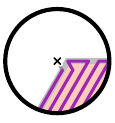}
            }{}
\node [anchor=center] () at (0,0) {\includegraphics[scale =\s]{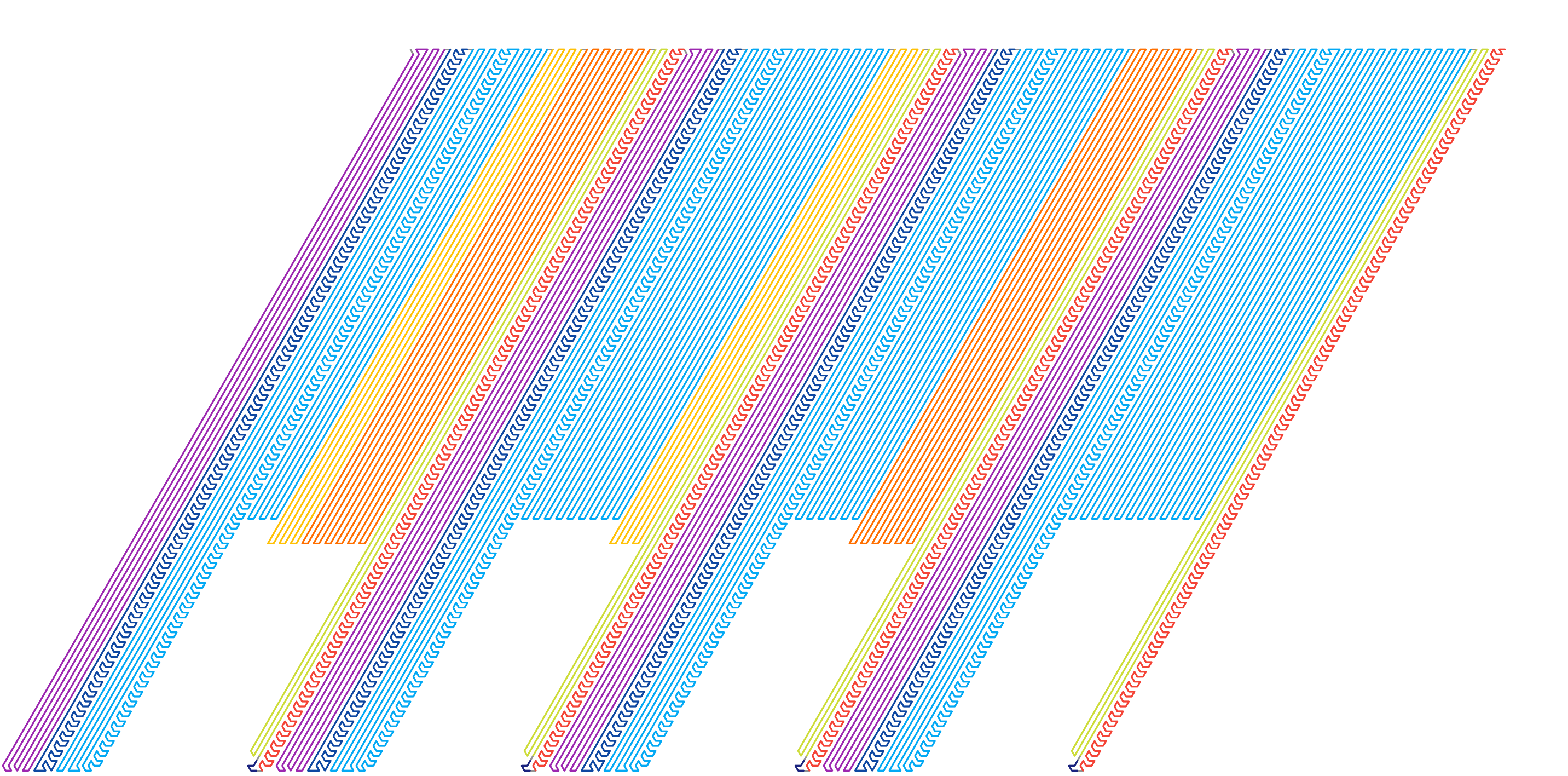}};
            \ifthenelse{\equal{#1}{}}{
\TIKZBLOCKZAGCOPY{1}{1}{\ensuremath{\epsilon}}{-30.000}{-19.486}
            }{}
    \end{tikzpicture}
}

%% file: OritatamiTuringPatterns/n=4-L=3-P=1/ZigCopy0_0__n=4-L=3-P=1.tex
\newcommand{\ZigCopyZBlockZ}[1]{
    \begin{tikzpicture}
        \node [anchor=center] () at (0,0) {\includegraphics[scale =\s]{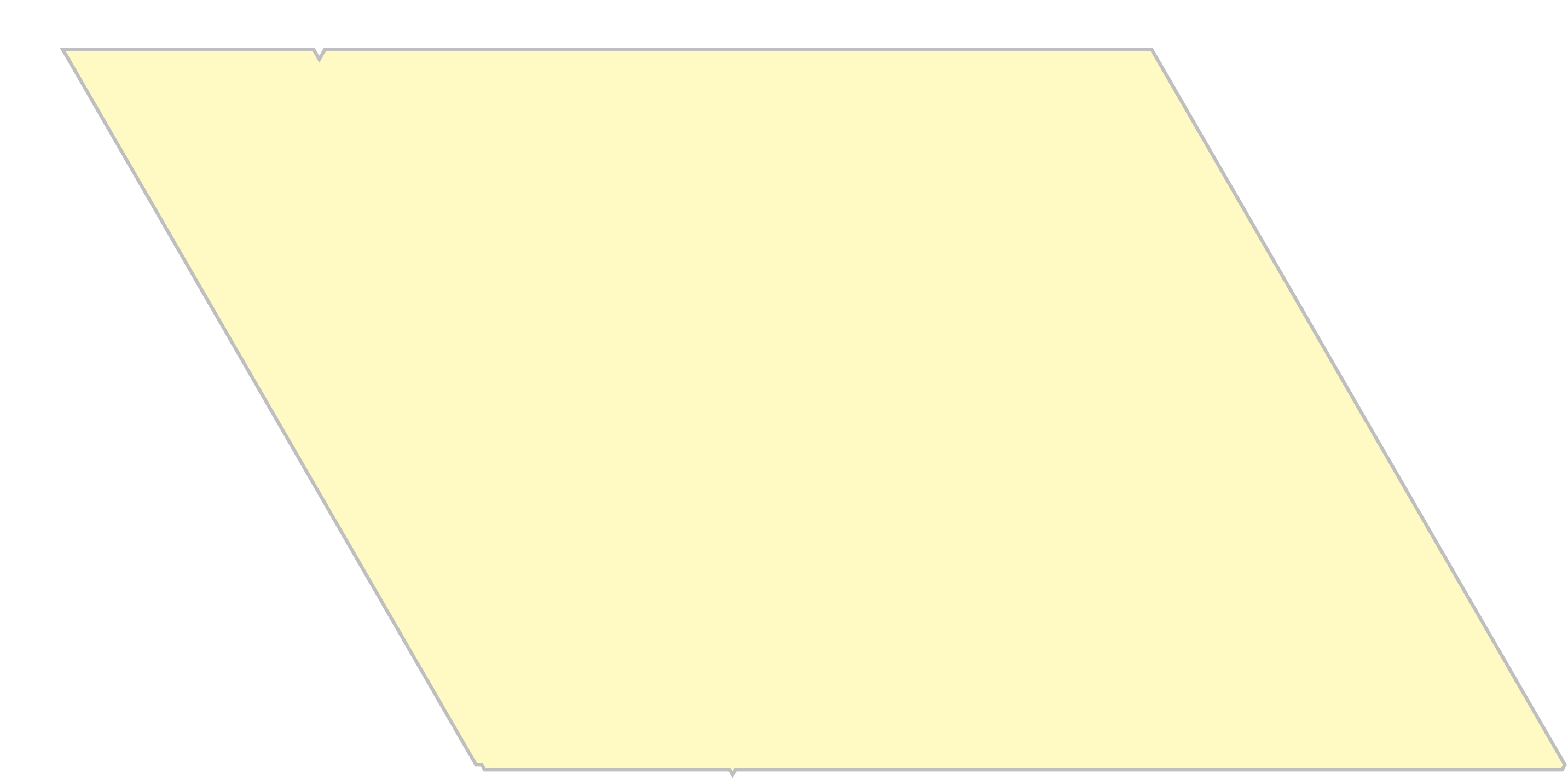}};
        \node [anchor=center] () at (0,0) {\includegraphics[scale =\s]{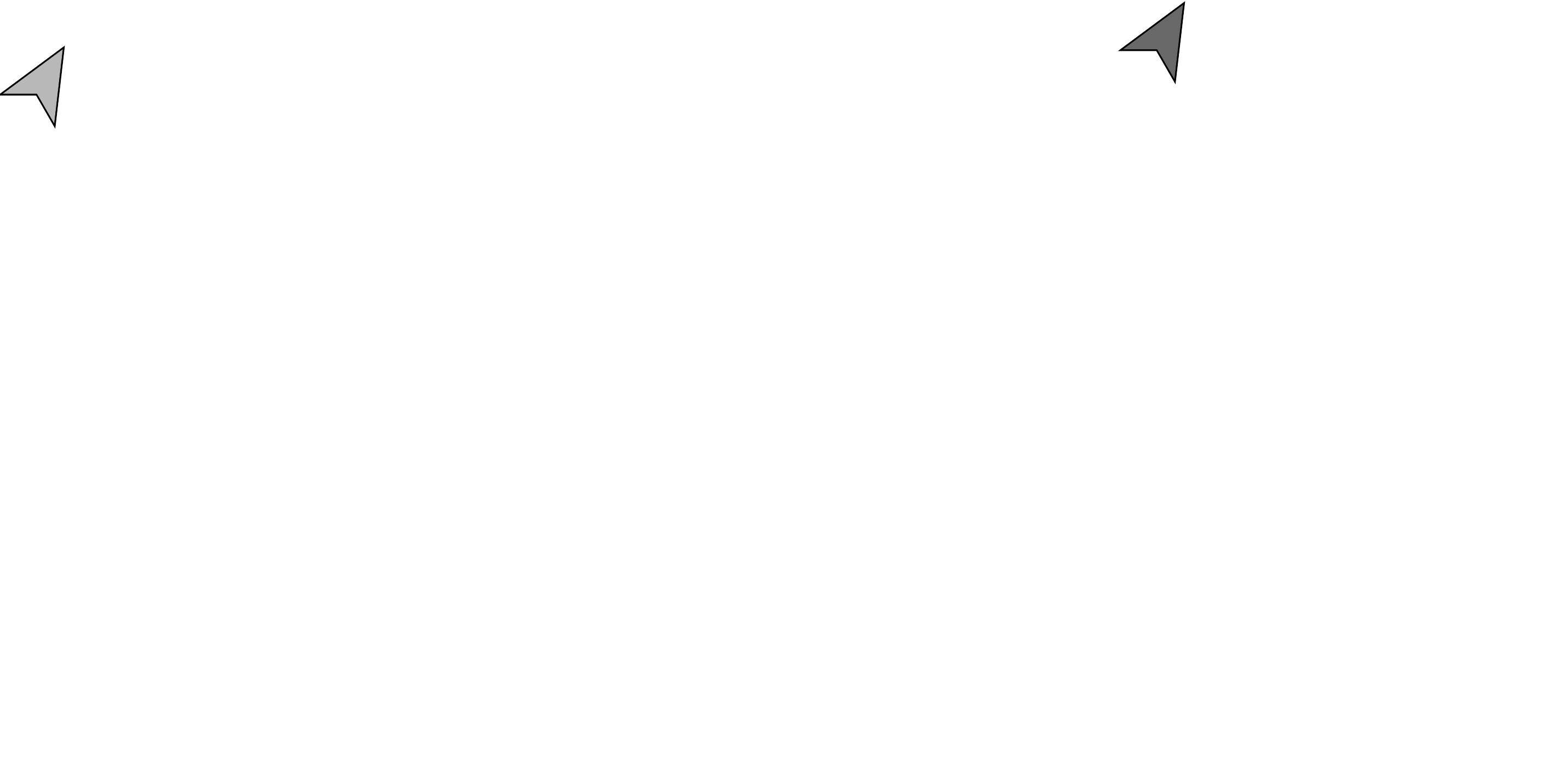}};
        \ifthenelse{\equal{#1}{}}{
\TIKZPointOfInterest{(-632.50000000*\st pt, 298.77876431*\st pt)}{a}{$(-1,1-h)$}{W}{100.0*\st pt}{Start}{(-1500.00000000*\st pt, -837.74990748*\st pt)}{8.0}{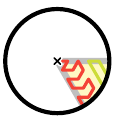}
\TIKZPointOfInterest{(-267.50000000*\st pt, -333.41978046*\st pt)}{b}{$(h-2,0)$}{W}{100.0*\st pt}{}{(-900.00000000*\st pt, -837.74990748*\st pt)}{8.0}{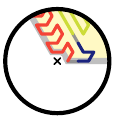}
\TIKZPointOfInterest{(-407.50000000*\st pt, 298.77876431*\st pt)}{c}{$(w+2,1-h)$}{NW}{100.0*\st pt}{Read \word0}{(-300.00000000*\st pt, -837.74990748*\st pt)}{8.0}{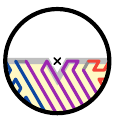}
\TIKZPointOfInterest{(-45.00000000*\st pt, -337.74990748*\st pt)}{d}{$(h+w+1,1)$}{SE}{100.0*\st pt}{Copy \word0}{(300.00000000*\st pt, -837.74990748*\st pt)}{8.0}{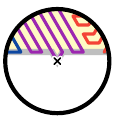}
\TIKZPointOfInterest{(-392.50000000*\st pt, 298.77876431*\st pt)}{e}{$(w+5,0)$}{NE}{100.0*\st pt}{\begin{minipage}[t]{2.5cm}\raggedright Start of 2$^{\text{nd}}$ copy unit\end{minipage}}{(900.00000000*\st pt, -837.74990748*\st pt)}{8.0}{ZigCopy0_0__n=4-L=3-P=1-PI-E-_48,0_-2nd-unit.pdf}
\TIKZPointOfInterest{(87.50000000*\st pt, 298.77876431*\st pt)}{f}{$((n-1)(w+6_-1,0)$}{NE}{100.0*\st pt}{\begin{minipage}[t]{2.5cm}\raggedright Start of last copy unit\end{minipage}}{(1500.00000000*\st pt, -837.74990748*\st pt)}{8.0}{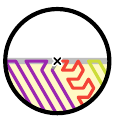}
\TIKZPointOfInterest{(687.50000000*\st pt, -333.41978046*\st pt)}{g}{$(W+h-3,0)$}{E}{100.0*\st pt}{}{(-1500.00000000*\st pt, -1437.74990748*\st pt)}{8.0}{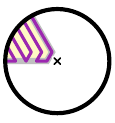}
\TIKZPointOfInterest{(327.50000000*\st pt, 298.77876431*\st pt)}{h}{$(W-1,1-h)$}{E}{100.0*\st pt}{Next}{(-900.00000000*\st pt, -1437.74990748*\st pt)}{8.0}{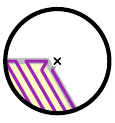}
            }{}
\node [anchor=center] () at (0,0) {\includegraphics[scale =\s]{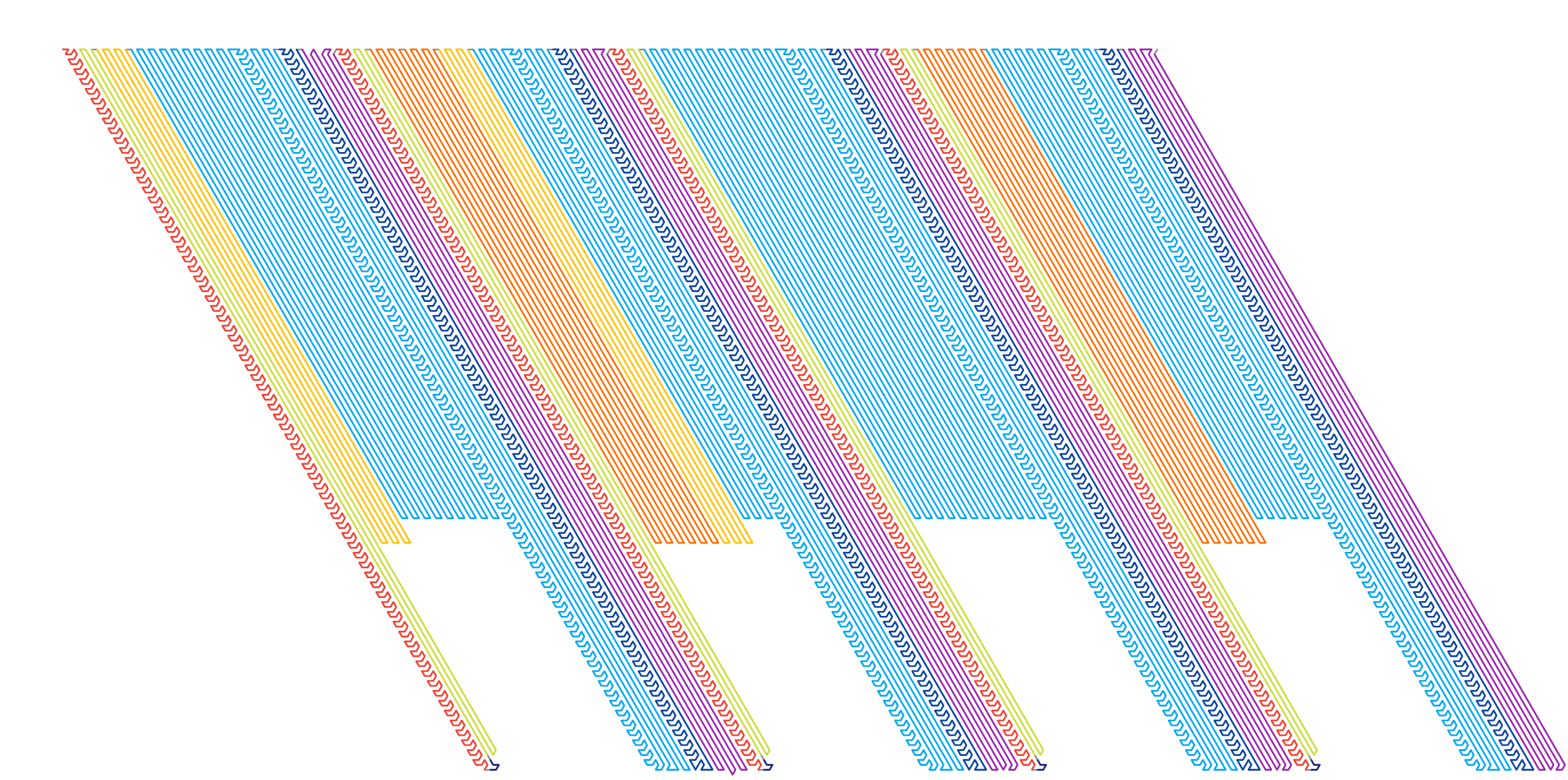}};
            \ifthenelse{\equal{#1}{}}{
\TIKZBLOCKZIGCOPY{0}{3}{0}{30.000}{-17.321}
            }{}
    \end{tikzpicture}
}

%% file: OritatamiTuringPatterns/n=4-L=3-P=1/ZigCopy0_110__n=4-L=3-P=1.tex
\newcommand{\ZigCopyZBlockUUZ}[1]{
    \begin{tikzpicture}
        \node [anchor=center] () at (0,0) {\includegraphics[scale =\s]{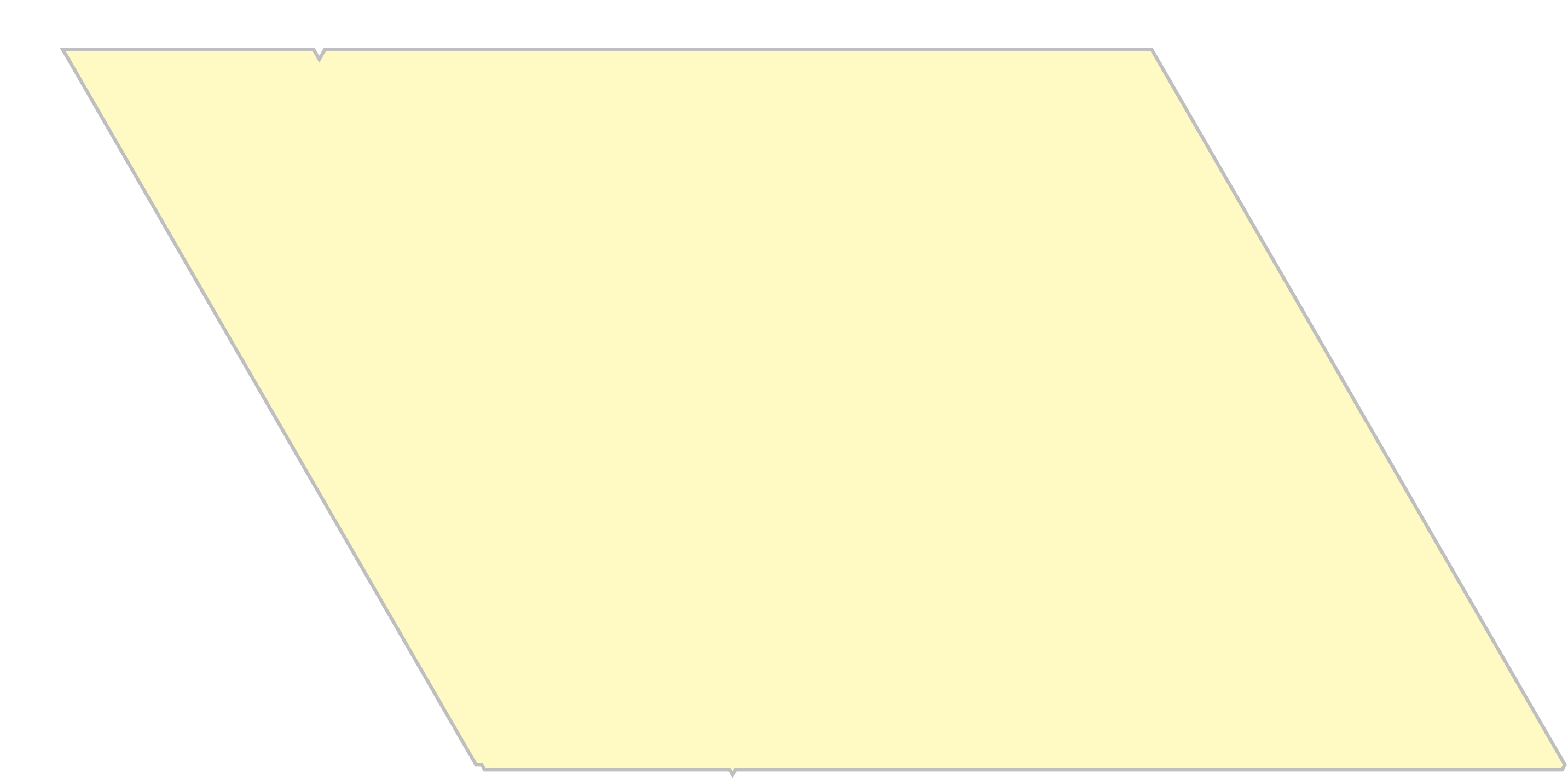}};
        \node [anchor=center] () at (0,0) {\includegraphics[scale =\s]{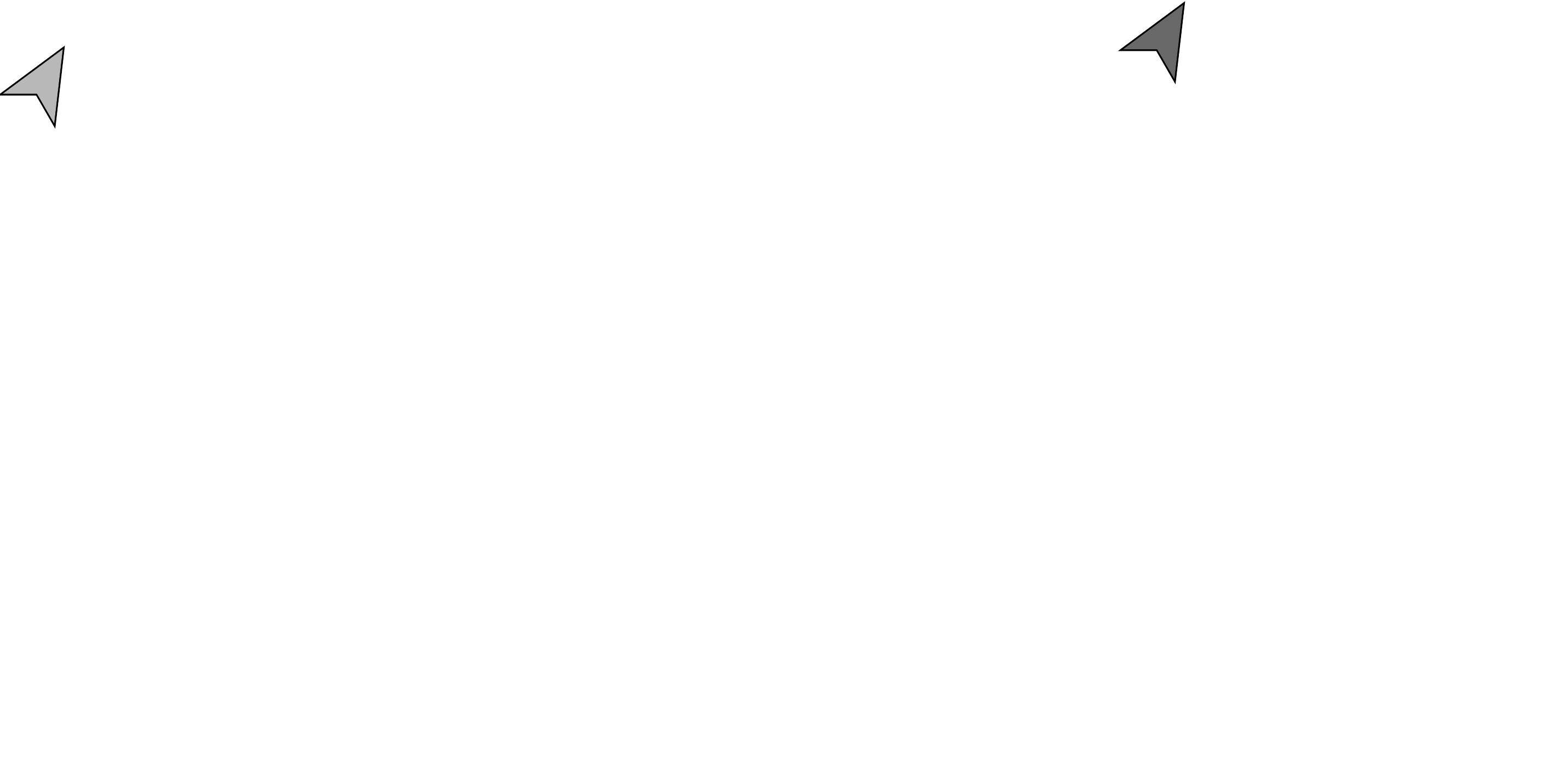}};
        \ifthenelse{\equal{#1}{}}{
\TIKZPointOfInterest{(-632.50000000*\st pt, 298.77876431*\st pt)}{a}{$(-1,1-h)$}{W}{100.0*\st pt}{Start}{(-1500.00000000*\st pt, -837.74990748*\st pt)}{8.0}{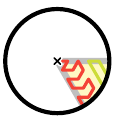}
\TIKZPointOfInterest{(-267.50000000*\st pt, -333.41978046*\st pt)}{b}{$(h-2,0)$}{W}{100.0*\st pt}{}{(-900.00000000*\st pt, -837.74990748*\st pt)}{8.0}{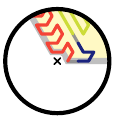}
\TIKZPointOfInterest{(-407.50000000*\st pt, 298.77876431*\st pt)}{c}{$(w+2,1-h)$}{NW}{100.0*\st pt}{Read \word0}{(-300.00000000*\st pt, -837.74990748*\st pt)}{8.0}{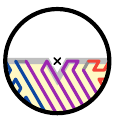}
\TIKZPointOfInterest{(-45.00000000*\st pt, -337.74990748*\st pt)}{d}{$(h+w+1,1)$}{SE}{100.0*\st pt}{Copy \word0}{(300.00000000*\st pt, -837.74990748*\st pt)}{8.0}{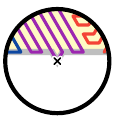}
\TIKZPointOfInterest{(-392.50000000*\st pt, 298.77876431*\st pt)}{e}{$(w+5,0)$}{NE}{100.0*\st pt}{\begin{minipage}[t]{2.5cm}\raggedright Start of 2$^{\text{nd}}$ copy unit\end{minipage}}{(900.00000000*\st pt, -837.74990748*\st pt)}{8.0}{ZigCopy0_110__n=4-L=3-P=1-PI-E-_48,0_-2nd-unit.pdf}
\TIKZPointOfInterest{(87.50000000*\st pt, 298.77876431*\st pt)}{f}{$((n-1)(w+6_-1,0)$}{NE}{100.0*\st pt}{\begin{minipage}[t]{2.5cm}\raggedright Start of last copy unit\end{minipage}}{(1500.00000000*\st pt, -837.74990748*\st pt)}{8.0}{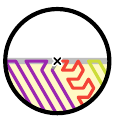}
\TIKZPointOfInterest{(687.50000000*\st pt, -333.41978046*\st pt)}{g}{$(W+h-3,0)$}{E}{100.0*\st pt}{}{(-1500.00000000*\st pt, -1437.74990748*\st pt)}{8.0}{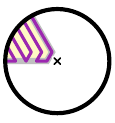}
\TIKZPointOfInterest{(327.50000000*\st pt, 298.77876431*\st pt)}{h}{$(W-1,1-h)$}{E}{100.0*\st pt}{Next}{(-900.00000000*\st pt, -1437.74990748*\st pt)}{8.0}{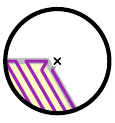}
            }{}
\node [anchor=center] () at (0,0) {\includegraphics[scale =\s]{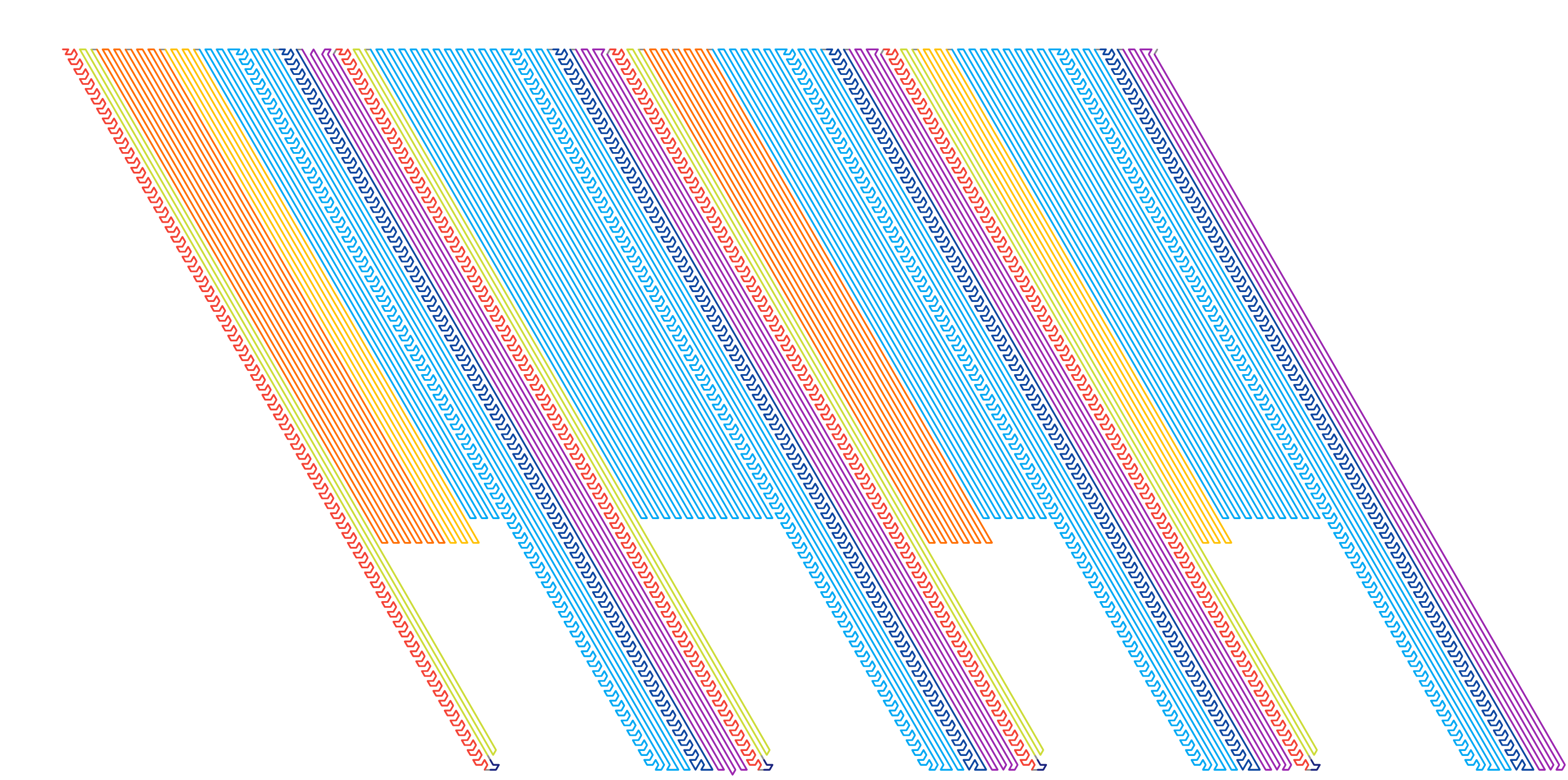}};
            \ifthenelse{\equal{#1}{}}{
\TIKZBLOCKZIGCOPY{0}{0}{110}{30.000}{-17.321}
            }{}
    \end{tikzpicture}
}

%% file: OritatamiTuringPatterns/n=4-L=3-P=1/ZigCopy0_11__n=4-L=3-P=1.tex
\newcommand{\ZigCopyZBlockUU}[1]{
    \begin{tikzpicture}
        \node [anchor=center] () at (0,0) {\includegraphics[scale =\s]{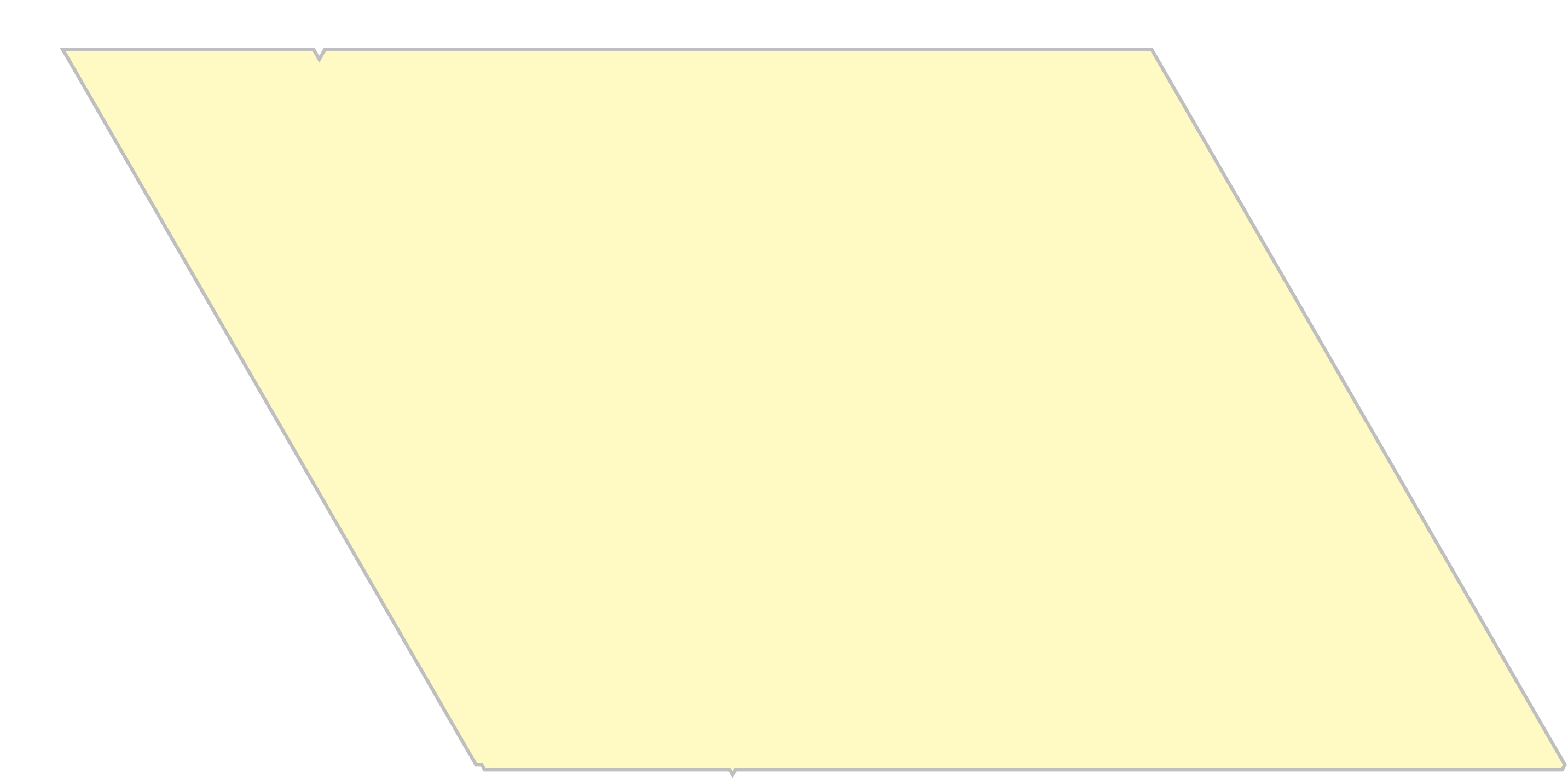}};
        \node [anchor=center] () at (0,0) {\includegraphics[scale =\s]{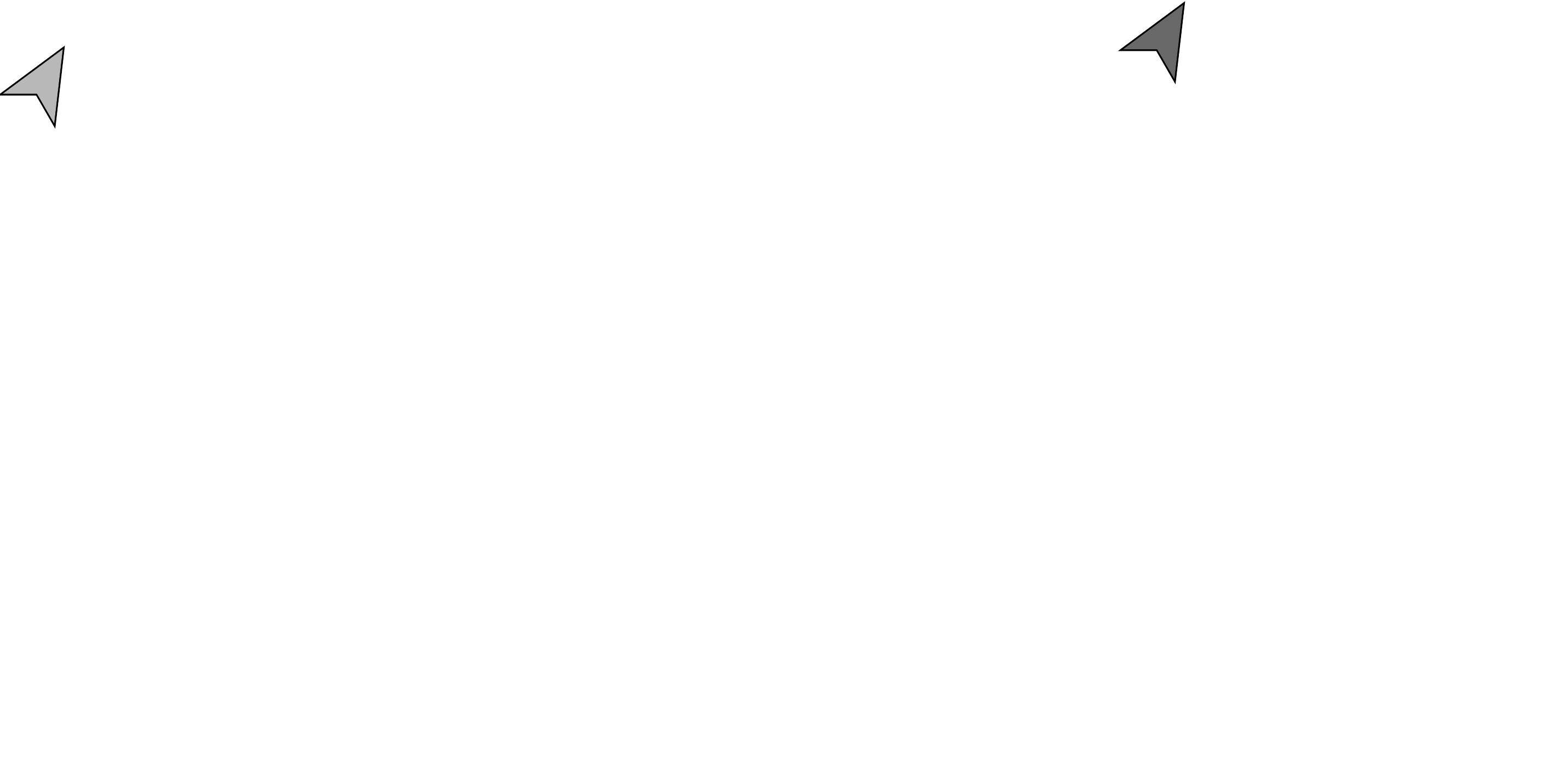}};
        \ifthenelse{\equal{#1}{}}{
\TIKZPointOfInterest{(-632.50000000*\st pt, 298.77876431*\st pt)}{a}{$(-1,1-h)$}{W}{100.0*\st pt}{Start}{(-1500.00000000*\st pt, -837.74990748*\st pt)}{8.0}{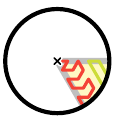}
\TIKZPointOfInterest{(-267.50000000*\st pt, -333.41978046*\st pt)}{b}{$(h-2,0)$}{W}{100.0*\st pt}{}{(-900.00000000*\st pt, -837.74990748*\st pt)}{8.0}{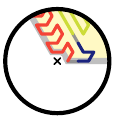}
\TIKZPointOfInterest{(-407.50000000*\st pt, 298.77876431*\st pt)}{c}{$(w+2,1-h)$}{NW}{100.0*\st pt}{Read \word0}{(-300.00000000*\st pt, -837.74990748*\st pt)}{8.0}{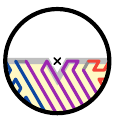}
\TIKZPointOfInterest{(-45.00000000*\st pt, -337.74990748*\st pt)}{d}{$(h+w+1,1)$}{SE}{100.0*\st pt}{Copy \word0}{(300.00000000*\st pt, -837.74990748*\st pt)}{8.0}{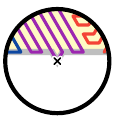}
\TIKZPointOfInterest{(-392.50000000*\st pt, 298.77876431*\st pt)}{e}{$(w+5,0)$}{NE}{100.0*\st pt}{\begin{minipage}[t]{2.5cm}\raggedright Start of 2$^{\text{nd}}$ copy unit\end{minipage}}{(900.00000000*\st pt, -837.74990748*\st pt)}{8.0}{ZigCopy0_11__n=4-L=3-P=1-PI-E-_48,0_-2nd-unit.pdf}
\TIKZPointOfInterest{(87.50000000*\st pt, 298.77876431*\st pt)}{f}{$((n-1)(w+6_-1,0)$}{NE}{100.0*\st pt}{\begin{minipage}[t]{2.5cm}\raggedright Start of last copy unit\end{minipage}}{(1500.00000000*\st pt, -837.74990748*\st pt)}{8.0}{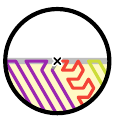}
\TIKZPointOfInterest{(687.50000000*\st pt, -333.41978046*\st pt)}{g}{$(W+h-3,0)$}{E}{100.0*\st pt}{}{(-1500.00000000*\st pt, -1437.74990748*\st pt)}{8.0}{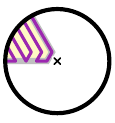}
\TIKZPointOfInterest{(327.50000000*\st pt, 298.77876431*\st pt)}{h}{$(W-1,1-h)$}{E}{100.0*\st pt}{Next}{(-900.00000000*\st pt, -1437.74990748*\st pt)}{8.0}{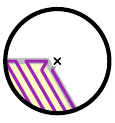}
            }{}
\node [anchor=center] () at (0,0) {\includegraphics[scale =\s]{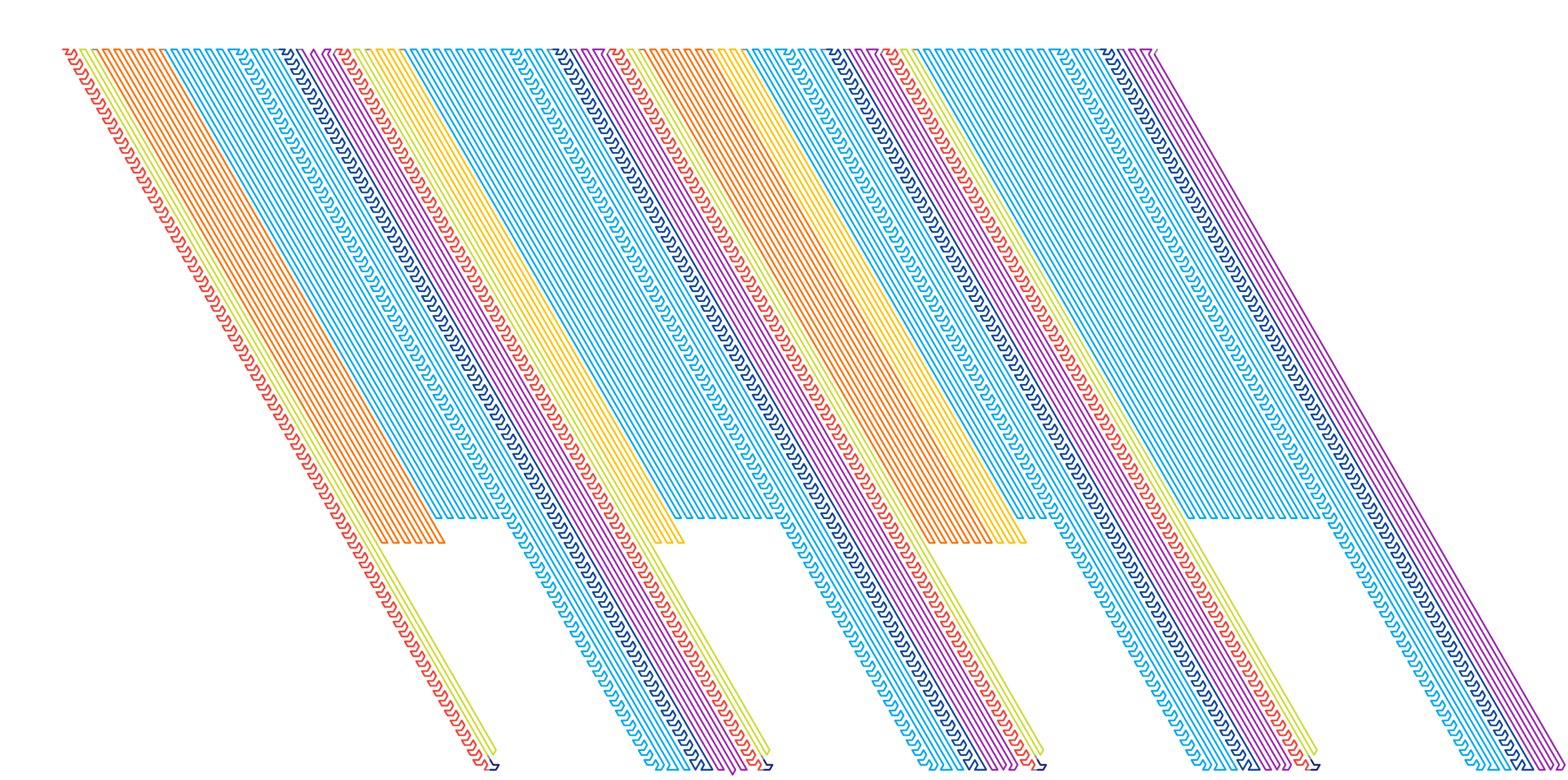}};
            \ifthenelse{\equal{#1}{}}{
\TIKZBLOCKZIGCOPY{0}{2}{11}{30.000}{-17.321}
            }{}
    \end{tikzpicture}
}

%% file: OritatamiTuringPatterns/n=4-L=3-P=1/ZigCopy0_____n=4-L=3-P=1.tex
\newcommand{\ZigCopyZBlockEpsilon}[1]{
    \begin{tikzpicture}
        \node [anchor=center] () at (0,0) {\includegraphics[scale =\s]{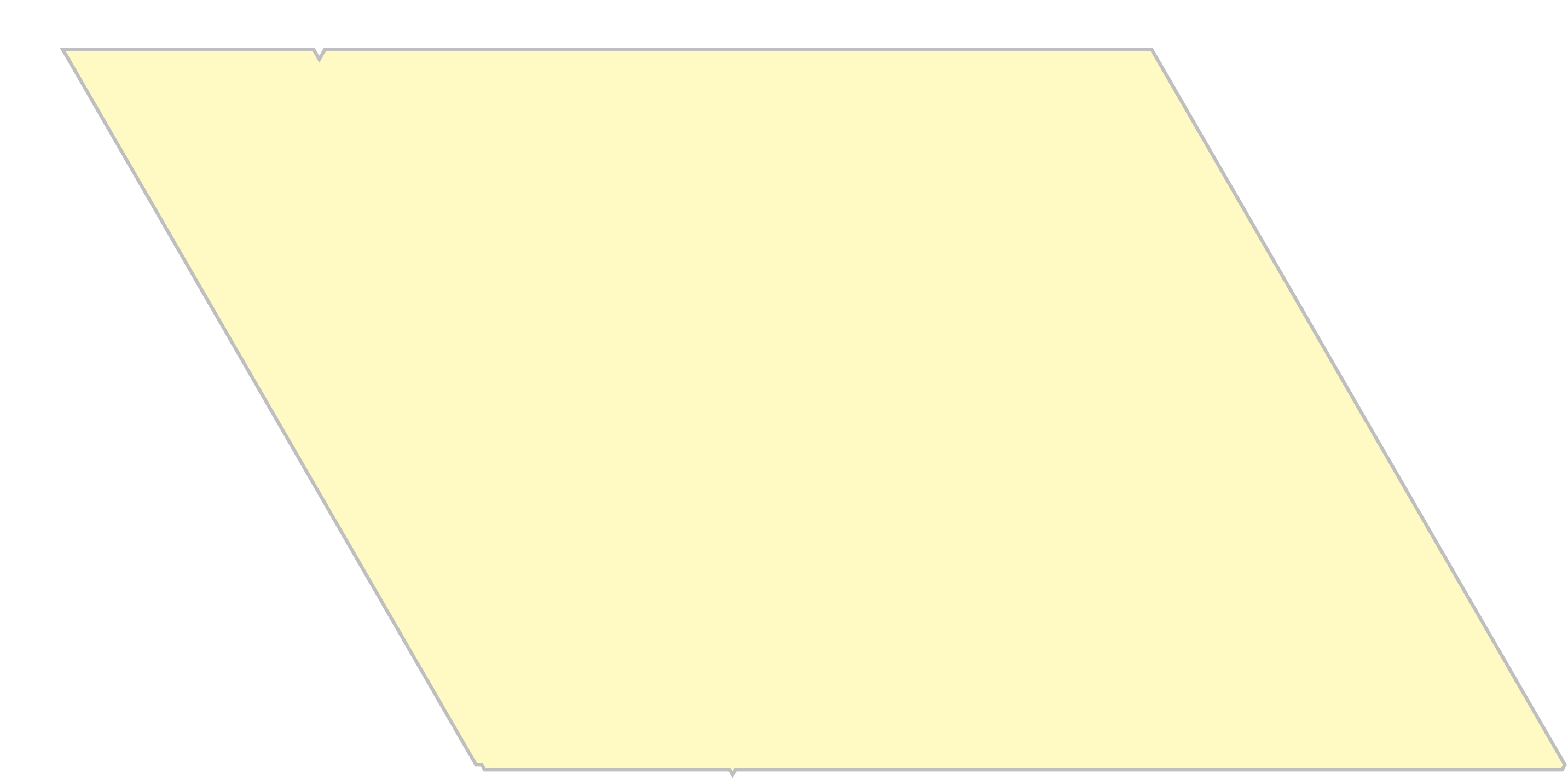}};
        \node [anchor=center] () at (0,0) {\includegraphics[scale =\s]{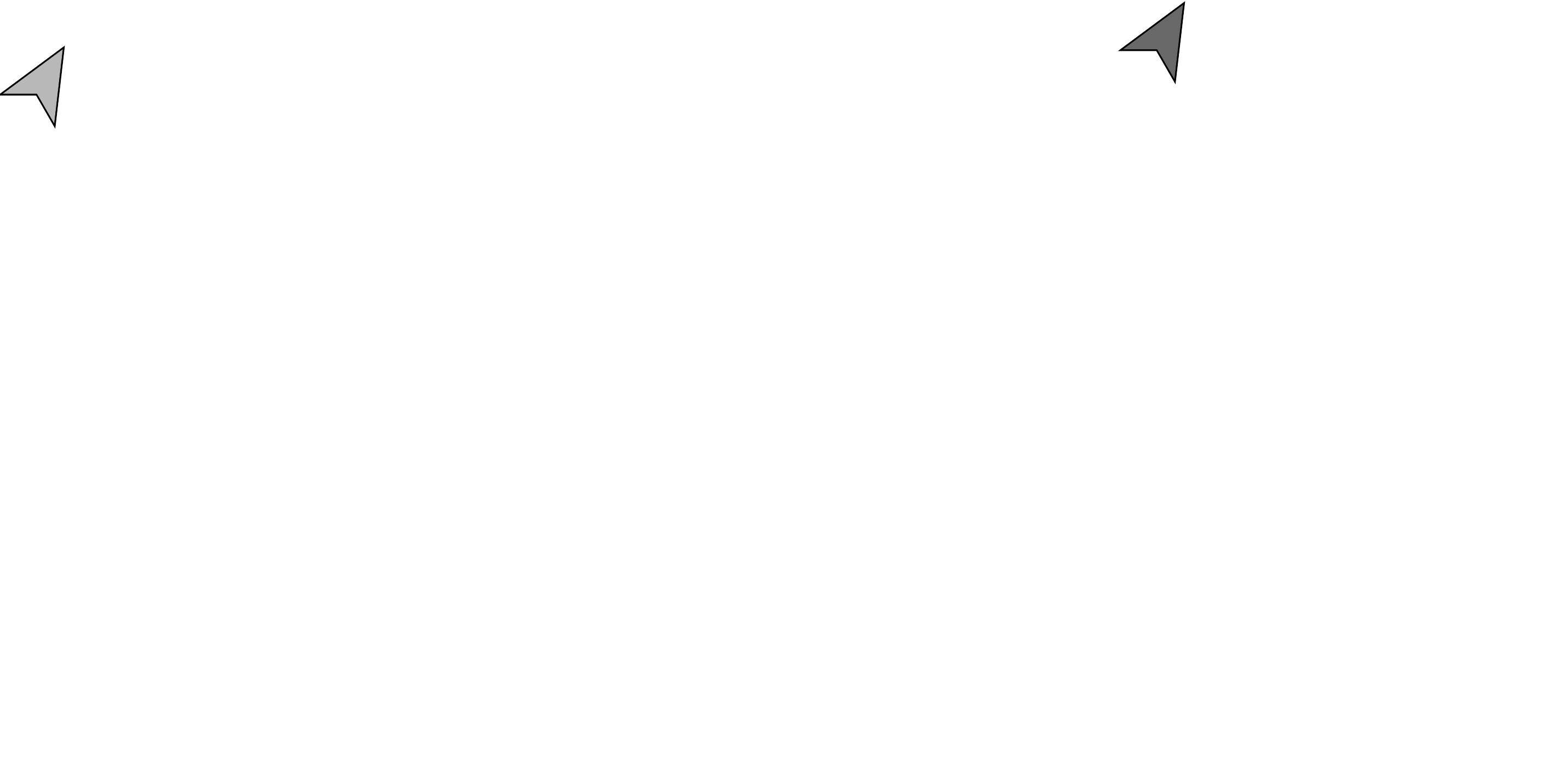}};
        \ifthenelse{\equal{#1}{}}{
\TIKZPointOfInterest{(-632.50000000*\st pt, 298.77876431*\st pt)}{a}{$(-1,1-h)$}{W}{100.0*\st pt}{Start}{(-1500.00000000*\st pt, -837.74990748*\st pt)}{8.0}{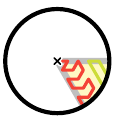}
\TIKZPointOfInterest{(-267.50000000*\st pt, -333.41978046*\st pt)}{b}{$(h-2,0)$}{W}{100.0*\st pt}{}{(-900.00000000*\st pt, -837.74990748*\st pt)}{8.0}{ZigCopy0____n=4-L=3-P=1-PI-B-_146,146_-bottom-left.pdf}
\TIKZPointOfInterest{(-407.50000000*\st pt, 298.77876431*\st pt)}{c}{$(w+2,1-h)$}{NW}{100.0*\st pt}{Read \word0}{(-300.00000000*\st pt, -837.74990748*\st pt)}{8.0}{ZigCopy0____n=4-L=3-P=1-PI-C-_45,0_-read0.pdf}
\TIKZPointOfInterest{(-45.00000000*\st pt, -337.74990748*\st pt)}{d}{$(h+w+1,1)$}{SE}{100.0*\st pt}{Copy \word0}{(300.00000000*\st pt, -837.74990748*\st pt)}{8.0}{ZigCopy0____n=4-L=3-P=1-PI-D-_191,147_-copy0.pdf}
\TIKZPointOfInterest{(-392.50000000*\st pt, 298.77876431*\st pt)}{e}{$(w+5,0)$}{NE}{100.0*\st pt}{\begin{minipage}[t]{2.5cm}\raggedright Start of 2$^{\text{nd}}$ copy unit\end{minipage}}{(900.00000000*\st pt, -837.74990748*\st pt)}{8.0}{ZigCopy0____n=4-L=3-P=1-PI-E-_48,0_-2nd-unit.pdf}
\TIKZPointOfInterest{(87.50000000*\st pt, 298.77876431*\st pt)}{f}{$((n-1)(w+6_-1,0)$}{NE}{100.0*\st pt}{\begin{minipage}[t]{2.5cm}\raggedright Start of last copy unit\end{minipage}}{(1500.00000000*\st pt, -837.74990748*\st pt)}{8.0}{ZigCopy0____n=4-L=3-P=1-PI-F-_144,0_-last-unit.pdf}
\TIKZPointOfInterest{(687.50000000*\st pt, -333.41978046*\st pt)}{g}{$(W+h-3,0)$}{E}{100.0*\st pt}{}{(-1500.00000000*\st pt, -1437.74990748*\st pt)}{8.0}{ZigCopy0____n=4-L=3-P=1-PI-G-_337,146_-bottom-right.pdf}
\TIKZPointOfInterest{(327.50000000*\st pt, 298.77876431*\st pt)}{h}{$(W-1,1-h)$}{E}{100.0*\st pt}{Next}{(-900.00000000*\st pt, -1437.74990748*\st pt)}{8.0}{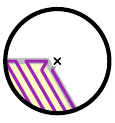}
            }{}
\node [anchor=center] () at (0,0) {\includegraphics[scale =\s]{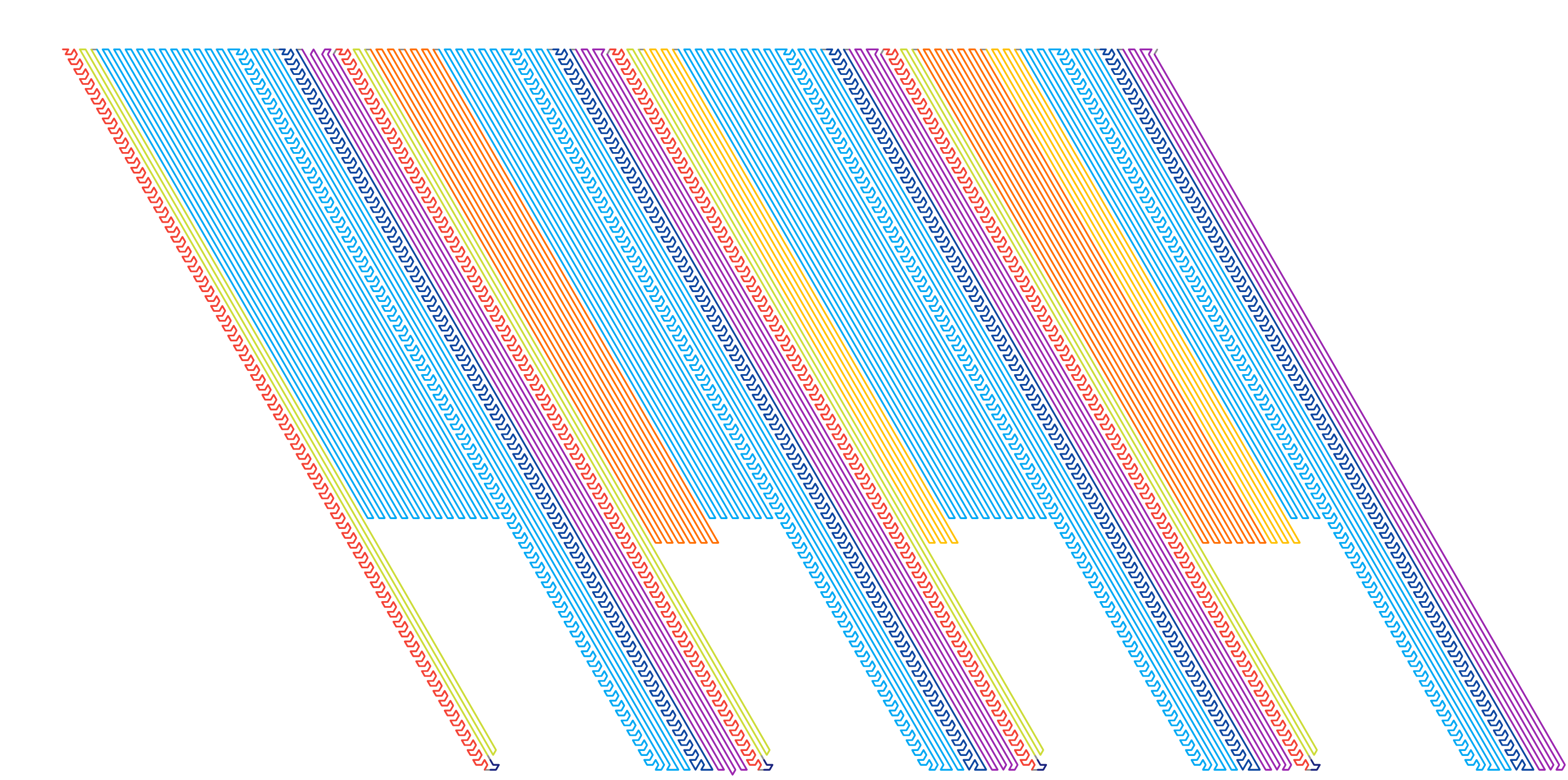}};
            \ifthenelse{\equal{#1}{}}{
\TIKZBLOCKZIGCOPY{0}{1}{\ensuremath{\epsilon}}{30.000}{-17.321}
            }{}
    \end{tikzpicture}
}

%% file: OritatamiTuringPatterns/n=4-L=3-P=1/ZigCopy1_0__n=4-L=3-P=1.tex
\newcommand{\ZigCopyUBlockZ}[1]{
    \begin{tikzpicture}
        \node [anchor=center] () at (0,0) {\includegraphics[scale =\s]{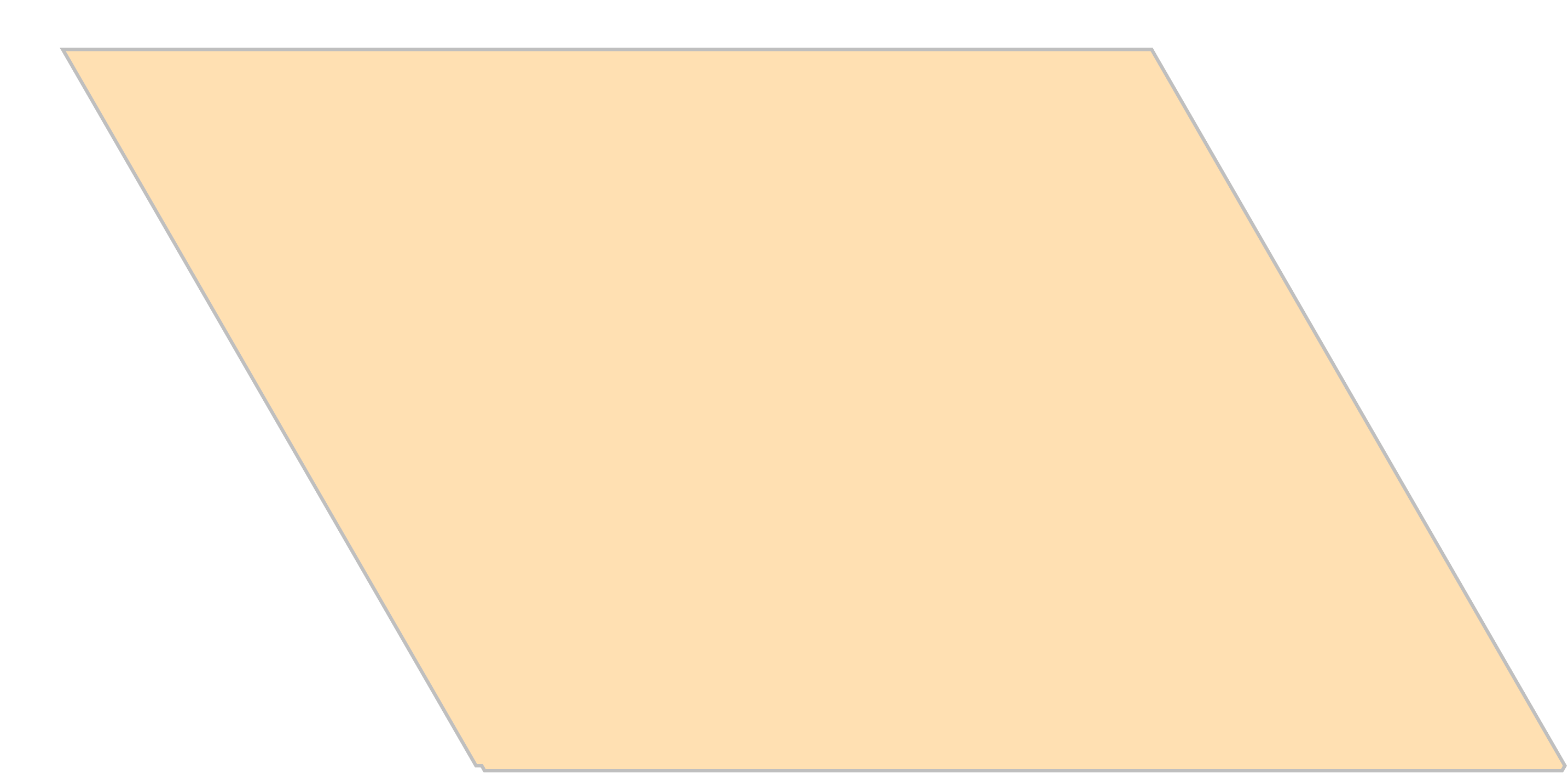}};
        \node [anchor=center] () at (0,0) {\includegraphics[scale =\s]{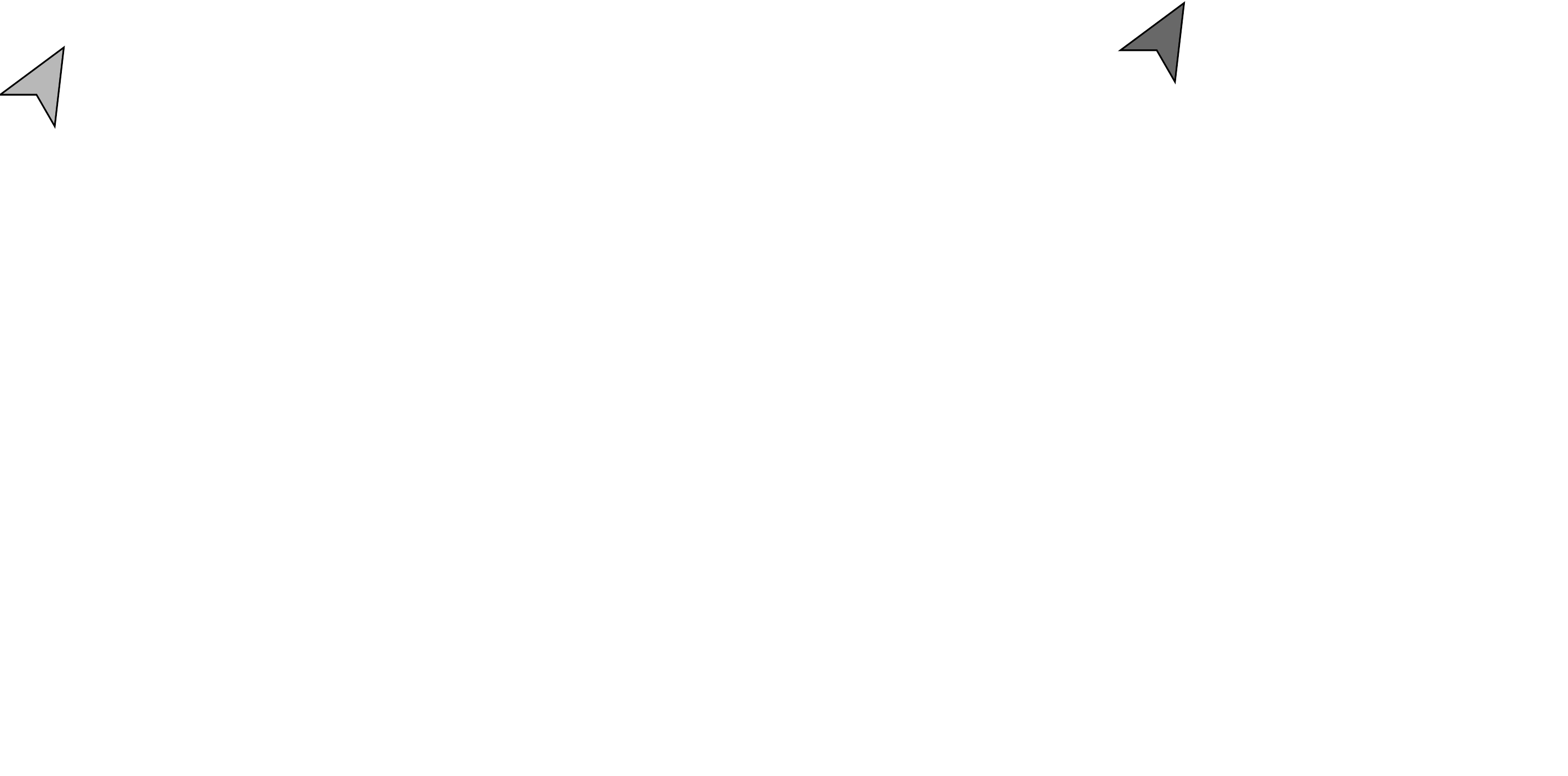}};
        \ifthenelse{\equal{#1}{}}{
\TIKZPointOfInterest{(-632.50000000*\st pt, 296.61370080*\st pt)}{a}{$(-1,1-h)$}{W}{100.0*\st pt}{Start}{(-1500.00000000*\st pt, -835.58484397*\st pt)}{8.0}{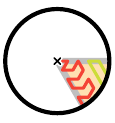}
\TIKZPointOfInterest{(-267.50000000*\st pt, -335.58484397*\st pt)}{b}{$(h-2,0)$}{W}{100.0*\st pt}{}{(-900.00000000*\st pt, -835.58484397*\st pt)}{8.0}{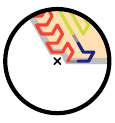}
\TIKZPointOfInterest{(-407.50000000*\st pt, 296.61370080*\st pt)}{c}{$(w+2,1-h)$}{NW}{100.0*\st pt}{Read \word1}{(-300.00000000*\st pt, -835.58484397*\st pt)}{8.0}{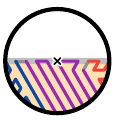}
\TIKZPointOfInterest{(-47.50000000*\st pt, -335.58484397*\st pt)}{d}{$(h+w,0)$}{SE}{100.0*\st pt}{Copy \word1}{(300.00000000*\st pt, -835.58484397*\st pt)}{8.0}{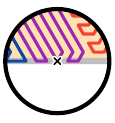}
\TIKZPointOfInterest{(687.50000000*\st pt, -335.58484397*\st pt)}{e}{$(W+h-3,0)$}{E}{100.0*\st pt}{}{(900.00000000*\st pt, -835.58484397*\st pt)}{8.0}{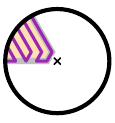}
\TIKZPointOfInterest{(327.50000000*\st pt, 296.61370080*\st pt)}{f}{$(W-1,1-h)$}{E}{100.0*\st pt}{Next}{(1500.00000000*\st pt, -835.58484397*\st pt)}{8.0}{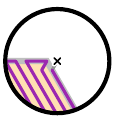}
            }{}
\node [anchor=center] () at (0,0) {\includegraphics[scale =\s]{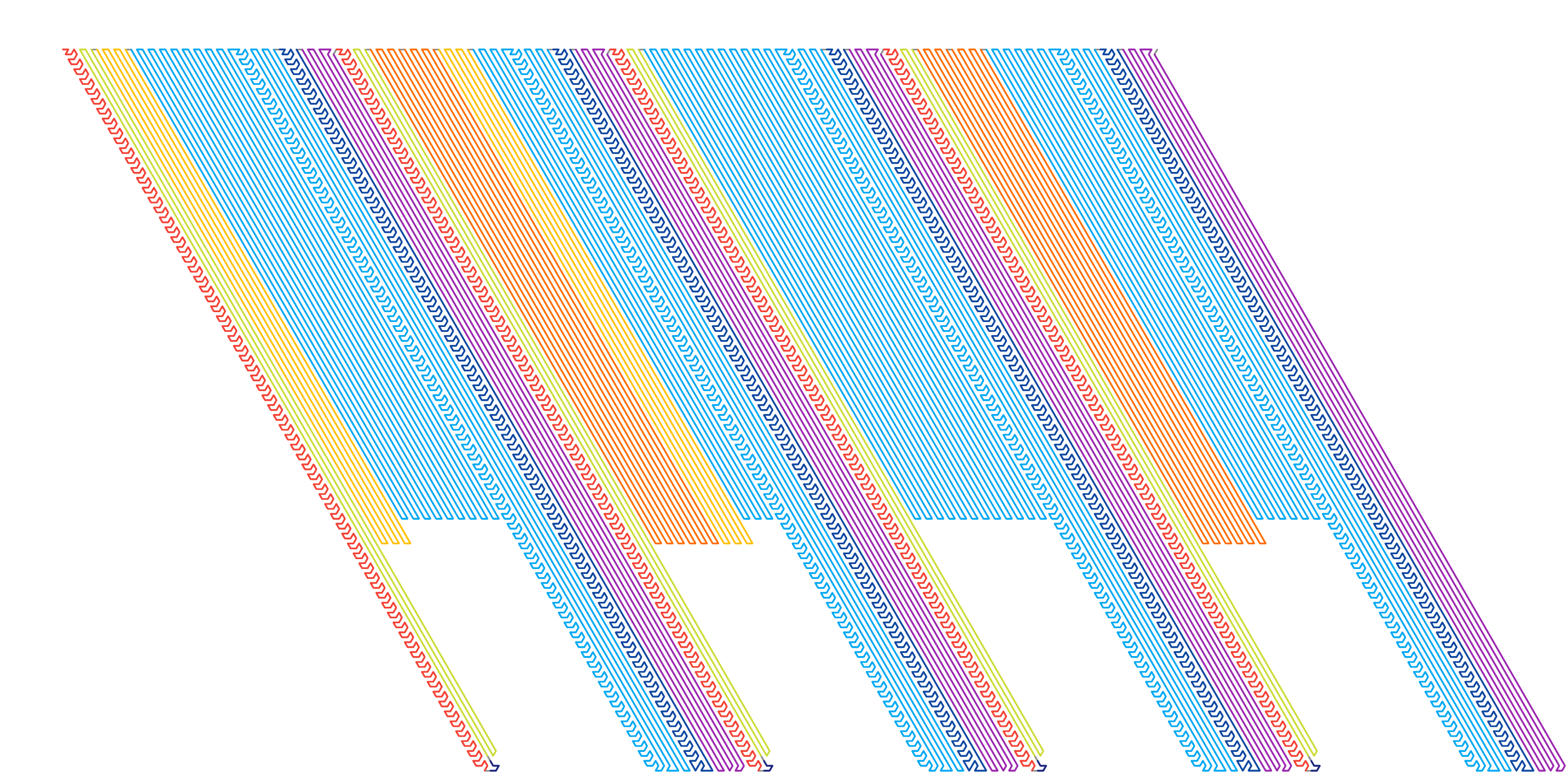}};
            \ifthenelse{\equal{#1}{}}{
\TIKZBLOCKZIGCOPY{1}{3}{0}{30.000}{-19.486}
            }{}
    \end{tikzpicture}
}

%% file: OritatamiTuringPatterns/n=4-L=3-P=1/ZigCopy1_110__n=4-L=3-P=1.tex
\newcommand{\ZigCopyUBlockUUZ}[1]{
    \begin{tikzpicture}
        \node [anchor=center] () at (0,0) {\includegraphics[scale =\s]{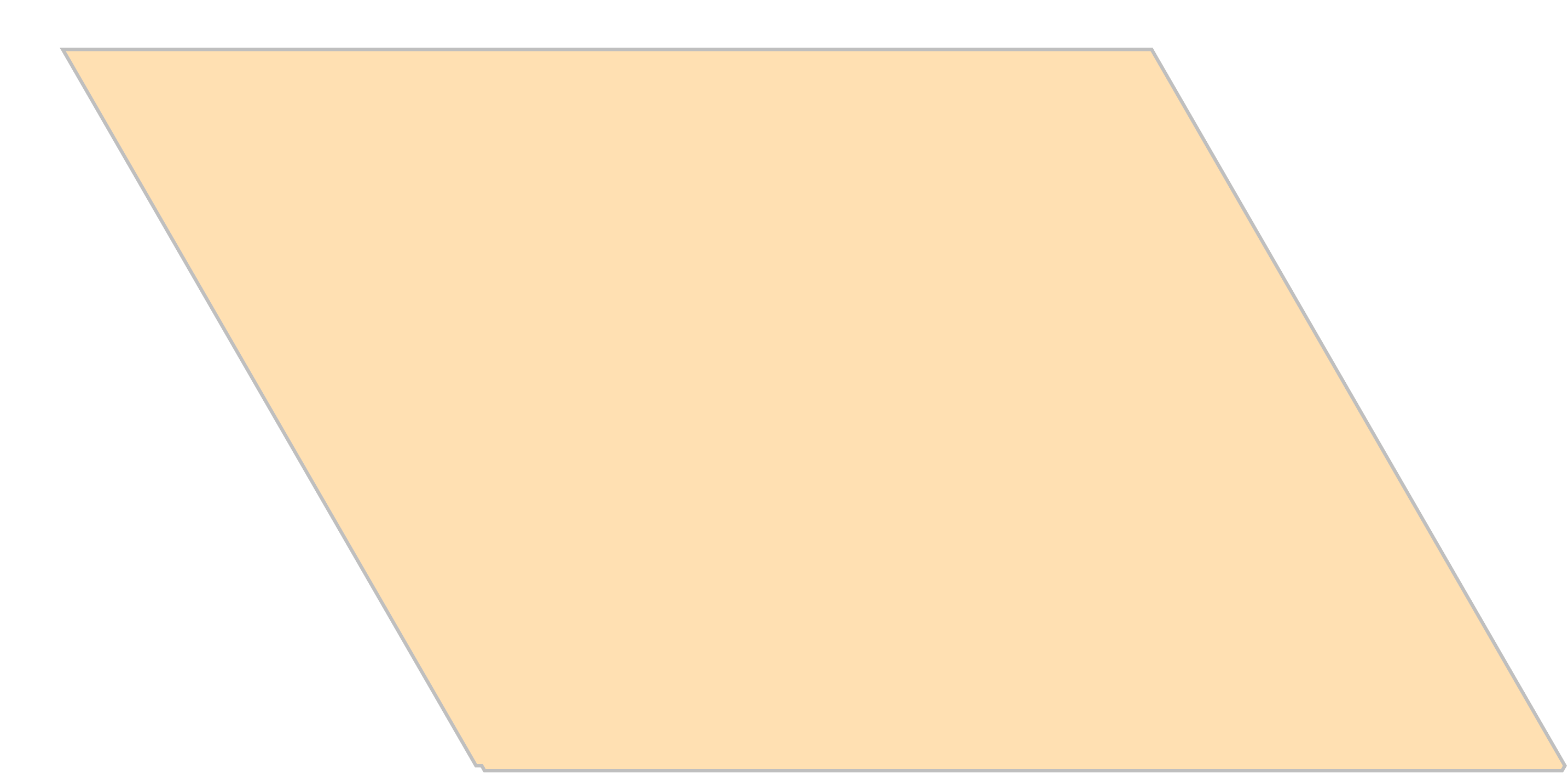}};
        \node [anchor=center] () at (0,0) {\includegraphics[scale =\s]{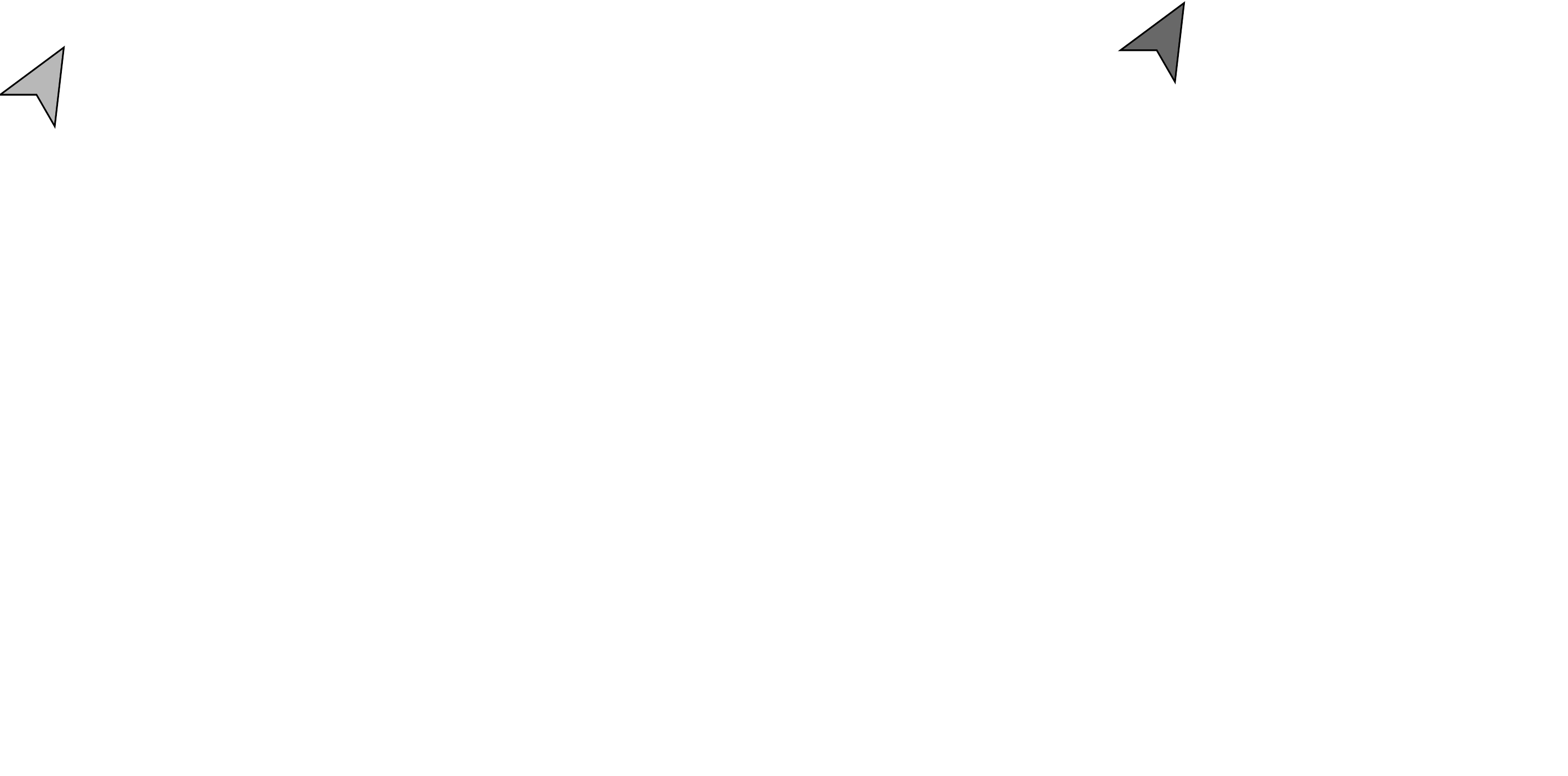}};
        \ifthenelse{\equal{#1}{}}{
\TIKZPointOfInterest{(-632.50000000*\st pt, 296.61370080*\st pt)}{a}{$(-1,1-h)$}{W}{100.0*\st pt}{Start}{(-1500.00000000*\st pt, -835.58484397*\st pt)}{8.0}{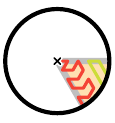}
\TIKZPointOfInterest{(-267.50000000*\st pt, -335.58484397*\st pt)}{b}{$(h-2,0)$}{W}{100.0*\st pt}{}{(-900.00000000*\st pt, -835.58484397*\st pt)}{8.0}{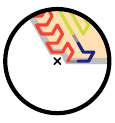}
\TIKZPointOfInterest{(-407.50000000*\st pt, 296.61370080*\st pt)}{c}{$(w+2,1-h)$}{NW}{100.0*\st pt}{Read \word1}{(-300.00000000*\st pt, -835.58484397*\st pt)}{8.0}{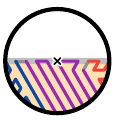}
\TIKZPointOfInterest{(-47.50000000*\st pt, -335.58484397*\st pt)}{d}{$(h+w,0)$}{SE}{100.0*\st pt}{Copy \word1}{(300.00000000*\st pt, -835.58484397*\st pt)}{8.0}{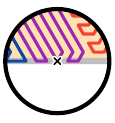}
\TIKZPointOfInterest{(687.50000000*\st pt, -335.58484397*\st pt)}{e}{$(W+h-3,0)$}{E}{100.0*\st pt}{}{(900.00000000*\st pt, -835.58484397*\st pt)}{8.0}{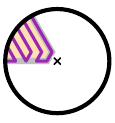}
\TIKZPointOfInterest{(327.50000000*\st pt, 296.61370080*\st pt)}{f}{$(W-1,1-h)$}{E}{100.0*\st pt}{Next}{(1500.00000000*\st pt, -835.58484397*\st pt)}{8.0}{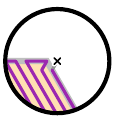}
            }{}
\node [anchor=center] () at (0,0) {\includegraphics[scale =\s]{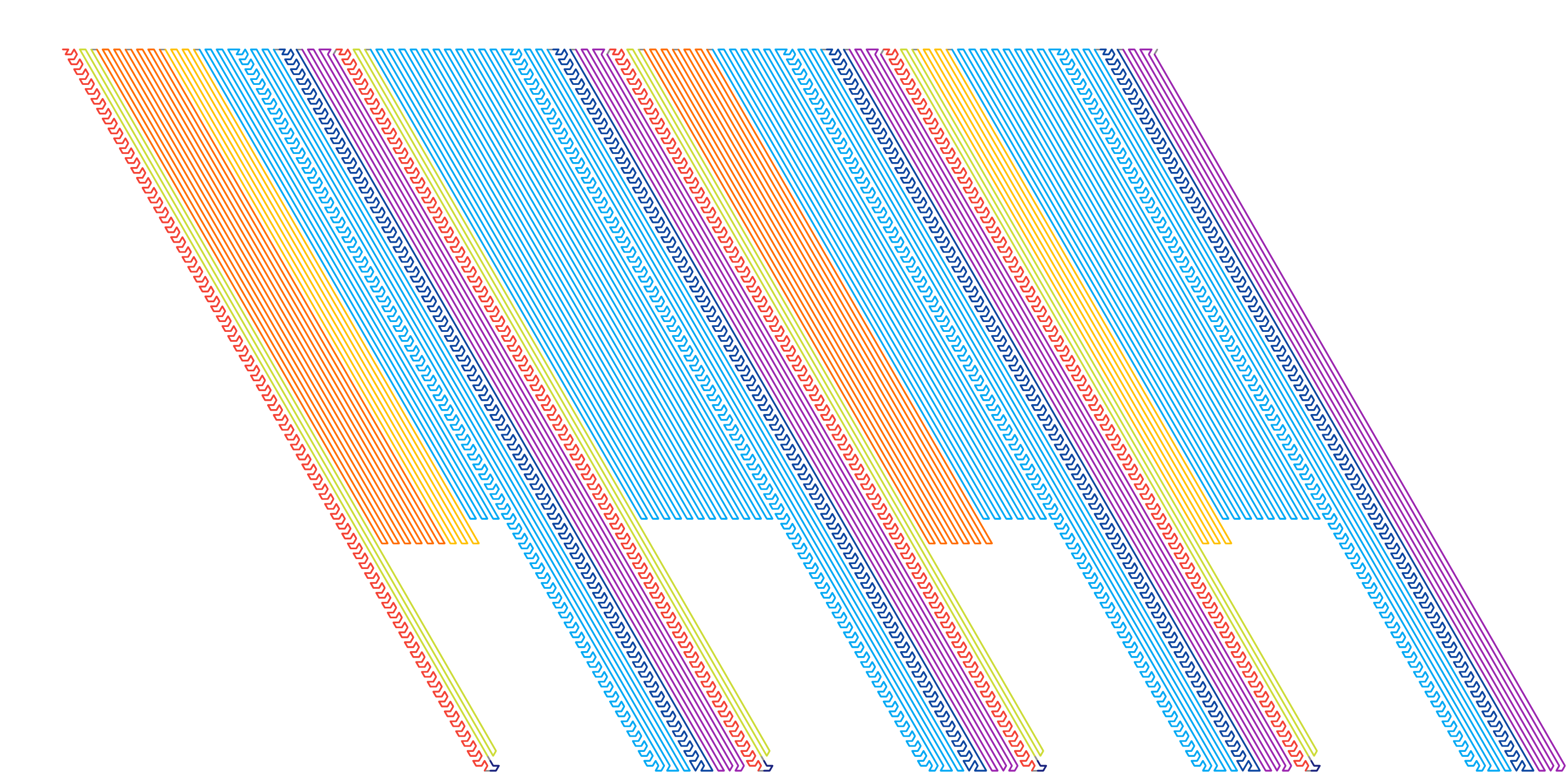}};
            \ifthenelse{\equal{#1}{}}{
\TIKZBLOCKZIGCOPY{1}{0}{110}{30.000}{-19.486}
            }{}
    \end{tikzpicture}
}

%% file: OritatamiTuringPatterns/n=4-L=3-P=1/ZigCopy1_11__n=4-L=3-P=1.tex
\newcommand{\ZigCopyUBlockUU}[1]{
    \begin{tikzpicture}
        \node [anchor=center] () at (0,0) {\includegraphics[scale =\s]{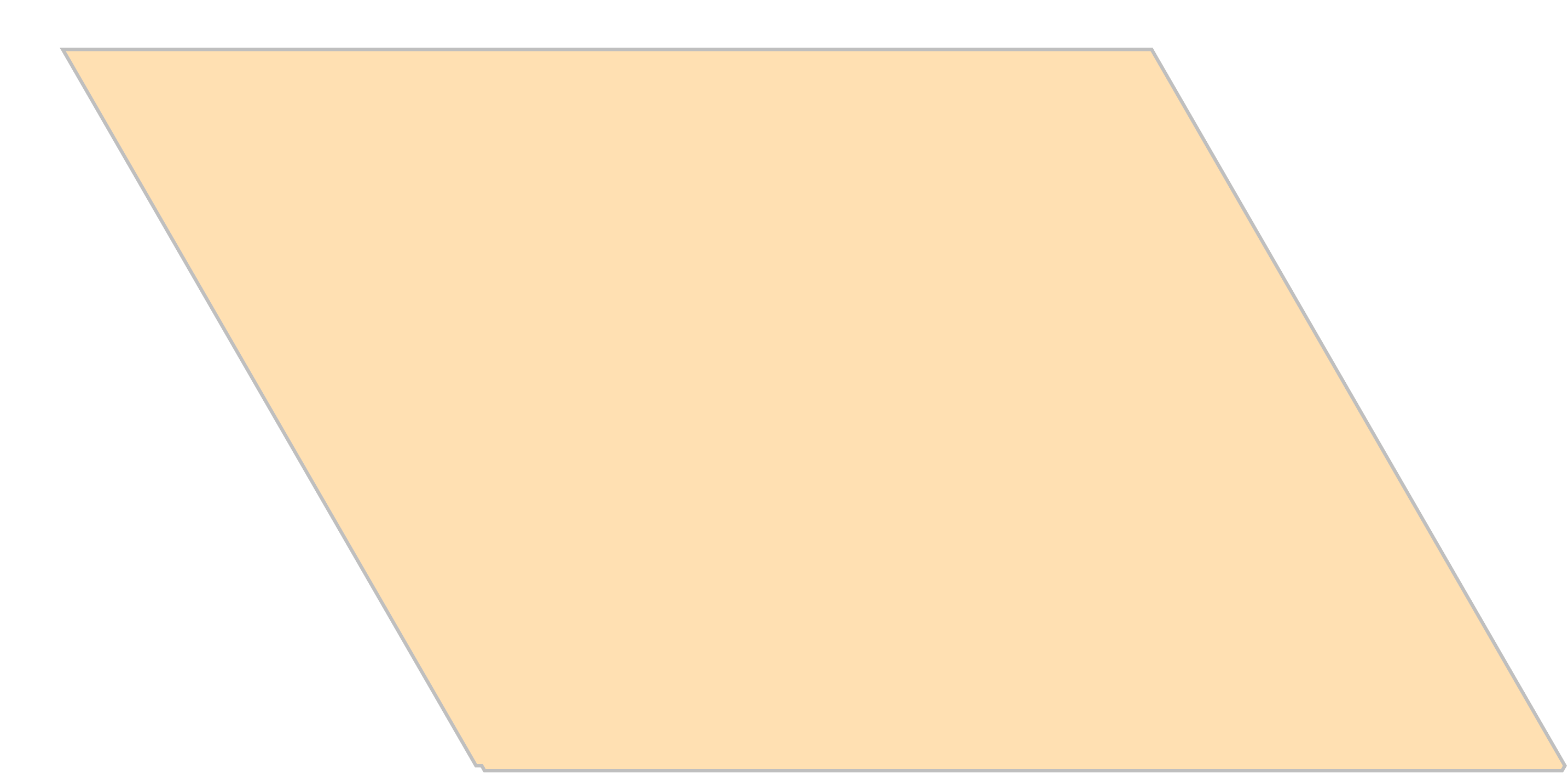}};
        \node [anchor=center] () at (0,0) {\includegraphics[scale =\s]{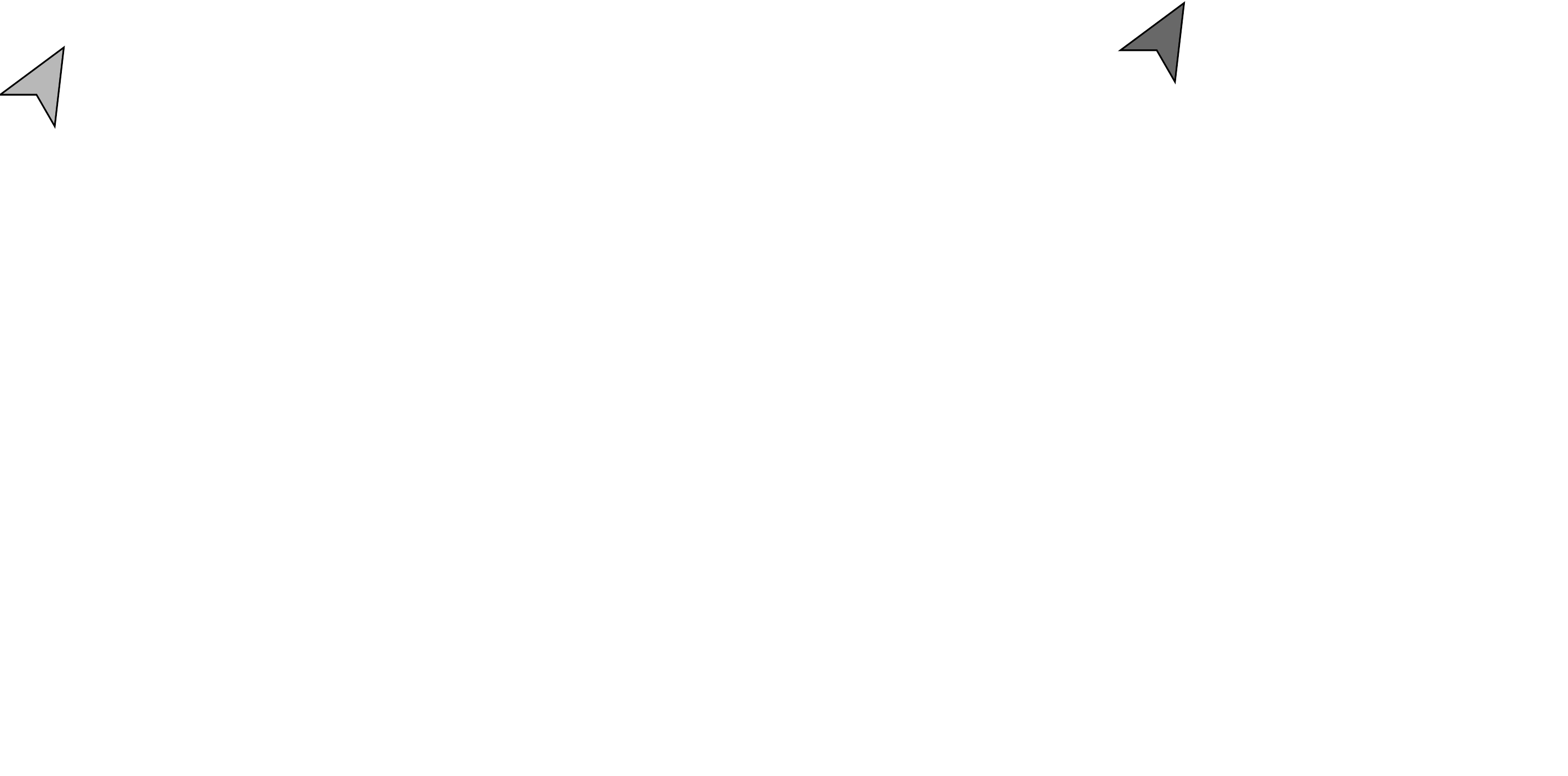}};
        \ifthenelse{\equal{#1}{}}{
\TIKZPointOfInterest{(-632.50000000*\st pt, 296.61370080*\st pt)}{a}{$(-1,1-h)$}{W}{100.0*\st pt}{Start}{(-1500.00000000*\st pt, -835.58484397*\st pt)}{8.0}{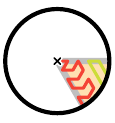}
\TIKZPointOfInterest{(-267.50000000*\st pt, -335.58484397*\st pt)}{b}{$(h-2,0)$}{W}{100.0*\st pt}{}{(-900.00000000*\st pt, -835.58484397*\st pt)}{8.0}{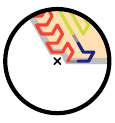}
\TIKZPointOfInterest{(-407.50000000*\st pt, 296.61370080*\st pt)}{c}{$(w+2,1-h)$}{NW}{100.0*\st pt}{Read \word1}{(-300.00000000*\st pt, -835.58484397*\st pt)}{8.0}{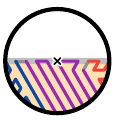}
\TIKZPointOfInterest{(-47.50000000*\st pt, -335.58484397*\st pt)}{d}{$(h+w,0)$}{SE}{100.0*\st pt}{Copy \word1}{(300.00000000*\st pt, -835.58484397*\st pt)}{8.0}{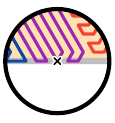}
\TIKZPointOfInterest{(687.50000000*\st pt, -335.58484397*\st pt)}{e}{$(W+h-3,0)$}{E}{100.0*\st pt}{}{(900.00000000*\st pt, -835.58484397*\st pt)}{8.0}{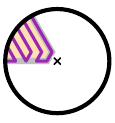}
\TIKZPointOfInterest{(327.50000000*\st pt, 296.61370080*\st pt)}{f}{$(W-1,1-h)$}{E}{100.0*\st pt}{Next}{(1500.00000000*\st pt, -835.58484397*\st pt)}{8.0}{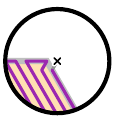}
            }{}
\node [anchor=center] () at (0,0) {\includegraphics[scale =\s]{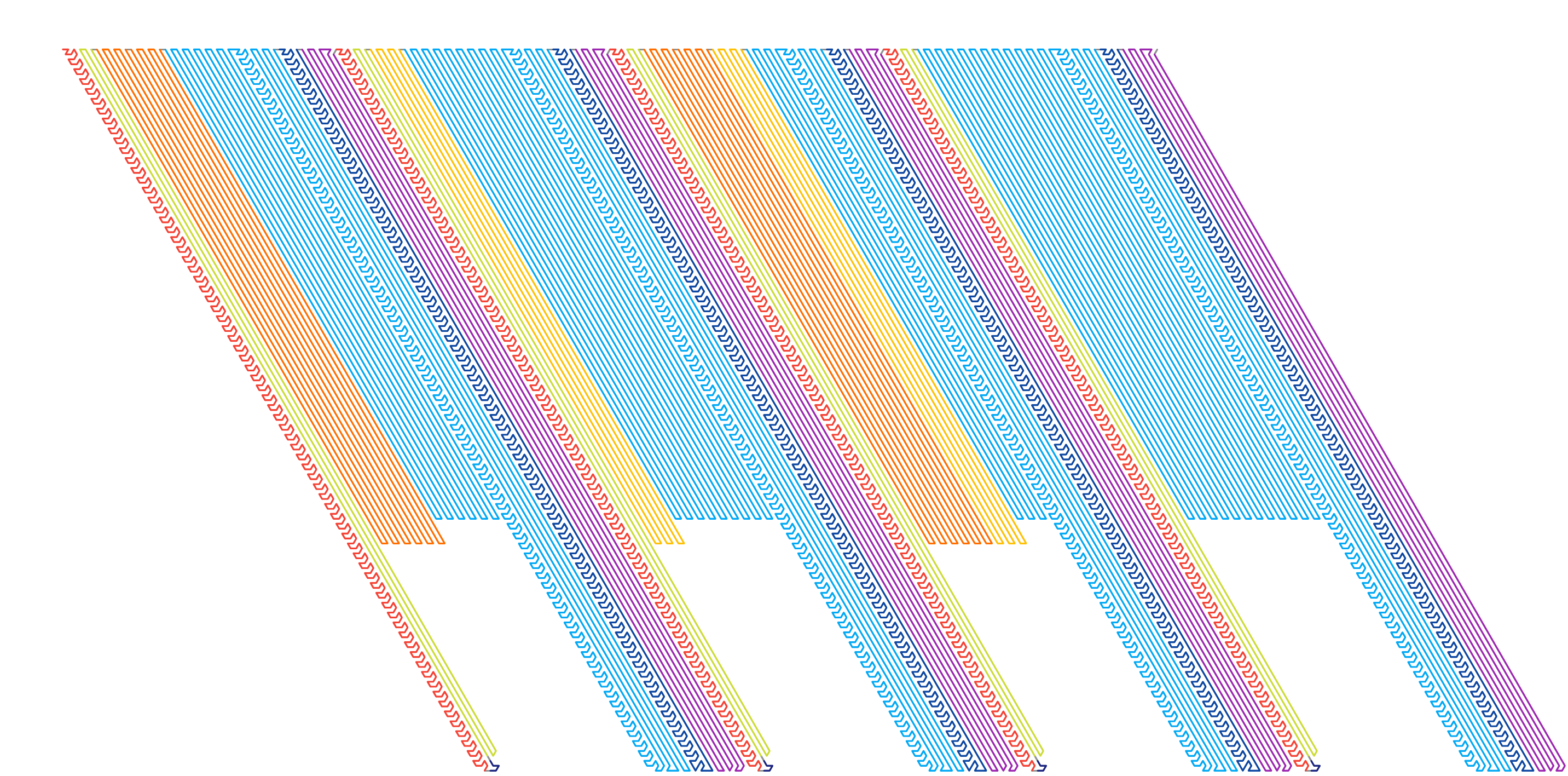}};
            \ifthenelse{\equal{#1}{}}{
\TIKZBLOCKZIGCOPY{1}{2}{11}{30.000}{-19.486}
            }{}
    \end{tikzpicture}
}

%% file: OritatamiTuringPatterns/n=4-L=3-P=1/ZigCopy1_____n=4-L=3-P=1.tex
\newcommand{\ZigCopyUBlockEpsilon}[1]{
    \begin{tikzpicture}
        \node [anchor=center] () at (0,0) {\includegraphics[scale =\s]{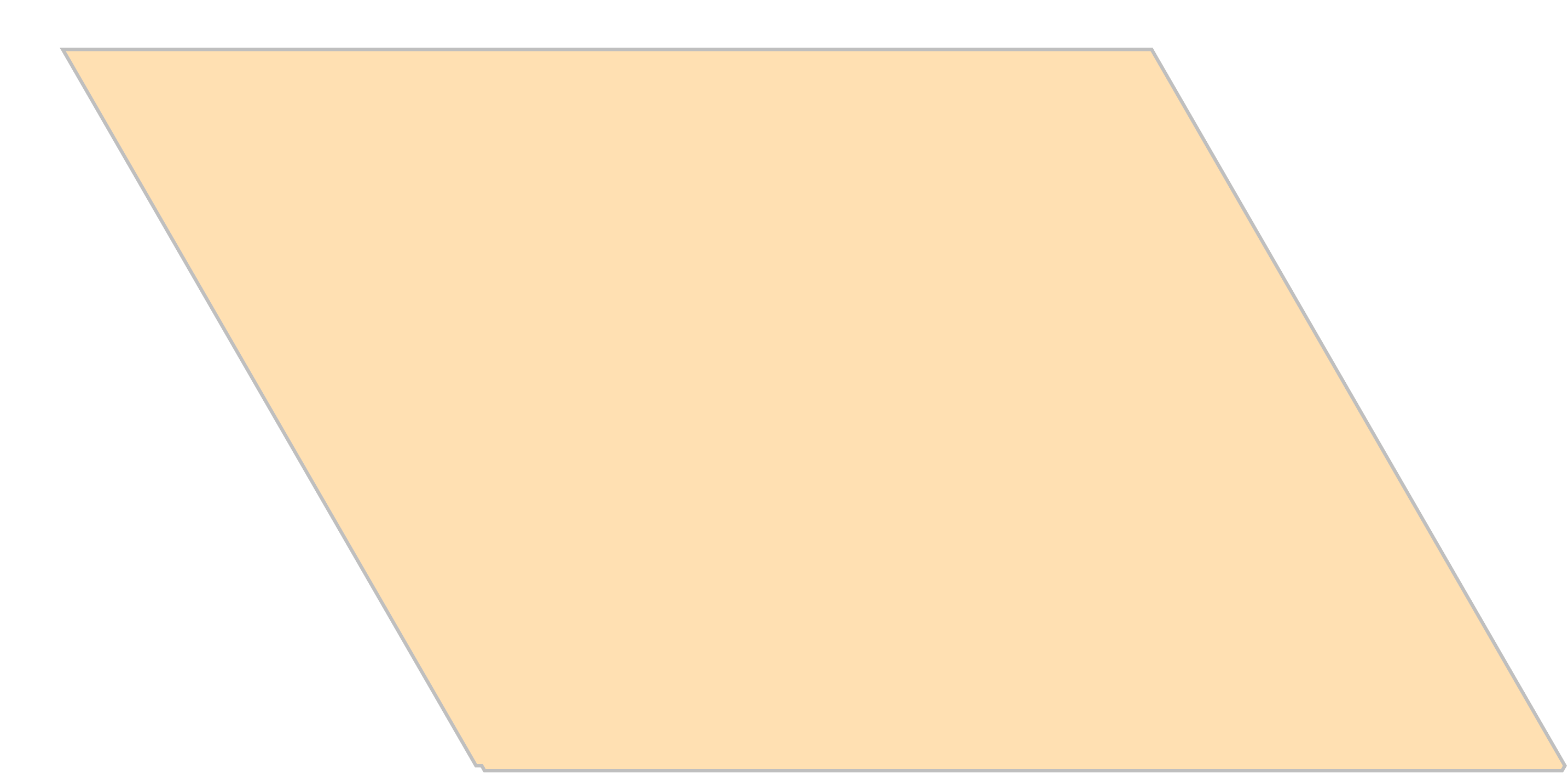}};
        \node [anchor=center] () at (0,0) {\includegraphics[scale =\s]{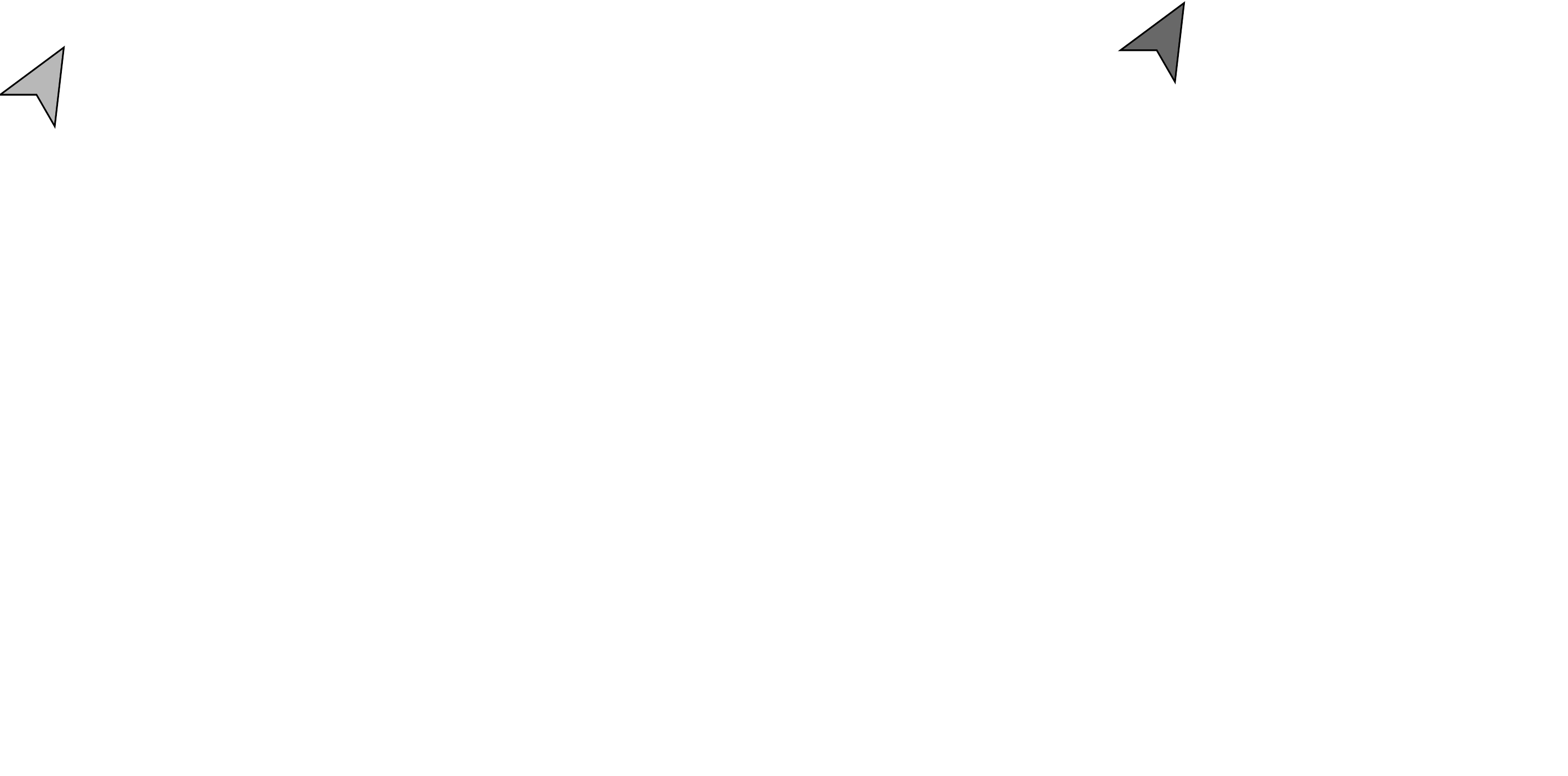}};
        \ifthenelse{\equal{#1}{}}{
\TIKZPointOfInterest{(-632.50000000*\st pt, 296.61370080*\st pt)}{a}{$(-1,1-h)$}{W}{100.0*\st pt}{Start}{(-1500.00000000*\st pt, -835.58484397*\st pt)}{8.0}{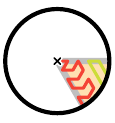}
\TIKZPointOfInterest{(-267.50000000*\st pt, -335.58484397*\st pt)}{b}{$(h-2,0)$}{W}{100.0*\st pt}{}{(-900.00000000*\st pt, -835.58484397*\st pt)}{8.0}{ZigCopy1____n=4-L=3-P=1-PI-B-_146,146_-bottom-left.pdf}
\TIKZPointOfInterest{(-407.50000000*\st pt, 296.61370080*\st pt)}{c}{$(w+2,1-h)$}{NW}{100.0*\st pt}{Read \word1}{(-300.00000000*\st pt, -835.58484397*\st pt)}{8.0}{ZigCopy1____n=4-L=3-P=1-PI-C-_45,0_-read1.pdf}
\TIKZPointOfInterest{(-47.50000000*\st pt, -335.58484397*\st pt)}{d}{$(h+w,0)$}{SE}{100.0*\st pt}{Copy \word1}{(300.00000000*\st pt, -835.58484397*\st pt)}{8.0}{ZigCopy1____n=4-L=3-P=1-PI-D-_190,146_-copy1.pdf}
\TIKZPointOfInterest{(687.50000000*\st pt, -335.58484397*\st pt)}{e}{$(W+h-3,0)$}{E}{100.0*\st pt}{}{(900.00000000*\st pt, -835.58484397*\st pt)}{8.0}{ZigCopy1____n=4-L=3-P=1-PI-E-_337,146_-bottom-right.pdf}
\TIKZPointOfInterest{(327.50000000*\st pt, 296.61370080*\st pt)}{f}{$(W-1,1-h)$}{E}{100.0*\st pt}{Next}{(1500.00000000*\st pt, -835.58484397*\st pt)}{8.0}{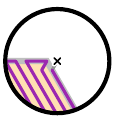}
            }{}
\node [anchor=center] () at (0,0) {\includegraphics[scale =\s]{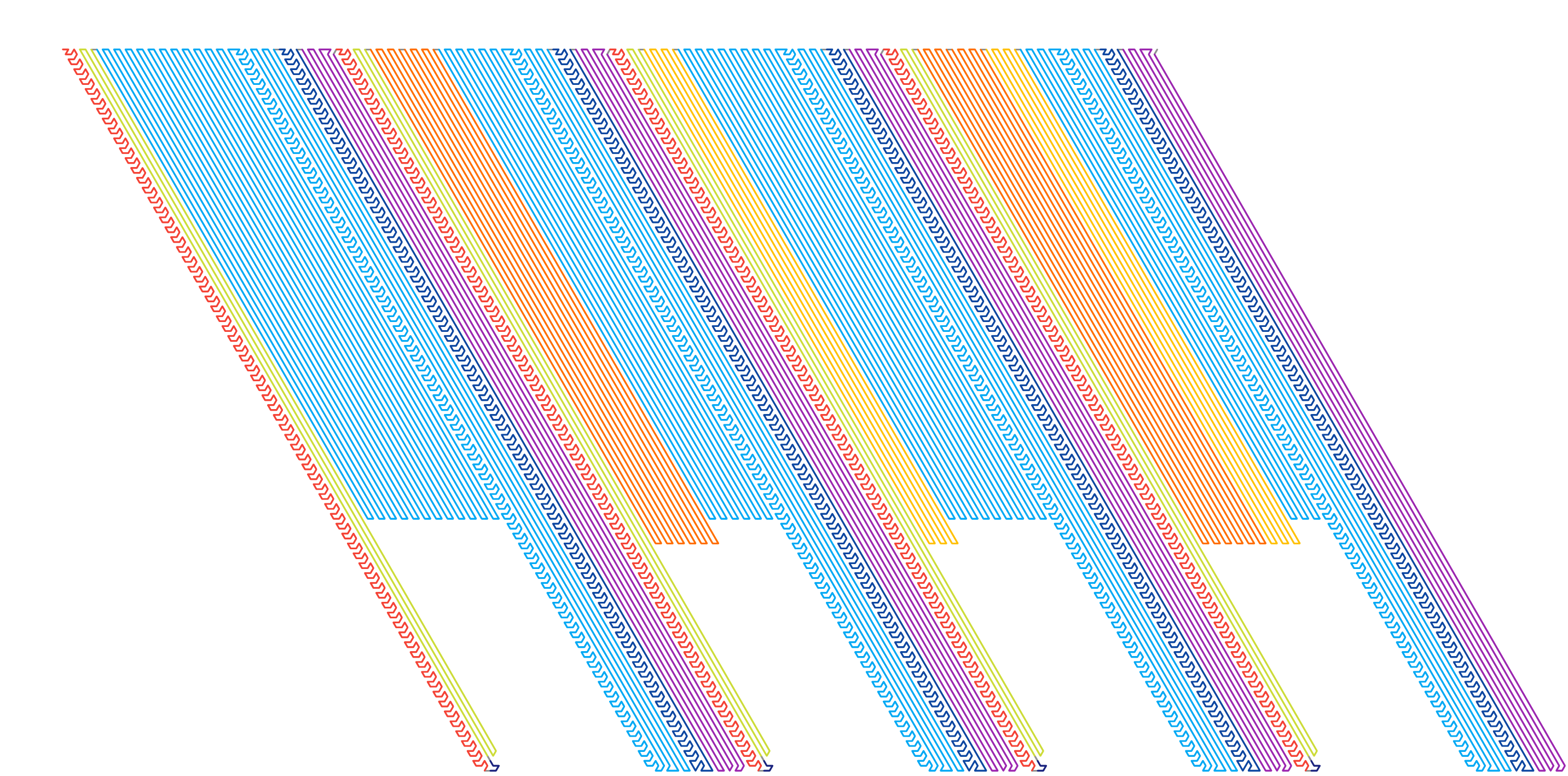}};
            \ifthenelse{\equal{#1}{}}{
\TIKZBLOCKZIGCOPY{1}{1}{\ensuremath{\epsilon}}{30.000}{-19.486}
            }{}
    \end{tikzpicture}
}

%% file: OritatamiTuringPatterns/n=4-L=3-P=1/execution_n=4-L=3-P=1-outlined.tex
\newcommand{\executionnQLTPUoutlined}[1]{
    \begin{tikzpicture}
        \node [anchor=center] () at (0,0) {\includegraphics[scale =\s]{execution_n=4-L=3-P=1-outlined-outline.pdf}};
        \node [anchor=center] () at (0,0) {\includegraphics[scale =\s]{execution_n=4-L=3-P=1-outlined-arrow.pdf}};
        \ifthenelse{\equal{#1}{}}{
            }{}
\node [anchor=center] () at (0,0) {\includegraphics[scale =\s]{execution_n=4-L=3-P=1-outlined-path.pdf}};
            \ifthenelse{\equal{#1}{}}{
\TIKZBLOCKSEED{010}{-2475.500}{2302.762}
\TIKZBLOCKSEEDLETTER{0}{-1569.975}{2243.049}
\TIKZBLOCKSEEDLETTER{1}{-2529.975}{2243.049}
\TIKZBLOCKSEEDLETTER{0}{-3489.975}{2243.049}
\TIKZBLOCKREAD{0}{0}{110}{-3786.250}{1712.565}
\TIKZBLOCKREAD{1}{1}{\ensuremath{\epsilon}}{-2826.250}{1712.565}
\TIKZBLOCKZIGCOPY{0}{2}{11}{-1506.250}{1712.565}
\TIKZBLOCKAPPENDCARRIAGERETURN{2}{858.250}{1774.053}
\TIKZBLOCKAPPENDLETTER{1}{0,0}{-266.500}{1727.721}
\TIKZBLOCKAPPENDLETTER{1}{1,1}{666.000}{1732.051}
\TIKZBLOCKZAGCOPY{1}{3}{0}{381.250}{1076.037}
\TIKZBLOCKZAGCOPY{1}{3}{0}{-578.750}{1076.037}
\TIKZBLOCKZAGCOPYLINEFEED{0}{3}{0}{-1538.750}{1076.037}
\TIKZBLOCKREAD{0}{3}{0}{-1866.250}{439.508}
\TIKZBLOCKREAD{1}{0}{110}{-906.250}{439.508}
\TIKZBLOCKZIGCOPY{1}{1}{\ensuremath{\epsilon}}{413.750}{439.508}
\TIKZBLOCKAPPENDCARRIAGERETURN{1}{2925.250}{246.384}
\TIKZBLOCKZAGCOPYLINEFEED{1}{2}{11}{381.250}{-197.021}
\TIKZBLOCKREAD{1}{2}{11}{53.750}{-833.549}
\TIKZBLOCKAPPENDCARRIAGERETURN{3}{2778.250}{-772.062}
\TIKZBLOCKAPPENDLETTER{0}{0,1}{1653.500}{-818.394}
\TIKZBLOCKZAGCOPYLINEFEED{0}{0}{110}{1341.250}{-1470.078}
\TIKZBLOCKREAD{0}{0}{110}{1013.750}{-2106.607}
\TIKZBLOCKHALT{1}{1591.000}{-2104.442}
            }{}
    \end{tikzpicture}
}

%% file: tikz-exploded.tex
\input{./OritatamiTuringPatterns/n=4-L=3-P=1/Append_0_n=4-L=3-P=1-exploded.tex}
\input{./OritatamiTuringPatterns/n=4-L=3-P=1/Append_10_n=4-L=3-P=1-exploded.tex}
\input{./OritatamiTuringPatterns/n=4-L=3-P=1/Append_110_n=4-L=3-P=1-exploded.tex}
\input{./OritatamiTuringPatterns/n=4-L=3-P=1/Append_11_n=4-L=3-P=1-exploded.tex}
\input{./OritatamiTuringPatterns/n=4-L=3-P=1/Append____n=4-L=3-P=1-exploded.tex}
\input{./OritatamiTuringPatterns/n=4-L=3-P=1/Halt_0_n=4-L=3-P=1-exploded.tex}
\input{./OritatamiTuringPatterns/n=4-L=3-P=1/Halt_110_n=4-L=3-P=1-exploded.tex}
\input{./OritatamiTuringPatterns/n=4-L=3-P=1/Halt_11_n=4-L=3-P=1-exploded.tex}
\input{./OritatamiTuringPatterns/n=4-L=3-P=1/Halt____n=4-L=3-P=1-exploded.tex}
\input{./OritatamiTuringPatterns/n=4-L=3-P=1/Read0_0_n=4-L=3-P=1-exploded.tex}
\input{./OritatamiTuringPatterns/n=4-L=3-P=1/Read0_110_n=4-L=3-P=1-exploded.tex}
\input{./OritatamiTuringPatterns/n=4-L=3-P=1/Read0_11_n=4-L=3-P=1-exploded.tex}
\input{./OritatamiTuringPatterns/n=4-L=3-P=1/Read0____n=4-L=3-P=1-exploded.tex}
\input{./OritatamiTuringPatterns/n=4-L=3-P=1/Read1_0_n=4-L=3-P=1-exploded.tex}
\input{./OritatamiTuringPatterns/n=4-L=3-P=1/Read1_110_n=4-L=3-P=1-exploded.tex}
\input{./OritatamiTuringPatterns/n=4-L=3-P=1/Read1_11_n=4-L=3-P=1-exploded.tex}
\input{./OritatamiTuringPatterns/n=4-L=3-P=1/Read1____n=4-L=3-P=1-exploded.tex}
\input{./OritatamiTuringPatterns/n=4-L=3-P=1/ZigCopy0_0_n=4-L=3-P=1-exploded.tex}
\input{./OritatamiTuringPatterns/n=4-L=3-P=1/ZigCopy0_110_n=4-L=3-P=1-exploded.tex}
\input{./OritatamiTuringPatterns/n=4-L=3-P=1/ZigCopy0_11_n=4-L=3-P=1-exploded.tex}
\input{./OritatamiTuringPatterns/n=4-L=3-P=1/ZigCopy0____n=4-L=3-P=1-exploded.tex}
\input{./OritatamiTuringPatterns/n=4-L=3-P=1/ZigCopy1_0_n=4-L=3-P=1-exploded.tex}
\input{./OritatamiTuringPatterns/n=4-L=3-P=1/ZigCopy1_110_n=4-L=3-P=1-exploded.tex}
\input{./OritatamiTuringPatterns/n=4-L=3-P=1/ZigCopy1_11_n=4-L=3-P=1-exploded.tex}
\input{./OritatamiTuringPatterns/n=4-L=3-P=1/ZigCopy1____n=4-L=3-P=1-exploded.tex}
\input{./OritatamiTuringPatterns/n=4-L=3-P=1/Read0_10_n=4-L=3-P=1-exploded.tex}
\input{./OritatamiTuringPatterns/n=4-L=3-P=1/Read1_10_n=4-L=3-P=1-exploded.tex}
\input{./OritatamiTuringPatterns/n=4-L=3-P=1/ZigCopy0_10_n=4-L=3-P=1-exploded.tex}
\input{./OritatamiTuringPatterns/n=4-L=3-P=1/ZigCopy1_10_n=4-L=3-P=1-exploded.tex}
\input{./OritatamiTuringPatterns/n=4-L=3-P=1/ZagCopy1_0_n=4-L=3-P=1-exploded.tex}
\input{./OritatamiTuringPatterns/n=4-L=3-P=1/ZagCopy1_10_n=4-L=3-P=1-exploded.tex}
\input{./OritatamiTuringPatterns/n=4-L=3-P=1/ZagCopy1_110_n=4-L=3-P=1-exploded.tex}
\input{./OritatamiTuringPatterns/n=4-L=3-P=1/ZagCopy1_11_n=4-L=3-P=1-exploded.tex}
\input{./OritatamiTuringPatterns/n=4-L=3-P=1/ZagCopy1____n=4-L=3-P=1-exploded.tex}

%% file: OritatamiTuringPatterns/n=4-L=3-P=1/Append_0_n=4-L=3-P=1-exploded.tex
\newcommand{\AppendBlockZnQLTPUexploded}[1]{
\begin{tikzpicture}
\ifthenelse{\equal{#1}{}}{
\TIKZlabel{(-1736.25000000*\st pt, 173.20508076*\st pt)}{A\explodedZigZagSize\ZigDown}{SW}{40.0*\st pt}
\TIKZlabel{(-1518.75000000*\st pt, -142.89419162*\st pt)}{B\explodedZigZagSize\ZigDown}{SW}{40.0*\st pt}
\TIKZlabel{(-1748.75000000*\st pt, 662.50943390*\st pt)}{C\explodedZigZagSize\ZigDown}{SE}{40.0*\st pt}
\TIKZlabel{(-996.25000000*\st pt, -138.56406461*\st pt)}{D0\explodedZigZagSize\Append}{SE}{80.0*\st pt}
\TIKZlabel{(-1.25000000*\st pt, -138.56406461*\st pt)}{E\explodedZigZagSize\bZigZag}{NE}{80.0*\st pt}
\TIKZlabel{(-681.25000000*\st pt, -484.97422612*\st pt)}{F\explodedZigZagSize\Zag}{SE}{40.0*\st pt}
\TIKZlabel{(-746.25000000*\st pt, -484.97422612*\st pt)}{G\explodedZigZagSize\Zag}{NW}{40.0*\st pt}
}{}
\node [anchor=center] () at (0,0) {\includegraphics[scale =\s]{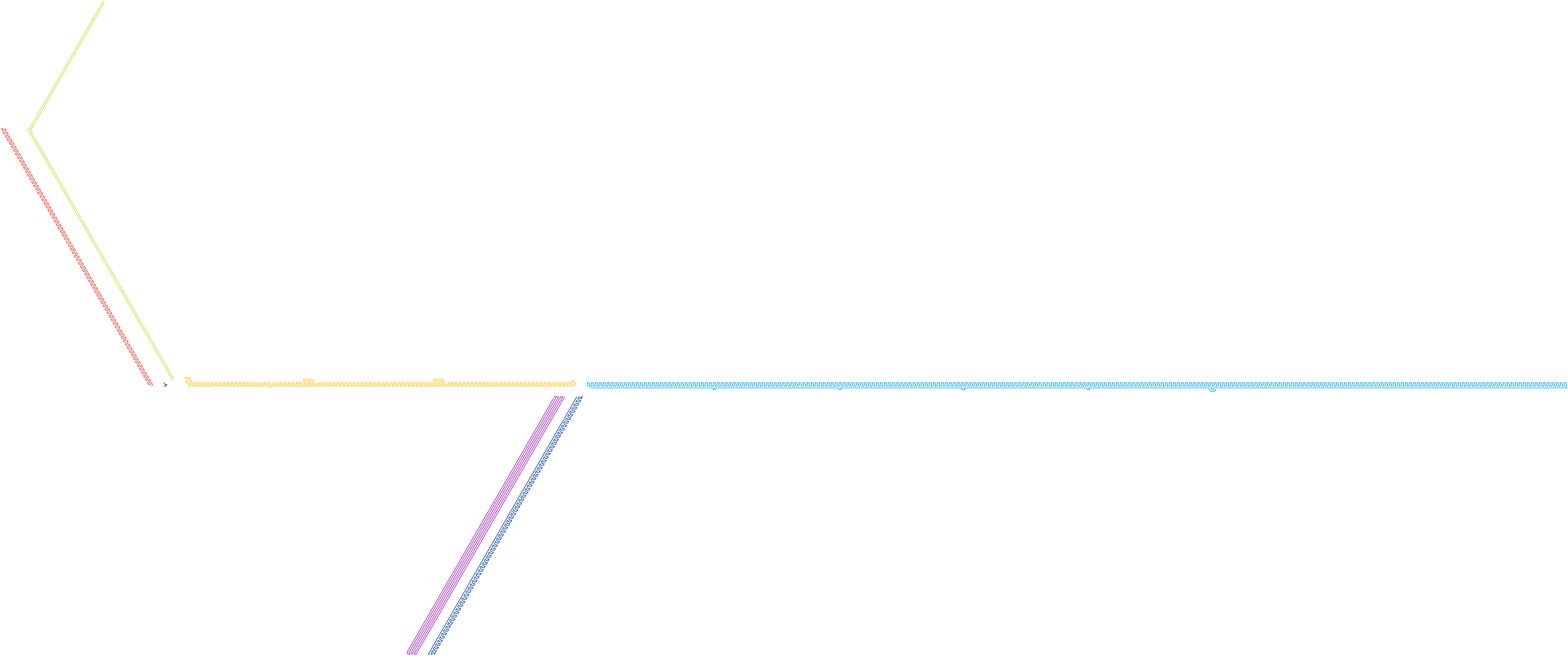}};
\end{tikzpicture}
}

%% file: OritatamiTuringPatterns/n=4-L=3-P=1/Append_10_n=4-L=3-P=1-exploded.tex
\newcommand{\AppendBlockUZnQLTPUexploded}[1]{
\begin{tikzpicture}
\ifthenelse{\equal{#1}{}}{
\TIKZlabel{(-1868.75000000*\st pt, 173.20508076*\st pt)}{A\explodedZigZagSize\ZigDown}{SW}{40.0*\st pt}
\TIKZlabel{(-1651.25000000*\st pt, -142.89419162*\st pt)}{B\explodedZigZagSize\ZigDown}{SW}{40.0*\st pt}
\TIKZlabel{(-1881.25000000*\st pt, 662.50943390*\st pt)}{C\explodedZigZagSize\ZigDown}{SE}{40.0*\st pt}
\TIKZlabel{(-1128.75000000*\st pt, -138.56406461*\st pt)}{D1\explodedZigZagSize\Append}{SE}{80.0*\st pt}
\TIKZlabel{(-173.75000000*\st pt, -138.56406461*\st pt)}{D0\explodedZigZagSize\Append}{SE}{80.0*\st pt}
\TIKZlabel{(851.25000000*\st pt, -138.56406461*\st pt)}{E\explodedZigZagSize\bZigZag}{NE}{80.0*\st pt}
\TIKZlabel{(171.25000000*\st pt, -484.97422612*\st pt)}{F\explodedZigZagSize\Zag}{SE}{40.0*\st pt}
\TIKZlabel{(106.25000000*\st pt, -484.97422612*\st pt)}{G\explodedZigZagSize\Zag}{NW}{40.0*\st pt}
}{}
\node [anchor=center] () at (0,0) {\includegraphics[scale =\s]{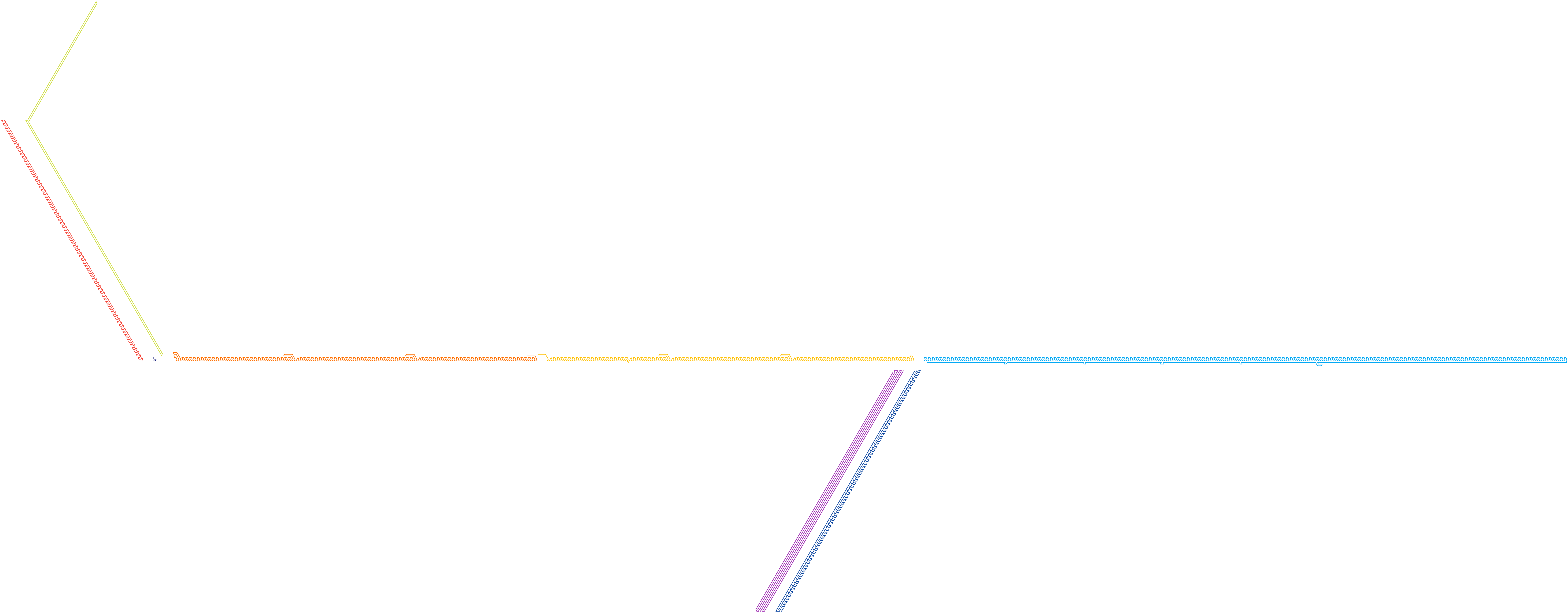}};
\end{tikzpicture}
}

%% file: OritatamiTuringPatterns/n=4-L=3-P=1/Append_110_n=4-L=3-P=1-exploded.tex
\newcommand{\AppendBlockUUZnQLTPUexploded}[1]{
\begin{tikzpicture}
\ifthenelse{\equal{#1}{}}{
\TIKZlabel{(-2001.25000000*\st pt, 173.20508076*\st pt)}{A\explodedZigZagSize\ZigDown}{SW}{40.0*\st pt}
\TIKZlabel{(-1783.75000000*\st pt, -142.89419162*\st pt)}{B\explodedZigZagSize\ZigDown}{SW}{40.0*\st pt}
\TIKZlabel{(-2013.75000000*\st pt, 662.50943390*\st pt)}{C\explodedZigZagSize\ZigDown}{SE}{40.0*\st pt}
\TIKZlabel{(-1261.25000000*\st pt, -138.56406461*\st pt)}{D1\explodedZigZagSize\Append}{SE}{80.0*\st pt}
\TIKZlabel{(-306.25000000*\st pt, -138.56406461*\st pt)}{D1\explodedZigZagSize\Append}{SE}{80.0*\st pt}
\TIKZlabel{(678.75000000*\st pt, -138.56406461*\st pt)}{D0\explodedZigZagSize\Append}{SE}{80.0*\st pt}
\TIKZlabel{(1703.75000000*\st pt, -138.56406461*\st pt)}{E\explodedZigZagSize\bZigZag}{NE}{80.0*\st pt}
\TIKZlabel{(1023.75000000*\st pt, -484.97422612*\st pt)}{F\explodedZigZagSize\Zag}{SE}{40.0*\st pt}
\TIKZlabel{(958.75000000*\st pt, -484.97422612*\st pt)}{G\explodedZigZagSize\Zag}{NW}{40.0*\st pt}
}{}
\node [anchor=center] () at (0,0) {\includegraphics[scale =\s]{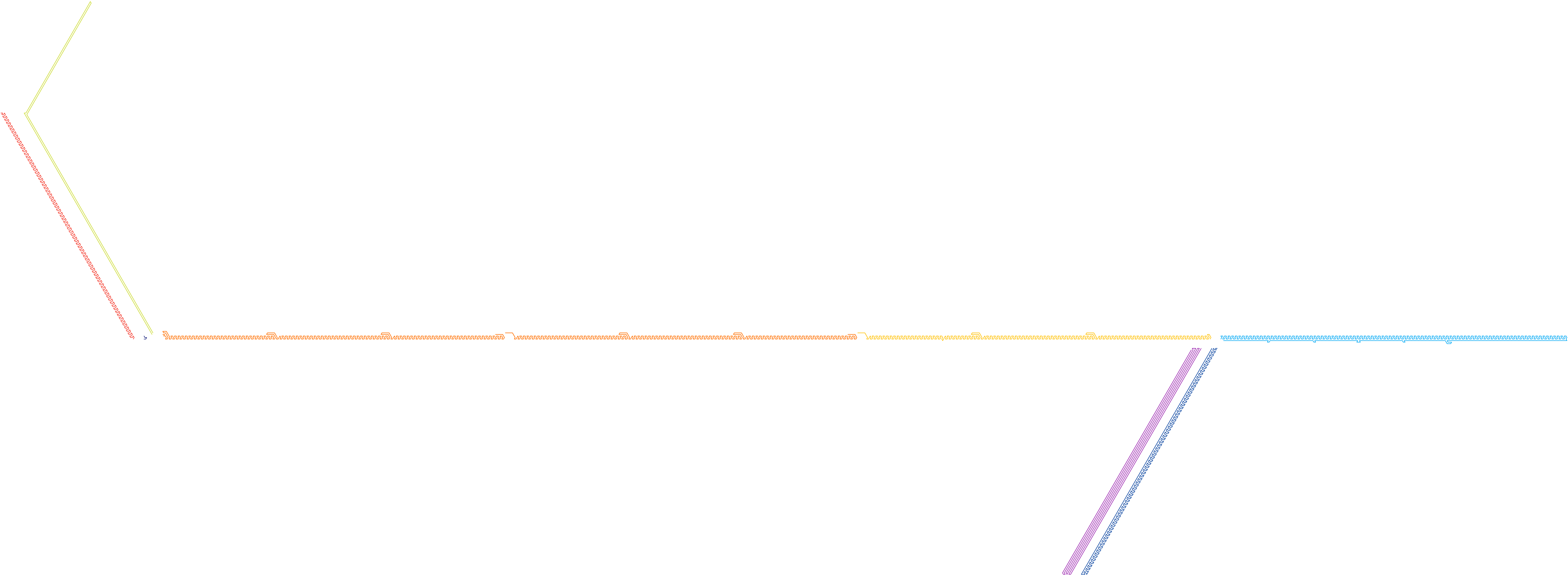}};
\end{tikzpicture}
}

%% file: OritatamiTuringPatterns/n=4-L=3-P=1/Append_11_n=4-L=3-P=1-exploded.tex
\newcommand{\AppendBlockUUnQLTPUexploded}[1]{
\begin{tikzpicture}
\ifthenelse{\equal{#1}{}}{
\TIKZlabel{(-1868.75000000*\st pt, 173.20508076*\st pt)}{A\explodedZigZagSize\ZigDown}{SW}{40.0*\st pt}
\TIKZlabel{(-1651.25000000*\st pt, -142.89419162*\st pt)}{B\explodedZigZagSize\ZigDown}{SW}{40.0*\st pt}
\TIKZlabel{(-1881.25000000*\st pt, 662.50943390*\st pt)}{C\explodedZigZagSize\ZigDown}{SE}{40.0*\st pt}
\TIKZlabel{(-1128.75000000*\st pt, -138.56406461*\st pt)}{D1\explodedZigZagSize\Append}{SE}{80.0*\st pt}
\TIKZlabel{(-173.75000000*\st pt, -138.56406461*\st pt)}{D1\explodedZigZagSize\Append}{SE}{80.0*\st pt}
\TIKZlabel{(851.25000000*\st pt, -138.56406461*\st pt)}{E\explodedZigZagSize\bZigZag}{NE}{80.0*\st pt}
\TIKZlabel{(171.25000000*\st pt, -484.97422612*\st pt)}{F\explodedZigZagSize\Zag}{SE}{40.0*\st pt}
\TIKZlabel{(106.25000000*\st pt, -484.97422612*\st pt)}{G\explodedZigZagSize\Zag}{NW}{40.0*\st pt}
}{}
\node [anchor=center] () at (0,0) {\includegraphics[scale =\s]{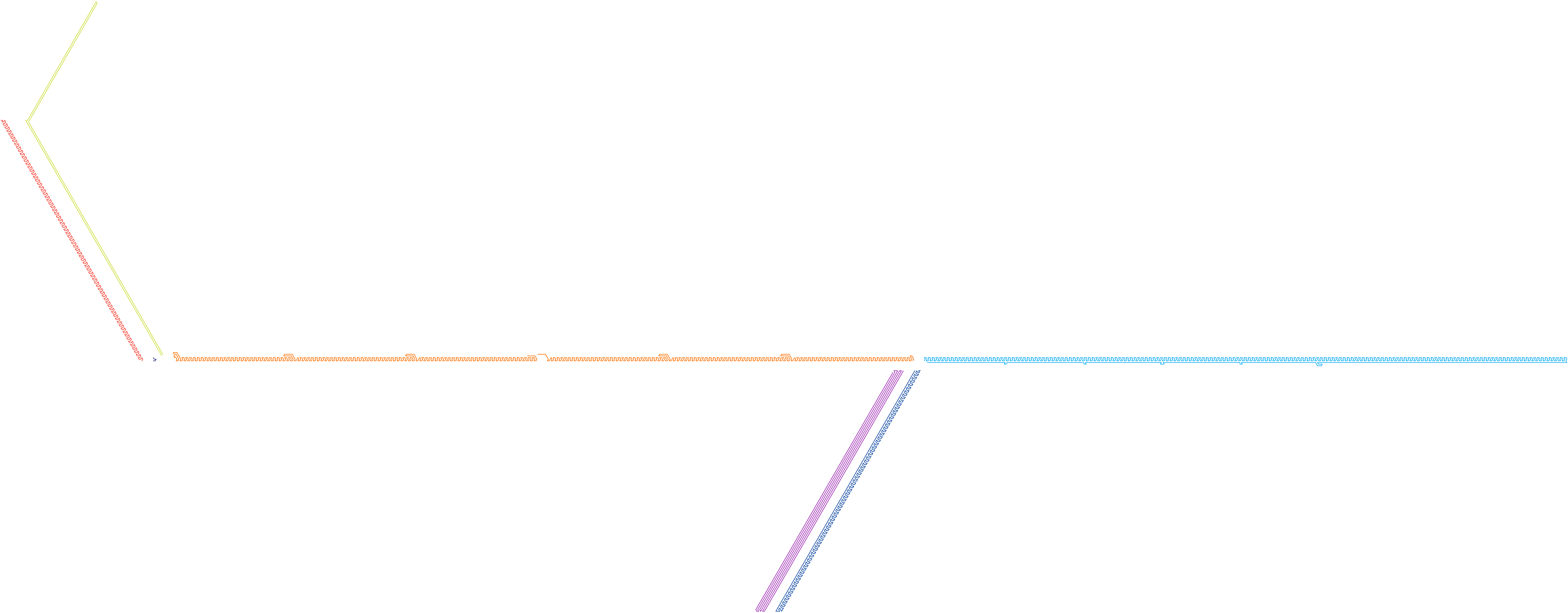}};
\end{tikzpicture}
}

%% file: OritatamiTuringPatterns/n=4-L=3-P=1/Append____n=4-L=3-P=1-exploded.tex
\newcommand{\AppendBlockEpsilonnQLTPUexploded}[1]{
\begin{tikzpicture}
\ifthenelse{\equal{#1}{}}{
\TIKZlabel{(-1603.75000000*\st pt, 173.20508076*\st pt)}{A\explodedZigZagSize\ZigDown}{SW}{40.0*\st pt}
\TIKZlabel{(-1386.25000000*\st pt, -142.89419162*\st pt)}{B\explodedZigZagSize\ZigDown}{SW}{40.0*\st pt}
\TIKZlabel{(-1616.25000000*\st pt, 662.50943390*\st pt)}{C\explodedZigZagSize\ZigDown}{SE}{40.0*\st pt}
\TIKZlabel{(-863.75000000*\st pt, -138.56406461*\st pt)}{E\explodedZigZagSize\bZigZag}{NE}{80.0*\st pt}
\TIKZlabel{(-1533.75000000*\st pt, -484.97422612*\st pt)}{F\explodedZigZagSize\Zag}{SE}{40.0*\st pt}
\TIKZlabel{(-1598.75000000*\st pt, -484.97422612*\st pt)}{G\explodedZigZagSize\Zag}{NW}{40.0*\st pt}
}{}
\node [anchor=center] () at (0,0) {\includegraphics[scale =\s]{Append___n=4-L=3-P=1-exploded.pdf}};
\end{tikzpicture}
}

%% file: OritatamiTuringPatterns/n=4-L=3-P=1/Halt_0_n=4-L=3-P=1-exploded.tex
\newcommand{\HaltBlockZnQLTPUexploded}[1]{
\begin{tikzpicture}
\ifthenelse{\equal{#1}{}}{
\TIKZlabel{(1.25000000*\st pt, 0.00000000*\st pt)}{A\explodedZigZagSize\ZigUp}{NW}{40.0*\st pt}
\TIKZlabel{(141.25000000*\st pt, 311.76914536*\st pt)}{B\explodedZigZagSize\Halt}{SW}{40.0*\st pt}
}{}
\node [anchor=center] () at (0,0) {\includegraphics[scale =\s]{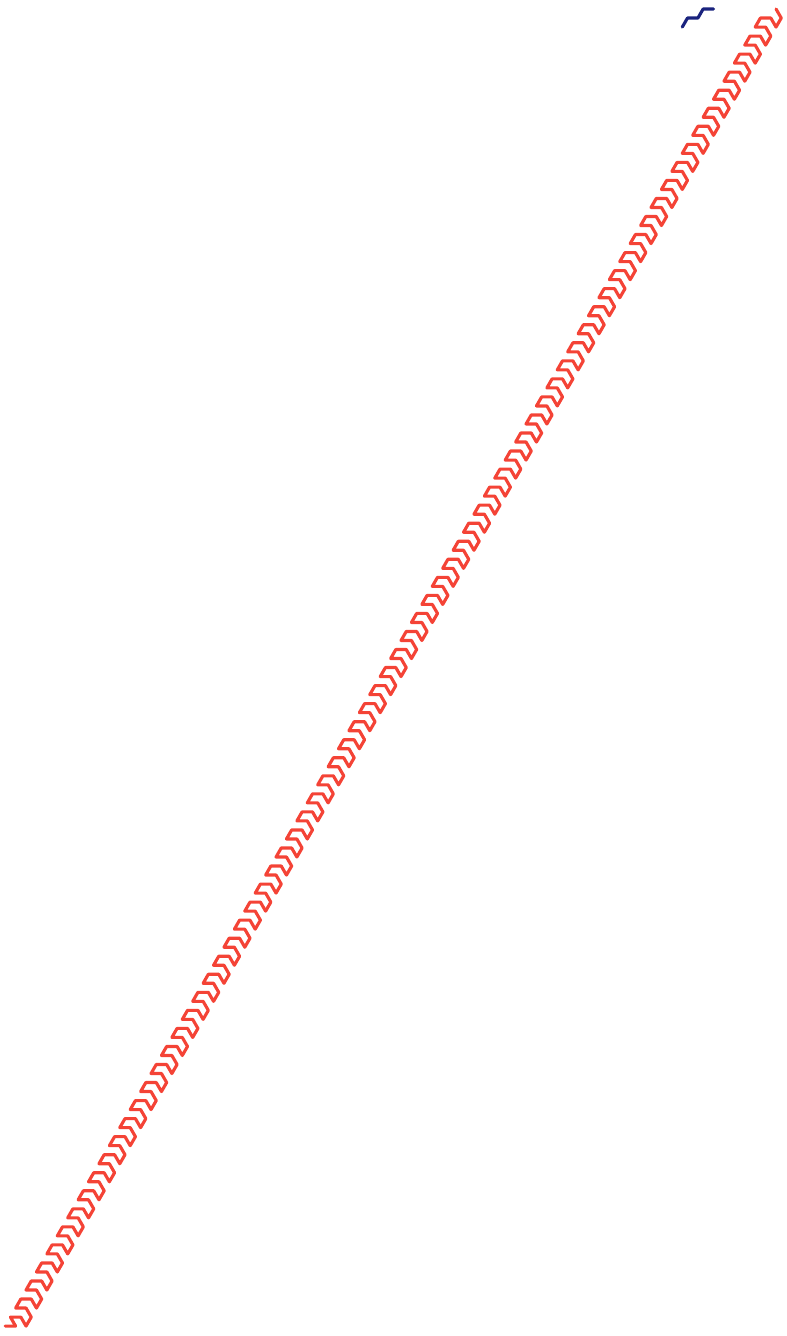}};
\end{tikzpicture}
}

%% file: OritatamiTuringPatterns/n=4-L=3-P=1/Halt_110_n=4-L=3-P=1-exploded.tex
\newcommand{\HaltBlockUUZnQLTPUexploded}[1]{
\begin{tikzpicture}
\ifthenelse{\equal{#1}{}}{
\TIKZlabel{(1.25000000*\st pt, 0.00000000*\st pt)}{A\explodedZigZagSize\ZigUp}{NW}{40.0*\st pt}
\TIKZlabel{(141.25000000*\st pt, 311.76914536*\st pt)}{B\explodedZigZagSize\Halt}{SW}{40.0*\st pt}
}{}
\node [anchor=center] () at (0,0) {\includegraphics[scale =\s]{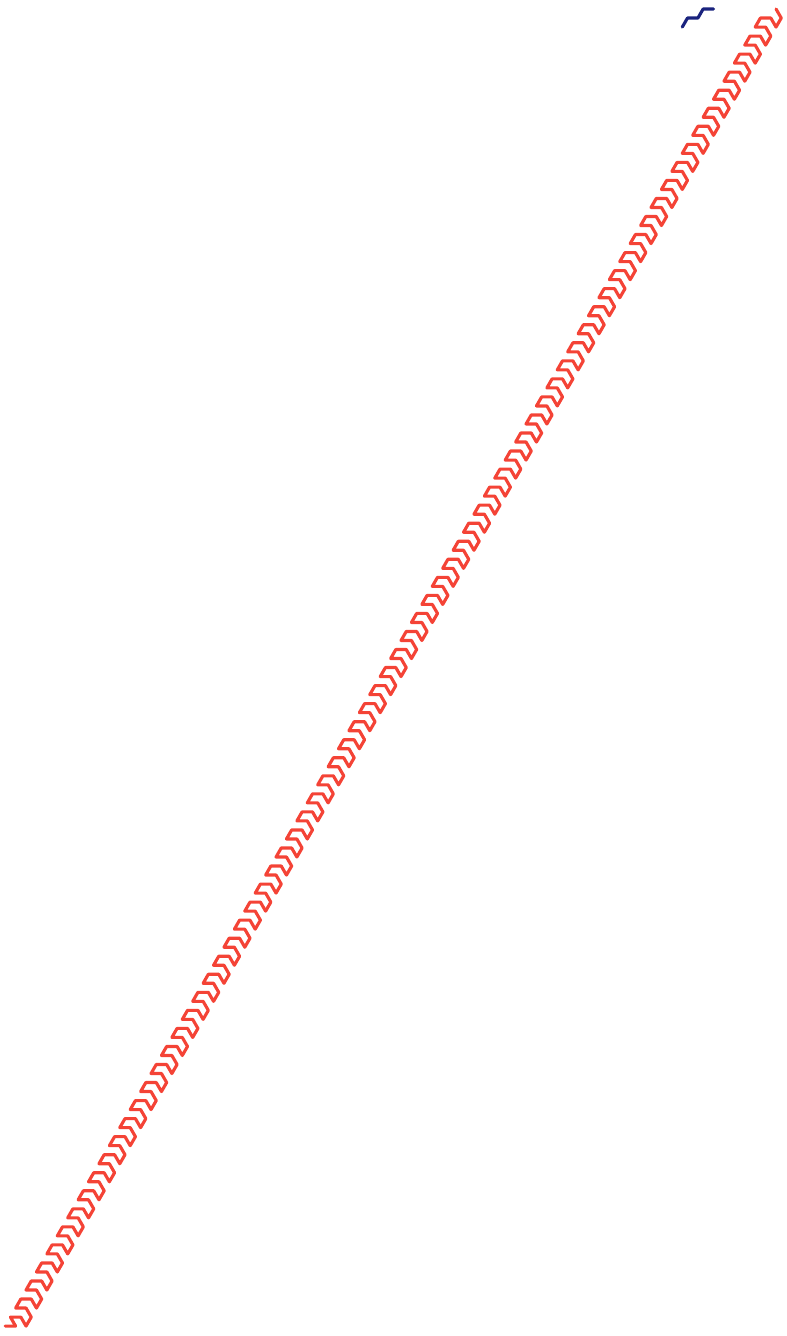}};
\end{tikzpicture}
}

%% file: OritatamiTuringPatterns/n=4-L=3-P=1/Halt_11_n=4-L=3-P=1-exploded.tex
\newcommand{\HaltBlockUUnQLTPUexploded}[1]{
\begin{tikzpicture}
\ifthenelse{\equal{#1}{}}{
\TIKZlabel{(1.25000000*\st pt, 0.00000000*\st pt)}{A\explodedZigZagSize\ZigUp}{NW}{40.0*\st pt}
\TIKZlabel{(141.25000000*\st pt, 311.76914536*\st pt)}{B\explodedZigZagSize\Halt}{SW}{40.0*\st pt}
}{}
\node [anchor=center] () at (0,0) {\includegraphics[scale =\s]{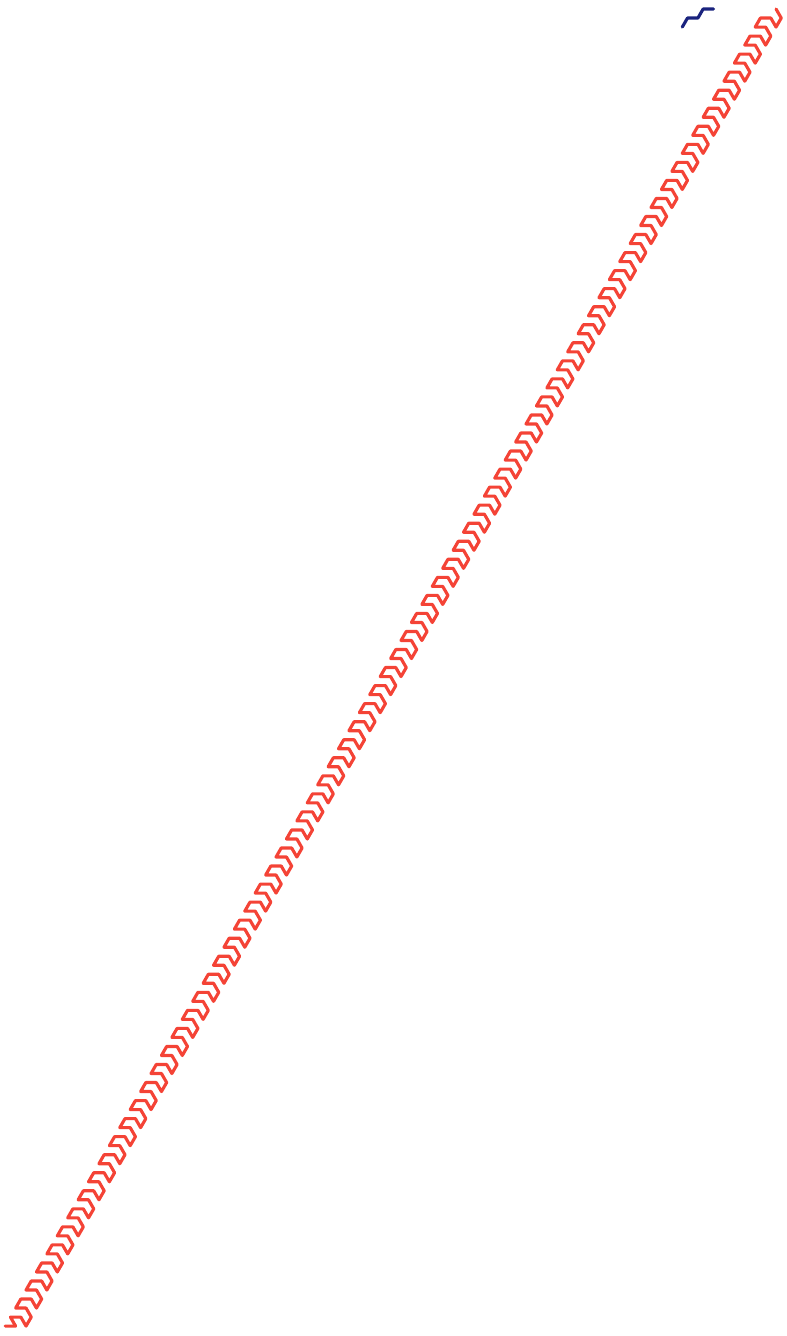}};
\end{tikzpicture}
}

%% file: OritatamiTuringPatterns/n=4-L=3-P=1/Halt____n=4-L=3-P=1-exploded.tex
\newcommand{\HaltBlockEpsilonnQLTPUexploded}[1]{
\begin{tikzpicture}
\ifthenelse{\equal{#1}{}}{
\TIKZlabel{(1.25000000*\st pt, 0.00000000*\st pt)}{A\explodedZigZagSize\ZigUp}{NW}{40.0*\st pt}
\TIKZlabel{(141.25000000*\st pt, 311.76914536*\st pt)}{B\explodedZigZagSize\Halt}{SW}{40.0*\st pt}
}{}
\node [anchor=center] () at (0,0) {\includegraphics[scale =\s]{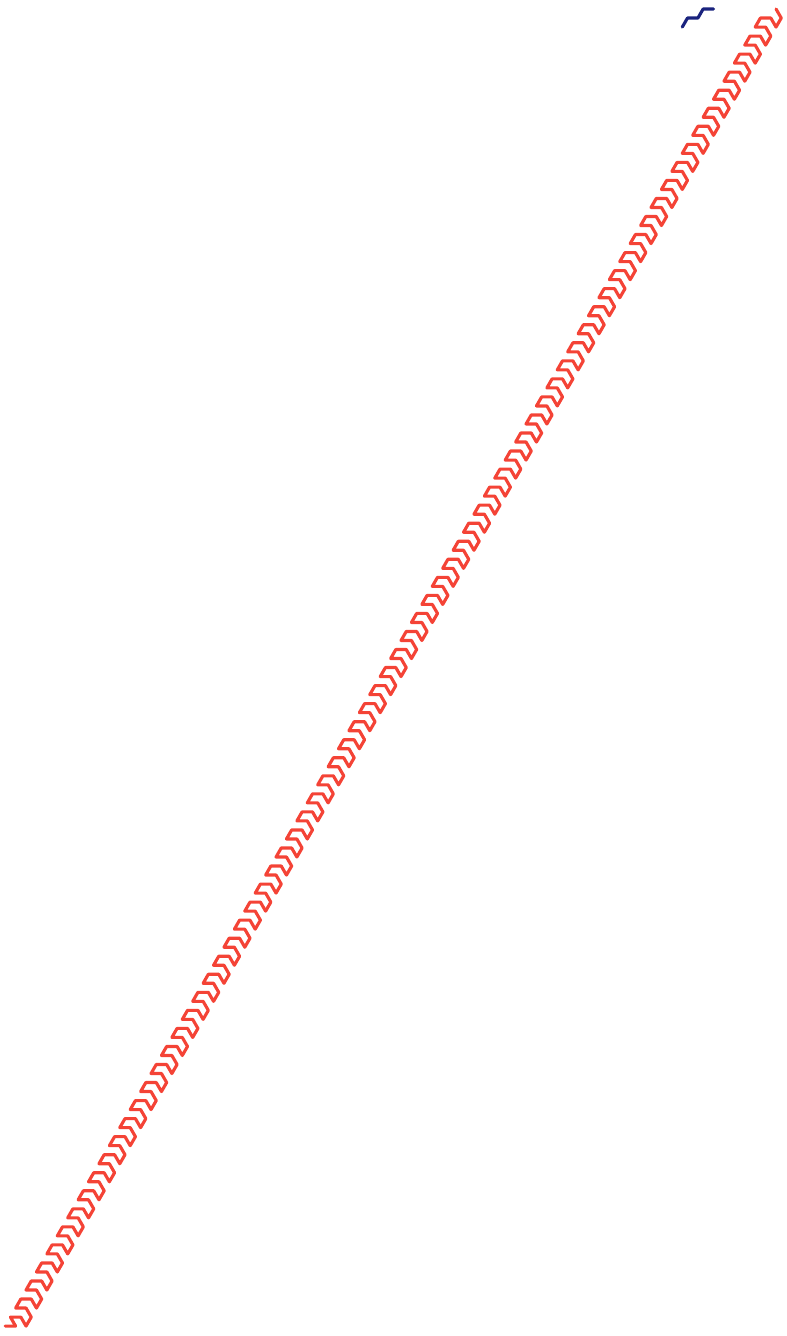}};
\end{tikzpicture}
}

%% file: OritatamiTuringPatterns/n=4-L=3-P=1/Read0_0_n=4-L=3-P=1-exploded.tex
\newcommand{\ReadZBlockZnQLTPUexploded}[1]{
\begin{tikzpicture}
\ifthenelse{\equal{#1}{}}{
\TIKZlabel{(-380.00000000*\st pt, 4.33012702*\st pt)}{A\explodedZigZagSize\ZigUp}{NW}{40.0*\st pt}
\TIKZlabel{(-162.50000000*\st pt, 320.42939940*\st pt)}{B\explodedZigZagSize\ZigUp}{NW}{40.0*\st pt}
\TIKZlabel{(-140.00000000*\st pt, 307.43901834*\st pt)}{C\explodedZigZagSize\ZigUp}{NE}{40.0*\st pt}
\TIKZlabel{(-447.50000000*\st pt, -311.76914536*\st pt)}{D0\explodedZigZagSize\ZigUp}{SW}{40.0*\st pt}
\TIKZlabel{(-377.50000000*\st pt, -311.76914536*\st pt)}{E\explodedZigZagSize\ZigUp}{SE}{80.0*\st pt}
\TIKZlabel{(-247.50000000*\st pt, -311.76914536*\st pt)}{F\explodedZigZagSize\ZigUp}{SE}{40.0*\st pt}
\TIKZlabel{(-22.50000000*\st pt, 0.00000000*\st pt)}{G\bZagZig}{SE}{40.0*\st pt}
}{}
\node [anchor=center] () at (0,0) {\includegraphics[scale =\s]{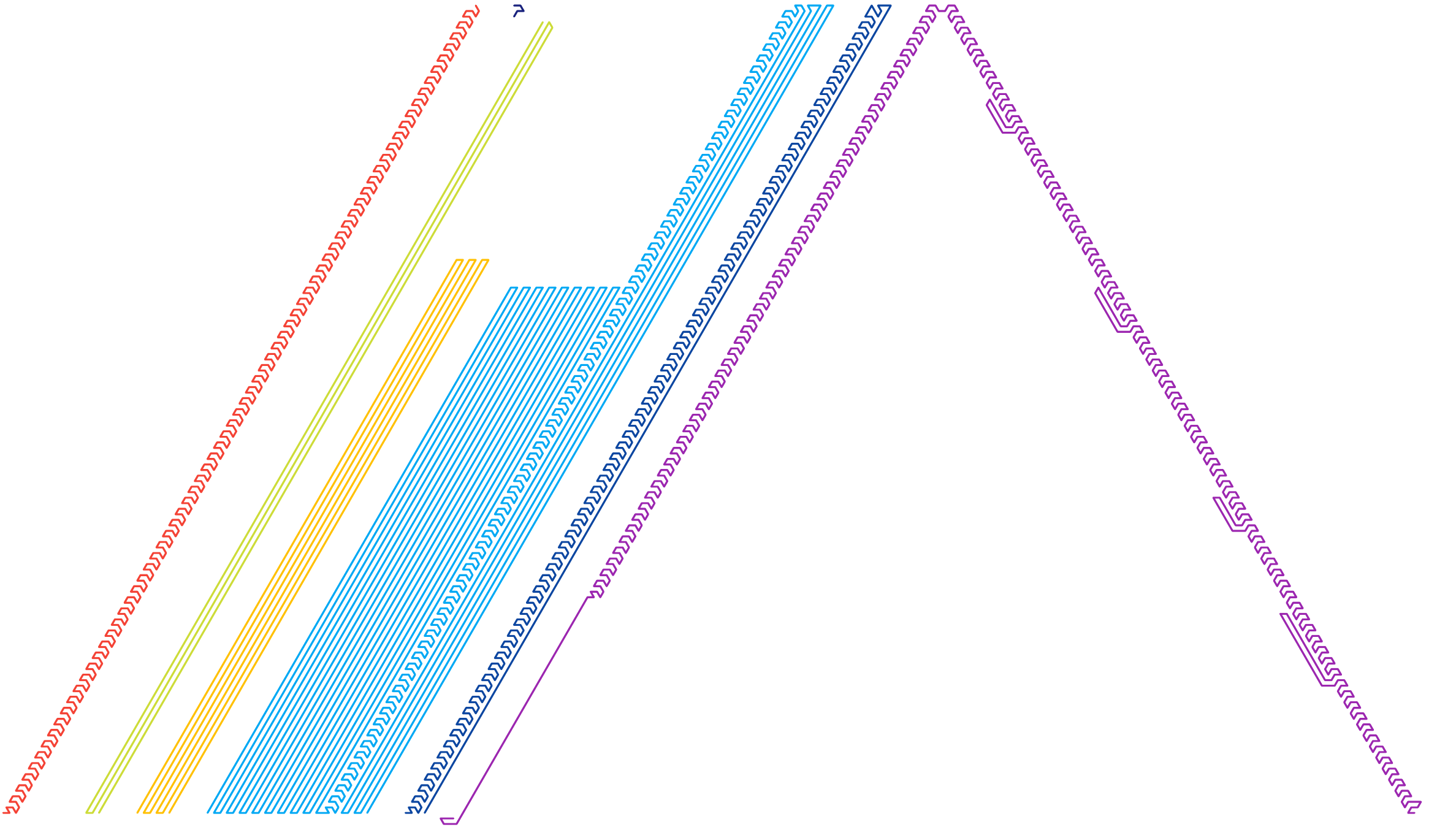}};
\end{tikzpicture}
}

%% file: OritatamiTuringPatterns/n=4-L=3-P=1/Read0_110_n=4-L=3-P=1-exploded.tex
\newcommand{\ReadZBlockUUZnQLTPUexploded}[1]{
\begin{tikzpicture}
\ifthenelse{\equal{#1}{}}{
\TIKZlabel{(-405.00000000*\st pt, 4.33012702*\st pt)}{A\explodedZigZagSize\ZigUp}{NW}{40.0*\st pt}
\TIKZlabel{(-187.50000000*\st pt, 320.42939940*\st pt)}{B\explodedZigZagSize\ZigUp}{NW}{40.0*\st pt}
\TIKZlabel{(-165.00000000*\st pt, 307.43901834*\st pt)}{C\explodedZigZagSize\ZigUp}{NE}{40.0*\st pt}
\TIKZlabel{(-472.50000000*\st pt, -311.76914536*\st pt)}{D1\explodedZigZagSize\ZigUp}{SW}{40.0*\st pt}
\TIKZlabel{(-417.50000000*\st pt, -311.76914536*\st pt)}{D1\explodedZigZagSize\ZigUp}{SW}{120.0*\st pt}
\TIKZlabel{(-362.50000000*\st pt, -311.76914536*\st pt)}{D0\explodedZigZagSize\ZigUp}{SW}{200.0*\st pt}
\TIKZlabel{(-292.50000000*\st pt, -311.76914536*\st pt)}{E\explodedZigZagSize\ZigUp}{SE}{80.0*\st pt}
\TIKZlabel{(-222.50000000*\st pt, -311.76914536*\st pt)}{F\explodedZigZagSize\ZigUp}{SE}{40.0*\st pt}
\TIKZlabel{(2.50000000*\st pt, 0.00000000*\st pt)}{G\bZagZig}{SE}{40.0*\st pt}
}{}
\node [anchor=center] () at (0,0) {\includegraphics[scale =\s]{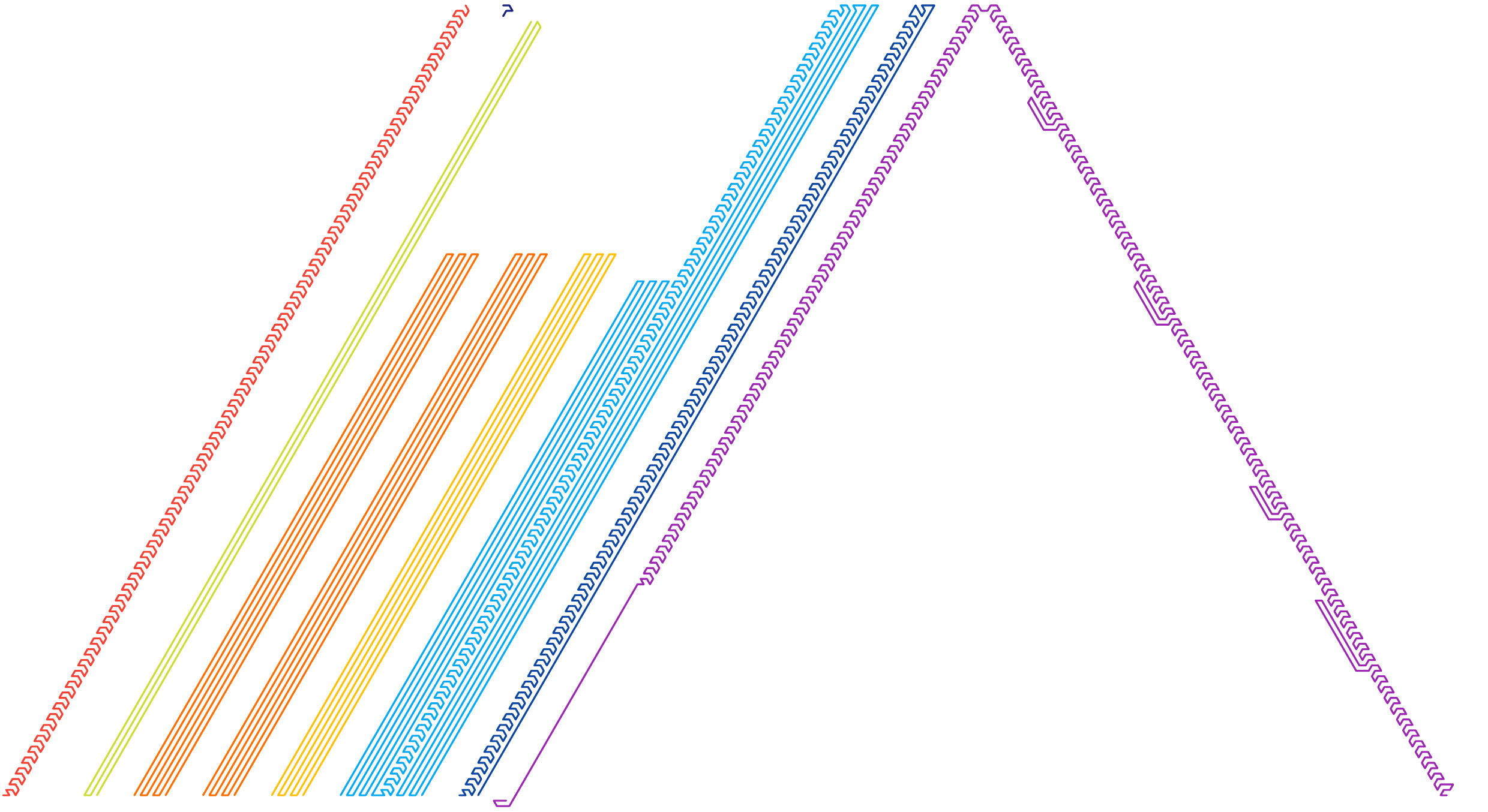}};
\end{tikzpicture}
}

%% file: OritatamiTuringPatterns/n=4-L=3-P=1/Read0_11_n=4-L=3-P=1-exploded.tex
\newcommand{\ReadZBlockUUnQLTPUexploded}[1]{
\begin{tikzpicture}
\ifthenelse{\equal{#1}{}}{
\TIKZlabel{(-392.50000000*\st pt, 4.33012702*\st pt)}{A\explodedZigZagSize\ZigUp}{NW}{40.0*\st pt}
\TIKZlabel{(-175.00000000*\st pt, 320.42939940*\st pt)}{B\explodedZigZagSize\ZigUp}{NW}{40.0*\st pt}
\TIKZlabel{(-152.50000000*\st pt, 307.43901834*\st pt)}{C\explodedZigZagSize\ZigUp}{NE}{40.0*\st pt}
\TIKZlabel{(-460.00000000*\st pt, -311.76914536*\st pt)}{D1\explodedZigZagSize\ZigUp}{SW}{40.0*\st pt}
\TIKZlabel{(-405.00000000*\st pt, -311.76914536*\st pt)}{D1\explodedZigZagSize\ZigUp}{SW}{120.0*\st pt}
\TIKZlabel{(-335.00000000*\st pt, -311.76914536*\st pt)}{E\explodedZigZagSize\ZigUp}{SE}{80.0*\st pt}
\TIKZlabel{(-235.00000000*\st pt, -311.76914536*\st pt)}{F\explodedZigZagSize\ZigUp}{SE}{40.0*\st pt}
\TIKZlabel{(-10.00000000*\st pt, 0.00000000*\st pt)}{G\bZagZig}{SE}{40.0*\st pt}
}{}
\node [anchor=center] () at (0,0) {\includegraphics[scale =\s]{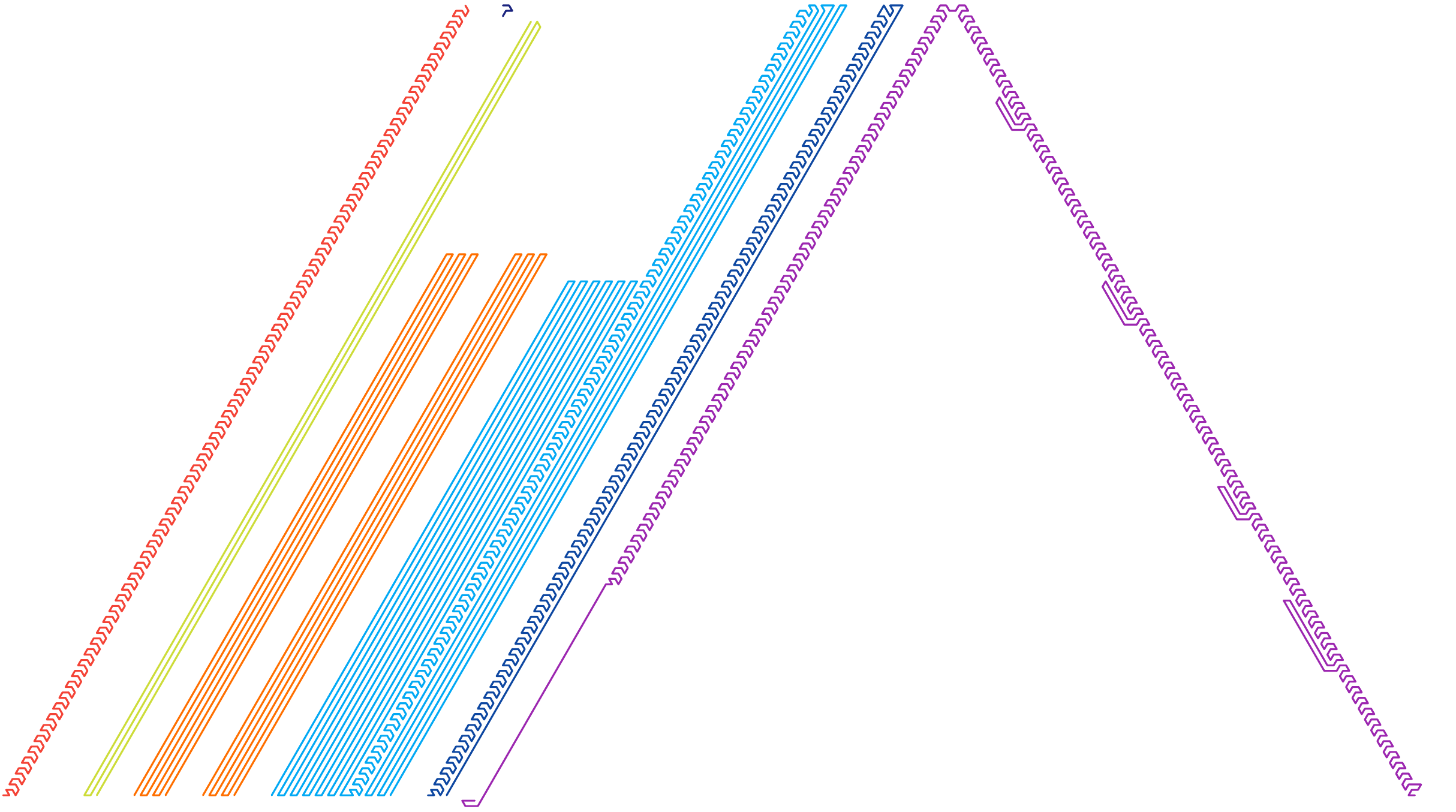}};
\end{tikzpicture}
}

%% file: OritatamiTuringPatterns/n=4-L=3-P=1/Read0____n=4-L=3-P=1-exploded.tex
\newcommand{\ReadZBlockEpsilonnQLTPUexploded}[1]{
\begin{tikzpicture}
\ifthenelse{\equal{#1}{}}{
\TIKZlabel{(-367.50000000*\st pt, 4.33012702*\st pt)}{A\explodedZigZagSize\ZigUp}{NW}{40.0*\st pt}
\TIKZlabel{(-150.00000000*\st pt, 320.42939940*\st pt)}{B\explodedZigZagSize\ZigUp}{NW}{40.0*\st pt}
\TIKZlabel{(-127.50000000*\st pt, 307.43901834*\st pt)}{C\explodedZigZagSize\ZigUp}{NE}{40.0*\st pt}
\TIKZlabel{(-420.00000000*\st pt, -311.76914536*\st pt)}{E\explodedZigZagSize\ZigUp}{SE}{80.0*\st pt}
\TIKZlabel{(-260.00000000*\st pt, -311.76914536*\st pt)}{F\explodedZigZagSize\ZigUp}{SE}{40.0*\st pt}
\TIKZlabel{(-35.00000000*\st pt, 0.00000000*\st pt)}{G\bZagZig}{SE}{40.0*\st pt}
}{}
\node [anchor=center] () at (0,0) {\includegraphics[scale =\s]{Read0___n=4-L=3-P=1-exploded.pdf}};
\end{tikzpicture}
}

%% file: OritatamiTuringPatterns/n=4-L=3-P=1/Read1_0_n=4-L=3-P=1-exploded.tex
\newcommand{\ReadUBlockZnQLTPUexploded}[1]{
\begin{tikzpicture}
\ifthenelse{\equal{#1}{}}{
\TIKZlabel{(-561.25000000*\st pt, 4.33012702*\st pt)}{A\explodedZigZagSize\ZigUp}{NW}{40.0*\st pt}
\TIKZlabel{(-343.75000000*\st pt, 320.42939940*\st pt)}{B\explodedZigZagSize\ZigUp}{NW}{40.0*\st pt}
\TIKZlabel{(-321.25000000*\st pt, 307.43901834*\st pt)}{C\explodedZigZagSize\ZigUp}{NE}{40.0*\st pt}
\TIKZlabel{(-628.75000000*\st pt, -311.76914536*\st pt)}{D0\explodedZigZagSize\ZigUp}{SW}{40.0*\st pt}
\TIKZlabel{(-558.75000000*\st pt, -311.76914536*\st pt)}{E\explodedZigZagSize\ZigUp}{SE}{80.0*\st pt}
\TIKZlabel{(-428.75000000*\st pt, -311.76914536*\st pt)}{F\explodedZigZagSize\ZigUp}{SE}{40.0*\st pt}
\TIKZlabel{(-203.75000000*\st pt, 0.00000000*\st pt)}{G\explodedZigZagSize\ZigUp}{SE}{40.0*\st pt}
}{}
\node [anchor=center] () at (0,0) {\includegraphics[scale =\s]{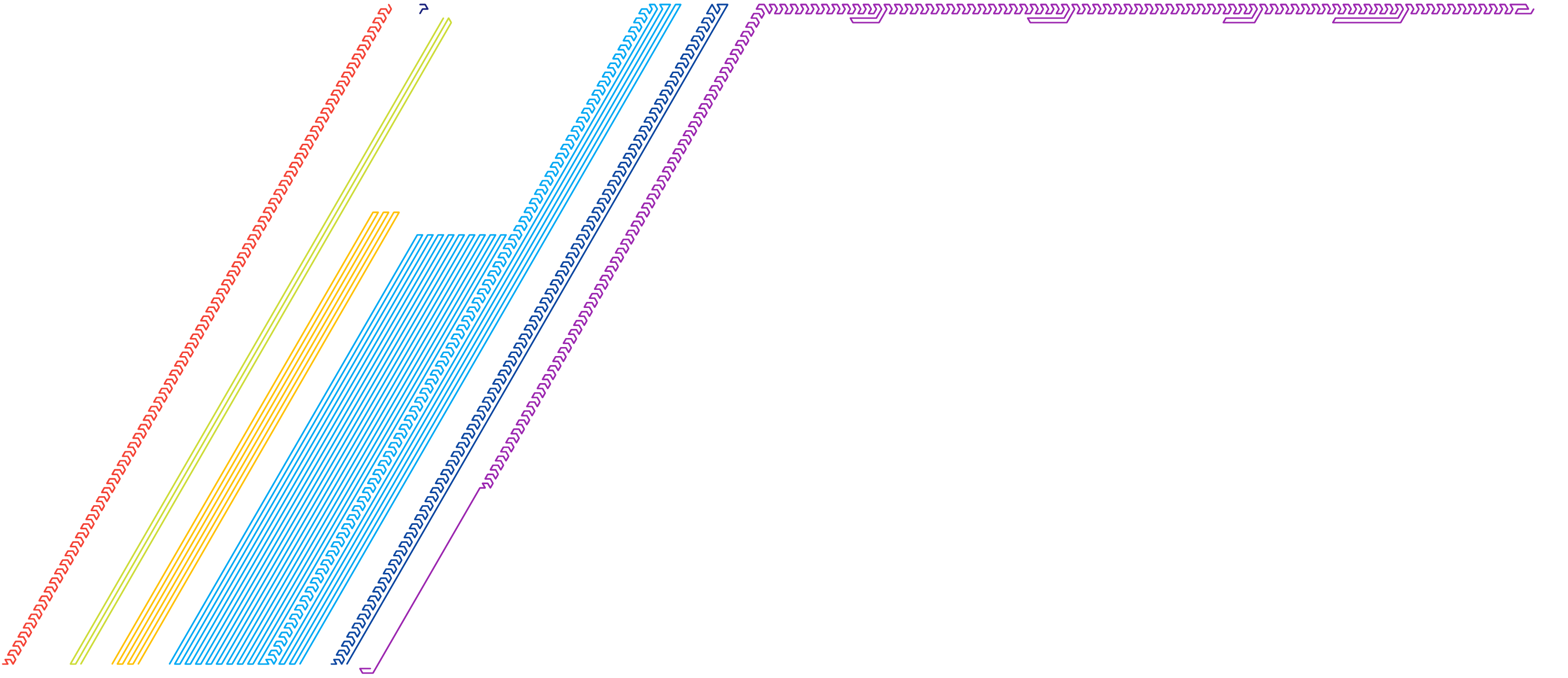}};
\end{tikzpicture}
}

%% file: OritatamiTuringPatterns/n=4-L=3-P=1/Read1_110_n=4-L=3-P=1-exploded.tex
\newcommand{\ReadUBlockUUZnQLTPUexploded}[1]{
\begin{tikzpicture}
\ifthenelse{\equal{#1}{}}{
\TIKZlabel{(-586.25000000*\st pt, 4.33012702*\st pt)}{A\explodedZigZagSize\ZigUp}{NW}{40.0*\st pt}
\TIKZlabel{(-368.75000000*\st pt, 320.42939940*\st pt)}{B\explodedZigZagSize\ZigUp}{NW}{40.0*\st pt}
\TIKZlabel{(-346.25000000*\st pt, 307.43901834*\st pt)}{C\explodedZigZagSize\ZigUp}{NE}{40.0*\st pt}
\TIKZlabel{(-653.75000000*\st pt, -311.76914536*\st pt)}{D1\explodedZigZagSize\ZigUp}{SW}{40.0*\st pt}
\TIKZlabel{(-598.75000000*\st pt, -311.76914536*\st pt)}{D1\explodedZigZagSize\ZigUp}{SW}{120.0*\st pt}
\TIKZlabel{(-543.75000000*\st pt, -311.76914536*\st pt)}{D0\explodedZigZagSize\ZigUp}{SW}{200.0*\st pt}
\TIKZlabel{(-473.75000000*\st pt, -311.76914536*\st pt)}{E\explodedZigZagSize\ZigUp}{SE}{80.0*\st pt}
\TIKZlabel{(-403.75000000*\st pt, -311.76914536*\st pt)}{F\explodedZigZagSize\ZigUp}{SE}{40.0*\st pt}
\TIKZlabel{(-178.75000000*\st pt, 0.00000000*\st pt)}{G\explodedZigZagSize\ZigUp}{SE}{40.0*\st pt}
}{}
\node [anchor=center] () at (0,0) {\includegraphics[scale =\s]{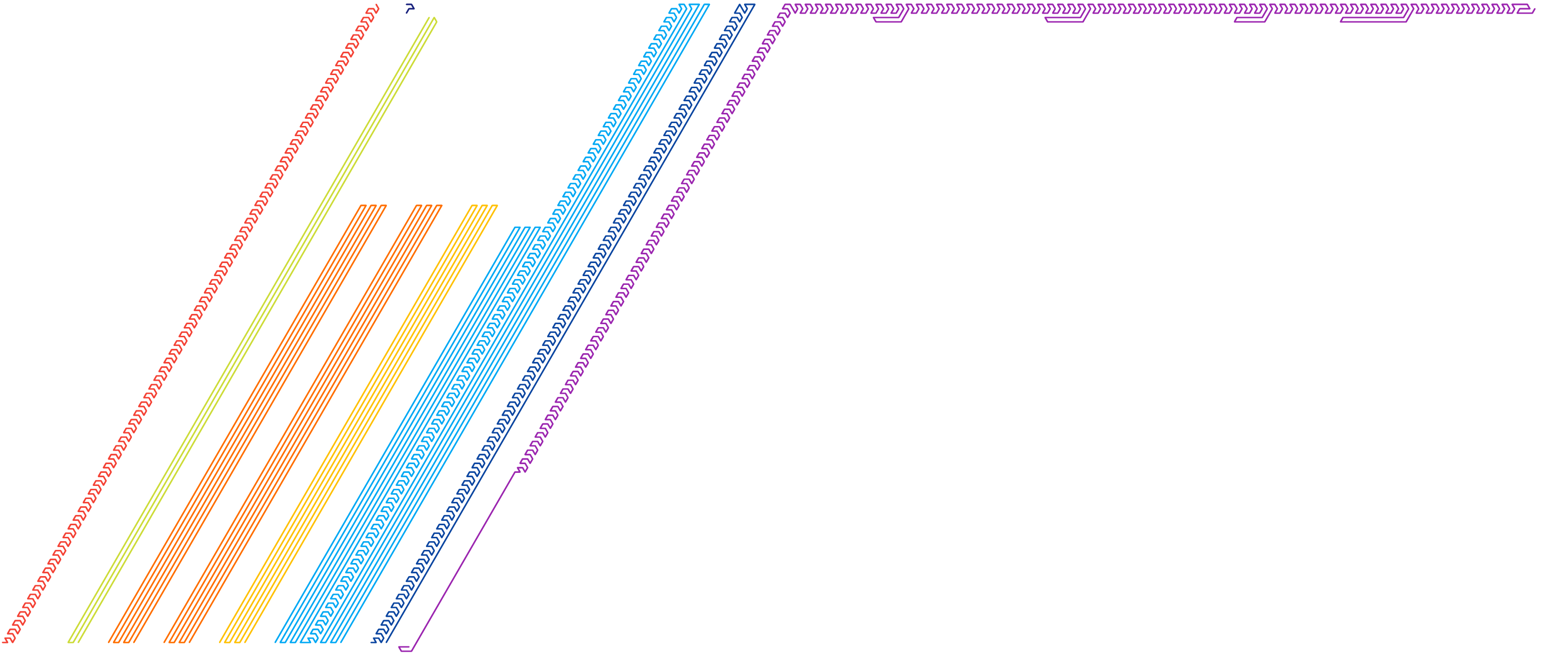}};
\end{tikzpicture}
}

%% file: OritatamiTuringPatterns/n=4-L=3-P=1/Read1_11_n=4-L=3-P=1-exploded.tex
\newcommand{\ReadUBlockUUnQLTPUexploded}[1]{
\begin{tikzpicture}
\ifthenelse{\equal{#1}{}}{
\TIKZlabel{(-573.75000000*\st pt, 4.33012702*\st pt)}{A\explodedZigZagSize\ZigUp}{NW}{40.0*\st pt}
\TIKZlabel{(-356.25000000*\st pt, 320.42939940*\st pt)}{B\explodedZigZagSize\ZigUp}{NW}{40.0*\st pt}
\TIKZlabel{(-333.75000000*\st pt, 307.43901834*\st pt)}{C\explodedZigZagSize\ZigUp}{NE}{40.0*\st pt}
\TIKZlabel{(-641.25000000*\st pt, -311.76914536*\st pt)}{D1\explodedZigZagSize\ZigUp}{SW}{40.0*\st pt}
\TIKZlabel{(-586.25000000*\st pt, -311.76914536*\st pt)}{D1\explodedZigZagSize\ZigUp}{SW}{120.0*\st pt}
\TIKZlabel{(-516.25000000*\st pt, -311.76914536*\st pt)}{E\explodedZigZagSize\ZigUp}{SE}{80.0*\st pt}
\TIKZlabel{(-416.25000000*\st pt, -311.76914536*\st pt)}{F\explodedZigZagSize\ZigUp}{SE}{40.0*\st pt}
\TIKZlabel{(-191.25000000*\st pt, 0.00000000*\st pt)}{G\explodedZigZagSize\ZigUp}{SE}{40.0*\st pt}
}{}
\node [anchor=center] () at (0,0) {\includegraphics[scale =\s]{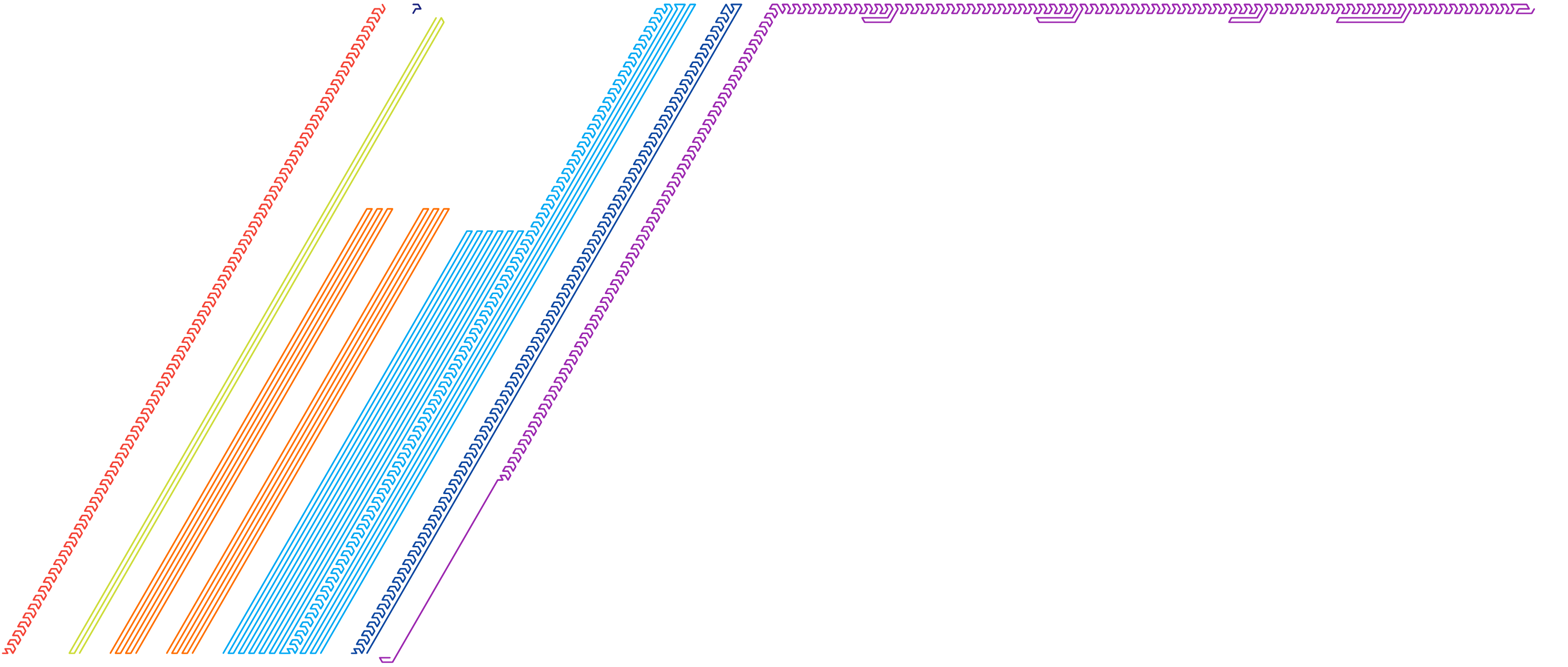}};
\end{tikzpicture}
}

%% file: OritatamiTuringPatterns/n=4-L=3-P=1/Read1____n=4-L=3-P=1-exploded.tex
\newcommand{\ReadUBlockEpsilonnQLTPUexploded}[1]{
\begin{tikzpicture}
\ifthenelse{\equal{#1}{}}{
\TIKZlabel{(-548.75000000*\st pt, 4.33012702*\st pt)}{A\explodedZigZagSize\ZigUp}{NW}{40.0*\st pt}
\TIKZlabel{(-331.25000000*\st pt, 320.42939940*\st pt)}{B\explodedZigZagSize\ZigUp}{NW}{40.0*\st pt}
\TIKZlabel{(-308.75000000*\st pt, 307.43901834*\st pt)}{C\explodedZigZagSize\ZigUp}{NE}{40.0*\st pt}
\TIKZlabel{(-601.25000000*\st pt, -311.76914536*\st pt)}{E\explodedZigZagSize\ZigUp}{SE}{80.0*\st pt}
\TIKZlabel{(-441.25000000*\st pt, -311.76914536*\st pt)}{F\explodedZigZagSize\ZigUp}{SE}{40.0*\st pt}
\TIKZlabel{(-216.25000000*\st pt, 0.00000000*\st pt)}{G\explodedZigZagSize\ZigUp}{SE}{40.0*\st pt}
}{}
\node [anchor=center] () at (0,0) {\includegraphics[scale =\s]{Read1___n=4-L=3-P=1-exploded.pdf}};
\end{tikzpicture}
}

%% file: OritatamiTuringPatterns/n=4-L=3-P=1/ZigCopy0_0_n=4-L=3-P=1-exploded.tex
\newcommand{\ZigCopyZBlockZnQLTPUexploded}[1]{
\begin{tikzpicture}
\ifthenelse{\equal{#1}{}}{
\TIKZlabel{(-187.50000000*\st pt, 2.16506351*\st pt)}{A\explodedZigZagSize\ZigDown}{SW}{40.0*\st pt}
\TIKZlabel{(30.00000000*\st pt, -313.93420887*\st pt)}{B\explodedZigZagSize\ZigDown}{SW}{40.0*\st pt}
\TIKZlabel{(52.50000000*\st pt, -300.94382782*\st pt)}{C\explodedZigZagSize\ZigDown}{SE}{40.0*\st pt}
\TIKZlabel{(-255.00000000*\st pt, 318.26433589*\st pt)}{D0\explodedZigZagSize\ZigDown}{NW}{40.0*\st pt}
\TIKZlabel{(-185.00000000*\st pt, 318.26433589*\st pt)}{E\explodedZigZagSize\ZigDown}{NE}{120.0*\st pt}
\TIKZlabel{(-55.00000000*\st pt, 318.26433589*\st pt)}{F\explodedZigZagSize\ZigDown}{NE}{40.0*\st pt}
\TIKZlabel{(192.50000000*\st pt, 2.16506351*\st pt)}{G\explodedZigZagSize\ZigDown}{NE}{40.0*\st pt}
}{}
\node [anchor=center] () at (0,0) {\includegraphics[scale =\s]{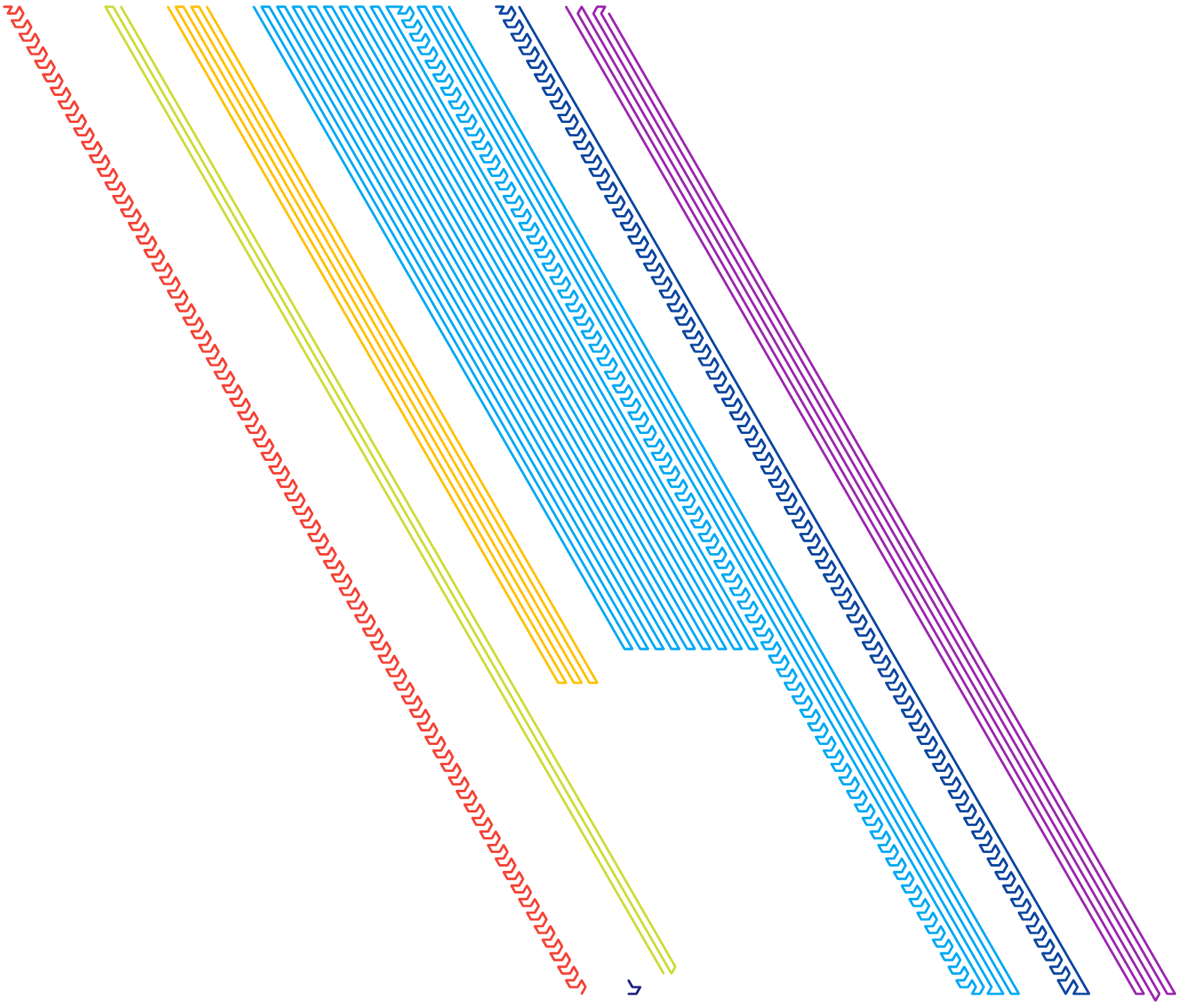}};
\end{tikzpicture}
}

%% file: OritatamiTuringPatterns/n=4-L=3-P=1/ZigCopy0_110_n=4-L=3-P=1-exploded.tex
\newcommand{\ZigCopyZBlockUUZnQLTPUexploded}[1]{
\begin{tikzpicture}
\ifthenelse{\equal{#1}{}}{
\TIKZlabel{(-212.50000000*\st pt, 2.16506351*\st pt)}{A\explodedZigZagSize\ZigDown}{SW}{40.0*\st pt}
\TIKZlabel{(5.00000000*\st pt, -313.93420887*\st pt)}{B\explodedZigZagSize\ZigDown}{SW}{40.0*\st pt}
\TIKZlabel{(27.50000000*\st pt, -300.94382782*\st pt)}{C\explodedZigZagSize\ZigDown}{SE}{40.0*\st pt}
\TIKZlabel{(-280.00000000*\st pt, 318.26433589*\st pt)}{D1\explodedZigZagSize\ZigDown}{NW}{40.0*\st pt}
\TIKZlabel{(-225.00000000*\st pt, 318.26433589*\st pt)}{D1\explodedZigZagSize\ZigDown}{NW}{120.0*\st pt}
\TIKZlabel{(-170.00000000*\st pt, 318.26433589*\st pt)}{D0\explodedZigZagSize\ZigDown}{NW}{200.0*\st pt}
\TIKZlabel{(-100.00000000*\st pt, 318.26433589*\st pt)}{E\explodedZigZagSize\ZigDown}{NE}{120.0*\st pt}
\TIKZlabel{(-30.00000000*\st pt, 318.26433589*\st pt)}{F\explodedZigZagSize\ZigDown}{NE}{40.0*\st pt}
\TIKZlabel{(217.50000000*\st pt, 2.16506351*\st pt)}{G\explodedZigZagSize\ZigDown}{NE}{40.0*\st pt}
}{}
\node [anchor=center] () at (0,0) {\includegraphics[scale =\s]{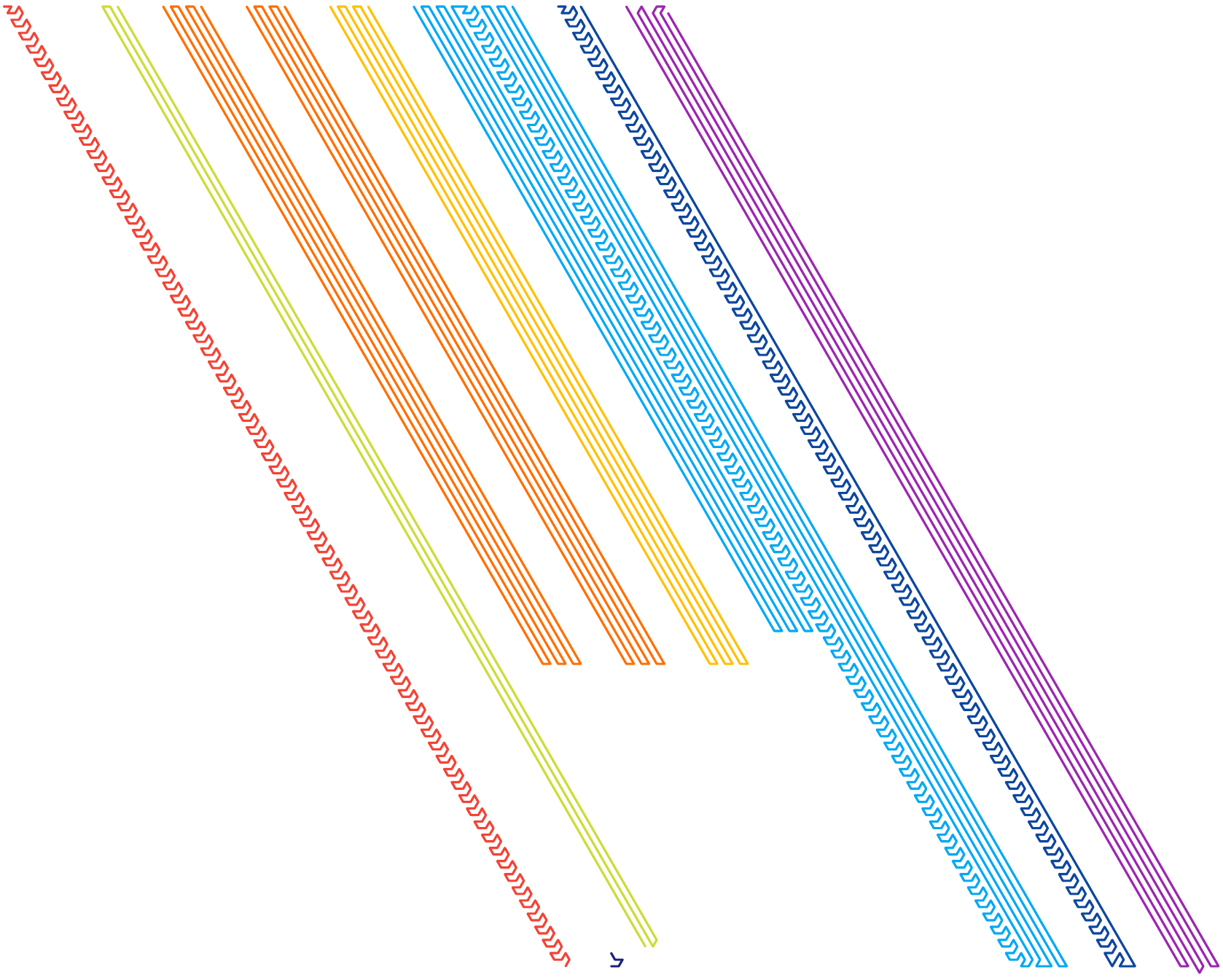}};
\end{tikzpicture}
}

%% file: OritatamiTuringPatterns/n=4-L=3-P=1/ZigCopy0_11_n=4-L=3-P=1-exploded.tex
\newcommand{\ZigCopyZBlockUUnQLTPUexploded}[1]{
\begin{tikzpicture}
\ifthenelse{\equal{#1}{}}{
\TIKZlabel{(-200.00000000*\st pt, 2.16506351*\st pt)}{A\explodedZigZagSize\ZigDown}{SW}{40.0*\st pt}
\TIKZlabel{(17.50000000*\st pt, -313.93420887*\st pt)}{B\explodedZigZagSize\ZigDown}{SW}{40.0*\st pt}
\TIKZlabel{(40.00000000*\st pt, -300.94382782*\st pt)}{C\explodedZigZagSize\ZigDown}{SE}{40.0*\st pt}
\TIKZlabel{(-267.50000000*\st pt, 318.26433589*\st pt)}{D1\explodedZigZagSize\ZigDown}{NW}{40.0*\st pt}
\TIKZlabel{(-212.50000000*\st pt, 318.26433589*\st pt)}{D1\explodedZigZagSize\ZigDown}{NW}{120.0*\st pt}
\TIKZlabel{(-142.50000000*\st pt, 318.26433589*\st pt)}{E\explodedZigZagSize\ZigDown}{NE}{120.0*\st pt}
\TIKZlabel{(-42.50000000*\st pt, 318.26433589*\st pt)}{F\explodedZigZagSize\ZigDown}{NE}{40.0*\st pt}
\TIKZlabel{(205.00000000*\st pt, 2.16506351*\st pt)}{G\explodedZigZagSize\ZigDown}{NE}{40.0*\st pt}
}{}
\node [anchor=center] () at (0,0) {\includegraphics[scale =\s]{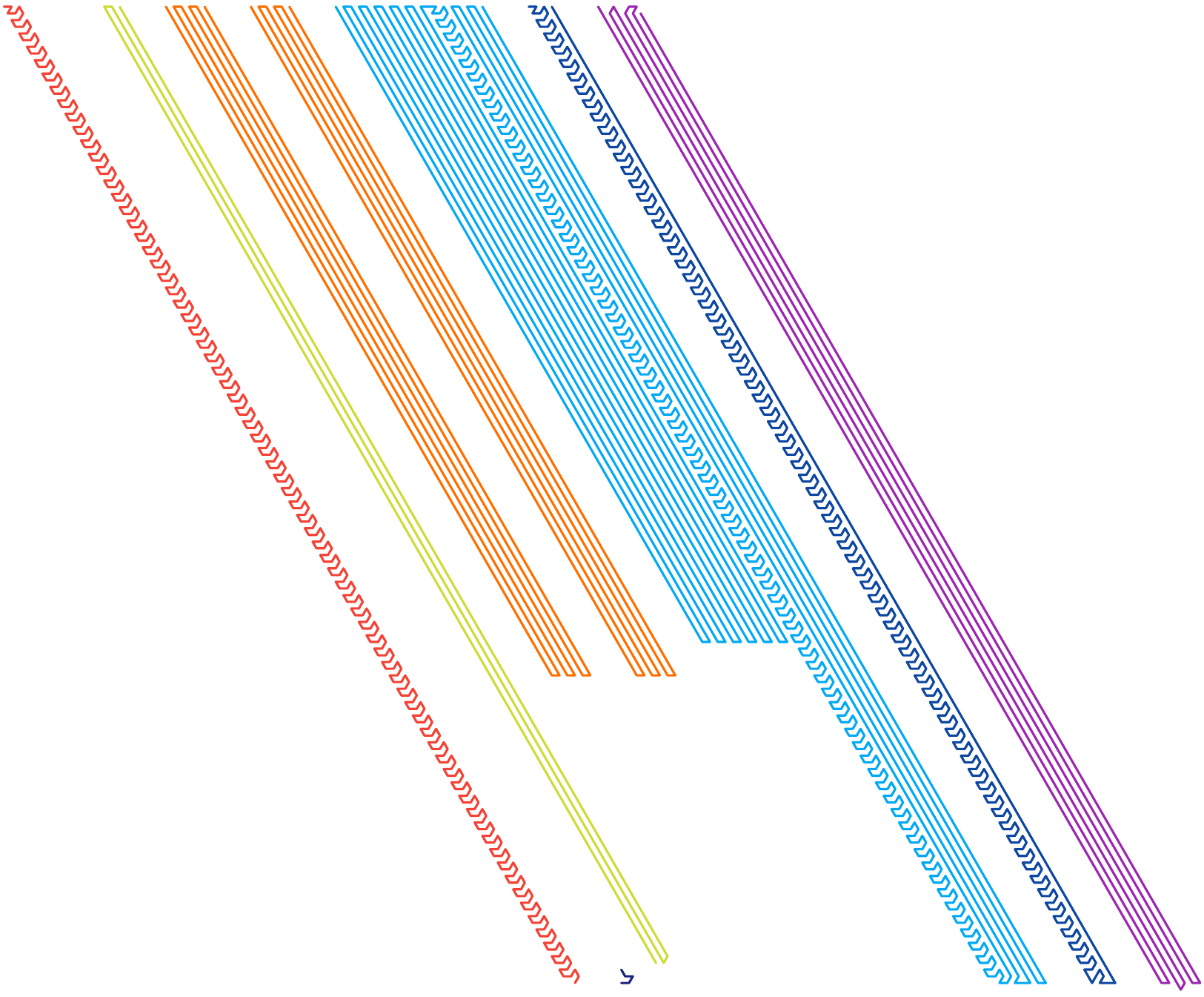}};
\end{tikzpicture}
}

%% file: OritatamiTuringPatterns/n=4-L=3-P=1/ZigCopy0____n=4-L=3-P=1-exploded.tex
\newcommand{\ZigCopyZBlockEpsilonnQLTPUexploded}[1]{
\begin{tikzpicture}
\ifthenelse{\equal{#1}{}}{
\TIKZlabel{(-175.00000000*\st pt, 2.16506351*\st pt)}{A\explodedZigZagSize\ZigDown}{SW}{40.0*\st pt}
\TIKZlabel{(42.50000000*\st pt, -313.93420887*\st pt)}{B\explodedZigZagSize\ZigDown}{SW}{40.0*\st pt}
\TIKZlabel{(65.00000000*\st pt, -300.94382782*\st pt)}{C\explodedZigZagSize\ZigDown}{SE}{40.0*\st pt}
\TIKZlabel{(-227.50000000*\st pt, 318.26433589*\st pt)}{E\explodedZigZagSize\ZigDown}{NE}{120.0*\st pt}
\TIKZlabel{(-67.50000000*\st pt, 318.26433589*\st pt)}{F\explodedZigZagSize\ZigDown}{NE}{40.0*\st pt}
\TIKZlabel{(180.00000000*\st pt, 2.16506351*\st pt)}{G\explodedZigZagSize\ZigDown}{NE}{40.0*\st pt}
}{}
\node [anchor=center] () at (0,0) {\includegraphics[scale =\s]{ZigCopy0___n=4-L=3-P=1-exploded.pdf}};
\end{tikzpicture}
}

%% file: OritatamiTuringPatterns/n=4-L=3-P=1/ZigCopy1_0_n=4-L=3-P=1-exploded.tex
\newcommand{\ZigCopyUBlockZnQLTPUexploded}[1]{
\begin{tikzpicture}
\ifthenelse{\equal{#1}{}}{
\TIKZlabel{(-186.25000000*\st pt, 0.00000000*\st pt)}{A\explodedZigZagSize\ZigDown}{SW}{40.0*\st pt}
\TIKZlabel{(31.25000000*\st pt, -316.09927238*\st pt)}{B\explodedZigZagSize\ZigDown}{SW}{40.0*\st pt}
\TIKZlabel{(53.75000000*\st pt, -303.10889132*\st pt)}{C\explodedZigZagSize\ZigDown}{SE}{40.0*\st pt}
\TIKZlabel{(-253.75000000*\st pt, 316.09927238*\st pt)}{D0\explodedZigZagSize\ZigDown}{NW}{40.0*\st pt}
\TIKZlabel{(-183.75000000*\st pt, 316.09927238*\st pt)}{E\explodedZigZagSize\ZigDown}{NE}{120.0*\st pt}
\TIKZlabel{(-53.75000000*\st pt, 316.09927238*\st pt)}{F\explodedZigZagSize\ZigDown}{NE}{40.0*\st pt}
\TIKZlabel{(193.75000000*\st pt, 0.00000000*\st pt)}{G\explodedZigZagSize\ZigDown}{NE}{40.0*\st pt}
}{}
\node [anchor=center] () at (0,0) {\includegraphics[scale =\s]{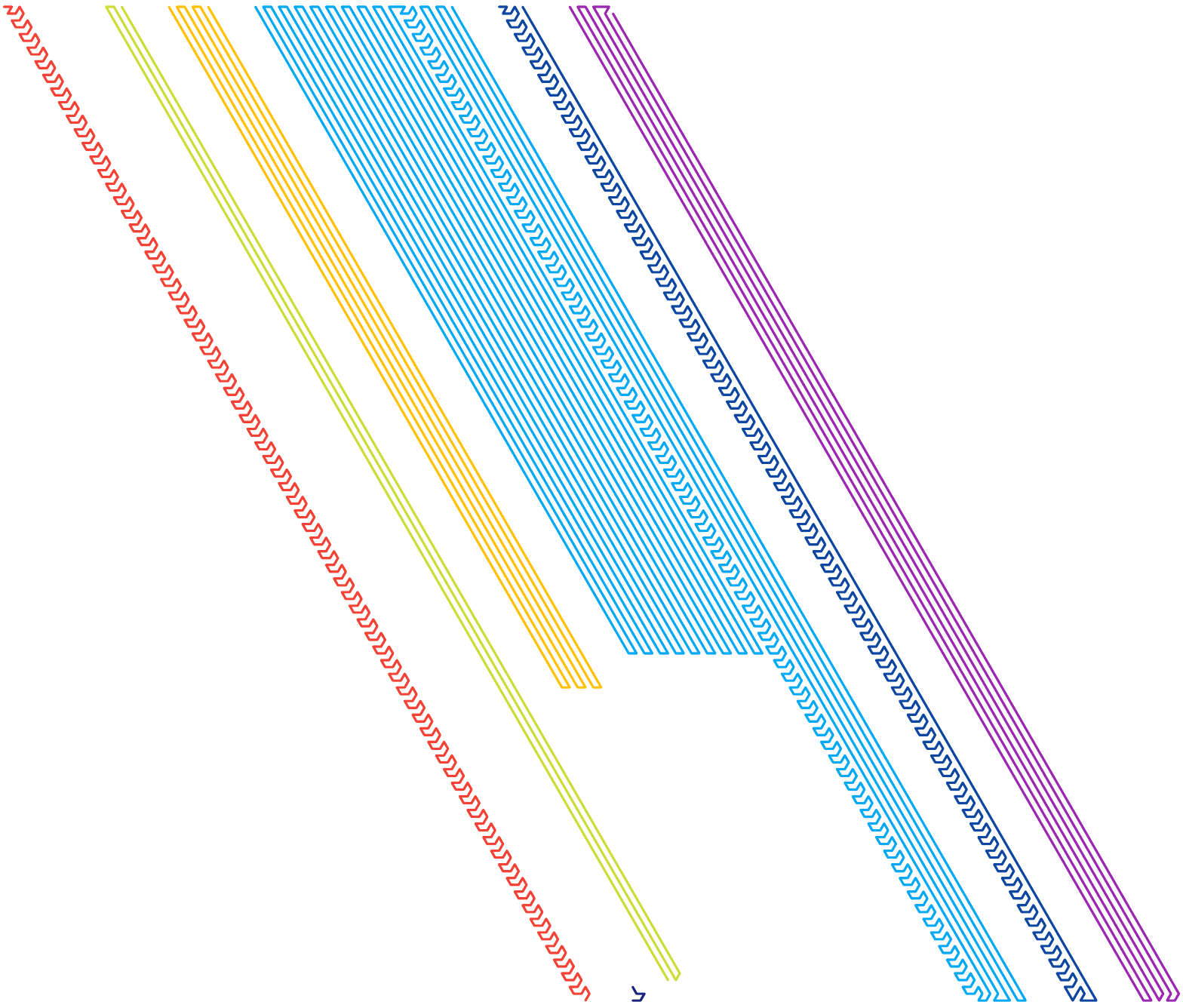}};
\end{tikzpicture}
}

%% file: OritatamiTuringPatterns/n=4-L=3-P=1/ZigCopy1_110_n=4-L=3-P=1-exploded.tex
\newcommand{\ZigCopyUBlockUUZnQLTPUexploded}[1]{
\begin{tikzpicture}
\ifthenelse{\equal{#1}{}}{
\TIKZlabel{(-211.25000000*\st pt, 0.00000000*\st pt)}{A\explodedZigZagSize\ZigDown}{SW}{40.0*\st pt}
\TIKZlabel{(6.25000000*\st pt, -316.09927238*\st pt)}{B\explodedZigZagSize\ZigDown}{SW}{40.0*\st pt}
\TIKZlabel{(28.75000000*\st pt, -303.10889132*\st pt)}{C\explodedZigZagSize\ZigDown}{SE}{40.0*\st pt}
\TIKZlabel{(-278.75000000*\st pt, 316.09927238*\st pt)}{D1\explodedZigZagSize\ZigDown}{NW}{40.0*\st pt}
\TIKZlabel{(-223.75000000*\st pt, 316.09927238*\st pt)}{D1\explodedZigZagSize\ZigDown}{NW}{120.0*\st pt}
\TIKZlabel{(-168.75000000*\st pt, 316.09927238*\st pt)}{D0\explodedZigZagSize\ZigDown}{NW}{200.0*\st pt}
\TIKZlabel{(-98.75000000*\st pt, 316.09927238*\st pt)}{E\explodedZigZagSize\ZigDown}{NE}{120.0*\st pt}
\TIKZlabel{(-28.75000000*\st pt, 316.09927238*\st pt)}{F\explodedZigZagSize\ZigDown}{NE}{40.0*\st pt}
\TIKZlabel{(218.75000000*\st pt, 0.00000000*\st pt)}{G\explodedZigZagSize\ZigDown}{NE}{40.0*\st pt}
}{}
\node [anchor=center] () at (0,0) {\includegraphics[scale =\s]{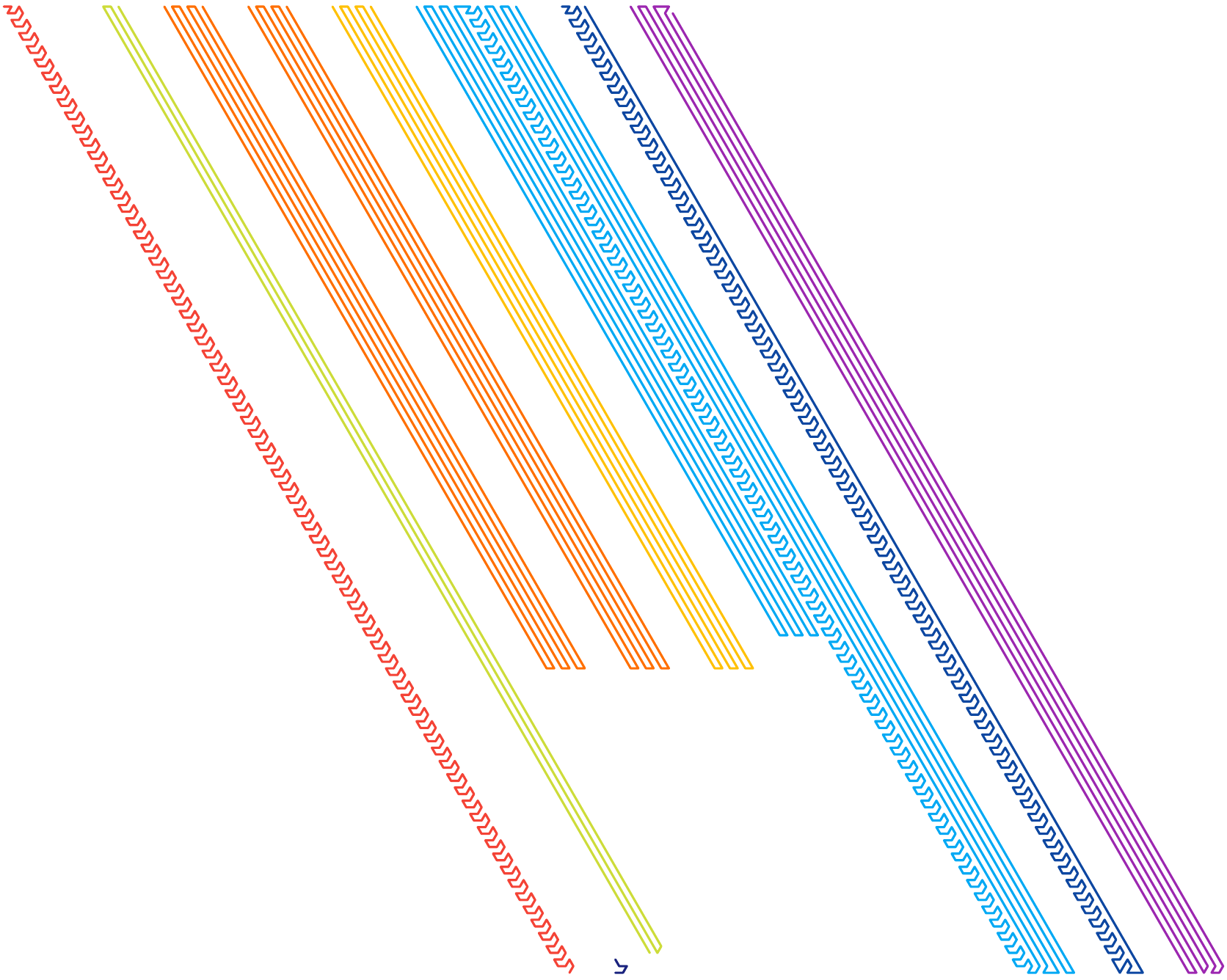}};
\end{tikzpicture}
}

%% file: OritatamiTuringPatterns/n=4-L=3-P=1/ZigCopy1_11_n=4-L=3-P=1-exploded.tex
\newcommand{\ZigCopyUBlockUUnQLTPUexploded}[1]{
\begin{tikzpicture}
\ifthenelse{\equal{#1}{}}{
\TIKZlabel{(-198.75000000*\st pt, 0.00000000*\st pt)}{A\explodedZigZagSize\ZigDown}{SW}{40.0*\st pt}
\TIKZlabel{(18.75000000*\st pt, -316.09927238*\st pt)}{B\explodedZigZagSize\ZigDown}{SW}{40.0*\st pt}
\TIKZlabel{(41.25000000*\st pt, -303.10889132*\st pt)}{C\explodedZigZagSize\ZigDown}{SE}{40.0*\st pt}
\TIKZlabel{(-266.25000000*\st pt, 316.09927238*\st pt)}{D1\explodedZigZagSize\ZigDown}{NW}{40.0*\st pt}
\TIKZlabel{(-211.25000000*\st pt, 316.09927238*\st pt)}{D1\explodedZigZagSize\ZigDown}{NW}{120.0*\st pt}
\TIKZlabel{(-141.25000000*\st pt, 316.09927238*\st pt)}{E\explodedZigZagSize\ZigDown}{NE}{120.0*\st pt}
\TIKZlabel{(-41.25000000*\st pt, 316.09927238*\st pt)}{F\explodedZigZagSize\ZigDown}{NE}{40.0*\st pt}
\TIKZlabel{(206.25000000*\st pt, 0.00000000*\st pt)}{G\explodedZigZagSize\ZigDown}{NE}{40.0*\st pt}
}{}
\node [anchor=center] () at (0,0) {\includegraphics[scale =\s]{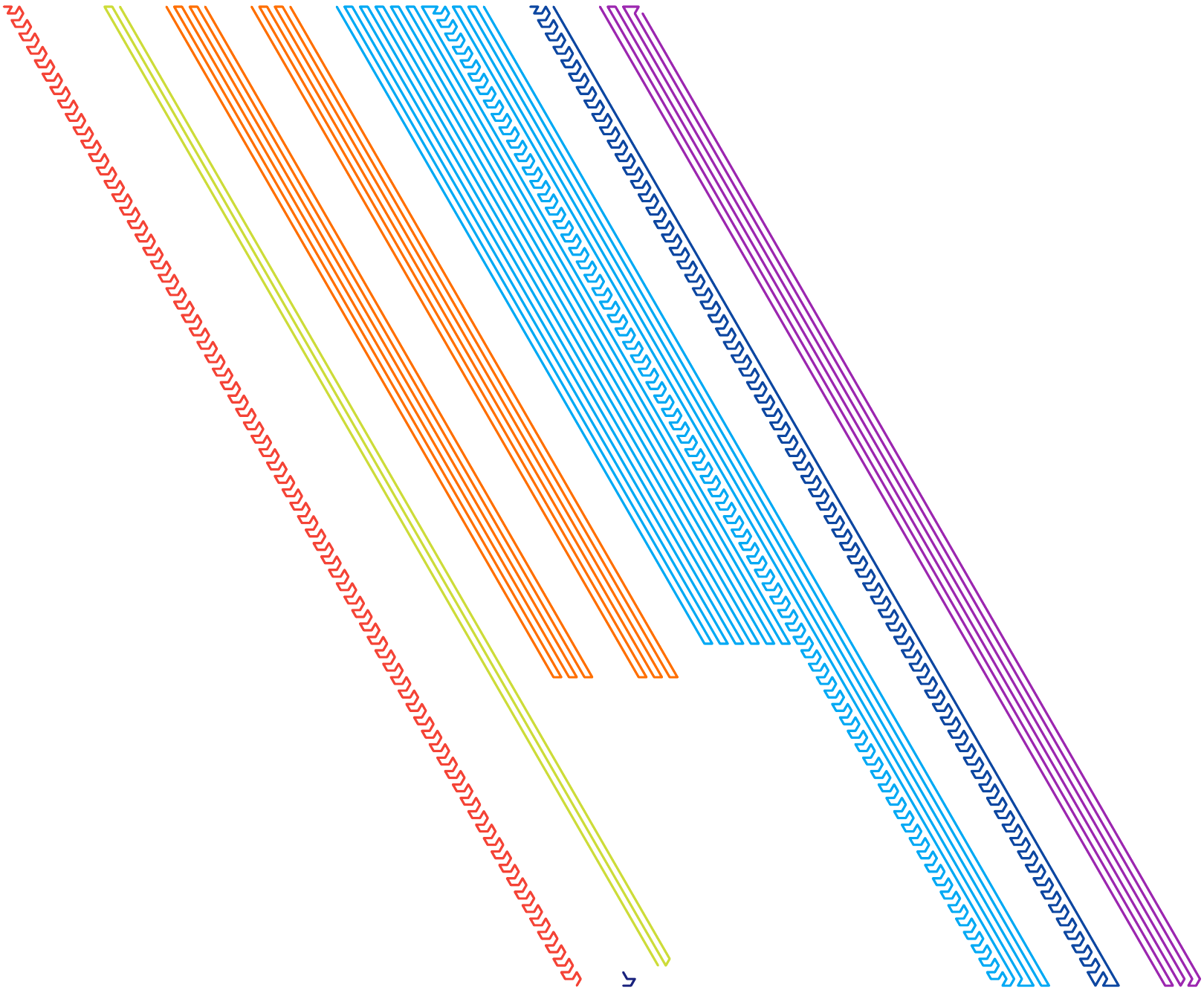}};
\end{tikzpicture}
}

%% file: OritatamiTuringPatterns/n=4-L=3-P=1/ZigCopy1____n=4-L=3-P=1-exploded.tex
\newcommand{\ZigCopyUBlockEpsilonnQLTPUexploded}[1]{
\begin{tikzpicture}
\ifthenelse{\equal{#1}{}}{
\TIKZlabel{(-173.75000000*\st pt, 0.00000000*\st pt)}{A\explodedZigZagSize\ZigDown}{SW}{40.0*\st pt}
\TIKZlabel{(43.75000000*\st pt, -316.09927238*\st pt)}{B\explodedZigZagSize\ZigDown}{SW}{40.0*\st pt}
\TIKZlabel{(66.25000000*\st pt, -303.10889132*\st pt)}{C\explodedZigZagSize\ZigDown}{SE}{40.0*\st pt}
\TIKZlabel{(-226.25000000*\st pt, 316.09927238*\st pt)}{E\explodedZigZagSize\ZigDown}{NE}{120.0*\st pt}
\TIKZlabel{(-66.25000000*\st pt, 316.09927238*\st pt)}{F\explodedZigZagSize\ZigDown}{NE}{40.0*\st pt}
\TIKZlabel{(181.25000000*\st pt, 0.00000000*\st pt)}{G\explodedZigZagSize\ZigDown}{NE}{40.0*\st pt}
}{}
\node [anchor=center] () at (0,0) {\includegraphics[scale =\s]{ZigCopy1___n=4-L=3-P=1-exploded.pdf}};
\end{tikzpicture}
}

%% file: OritatamiTuringPatterns/n=4-L=3-P=1/Read0_10_n=4-L=3-P=1-exploded.tex
\newcommand{\ReadZBlockUZnQLTPUexploded}[1]{
\begin{tikzpicture}
\ifthenelse{\equal{#1}{}}{
\TIKZlabel{(-392.50000000*\st pt, 4.33012702*\st pt)}{A\explodedZigZagSize\ZigUp}{NW}{40.0*\st pt}
\TIKZlabel{(-175.00000000*\st pt, 320.42939940*\st pt)}{B\explodedZigZagSize\ZigUp}{NW}{40.0*\st pt}
\TIKZlabel{(-152.50000000*\st pt, 307.43901834*\st pt)}{C\explodedZigZagSize\ZigUp}{NE}{40.0*\st pt}
\TIKZlabel{(-460.00000000*\st pt, -311.76914536*\st pt)}{D1\explodedZigZagSize\ZigUp}{SW}{40.0*\st pt}
\TIKZlabel{(-405.00000000*\st pt, -311.76914536*\st pt)}{D0\explodedZigZagSize\ZigUp}{SW}{120.0*\st pt}
\TIKZlabel{(-335.00000000*\st pt, -311.76914536*\st pt)}{E\explodedZigZagSize\ZigUp}{SE}{80.0*\st pt}
\TIKZlabel{(-235.00000000*\st pt, -311.76914536*\st pt)}{F\explodedZigZagSize\ZigUp}{SE}{40.0*\st pt}
\TIKZlabel{(-10.00000000*\st pt, 0.00000000*\st pt)}{G\explodedZigZagSize{Read\word0}}{SE}{40.0*\st pt}
}{}
\node [anchor=center] () at (0,0) {\includegraphics[scale =\s]{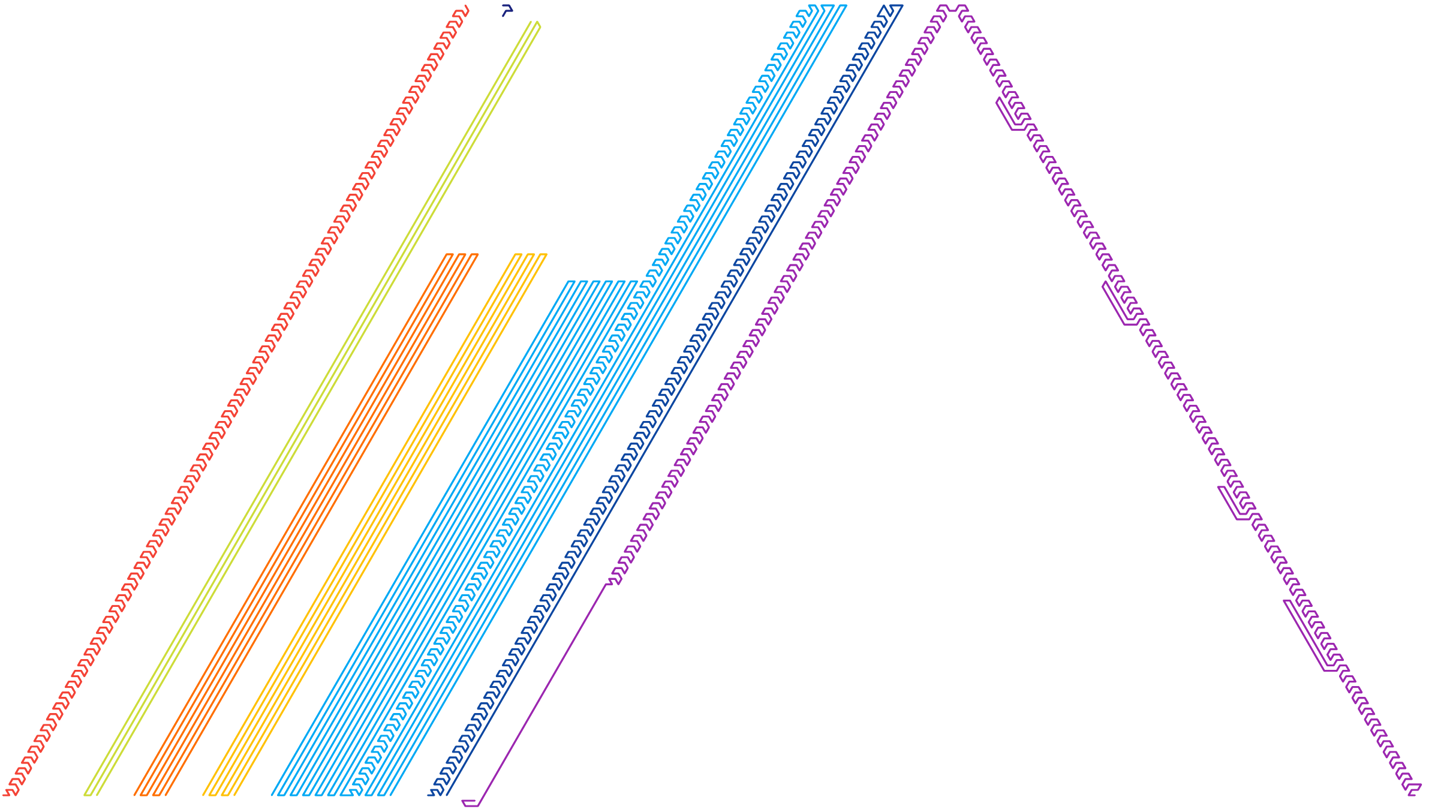}};
\end{tikzpicture}
}

%% file: OritatamiTuringPatterns/n=4-L=3-P=1/Read1_10_n=4-L=3-P=1-exploded.tex
\newcommand{\ReadUBlockUZnQLTPUexploded}[1]{
\begin{tikzpicture}
\ifthenelse{\equal{#1}{}}{
\TIKZlabel{(-573.75000000*\st pt, 4.33012702*\st pt)}{A\explodedZigZagSize\ZigUp}{NW}{40.0*\st pt}
\TIKZlabel{(-356.25000000*\st pt, 320.42939940*\st pt)}{B\explodedZigZagSize\ZigUp}{NW}{40.0*\st pt}
\TIKZlabel{(-333.75000000*\st pt, 307.43901834*\st pt)}{C\explodedZigZagSize\ZigUp}{NE}{40.0*\st pt}
\TIKZlabel{(-641.25000000*\st pt, -311.76914536*\st pt)}{D1\explodedZigZagSize\ZigUp}{SW}{40.0*\st pt}
\TIKZlabel{(-586.25000000*\st pt, -311.76914536*\st pt)}{D0\explodedZigZagSize\ZigUp}{SW}{120.0*\st pt}
\TIKZlabel{(-516.25000000*\st pt, -311.76914536*\st pt)}{E\explodedZigZagSize\ZigUp}{SE}{80.0*\st pt}
\TIKZlabel{(-416.25000000*\st pt, -311.76914536*\st pt)}{F\explodedZigZagSize\ZigUp}{SE}{40.0*\st pt}
\TIKZlabel{(-191.25000000*\st pt, 0.00000000*\st pt)}{G\explodedZigZagSize{Read\word1}}{SE}{40.0*\st pt}
}{}
\node [anchor=center] () at (0,0) {\includegraphics[scale =\s]{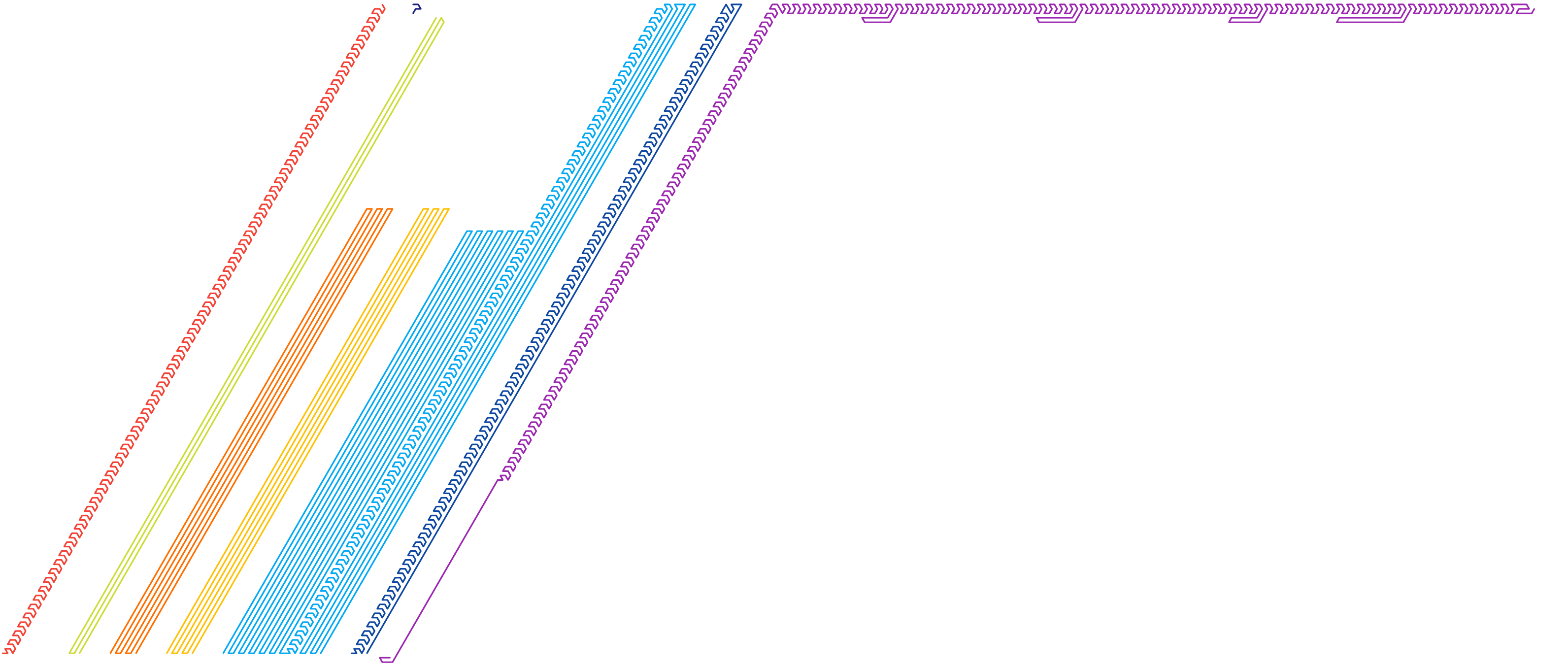}};
\end{tikzpicture}
}

%% file: OritatamiTuringPatterns/n=4-L=3-P=1/ZigCopy0_10_n=4-L=3-P=1-exploded.tex
\newcommand{\ZigCopyZBlockUZnQLTPUexploded}[1]{
\begin{tikzpicture}
\ifthenelse{\equal{#1}{}}{
\TIKZlabel{(-200.00000000*\st pt, 2.16506351*\st pt)}{A\explodedZigZagSize\ZigDown}{SW}{40.0*\st pt}
\TIKZlabel{(17.50000000*\st pt, -313.93420887*\st pt)}{B\explodedZigZagSize\ZigDown}{SW}{40.0*\st pt}
\TIKZlabel{(40.00000000*\st pt, -300.94382782*\st pt)}{C\explodedZigZagSize\ZigDown}{SE}{40.0*\st pt}
\TIKZlabel{(-267.50000000*\st pt, 318.26433589*\st pt)}{D1\explodedZigZagSize\ZigDown}{NW}{40.0*\st pt}
\TIKZlabel{(-212.50000000*\st pt, 318.26433589*\st pt)}{D0\explodedZigZagSize\ZigDown}{NW}{120.0*\st pt}
\TIKZlabel{(-142.50000000*\st pt, 318.26433589*\st pt)}{E\explodedZigZagSize\ZigDown}{NE}{120.0*\st pt}
\TIKZlabel{(-42.50000000*\st pt, 318.26433589*\st pt)}{F\explodedZigZagSize\ZigDown}{NE}{40.0*\st pt}
\TIKZlabel{(205.00000000*\st pt, 2.16506351*\st pt)}{G\explodedZigZagSize\ZigDown}{NE}{40.0*\st pt}
}{}
\node [anchor=center] () at (0,0) {\includegraphics[scale =\s]{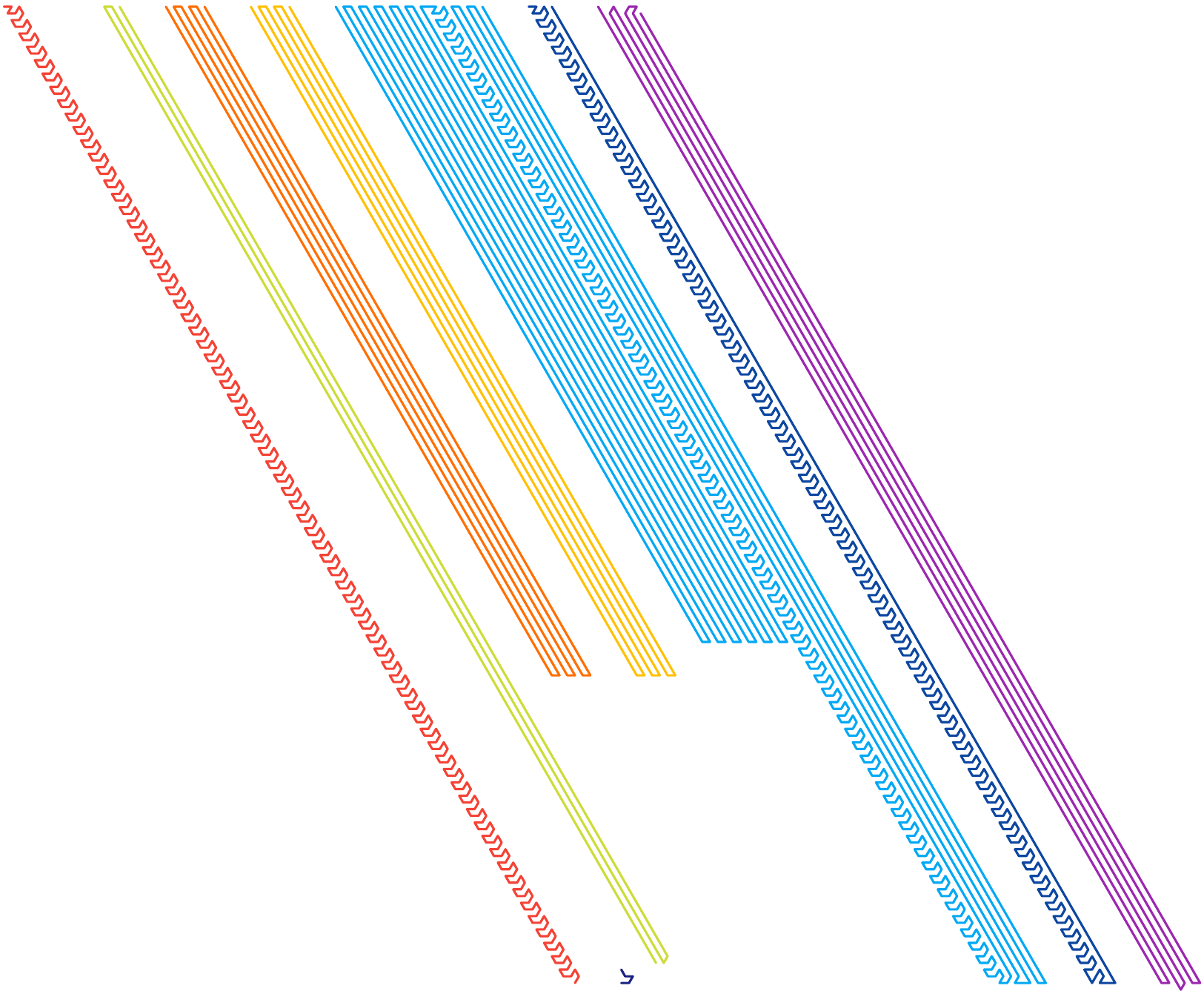}};
\end{tikzpicture}
}

%% file: OritatamiTuringPatterns/n=4-L=3-P=1/ZigCopy1_10_n=4-L=3-P=1-exploded.tex
\newcommand{\ZigCopyUBlockUZnQLTPUexploded}[1]{
\begin{tikzpicture}
\ifthenelse{\equal{#1}{}}{
\TIKZlabel{(-198.75000000*\st pt, 0.00000000*\st pt)}{A\explodedZigZagSize\ZigDown}{SW}{40.0*\st pt}
\TIKZlabel{(18.75000000*\st pt, -316.09927238*\st pt)}{B\explodedZigZagSize\ZigDown}{SW}{40.0*\st pt}
\TIKZlabel{(41.25000000*\st pt, -303.10889132*\st pt)}{C\explodedZigZagSize\ZigDown}{SE}{40.0*\st pt}
\TIKZlabel{(-266.25000000*\st pt, 316.09927238*\st pt)}{D1\explodedZigZagSize\ZigDown}{NW}{40.0*\st pt}
\TIKZlabel{(-211.25000000*\st pt, 316.09927238*\st pt)}{D0\explodedZigZagSize\ZigDown}{NW}{120.0*\st pt}
\TIKZlabel{(-141.25000000*\st pt, 316.09927238*\st pt)}{E\explodedZigZagSize\ZigDown}{NE}{120.0*\st pt}
\TIKZlabel{(-41.25000000*\st pt, 316.09927238*\st pt)}{F\explodedZigZagSize\ZigDown}{NE}{40.0*\st pt}
\TIKZlabel{(206.25000000*\st pt, 0.00000000*\st pt)}{G\explodedZigZagSize\ZigDown}{NE}{40.0*\st pt}
}{}
\node [anchor=center] () at (0,0) {\includegraphics[scale =\s]{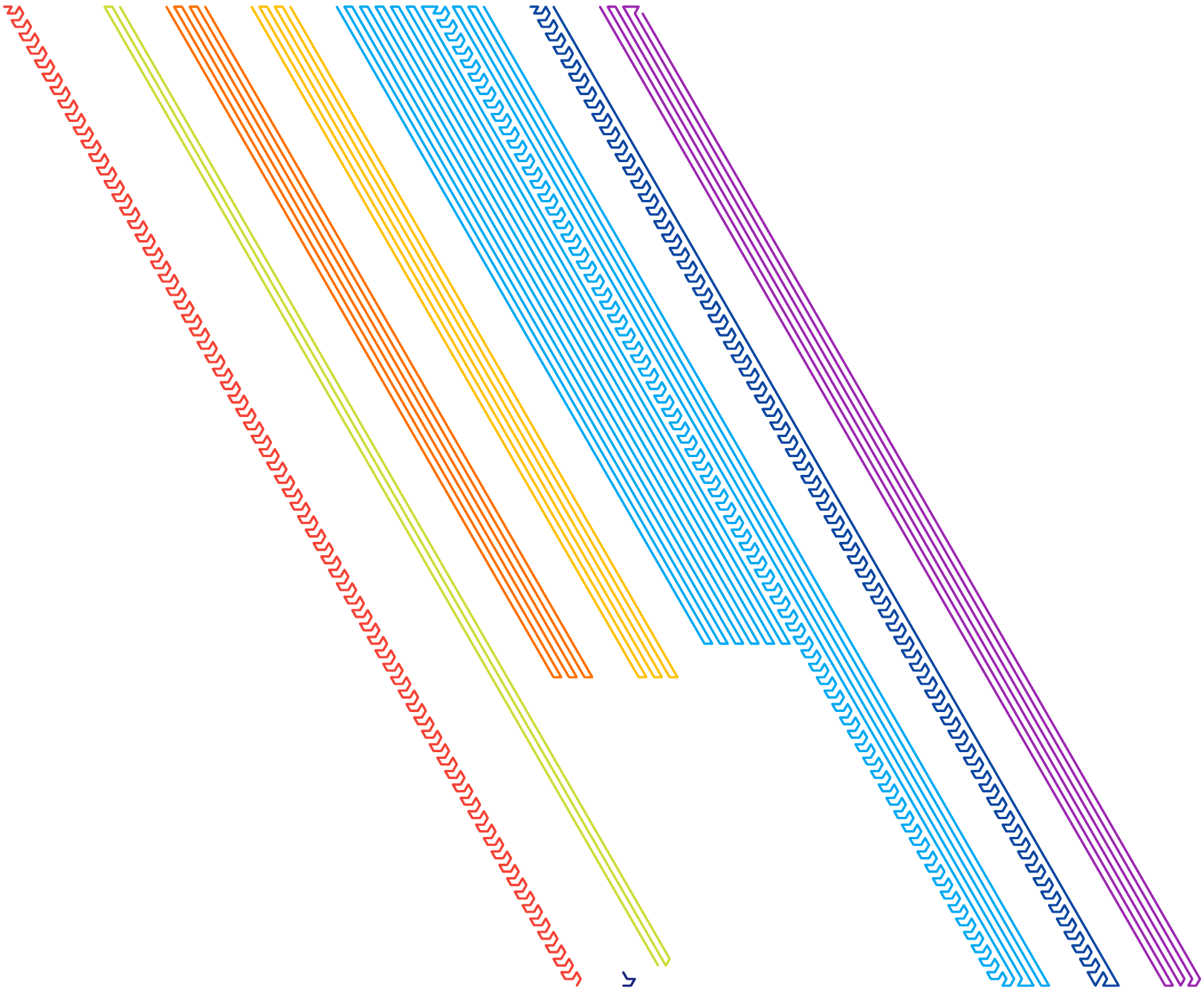}};
\end{tikzpicture}
}

%% file: OritatamiTuringPatterns/n=4-L=3-P=1/ZagCopy1_0_n=4-L=3-P=1-exploded.tex
\newcommand{\ZagCopyUBlockZnQLTPUexploded}[1]{
\begin{tikzpicture}
\ifthenelse{\equal{#1}{}}{
\TIKZlabel{(360.00000000*\st pt, 311.76914536*\st pt)}{A\explodedZigZagSize\Zag}{SE}{40.0*\st pt}
\TIKZlabel{(142.50000000*\st pt, -4.33012702*\st pt)}{B\explodedZigZagSize\Zag}{SE}{40.0*\st pt}
\TIKZlabel{(120.00000000*\st pt, 8.66025404*\st pt)}{C\explodedZigZagSize\Zag}{SW}{40.0*\st pt}
\TIKZlabel{(427.50000000*\st pt, 627.86841774*\st pt)}{D0\explodedZigZagSize\Zag}{NE}{40.0*\st pt}
\TIKZlabel{(357.50000000*\st pt, 627.86841774*\st pt)}{E\explodedZigZagSize\Zag}{NW}{120.0*\st pt}
\TIKZlabel{(227.50000000*\st pt, 627.86841774*\st pt)}{F\explodedZigZagSize\Zag}{NW}{40.0*\st pt}
\TIKZlabel{(0.00000000*\st pt, 311.76914536*\st pt)}{G\explodedZigZagSize\bZagZig}{NW}{40.0*\st pt}
}{}
\node [anchor=center] () at (0,0) {\includegraphics[scale =\s]{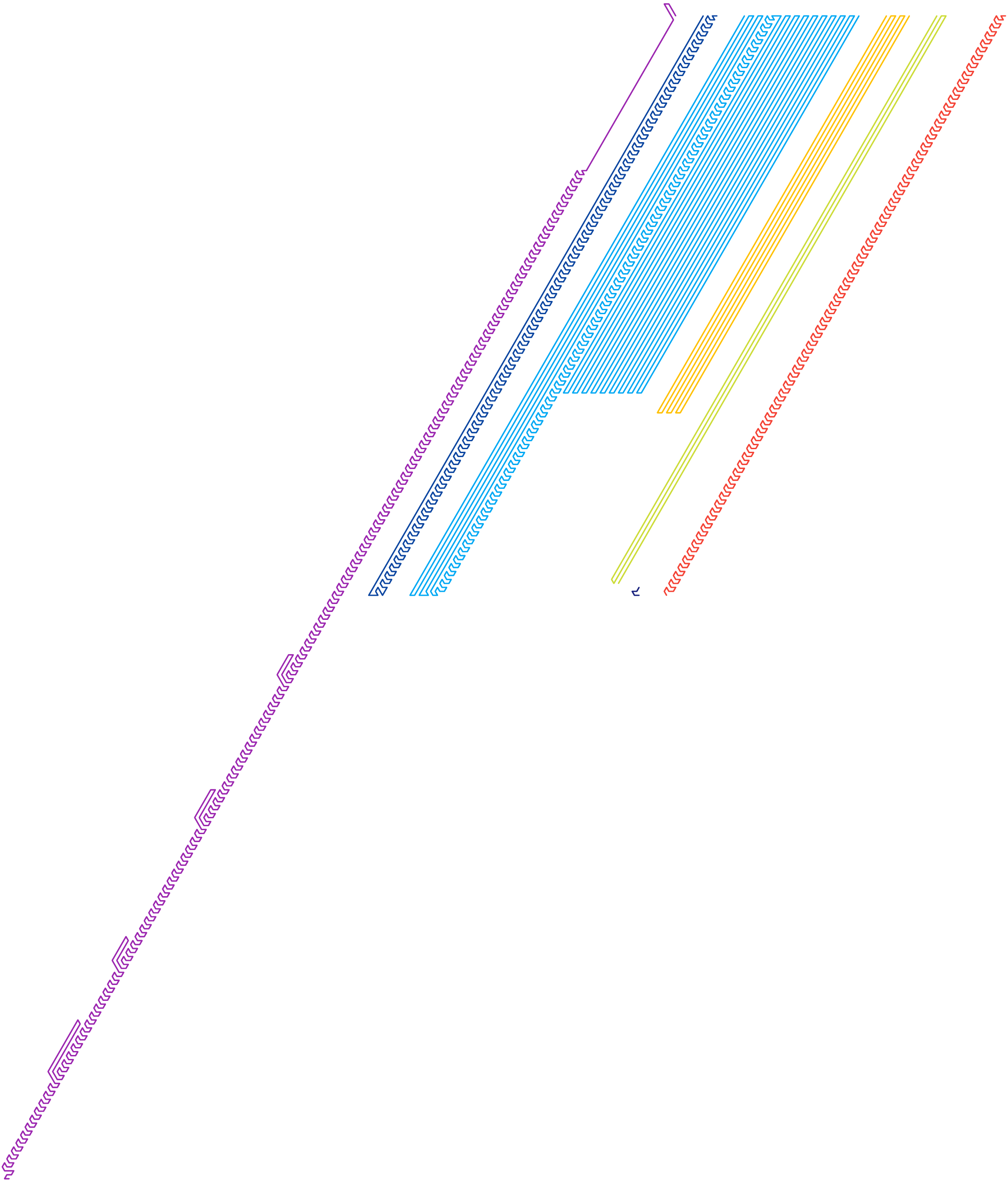}};
\end{tikzpicture}
}

%% file: OritatamiTuringPatterns/n=4-L=3-P=1/ZagCopy1_10_n=4-L=3-P=1-exploded.tex
\newcommand{\ZagCopyUBlockUZnQLTPUexploded}[1]{
\begin{tikzpicture}
\ifthenelse{\equal{#1}{}}{
\TIKZlabel{(372.50000000*\st pt, 311.76914536*\st pt)}{A\explodedZigZagSize\Zag}{SE}{40.0*\st pt}
\TIKZlabel{(155.00000000*\st pt, -4.33012702*\st pt)}{B\explodedZigZagSize\Zag}{SE}{40.0*\st pt}
\TIKZlabel{(132.50000000*\st pt, 8.66025404*\st pt)}{C\explodedZigZagSize\Zag}{SW}{40.0*\st pt}
\TIKZlabel{(440.00000000*\st pt, 627.86841774*\st pt)}{D1\explodedZigZagSize\Zag}{NE}{40.0*\st pt}
\TIKZlabel{(385.00000000*\st pt, 627.86841774*\st pt)}{D0\explodedZigZagSize\Zag}{NE}{120.0*\st pt}
\TIKZlabel{(315.00000000*\st pt, 627.86841774*\st pt)}{E\explodedZigZagSize\Zag}{NW}{120.0*\st pt}
\TIKZlabel{(215.00000000*\st pt, 627.86841774*\st pt)}{F\explodedZigZagSize\Zag}{NW}{40.0*\st pt}
\TIKZlabel{(-12.50000000*\st pt, 311.76914536*\st pt)}{G\explodedZigZagSize\bZagZig}{NW}{40.0*\st pt}
}{}
\node [anchor=center] () at (0,0) {\includegraphics[scale =\s]{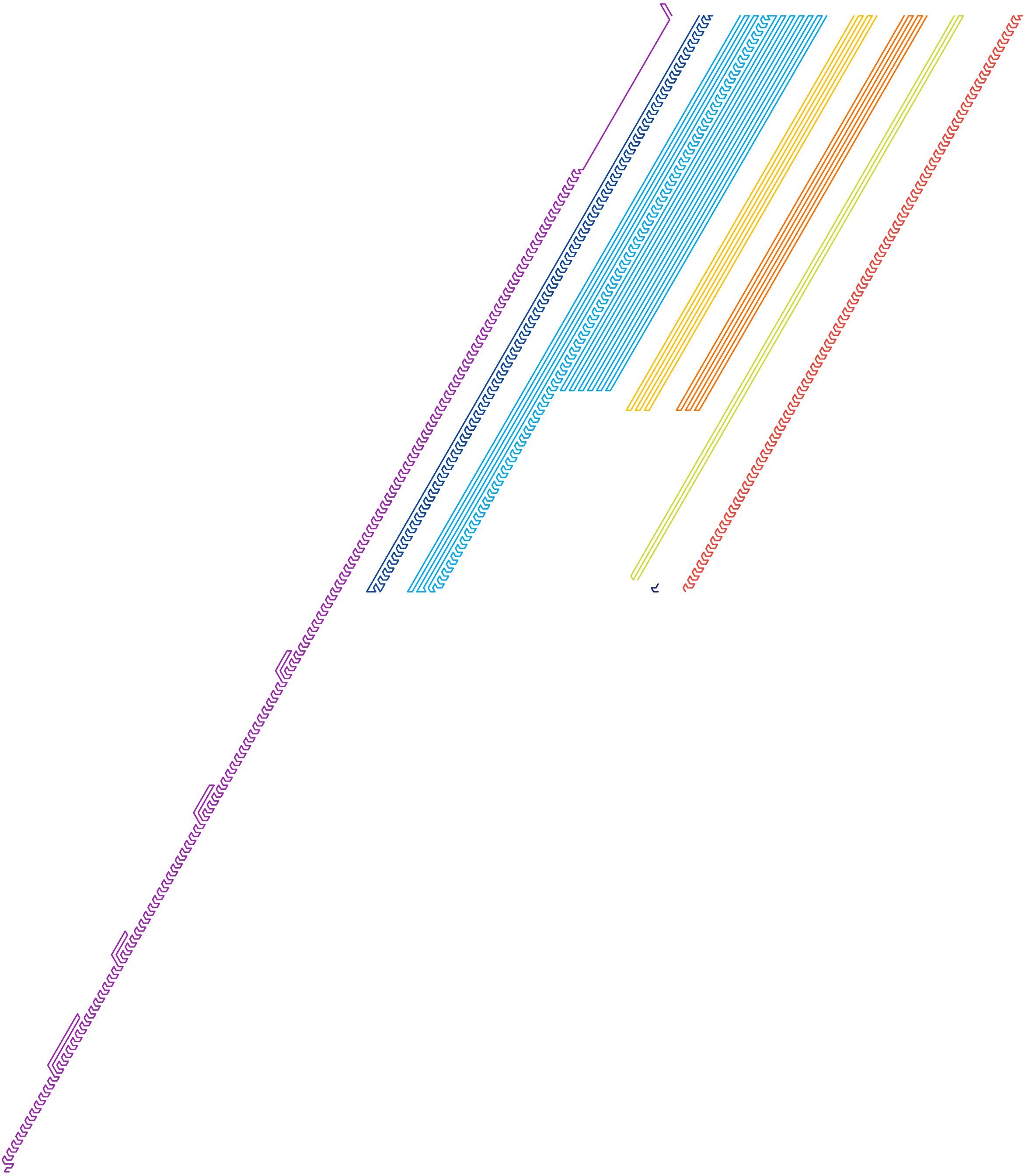}};
\end{tikzpicture}
}

%% file: OritatamiTuringPatterns/n=4-L=3-P=1/ZagCopy1_110_n=4-L=3-P=1-exploded.tex
\newcommand{\ZagCopyUBlockUUZnQLTPUexploded}[1]{
\begin{tikzpicture}
\ifthenelse{\equal{#1}{}}{
\TIKZlabel{(385.00000000*\st pt, 311.76914536*\st pt)}{A\explodedZigZagSize\Zag}{SE}{40.0*\st pt}
\TIKZlabel{(167.50000000*\st pt, -4.33012702*\st pt)}{B\explodedZigZagSize\Zag}{SE}{40.0*\st pt}
\TIKZlabel{(145.00000000*\st pt, 8.66025404*\st pt)}{C\explodedZigZagSize\Zag}{SW}{40.0*\st pt}
\TIKZlabel{(452.50000000*\st pt, 627.86841774*\st pt)}{D1\explodedZigZagSize\Zag}{NE}{40.0*\st pt}
\TIKZlabel{(397.50000000*\st pt, 627.86841774*\st pt)}{D1\explodedZigZagSize\Zag}{NE}{120.0*\st pt}
\TIKZlabel{(342.50000000*\st pt, 627.86841774*\st pt)}{D0\explodedZigZagSize\Zag}{NE}{200.0*\st pt}
\TIKZlabel{(272.50000000*\st pt, 627.86841774*\st pt)}{E\explodedZigZagSize\Zag}{NW}{120.0*\st pt}
\TIKZlabel{(202.50000000*\st pt, 627.86841774*\st pt)}{F\explodedZigZagSize\Zag}{NW}{40.0*\st pt}
\TIKZlabel{(-25.00000000*\st pt, 311.76914536*\st pt)}{G\explodedZigZagSize\bZagZig}{NW}{40.0*\st pt}
}{}
\node [anchor=center] () at (0,0) {\includegraphics[scale =\s]{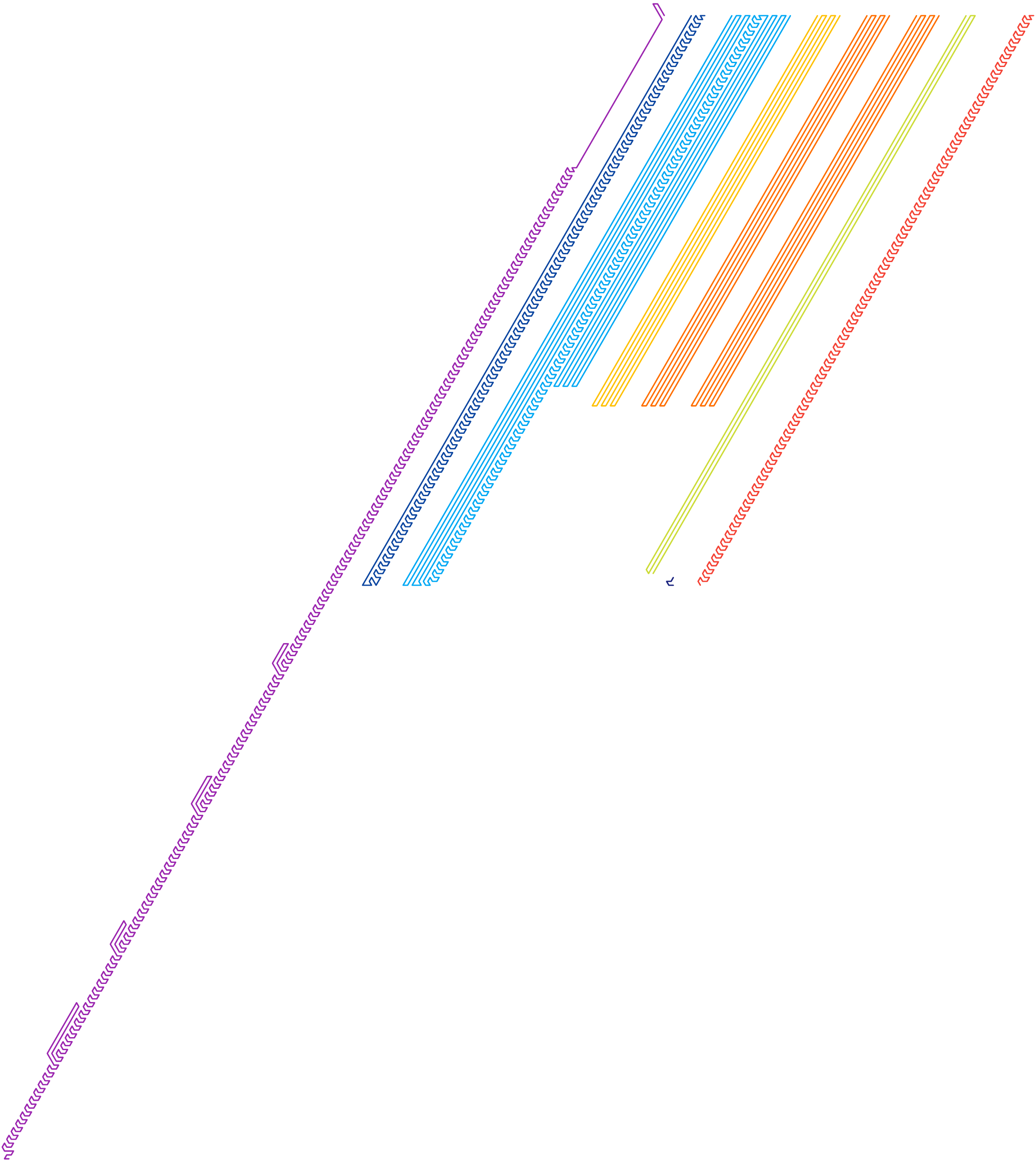}};
\end{tikzpicture}
}

%% file: OritatamiTuringPatterns/n=4-L=3-P=1/ZagCopy1_11_n=4-L=3-P=1-exploded.tex
\newcommand{\ZagCopyUBlockUUnQLTPUexploded}[1]{
\begin{tikzpicture}
\ifthenelse{\equal{#1}{}}{
\TIKZlabel{(372.50000000*\st pt, 311.76914536*\st pt)}{A\explodedZigZagSize\Zag}{SE}{40.0*\st pt}
\TIKZlabel{(155.00000000*\st pt, -4.33012702*\st pt)}{B\explodedZigZagSize\Zag}{SE}{40.0*\st pt}
\TIKZlabel{(132.50000000*\st pt, 8.66025404*\st pt)}{C\explodedZigZagSize\Zag}{SW}{40.0*\st pt}
\TIKZlabel{(440.00000000*\st pt, 627.86841774*\st pt)}{D1\explodedZigZagSize\Zag}{NE}{40.0*\st pt}
\TIKZlabel{(385.00000000*\st pt, 627.86841774*\st pt)}{D1\explodedZigZagSize\Zag}{NE}{120.0*\st pt}
\TIKZlabel{(315.00000000*\st pt, 627.86841774*\st pt)}{E\explodedZigZagSize\Zag}{NW}{120.0*\st pt}
\TIKZlabel{(215.00000000*\st pt, 627.86841774*\st pt)}{F\explodedZigZagSize\Zag}{NW}{40.0*\st pt}
\TIKZlabel{(-12.50000000*\st pt, 311.76914536*\st pt)}{G\explodedZigZagSize\bZagZig}{NW}{40.0*\st pt}
}{}
\node [anchor=center] () at (0,0) {\includegraphics[scale =\s]{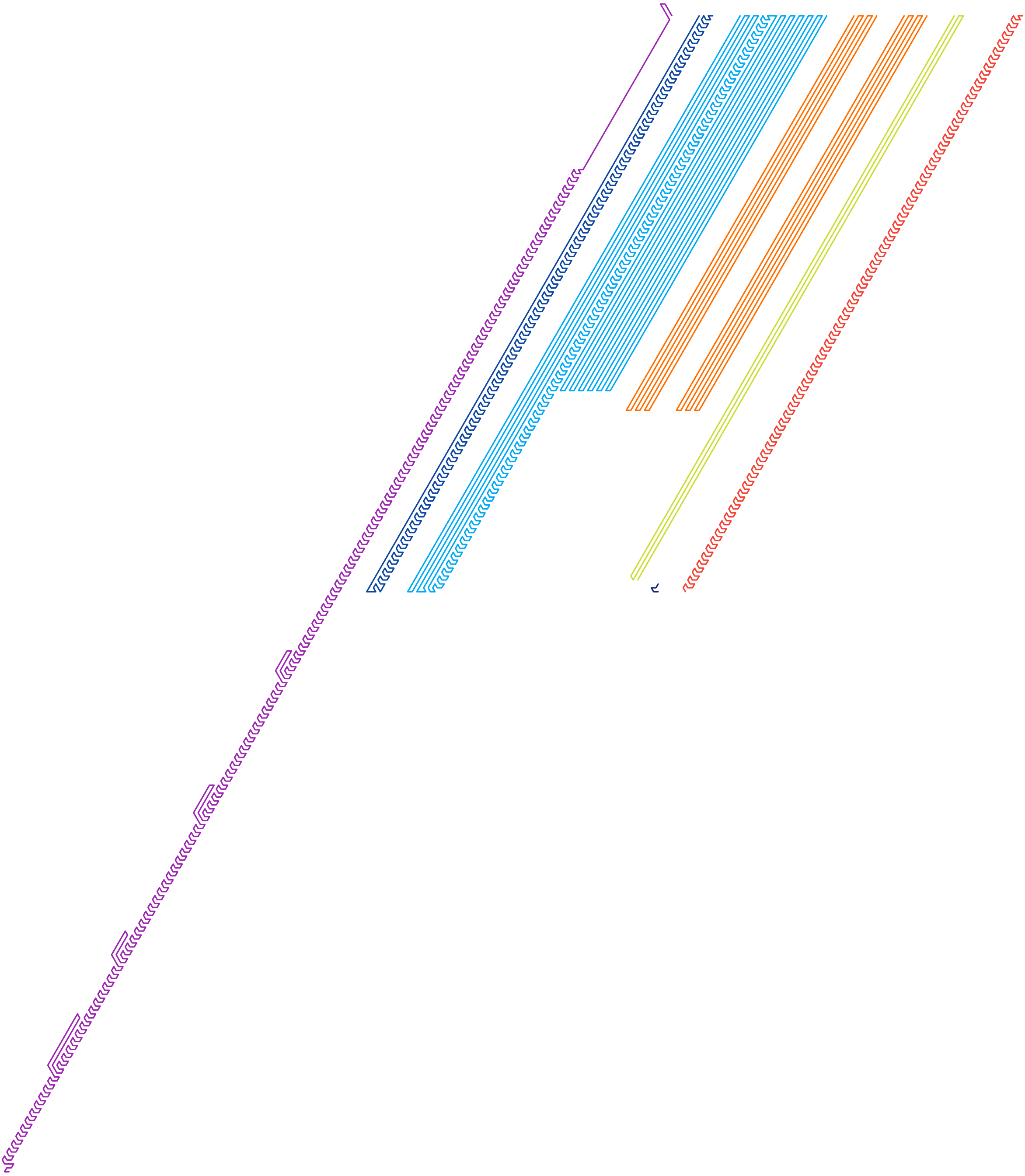}};
\end{tikzpicture}
}

%% file: OritatamiTuringPatterns/n=4-L=3-P=1/ZagCopy1____n=4-L=3-P=1-exploded.tex
\newcommand{\ZagCopyUBlockEpsilonnQLTPUexploded}[1]{
\begin{tikzpicture}
\ifthenelse{\equal{#1}{}}{
\TIKZlabel{(347.50000000*\st pt, 311.76914536*\st pt)}{A\explodedZigZagSize\Zag}{SE}{40.0*\st pt}
\TIKZlabel{(130.00000000*\st pt, -4.33012702*\st pt)}{B\explodedZigZagSize\Zag}{SE}{40.0*\st pt}
\TIKZlabel{(107.50000000*\st pt, 8.66025404*\st pt)}{C\explodedZigZagSize\Zag}{SW}{40.0*\st pt}
\TIKZlabel{(400.00000000*\st pt, 627.86841774*\st pt)}{E\explodedZigZagSize\Zag}{NW}{120.0*\st pt}
\TIKZlabel{(240.00000000*\st pt, 627.86841774*\st pt)}{F\explodedZigZagSize\Zag}{NW}{40.0*\st pt}
\TIKZlabel{(12.50000000*\st pt, 311.76914536*\st pt)}{G\explodedZigZagSize\bZagZig}{NW}{40.0*\st pt}
}{}
\node [anchor=center] () at (0,0) {\includegraphics[scale =\s]{ZagCopy1___n=4-L=3-P=1-exploded.pdf}};
\end{tikzpicture}
}

%% file: OritatamiTuringPatterns/n=4-L=3-P=1/blockColors.tex
\definecolor{blockReadOneOutlineColor}{rgb}{0.882353,0.745098,0.905882}
\definecolor{moduleSeedOneColor}{rgb}{0.243137,0.152941,0.137255}
\definecolor{moduleAColor}{rgb}{0.956863,0.262745,0.211765}
\definecolor{blockZigCopyOneOutlineColor}{rgb}{1.000000,0.878431,0.698039}
\definecolor{blockAppendZeroOutlineColor}{rgb}{0.784314,0.901961,0.788235}
\definecolor{moduleSeedBeginColor}{rgb}{0.364706,0.250980,0.215686}
\definecolor{moduleDOneColor}{rgb}{1.000000,0.435294,0.000000}
\definecolor{moduleEColor}{rgb}{0.011765,0.662745,0.956863}
\definecolor{blockSeedZeroOutlineColor}{rgb}{0.843137,0.800000,0.784314}
\definecolor{moduleSeedZeroColor}{rgb}{0.552941,0.431373,0.388235}
\definecolor{blockSeedBeginOutlineColor}{rgb}{0.631373,0.533333,0.498039}
\definecolor{blockZigCopyZeroOutlineColor}{rgb}{1.000000,0.976471,0.768627}
\definecolor{blockReadZeroOutlineColor}{rgb}{0.972549,0.733333,0.815686}
\definecolor{blockZagCopyOneOutlineColor}{rgb}{1.000000,0.800000,0.737255}
\definecolor{moduleBColor}{rgb}{0.101961,0.137255,0.494118}
\definecolor{blockAppendOneOutlineColor}{rgb}{0.819608,0.768627,0.913725}
\definecolor{moduleDZeroColor}{rgb}{1.000000,0.756863,0.027451}
\definecolor{moduleSeedEndColor}{rgb}{0.364706,0.250980,0.215686}
\definecolor{blockAppendOutlineColor}{rgb}{0.701961,0.898039,0.988235}
\definecolor{blockOutlineLineColor}{rgb}{0.700000,1.000000,0.988235}
\definecolor{moduleCColor}{rgb}{0.803922,0.862745,0.223529}
\definecolor{blockSeedEndOutlineColor}{rgb}{0.631373,0.533333,0.498039}
\definecolor{blockZagCopyZeroOutlineColor}{rgb}{1.000000,0.925490,0.701961}
\definecolor{blockHaltOutlineColor}{rgb}{1.000000,0.803922,0.823529}
\definecolor{moduleFColor}{rgb}{0.050980,0.278431,0.631373}
\definecolor{bondColor}{rgb}{0.500000,1.000000,0.631373}
\definecolor{blockSeedOneOutlineColor}{rgb}{0.737255,0.666667,0.643137}
\definecolor{moduleGColor}{rgb}{0.611765,0.152941,0.690196}

%% file: intro.tex

\section{Introduction}

Oritatami model was introduced in~\cite{GeMeScSe2016}  to try to understand the kinetics of co-transcriptional folding. This process has been shown to play an important role in the final shape of biomolecules~\cite{cody1}, especially in the case of RNA~\cite{cody2}.
The rationale of this choice is that the wetlab version of Oritatami already exists, and has been successfully used to engineer shapes with RNA in the wetlab~\cite{GearyRothemundAndersen2014}. 

\noindent\captionpar{In oritatami,} we consider a finite set of \emph{bead types}, and a periodic sequence of \emph{beads}, each of a specific bead type. Beads are attracted to each other according to a fixed symmetric relation, and in any folding (a folding is also called a \emph{configuration}), whenever two beads attracted to each other are found at adjacent positions, a \emph{bond} is formed between them.

At each step, the latest few beads in the sequence are allowed to explore all possible positions, and we keep only those positions that minimise the energy, or otherwise put, those positions that maximise the number of bonds in the folding.
``Beads'' are a metaphor for domains, i.e. subsequences, in RNA and DNA.

\noindent\captionpar{Previous work} on oritatami includes the implementation of a binary counter~\cite{GeMeScSe2016}, the Heighway dragon fractal \cite{MaShUb2018}, folding of shapes at small scale \cite{shape2018DNA}, and NP-hardness of the rule minimization \cite{OtaSeki2017,HanKim2017} and of the equivalence of non-deterministic oritatami systems \cite{HaKiOtSe2016}.

\noindent\captionpar{Main result.}
In this paper, we construct a ``universal'' set of 542 bead types, along with a single universal attraction rule for these bead types, with which we can simulate any tag system, and therefore any Turing machine $\mathcal M$, within a polynomial factor of the running time $\mathcal M$.
The reduction proceeds as follows:
$$\small
	\textsf{Turing machine} 
\xrightarrow{\text{\cite{WoodsNeary2006,NearyPhD}}} 
	\textsf{Cyclic tag system} 
\xrightarrow{\text{Prop.~\ref{prop:TM>SCTS}}} 
	\textsf{Skipping cyclic tag system} 
\xrightarrow{\text{Thm.~\ref{thm:SCTS>oritatami}}}
	\textsf{Oritatami system}
$$
Our result relies on the development of a generic toolbox, which is easily reusable for future work to design complex functions in oritatami systems. 

%


\noindent\captionpar{Proving our designs.}
\label{challenge1}
The main challenge we faced in this paper was the size of our constructions: indeed, while we developed higher-level geometric constructs to program useful shapes, there is a large number of possible interactions between all different parts of the sequence.
Getting solid proofs on large objects is a common problem in discrete dynamical systems, for instance on cellular automata~\cite{gacs1997reliable,Cook2004} or tile assembly systems~\cite{DBLP:conf/icalp/KariKMPS15}.
In this paper, we introduce a general framework to deal with that complexity, and prove our constructions rigorously. This method proceeds by decomposing the sequence into different \emph{modules}, and the space into different areas: \emph{blocks}, where exactly one step of the simulation is performed, which are composed of \emph{bricks}, where exactly one module grows. We can then reason on the modules separately, and only deal with interactions at the border between all possible modules that can have a common border.

%% file: block-simulation.tex

\section{The block simulation of SCTS: Proving the correctness of local folding is enough}
\label{sec:block:simul}

Given a SCTS \TMS, we design an oritatami system $\TMO_\TMS$ that folds into a version, at a larger scale, of the \emph{annotated trimmed space-time diagram} of $\TMS$ (or \emph{trimmed diagram} for short) defined as follows:  

\subparagraph{Trimmed diagram of SCTS.} Any SCTS proceeds as follows: it trims all the leading \word0s in the data word and then appends the currently marked appendant when it reads the first \word1 (if any; otherwise it halts). It is thus natural to group all these steps (trim leading \word0s and process the leading \word1) as one single macro step. This motivates the following representation. Given a SCTS $(\alpha^0,\ldots,\alpha^{n-1}; u^0)$, we denote by ${0\leq t_1 < t_2 <\cdots}$ all the times $t$ such that the dataword $u^t$ starts with letter $\word1$ and set $t_0 = -1$ by convention. Let us now group all deletion steps occurring during steps $t_{i}+1$ to $t_{i+1}-1$ by simply indicating in exponent the marker $m^t$ before each letter read. In the case of our STCS \TME, we have $t_0=-1, t_1=1, t_2=3, t_3=4$ and its execution is now represented as:
$$\pQ0\word0\pQ1\word{10}
\tagTransition{Append}{[2:\word{11}]} 
\pQ3\word0\pQ0\word1\word1
\tagTransition{Append}{[1:\epsilon]} 
\pQ2\word{1} 
\tagTransition{Append}{[3:\word{0}]} 
\pQ0\word0\pQ1
~ \texttt{Halt}
$$
Now, let's align the resulting datawords in a 2D diagram according to their common parts:
\begin{center}
\input{\texInputDir/scts-diagram}
\end{center}
\noindent This defines the \emph{annotated trimmed space-time diagram} for the SCTS \TME.  Lemma~\ref{lem:SCTS:diagram} in appendix provides the formal definition for an arbitrary SCTS. 

\subparagraph{The transcript.}
The proof of Theorem~\ref{thm:SCTS>oritatami} relies on constructing a transcript (and a fixed rule) that will reproduce faithfully the trimmed diagram of the simulated STCS.  Figure~\ref{fig:block:example} illustrates the folded configuration of the transcript corresponding to SCTS \TME.
Macroscopically, the transcript folds into a zig-zag sequence of \emph{blocks}, each performing a specific operation.  
\begin{description}[topsep=1mm]
\item[The current dataword]  is encoded at the bottom of each row of blocks: \word0s are encoded by a spike, and \word1s are encoded by a flat surface.
\item[the seed configuration] encodes the initial dataword and opens the first zig row at which the folding of the transcript starts. Letters \word0 and \word1 are encoded by a \emph{spike} (see Fig.~\ref{fig:G:bounce:read0})
 and a \emph{flat surface}  (see Fig.~\ref{fig:G:bounce:read1}) respectively.
\item[in each zig row (left to right),]  the transcript folds into a series of \bnREAD0{} blocks (trimming the leading \word0s from the dataword encoded above), then into a \bnREAD1{} block, if the dataword contains a \word1, or into a \bnHALT{} block terminating the folding, otherwise; this is the \emph{zig-up phase}. Then, the transcript starts the \emph{zig-down phase} which consists in folding into \bnZIGCOPY{}{} block copying the letters encoded above to the bottom of the row;  once the end of the dataword is reached, the transcript folds into an \bnAPPENDRETURN{} block which encodes, at the bottom of the row, the currently marked appendant, and finally, opens the next zag row.
\item[in each zag row (right to left),] the transcript folds into \bnZAGCOPY{}{} blocks copying the dataword encoded above to the bottom of the row. For the leftmost letter, the transcript folds into the special \bnCOPYLINEFEED{}{} block which also opens the next zig row.
\end{description} 

\input{\texInputDir/block-example}

 The transcript is a periodic sequence whose period is the concatenation of  $n$ bead type sequences $\APPENDANTa{0},\ldots,\APPENDANTa{n-1}$ called segments, each encoding one appendant. 

\subparagraph{Encoding of the marker.} \bnREAD{}{} and \bnAPPENDRETURN{} blocks consist of the folding of \emph{exactly one} segment, whereas \bnZIGCOPY{}{}, \bnZAGCOPY{}{} and \bnCOPYLINEFEED{}{} consist of the folding of \emph{exactly $n$} segments. It follows that the appendant encoded in the \emph{leading} segment folded inside each block corresponds to the \emph{currently marked} appendant in the simulated SCTS. As a consequence, the appendant contained in the folded \bnAPPENDRETURN{} block is indeed the appendant to be appended to the dataword.

\subparagraph{The segment sequence.}
Each segment $\APPENDANTa{i}$ encodes the appendant $\alpha^i$ as a sequence of $6+|\alpha^i|$ modules: one of each \modA, $\ModuleB$, and $\ModuleC$, then $|\alpha^i|$ of \modD, then one of each \modE, $\ModuleF$ and $\ModuleG$. Each module is a bead type sequence that plays a particular role in the design:
\begin{description}[topsep=1mm]
\item[\ModA] folds into the initial scaffold upon which the next modules rely.
\item[\ModB] detects if the dataword is empty: if so, it folds to the left and the folding gets trapped in a closed space and halts; otherwise, it folds to the right and the folding continues.
\item[\ModC] detects the end of the dataword and triggers the appending of the marked appendant accordingly.
\item[\ModD]  encodes each letter of the appendant.
\item[\ModEa{}] ensures by padding that all appendant sequences have the same length when folded (even if the appendant have different length). It serves two other purposes: \ModB\/ senses its presence to detect if the dataword is empty; and its folding initiates the opening the zag row once the marked appendant has been appended to the dataword.
\item[\ModF] is the scaffold upon which \ModG\/ folds. It is specially designed to induce two very distinct shapes on $\ModuleG$ depending on the initial shift of $\ModuleG$. Furthermore, when \ModF\/ is exposed, \ModC\/ folds along $\ModuleF$ which triggers the appending of the marked appendant encoded by the modules $\ModuleDx{}$ following $\ModuleC$. 
\item[\ModG] is the ``logical unit'' of the transcript. It implements three distinct functions which are triggered by its interactions with its environment: Reading the leading letter of the dataword, Copying a letter of the dataword, and Opening the next zig row at the leftmost end of a zag row.
\end{description}

We call \emph{bricks} the folding of each of these modules. The blocks into which the transcript folds, depend on the bricks in which its modules fold, as illustrated in Fig.~\ref{fig:block:exploded}. Please refer to sections~\ref{sec:all:bricks:paths} to \ref{sec:all:bricks:full} in appendix for the description of blocks in terms of bricks and of how they articulate with each other to produce the desired macroscopic folding pattern.   

The full description of each of these sequences is given in Section~\ref{sec:all:bricks:full} in appendix.

\medskip

Let $\TMS = (\alpha^0,\alpha^1,\ldots, \alpha^{n-1}; u^0)$ be a skipping cyclic tag system, and, as before, let for all integer $i \geq 0$, $t_i$ be the $i^{\text{th}}$ step where $u^{t}$ starts with $\word1$ (starting from $0$, i.e. $t_0$ is the first step where $u^{t_0}$ starts with $\word1$). The following lemma shows that the transcript described above folds indeed into blocks that simulates the trimmed diagram of \TMS. Proposition~\ref{prop:TM>SCTS} and Theorem~\ref{thm:key} are direct corollaries of this lemma. 

\begin{lemma}[Key lemma] \label{lem:actual:ith:row} \label{lem:key}
There is a bead type set $B$ and a rule $\heart$ such that: for every SCTS $\TMS$, there are $\pi_\TMS$ and $(\sigma_\TMS,s_\TMS)$ defined as in Theorem~\ref{thm:main} such that, for every initial dataword $u^0$, the (possibly infinite) final folded path of the oritatami system $\TMO_\TMS = ((\pi_\TMS)^\infty,\heart,\delay=3)$ from the seed configuration $(\sigma(u ^0), s(u^0))$ is exactly structured as the following sequence of blocks organized in zig and zag rows as follows: (recall Fig.~\ref{fig:block:example:execution})

\newcommand{\theScale}{.7}
\begin{itemize}[nosep]

\item First, the block \bSEED{$u^0$} ending at coordinates $(-1,0)$.

\item Then, for $i\geq0$, the $i$-th row consists of a zig row located between $y=2(i-1)h+1$ and $y=2ih$, and a zag row located between $y=2ih+1$ and $y=2(i+1)h$, composed as follows:
  \begin{description}[topsep=0pt]
  \item[$\bullet$ \textup{(Compute)} if $u^{1+t_i} = \word0^r1\cdot s$ and if $s\neq\epsilon$ or $\alpha^{1+i+t_{i+1}}\neq\epsilon$:] then ${r = t_{i+1}-t_i-1}$ and:
    \begin{itemize}[itemsep=4pt]
    \item the $i$-th zig-row consists from left to right of the following sequence of blocks whose origins are located at the following coordinates:\\[4pt]
      \centerline{\hspace*{-1.5cm}\scalebox{\theScale}{
          $\begin{array}{cccccccccc}
             \swarrow y 	&	\multicolumn{4}{c}{2ih} && \multicolumn{4}{c}{(2i-1)h+1}\\ \cmidrule{1-5} \cmidrule{7-10}
             \rightarrow  x	&	ih+(1+t_i)W & \cdots& ih+(t_{i+1}-1)W & ih+t_{i+1}W &  &ih+(1+t_{i+1})W-1  & \cdots & ih+(|s|+t_{i+1})W-1 & ih+(1+|s|+t_{i+1})W-1 \\ \toprule
             \textup{Blocks} & \bnREAD{\word0}{} & \cdots & \bnREAD{\word0}{} & \bnREAD{\word1}{} & & \bnZIGCOPY{$(s_0)$}{} & \cdots & \bnZIGCOPY{$(s_{|s|-1})$}{} & \bnAPPENDRETURN{[\alpha^{1+i+t_{i+1}}]}{}\\
              \midrule
              \textup{Marker}	&	i+1+t_i & \cdots& i+r+t_i & i+t_{i+1} &  & i+1+t_{i+1}  & \cdots & i+1+t_{i+1} & i+1+t_{i+1} 
           \end{array}
           $}}
       \smallskip\\
       This row ends at position $((i+1)h+(1+|s|+|\alpha^{i+1+t_{i+1}}|+t_{i+1})W-7,2ih+2)$.

     \item the $i$-th zag-row consists from right to left of the following sequence of blocks whose origins are  located at the following coordinates:\\[4pt]
       \centerline{\hspace*{-1.5cm}\scalebox{\theScale}{
           $\begin{array}{ccccc}
              \swarrow y	&	\multicolumn{4}{c}{2ih+1}\\ \midrule
              \rightarrow x	&	(i+1)h+(2+t_{i+1})W-8  & (i+1)h+(3+t_{i+1})W-8  & \cdots & (i+1)h+(1+|v|+t_{i+1})W-8
              \\ \toprule
              \textup{Blocks} & \bnCOPYLINEFEED{$(v_0$)}{} & \bnZAGCOPY{$(v_1)$}{} & \cdots & \bnZAGCOPY{$(v_{|v|-1})$}{}
              \\ 
              \midrule
              \textup{Marker}	&	i+2+t_{i+1} &	i+2+t_{i+1} & \cdots& i+2+t_{i+1}
            \end{array}
            $}}\\[4pt]	
        where ${v = u^{1+t_{i+1}} = s\cdot p_{i+1+t_{i+1}}}\neq\epsilon$ (as $s$ and $\alpha^{i+1+t_{i+1}}$ are not both $\epsilon$). 
        This row ends at position $((i+1)h+(1+t_{i+1})W-1,2(i+1)h)$.
      \end{itemize} 
    \item[$\bullet$ \textup{(Halt 1)} if $u^{1+t_i} = \word0^r1$ and $\alpha^{1+i+t_{i+1}} = \epsilon$:] then $r = t_{i+1}-t_i-1$ and the last rows of the configuration consists from left to right of the following sequence of blocks located at the following coordinates:\\[4pt] 
      \centerline{\hspace*{-1.5cm}\scalebox{\theScale}{
          $\begin{array}{ccccccccc}
             \swarrow y	&	\multicolumn{4}{c}{2ih} & & (2i-1)h+1 & & 2(i+1)h
             \\ \cmidrule{1-5} \cmidrule{7-7} \cmidrule{9-9}
             \rightarrow x	&	ih+(1+t_{i})W  & \cdots & ih+(t_{i+1}-1)W & ih+t_{i+1}W & & ih+(1+t_{i+1})W-1&  & (i+1)h+(1+t_{i+1})W
             \\ \toprule
             \textup{Blocks} & \bnREAD{\word0}{} & \cdots & \bnREAD{\word0}{} & \bnREAD{\word1}{} & & \bnRETURNLINEFEED{} & &  \bnHALT
             \\ \midrule
             \textup{Marker}	&	i+1+t_i &	\cdots & i+t_{i+1}-1 & i+t_{i+1}  &&i+1+t_{i+1}  & & i+2+t_{i+1}
           \end{array}
           $}}\\[4pt]	
     \item[$\bullet$ finally, \textup{(Halt 2)} if $u^{1+t_i} = \word0^r$ for some $r\geq0$:] then the $i$-th zig-row is last row of the configuration and consists of the following sequence of blocks located at the following coordinates:\\[4pt] 
       \centerline{\hspace*{-1.5cm}\scalebox{\theScale}{
           $\begin{array}{ccccc}
              \swarrow y	&	\multicolumn{4}{c}{2ih} 
              \\ \midrule
              \rightarrow x	&	ih+(1+t_i)W & \cdots & ih+(r+t_{i})W & ih+(1+r+t_{i})W
              \\ \toprule
              \textup{Blocks} & \bnREAD{\word0}{} & \cdots & \bnREAD{\word0}{} & \bnHALT{}
               \\ \midrule
               \textup{Marker}	&	i+1+t_i &	\cdots & i+r+t_i & i+r+1+t_i
            \end{array}
            $}}
      \end{description}
    \end{itemize}
\end{lemma}

The following sections are dedicated to the proof of Key Lemma~\ref{lem:key}.

%% file: scts-diagram.tex

{\setlength{\extrarowheight}{3pt}
\ensuremath{
\begin{array}{r@{}r@{}r@{\,}r@{\,}r@{\,}r@{\,}r@{\,}rr@{{~~\longrightarrow~~}}c@{~}r@{\,:\,}l}
\pointHere{t_0}&&\pointHere{t_1}&&\pointHere{t_2}&\pointHere{t_3}\\
\cline{2-4}
&\kase{\pQ0\word0}&\kase{\pQ1\word1}&\kase{\word0}&\kase{}&&&&&{\texttt{Append}}&[2&\word{11}]\\[2pt]
\cline{2-6}
&&&\kase{\pQ3\word0}&\kase{\pQ0\word1}&\kase{\word1}&\kase{}&&&
{\texttt{Append}} &[1&\epsilon]\\[2pt]
\cline{4-6}
&&&&&\kase{\pQ2\word1}&\kase{}&&&{\texttt{Append}}&[3&\word{0}]\\
\cline{6-7}
&&&&&&\kase{\pQ0\word0}&\kase{\pQ1}&&{\texttt{Halt}} &[1&\epsilon]\\[2pt]
\cline{7-7}
\end{array}
}
}%

%% file: block-example.tex

\setlength{\TIKZtextwidth}{\textwidth}
\setlength{\TIKZtextheight}{\textheight}

\begin{figure}

\begin{subfigure}{\textwidth}
\hspace*{-3cm}
\begin{tikzpicture}
\node[anchor=north west] (blocks) at (0,0) {\adjustbox{scale=10,max width =1.4\TIKZtextwidth, max height = .9\TIKZtextheight}
{\executionnQLTPUoutlined{}}};
\node[draw, anchor=north west,shift={(2mm,-1.5cm)}] () at (blocks.west) {\adjustbox{max width =8cm, max height = 8cm}
{\input{\texInputDir/scts-diagram}}};
\end{tikzpicture}
\caption{Folding of the oritatami system simulating the STCS \TME.}
\label{fig:block:example:execution}
\end{subfigure}

\vspace*{2mm}

\begin{subfigure}{\textwidth}
\hspace*{-3cm}
\adjustbox{scale=1.5,max width =1.4\TIKZtextwidth, max height = .9\TIKZtextheight}{%
	\begin{tabular}{c@{}c@{}c@{}c@{}c}
	\bnREAD{0}{}
	&	\bnREAD{1}{}
	&	\bnZIGCOPYUNIT{0}{}
	&	\bnZIGCOPYUNIT{1}{}
	&	\bnHALT{}
	\\
	\ReadZBlockUZnQLTPUexploded{}
	&	\ReadUBlockUZnQLTPUexploded{}
	&	\ZigCopyZBlockUZnQLTPUexploded{}
	&	\ZigCopyUBlockUZnQLTPUexploded{}
	&	\HaltBlockEpsilonnQLTPUexploded{}
	\\[1mm]	
	\multicolumn{5}{c}{
		\begin{tabular}{@{}c@{\hspace*{-1cm}}c@{}}
		\raisebox{-1cm}[0cm][0cm]{\bnAPPENDRETURN{}}
		& 	\raisebox{-3cm}[0cm][0cm]{~\bnCOPYLINEFEEDUNIT{}{}{}}
		\\[-8mm]
		\AppendBlockUZnQLTPUexploded{}
		&  \ZagCopyUBlockUZnQLTPUexploded{}
		\end{tabular}
	}
	\\
	\end{tabular}
}
\caption{Exploded view of the bricks and modules inside the blocks involved in the simulation above.}
\label{fig:block:exploded}
\end{subfigure}
\caption{Folding of the transcript simulating the STCS \TME, and some block internal structures.}
\label{fig:block:example}
\end{figure}

%% file: design-toolbox.tex

	\label{sec:toolbox}

In this section, we present several key tools to program Oritatami design and which we believe to be generic as they allowed us to get a lot of freedom in our design.

%
%

\subsection{Implementing the logic}

As in \cite{GeMeScSe2016}, the internal state of our ``molecular computing machinery'' consists essentially of two parameters: 1) the \emph{position inside the transcript} of the part currently folding; and 2) the \emph{entry point} of transcript inside the environment. Indeed, depending on the entry point or the position inside the transcript, different beads will be in contact with the environment and thus different \emph{functions} will be applied as a result of their interactions. 
This happens during the zig phase: in the first (zig-up) part, the transcript starts folding at the bottom, forcing the modules $\ModuleG$ to fold into $\GRead{}$ bricks; whereas during the second (zig-down) part, the transcript starts folding at the top, forcing  the modules $\ModuleG$ to fold into $\GZigCopy{}$ bricks instead. 
Similarly, the \emph{memory} of the system consists of the beads already placed on the surrounding of the area currently visited (the \emph{environment}). This happens in every row of the folding: depending on the letter encoded at the bottom of the row above, the modules $\ModuleG$ fold into $\GReadZ$ or $\GReadU$ bricks (zig-up phase),  $\GZigCopyZ$ or $\GZigCopyU$ bricks (zig-down phase), and $\GZagCopyZ$ or $\GZagCopyU$ bricks (zag phase).

At different places, we need the transcript to read information from the environment and trigger the appropriate folding. This is obtained through different mechanisms.

\begin{figure}
	\begin{subfigure}{.32\textwidth}
	\center
	\includegraphics[scale=.3, trim={12cm 5cm 10cm 7cm},clip]{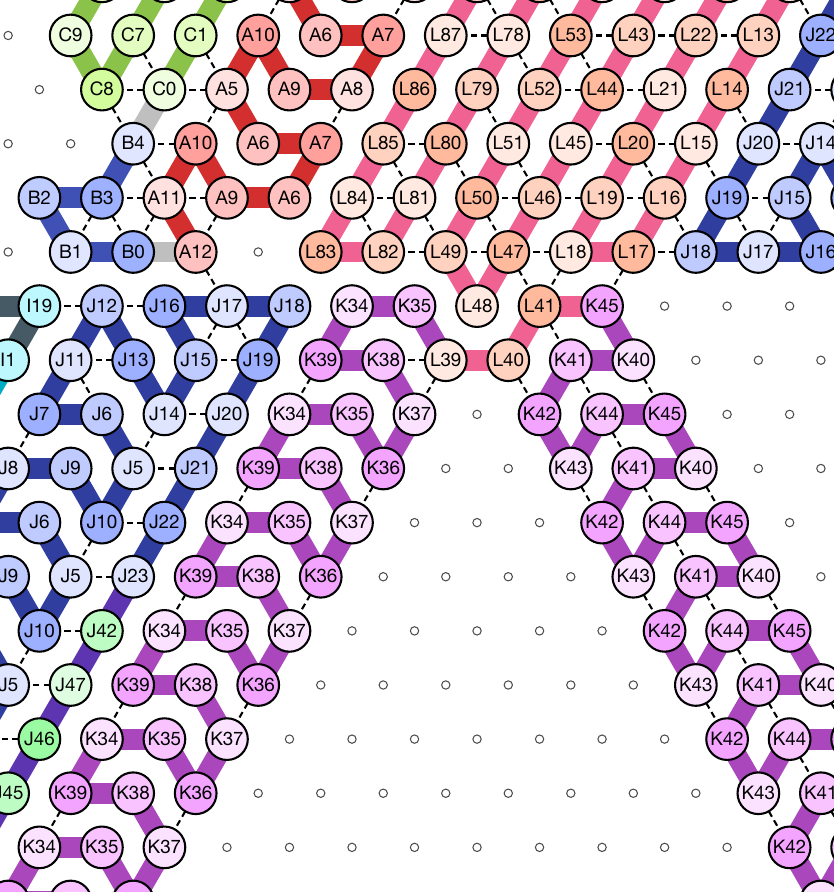}
	\caption{$\ModuleG$ bounces southeastward in presence of a spike encoding a \word0 and folds into $\GReadZ$.}
	\label{fig:G:bounce:read0}
	\end{subfigure}
	\hfill
	\begin{subfigure}{.32\textwidth}
	\center
	\includegraphics[scale=.3, trim={10cm 5cm 9cm 7cm},clip]{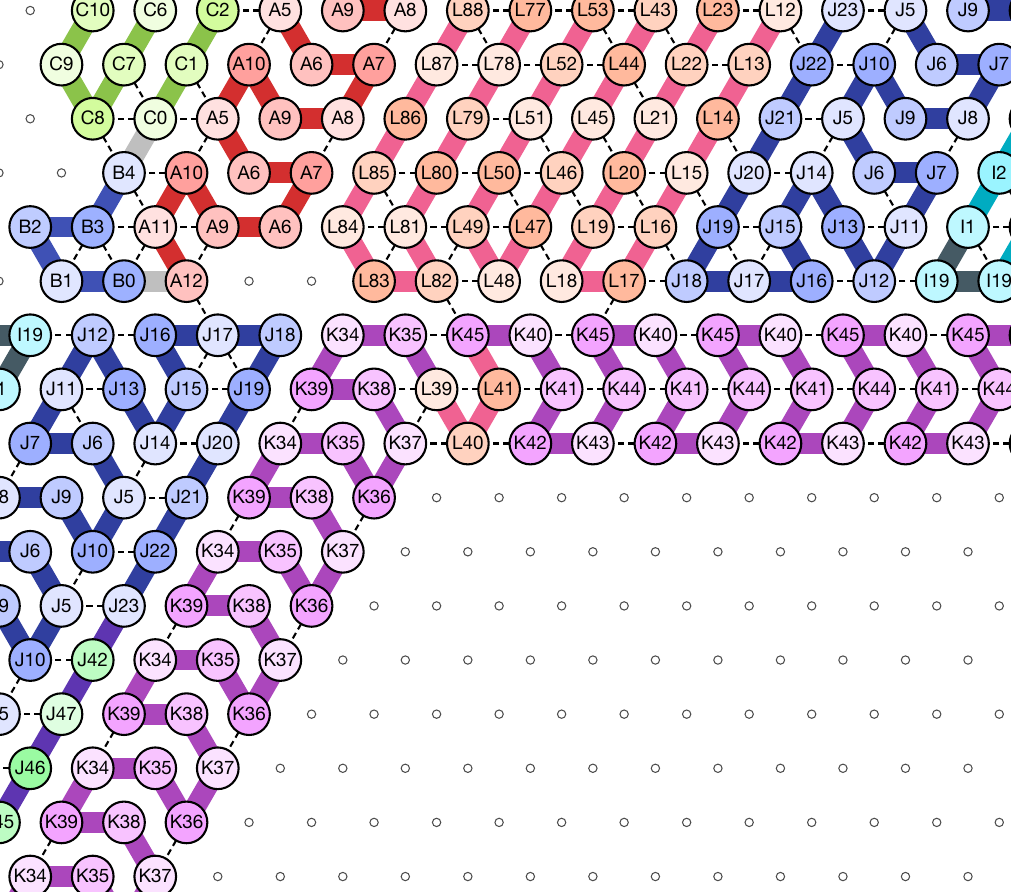}
	\caption{$\ModuleG$ bounces eastward on a flat surface encoding a \word1, and folds into $\GReadU$.}
	\label{fig:G:bounce:read1}
	\end{subfigure}
	\hfill
	\begin{subfigure}{.32\textwidth}
	\center
	\includegraphics[scale=.3, trim={12cm 7cm 9cm 5cm},clip]{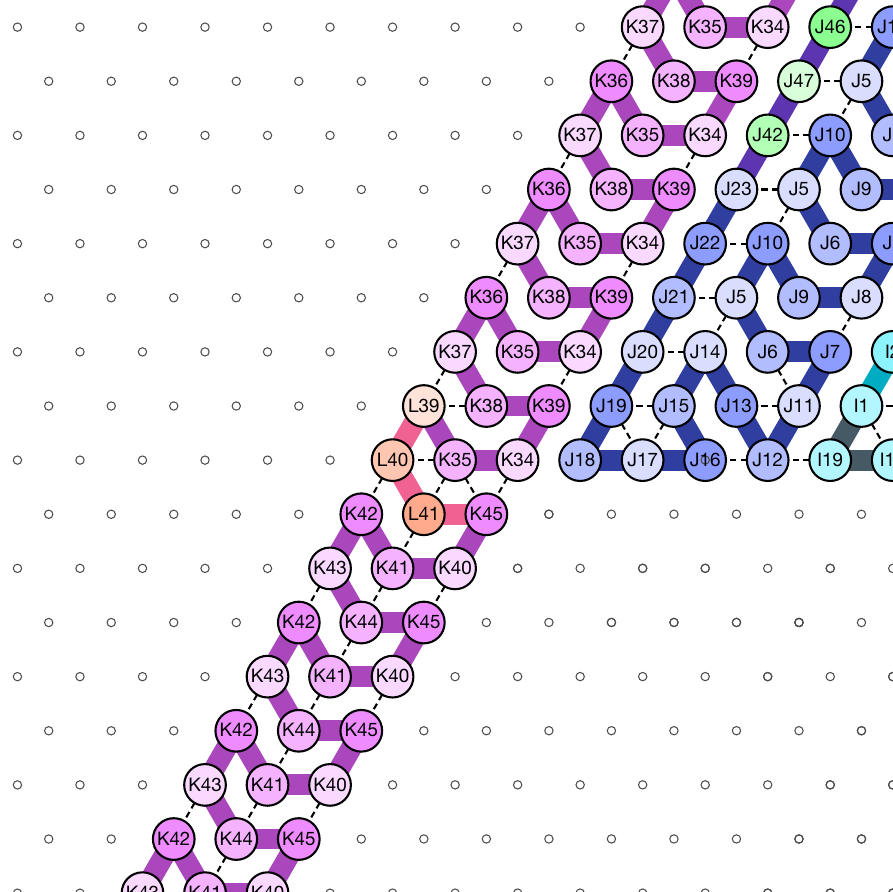}
	\caption{$\ModuleG$ goes straight southwestward in absence of obstacle, and folds into $\GLF$.}
	\label{fig:G:bounce:LF}
	\end{subfigure}
	\caption{The interactions of module $\ModuleG$ in ``glider''-mode with different environments result in heading to different entry points to the next area of the folding.}
	\label{fig:G:bounce}
\end{figure}

\begin{description}
\item[Default folding.] By default, during the zig-up phase, $\ModuleB$ is attracted to the left by $\ModuleF$ and folds to the right only in presence of $\ModuleEa{}$ above. This allows to continue the folding only if the tape word is not empty or to halt it otherwise (see Figure~\ref{fig:B:brick:ZigUp} in appendix).
\item[Geometry obstruction.] An typical example is illustrated by $\ModuleG$. During the zig-up phase where the absence of environment  below the block $\bnREAD{}{}$ allows $\ModuleG$ to fold downward at the beginning (see Figure~\ref{fig:G:brick:Read0}) which shift the transcript by $7$ beads along $\ModuleF$ resulting in $\ModuleG$ to adopt the glider-shape (more details on this mechanism in the next section). Whereas during the zig-down phase, $\ModuleG$ cannot make this loop because it is occupied by a previously placed $\ModuleG$. This results in a perfect alignment of $\ModuleG$ with $\ModuleF$ whose strong attraction forces $\ModuleG$ to adopt the switchback shape (see Figure~\ref{fig:G:brick:ZigCopy0}).     
\item[Geometry of the environment.]  Figure~\ref{fig:G:bounce} shows how the shape of the environment is used to change the direction of $\ModuleG$ in glider-shape. This results in modifying the entry point in the environment and allows the Oritatami system to trim the leading $\word0$s in the tape word by going back to the same entry point (Fig.~\ref{fig:G:bounce:read0}), switch from zip-up to zig-down phase when reading a $\word1$ by opening the next block from the top (Fig.~\ref{fig:G:bounce:read1}), and from zag to zig-up phase when it has rewind to the beginning of the tape word, by getting down to the bottom of the next zig row (Fig.~\ref{fig:G:bounce:LF}).
\end{description}

\subsection{Easing the design: getting the freedom you need}

Several key tools allowed to ease considerably our design, and even in some cases to make it feasible. These tools are generic enough to be considered as \emph{programming paradigms}. One main difficulty we faced is that the different functions one wants to implement tend to concentrate at the same ``hot-spots'' in the transcript. A typical example is the midpoint of $\ModuleG$ where one wants to implement all the functions:
Read, Copy and Line Feed. The following powerful tools allow to overcome these difficulties:

\label{separating-functions}
\begin{description}
\item[Socks] work by letting a glider/switchback module fold into a switchback turn conformation for some time when it would otherwise fold into a glider. Examples are given in Figure~\ref{fig:socks}. They are easy to implement: indeed, the socks naturally adopt the same shape as the corresponding switchback turn and require thus \emph{no extra interfering bonds}. They allow a lot of freedom in the design, for several reasons:
  \begin{itemize}
  \item First, they simplify the design of important switchback part by \emph{lifting the need for implementing the glider configuration} for that part, as shown in Figure~\ref{fig:socks:ease}.
  \item Second, a glider naturally progresses at speed $1/3$. Adding a sock allows us to \emph{slow its progression down} to speed $1/5$ for some time (see Fig.~\ref{fig:socks:delay}) and therefore to realign them.
    We used that feature repeatedly to ``shift'' some modules: starting the folding at an initial speed-$1$ (i.e. straight line) and then compensating for that speed later on by introducing socks (see Fig.~\ref{fig:socks:delay}).
    This is a key point in our design, as it allowed us to \emph{spread apart} the Read and Copy functions into different subsequences of module $\ModuleG$, and therefore to get less constraints on our rule design. In the specific case of module \ModuleG, the Copy-function occurs at the center of the module, while the Read-function is implemented earlier in module! (see section~\ref{sec:full:description:G} for full details)
  \item Finally, socks allow to prevent unwanted interactions between beads by \emph{concealing} potentially harmful beads in unreachable area as in Figure~\ref{fig:socks:unwanted}.
  \end{itemize}
  \begin{figure}
	\begin{subfigure}{.32\textwidth}
	\center
	\includegraphics[scale=.25, trim={19.5cm 9.5cm 13.5cm 12cm},clip]{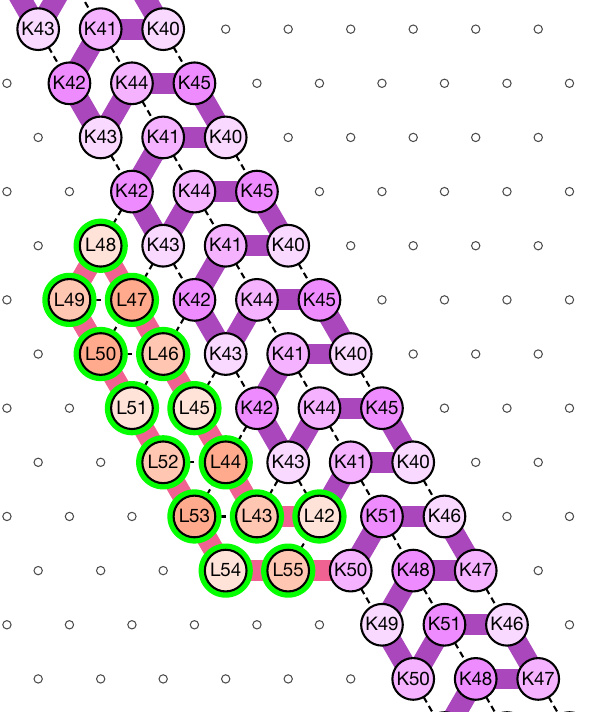}
	\includegraphics[scale=.25, trim={15cm 13cm 19cm 8.5cm},clip]{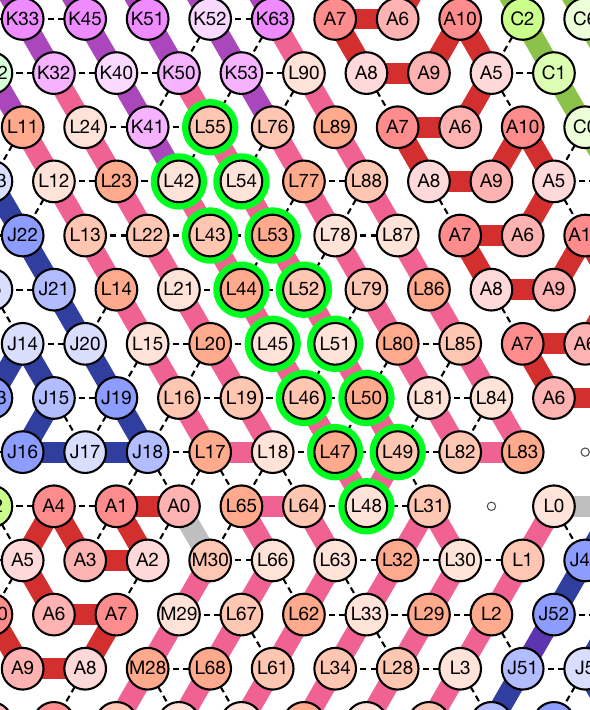}
	\caption{Easing the design of switchback/glider by letting the switchback (in green) folds in its natural shape at its extremities even in ``glider''-mode.}
	\label{fig:socks:ease}
	\end{subfigure}
	\hfill
	\begin{subfigure}{.32\textwidth}
	\center
	\includegraphics[width=\textwidth]{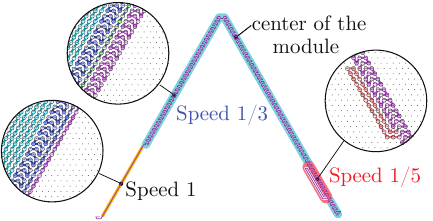}
	\caption{Module $\ModuleG$: Realigning a pattern by slowing its folding down at the end to compensate speeding it at the beginning.}
	\label{fig:socks:delay}
	\end{subfigure}
	\hfill
	\begin{subfigure}{.32\textwidth}
	\center
	\includegraphics[scale=.3, trim={5.5cm 9.5cm 13cm 3.5cm},clip]{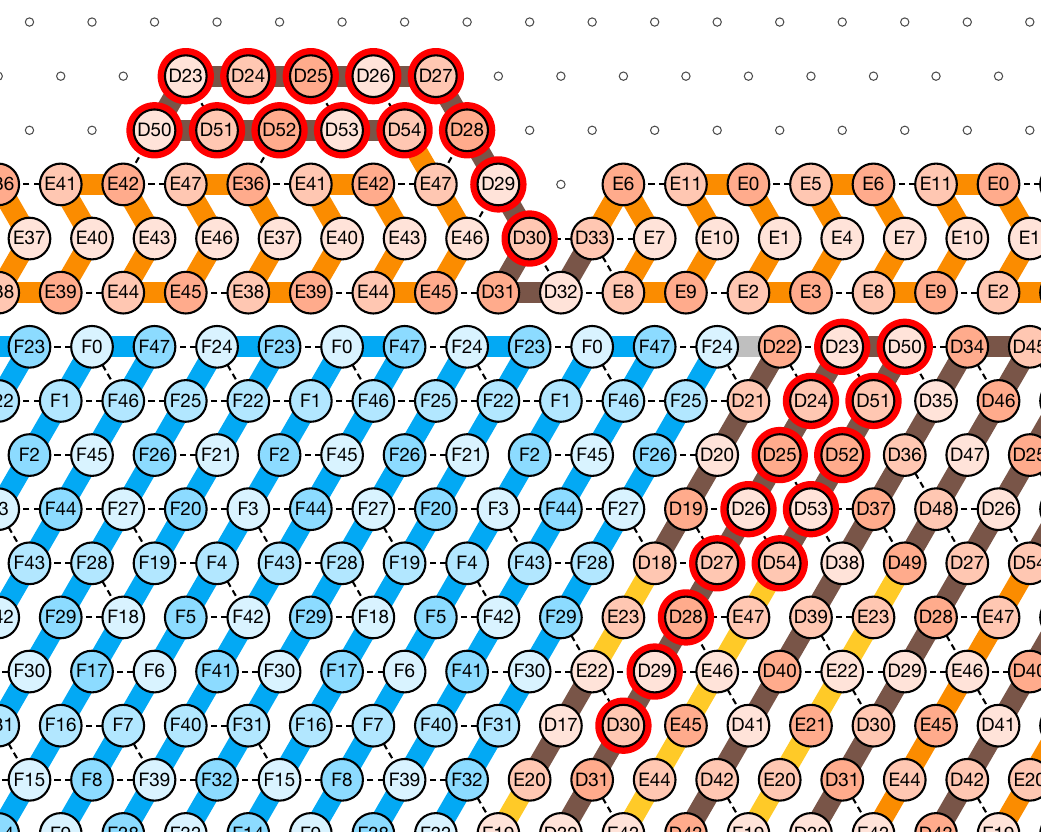}
	\caption{Module $\ModuleDx{}$: Preventing unwanted interactions between the beads outlined in red and the ones below, by concealing the red ones on top of the glider.}
	\label{fig:socks:unwanted}
	\end{subfigure}
\caption{Different uses of socks: (a) Easing bond design; (b) Delaying; (c)  Preventing unwanted interactions.} 
\label{fig:socks}
\end{figure}

\item[Exponential bead type coloring] is a key tool to allow module $\ModuleG$ to fold into different shapes, glider or switchback, along module $\ModuleF$, when folding in the $\bnREAD{}{}$ configuration. The problem it solves is that in order for $\ModuleG$ to fold into switchbacks, we need strong interactions between $\ModuleG$ and its neighboring module $\ModuleF$ (see Fig.~\ref{fig:G:brick:Read0}), whereas in order for $\ModuleG$ to fold as glider, we want to avoid those interactions (see Fig.~\ref{fig:G:brick:ZigCopy0}). This is made possible because gliders progress at speed $1/3$ while switchbacks progress at speed~$1$. Using a power-of-3 coloring, we manage to easily achieve these contradicting goals altogether (the construction is analysed in Lemma~\ref{lem:exp:col} in Section~\ref{sec:enum:env}).
\end{description}

%% file: omitted.tex

\noindent Please find next the omitted part of article due to space constraints.

\section{Skipping Cyclic Tag Systems}
 
\subparagraph{Notations.}
We index the letters of every word $u=u_0\ldots u_{|u|-1}$ from $0$ to $|u|-1$. Given two words $u$ and $v$, we denote by $u\cdot v$ their concatenation: ${u\cdot v = u_0\ldots u_{|u|-1} v_0\ldots v_{|v|-1}}$. We denote by $u^\infty$ the oneway infinite periodic word $u\cdot u\cdots$. For all $i\leq j$, we denote by $u_{i..j}$ the (possibly empty) factor $u_{\max(0,i)}\ldots u_{\min(j,|u|-1)}$. The empty word is denoted by~$\epsilon$. The indices in the notation $O_{L}()$ where $L$ is a list of variables (for instance $L=A,B$) indicates that the constant in the $O()$ only depends on the variable in $L$ (for instance $A$ and $B$) and on no other values.

\subsection{Trimmed diagram} 
 
The following lemma gives the formal description of the trimmed diagram of a SCTS $(\alpha; u^0)$ with marker $m^t$. Recall that $t_i$ is the $i$-th time $t$ such that the dataword $u^t$ starts with letter $\word1$ ($t_0 = -1$ by convention).

\begin{lemma} \label{lem:SCTS:diagram}
 The annotated word on the row~$i$ (indexed from $i=0$) of the trimmed diagram is: (the markers in exponent are computed modulo $n$)
\begin{description}[topsep=0pt]
\item[\labelitemi~ if $u^{1+t_i} = \word0^r\word1\cdot s$ for some $r\geq0$ and $s\in\letterSet^*$:] then, $r={t_{i+1}-t_i-1}$ and the annotated word on row~$i$ is $\pQ{i+1+t_i}\word0\cdots\pQ{i-1+t_{i+1}}\word0\pQ{i+t_{i+1}}\word1\cdot s$ whose first letter is placed  in column $t_i+1$ (assuming the leftmost column is indexed by $0$); 
\item[\labelitemi~ if $u^{1+t_i} = \word0^r$ for some $r>0$:] then, row~$i$ is the last row of the diagram and its annotated word is $\pQ{i+1+t_i}\word0\cdots\pQ{i+t_i+r}\word0\pQ{i+t_i+r+1}$ and starts at column $t_i+1$.
\end{description}
\end{lemma}

\begin{proof}[Proof sketch]
Simply observe that ${m^{t_i} = i+t_i\mod n}$, as indeed exactly $t_i$ letters have been read and exactly $i$ appending steps have occurred before reading the $i$-th \word1. 
\end{proof}

\subsection{Turing-universality of Skipping Cyclic Tag Systems}

This proof makes use of the time-efficient reduction from Turing machines to cyclic tag systems (CTS) 
designed in \cite[Theorem~4.3.2 p.~65]{NearyPhD}, improving on \cite{Cook2004,WoodsNeary2006}.
\subparagraph{A cyclic tag system} $\TMC = (\alpha^0,\ldots,\alpha^{n-1};v^0)$ consists of a list of $n$ appendants $\alpha^0,\ldots,\alpha^{n-1}\in\letterSet^*$ and an initial dataword $v^0\in\letterSet^*$. Its configuration at time $t$ consists of a \emph{marker} $m^t=t\mod n$, recording the index of the current appendant at time~$t$, and a dataword $v^t$.  Initially, $m^0 = 0$ and the dataword is $v^0$. At each time step~$t$, the CTS acts deterministically on configuration $(m^t,v^t)$ in one of three ways:
\begin{description}[topsep=2pt,itemsep=-1pt]
\item[\mdseries (Halt step) If $v^t$ is the empty word $\epsilon$,] then the CTS halts;
%
%
\item[\mdseries (Nop step) If the first letter $v^t_0$ of $v^t$ is $\word 0$,] then $v^t_0$ is deleted and the marker moves to the next appendant cyclically: i.e., $m^{t+1} = (m^t+1) \mod n$ and ${v^{t+1} = v^t_1\cdots v^t_{|v^t|-1}}$; 
\item[\mdseries (Append step) If $v^t_0 = \word1$,] then $v^t_0$ is deleted, the currently marked appendant  $\alpha^{(m^t~\mod~n)}$ is appended onto the right end of $v^t$: i.e., $v^{t+1} = v^t_1\cdots v^t_{|v^t|-1}\cdot \alpha^{(m^t~\mod~n)}$ and ${m^{t+1} = (m^t+1)\mod n}$ (no skipping). 
\end{description}

According to the definition in \cite{NearyPhD, Cook2004,WoodsNeary2006}, the computation of a CTS is said to end if either the dataword is $\epsilon$ or if it repeats a configuration. This relaxed definition of termination was introduced for the purpose of reducing any Turing machine to cellular automaton rule 110, whose computation never halts.  Precisely, \cite[Theorem~4.3.2, p.~65]{NearyPhD} states the following:  let $\TMM$ be any deterministic Turing machine using a single tape; there is a cyclic tag system $\TMC_\TMM$ with appendants $\alpha^0,\ldots,\alpha^{n-1}$ and a linear-time encoding $v_\TMM$ of the input $x$ of $\TMM$, such that for all input $x$: (1) $\TMC_\TMM$ halts from initial dataword $v_\TMM(x)$ if and only if $\TMM$ halts from input $x$; and (2) for all $t$, if $\TMM$ halts after $t$ steps on $x$, then $\TMT_\TMM$ halts after $O_\TMM(t^2\log t)$ steps on $v_\TMM(x)$. For our purpose, we need the computation of the CTS to \emph{stop with an empty dataword} (and not to enter a cycle) if the simulated Turing machine stops, precisely: 

\begin{lemma}[Corollary of Theorem~4.3.2 in \cite{NearyPhD}]
For every Turing machine \TMM, there is CTS $\hat\TMC_\TMM$ and a linear time encoding $\hat v_\TMM(x) = \homo(v_\TMM(x))$ that encodes any input $x$ of \TMM\/ into an initial dataword $\hat v^0 = \hat v_\TMM(x)$ such that: (1) $\hat\TMC_\TMM$ halts from $\hat v^0$ \empty{with an empty dataword} iff $\TMM$ halts from input $x$; and (2) if \TMM\/ halts after $t$ steps from $x$, then  $\hat\TMC_\TMM$ after $O(t^2\log t)$ steps from $\hat v^0$. 
\end{lemma}

\begin{proof}
We proceed as follows by defining a CTS $\hat\TMC_\TMM$ ``with two processing modes'': the first mode emulates $\TMC_\TMM$, the second mode just erases the data word; switching from one mode to the other just requires inserting a single letter \word0 in the dataword. Consider $\homo$ the homomorphism on $\letterSet^*$ such that $\homo(\word0) = \word{00}$ and $\homo(\word1) = \word{01}$, i.e. $\homo$ inserts a \word0 before every letter of a word.  Then, consider the CTS $\hat\TMC_\TMM$ with $2n$ appendants: $\hat\alpha^{2i+1} = \homo(\hat\alpha^i)$ and ${\hat\alpha}_\TMM^{2i} = \epsilon$, for $0\leq i < n$. An immediate induction shows that $\hat\TMC_\TMM$ simulates $\TMC_\TMM$ exactly twice slower, indeed: if $v^t$ and ${\hat v}^t$ denote the datawords of $\TMC_\TMM$ and $\hat\TMC_\TMM$ starting from the initial datawords $v^0$ and ${\hat v^0} = \homo(v^0)$ respectively, then for all time $t$, $\hat v^{2t} = \homo(v^t)$.  Now, if we shift the dataword of $\hat\TMC_\TMM$ by one letter, the appendants that will be appended next, are all the empty word $\epsilon$, and the dataword will be completely erased, yielding to the desired terminaison. Without loss of generality, we assume that $\TMM$ has an unique final state $q_F$. According to the design of $\TMC_\TMM$ in the proof Theorem~4.3.2 p.~65 in \cite{NearyPhD}, for every step $t$ of \TMM\/ where the configuration  is $\cdots B\cdots B \sigma_{1}\ldots \sigma_{j}[q,d]\sigma_{j+1}\sigma_{j+2}\cdots \sigma_{s-2}B\cdots B\cdots$ (i.e., where the head is over position $\sigma_{j+1}$, the current state is $q$ and the next head movement is $d$, $B$ denotes the blank symbol), the dataword of $\TMC_\TMM$ is,  at the step $O(t^2\log t)$ corresponding to first stage of the processing of this configuration by $\TMC_\TMM$:
$$
\tuple{1,q,d}\tuple{\sigma_{j+1}}\cdots\tuple{\sigma_{s-2}}\tuple{\sigma_{B}}\mu^{s'}\tuple{\sigma_{1}}\cdots\tuple{\sigma_{j}}
$$
and the marker is $0$. Each $\tuple{\cdot}$ and $\mu$ stands for a binary encoding containing a single \word1. Furthermore, there is at most one pattern $\tuple{1,q,d}$ in the dataword of $\TMC_\TMM$ at all time. Let $i$ be the index of the only \word1 in $\tuple{1,q_F,d}$. We then change the appendant $\hat\alpha^{2i+1}$ to $\word0$.  It follows that, the first time the pattern $\chi(\tuple{1,q_F,d})$ appears, i.e. the first time the simulated Turing machine \TMM\/ enters the final state, the CTS $\hat\TMC_\TMM$ switches to the even-indexed empty-appendants-mode, then erases the whole dataword, and halts with an empty dataword, as desired. Furthermore, if \TMM\/ halts after $t$ steps, $\hat\TMC_\TMM$ halts after $O(t^2\log t)$ steps. 
\end{proof}

We now show how to simulate the CTS $\hat\TMC_\TMM$ with a SCTS. 

\begin{omittedproof}{Proposition}{prop:TM>SCTS}
 The original cyclic tag system by Cook \cite{Cook2004} differs from the skipping cyclic tag system only in that in the original, the list rotates by 1 no matter which letter the current word begins with. Consider the CTS $\hat\TMC_\TMM$, given by the lemma above, with $2n$ appendants $\hat\alpha^0,\ldots,\hat\alpha^{2n-1}$,  together with its linear-time input encoding $\hat v_\TMM$. Let $\homo'$ be the homomorphism over $\letterSet^*$ defined as $\homo'(\word 0) = \word{00}$ and $\homo'(\word1) = \word1$.  Let $\TMS_\TMM$ be the SCTS with $4n$ appendants: ${\beta^{2i} = \epsilon}$ and ${\beta^{2i+1} = \homo'(\hat\alpha^i)}$ for $0\leq i< 2n$. An immediate recurrence shows that $\TMS_\TMM$  simulates $\hat\TMC_\TMM$, precisely: if $v^t$ and $u^t$ denote respectively the datawords of $\hat\TMC_\TMM$  and $\TMS_\TMM$ with initial datawords $v^0\in\letterSet^*$ and $u^0 = \homo'(v^0)$ then, for all time $t$, $u^{t+r_t} = v^t$ where $r_t$ is the number of \word0s read by $\hat\TMC_\TMM$ up to time $t$ (note that $r_t\leq t$). Let $u_\TMM(x) =  \homo'(\hat v_\TMM(x))$ denote the linear time encoding of the input $x$ of \TMM\/ as the initial dataword of $\TMS_\TMM$. It follows that (1) $\TMS_\TMM$ halts from input dataword $u_\TMM(x)$ iff $\TMM$ halts from input $x$; and (2) if $\TMM$ halts from input $x$ after $t$ steps, then $\TMS_\TMM$ halts from $u_\TMM(x)$ with an empty dataword after $O(t^2\log t)$ steps. Note that moreover, the number of appendants of $\TMS_\TMM$ is a multiple of $4$.
\end{omittedproof}

\section{Proof of main Theorem~\ref{thm:main} as a consequence of key Theorem~\ref{thm:SCTS>oritatami}}

The remaining of this article is dedicated to prove the following theorem which implies Theorem~\ref{thm:main} by the proposition above (see appendix \vpageref{proof:thm:main}).

\begin{theorem}[Key theorem]
\label{thm:SCTS>oritatami}
\label{thm:key}
There is a fixed set $B$ of 542 bead types and a fixed attraction rule~$\heart$ on $B$ together with two polynomial-time encodings:
\begin{itemize}[topsep=2pt]
\item $\pi$ that maps any SCTS $\TMS$ with $n\geq 8$ appendants $\alpha = \tuple{\alpha^0,\ldots,\alpha^{n-1}}$ where $n$ is a multiple of~$4$, to a bead-type sequence $\pi_\TMS \in B^*$ of exact length:\\[1mm]
\centerline{$|\pi_\TMS| = 18Kn(Kn+12n-8)+3n(192n-171)+30\sum_{i=0}^{n-1}|\alpha^i| = O(|\alpha|^4)$
\hspace*{2mm}}\\[1mm]
where $K = L + 12 - (L\mod 2)\leq L+12$ with $L = \max_i |\alpha^i|$ being the length of the longest appendant, and $|\alpha|=\sum_{i=0}^{n-1}|\alpha^i|$. Note that $\pi_\TMS$ only depends on the appendants of~$\TMS$.   
\item $(s_\TMS,\sigma_\TMS)$ that maps any input dataword $u$ of $\TMS$ to a seed configuration $\sigma_\TMS(u)$ of a bead type sequence $s_\TMS(u)$ of length $O_\TMS(|u|)$, precisely:\\[1mm]
\centerline{$|\sigma_\TMS(u)| = 2|u|(3K+16) + |u|_{\word0} + 9K(n-1)+36n-21 = O_\TMS(|u|)$
\hspace*{2mm}}\\[1mm]
where $|u|_{\word0} = \#\{i:u_i=\word0\}$ is the number of $\word0$s in $u$,   
\end{itemize}
such that: For any SCTS $\TMS$ with $n\geq 8$ appendants, where $n$ is a multiple of~$4$, and every input dataword $u$ of \TMS, the deterministic and periodic oritatami system $\TMO_\TMS = ((\pi_\TMS)^\infty, \heart, 3)$ with bead type sequence $(\pi_\TMS)^\infty$ and delay $\delta = 3$, halts when folding from seed configuration $\sigma_\TMS(u)$ if and only if $\TMS$ halts on input dataword $u$. Furthermore, for all $t$, if $\TMS$ halts on $u$ after $t$ steps, then the folding $\TMO_\TMS$ from seed configuration $\sigma_\TMS(s)$ halts after folding $O_\TMS(t^2)$ beads. 
\end{theorem}

Note that requiring that $n\geq 8$ and $n$ being a multiple of $4$ does not restrict this result. Indeed, repeating the appendants sequence $k$ times in a SCTS, yields a strictly identical SCTS with $k$ times the number of appendants. These requirements are however necessary to ensure the proper folding alignment in the design of our oritatami system.

\begin{omittedproof}{Theorem}{thm:main}
Consider a universal Turing machine $\TMM$ and the skipping cyclic tag system $\TMS_\TMM$ provided by Proposition~\ref{prop:TM>SCTS} together with its linear-time input encoder $u_\TMM$. Consider the set of 542 bead types $B$, the rule $\heart$, the oritatami system $\TMO_\TMM=((\pi_\TMM)^\infty,\heart,3)$ whose primary structure has period $\pi_\TMM = \pi_{\TMS_\TMM}$, and the linear-time seed encodings $(s_{\TMS_\TMM},\sigma_{\TMS_\TMM})$ provided by Theorem~\ref{thm:SCTS>oritatami} when applied to $\TMS_\TMM$. Let us define for short $s_\TMM(x) = s_{\TMS_\TMM}(u_\TMM(x))$ and $\sigma_\TMM(x) = \sigma_{\TMS_\TMM}(u_\TMM(x))$, the seed bead types sequence and the seed conformation of $\TMO_\TMM$ corresponding to the input $x$ of~$\TMM$.  Then, by construction:
\begin{enumerate}
\item For all input $x$ of $\TMM$,  $\TMM$ halts on input $x$, if and only if $\TMS_\TMM$ halts on input dataword $u_\TMM(x)$, if and only if $\TMO_\TMM$ halts its folding from seed conformation $\sigma_\TMM(x) = \sigma_{\TMS_\TMM}(u_\TMM(x))$.
\item For all input $x$ of $\TMM$ and all time~$t$, if $\TMM$ halts on $x$ after $t$ steps, then $\TMS_\TMM$ halts on input dataword $u_\TMM(x)$ after $T=O_\TMM(t^2\log t)$ steps, and thus $\TMO_\TMM$ halts its folding from seed conformation $\sigma_\TMM(x)$ after $O_{\TMS_\TMM}(T^2) = O_\TMM(t^4\log^2 t)$ steps.
\item For all input $x$ of $\TMM$, the length of the seed conformation encoding $x$ in $\TMO_\TMM$ is $O_{\TMS_\TMM}(|u_\TMM(x)|) =O_\TMM(|x|)$, linear in $|x|$.
\item Finally, the oritatami system $\TMO_\TMM$ and seed encoding $(s_\TMM,\sigma_\TMM)$ are obtained in polynomial time from the skipping tag system $\TMS_\TMM$, which is also obtained in polynomial time from $\TMM$. The reduction is thus computed in polynomial time.
\end{enumerate}

\end{omittedproof}

%% file: all-bricks.tex

{
\newcommand{\scaleBrick}{.3} 
\newcommand{\myN}{4}
\newcommand{\myL}{3}
\newcommand{\myP}{1}
\newcommand{\nameBrickFile}[2]{Module-#1-#2_n=\myN-L=\myL-P=\myP-arrow.pdf}
\newcommand{\newfigBrick}[2]{\includegraphics{\nameBrickFile{#1}{#2}}}
\newcommand{\scaledbrickfig}[2]{\includegraphics[scale=\scaleBrick]{\nameBrickFile{#1}{#2}}}
\newcommand{\scaledbrickfigOptions}[3]{\includegraphics[scale=\scaleBrick,#3]{\nameBrickFile{#1}{#2}}}
%

\section{Folding paths of all the bricks of our design}
\label{sec:all:bricks:paths}

This section presents  all the bricks, i.e. all the folding paths of the 7 modules (i.e. subsequences of the transcript) composing each unit of the transcript. The folding of each module into one of these bricks depending on the context, is the key to the correctness of the folding of our transcript design into the shape of the trimmed diagram the simulated STCS as stated in Lemma~\ref{lem:key}. This section just presents the shape of each brick for each module together with an illustration to scale. Its purpose is to provide a guideline to the description of the blocks in the next section. The full description of each brick will be given in section~\ref{sec:all:bricks:full}.

To ease the reading, the brick name contains a specific symbol indicating to which kind of phase of the folding the brick belongs:
\begin{itemize}
\item Zig-up bricks are annotated by $\ZigUp$
\item Zig-down bricks are annotated by $\ZigDown$
\item Appending bricks are annotated by $\Append$
\item Carriage-return bricks (spanning from a zig row to the next zag row)  are annoted by $\bZigZag$
\item Zag bricks are annotated by $\Zag$
\item Line-feed bricks (spanning from a zag row to the next zig row)  are anoted by $\bZagZig$
\item Halting bricks are annotated by $\Halt$
\end{itemize}

The subsequences corresponding to each of the 7 modules are referred by $\ModuleA$, $\ModuleB$, $\ModuleC$, $\ModuleDxrt{$(x)$}{r,t}$, $\ModuleEa{a}$, $\ModuleF$, $\ModuleG$. Some modules ($\ModuleDx{}$ and $\ModuleEa{}$) have  parameters that will be explained later, in the next sections. Their bricks, i.e. their possible folding depending on the context are referred by:

\begin{description}[topsep=2mm,itemsep=2mm]
\item[\ModA\/ (see section~\ref{sec:A:bricks}):] \AZigUp (zig-up), \AZigDown (zig-down) and \AZag (zag).
\item[\ModB\/ (see section~\ref{sec:B:bricks}):]  \BZigUp (zig-up), \BZigDown (zig-down), \BZag (zag) and \BHalt (halt).
\item[\ModC\/ (see section~\ref{sec:C:bricks}):]  \CZigUp (zig-up), \CZigDown (zig-down), \CZag (zag) and \CEnd (zig-up, enf of dataword detected).
\item[\ModDxrt{$(x)$}{r,t}\/ (see section~\ref{sec:D:bricks}):]  \DxrtZigUp{$(x)$}{r,t} (zig-up), \DxrtZigDown{$(x)$}{r,t} (zig-down), {\DxrtZag{$(x)$}{r,t}}~(zag) and \DxrtAppend{$(x)$}{r,t} (append).
\item[\ModEa{a}\/ (see section~\ref{sec:E:bricks}):]  \EaZigUp{a} (zig-up), \EaZigDown{a} (zig-down), \EaZag{a} (zag) and {\EaCR{a}}~(carriage return).
\item[\ModF (see section~\ref{sec:F:bricks}):]  \FZigUp (zig-up), \FZigDown (zig-down) and \FZag (zag).
\item[\ModG\/ (see section~\ref{sec:G:bricks}):]  \GReadZ (read \word0), \GReadU (read \word1), \GZigCopyZ (zig-down), \GZigCopyU (zig-down), \GZagCopyZ (zag), \GZagCopyU (zag),  and \GLF (line feed).
\end{description}

In the following figures, a lighter and a darker grey arrow indicates the beginning and the end of the folding path of each brick respectively. The parameter $h$ will be defined later and refers to the height of the blocks composing the folded shape of our design. 

\subsection{All bricks for Module A}
\label{sec:A:bricks}

Module \ModuleA\/ always folds as a glider of height $h$ and width $3$, pointing to NE in zig-up phase, SE in zig-down phase and SW in zag phase. The folding of module $\ModuleA$ serves as a scaffold for the folding of the next modules in the zig-up and zig-down phases.

All its possible bricks are displayed in Fig.~\vref{fig:A:all:bricks}.

\begin{figure}[h]
	\begin{subfigure}{.49\textwidth}
	\scaledbrickfig{A}{ZigUp}
	\caption{The brick $\AZigUp$.}
	\end{subfigure}
	\quad
	\begin{subfigure}{.49\textwidth}
	\scaledbrickfig{A}{ZigDown}
	\caption{The brick $\AZigDown = \mirror(\AZigUp)$.}
	\end{subfigure}
	\\[5mm]
	\centerline{
	\begin{subfigure}{.49\textwidth}
	\scaledbrickfig{A}{Zag}
	\caption{The brick $\AZag = \rot(\AZigUp)$.}
	\end{subfigure}}
\caption{Folding paths to scale of all the bricks for Module $\ModuleA$ (see section~\ref{sec:full:description:A} for full description).}
\label{fig:A:all:bricks}
\end{figure}

\afterpage{\clearpage}
\newpage

\subsection{All bricks for Module B}
\label{sec:B:bricks}

Module \ModuleB\/ is 5 beads long. It folds along the preceding brick of Module \ModuleA\/:
\begin{itemize}
\item to the right in zig-down; 
\item to the left in zag phase; 
\item to right in zig-up phase if the dataword encoded in the zag-row above is not empty;
\item  but to the left in the zig-up phase if the dataword encoded above is empty (terminating the folding as it is now trapped in a closed area).
\end{itemize}
All its possible bricks are displayed in Fig.~\vref{fig:B:all:bricks}. 
Note that $\BZigDown = \mirror(\BZigUp)$ and $\BZag = \rot(\BZigUp)$.

\begin{figure}[h]
	\center
	\begin{subfigure}[b]{.21\textwidth}
	\scaledbrickfig{B}{ZigUp}
	\caption{The brick $\BZigUp$.}
	\end{subfigure}
	\quad
	\begin{subfigure}[b]{.21\textwidth}
	\scaledbrickfig{B}{ZigDown}
	\caption{The brick $\BZigDown$.}
	\end{subfigure}
	\quad
	\begin{subfigure}[b]{.21\textwidth}
	\scaledbrickfig{B}{Zag}
	\caption{The brick $\BZag$.}
	\end{subfigure}
	\quad
	\begin{subfigure}[b]{.21\textwidth}
	\scaledbrickfig{B}{Halt}
	\caption{The brick $\BHalt$.}
	\end{subfigure}
\caption{Folding paths to scale of all the bricks for Module $\ModuleB$ (see section~\ref{sec:full:description:B} for full description).}
\label{fig:B:all:bricks}
\end{figure}

\afterpage{\clearpage}
\newpage

\subsection{All bricks for Module C}
\label{sec:C:bricks}

Module \ModuleC\/ folds in switchbacks of height almost $h$, along the brick of the preceding module \ModuleA\/: 
\begin{itemize}
\item in 3 switchbacks in the zag phases;
\item in 3 switchbacks in the zig-up or zig-down phases if the current folding did not reach the end of the dataword encoded in the zag-row above yet;
\item but in 2 switchbacks ($\CEnd$) if this end is reached, creating the initial condition for folding the next modules as the encoding of the letters of the appendant to be appended at this stage of the simulation.  
\end{itemize}
All its possible bricks are displayed in Fig.~\vref{fig:C:all:bricks}. 

\begin{figure}[h]
	\begin{subfigure}{.49\textwidth}
	\scaledbrickfig{C}{ZigUp}
	\caption{The brick $\CZigUp$.}
	\end{subfigure}
	\quad
	\begin{subfigure}{.49\textwidth}
	\scaledbrickfig{C}{ZigDown}
	\caption{The brick $\CZigDown = \mirror(\CZigUp)$.}
	\end{subfigure}
	\\[5mm]
	\begin{subfigure}{.49\textwidth}
	\scaledbrickfig{C}{Zag}
	\caption{The brick $\CZag = \rot(\CZigUp)$.}
	\end{subfigure}
	\quad
	\begin{subfigure}{.49\textwidth}
	\scaledbrickfig{C}{End}
	\caption{The brick $\CEnd$.}
	\end{subfigure}
\caption{Folding paths to scale of all the bricks for Module $\ModuleC$ (see section~\ref{sec:full:description:C} for full description).}
\label{fig:C:all:bricks}
\end{figure}

\afterpage{\clearpage}
\newpage

\subsection{All bricks for Module D}
\label{sec:D:bricks}

\ModDxrt{$(x)$}{r,t}\/ is used to encode each letter of the appendant stored in each block unit. Its parameters $x,r,t$ stands for the letter $x\in\letterSet$ and the position index of that letter in the encoded appendant ($r$ says if it is either at the first, odd or even, and $t$ if it is the last or not). All these variants of \modD\/ fold slightly differently. \ModDxrt{$(x)$}{r,t}\/ folds either:
\begin{itemize}
\item in 6 switchbacks of height approximatively $h/2$, along the preceding brick of Module $\ModuleC$ in the zig-up, zig-down and zag phases; 
\item or as a glider in the append phase where it is forced by the preceding brick of module $\ModuleB$ to adopt this shape.
\end{itemize}
All its possible bricks are displayed in Fig.~\vref{fig:D:all:bricks}. 
\\
Note that: $\DZigDown = \mirror(\DZigUp)$ and $\DZag = \rot(\DZigUp)$.

\enlargethispage*{10cm}

\begin{figure}[h]
	\center
	\begin{subfigure}{.3\textwidth}
	\scaledbrickfig{D000}{ZigUp}
	\caption{The brick $\DZigUp$.}
	\end{subfigure}
	\quad
	\begin{subfigure}{.3\textwidth}
	\scaledbrickfig{D000}{ZigDown}
	\caption{The brick $\DZigDown$.}
	\end{subfigure}
	\begin{subfigure}{.3\textwidth}
	\scaledbrickfig{D000}{Zag}
	\caption{The brick $\DZag$.}
	\end{subfigure}
	\\[5mm]
	\center
	\begin{subfigure}{\textwidth}
	\begin{tabular}{rc}
		\raisebox{4mm}{$\DZAppend{0,1}$}
	&	\scaledbrickfig{D001}{Append}\\
		\raisebox{4mm}{$\DZAppend{0,0}$}
	&	\scaledbrickfig{D000}{Append}\\
		\raisebox{4mm}{$\DZAppend{1,0}$}
	&	\scaledbrickfig{D010}{Append}\\
		\raisebox{4mm}{$\DZAppend{1,1}$}
	&	\scaledbrickfig{D011}{Append}\\
		\raisebox{4mm}{$\DUAppend{0,1}$}
	&	\scaledbrickfig{D101}{Append}\\
		\raisebox{4mm}{$\DUAppend{0,0}$}
	&	\scaledbrickfig{D100}{Append}\\
		\raisebox{4mm}{$\DUAppend{1,0}$}
	&	\scaledbrickfig{D110}{Append}\\
		\raisebox{4mm}{$\DUAppend{1,1}$}
	&	\scaledbrickfig{D111}{Append}
	\end{tabular}\\
	\caption{The bricks $\DZAppend{r,t}$ and $\DUAppend{r,t}$.}
	\end{subfigure}
\caption{Folding paths to scale of all the bricks for Module $\ModuleDx{}$ (see section~\ref{sec:full:description:D} for full description).}
\label{fig:D:all:bricks}
\end{figure}

\afterpage{\clearpage}
\newpage

\subsection{All bricks for Module E}
\label{sec:E:bricks}

Module \ModuleEa{a}\/ is used for padding and carriage return. Its parameter $a$ stands for the number of letter of the appendant it needs to pad so as the block units of all appendants have the same dimensions. The length of \modEa{a}\/ decreases with $a$ accordingly. \ModEa{a}\/ folds either:
\begin{itemize}
\item in many short switchbacks of height approximatively $h/2$ followed by a glider and 5 large switchbacks of height $h$ in zig-up, zig-down and zag phases. The many short switchbacks are used to pad the appendants to a fixed length in the block units.
\item or as a glider in append phase, achieving the carriage return from zig to zag phase. The glider folding is triggered as for module $\ModuleDx{}$, either by the preceding brick of module $\ModuleB$ or the preceding glider append brick of a module $\ModuleDx{}$.
\end{itemize}
All its possible bricks are displayed in Fig.~\vref{fig:E:all:bricks}. 
\\
Note that: $\EZigDown = \mirror(\EZigUp)$ and $\EZag = \rot(\EZigUp)$.

\enlargethispage*{10cm}

\begin{figure}[h]
	\center
	\begin{subfigure}{.45\textwidth}
	\scaledbrickfig{E0}{ZigUp}
	\caption{The bricks $\EaZigUp{a}$.}
	\end{subfigure}
	\quad
	\begin{subfigure}{.45\textwidth}
	\scaledbrickfig{E0}{ZigDown}
	\caption{The bricks $\EaZigDown{a}$.}
	\end{subfigure}
	\\[5mm]
	\center
	\begin{subfigure}{.3\textwidth}
	\scaledbrickfig{E0}{Zag}
	\caption{The bricks $\EaZag{a}$.}
	\end{subfigure}
	\\[5mm]
	\center
	\begin{subfigure}{\textwidth}
	\begin{tabular}{c}
	$\EaCR{0}$\\[-5mm]
	\scaledbrickfigOptions{E0}{Append}{trim={0cm 0cm 89cm 0cm},clip}
	\raisebox{3mm}{$\cdots$}
	\scaledbrickfigOptions{E0}{Append}{trim={89cm 0cm 0cm 0cm},clip}
	\\[3mm]
	$\EaCR{a>0}$\\
	\scaledbrickfigOptions{E1}{Append}{trim={0cm 0cm 70cm 0cm},clip}
	\raisebox{3mm}{$\cdots$}
	\scaledbrickfigOptions{E1}{Append}{trim={70cm 0cm 0cm 0cm},clip}
	\end{tabular}\\
	\caption{The bricks $\EaCR{a}$.}
	\end{subfigure}
\caption{Folding paths to scale of all the bricks for Module $\ModuleEa{a}$ (see section~\ref{sec:full:description:E} for full description).}
\label{fig:E:all:bricks}
\end{figure}

\afterpage{\clearpage}
\newpage

\subsection{All bricks for Module F}
\label{sec:F:bricks}

Module \ModuleF\/ always folds as a glider of height $h$ and width $4$, pointing to NE in zig-up phase, SE in zig-down phase and SW in zag phase. As for module $\ModuleA$ in the zig phases, the folding of module $\ModuleF$ serves as a scaffold for the folding of the next modules in the zag phases.

All its possible bricks are displayed in Fig.~\vref{fig:F:all:bricks}.

\begin{figure}[h]
	\begin{subfigure}{.49\textwidth}
	\scaledbrickfig{F}{ZigUp}
	\caption{The brick $\FZigUp$.}
	\end{subfigure}
	\quad
	\begin{subfigure}{.49\textwidth}
	\scaledbrickfig{F}{ZigDown}
	\caption{The brick $\FZigDown = \mirror(\FZigUp)$.}
	\end{subfigure}
	\\[5mm]
	\centerline{
	\begin{subfigure}{.49\textwidth}
	\scaledbrickfig{F}{Zag}
	\caption{The brick $\FZag = \rot(\FZigUp)$.}
	\end{subfigure}}
\caption{Folding paths to scale of all the bricks for Module $\ModuleF$ (see section~\ref{sec:full:description:F} for full description).}
\label{fig:F:all:bricks}
\end{figure}

\afterpage{\clearpage}
\newpage

\subsection{All bricks for Module G}
\label{sec:G:bricks}

\ModG\/ is the most sophisiticated module as it can fold, depending of the context, into very different shapes:
\begin{description}
\item[Read bricks (Fig.~\ref{fig:G:all:bricks:read}):] In zig-up phase, \modG\/ folds as a glider of total length approximatively $2h$, heading first to NE and then bouncing to SE ($\GReadZ$) or E ($\GReadU$) depending on whether it hits the encoding of letter \word0\/ or \word1\/ in the zag row above respectively. 
\item[Copy bricks (Fig.~\ref{fig:G:all:bricks:copy}):] In zig-down and zag phases, \modG\/ folds into 6 switchbacks of height $h$, copying the letter, \word0\/ or \word1, encoded at the bottom of the row above, to the top of the row bellow. 
\item[Line feed brick (Fig.~\ref{fig:G:all:bricks:linefeed}):] At the end of the zag row, \modG\/ folds into a glider of length $2h$ heading SW opening the next zig row.
\end{description}
The design of module $\ModuleG$ requires a lot of care and was made possible thanks to the advanced design tools we developed for that purpose.

All its possible bricks are displayed in Fig.~\ref{fig:G:all:bricks:read}, \ref{fig:G:all:bricks:copy} and~\ref{fig:G:all:bricks:linefeed}.

\begin{figure}[h]
	\begin{subfigure}{\textwidth}
	\center \scaledbrickfig{G}{Read0}
	\caption{The brick $\GReadZ$.}
	\end{subfigure}
	\\[5mm]
	\begin{subfigure}{\textwidth}
	\center\scaledbrickfig{G}{Read1}
	\caption{The brick $\GReadU$.}
	\end{subfigure}
\caption{Folding paths to scale of all the Read bricks for Module $\ModuleG$ (see section~\ref{sec:full:description:G} for full description).}
\label{fig:G:all:bricks:read}
\end{figure}

\begin{figure}[h]
	\center 
	\begin{subfigure}{.4\textwidth}
	\scaledbrickfig{G}{ZigCopy0}
	\caption{The brick $\GZigCopyZ$.}
	\end{subfigure}
	\begin{subfigure}{.4\textwidth}
	\scaledbrickfig{G}{ZigCopy1}
	\caption{The brick $\GZigCopyU$.}
	\end{subfigure}
	\\[5mm]
	\center 
	\begin{subfigure}{.4\textwidth}
	\scaledbrickfig{G}{ZagCopy0}
	\caption{The brick $\GZagCopyZ$.}
	\end{subfigure}
	\begin{subfigure}{.4\textwidth}
	\center \scaledbrickfig{G}{ZagCopy1}
	\caption{The brick $\GZigCopyU$.}
	\end{subfigure}
\caption{Folding paths to scale of all the Copy bricks for Module $\ModuleG$ (see section~\ref{sec:full:description:G} for full description).}
\label{fig:G:all:bricks:copy}
\end{figure}

\begin{figure}[h]
	\begin{subfigure}{\textwidth}
	\scaledbrickfig{G}{LineFeed}
	\caption{The brick $\GLF$.}
	\end{subfigure}
\caption{Folding paths to scale of all the Line Feed brick for Module $\ModuleG$ (see section~\ref{sec:full:description:G} for full description).}
\label{fig:G:all:bricks:linefeed}
\end{figure}

\afterpage{\clearpage}
\newpage

}

%% file: blocks.tex

\section{Inside of the blocks}
\label{sec:blocks}

This section presents the exact content of each block, i.e. the bricks they are composed of.

\begin{figure}[hb]
\centerline{\includegraphics[width=.3\textwidth]{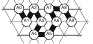}}
\caption{The conformation encoded by \protect{$\bA0 \SEbond\bA1 \NEbond\bA2 \SEbond\bA3 \SWbond\bA4 \Ebond\bA5 \NEbond\bA6 \NWbond\bA7 \Ebond\bA8 \SEbond\bA9 $}.}
\label{fig:conformation:example}
\end{figure}

\subparagraph{Notation.}
We describe conformations using the following notations: given a conformation $\sigma$ and a bead type $b$, we denote by $\sigma\Ebond b$, $\sigma\SEbond b$, $\sigma\SWbond b$, $\sigma\Wbond b$, $\sigma\NWbond b$, and $\sigma\NEbond b$ the elongations of the conformation~$\sigma$ by one bead of bead type $b$ located respectively to the east, south-east, south-west, west, north-west, and north-east of the last bead of $\sigma$. We refer by $b$ the conformation that consists of a single bead of bead type $b$ located at $(0,0)$. Fig.~\ref{fig:conformation:example} illustrates this notation.

We extend naturally this notation to two conformations: for instance, $\sigma_1 \Eb \sigma_2$ is the conformation beginning with conformation~$\sigma_1$ followed by conformation $\sigma_2$ whose origin has been translated to the vertex at the east of the last bead of $\sigma_1$.  

We denote by $\mirror(\sigma)$, $\vmirror(\sigma)$ and $\rot(\sigma)$ the conformations obtained by respectively mirroring horizontally, vertically and rotating by $180^\circ$ the conformation~$\sigma$.

Tables~\ref{tab:zig:block:inside} and~\ref{tab:zag:block:inside} on pages~\pageref {tab:zig:block:inside} and~\pageref{tab:zag:block:inside}, present an exploded view of the bricks inside each zig- and zag-block respectively.

The folded paths of the bricks composing the blocks in the definitions bellow have been presented in Section~\ref{sec:all:bricks:paths}. Their full description will be given in  section~\ref{sec:all:bricks:full}.  

These blocks are composed of one or $n$ \emph{block units} where $n$ is the number of appendants in the simulated STCS. Each block unit encodes an appendant $\alpha^i$ inside and consists of a sequence of $6+|\alpha^i|$ bricks: one of each \modA, \modB, and \modC, then $|\alpha^i|$ of \modD, then one of each \modE, \modF\/ and \modG. Read and Append blocks are composed of one block unit, whereas Copy blocks are composed of $n$ block units. 

We annotate each block by $[i]$ where $i$ is the index of the appendant $\alpha^i$ it contains. A block composed of $n$ block units is annotated by the index of its leading unit (i.e., the leftmost block unit of zig-block and the rightmost block unit of a zag-block).  These indices are computed modulo $n$, thus in the following $[i]$ refers to $[i \mod n]$.

As for the bricks before, specific symbols indicate to which kind of phase the blocks belong:
\begin{itemize}
\item Zig-up blocks are annotated by $\ZigUp$
\item Zig-down blocks are annotated by $\ZigDown$
\item Appending blocks are annotated by $\Append$
\item Carriage-return blocks (spanning from a zig row to the next zag row)  are annotated by~$\bZigZag$
\item Zag blocks are annotated by $\Zag$
\item Line-feed blocks (spanning from a zag row to the next zig row)  are annotated by $\bZagZig$
\item Halting blocks  are annotated by $\Halt$
\end{itemize}

This section presents the exact composition in terms of bricks of each block. It is more like a reference section.  The next section presents with illustration the exact geometry of each block. We encourage the reader to use the figures of the next section to picture each block when reading the block definition. We will refer to the appropriate figure after each block definition. 

\newcommand{\seeFigRight}[1]{\makebox[0em][l]{\small (See fig.~\ref{#1} p.~\pageref{#1})}}

\subsection{The bricks encoding the appendants}

Let us first describe how the appendants are encoded by \ModuleDx{}- and \ModuleEa{}-bricks inside the blocks.

Each letter of an appendant $\alpha^i$ is encoded by a \modD. There are 8 variants $\ModuleDxrt{$(x)$}{r,t}$ of modules $\ModuleDx{}$ depending on:
\begin{itemize}
\item the letter $x\in\letterSet$
\item the rank $r\in\{0,1,2\}$ of the letter in $\alpha^i$: $r=0$ for the first letter, $r=1$ if the index of letter is odd, and $r=2$ if the index of the letter is positive and even.
\item $t\in\{0,1\}$ where $t=1$ iff it is the last letter of $\alpha^i$.    
\end{itemize}
Each of these variant folds into a slightly different brick. This is why we must take the index of the letter into account in the definitions bellow. 
\begin{align*}
\WordZigUp{\epsilon}
&	= \EaZigUp{L}
\\
\WordZigUp{x}
&	= \DxrtZigUp{$(x)$}{0,1}\Eb \EaZigUp{L-1},
&	\makebox[0em][r]{for letter $x\in\{\word0,\word1\}$}
\\
\WordZigUp{v}
&	= \DxrtZigUp{$(v_0)$}{0,0}\Eb 
		\concatup{i=1}{|v|-2} \left(\DxrtZigUp{$(v_i)$}{2-(i\,\mod\,2),0} \,\Eb \right)
		\DxrtZigUp{$(v_{|v|-1})$}{2-((|v|-1)\,\mod\,2),1}\Eb 
		\EaZigUp{L-|v|},
\\[-2mm]
&&\makebox[0em][r]{for $v\in\{\word0,\word1\}^*$ with $|v|\geq 2$}
\\[2mm]
\WordZigDown{v}
&	= \mirror\!\left(\WordZigUp{v}\right)
&	\makebox[0em][r]{for all $v\in\{\word0,\word1\}^*$}
\\
\WordZag{v}
&	= \rot\!\left(\WordZigUp{v}\right)
&	\makebox[0em][r]{for all $v\in\{\word0,\word1\}^*$}
\\
\end{align*}

\subsection{The Zig-Blocks}

Let us now describe the bricks inside each of the blocks present on a zig row, that is to say: \bnREAD{}{} and \bnZIGCOPY{}{} blocks.
\begin{align*}
\bnREAD{0}{[i]} 
&	= \AZigUp \Eb \BZigUp \SWb \CZigUp \Eb \WordZigUp{\alpha^i} \Eb \FZigUp \SWb \GReadZ
& \seeFigRight{fig:geom:READ0}
\\
\bnREAD{1}{[i]} 
&	= \AZigUp \Eb \BZigUp \SWb \CZigUp \Eb \WordZigUp{\alpha^i} \Eb \FZigUp \SWb \GReadU
& \seeFigRight{fig:geom:READ1}
\\[2mm]
\bnZIGCOPYUNIT{0}{[i]}
&	= \AZigDown \Eb \BZigDown \NWb \CZigDown \Eb \WordZigDown{\alpha^i} \Eb \FZigDown \Eb \GZigCopyZ
\\
\bnZIGCOPYUNIT{1}{[i]}
&	= \AZigDown \Eb \BZigDown \NWb \CZigDown \Eb \WordZigDown{\alpha^i} \Eb \FZigDown \Eb \GZigCopyU
\\[1mm]
\bnZIGCOPY{0}{[i]}
&	= \bnZIGCOPYUNIT{0}{[i]} \, \concatup{j=1}{n-1} \left(\Eb\, \bnZIGCOPYUNIT{1}{[i+j]}\right)
& \seeFigRight{fig:geom:ZIGCOPY0}
\\
\bnZIGCOPY{1}{[i]}
&	= \concatup{j=0}{n-1} \left( \bnZIGCOPYUNIT{1}{[i+j]} \Eb\right)
& \seeFigRight{fig:geom:ZIGCOPY1}
\end{align*}

Note that the \bnZIGCOPY{}{}\/ blocks are composed of $n$ copy block units with consecutive appendant indices modulo $n$. They are indexed by the index of their leading copy block unit. Recall that the index $[i+j]$ is computed modulo $n$ and refers thus to $[(i+j)\,\mod\,n]$.

\subsection{Append and Carriage return blocks}

Let us define the following convenience block:
\begin{align*}
\WordAppend{\epsilon}
&	= \EaCR{L}
\\
\WordAppend{x}
&	= \DxrtAppend{$(x)$}{0,1}\Eb \EaCR{L-1},
&	\makebox[0em][r]{for letter $x\in\{\word0,\word1\}$}
\\
\WordAppend{v}
&	= \DxrtAppend{$(v_0)$}{0,0}\Eb 
		\concatup{i=1}{|v|-2} \left(\DxrtAppend{$(v_i)$}{2-(i\,\mod\,2),0} \,\Eb \right) 
\\ &	\hspace*{4cm}
		\DxrtAppend{$(v_{|v|-1})$}{2-((|v|-1)\,\mod\,2),1}\Eb 
		\EaCR{L-1},
\\
&&\makebox[0em][r]{for $v\in\{\word0,\word1\}^*$ with $|v|\geq 2$}
\end{align*}

Now,

\begin{align*}
\bnAPPENDRETURN{[i]}
&	= \AZigDown \Eb \BZigDown \NWb \CEnd \SEb \WordAppend{\alpha^i} \Wb \FZag \Wb \GZagCopyU
& \seeFigRight{fig:geom:Append-p-a}
\end{align*}

\subsection{Zag row}

Naturally, by symmetry, let:

\begin{align*}
\bnZAGCOPYUNIT{0}{[i]} 
&	= \vmirror\left(\bnZIGCOPYUNIT{0}{[i]}\right)
\\
\bnZAGCOPYUNIT{1}{[i]} 
&	= \vmirror\left(\bnZIGCOPYUNIT{1}{[i]}\right)
\\
\bnCOPYLINEFEEDUNIT{[i]} 
&	= \AZag \Wb \BZag \NEb \CZag \Wb \WordZag{\alpha^i} \Wb \FZag \Wb \GLF
\end{align*}

Then,

\begin{align*}
\bnZAGCOPY{0}{[i]} 
&	= \concatup{j=0}{n-3} \left( \bnZAGCOPYUNIT{1}{[i+j]}\, \Wb\right)
	\bnZAGCOPYUNIT{0}{[i+n-2]} ~	\Wb 
\\[-2mm]	&	\hspace*{3cm}
	\bnZAGCOPYUNIT{1}{[i+n-1]} 
& \seeFigRight{fig:geom:ZAGCOPY0}
\\
\bnZAGCOPY{1}{[i]} 
&	= \concatup{j=0}{n-1} \left( \bnZAGCOPYUNIT{1}{[i+j]}\, \Wb\right)
& \seeFigRight{fig:geom:ZAGCOPY1}
\\[2mm]
\bnCOPYLINEFEED{0}{[i]} 
&	= \concatup{j=0}{n-3} \left( \bnZAGCOPYUNIT{1}{[i+j]}\, \Wb\right) 
	\bnZAGCOPYUNIT{0}{[i+n-2]} ~ \Wb  
\\[-2mm]	&	\hspace*{3cm}
	\bnCOPYLINEFEEDUNIT{[i+n-1]} 
& \seeFigRight{fig:geom:ZAGCOPY0LF}
\\
\bnCOPYLINEFEED{1}{[i]} 
&	= \concatup{j=0}{n-2} \left( \bnZAGCOPYUNIT{1}{[i+j]}\, \Wb\right) 
\\[-2mm]	&	\hspace*{3cm}
	\bnCOPYLINEFEEDUNIT{[i+n-1]} 
& \seeFigRight{fig:geom:ZAGCOPY1LF}
\end{align*}

\subsection{The special case: appending $\epsilon$ to an empty dataword}

\begin{align*}
\bnRETURNLINEFEED
&	= \AZigDown \Eb \BZigDown \NWb \CEnd \SEb \WordAppend{\epsilon} \Wb \FZag \Wb \GLF
\\[1mm]
\bnRETURNLINEFEEDHALT
&	= \bnRETURNLINEFEED \Eb \AZigUp \Wb \BHalt
\hspace*{.75cm} 
\seeFigRight{fig:geom:CRLFHalt}
\end{align*}

\input{\texInputDir/block-inside}

\afterpage{\clearpage}
\newpage

%% file: block-inside.tex



\begin{sidewaystable}
\adjustbox{max size={.95\textwidth}{\textheight}}{\input{\texInputDir/newbricksInsideZigBlocks.tex}}
\caption{The bricks inside the zig\,\bZig-blocks.}
\label{tab:zig:block:inside}
\end{sidewaystable}

\begin{sidewaystable}
\adjustbox{max size={.95\textwidth}{\textheight}}{\input{\texInputDir/newbricksInsideZagBlocks.tex}}
\caption{The bricks inside the \bZag\,zag-blocks.}
\label{tab:zag:block:inside}
\end{sidewaystable}

%% file: newbricksInsideZigBlocks.tex

{
\newcommand{\scaleBlock}{.5} 
\newcommand{\scaleBrick}{.5} 
\newcommand{\figBricks}[1]{\includegraphics[scale=\scaleBrick]{bricks/#1}}
\newcommand{\figBlocks}[1]{\includegraphics[scale=\scaleBlock]{blocks/#1}}
\newcommand{\figBlocksColor}[1]{\includegraphics[scale=\scaleBlock]{blocks-colored/#1}}
\newcommand{\newscaleBlockBrick}{.075} 
\pgfmathsetmacro{\newscaleBlock}{\newscaleBlockBrick/\s}
\newcommand{\myN}{4}
\newcommand{\myL}{3}
\newcommand{\myP}{1}
\newcommand{\nameBrickFile}[2]{Module-#1-#2_n=\myN-L=\myL-P=\myP-arrow.pdf}
\newcommand{\newfigBrick}[2]{\includegraphics[scale=\newscaleBlockBrick]{\nameBrickFile{#1}{#2}}}
\newcommand{\newfigBlock}[1]{\scalebox{\newscaleBlock}{#1}}
\newcommand{\figFontSize}{\small}
\newcommand{\flipFig}[3]{%
	\settoheight{\figHeight}{#3}
	\raisebox{\figHeight}{\scalebox{#1}[#2]{#3}}
}%
\newcommand{\NotApplicable}[1]{\raisebox{#1}{\textit{\figFontSize\sffamily N\!/\!A}}}
\newcommand{\figModule}[2]{\raisebox{#1}{\figFontSize #2}}
\newcommand{\figLegend}[1]{\scalebox{.8}{\figFontSize #1}}
\newcommand{\figBrickRef}[2]{\makebox[0em][r]{\scriptsize\sffamily(see Figure~\ref{#1})\hspace*{#2}}}
\newcommand{\nameit}[2]{\hspace*{#2}{\scriptsize #1}}
\newcommand{\figBrickReff}[3]{\makebox[0em][r]{\sffamily\begin{tabular}[b]{r}\nameit{#3}{0mm}\\[-1mm]\scriptsize (see Figure~\ref{#1})\end{tabular}\hspace*{#2}}}


\begin{tabular}{cccccccc}
%
\toprule
Zig\,\bZig-Block
&	\makebox[0em][r]{Modules:\quad} \ModuleA
&	\ModuleB
&	\ModuleC
&	\ModuleDx{}
&	\ModuleE
&	\ModuleF
&	\ModuleG
\\[1mm] 
\toprule
\newfigBlock{\ReadZBlockUUZ{no}}
&	\newfigBrick{A}{ZigUp}
&	\raisebox{16.5mm}{\newfigBrick{B}{ZigUp}}
&	\newfigBrick{C}{ZigUp}
&	\newfigBrick{D000}{ZigUp} \newfigBrick{D100}{ZigUp}
&	\newfigBrick{E0}{ZigUp}
&	\newfigBrick{F}{ZigUp}
&	\newfigBrick{G}{Read0}
\\ 
\multicolumn{1}{c}{\bkREAD{\word0}} 
&	$\AZigUp$
&	$\BZigUp$
&	$\CZigUp$
&	$\DZZigUp{r,t}$ $\DUZigUp{r,t}$
&	$\EaZigUp{\ensuremath{a}}$
&	$\FZigUp$
&	$\GReadZ$
\\[2mm]
\midrule
\newfigBlock{\ReadUBlockZ{no}}
&	\newfigBrick{A}{ZigUp}
&	\raisebox{16.5mm}{\newfigBrick{B}{ZigUp}}
&	\newfigBrick{C}{ZigUp}
&	\newfigBrick{D000}{ZigUp} \newfigBrick{D100}{ZigUp}
&	\newfigBrick{E0}{ZigUp}
&	\newfigBrick{F}{ZigUp}
&	\newfigBrick{G}{Read1}
\\ 
\multicolumn{1}{c}{\bkREAD{\word1}} 
&	$\AZigUp$
&	$\BZigUp$
&	$\CZigUp$
&	$\DZZigUp{r,t}$ $\DUZigUp{r,t}$
&	$\EaZigUp{\ensuremath{a}}$
&	$\FZigUp$
&	$\GReadU$
\\[2mm]
\midrule
\newfigBlock{\ZigCopyZBlockUU{no}}
&	\newfigBrick{A}{ZigDown}
&	\newfigBrick{B}{ZigDown}
&	\newfigBrick{C}{ZigDown}
&	\raisebox{6mm}{\newfigBrick{D000}{ZigDown} \newfigBrick{D100}{ZigDown}}
&	\newfigBrick{E0}{ZigDown}
&	\newfigBrick{F}{ZigDown}
&	\newfigBrick{G}{ZigCopy0} \newfigBrick{G}{ZigCopy1}
\\ 
\multicolumn{1}{c}{\bkZIGCOPY{\word0}} 
&	$\AZigDown$
&	$\BZigDown$
&	$\CZigDown$
&	$\DZZigDown{r,t}$ $\DUZigDown{r,t}$
&	$\EaZigDown{\ensuremath{a}}$
&	$\FZigDown$
&	$\GZigCopyZ$ $\GZigCopyU$
\\[2mm]
\midrule
\newfigBlock{\ZigCopyZBlockEpsilon{no}}
&	\newfigBrick{A}{ZigDown}
&	\newfigBrick{B}{ZigDown}
&	\newfigBrick{C}{ZigDown}
&	\raisebox{6mm}{\newfigBrick{D000}{ZigDown} \newfigBrick{D100}{ZigDown}}
&	\newfigBrick{E0}{ZigDown}
&	\newfigBrick{F}{ZigDown}
&	\newfigBrick{G}{ZigCopy1}
\\
\multicolumn{1}{c}{\bkZIGCOPY{\word1}} 
&	$\AZigDown$
&	$\BZigDown$
&	$\CZigDown$
&	$\DZZigDown{r,t}$ $\DUZigDown{r,t}$
&	$\EaZigDown{\ensuremath{a}}$
&	$\FZigDown$
&	$\GZigCopyU$
\\[2mm]
\midrule
\multicolumn{1}{r}{\newfigBlock{\AppendBlockUUZ{no}}}
&	\raisebox{16.5mm}{\newfigBrick{A}{ZigDown}}
&	\raisebox{16.5mm}{\newfigBrick{B}{ZigDown}}
&	\raisebox{16.5mm}{\newfigBrick{C}{End}}
&	\raisebox{9mm}{\begin{tabular}[b]{c}
		\newfigBrick{D001}{Append}\\
		\newfigBrick{D000}{Append}\\
		\newfigBrick{D010}{Append}\\
		\newfigBrick{D011}{Append}\\
		\newfigBrick{D101}{Append}\\
		\newfigBrick{D100}{Append}\\
		\newfigBrick{D110}{Append}\\
		\newfigBrick{D111}{Append}
	\end{tabular}}
&	\raisebox{12.5mm}{\begin{tabular}[b]{c}
		\newfigBrick{E0}{Append}\\
		\newfigBrick{E1}{Append}
	\end{tabular}}
&	\newfigBrick{F}{Zag}
&	\newfigBrick{G}{ZagCopy1}
\\ 
\multicolumn{1}{c}{\bkAPPENDRETURN{(\alpha)}} 
&	$\AZigDown$
&	$\BZigDown$
&	$\CEnd$
&	$\DZAppend{r,t}$ $\DUAppend{r,t}$
&	$\EaCR{\ensuremath{a}}$
&	$\FZag$
&	$\GZagCopyU$
\\[2mm]
\midrule
\newfigBlock{\HaltBlockZ{no}}
&	\newfigBrick{A}{ZigUp}
&	\raisebox{16.5mm}{\newfigBrick{B}{Halt}}
\\ 
\bkHALT
&	$\AZigUp$
&	$\BHalt$
\\[2mm]
\midrule
\multicolumn{1}{r}{\newfigBlock{\AppendLineFeedHaltBlockEpsilon{no}}}
&	\raisebox{37mm}{\newfigBrick{A}{ZigDown}} \newfigBrick{A}{ZigUp}
&	\raisebox{16.5mm}{\begin{tabular}[b]{c}\newfigBrick{B}{ZigDown}\\[16mm] \newfigBrick{B}{HaltStraight}\end{tabular}}
&	\raisebox{37mm}{\newfigBrick{C}{End}} \newfigBrick{C}{HaltStraight}
&	\raisebox{30mm}{\begin{tabular}[b]{c}
		\newfigBrick{D001}{Append}\\
		\newfigBrick{D000}{Append}\\
		\newfigBrick{D010}{Append}\\
		\newfigBrick{D011}{Append}\\
		\newfigBrick{D101}{Append}\\
		\newfigBrick{D100}{Append}\\
		\newfigBrick{D110}{Append}\\
		\newfigBrick{D111}{Append}
	\end{tabular}}
&	\raisebox{30mm}{\begin{tabular}[b]{c}
		\newfigBrick{E0}{Append}\\
		\newfigBrick{E1}{Append}
	\end{tabular}}
&	\raisebox{16.5mm}{\newfigBrick{F}{Zag}}
&	\newfigBrick{G}{LineFeed}
\\
\bkRETURNLINEFEEDHALT 
&	$\AZigDown$ $\AZigUp$
&	$\BZigDown$ $\BHalt$
&	$\CEnd$
&	$\DZAppend{r,t}$ $\DUAppend{r,t}$
&	$\EaCR{\ensuremath{a}}$
&	$\FZag$
&	$\GLF$
\\[2mm]
\bottomrule
\end{tabular}
}

%% file: newbricksInsideZagBlocks.tex

{
\newcommand{\scaleBlock}{.5} 
\newcommand{\scaleBrick}{.5} 
\newcommand{\figBricks}[1]{\includegraphics[scale=\scaleBrick]{bricks/#1}}
\newcommand{\figBlocks}[1]{\includegraphics[scale=\scaleBlock]{blocks/#1}}
\newcommand{\figBlocksColor}[1]{\includegraphics[scale=\scaleBlock]{blocks-colored/#1}}
\newcommand{\figFontSize}{\small}
\newcommand{\flipFig}[3]{%
	\settoheight{\figHeight}{#3}
	\raisebox{\figHeight}{\scalebox{#1}[#2]{#3}}
}%
\newcommand{\newscaleBlockBrick}{.075} 
\pgfmathsetmacro{\newscaleBlock}{\newscaleBlockBrick/\s}
\newcommand{\myN}{4}
\newcommand{\myL}{3}
\newcommand{\myP}{1}
\newcommand{\nameBrickFile}[2]{Module-#1-#2_n=\myN-L=\myL-P=\myP-arrow.pdf}
\newcommand{\newfigBrick}[2]{\includegraphics[scale=\newscaleBlockBrick]{\nameBrickFile{#1}{#2}}}
\newcommand{\newfigBlock}[1]{\scalebox{\newscaleBlock}{#1}}
\newcommand{\NotApplicable}[1]{\raisebox{#1}{\textit{\figFontSize\sffamily N\!/\!A}}}
\newcommand{\figModule}[2]{\raisebox{#1}{\figFontSize #2}}
\newcommand{\figLegend}[1]{\scalebox{.8}{\figFontSize #1}}
\newcommand{\figBrickRef}[2]{\makebox[0em][r]{\scriptsize\sffamily(see Figure~\ref{#1})\hspace*{#2}}}
\newcommand{\nameit}[2]{\hspace*{#2}{\scriptsize #1}}
\newcommand{\figBrickReff}[3]{\makebox[0em][r]{\sffamily\begin{tabular}[b]{r}\nameit{#3}{0mm}\\[-1mm]\scriptsize (see Figure~\ref{#1})\end{tabular}\hspace*{#2}}}


\begin{tabular}{rrcccccc}
%
\toprule
\multicolumn{1}{c}{\bZag\,Zag-Block}
&	\multicolumn{1}{c}{\makebox[0em][r]{Modules:\quad} \ModuleG}
&	\ModuleF
&	\ModuleE
&	\ModuleDx{}
&	\ModuleC
&	\ModuleB
&	\ModuleA
\\[1mm] 
\toprule
\newfigBlock{\ZagCopyZBlockUUZ{no}}
&	\newfigBrick{G}{ZagCopy0} \newfigBrick{G}{ZagCopy1}
&	\newfigBrick{F}{Zag}
&	\newfigBrick{E0}{Zag}
&	\raisebox{6mm}{\newfigBrick{D000}{Zag} \newfigBrick{D100}{Zag}}
&	\newfigBrick{C}{Zag}
&	\newfigBrick{B}{Zag}
&	\newfigBrick{A}{Zag}
\\ 
\multicolumn{1}{c}{\bkZAGCOPY{\word0}} 
&	\multicolumn{1}{c}{$\GZagCopyZ$ $\GZagCopyU$}
&	$\FZag$
&	$\EaZag{\ensuremath{a}}$
&	$\DZZag{r,t}$ $\DUZag{r,t}$
&	$\CZag$
&	$\BZag$
&	$\AZag$
\\[2mm]
\midrule
\newfigBlock{\ZagCopyUBlockZ{no}}
&	\newfigBrick{G}{ZagCopy1}
&	\newfigBrick{F}{Zag}
&	\newfigBrick{E0}{Zag}
&	\raisebox{6mm}{\newfigBrick{D000}{Zag} \newfigBrick{D100}{Zag}}
&	\newfigBrick{C}{Zag}
&	\newfigBrick{B}{Zag}
&	\newfigBrick{A}{Zag}
\\ 
\multicolumn{1}{c}{\bkZAGCOPY{\word1}}  
&	$\GZagCopyU$
&	$\FZag$
&	$\EaZag{\ensuremath{a}}$
&	$\DZZag{r,t}$ $\DUZag{r,t}$
&	$\CZag$
&	$\BZag$
&	$\AZag$
\\[2mm]
\midrule
\raisebox{-17mm}{\newfigBlock{\ZagCopyZLineFeedBlockUU{no}}}
&	\raisebox{-17mm}{\newfigBrick{G}{LineFeed}} \newfigBrick{G}{ZagCopy0} \newfigBrick{G}{ZagCopy1}
&	\newfigBrick{F}{Zag}
&	\newfigBrick{E0}{Zag}
&	\raisebox{6mm}{\newfigBrick{D000}{Zag} \newfigBrick{D100}{Zag}}
&	\newfigBrick{C}{Zag}
&	\newfigBrick{B}{Zag}
&	\newfigBrick{A}{Zag}
\\ 
\multicolumn{1}{c}{\bkZAGCOPYLF{\word0}}  
&	\multicolumn{1}{c}{$\GLF$ $\GZagCopyZ$ $\GZagCopyU$}
&	$\FZag$
&	$\EaZag{\ensuremath{a}}$
&	$\DZZag{r,t}$ $\DUZag{r,t}$
&	$\CZag$
&	$\BZag$
&	$\AZag$
\\[2mm] 
\midrule
\raisebox{-17mm}{\newfigBlock{\ZagCopyULineFeedBlockEpsilon{no}}}
&	\raisebox{-17mm}{\newfigBrick{G}{LineFeed}} \newfigBrick{G}{ZagCopy1}
&	\newfigBrick{F}{Zag}
&	\newfigBrick{E0}{Zag}
&	\raisebox{6mm}{\newfigBrick{D000}{Zag} \newfigBrick{D100}{Zag}}
&	\newfigBrick{C}{Zag}
&	\newfigBrick{B}{Zag}
&	\newfigBrick{A}{Zag}
\\
\multicolumn{1}{c}{\bkZAGCOPYLF{\word1}}  
&	\multicolumn{1}{c}{$\GLF$ $\GZagCopyU$}
&	$\FZag$
&	$\EaZag{\ensuremath{a}}$
&	$\DZZag{r,t}$ $\DUZag{r,t}$
&	$\CZag$
&	$\BZag$
&	$\AZag$
\\[2mm]
\bottomrule
\end{tabular}

}

%% file: geometry.tex

\section{Geometry of the blocks}
\label{sec:block:geom}

We give here the geometrical description of each block. We describe in particular the positions of the important features of each block, such as the positions where a letter is read, copied or written.
 
\subparagraph{Convenience variables.} 
Let $\TMS = (\alpha; u^0)$ be the SCTS to be simulated. 
In order to simplify the description, we introduce the following variables which correspond to key geometrical parameters of the different blocks:  

\noindent
\begin{align*}
L	
&	= \max_{0\leq i< n} |\alpha_i|
&	\text{the maximum length of an appendant in $\TMS$}
\\
P
&	 = 12-(L\mod 2)
&	\text{ the \emph{padding} constant, such that $L+P$ is even and at least $12$}
\\
w
&	= 6(L+P)+18
&	\begin{array}[t]{r@{}}
	\text{the width of the appendant module excluding its}\\
 	\text{read/copy/linefeed part}
	\end{array}
\\
W
&	= n\cdot (w+6)
&	\text{the width of the \bnREAD{}{}, \bnZIGCOPY{}{}, and \bnZAGCOPY{}{} blocks}
\\
h
&	= W-(w+3)
&	\text{the height of the zig and zag rows}
\\
c(a)
&	= \makebox[0em][l]{$(6\lambda(L-a+P)+8h-16)/4$}
&	\text{the width of the brick $\EaCR{a}$ }
\end{align*}

As $n$ is a multiple of $4$ and as $L+P$ is even, we have: 
\begin{fact}
All convenience variables $L, P, w, W, h, c(a)$ are integers. Furthermore, 
$$
\begin{array}{r@{~}c@{~}lr@{~}c@{~}lr@{~}c@{~}lr@{~}c@{~}l}
	w& =& 6 \mod 12,
&	W& = &0 \mod 48,
&	 h &= &3 \mod 12,
&	c(a)& =& 2 \mod 12 \text{, for all $0\leq a\leq L$.}
\end{array}
$$
\end{fact}

These relations ensure for instance that all gliders finish in the correct position and that all patterns are properly aligned.

\begin{sidewaysfigure}
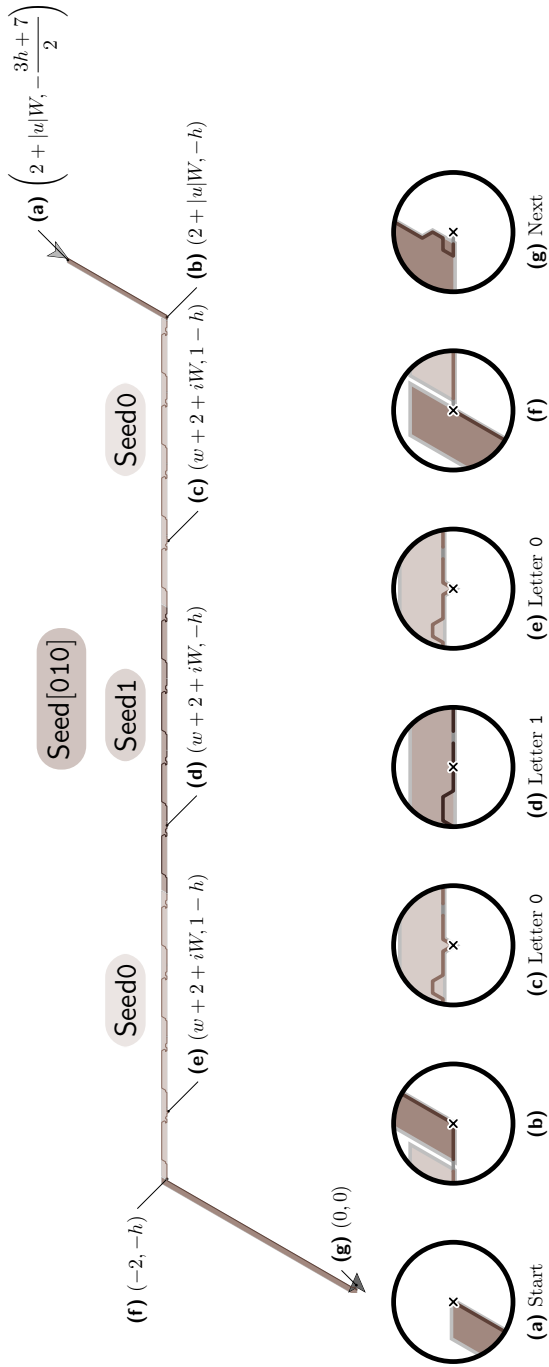

	\caption{\textsf{\bfseries Geometry of the \bnSEED{} block.} This block encodes the initial word so that the oritatami system simulates properly the corresponding tag system.  It consists of placing the different letter at the expected Write positions. Its rightmost part consists in a northeast-bound segment signalling the end of the (initial) word. Its leftmost part ends at the position $(-1,0)$ where the transcript will start folding the first zig-row. }
	\label{fig:geom:SEED}
	\blockFigs{}{\SeedZUZ{}}
\end{sidewaysfigure}

\begin{itemize}
\item \bnREAD{}{}{} blocks are described in figure~\ref{fig:geom:READ}
\item \bnZIGCOPY{}{}{} blocks are described in figure~\ref{fig:geom:ZIGCOPY} 
\item \bnSEED{} blocks are described in figure~\ref{fig:geom:SEED} 
\item \bnZAGCOPY{}{}{} blocks are described in figure~\ref{fig:geom:ZAGCOPY} 
\item \bnCOPYLINEFEED{}{}{} blocks are described in figure~\ref{fig:geom:ZAGCOPYLF} 
\item \bnAPPENDRETURN{} blocks are described in figure~\ref{fig:geom:Append-p-a} 
\item \bnRETURNLINEFEEDHALT{} and \bHALT{}{}{} blocks are described in figures~\ref{fig:geom:CRLFHalt} and~\ref{fig:geom:HALT} 
\end{itemize}

Intuitively, each zig block corresponds to one cell of the trimmed diagram of the simulated SCTS, upscaled to a parallelogram of width $W$ and heigth $h$.  Zag-blocks are just used to copy these cells while returning at the beginning of the current datawork. 

Recall that the coordinates are expressed according to the east and south-west axis: every position $(x,y)$ in $\Tlat$ is mapped in the euclidean plane to $x\cdot \vec E + y \cdot \vec{SW}$ using the vector basis $\vec E = (1,0)$ and $\vec{SW} = \rotateClockwiseAA{\vec E} = (-\frac12, -\frac{\sqrt3}2)$.

\begin{figure}
\caption{\textsf{\bfseries Geometry of the \protect\bnREAD{}{} blocks.} Note that the internal structures (the lines in white) of both blocks \bnREAD0{} and \bnREAD1{} agree until position $(w+2,1-h)$ where the presence or absence of a spike, encoding a \word0\/, at the bottom of the row above forces the block to adopt the shape \bnREAD0\/ or \bnREAD1 respectively.}
\label{fig:geom:READ}
	\begin{subfigure}{\textwidth}
	\blockFigs{Block-Read-0-a.pdf}{\ReadZBlockUU{}}
	\caption{The \bnREAD0{} block has the shape of a trapezium whose bottom basis has length $W$ and top basis has length $w+5$, with height $h$. It has a dent (an empty position) located at $(w+2,-h+1)$ (w.r.t. to its origin at the bottom left corner), in which plugs the spike of the block from the row above it, encoding the letter~\word0. The next block will start folding at the bottom right corner, at~$(W,0)$.}
	\label{fig:geom:READ0}
	\end{subfigure}\\[\figSkip]
	\begin{subfigure}{\textwidth}
	\blockFigs{Block-Read-1-a.pdf}{\ReadUBlockEpsilon{}}
	\caption{The \bnREAD1{} block has the shape of a parallelogram with horizontal side length $W$ and vertical side length $h$. The red rectangle area at position $(w+2,1-h)$ (w.r.t. its origin at the bottom left corner) aligns with the flat bottom block above encoding the letter \word1 (as opposed to a spiked-block encoding a \word0). The next block will start folding at the top right corner, at~$(W-1,1-h)$.}
	\label{fig:geom:READ1}
	\end{subfigure}
\end{figure}


\begin{figure}
\caption{\textsf{\bfseries Geometry of the \bnZIGCOPY{}{} blocks.} The \bnZIGCOPY0{}{} and \bnZIGCOPY1{}{} blocks have both the shape of a parallelogram with horizontal side length $W$ and vertical side length $h$. For both, the next block will start folding at the top right corner, at~$(W-1,1-h)$. Note that the \bnZIGCOPY0{}{} and  \bnZIGCOPY1{}{} blocks have identical internal structure apart from the line joining  the two purple areas at $(w+2,1-h)$ and $(h+w+1,1)$. Indeed, when folding, the part of the transcript located in the red area, either:  (1) detects a spike on top (encoding a \word0) and then folds into a dent on top which induces spike at the bottom (copying the \word0 below, the block \bnZIGCOPY0{}{}); or (2) folds flat (encoding a \word1) on top which induces a flat folding at the bottom, copying the \word1 from the top to the bottom of the Zig-row (the block \bnZIGCOPY1{}). Furthermore, this block is made of $n$ \bnZIGCOPYUNIT{}{}{} blocks (from left to right), each of width $w+6$ and height $h$.}
\label{fig:geom:ZIGCOPY}
	\begin{subfigure}{\textwidth}
	\blockFigs{Block-Copy-0-Zig-a.pdf}{\ZigCopyZBlockUUZ{}}
	\caption{The \bnZIGCOPY0{}{} block has a dent (an empty position) located at $(w+2,1-h)$, in which plugs the spike of the block from the row above it, and which induces (when folding) a spike at the bottom at $(h+w+1,1)$, copying the letter~\word0 from the top to the bottom of the Zig-row. Note that this block is made of $n$ \bnZIGCOPYUNIT{}{}{} blocks (from left to right): one \bnZIGCOPYUNIT0{}{} followed by $n-1$ \bnZIGCOPYUNIT1{}{}, each of width $w+6$ and height $h$.}
	\label{fig:geom:ZIGCOPY0}
	\end{subfigure}\\[\figSkip]
	\begin{subfigure}{\textwidth}
	\blockFigs{Block-Copy-1-Zig-a.pdf}{\ZigCopyUBlockZ{}}
	\caption{The \bnZIGCOPY1{}{} block is flat at $(w+2,1-h)$, which, aligned with a flat block above (encoding a \word1), induces (when folding) a flat bottom at $(h+w,0)$, copying the letter~\word1 from the top to the bottom of the Zig-row. Note that this block is made of $n$ \bnZIGCOPYUNIT1{}{} blocks (from left to right), each of width $w+6$ and height $h$.}
	\label{fig:geom:ZIGCOPY1}
	\end{subfigure}
\end{figure}

\begin{figure}
\caption{\textsf{\bfseries Geometry of the $\bnZAGCOPY{}{}$ blocks.} The \bnZAGCOPY0{}{} and \bnZAGCOPY1{}{} blocks are the horizontal mirror images of the \bnZIGCOPY0{}{} and \bnZIGCOPY1{}{} blocks  (see Figure~\ref{fig:geom:ZIGCOPY}).}
\label{fig:geom:ZAGCOPY}
	\begin{subfigure}{\textwidth}
	\blockFigs{Block-Copy-0-Zag-a.pdf}{\ZagCopyZBlockUUZ{}}
	\caption{The \bnZAGCOPY0{}{} block is the horizontal mirror image of the \bnZIGCOPY0{}{} block  (see Figure~\ref{fig:geom:ZIGCOPY0}).}
	\label{fig:geom:ZAGCOPY0}
	\end{subfigure}\\[\figSkip]
	\begin{subfigure}{\textwidth}
	\blockFigs{Block-Copy-1-Zag-a.pdf}{\ZagCopyUBlockZ{}}
	\caption{The \bnZAGCOPY1{}{} block is the horizontal mirror image of the \bnZIGCOPY1{}{} block  (see Figure~\ref{fig:geom:ZIGCOPY1})} 
	\label{fig:geom:ZAGCOPY1}
	\end{subfigure}
\end{figure}

\begin{figure}
\caption{\textsf{\bfseries Geometry of the $\bnCOPYLINEFEED{}{}$ blocks.} These blocks adopt the shape of a ${(W-6)\times h}$-parallelogram prolongated by an southwestbound ``arm'' hoping to the beginning of the next zig-row. Folding from right to left, the \bnCOPYLINEFEED{}{} blocks are identical to the \bnZAGCOPY{}{} blocks until position $(h-W-2,0)$ where it detects that there are no more blocks (encoding letter) in the row above (the detection of the absence of a block on top is made possible by the horizontal offset of $7$ beads between the zig- and zag-rows). Then, instead of completing a parallelogram, the end of the $\bnCOPYLINEFEED{}{}$ blocks is attracted upwards and then folds into a southwestbound glider pattern to reach the opening position of the next zig-row. The next block will start folding at $(h-W,2h)$. Furthermore, this block is made (from right to left) of $n-2$ \bnZAGCOPYUNIT1{}{} blocks followed by a  \bnZAGCOPYUNIT{$(x)$}{}{} and a \bnCOPYLINEFEEDUNIT{}{}, where $x$ is the letter copied.}
\label{fig:geom:ZAGCOPYLF}
	\begin{subfigure}{\textwidth}
	\blockFigs{Block-Copy-0-LineFeed-a.pdf}{\ZagCopyZLineFeedBlockUU{}}
	\caption{The \bnCOPYLINEFEED0{}{} block proceeds as \bnZAGCOPY0{}{} to copy the spike encoding a \word0 from the row above to the row below. It has a dent (an empty position) at $(h-W+w+1,1)$ in which plugs the spike (encoding a \word0) of the block above. When folding, this dent induces a spike at the bottom at position $(h-W+w+2,1+h)$. Note that the spike below is at position $(h-W+w+2,1+h)$, which is consistent with the position of the dent in the \bnREAD0{}{} block that will fold from $(h-W,2h)$ (see Figure~\ref{fig:geom:READ0}).}
	\label{fig:geom:ZAGCOPY0LF}
	\end{subfigure}
\end{figure}

\addtocounter{figure}{-1}

\begin{figure}
\caption{\textsf{\bfseries Geometry of the $\bnCOPYLINEFEED{}{}$ blocks.} (Continued)}
	\addtocounter{subfigure}{1}
	\begin{subfigure}{\textwidth}
	\blockFigs{Block-Copy-1-LineFeed-a.pdf}{\ZagCopyULineFeedBlockZ{}}
	\caption{The \bnCOPYLINEFEED1{} block.}
	\label{fig:geom:ZAGCOPY1LF}
	\end{subfigure}
\end{figure}

\begin{sidewaysfigure}
	\caption{\textsf{\bfseries Geometry of the \protect\bnAPPENDRETURN{(u)} blocks.}   The folding into this block is triggered by the absence of a block in row above (indicating the end of the word). It has one northeastbound 2-beads wide arm climbing along the east side of the block in the row above then a southeastbound 4-beads wide arm stopping at the bottom of the current zig-row. Then, the block consists in an 3-beads high $|u|W$-beads long eastbound glider path going along the bottom of the current zig-row and encoding each letter of $u$: the path contains a spike (below, and a dent on top) for each $u_j = \word 0$ at position $(jW+w+h+1,1)$ (\word1s  are encoded by the absence of spike). It then expands upto position $(|u|W+h+c(|u|),0)$ and go back to its origin and grows a $10$-beads wide $h$-beads high southwestbound arm opening the next zag-row to end at the position $(|u|W+h-9,2)$, at the top right corner of the upcoming zag-row. The next block will start at $(|u|W+h-8,1)$.}
	\label{fig:geom:Append-p-a}
	\blockFigs{Block-Copy-1-LineFeed-a.pdf}{\AppendBlockUUZ{}}
\end{sidewaysfigure}

\begin{sidewaysfigure}
	\caption{\textsf{\bfseries Geometry of the \protect\bnAPPENDRETURN{(\epsilon)} block.}   This block is the special case of Figure~\ref{fig:geom:Append-p-a} where $u=\epsilon$. It is given for clarity.}
	\blockFigs{Block-Copy-1-LineFeed-a.pdf}{\AppendBlockEpsilon{}}
\end{sidewaysfigure}

\begin{sidewaysfigure}
	\caption{\protect\textsf{\bfseries Geometry of the \protect\bnRETURNLINEFEEDHALT{}\/ block.} This block is identical to the \bnAPPENDRETURN{(\epsilon)}{} block until it reaches position $(h-1,1)$. Then, when, folding, it detects the absence of a block above which indicates that the current word is empty. It then folds as the lefttmost part of the \bnCOPYLINEFEED{}{}  blocks (see Figure~\ref{fig:geom:ZAGCOPYLF}) to open a new zig-row at $(h-1,2h)$. It then goes up to $(h,h-1)$. And as there are no block on the zag-row above, it is attracted inside itself and gets blocked at $(h-1,2h-3)$.}
\label{fig:geom:CRLFHalt}
	\blockFigs{Block-Copy-1-LineFeed-a.pdf}{\AppendLineFeedHaltBlockEpsilon{}}
\end{sidewaysfigure}

\begin{figure}
	\caption{\textsf{\bfseries Geometry of the $\bnHALT{}$ block.} This block appears at the end of the computation. It starts as a \bnREAD{}{} block with a $3$-beads wide $h$-beads high southeastbound glider until it reaches position $(2,1-h)$. But, as there are no block in the zag-row above, the next beads are attracted to the left and the construction stops there.}
	\label{fig:geom:HALT}
	\hfill \blockFigs{Block-Halt-a.pdf}{\HaltBlockEpsilon{}} \hfill ~
\end{figure}

%% file: modules+bricks.tex

\section{Full description of the SCTS oritatami simulator}
\label{sec:all:bricks:full}

Consider a SCTS $\TMS$ with $n$ appendants $\alpha_0, \ldots, \alpha_{n-1}\in\{\word0, \word1\}^*$ , where $n$ is at least $8$ and a multiple of~$4$, together with an input dataword $u\in\{\word0, \word1\}^*$.

We give here the full description of the primary structure $\pi_\TMS$ of the oritatami systems $\TMO_\TMS=((\pi_\TMS)^\infty, \heart, 3)$ together with its seed conformation $\sigma_\TMS(u)$, which simulates step by step the computation of~$\TMS$ on input dataword~$u$.
 
\subparagraph{Convenience variables.} 
In order to simplify the description, we introduce the following variables which correspond to key geometrical parameters of the different modules: (note that some of the variables were already introduced in Section~\ref{sec:block:geom})  

\noindent
\begin{align*}
L	
&	= \max_{0\leq i< n} |\alpha_i|
&	\text{the maximum length of an appendant in $\TMS$}
\\
P
&	 = 12-(L\mod 2)
&	\text{ the \emph{padding} constant, such that $L+P$ is even and at least $12$}
\\
w
&	= 6(L+P)+18
&	\begin{array}[t]{r@{}}
	\text{the width of the appendant module excluding its}\\
 	\text{read/copy/linefeed part}
	\end{array}
\\
W
&	= n\cdot (w+6)
&	\text{the width of the \bnREAD{}{}, \bnZIGCOPY{}{}, and \bnZAGCOPY{}{} blocks}
\\
h
&	= W-(w+3)
&	\text{the height of the zig and zag rows}
\\
k
&	= (h-3)/6
&	\text{the number of periods of a glider of length $h$}
\\
\lambda
&	= W/2
&	\begin{array}[t]{r@{}}
	\text{\makebox[0em][r]{the height} ($\lambda+5$) of the letter modules inside the}\\
	\text{\bnREAD{}{}, \bnZIGCOPY{}{}, and \bnZAGCOPY{}{} blocks}
	\end{array}
\\
\kappa
&	= W/24
&	\begin{array}[t]{r@{}}
		\text{the number of periods of the glider/switchback pattern}\\
		\text{in the letter and padding modules}
	\end{array}
\\
q
&	= (h-3)/4
&	\text{\makebox[0em][r]{the number} of periods of the glider in the backbone of module $\ModuleE$}
\\
c(a)
&	= \makebox[0em][l]{$(6\lambda(L-a+P)+8h-16)/4$}
&	\text{the width of the brick $\EaCR{a}$ }
\end{align*}

As $n$ is a multiple of $4$ and as $L+P$ is even, we have: 
\begin{fact}
\label{fact:variable:modulo}
All convenience variables $L, P, w, W, h, k, \lambda, \kappa, q, c(a)$ are integers. Furthermore, 
$$
\begin{array}{r@{~}c@{~}lr@{~}c@{~}lr@{~}c@{~}lr@{~}c@{~}l}
	w& =& 6 \mod 12,
&	W& = &0 \mod 48,
&	 h &= &3 \mod 12,
&	k &=& 0\mod 2,
\\
	\lambda &= &0 \mod 24,
&	\kappa &=& 0 \mod 2,
&	q &=& 0 \mod 3,
&	c(a)& =& 2 \mod 12 \text{, for all $0\leq a\leq L$.}
\end{array}
$$
\end{fact}

\subparagraph{Notations for describing of the bead type sequences.}
if $u$ and $v$ are two finite bead type sequences, we write their concatenation as $u\cdot v$. For any two integers $0\leq i\leq j<|u|$, we write $u_{i\ldots j}$ for $u_iu_{i+1}\ldots u_j$. The \emph{reverse sequence} of $u$, written as $u^R$, is $u_{|u|-1}u_{|u|-2}\ldots u_1 u_0$.

  Finally, given a sequence $u$, we write $u\subst{a_1\!@i_1,\ldots, a_k\!@i_k}$ for the sequence $w$ where  the bead type indexed by $i_j$ in $u$ has been replaced by $a_j$ for $j=1,\ldots,k$:
$$
w_i = \left\{
  \begin{array}{cl}
    a_j	&\text{if $i = i_j$ for some $j$}\\
    u_i	&\text{otherwise} 
  \end{array}
\right.
$$

By extension, we write $u\subst{v@{k..l}}$ for the sequence $w$ where for all $i\in\{k,k+1, \ldots, l\}$, the beads at indices $k$ to $l$ of $u$ have been replaced by the word $v$ (of length $l-k+1$):
$$
w_i = \left\{
  \begin{array}{cl}
    v_{i-k}	&\text{if $k\leq i \leq l$}\\
    u_i	&\text{otherwise} 
  \end{array}
\right.
$$

For an infinite sequence of (finite) words $(u_i)_{i\geq1}$, we denote by $\concatop_{i\geq1}u_i$ the infinite word $u_1 u_2\cdots u_i\cdots$ obtained by containing all the words $u_1,u_2,\ldots$

\subparagraph{Notation for describing conformations.}
Given an infinite sequence of directions $(d_i)\in\{\NEb,  \Eb,  \SEb,  \SWb, \Wb, \NWb\}^{\mathbb N}$, and a finite bead type sequence $b\in B^*$, we denote by $\conformation(b, d)$ the conformation $b_0 d_0 b_1 d_1\cdots d_{|b|-2} b_{|b|-1}$ that maps $b$ along the path $d$. We will use the following convenience functions which will ease the description of the bricks:
\begin{itemize}     
\item $\Epath(b) = \conformation(b, (\Eb)^\infty)$ and similarly, $\SEpath$, \SWpath$, \Wpath$, $\NWpath$ and $\NEpath$ that map a bead type sequence along the paths $(\SEb)^\infty$, $(\SWb)^\infty$, $(\Wb)^\infty$, $(\NWb)^\infty$, and $(\NEb)^\infty$ respectively.
\item $\Eglider(b) = \conformation\left(b, \left(\NEb, \NWb, \Eb, \SEb, \SWb, \Eb\right)^\infty\right)$ 
\item $\Eglider'(b) = \conformation\left(b, \left(\SEb, \SWb, \Eb, \NEb, \NWb, \Eb\right)^\infty\right)$ 
\item $\SERglider(b) = \conformation\left(b, \left(\NEb, \Eb, \SEb, \Wb, \SWb, \SEb \right)^\infty\right)$ 
\item $\NEglider(b) = \conformation\left(b, \left(\NWb, \Wb, \NEb, \Eb, \SEb, \NEb \right)^\infty\right)$ 
\end{itemize}

Recall that the coordinates are expressed according to the east and south-west axis: every position $(x,y)$ in $\Tlat$ is mapped in the euclidean plane to $x\cdot \vec E + y \cdot \vec{SW}$ using the vector basis $\vec E = (1,0)$ and $\vec{SW} = \rotateClockwiseAA{\vec E} = (-\frac12, -\frac{\sqrt3}2)$.

\subsection{The periodic primary structure  $(\pi_\TMS)^\infty$}

The period of the primary structure consists in the concatenation of one appendant bead type sequence for each appendant:
$$ 
\pi_\TMS 
= \APPENDANTa{0}\conc\APPENDANTa{1}\conc \,\cdots \, \conc \APPENDANTa{n-1}
$$

\subsection{The appendant bead type sequences}

Each appendant bead type sequence $\APPENDANTa{i}$ has the exact same structure: it is the concatenation of 6 sequences: Modules \ModuleA, \ModuleB, and \ModuleC, followed a sequence \WORD{\alpha_i} encoding the appendant~$\alpha_i$ itself, followed by modules  \ModuleF\/ and \ModuleG:
$$
\APPENDANTa{i}
=	\ModuleA \conc \ModuleB \conc \ModuleC \conc \WORD{\alpha_i} \conc \ModuleF \conc \ModuleG
$$
For each word $v\in\{\word0,\word1\}^*$ with $|v|\leq L$, \WORD{v} encodes each letter of $v$ using one of the 6 variants of module \ModuleDx{} and terminates with a padding sequence \ModuleEa{L-|v|} which ensures that the folded size of $\WORD{v}$ is independent of the length of~$v$. The 6 variants of the module \ModuleDx{} are \ModuleDxrt{$(x)$}{r,t} where:
\begin{itemize}[topsep=2pt,itemsep=0pt]
\item $x\in\{\word0,\word1\}$  is the encoded letter; 
\item $r\in\{0,1,2\}$ is the rank of the letter inside the encoded word $v$: $r=0$ if it is the first letter of $v$; $r=1$ if its index in $v$ is odd; and $r=2$ if its index in $v$ is even but not $0$;
\item $t\in\{0,1\}$ is $1$ if the encoded letter is the last letter of $v$, and $0$ otherwise.
\end{itemize}
The definition of the sequence \WORD{v} follows as: 
\begin{align*}
\WORD\epsilon
&	= \ModuleEa{L}
\\
\WORD{\word0}
&	= \ModuleDZ{0,1} \conc \ModuleEa{L-1}
\\
\WORD{\word1}
&	= \ModuleDU{0,1} \conc \ModuleEa{L-1}
\\
\text{and for $|v|\geq 2$,\quad}
\WORD{v}
&	= \ModuleDxrt{$(v_0)$}{0,0} 
	\conc \left( \concatup{i=1}{|v|-2} \ModuleDxrt{$(v_i)$}{2-(i\,\mod\,2), 0}\right) 
	\conc \ModuleDxrt{$(v_{n-1})$}{2-((|v|-1)\,\mod\,2), 1} 
	\conc \ModuleEa{L-|v|}
\end{align*}

The next section concludes the full description of the primary structure by giving the sequences for modules \ModuleA, \ModuleB, \ModuleC, \ModuleDxrt{$(x)$}{r,t}, \ModuleEa{a}, \ModuleF, and \ModuleG.

\subsection{Modules bead type sequences and brick conformations}

The modules are given using $546$ bead types. However, 5 of them, $\bA6$, $\bC{16}$, $\bJ{9}$, $\bL{86}$, and $\bL{89}$, are \emph{neutral}, i.e. do not have any interaction with any other bead types (see Section~\ref{sec:rule}) and can thus be all substituted by one single \emph{neutral} bead type $\bN0$. The total number of distinct bead types is thus $542$ ($541$ + $1$ neutral bead type $\bN0$).  However, in the description bellow, we prefer to use $\bA6$, $\bC{16}$, $\bJ{9}$, $\bL{86}$, and $\bL{89}$ (and not $\bN0$) as it keeps the bead types homogenous  and continuously numbered in each module. 

\subsection{Module A} 
\label{sec:full:description:A}

\subsubsection{Bead type sequence of Module A} 
Length: $3h-2$; 13 bead types used: $\bbA{0}{12}$
$$
\ModuleA =  \bSN{\bbA{0}{4}}\conc (\bbA{5}{10})^{3k-1}\conc \bbA{5}{7} \conc \bSN{\bA{6}}\conc \bbA{9}{10} \conc \bSN{\bbA{11}{12}}
$$
	
\subsubsection{The bricks for module A}

Module $\ModuleA$ adopts three brick conformations: \AZigUp, \AZigDown, and \AZag, where the two last ones are just obtained by mirroring and rotating the first:
\begin{align*}
\AZigUp 
&	=  \bA0 \Eb \bA1\NWb \bA2 \Eb \bA3 \SEb\bA4 \NEb (\bA5 \NWb \bA6 \Wb \bA7 \NEb \bA8 \Eb \bA9 \SEb \bA{10} \NEb)^{3k-1} \bA5 \NWb \\
& 	\retraitAlign  
	\bA6 \Wb \bA7 \NEb \bA6 \Eb \bA9 \SEb \bA{10} \NEb \bA{11} \NWb \bA{12}
\\[2mm]
\AZigDown 
&	= \mirror(\AZigUp)
\\[1mm]
\AZag 
&	= \rot(\AZigUp)
\end{align*}

Figure~\ref{fig:A:brick:ZigUp} displays brick $\AZigUp$.

\figBrickHere%
	{A-Zig_Init-UPRIGHT}%
	{Module \ModuleA: Brick $\AZigUp$.}%
	{fig:A:brick:ZigUp}%
	
\subsubsection{Subrule for Module A}	

Module \ModuleA\/ interacts with \ModuleA, \ModuleB, \ModuleC, \ModuleDx{}, \ModuleF\/ and \ModuleG. Figure~\ref{fig:A:subrule} presents the subrule for the interactions between the beads of $\ModuleA$ and the beads of the other modules.

\figSubRuleHere%
	{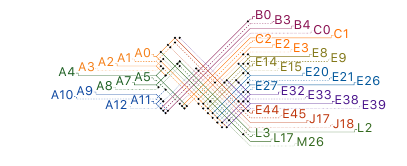}%
	{Subrule for Module \ModuleA.}%
	{fig:A:subrule}%
			
\newpage
	
\subsection{Module B} 
\label{sec:full:description:B}

\subsubsection{Bead type sequence for Module B} 
Length: $5$; 5 bead types used: $\bbB{0}{4}$
$$
\ModuleB =  \bbB{0}{4}
$$

\subsubsection{The bricks for module B}

Module $\ModuleB$ adopts four brick conformations plus some incomplete ones if the folding stops because the dataword in the simumlated SCTS is empty.

\begin{align*}
\BZigUp 
&	=  \bB0 \Eb \bB1 \SEb \bB2 \Wb \bB3 \SWb \bB4
\\[2mm]
\BZigDown 
&	=  \mirror(\BZigUp)
\\[1mm]
\BZag 
&	=  \rot(\BZigUp)
\\[2mm]
\BHalt 
&	=  \bB0 \Wb \bB1 \SWb \bB2 \SWb \bB3 \SWb \bB4
\end{align*}

Figure~\ref{fig:B:brick:ZigUp} displays bricks $\BZigUp$ and $\BHalt$ (shaded).

\figBrickHere%
	{B-Empty_Word_Tape}%
	{Module \ModuleB: Brick $\BZigUp$ to the right, and brick $\BHalt$ shaded to the left.}%
	{fig:B:brick:ZigUp}

\subsubsection{Subrule for Module B}	

Module \ModuleB\/ interacts with \ModuleA, \ModuleC, \ModuleDx{}, \ModuleE\/ and \ModuleF.
Figure~\ref{fig:B:subrule} presents the subrule for the interactions between the beads of $\ModuleB$ and the beads of the other modules.

\figSubRuleHere%
	{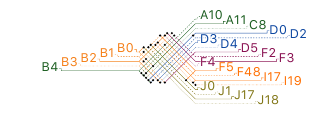}%
	{Subrule for Module \ModuleB.}%
	{fig:B:subrule}%
			
\afterpage{\clearpage}
\newpage

\subsection{Module C} 
\label{sec:full:description:C}

\subsubsection{Bead type sequence for Module C} 
Length: $3h-10$; 17 bead types used: $\bbC{0}{16}$
$$
 \ModuleC = (\bbC{0}{2})^{2k}\conc (\bbC{3}{5})^{k}\conc \bC3\conc \bbC{7}{8}\conc (\bbC{6}{8})^{k-1}\conc (\bbC{9}{14})^{k-1}\conc \bbC{9}{10}\conc \bSN{\bbC{15}{16}} \conc \bC{13}
 $$

\subsubsection{The bricks for module C}

Module $\ModuleC$ adopts four brick conformations:

\begin{align*}
\CZigUp 
&	= 	(\bC0 \SWb \bC1 \SWb \bC2\, \SWb )^{2k-1} \bC0 \SWb \bC1 \SWb \bC2 \, \Eb 
\\
&		\retraitAlign 
		(\bC3 \NEb \bC4 \NEb \bC5\, \NEb)^{k} \, \bC3 \NEb \bC7 \NEb 
		(\bC8 \NEb \bC6 \NEb \bC7\, \NEb) ^{k-1} \, \bC8 \, \SEb
\\
&		\retraitAlign 
		(\bC9 \SWb \bC{10} \SWb \bC{11} \SWb \bC{12} \SWb \bC{13} \SWb \bC{14} \, \SWb)^{k-1}
		\bC9 \SWb \bC{10} \SWb \bC{15} \SWb \bC{16} \SWb \bC{13}
\\[2mm]
\CZigDown 
&	=  \mirror(\CZigUp)
\\[1mm]
\CZag 
&	=  \rot(\CZigUp)
\\[2mm]
\CEnd
&	=  (\bC0 \NWb \bC1 \NWb \bC2 \, \NWb )^{2k-1} \bC0 \NWb \bC1 \NWb \bC2 \, \Eb 
\\
&		\retraitAlign 
		(\bC3 \NEb \bC4 \NEb \bC5 \, \NEb)^{k} \, \bC3 \, \SEb 	
\\
&		\retraitAlign 
		\bC7 \SWb  \bC8 \SWb (\bC6 \SWb \bC7 \SWb \bC8 \, \SWb ) ^{k-1} \, \bC9 \, \SWb
\\
&		\retraitAlign 
		(\bC{10} \SEb \bC{11} \SEb \bC{12} \SEb \bC{13} \SEb \bC{14} \SEb \bC9 \, \SEb)^{k-1}
		\SEb \bC{10} \SEb \bC{15} \SEb \bC{16} \SEb \bC{13}
\end{align*}

Figures~\ref{fig:C:brick:ZigUp} and Figure~\ref{fig:C:brick:End} display bricks $\CZigUp$ and $\CEnd$ respectively.

\figBrickHere%
	{C-End_Word_Tape-Upright}%
	{Module \ModuleC: Brick $\CZigUp$.}%
	{fig:C:brick:ZigUp}
	
\figBrickHere%
	{C-End_of_Tape-Expanded}%
	{Module \ModuleC: Brick $\CEnd$.}%
	{fig:C:brick:End}
	
\subsubsection{Subrule for Module C}	

Module \ModuleC\/ interacts with \ModuleA, \ModuleB, \ModuleDx{}, \ModuleE\/ and \ModuleF\/ and might interact with \ModuleG.
Figure~\ref{fig:C:subrule} presents the subrule for the interactions between the beads of $\ModuleC$ and the beads of the other modules.

\figSubRuleHere%
	{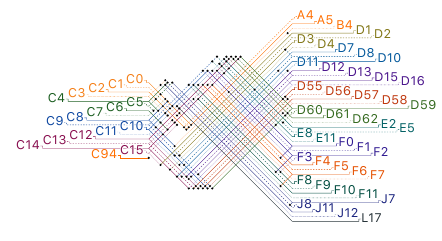}%
	{Subrule for Module \ModuleC.}%
	{fig:C:subrule}%
			
\afterpage{\clearpage}
	
\newpage
\subsection{Modules D} 
\label{sec:full:description:D}

\subsubsection{Bead type sequence for Modules D} 
Length: $3W+30 = 6(\lambda+5)$;\\ uses: 111 proper bead types, $\bbD{0}{62}$ and $\bbE{0}{47}$; plus 2 special bead types from~\ModuleG, $\bbL{17}{18}$

\begin{align*}
\SEGD0 & =  \bSB{\bbD{23}{33}}\conc \bbE{6}{11}\conc (\bbE{0}{11})^{\kappa-1}
\\
\SEGD1 & =  (\bbE{12}{23})^{\kappa}\conc \bSB{\bbD{49}{45}}
\\
\SEGD2 & =  \bSB{\bbD{34}{44}}\conc \bbE{30}{35}\conc (\bbE{24}{35})^{\kappa-1}
\\
\SEGD3 & =  (\bbE{36}{47})^{\kappa}\conc \bSB{\bbD{54}{50}}
\\
\ModuleDU{2,0} 
&	= \SEGD0\conc \SEGD1\conc  \SEGD2\conc \SEGD3\conc  \SEGD0\conc \SEGD1
\\
\ModuleDU{1,0} 
&	= \SEGD2\conc \SEGD3\conc  \SEGD0\conc \SEGD1\conc \SEGD2\conc \SEGD3
\\ 
\ModuleDU{0,0} 
&	= \ModuleDU{2,0}\subst{\bS{\bbD{0}{16}}@0..16}
\\
\ModuleDU{r,1} 
&	= \ModuleDU{r,0}\subst{\bS{\bD{17}}@(3W+22), \bS{\bbD{18}{22}}@(3W+25)..(3W+29)} 
& 	\text{for $r\in\{0,1,2\}$}
\\
\ModuleDZ{r,t} 
&	= \ModuleDU{r,t}\subst{\bS{\bL{17}}@(3w+1), \bS{\bL{18}}@(3w+2), \bSP{\bbD{55}{62}}@(3w+6)..(3w+13)}
& 	\begin{array}[t]{r} 
		\text{for $r\in\{0,1,2\}$}
		\\
		\text{and $t\in\{0,1\}$}
	\end{array}
\end{align*}
	
\subsubsection{The bricks for the modules D}

\paragraph{The zig and zag brick conformations for Module D}

\begin{align*}
\SegDZigUp{0} 
&	= 	\NEpath\!\left(\bbD{23}{33}\conc\bbE{6}{11}\conc(\bbE{0}{11})^{\kappa-1}\right)
\\
\SegDZigUp{1} 
&	=	\SWpath\!\left( (\bbE{12}{23})^{\kappa}\conc \bbD{49}{45} \right)
\\
\SegDZigUp{2} 
&	=	\NEpath\!\left( \bSB{\bbD{34}{44}}\conc \bbE{30}{35}\conc (\bbE{24}{35})^{\kappa-1}\right)
\\
\SegDZigUp{3} 
&	=	\SWpath\!\left( (\bbE{36}{47})^{\kappa}\conc \bbD{54}{50} \right)
\\[2mm]
\DUZigUp{2,0} 
&	= \SegDZigUp{0} \Eb \SegDZigUp{1} \Eb \SegDZigUp{2} \Eb \SegDZigUp{3} \Eb \SegDZigUp{0} \Eb \SegDZigUp{1} 
\\
\DUZigUp{1,0} 
&	= \SegDZigUp{2} \Eb \SegDZigUp{3} \Eb \SegDZigUp{0} \Eb \SegDZigUp{1} \Eb \SegDZigUp{2} \Eb \SegDZigUp{3} 
\\
\DUZigUp{0,0}  
&	= \DUZigUp{2,0}\subst{(\NEpath(\bbD{0}{16})\, \NEb) @0..16}
\\
\DUZigUp{r,1} 
&	= \DUZigUp{r,0}\subst{(\bD{17}\,\SWb )@(3W+22), \SWpath(\bbD{18}{22})@(3W+25)..(3W+29)} 
& 	\text{for $r\in\{0,1,2\}$}
\\
\DZZigUp{r,t} 
&	= \DUZigUp{r,t}\substOpen{(\bL{17} \NEb \bL{18} \, \NEb)@(3w+1..3w+2),}
\\[-2mm]
&	\hspace*{2cm} \substClose{(\NEpath(\bbD{55}{62})\,\NEb)@(3w+6)..(3w+13)}
&	\begin{array}[t]{r} 
		\text{for $r\in\{0,1,2\}$}
		\\
		\text{and $t\in\{0,1\}$}
	\end{array}
\end{align*}
\\
The zig-down and zag brick conformations are obtained by mirroring and rotating the zig-up brick conformation: for $x\in\{\word0,\word1\}$, $r\in\{0,1,2\}$ and $t\in\{0,1\}$, we have 
\begin{align*}
\DxrtZigDown{$(x)$}{r,t}
&	= \mirror(\DxrtZigUp{$(x)$}{r,t})
\\
\DxrtZag{$(x)$}{r,t}
&	= \rot(\DxrtZigUp{$(x)$}{r,t})
\end{align*}


Figure~\ref{fig:D0:brick:ZigUp} displays the bricks $\DZZigUp{r,t}$.

\figBrickHereSize%
	{D-NewLetter-0-Zig}%
	{Modules \ModuleDZ{r,t}: Bricks $\DZZigUp{r,t}$.}%
	{fig:D0:brick:ZigUp}%
	{1.3\textwidth}%
	{\textheight}
	
\afterpage{\clearpage}
\newpage

\paragraph{The append brick conformations for Module D}

\begin{align*}
\SegDAppend{0} 
&	= 	\Epath(\bbD{23}{27}) \SEb \SEpath(\bbD{28}{30}) \SWb \bD{31} \Eb \bD{32} \NEb \bD{33} \NEb
\\
&	\retraitAlign
		\Eglider'\!\left(\bbE{6}{11}\conc(\bbE{0}{11})^{\kappa-1}\right)
\\
\SegDAppend{1} 
&	=	\Eglider'\!\left( (\bbE{12}{23})^{\kappa}\right) \NWb \Wpath(\bbD{49}{45})
\\
\SegDAppend{2} 
&	= 	\Epath(\bbD{34}{38}) \SEb \SEpath(\bbD{39}{41}) \SWb \bD{42} \Eb \bD{43} \NEb \bD{44} \NEb
\\
&	\retraitAlign
		\Eglider'\!\left(\bbE{30}{35}\conc(\bbE{24}{35})^{\kappa-1}\right)
\\
\SegDAppend{3} 
&	=	\Eglider'\!\left( (\bbE{36}{47})^{\kappa}\right)\NWb \Wpath(\bbD{54}{50})
\\
\SegDAppend{\textit{init}} 
&	=	\bD{0} \Eb \bD{1} \SWb \bD{2} \Eb \bD{3} \SEb \bD{4} \SWb \bD{5} \Eb \bD{6} \NEb \bD{7} \NWb \bD{8} \NWb \bD{9} \NWb \bD{10} \Wb
\\
&	\retraitAlign  \bD{11} \NWb \bD{12} \Eb \bD{13} \Eb \bD{14} \SEb \bD{15} \SEb \bD{16}
\\
\SegDAppend{\textit{end}} 
&	=	\bD{18} \Eb \bD{19} \SEb \bD{20} \SEb \bD{21} \SWb \bD{22}
\\
\SegDAppend{\textit{spike}} 
&	=	\bD{55} \SWb \bD{56} \SEb \bD{57} \NEb \bD{58} \NEb \bD{59} \NEb \bD{60} \SEb \bD{61} \SWb \bD{62}
\\
\\[2mm]
\DUAppend{2,0} 
&	= \SegDAppend{0} \Eb \SegDAppend{1} \NEb \SegDAppend{2} \Eb \SegDAppend{3}
\NEb \SegDAppend{0} \Eb \SegDAppend{1} 
\\
\DUAppend{1,0} 
&	= \SegDAppend{2} \Eb \SegDAppend{3} \NEb \SegDAppend{0} \Eb \SegDAppend{1}
\NEb \SegDAppend{2} \Eb \SegDAppend{3} 
\\
\DUAppend{0,0}  
&	= \DUAppend{2,0}\subst{(\SegDAppend{\textit{init}}\, \SEb) @0..16}
\\
\DUAppend{r,1} 
&	= \DUAppend{r,0}\substOpen{(\bD{17}\,\NEb )@(3W+22),}
\\[-2mm]
&	\hspace*{2.9cm} \substClose{(\NWb \SegDAppend{\textit{end}})@(3W+25)..(3W+29)} 
	\quad \text{for $r\in\{0,1,2\}$}
\\
\DZAppend{r,t} 
&	= \DUAppend{r,t}\substOpen{(\bL{17} \Eb \bL{18} \, \NEb)@(3w+1..3w+2),}
\\[-2mm]
&	\hspace*{2.9cm} \substClose{(\SegDAppend{\textit{spike}} \,\Eb)@(3w+6)..(3w+13)}
	\quad
	\begin{array}[t]{r} 
		\text{for $r\in\{0,1,2\}$}
		\\
		\text{and $t\in\{0,1\}$}
	\end{array}
\end{align*}

Figure~\ref{fig:brick:D:write} displays the bricks $\DZAppend{r,t}$.

\begin{sidewaysfigure}
  \subfigBrickLandscape{D-NewLetter0-0-Expanded.pdf}
    {The brick $\DZAppend{0,0}$.}%
    {fig:D:brick:000:Append}
  \subfigBrickLandscape{D-NewLetter0-1-Expanded.pdf}
    {The brick $\DZAppend{1,0}$.}%
    {fig:D:brick:010:Append}
  \subfigBrickLandscape{D-NewLetter0-2-Expanded.pdf}%
    {The brick $\DZAppend{2,0}$.}%
    {fig:D:brick:020:Append}

  \subfigBrickLandscape{D-NewLetter0-0Last-Expanded.pdf}%
    {The brick $\DZAppend{0,1}$.}%
    {fig:D:brick:001:Append}
  \subfigBrickLandscape{D-NewLetter0-1Last-Expanded.pdf}%
    {The brick $\DZAppend{1,1}$.}%
    {fig:D:brick:021:Append}
  \subfigBrickLandscape{D-NewLetter0-2Last-Expanded.pdf}%
    {The brick $\DZAppend{2,1}$.}%
    {fig:D:brick:021:Append}

  \caption{Module $\ModuleDZ{r,t}$: Bricks $\DWrite$.}
  \label{fig:brick:D:write}
\end{sidewaysfigure}
	
\subsubsection{Subrule for Module D}	

Module \ModuleDx{}\/ interacts with \ModuleA, \ModuleB, \ModuleC, \ModuleE\/ and \ModuleG.
Figure~\ref{fig:D:subrule} presents the subrule for the interactions between the beads of $\ModuleDx{}$ and the beads of the other modules.

\figSubRuleHere%
	{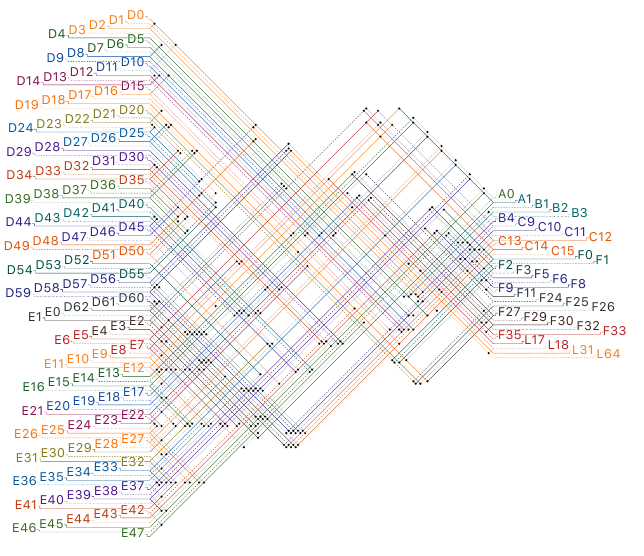}%
	{Subrule for Module \ModuleDx{}.}%
	{fig:D:subrule}%

\afterpage{\clearpage}
\newpage

\subsection{Modules E} 
\label{sec:full:description:E}

\subsubsection{Bead type sequences for Modules E} 
Length: $6\lambda(L-a+P)+8h-1$; \\ 
146 bead types used: $\bbF{0}{51}$, $\bbG{0}{48}$, $\bbH{0}{24}$, and $\bbI{0}{19}$.
\begin{align*}
\SEGE{A} 
&	= \bigl((\bbF{0}{11})^\kappa \conc (\bbF{12}{23})^\kappa \conc (\bbF{24}{35})^\kappa \conc (\bbF{36}{47})^\kappa\bigr)^\infty
\\
\SEGE{B} 
&	= \bigl((\bbG{0}{11})^\kappa \conc (\bbG{12}{23})^\kappa \conc (\bbG{24}{35})^\kappa \conc (\bbG{36}{47})^\kappa\bigr)^\infty
\\
\HEAD{E$_a$} 
&	= \SEGE{A}_{b..b+3c(a)-2} \conc \bS{\bF{51}} \conc \SEGE{B}_{b+3c(a)..b+K-2}
&	\begin{array}[t]{r} 
		\text{\quad where $b= 0$ if $a$ is even,}
		\\ 
		\text{and $b=2\lambda$ if $a$ is odd}
		\end{array}
\\[2mm]
\SEGE{C} 
&	= \bSN{\bbH{0}{4}} \conc (\bbH{5}{16})^{q-1} \conc \bbH{5}{10} \conc \bSC{\bbH{17}{24}}
\\
\SEGE{D$_0$}
&	= \bSC{\bI{15}} \conc \bbI{1}{5} \conc (\bbI{0}{5})^{k-1} \conc \bbI{0}{1} \conc \bSC{\bI{18}}
\\
\SEGE{D$_1$} 
&	= \bSC{\bI{19}} \conc \bbI{7}{8} \conc (\bbI{6}{8})^{2k-1}\conc \bbI{6}{7}\conc \bSC{\bbI{15}{16}}
\\
\SEGE{D$_2$}
&	= \bSC{\bI{17}}\conc \bbI{10}{11} \conc (\bbI{9}{11})^{2k-1} \conc \bbI{9}{10}\conc \bSC{\bI{19}}
\\
\SEGE{D$_3$}
&	= \bSC{\bI{18}} \conc \bbI{13}{14} \conc (\bbI{12}{14})^{2k-1} \conc \bbI{12}{13} \conc \bSC{\bI{19}}
\\
\SEGE{D$_4$}
&	= \bSC{\bI{19}} \conc \bbI{1}{2} \conc (\bbI{0}{2})^{2k}
\\
\TAIL{E} 
&	= \SEGE{C} \conc \SEGE{D$_0$} \conc \SEGE{D$_1$} \conc \SEGE{D$_2$} \conc \SEGE{D$_3$} \conc \SEGE{D$_4$}
\\[2mm]
\ModuleEa{0} 
&	= \HEAD{E$_0$}\subst{\bS{\bbF{48}{49}}@0..1, \bS{\bF{50}}@11} \conc \bS{\bG{48}} \conc \TAIL{E}
\\
\ModuleEa{a>0} 
&	= \HEAD{E$_{a}$}\conc \bS{\bG{48}} \conc \TAIL{E}
\end{align*}

\afterpage{\clearpage}
\newpage

\subsubsection{The bricks for the modules E}

\paragraph{Zig-up, zig-down and zag bricks for Module E}

\begin{align*}
\SEGE{A\ZigUp} 
&	= \Bigl(\NEpath\bigl((\bbF{0}{11})^\kappa\bigr) \Eb 
		 \SWpath\bigl((\bbF{12}{23})^\kappa\bigr) \Eb
\\[-2mm]	&	\retraitAlign
		\NEpath\bigl((\bbF{24}{35})^\kappa\bigr) 
		\Eb \SWpath\bigl((\bbF{36}{47})^\kappa\bigr)\Bigr)^\infty
\\
\SEGE{B\ZigUp} 
&	= \Bigl(\NEpath\bigl(\bbG{0}{11})^\kappa\bigr) \Eb 
		\SWpath\bigl((\bbG{12}{23})^\kappa\bigr) \Eb
\\[-2mm]	&	\retraitAlign
		 \NEpath\bigl((\bbG{24}{35})^\kappa\bigr) \Eb
		 \SWpath\bigl((\bbG{36}{47})^\kappa\bigr)\Bigr)^\infty
\\
\HEAD{E$_a$\ZigUp} 
&	= \SEGE{A\ZigUp}_{b..b+3c(a)-2} ~ d ~ \bS{\bF{51}} ~ d ~ \SEGE{B\ZigUp}_{b+3c(a)..b+K-2}
&	\begin{array}[t]{r} 
		\makebox[1cm][r]{\text{\quad where $b= 0$ if $a$ is even,}}
		\\ 
		\makebox[1cm][r]{\text{and $b=2\lambda$ if $a$ is odd;}}
		\\
		\makebox[1cm][r]{\text{and $d = \NEb$ if $\lfloor\frac{3c-2}{\lambda}\rfloor$ is even, else $d=\SWb$}}  
		\end{array}
\\[2mm]
\SEGE{C\ZigUp} 
&	= \bH{0} \Eb \bH{1} \NWb \bH{2} \Eb \bH{3} \SEb \bH{4} \NEb 
\\	&	\retraitAlign
		\NEglider\bigl((\bbH{5}{16})^{q-1} \conc \bbH{5}{10} \conc \bbH{17}{22}\bigr)
		\NEb \bH{23} \NWb \bH{24}
\\
\SEGE{D$_0$\ZigUp}
&	= \bI{15} \SEb \SWpath\bigl(\bbI{1}{5} \conc (\bbI{0}{5})^{k-1} \conc \bbI{0}{1} \conc \bI{18}\bigr)
\\
\SEGE{D$_1$\ZigUp} 
&	= \NEpath\bigl(\bI{19} \conc \bbI{7}{8} \conc (\bbI{6}{8})^{2k-1}\conc \bbI{6}{7}\bigr) \NWb \bI{15} \Eb \bI{16}
\\
\SEGE{D$_2$\ZigUp}
&	= \SWpath\bigl(\bI{17}\conc \bbI{10}{11} \conc (\bbI{9}{11})^{2k-1} \conc \bbI{9}{10}\conc \bI{19}\bigr)
\\
\SEGE{D$_3$\ZigUp}
&	= \NEpath\bigl(\bI{18} \conc \bbI{13}{14} \conc (\bbI{12}{14})^{2k-1} \conc \bbI{12}{13} \conc \bI{19}\bigr)
\\
\SEGE{D$_4$\ZigUp}
&	= \SWpath\bigl(\bI{19} \conc \bbI{1}{2} \conc (\bbI{0}{2})^{2k}\bigr)
\\
\TAIL{E\ZigUp} 
&	= \SEGE{C\ZigUp} \Eb \SEGE{D$_0$\ZigUp} \Eb \SEGE{D$_1$\ZigUp} \Eb \SEGE{D$_2$\ZigUp} \Eb \SEGE{D$_3$\ZigUp} \Eb \SEGE{D$_4$\ZigUp}
\\[2mm]
\EaZigUp{0} 
&	= \HEAD{E$_0$\ZigUp}\subst{(\bF{48}\NEb \bF{49}\,\NEb)@0..1, (\bF{50}\,\NEb)@11} \SWb \bG{48} 
		\Eb \TAIL{E\ZigUp}
\\
\EaZigUp{a>0} 
&	= \HEAD{E$_{a}$\ZigUp}\SWb \bG{48} \Eb \TAIL{E\ZigUp}
\\[2mm]
\EaZigDown{a\geq 0} 
&	= \mirror(\EaZigUp{a})
\\
\EaZag{a\geq 0} 
&	= \rot(\EaZigUp{a})
\end{align*}

\begin{sidewaysfigure}[h]
  \begin{subfigure}{\textheight}
    \centering
    \adjustbox{max size= {\textheight}{.4\textwidth}}{
    \begin{tikzpicture}
    \node () at (0,0){\includegraphics[width=15cm,height=15cm,keepaspectratio]{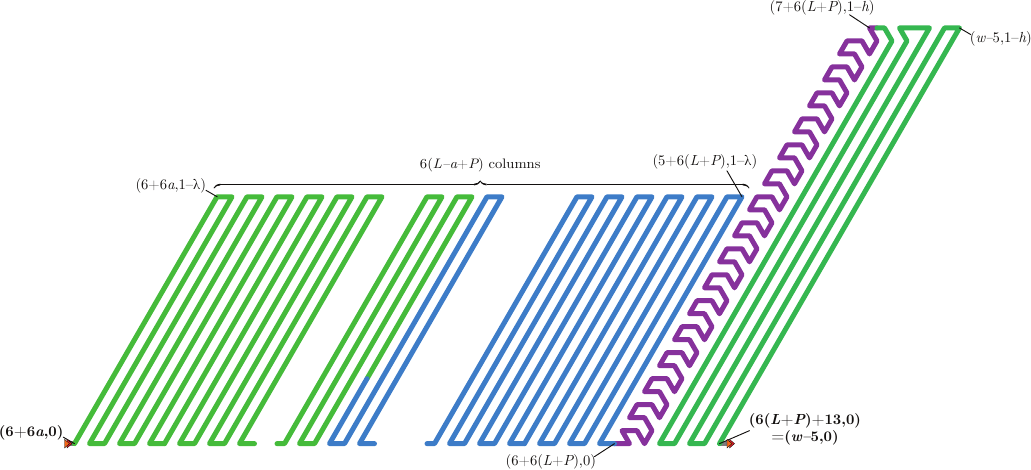}};
%
%
    \node () at (-3.5cm,-1cm) {\small\begin{tabular}{c} Segment $\SEGE{A}$\\ folded as brick $\SEGE{A\ZigUp}$\end{tabular}};
    \node () at (0.3cm,-1cm) {\small\begin{tabular}{c} Segment $\SEGE{B}$\\ folded as brick $\SEGE{B\ZigUp}$\end{tabular}};
    \node () at (2.65cm,2cm) {\small\begin{tabular}{r} Segment $\SEGE{C}$\\ folded as brick $\SEGE{C\ZigUp}$\hspace*{2.5mm}\end{tabular}};
    \node [anchor=west] () at (4cm,0cm) {\small\begin{tabular}{l} \hspace*{2.5mm} Segment $\SEGE{D\ensuremath{_{0..4}}}$\\ folded as brick $\SEGE{D\ensuremath{_{0..4}}\ZigUp}$\end{tabular}};
    \end{tikzpicture}
    }
    \caption{Module $\ModuleE$: Blueprint of the brick $\EaZigUp{a}$.}
    \label{fig:brick:E:upright:outline}
  \end{subfigure}
  \begin{subfigure}{\textheight}
    \hspace*{-1cm}\adjustbox{max size= {1.075\textheight}{.6\textwidth}}{
    \begin{tikzpicture}
    \node () at (0,0){\includegraphics[scale=1.25]{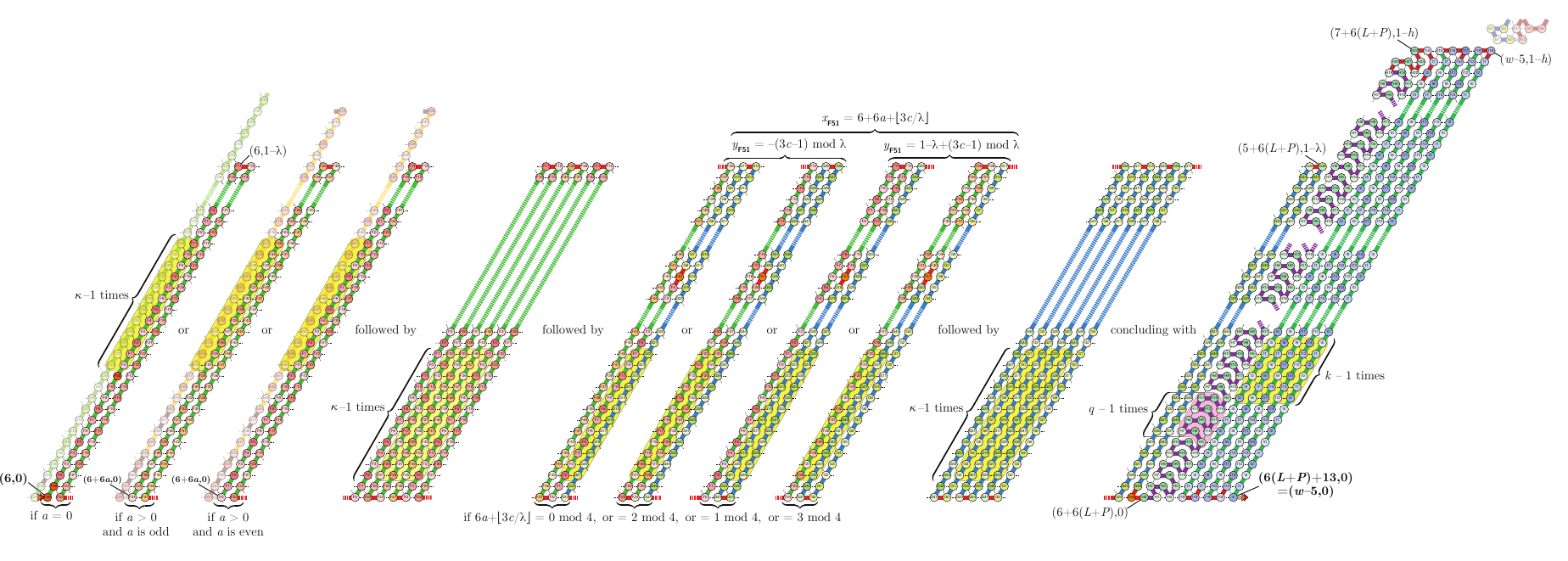}};
%
%
    \node () at (-16.5cm,-6.75cm) {\small\begin{tabular}{c} Beginning of brick $\SEGE{A\ZigUp}$\end{tabular}};
    \node () at (-10cm,-6.25cm) {\small\begin{tabular}{c} Repeated\\bond pattern\\ inside brick $\SEGE{A\ZigUp}$\end{tabular}};
    \node () at (-2.75cm,-6.5cm) {\small\begin{tabular}{r} Junction between bricks $\SEGE{A\ZigUp}$ and $\SEGE{B\ZigUp}$\end{tabular}};
    \node () at (4.5cm,-6.25cm) {\small\begin{tabular}{c} Repeated\\bond pattern\\ inside brick $\SEGE{B\ZigUp}$\end{tabular}};
    \node () at (9.5cm,-6.5cm) {\small\begin{tabular}{r} Bricks $\SEGE{C\ZigUp}$ and $\SEGE{D\ZigUp}$\end{tabular}};
    \end{tikzpicture}
    }
    \caption{Precise description of the brick $\EaZigUp{a}$.}
    \label{fig:brick:E:upright:details}
  \end{subfigure}
  \caption{Module $\ModuleEa{a}$: Bricks $\EaZigUp{a}$.}
  \label{fig:brick:E:upright}
\end{sidewaysfigure}

Figure~\ref{fig:brick:E:upright} displays the bricks $\EaZigUp{a}$.

\afterpage{\clearpage}
\newpage

\paragraph{Carriage-return bricks for Module E}

\begin{align*}
\SEGE{A\bZigZag} 
&	= \Eglider\bigl(\SEGE{A}\bigr)
\\
\SEGE{B\bZigZag} 
&	= \Wpath\bigl(\SEGE{B}\bigr)
\\
\HEAD{E$_a$\bZigZag} 
&	= \SEGE{A\bZigZag}_{b..b+3c(a)-2} \SWb \bF{51} \SEb \SEGE{B\bZigZag}_{b+3c(a)..b+K-2}
&	\begin{array}[t]{r} 
		\makebox[0em][r]{\text{\quad where $b= 0$ if $a$ is even,}}
		\\ 
		\makebox[0em][r]{\text{and $b=2\lambda$ if $a$ is odd;}}
		\end{array}
\\[2mm]
\SEGE{C\bZigZag} 
&	= \Wpath\bigl(\bbH{0}{4} \conc (\bbH{5}{16})^{q-1} \conc \bbH{5}{10} \conc \bbH{17}{19}\bigr)
		\SEb \bH{20} \Eb \bH{21} \Eb \bH{22} \SWb \bH{23} \Wb \bH{24}
\\
\SEGE{D$_0$\bZigZag}
&	= \bI{15} \NWb \bI{1} \NWb \Wpath\bigl( \bbI{2}{5} \conc (\bbI{0}{5})^{k-1} \conc \bbI{0}{1}\bigr)
		\SEb \bI{18}
\\
\SEGE{D$_1$\bZigZag} 
&	= \bI{19} \NWb \Wpath\bigl(\bbI{7}{8} \conc (\bbI{6}{8})^{2k-1}\conc \bbI{6}{7}\bigr) \SEb \bI{15} \Wb \bI{16}
\\
\SEGE{D$_2$\bZigZag}
&	= \bI{17}\NEb \Wpath\bigl(\bbI{10}{11} \conc (\bbI{9}{11})^{2k-1} \conc \bbI{9}{10}\bigr) \SEb \bI{19}
\\
\SEGE{D$_3$\bZigZag}
&	= \bI{18}  \NWb \Wpath\bigl(\bbI{13}{14} \conc (\bbI{12}{14})^{2k-1} \conc \bbI{12}{13}\bigr) \SWb \bI{19}
\\
\SEGE{D$_4$\bZigZag}
&	= \bI{19} \NEb \Wpath\bigl(\bbI{1}{2} \conc (\bbI{0}{2})^{2k}\bigr)
\\
\TAIL{E\bZigZag} 
&	= \SEGE{C\bZigZag} \Wb \SEGE{D$_0$\bZigZag} \Wb \SEGE{D$_1$\bZigZag} \Wb \SEGE{D$_2$\bZigZag} \Wb \SEGE{D$_3$\bZigZag} \Wb \SEGE{D$_4$\bZigZag}
\\[2mm]
\EaCR{0} 
&	= \HEAD{E$_0$\bZigZag}\subst{(\bF{48}\Eb \bF{49}\,\SWb)@0..1, (\bF{50}\,\Eb)@11} \Wb \bG{48} 
		\Wb \TAIL{E\bZigZag}
\\
\EaCR{a>0} 
&	= \HEAD{E$_{a}$\bZigZag}\Wb \bG{48} \Wb \TAIL{E\bZigZag}
\end{align*}

\begin{figure}[h]
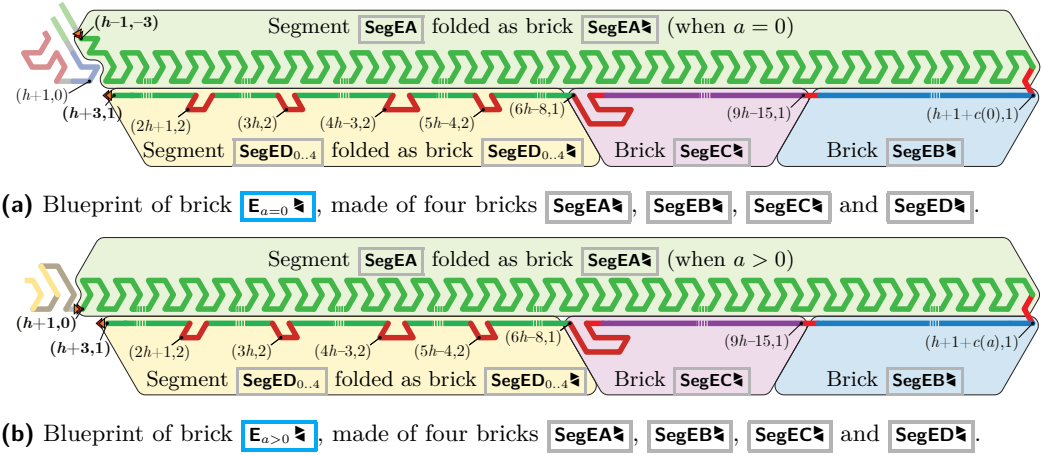

	\newSubFigBrick%
		{E-Padding-Outline0.pdf}
		{Blueprint of brick $\EaCR{a=0}$, made of four bricks $\SEGE{A\bZigZag}$, $\SEGE{B\bZigZag}$, $\SEGE{C\bZigZag}$ and $\SEGE{D\bZigZag}$.}
		{fig:brick:schema:EL:CR}
		{\textwidth}
		{%
			\nodeLabel{0}{0.8}{c}{Segment $\SEGE{A}$ folded as brick $\SEGE{A\bZigZag}$ (when $a=0$)}
			\nodeLabel{-2.45}{-1}{c}{Segment $\SEGE{D\ensuremath{_{0..4}}}$ folded as brick $\SEGE{D\ensuremath{_{0..4}}\bZigZag}$}
			\nodeLabel{2.2}{-1}{c}{Brick $\SEGE{C\bZigZag}$}
			\nodeLabel{5.3}{-1}{c}{Brick $\SEGE{B\bZigZag}$}
		}
		{} 

	\newSubFigBrick%
		{E-Padding-Outline1.pdf}
		{Blueprint of brick $\EaCR{a>0}$, made of four  bricks $\SEGE{A\bZigZag}$, $\SEGE{B\bZigZag}$, $\SEGE{C\bZigZag}$ and $\SEGE{D\bZigZag}$.}
		{fig:brick:schema:Ea<L:CR}
		{\textwidth}
		{%
			\nodeLabel{0}{0.8}{c}{Segment $\SEGE{A}$ folded as brick $\SEGE{A\bZigZag}$ (when $a>0$)}
			\nodeLabel{-2.4}{-0.95}{c}{Segment $\SEGE{D\ensuremath{_{0..4}}}$ folded as brick $\SEGE{D\ensuremath{_{0..4}}\bZigZag}$}
			\nodeLabel{2.2}{-0.95}{c}{Brick $\SEGE{C\bZigZag}$}
			\nodeLabel{5.3}{-0.95}{c}{Brick $\SEGE{B\bZigZag}$}
		}
		{} 
		
  \caption{Outline of the four different parts of module $\ModuleEa{a}$, when folded at the end of the appended appendant. See Figure~\ref{fig:brick:E:CR} for the detailed beads of each part.}
  \label{fig:E-outline}
\end{figure}

Figures~\ref{fig:E-outline} and~\ref{fig:brick:E:CR} display the blueprints and the description of bricks $\EaCR{a}$ respectively.

\begin{figure}[h]
  \subfigBrickPortrait{E-Padding-EA0.pdf}%
  {The $\SEGE{A}$-part of brick \HEAD{E$_{a=0}$\bZigZag}, when $a=0$.}%
  {fig:brick:EA0:CR}

  \subfigBrickPortrait{E-Padding-EA1.pdf}%
  {The $\SEGE{A}$-part of brick $\HEAD{E$_a$\bZigZag}$, when $a$ is odd.}%
  {fig:brick:EA1:CR}

  \subfigBrickPortrait{E-Padding-EA2.pdf}%
  {The $\SEGE{A}$-part of brick $\HEAD{E$_{a>0}$\bZigZag}$, when $a$ is even and positive.}%
  {fig:brick:EA2:CR}

  \subfigBrickPortrait{E-Padding-EB.pdf}%
  {The $\SEGE{B}$-part of brick $\HEAD{E$_a$\bZigZag}$.}%
  {fig:brick:EB:CR}\\[1mm]

  \subfigBrickPortrait{E-Padding-EC.pdf}%
  {The brick $\SEGE{C\bZigZag}$.}%
  {fig:brick:EC:CR}

  \subfigBrickPortrait{E-Padding-ED.pdf}%
  {The brick $\SEGE{D$_{0..4}$\bZigZag}$.}%
  {fig:brick:ED:CR}
  \caption{The bricks $\EaCR{a}$.}
  \label{fig:brick:E:CR}
\end{figure}

\subsubsection{Subrule for Module E}	

Module \ModuleE\/ interacts with \ModuleB, \ModuleC, \ModuleDx{} and \ModuleF.
Figure~\ref{fig:E:subrule} presents the subrule for the interactions between the beads of $\ModuleE$ and the beads of the other modules.

\figSubRuleHere%
	{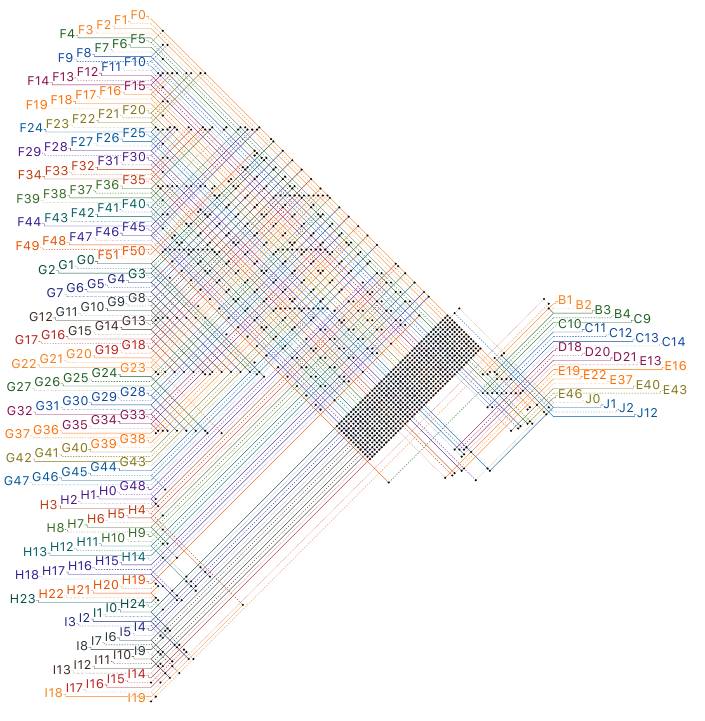}%
	{Subrule for Module \ModuleE.}%
	{fig:E:subrule}%

\afterpage{\clearpage}
\newpage

\subsection{Module F} 
\label{sec:full:description:F}

\subsubsection{Bead type sequence for Module F} 
Length: $4h$; 53 bead types used: $\bbJ{0}{52}$.
\begin{align*}
\HEAD F 
&	= \bSN{\bbJ{0}{4}} \conc (\bbJ{5}{10})^{3k-1} \conc \bbJ{5}{7} \conc \bSN{\bbJ{11}{23}}
\\
\TAIL F 
&	= \bJ{48} \conc (\bbJ{51}{48})^{9} \conc \bJ{51} \conc \bSN{\bJ{52}} \conc \bbJ{49}{48}
\\[2mm]
\SEG{Exp}{$(2i)$} 
&	= \bSE{\bbJ{24}{29}} \conc (\bbJ{30}{35})^{3^{2i-1}-1}
&	\text{for $i\geq 1$}
\\
\SEG{Exp$(2i+1)$}{} 
&	= \bSE{\bbJ{36}{41}} \conc (\bbJ{42}{47})^{3^{2i}-1}
&	\text{for $i\geq 1$}
\\
\SEG{Exp}{F} 
&	= \concat{i\geq2}  \SEG{Exp$(i)$}{}
\\
\ModuleF 
&	= \HEAD F \cdot \left(\bSE{\bbJ{39}{41}}\conc \SEG{Exp}{F}_{0..(h-51)}\right)^R \conc  \TAIL F
\end{align*}

\subsubsection{The bricks for the modules F}

Module $\ModuleF$ adopts three brick conformations: zig-up $\FZigUp$, zig-down $\FZigDown$, and zag $\FZag$. the two last are obtained by mirroring and rotating the first.
\begin{align*}
\HEAD{F\ZigUp} 
&	= \bJ0 \Eb \bJ1\NWb \bJ2 \Eb \bJ3 \SEb\bJ4 \NEb \NEglider\left((\bbJ{5}{10})^{3k-1}\right)
	 	\NEb \bJ{5} \NWb \bJ{6} \Wb \bJ{7} \NEb
\\	&	\retraitAlign
		 \bJ{11} \NEb \bJ{12} \SEb \bJ{13} \SEb \bJ{14}
		\NEb \bJ{15} \NWb \bJ{16}  \Eb \bJ{17} \Eb \SWpath(\bbJ{18}{23})
\\
\TAIL{F\ZigUp} 
&	= \SWpath(\bJ{48} \conc (\bbJ{51}{48})^{9} \conc \bJ{51} \conc \bJ{52} \conc \bbJ{49}{48})
\\[2mm]
\text{Recall that } \SEG{Exp}{F} 
&	= \concat{i\geq2}  \SEG{Exp$(i)$}{}
\\
\FZigUp 
&	= \HEAD{F\ZigUp} \, \SWb \, \SWpath\!\left(\bbJ{39}{41}\conc \SEG{Exp}{F}_{0..(h-51)}\right)^R \, \SWb  \, \TAIL{F\ZigUp}
\\[2mm]
\FZigDown 
&	= \mirror(\FZigUp)
\\[1mm]
\FZag 
&	= \rot(\FZigUp)
\end{align*}


Figure~\ref{fig:F:brick:ZigUp} displays the brick $\FZigUp$.

\figBrickHereSize%
	{F-Zag_init-Upright}%
	{Module \ModuleF: Brick $\FZigUp$.}%
	{fig:F:brick:ZigUp}
	{1.3\textwidth}
	{\textheight}

\subsubsection{Subrule for Module F}	

Module \ModuleF\/ interacts with \ModuleA, \ModuleB, \ModuleC, \ModuleE\/ and \ModuleG.
Figure~\ref{fig:F:subrule} presents the subrule for the interactions between the beads of $\ModuleF$ and the beads of the other modules.

\figSubRuleHere%
	{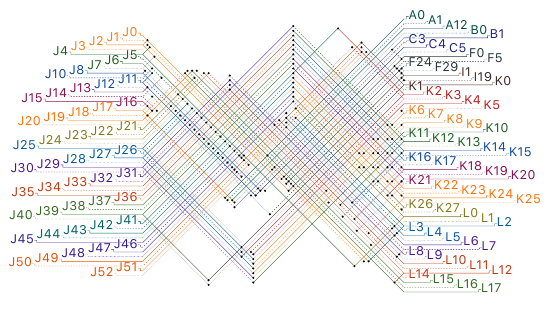}%
	{Subrule for Module \ModuleF.}%
	{fig:F:subrule}%

\afterpage{\clearpage}
\newpage

\subsection{Module G} 
\label{sec:full:description:G}

\subsubsection{Bead type sequence for Module G} 
Length: $6h-1$;  201 bead types used: $\bbK{0}{69}$, $\bbL{0}{99}$, and $\bbM{0}{30}$.

\begin{align*}
\SEG{Exp'}{$(2i)$} 
&	= \bSE{\bbK{4}{9}} \conc (\bbK{10}{15})^{3^{2i-1}-1}
&	\text{for $i\geq 1$}
\\
\SEG{Exp'}{$(2i+1)$} 
&	= \bSE{\bbK{16}{21}} \conc (\bbK{22}{27})^{3^{2i}-1}
&	\text{for $i\geq 1$}
\\
\SEG{Exp}{G} 
&	= \concat{i\geq2}  \SEG{Exp'}{$(i)$}  
\\[1mm]
\SEG G1 
&	= \bS{\bbL{0}{6}} \conc \bK3 \conc (\bbK{0}{3})^9 \conc \bbK{0}{2} \conc \bS{\bbL{7}{10}} \conc \SEG{Exp}{G}_{8..h-51}
\\
\SEG G2 
&	= \bK{32} \conc \bK{33}\conc (\bbK{28}{33})^{k-3} \conc \bbK{28}{32}
\\
\SEG G3 
&	= \begin{array}[t]{l} 
		\bbK{35}{39} \conc (\bbK{34}{39})^{k-14}\conc \bK{34}\conc \bK{35} \conc \bSR{\bbL{39}{41}} \conc \bK{45} \conc (\bbK{40}{45})^{10}\, \conc
		\\ \quad
		  \bK{40}\conc \bK{41}
		  \end{array}
\\
\SEG G4
&	= \bbK{50}{51} \conc  (\bbK{46}{51})^{k-3}\conc \bbK{46}{48}
\\
\SEG G5
&	=  \begin{array}[t]{l} 
		\bbK{55}{57} \conc (\bbK{52}{57})^{k-6} \conc \bbK{52}{53} \conc \bS{\bL{74}\conc \bL{75}}  \conc \bbK{56}{57} \,\conc
		\\ \quad
		(\bbK{52}{57})^2 \conc \bbK{52}{53}
		\end{array}
\\
\SEG G6 
&	= \begin{array}[t]{l} 
		\bK{63}\conc (\bbK{58}{63})^{k-19} \conc \bbK{58}{61} \conc \bSC{\bbL{91}{99}\conc\bbM{0}{19}} \conc \bbK{67}{69} \, \conc 
		\\ \quad
		(\bbK{64}{69})^{10} \conc \bS{\bbM{20}{30}}
		\end{array}
\\[2mm]
\ModuleG 
&	= \SEG G1 \conc  \bS{\bbL{11}{24}}  \conc \SEG G{2}\conc  \bSCW{\bbL{25}{38}} \conc \SEG G{3}\conc  \bSCW{\bbL{42}{55}} \conc \SEG G{4}\, \conc\\
&	\hspace*{1cm}  \bSC{\bbL{56}{73}} \conc \SEG G{5} \conc  \bSC{\bbL{76}{90}} \conc \SEG G{6}
\end{align*}

\afterpage{\clearpage}
\newpage

\subsubsection{The bricks for the modules G}

\paragraph{Zig-up bricks for Module G}

$\ModuleG$ adopts two different bricks in zig-up phase, depending on the letter encoded in the zag-phase above: $\GReadZ$ and $\GReadU$.


\subparagraph{The $\protect\GReadZ$ brick.} ~

\begin{align*}
\SEG{G1Start\ZigUp}{Read} 
&	= \bL{0} \Wb \bL{1} \Wb \bL{2} \SEb \bL{3} \Eb \bL{4} \Eb \bL{5} \NEb \bL{6} \NEb \bK3
\\
\SEG{G1GliderNE\ZigUp}{Read} 
&	=  \begin{array}[t]{@{}l@{}l@{}}
		\NEpath\Bigl((\bbK{0}{3})^9 \conc \bbK{0}{2}\Bigr) \Eb \bL{7}\NWb \bL{8} \Eb \bL{9} \SEb \bL{10} 
		\NEb 
		\\ \retraitAlign
		\NEglider\Bigl(\SEG{Exp}{G}_{8..h-51} \conc \bbL{11}{23}\Bigr)
		\end{array} 
\\
\SEG{G2GliderNE\ZigUp}{Read} 
&	= \NEglider\bigl(\bL{24} \conc \bbK{32}{33}\conc (\bbK{28}{33})^{k-3} \conc \bbK{28}{32} \conc \bbL{25}{34}\bigr)
\\
\SEG{G3GliderNE\ZigUp}{Read} 
&	= \NEglider\bigl(\bbL{35}{38} \conc \bbK{35}{39} \conc (\bbK{34}{39})^{k-14}\conc \bK{34}\conc \bK{35} \conc \bL{39}\bigr)
\\[1mm]
\BigSEG{G1..3GliderNE\ZigUp}{Read} 
&	= \SEG{G1GliderNE\ZigUp}{Read} \NEb  \SEG{G2GliderNE\ZigUp}{Read} \NEb \SEG{G3GliderNE\ZigUp}{Read}
\\[3mm]
\SEG{G3GliderSE\ZigUp}{Read\word0} 
&	= \SERglider\bigl(\bbL{40}{41} \conc \bK{45} \conc (\bbK{40}{45})^{10} \conc
		  \bbK{40}{41}\conc \bL{42}\bigr)
\\
\SEG{G3SockNW\ZigUp}{Read\word0} 
&	= \NWpath(\bbL{43}{48}) \SWb \SEpath(\bbL{49}{54}) \Eb \bL{55}
\\
\SEG{G4GliderSE\ZigUp}{Read\word0}
&	= \SERglider\bigl(\bbK{50}{51} \conc  (\bbK{46}{51})^{k-3}\conc \bbK{46}{48} \conc \bL{56}\bigr)
\\
\SEG{G4SockNW\ZigUp}{Read\word0}
&	= \NWpath(\bbL{57}{64}) \SWb \SEpath(\bbL{65}{72}) \Eb \bL{73} 
\\
\SEG{G5GliderSE\ZigUp}{Read\word0}
&	=  \begin{array}[t]{@{}l@{}l@{}}
		\SERglider\Bigl(& 
		\bbK{55}{57} \conc (\bbK{52}{57})^{k-6} \conc \bbK{52}{53} \conc \bbL{74}{75}  \conc \bbK{56}{57} \, \conc
		\\
		& (\bbK{52}{57})^2 \conc \bbK{52}{53} \conc \bL{76} \Bigr)
		\end{array}
\\
\SEG{G5SockNW\ZigUp}{Read\word0}
&	= \NWpath(\bbL{77}{82}) \Wb \SEpath(\bbL{83}{89}) \Eb \bL{90} 
\\
\SEG{G6AGliderSE\ZigUp}{Read\word0} 
&	= \SERglider\bigl(\bK{63}\conc (\bbK{58}{63})^{k-19} \conc \bbK{58}{61} \conc \bL{91}\bigr)
\\
\SEG{G6SockNW\ZigUp}{Read\word0}
&	= \NWpath(\bbL{92}{99} \conc \bbM{0}{4}) \Wb \SEpath(\bbM{5}{18}) \Eb \bM{19} 
\\
\SEG{G6BGlider\ZigUp}{Read\word0} 
&	= \SERglider\bigl(\bbK{67}{69} \conc (\bbK{64}{69})^{10} \conc \bbM{20}{22}\bigr) \SEb 
		\bM{23}
\\
\SEG{G6End\ZigUp}{Read\word0} 
&	= \bM{24} \Eb \bM{25} \Eb \bM{26} \SWb \bM{27} \Wb \bM{28} \SWb \bM{29} \Eb \bM{30}
\\[2mm]
\GReadZ 
&	= \SEG{G1Start\ZigUp}{Read} \NEb \BigSEG{G1..3GliderNE\ZigUp}{Read} 
		\Eb \SEG{G3GliderSE\ZigUp}{Read\word0} \, \Wb  
\\	&	\retraitAlign
		 \SEG{G3SockNW\ZigUp}{Read\word0} \Eb 
		 \SEG{G4GliderSE\ZigUp}{Read\word0} \Wb
		  \SEG{G4SockNW\ZigUp}{Read\word0}\, \Eb 
\\	&	\retraitAlign 
		\SEG{G5GliderSE\ZigUp}{Read\word0}  \Wb 
		 \SEG{G5SockNW\ZigUp}{Read\word0} \Eb
		\SEG{G6AGliderSE\ZigUp}{Read\word0} \,\Wb 
\\	&	\retraitAlign
		\SEG{G6SockNW\ZigUp}{Read\word0} \Eb 
		\SEG{G6BGlider\ZigUp}{Read\word0} \NEb 
		\SEG{G6End\ZigUp}{Read\word0}
\end{align*}


Figure~\ref{fig:G:brick:Read0} displays the brick $\GReadZ$.

\figBrickHereSize%
	{G-Read_0}%
	{Module \ModuleG: Brick $\GReadZ$.}%
	{fig:G:brick:Read0}%
	{1.35\textwidth}%
	{\textheight}
	
\afterpage{\clearpage}
\newpage


\subparagraph{The $\protect\GReadU$ brick.} ~

\begin{align*}
\SEG{G3GliderE\ZigUp}{Read\word1} 
&	= \bL{41} \NWb \bK{45} \Eb \Eglider'\bigl((\bbK{40}{45})^{10}\bigr) 
		\Eb \bK{40} \SEb \bK{41} \SWb \bL{42}
\\
\SEG{G3SockW\ZigUp}{Read\word1} 
&	= \Wpath(\bbL{43}{48}) \SEb \Epath(\bbL{49}{54}) \NEb \bL{55}
\\
\SEG{G4GliderE\ZigUp}{Read\word1}
&	= \Eglider\bigl(\bbK{50}{51} \conc  (\bbK{46}{51})^{k-3}\conc \bbK{46}{48} \conc \bL{56}\bigr)
\\
\SEG{G4SockW\ZigUp}{Read\word1}
&	= \Wpath(\bbL{57}{64}) \SEb \Epath(\bbL{65}{72}) \NEb \bL{73} 
\\
\SEG{G5GliderE\ZigUp}{Read\word1}
&	=  \begin{array}[t]{@{}l@{}l@{}}
		\Eglider\Bigl(& 
			\bbK{55}{57} \conc (\bbK{52}{57})^{k-6} \conc \bbK{52}{53} \conc \bbL{74}{75}  \conc \bbK{56}{57} \, \conc
		\\ 
			&\retraitAlign (\bbK{52}{57})^2 \conc \bbK{52}{53} \conc \bL{76} \Bigr)
		\end{array}
\\
\SEG{G5SockW\ZigUp}{Read\word1}
&	= \Wpath(\bbL{77}{82}) \SWb \Epath(\bbL{83}{89}) \NEb \bL{90} 
\\
\SEG{G6AGliderE\ZigUp}{Read\word1} 
&	= \Eglider\bigl(\bK{63}\conc (\bbK{58}{63})^{k-19} \conc \bbK{58}{61} \conc \bL{91}\bigr)
\\
\SEG{G6SockW\ZigUp}{Read\word1}
&	= \Wpath(\bbL{92}{99} \conc \bbM{0}{4}) \SWb \Epath(\bbM{5}{18}) \NEb \bM{19} 
\\
\SEG{G6BGliderE\ZigUp}{Read\word1} 
&	= \Eglider\bigl(\bbK{67}{69} \conc (\bbK{64}{69})^9 \conc \bbK{64}{66}\bigr) \Eb \bK{67} \NEb 
		\bK{68} \NWb \bK{69}
\\[2mm]
\BigSEG{G3..6GliderE\ZigUp}{Read\word1}
&	=	\SEG{G3GliderE\ZigUp}{Read\word1} \SWb
		\SEG{G3SockW\ZigUp}{Read\word1} \NEb
		\SEG{G4GliderE\ZigUp}{Read\word1}\SWb
\\	&	\retraitAlign
		\SEG{G4SockW\ZigUp}{Read\word1}\NEb
		\SEG{G5GliderE\ZigUp}{Read\word1}\SWb
		\SEG{G5SockW\ZigUp}{Read\word1}\NEb
\\	&	\retraitAlign
		\SEG{G6AGliderE\ZigUp}{Read\word1} \SWb
		\SEG{G6SockW\ZigUp}{Read\word1} \NEb
		\SEG{G6BGliderE\ZigUp}{Read\word1} 
\\[3mm]
\SEG{G6End\ZigUp}{Read\word1} 
&	= \Epath(\bbM{20}{22}) \SEb \Wpath(\bbM{23}{25}) \SWb \Epath(\bbM{26}{29}) \NEb \bM{30}
\\[2mm]
\GReadU 
&	= \SEG{G1Start\ZigUp}{Read} \NEb 
		\BigSEG{G1..3GliderNE\ZigUp}{Read} \SEb \bL{40} \NEb 
\\	&	\retraitAlign
		\BigSEG{G3..6GliderE\ZigUp}{Read\word1} \Eb
		 \SEG{G6End\ZigUp}{Read\word1}
\end{align*}


Figure~\ref{fig:G:brick:Read1} displays the brick $\GReadU$.

\figBrickHereLandscapeSize%
	{G-Read_1}%
	{Module \ModuleG: Brick $\GReadU$.}%
	{fig:G:brick:Read1}%
	{1.1\textheight}%
	{1.3\textwidth}
	
\afterpage{\clearpage}
\newpage

\paragraph{Zig-down and Zag bricks for Module G: Copy letter \word0\/ and \word1.}

\begin{align*}
\SEG{G1\ZigDown}{Copy} 
&	= \SEpath\bigl(\bbL{0}{6} \conc \bK3 \conc (\bbK{0}{3})^9 \conc \bbK{0}{2} \conc \bbL{7}{10} \conc \SEG{Exp}{G}_{8..h-51} \conc \bbL{11}{17}\bigr) 
\\
\SEG{G2\ZigDown}{Copy} 
&	= \NWpath\bigl(\bbL{18}{24} \conc \bbK{32}{33}\conc (\bbK{28}{33})^{k-3} \conc \bbK{28}{32} \conc \bbL{25}{30}\bigr)
\\
\SEG{G3\ZigDown}{Copy} 
&	= \begin{array}[t]{@{}l@{}l@{}}
		\SEpath\bigl(&\bbL{32}{38} \conc \bbK{35}{39} \conc (\bbK{34}{39})^{k-14}\conc \bbK{34}{35} \conc \bbL{39}{41} \conc \bK{45} \conc 
		\\	
		&(\bbK{40}{45})^{10} \conc \bbK{40}{41}\conc \bbL{42}{48}\bigr)
		\end{array}
\\
\SEG{G4\ZigDown}{Copy} 
&	= \NWpath\bigl(\bbL{49}{55}\conc \bbK{50}{51} \conc  (\bbK{46}{51})^{k-3}\conc \bbK{46}{48} \conc \bbL{56}{63}\bigr)
\\
\SEG{G5\ZigDown}{Copy}
&	= \begin{array}[t]{@{}l@{}l@{}}
		\SEpath(&\bbL{66}{73}\conc \bbK{55}{57} \conc (\bbK{52}{57})^{k-6} \conc \bbK{52}{53} \conc \bbL{74}{75}  \conc \bbK{56}{57}\conc
		\\ 
	 	&	(\bbK{52}{57})^2 \conc \bbK{52}{53} \conc \bbL{76}{81}\bigr)
		\end{array}
\\
\SEG{G6\ZigDown}{Copy} 
&	= \begin{array}[t]{@{}l@{}l@{}}
		\NWpath\Bigl(&\bbL{84}{90}\conc \bK{63}\conc (\bbK{58}{63})^{k-19} \conc \bbK{58}{61} \conc \bbL{91}{99} \conc \bbM{0}{19} \, \conc\\
		& \bbK{67}{69} \conc (\bbK{64}{69})^{10} \conc \bbM{20}{30}\Bigr)
	\end{array}
\\[2mm]
\GZigCopyZ 
&	= \begin{array}[t]{@{}l@{}l@{}}
		\SEG{G1\ZigDown}{Copy} \Eb  \SEG{G2\ZigUp}{Copy} \NEb \bL{31} \SEb  
		\SEG{G3\ZigDown}{Copy} \NEb \SEG{G4\ZigDown}{Copy} \NEb \bL{64} \Eb \bL{65} \, \SWb
		\\[1mm] \retraitAlign
		\SEG{G5\ZigDown}{Copy} \SEb \bL{82} \Eb  \bL{83} \NWb \SEG{G6\ZigDown}{Copy}
		\end{array}
\\
\GZigCopyU 
&	= \begin{array}[t]{@{}l@{}l@{}}
		\SEG{G1\ZigDown}{Copy} \Eb  \SEG{G2\ZigUp}{Copy} \NWb \bL{31} \Eb  
		\SEG{G3\ZigDown}{Copy} \NEb \SEG{G4\ZigDown}{Copy} \Eb \bL{64} \Eb \bL{65} \, \SWb
		\\[1mm] \retraitAlign
		\SEG{G5\ZigDown}{Copy} \SWb \bL{82} \Eb  \bL{83} \NEb  \SEG{G6\ZigDown}{Copy}
	\end{array}
\\[2mm]
\GZagCopyZ
&	= \vmirror\bigl(\GZigCopyZ\bigr) 
\\
\GZagCopyU
&	= \vmirror\bigl(\GZigCopyU\bigr) 
\end{align*}


Figures~\ref{fig:G:brick:ZigCopy0} and~\ref{fig:G:brick:ZigCopy1} display the bricks $\GZigCopyZ$ and $\GZigCopyU$.

\figBrickHereSize%
	{G-Copy_0-Zig}%
	{Module \ModuleG: Brick $\GZigCopyZ$.}%
	{fig:G:brick:ZigCopy0}%
	{1.3\textwidth}%
	{\textheight}

\figBrickHereSize%
	{G-Copy_1-Zig}%
	{Module \ModuleG: Brick $\GZigCopyU$.}%
	{fig:G:brick:ZigCopy1}%
	{1.3\textwidth}%
	{\textheight}
	
\afterpage{\clearpage}
\newpage

\paragraph{Line feed brick for Module G}

\begin{align*}
\SEG{G1A\bZagZig}{LF} 
&	= \NWpath(\bbL{0}{3}) \Wb \SEpath(\bbL{4}{6} \conc \bK{3})
\\
\GLF 
&	= \begin{array}[t]{@{}l}
		\SEG{G1A\bZagZig}{LF} \SWb 
		\rot\bigl(\BigSEG{G1..3GliderNE\ZigUp}{Read} \bigr) 
		\SWb \bL{40} \SEb
		\\ \retraitAlign
		\rotateClockwiseAA{\BigSEG{G3..6GliderE\ZigUp}{Read\word1}}
		\SWb \bM{20} \Wb \bM{21} \NWb \bM{22} 
		\\ \retraitAlign
		\SWb \bM{23} \SEb \SEG{G6End\ZigUp}{Read\word0}
	\end{array}
\end{align*}


Figure~\ref{fig:G:brick:LineFeed} displays the brick $\GLF$.

\figBrickHereSize%
	{G-Line_Feed}%
	{Module \ModuleG: Brick $\GLF$.}%
	{fig:G:brick:LineFeed}%
	{1.3\textwidth}%
	{\textheight}

\subsubsection{Subrule for Module G}	

Module \ModuleG\/ interacts with \ModuleA, \ModuleC, \ModuleDx{}, \ModuleF.
Figure~\ref{fig:G:subrule} presents the subrule for the interactions between the beads of $\ModuleG$ and the beads of the other modules.

\figSubRuleHere%
	{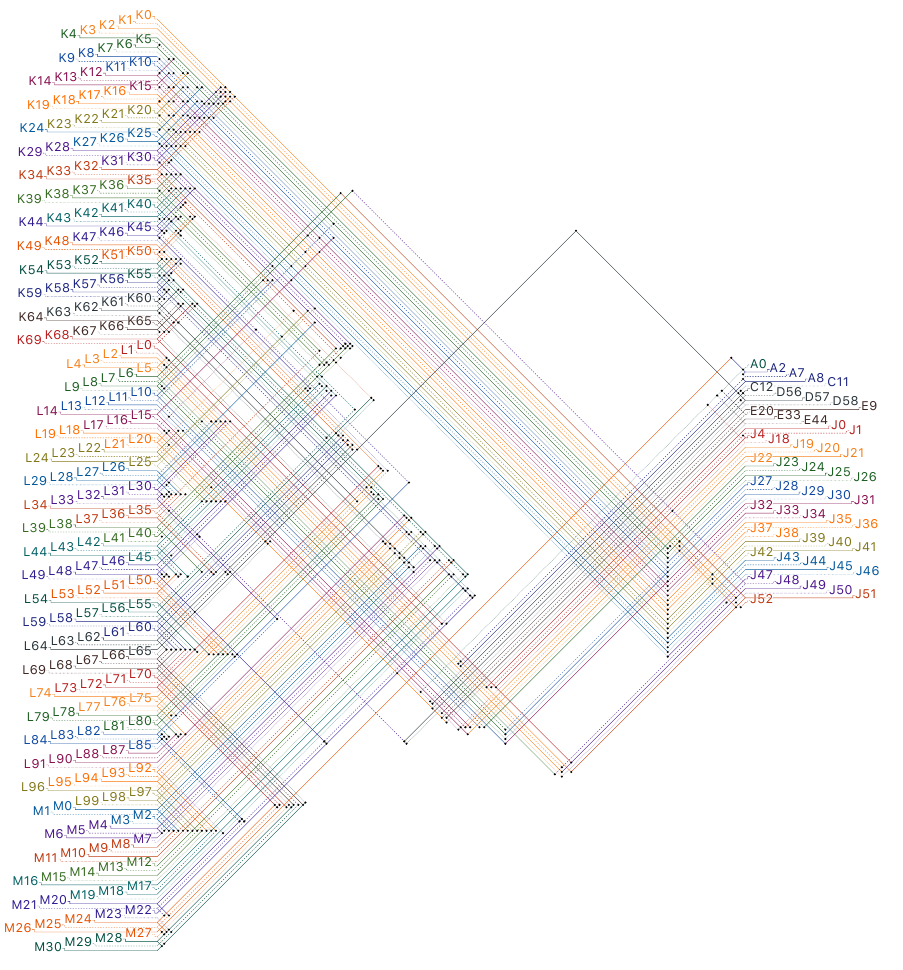}%
	{Subrule for Module \ModuleG.}%
	{fig:G:subrule}%

\afterpage{\clearpage}
\newpage


\subsection{The seed conformation}

The seed $\ModuleSeed{(u)}$ is a conformation encoding the input dataword $u$ of the simulated SCTS. It is made of 4 types of conformations, (see Figure~\ref{fig:brick:Seed:u} for an illustration):
\begin{align*}
\SEG{Seed}{Begin} 
&	= 
	{\left(\bJ{8}\SWbond\bJ{7}\,\SWbond\right)^{2+\frac{h-3}{4}}}
	\bJ{11}
	\SWbond\bJ{12}
	\Wbond\bJ{16}
	\Wbond\bJ{17}
	\Wbond\bJ{18}
\\
\SEG{Seed}{Suffix} 
&	= \begin{array}[t]{l}
	\bA9\SWbond
	\bA{12}\Wbond
	\bB0\Wbond
	\bB1\NWbond
	\bB2\Ebond
	\bB3\NEbond
	\bB4\NWbond
	\bC4\NWbond
	\bA0\left(\Wbond\,\bA0\right)^{w-17}\SWbond
	\bH{8}\SWbond
	\\ \quad 
	\bH{19}\SWbond
	\bH{20}\Wbond
	\bH{21}\SWbond
	\bH{24}\Wbond
	\bI{15}\Wbond
	\bI{15}\Wbond
	\bI{16}\Wbond
	\bI{17}\Wbond
	\bI{19}\Wbond
	\bI{19}\Wbond
	\bJ{12}\Wbond
	\\ \quad
	\bJ{16}\Wbond
	\bJ{17}\Wbond
	\bJ{18} 
	\end{array}
\\
\SEG{Seed}{$(\word 0)$} 
&	= 
	\bL{17}\Wbond
	\bL{18}\Wbond
	\bL{47}\SWbond
	\bL{48}\NWbond
	\bL{49}\Wbond
	\bL{82}\Wbond
	\bL{83}\NWbond
	\bA6
\\
\SEG{Seed}{$(\word 1)$} 
&	= 
	\bL{17}\Wbond
	\bL{18}\Wbond
	\bL{48}\Wbond
	\bL{82}\Wbond
	\bL{83}\NWbond
	\bL{84}\Wbond
	\bA6
\\
\SEG{Seed}{End} 
&	= \begin{array}[t]{l}
	\bK{34}\SWbond
	\left(\bK{45}\SWbond\bK{40}\,\SWbond\right)^{11}
	\left(\bK{46}\SWbond\bK{47}\,\SWbond\right)^{k-2}
	\left(\bK{57}\SWbond\bK{52\,\SWbond}\right)^{k-2}
	\\ \quad
	\left(\bK{59}\SWbond\bK{60}\,\SWbond\right)^{k-18}
	\left(\bK{69}\SWbond\bK{64}\,\SWbond\right)^{10}
	\bK{69} \SWbond
	\bM{20} \SEbond
	\bM{26} \SWbond
	\bM{27}
	\\ \quad 
	\Wbond\,
	\bM{28}\SWbond
	\bM{29}\Ebond
	\bM{30}.
	\end{array}
\end{align*}

\medskip 

\noindent
Each letter $a\in\{\word0,\word1\}$ is encoded in the seed by the conformation: (note that it heads westwards)
$$
\SEG{Seed}{Letter$(a)$} = \left(\SEG{Seed}{$(\word1)$}\Wbond\SEG{Seed}{Suffix}\,\Wbond\right)^{n-1} 
\SEG{Seed}{$(a)$}\Wbond\SEG{Seed}{Suffix}
$$
\medskip

Then, the conformation $\ModuleSeed{(u)}$ is: (note that it heads southwestwards, see Fig.~\ref{fig:brick:Seed:u})
$$
\ModuleSeed{(u)} = 
\SEG{Seed}{Begin} 
\Wbond
\left(\overset{|u|}{\concat{i=1}} 
  \SEG{Seed}{Letter$(u_{|u|-i})$}
  \,\Wbond
\right)
\SEG{Seed}{End}
$$

This completes the definition of the primary structure of the oritatami system ${\TMO_\TMS=((\pi_\TMS)^\infty,\heart,3)}$ whose folding from the seed configuration \ModuleSeed{(u)} simulates step-by-step the computation of SCTS $\TMS$ starting with input dataword~$u$.  Section~\ref{sec:rule} presents the full description of the attraction rule~$\heart$, completing the description of the oritatami system $\TMO_\TMS$.

\figBrickHereLandscapeSize%
	{Seed-Brick}%
	{The brick $\protect\ModuleSeed{(u)}$.}%
	{fig:brick:Seed:u}%
	{1.075\textheight}%
	{\textwidth}

%% file: proof-environments.tex

\label{sec:enum:env}

We will now resume and expand the explanations given in Section~\ref{sec:proof:local:folding}.
Here is how we proceeded to ensure the correctness of our design:
\begin{enumerate}
\item
Enumerate all the surrounding for each brick of each module
\item
Enumerate all possible modules following the module
\item
Generate automatically human-readable certificate of the correctness of the folding for each possibility, in the form of \emph{proof trees}.
\item\label{exceptions}
In the few cases where the surrounding may vary, prove that it has no incidence on the folding of the brick. This happens only for three bricks exactly:  when the brick $\GRead$\/ zig-folds along $\FZigUp$, when the top of the brick {$\GReadU$} folds, and when the zag-bricks folds under $\DWrite$.
\end{enumerate}

One can check using Fact~\ref{fact:variable:modulo} that the beads alignments in each brick do not change when $n$ and $L$ vary. This implies that the figures of the bricks are indeed generic. It follows that with the exception of the three cases listed in point~\ref{exceptions} above, and handled in Section~\ref{sec:exc}, it is enough to prove the folding of each brick only once. And as most of them are made of repeating patterns, only a finite number of environments have to be considered. That last case will be treated in Section~\ref{proof-trees} using an automatic procedure which produces human-readable certificates called \emph{proof-trees}.

\subsubsection{The three bricks with varying environments}
\label{sec:exc}

The following lemma show that it is enough to proof one folding of the zag bricks under a $\DWrite$, all the other are the same since there are no interaction between the $\DWrite$ brick and any zag brick folding immediately below it.

\begin{lemma}[Zag-folding under $\DWrite$] \label{lem:env:below:write}
The modules zag-folding under  the bricks $\DWrite$ have no interaction with $\DWrite$, with the only exceptions of:
\begin{itemize}
\item the beads $\bA0$ and $\bA1$ of module $\ModuleA$ which have bonds with beads \bE{$(2+12i)$} and $\bL{17}$ for~$\bA0$, and \bE{$(9+12i)$} for $\bA1$, for all $0\leq i\leq3$.  
\item the beads $\bL{17}, \bL{18}, \bD{57}, \bD{58}$ (the bump in module $\ModuleDZ{}$) which bond with the beads $\bL{65}, \bL{64},\bL{31}$ so that the corresponding module $\ModuleG$ folds into the expected brick $\GZagCopyZ$.
\end{itemize}
\end{lemma}

\begin{proof}
Figure~\ref{fig:rule:Dwrite:zagFold} lists all the possible \heart-interactions between the beads accessible from below the {$\DWrite$} bricks (to the left) with the beads at the top the modules zag-folding below it that can interact with them (to the right).  

\begin{figure}
\centerline{\includegraphics[width=.95\textwidth]{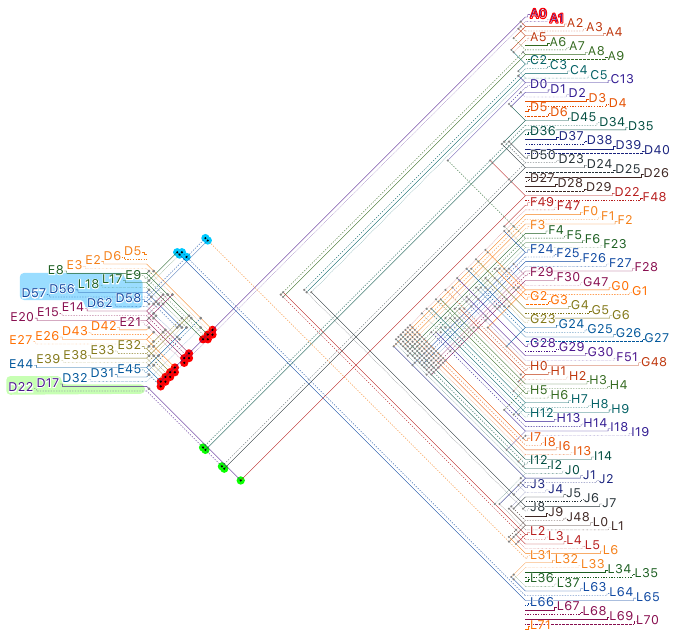}}
\caption{The \protect\heart-rule between the beads accessible from below of brick {\protect$\DWrite$} and the beads that will get in touch with them from all the modules Zag-folding below.} 
\label{fig:rule:Dwrite:zagFold}
\end{figure}

The only possible bonds are thus:
\begin{description}
\item[with beads \bD{17} and \bD{22}:] (in green on Figure~\ref{fig:rule:Dwrite:zagFold}) these are only present at the junction between the bricks $\DWrite$\/ and \ECR, at the end of the rightmost $\DWrite$ brick. The correctness of the zag-folding of the \FZag\/ brick below is given next in the proof-trees section. 
\item[with beads \bL{17}, \bL{18}, \bD{56}, \bD{57}, \bD{58}, \bD{62}:] (in blue on Figure~\ref{fig:rule:Dwrite:zagFold}) these beads are only present in the spike encoding a \word0\/ in the brick $\DWrite$, and these interactions are the one expected to ensure the copy of the encoding of \word0\/ by the module $\ModuleG$ that will Zag-fold below.
\item[and finally between beads \bA0 and \bA1, and 4 groups of beads:] $\bE 2$, $\bE 3$, $\bE 8$, $\bE 9$, then $\bE {14}$, $\bE {15}$, $\bE {20}$, $\bE {21}$, then  $\bE{26}$, $\bE{27}$, $\bE{32}$, $\bE{33}$, and finally $\bE{38}$, $\bE{39}$, $\bE{44}$, $\bE{45}$ (in red on Figure~\ref{fig:rule:Dwrite:zagFold}). As the width of a zag-folded production segment is $w+6=0\mod 12$, the beads $\bA 0$ and $\bA 1$ are always aligned with the same beads within each of these groups (see Figure~\ref{fig:brick:D:write}), namely $\bA 0$ with $\bE 2$, $\bE{14}$, $\bE{26}$ and $\bE{38}$, and $\bA 1$ with $\bE 9$, $\bE{21}$, $\bE{33}$ and $\bE{45}$. Furthermore as the interactions of $\bA 0$ and $\bA 1$ are the same with each of them, it is enough to prove that the module $\ModuleA$ zag-folds correctly between \emph{one} of these groups only, which is done next in the proof-trees section.
\end{description}

It follows that outside these three cases (each handled by a proof-tree, see later), no interactions are possible and the modules will zag-fold below the $\DWrite$\/ bricks independently of the exact beads that are present inside. It is thus enough to show that each module zag-folds correctly at any location to ensure that it zag-folds correctly anywhere below the $\DWrite$\/ brick.
\end{proof}

\begin{lemma}[Top of $\GReadU$] \label{lem:env:top:read1}
During the folding of the brick $\GReadU$, no bead in $\ModuleG$ interacts with the row above but at its two extremeties, i.e. the 82 top-leftmost beads and the 11 last ($\bK{34}$..$\bL{55}$ and $\bM{20}$..$\bM{30}$ resp. in Figure~\ref{fig:G:brick:Read1}).
\end{lemma}

\begin{proof}
Figure~\ref{fig:G-READ1:accessible} lists the only beads exposed and accessible from below above $\GReadU$. And Figure~\ref{fig:rule:G-READ1:above} lists all the possible \heart-interactions between them (to the left) and the beads of the brick $\GReadU$\/ zig-folding below (to the right).

\begin{figure}
	\begin{subfigure}{\textwidth}
	\centerline{\includegraphics[width=.95\textwidth]{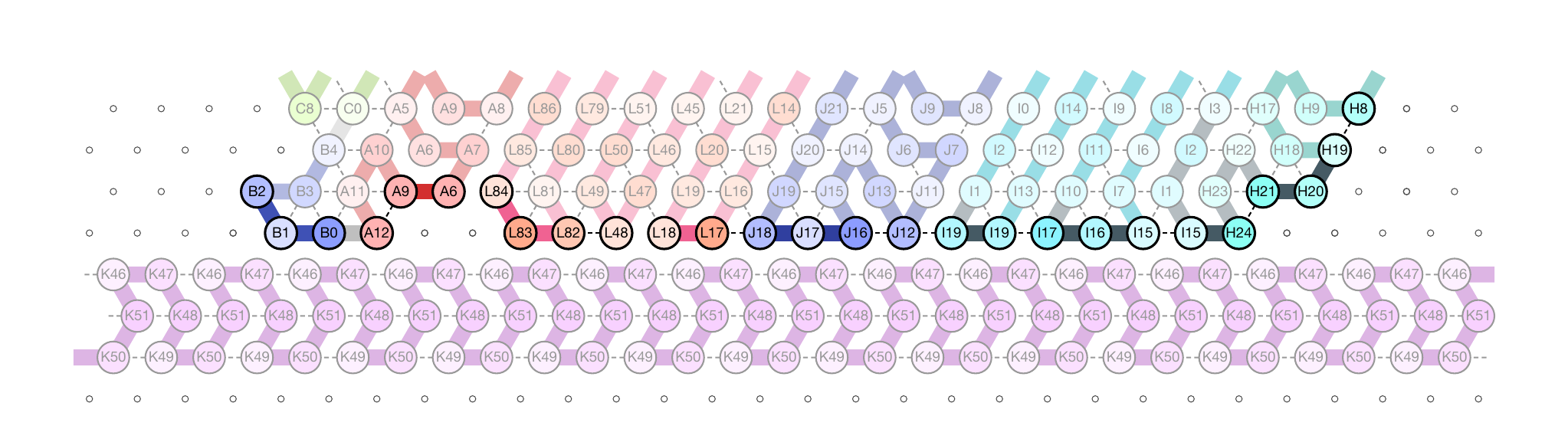}}
	\caption{The beads accessible when the brick $\GReadU$\/ zig-folds itself.}
	\label{fig:G-READ1:accessible}
	\end{subfigure}\\[5mm]

	\begin{subfigure}{\textwidth}
	\centerline{\includegraphics[width=.95\textwidth]{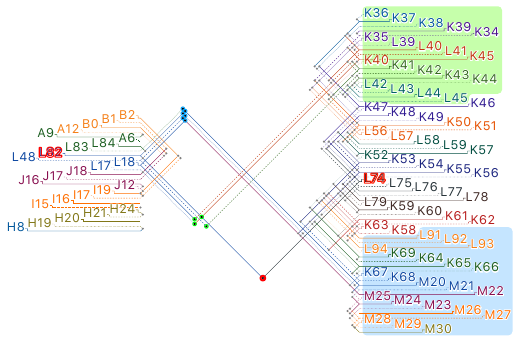}}
	\caption{The \heart-rule for the beads accessible by the beads in $\GReadU$\/ as it zig-folds.} 
	\label{fig:rule:G-READ1:above}
	\end{subfigure}\\[5mm]

	\begin{subfigure}{\textwidth}
	\centerline{\includegraphics[width=.95\textwidth]{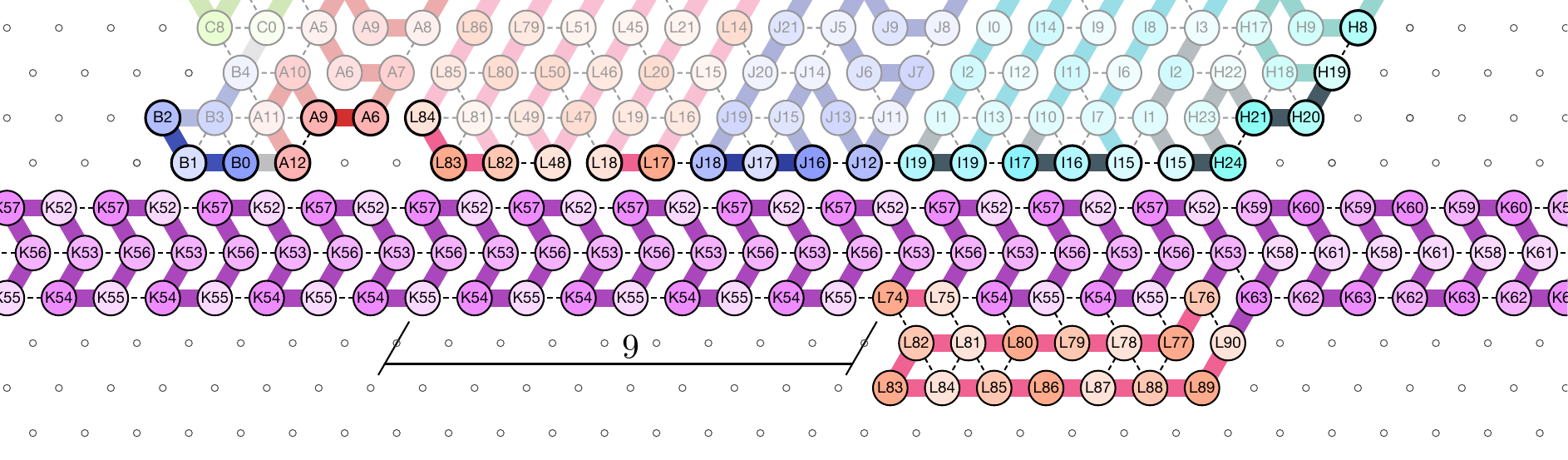}}
	\caption{The closest bead L74 in brick $\GReadU$\/ can get from one bead L82 above (case $n=1\mod 3$).} 
	\label{fig:G-READ1:L74:closest:L82}
	\end{subfigure}
\end{figure}

According to the rule in Figure~\ref{fig:rule:G-READ1:above}, besides the interactions at the 82 first beads at the very top-leftmost part of $\GReadU$\/ ($\bK{34}$..$\bL{55}$ in Figure~\ref{fig:G:brick:Read1}, interactions in green in Figure~\ref{fig:rule:G-READ1:above}) and the 11 beads at the very end of $\GReadU$\/ ($\bM{20}$..$\bM{30}$ in Figure~\ref{fig:G:brick:Read1}, interactions in blue in Figure~\ref{fig:rule:G-READ1:above}), the only possible interaction between $\GReadU$\/ and the already present beads above it is: $\bL{82}$\heart $\bL{74}$.  But $\bL{74}$ appears only once in $\GReadU$, at coordinates $(w+10+4k,1-h)$ (see Figure~\ref{fig:brick:G:read1}), while $\bL{82}$ appears above $\GReadU$\/ at coordinates $(w+1+i(w+6),2-h)$ for $i=0..n$. The minimal $x$-distance between $\bL{82}$ and $\bL{74}$ is thus $\min_{i=0..n} |9+4k-i(w+6)|$. But $9+4k-i(w+6) = 9+4(n-1)(w+6)/6-i(w+6) = 9+2(n-1-3i)(2(L+P)+8)$. It follows that the minimum difference in $x$-coordinate between $\bL{82}$ and $\bL{74}$ is:
\begin{itemize}
\item $17+2(L+P)\geq41$, if $n=0\mod 3$; 
\item $9$, if $n = 1\mod3$; and 
\item $1-2(L+P) \leq -23$, if $n=2\mod 3$. 
\end{itemize}
As a consequence, $\bL{74}$ never gets close enough to interact with $\bL{82}$ above (see Figure~\ref{fig:G-READ1:L74:closest:L82} for the closest situation). It follows that one only need to take into account the environnement for the folding of the top-leftmost and top-rightmost part of brick $\GReadU$ (which is done next using proof-trees), the glider between them, zig-folds regardless of the beads above in the environment.
\end{proof}

\begin{lemma}[{$\GReadU$} along $\FZigUp$] \label{lem:exp:col}
When $\ModuleG$ folds into the brick \GRead, no bead in $\SEG{Exp}G$ can make bonds with the beads in $\FZigUp$ nearby and thus folds regardless of the beads nearby  (as a glider). 
\end{lemma}

\begin{proof}
Figure~\ref{fig:rule:GReadExp:FZigExp} lists the interactions between the beads in $\SEG{Exp}G$ and the beads in $\SEG{Exp}F$: these are exactly \bK{$(4+i)$}\heart \bJ{$(24+i)$} for $i=0..23$; in particular red-shaded beads \bK{4}..\bK 9 in $\ModuleG$ (resp. yellow, \bK{10}..\bK{15}; blue, \bK{16}..\bK{21}; and green, \bK{22}..\bK{27})  can only bond with beads of the same shade \bJ{24}..\bJ{29} in $\ModuleF$ (resp. \bJ{30}..\bJ{35}; \bJ{36}..\bJ{41}; \bJ{42}..\bJ{47}).

\begin{figure}
\includegraphics[width=.9\textwidth]{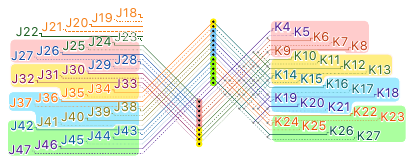}
\caption{\protect\heart-rule between the two exponential segments in $\protect\ModuleF$ and $\protect\ModuleG$. Note that each bead makes exactly one bond, with a bead of the same shade, red, blue, yellow or green (see Figure~\ref{fig:brick:G:read0} and~\ref{fig:brick:G:read1}) and of the same rank within the shade.}
\label{fig:rule:GReadExp:FZigExp}
\end{figure}

As shown on Figure~\ref{fig:F:brick:ZigUp} and~\ref{fig:G:brick:Read0} the $y$-coordinates explored by these beads are as follows when $\ModuleG$ zig-folds into $\GReadZ$ or $\GReadU$:
\begin{description}
\item[Red]: the $y$-coordinates of beads \bJ{24}..\bJ{29} in $\ModuleF$ belong to ${\{-40-3^{2j},\ldots,-35-3^{2j}\}}$ for $j\geq1$, while the corresponding beads \bK 4..\bK 9 in $\ModuleG$ explore $y$-coordinates in ${\{-38-3^{2j'+1}, \ldots, -34-3^{2j'+1}\}}$ for $j'\geq1$. 
\item[Yellow]: the $y$-coordinates of beads \bJ{30}..\bJ{35} in $\ModuleF$ belong to ${\{-34-3^{2j+1},\ldots,-41-3^{2j}\}}$ for $j\geq1$, while the corresponding beads \bK{10}..\bK{15} in $\ModuleG$ explore $y$-coordinates in ${\{-36-3^{2j'+2}, \ldots, -36-3^{2j'+1}\}}$ for $j'\geq1$. 
\item[Blue]: the $y$-coordinates of beads \bJ{36}..\bJ{41} in $\ModuleF$ belong to ${\{-40-3^{2j+1},\ldots,-35-3^{2j+1}\}}$ for $j\geq1$, while the corresponding beads \bK{16}..\bK{21} in $\ModuleG$ explore $y$-coordinates in ${\{-38-3^{2j'}, \ldots, -34-3^{2j'}\}}$ for $j'\geq1$. 
\item[Green]: the $y$-coordinates of beads \bJ{42}..\bJ{47} in $\ModuleF$ belong to ${\{-34-3^{2j+2},\ldots,-41-3^{2j+1}\}}$ for $j\geq1$, while the corresponding beads \bK 2..\bK{27} in $\ModuleG$ explore $y$-coordinates in ${\{-36-3^{2j'+1}, \ldots, -36-3^{2j'}\}}$ for $j'\geq1$. 
\end{description}
Now, as for all $j\geq 1$ (with the notation, $a\lessdot b$ iff $a\leq b-2$)
\begin{align*}
-35-3^{2j+2} \lessdot -38-3^{2j+1} & \lessdot -34-3^{2j+1} \lessdot -40-3^{2j}\\
\text{and }
-36-3^{2j+1} \lessdot -34-3^{2j+1} & \lessdot -41-3^{2j} \lessdot -36-3^{2j}\\
\text{and }
 -34-3^{2j+2} \lessdot -40-3^{2j+1} & \lessdot -35-3^{2j+1} \lessdot -38-3^{2j} \\
\text{and }
-41-3^{2j+1} \lessdot -36-3^{2j+1} & \lessdot -36-3^{2j} \lessdot -34-3^{2j}\\
\end{align*}
none of the (same-shade) interacting beads ever get close enough to each other and the beads in the segment \SEG{Exp}G folds without making any bond (into a glider), regardless of the beads next to them in $\FZigUp$ when $\ModuleG$ zig-folds into brick $\GRead$.
\end{proof}

\subsubsection{Proof-trees: An automated human-readable certificate for the correctness of oritatami system}
\label{sec:proof:trees}
\label{proof-trees}
\begin{figure}
\includegraphics[height=.9\textheight]{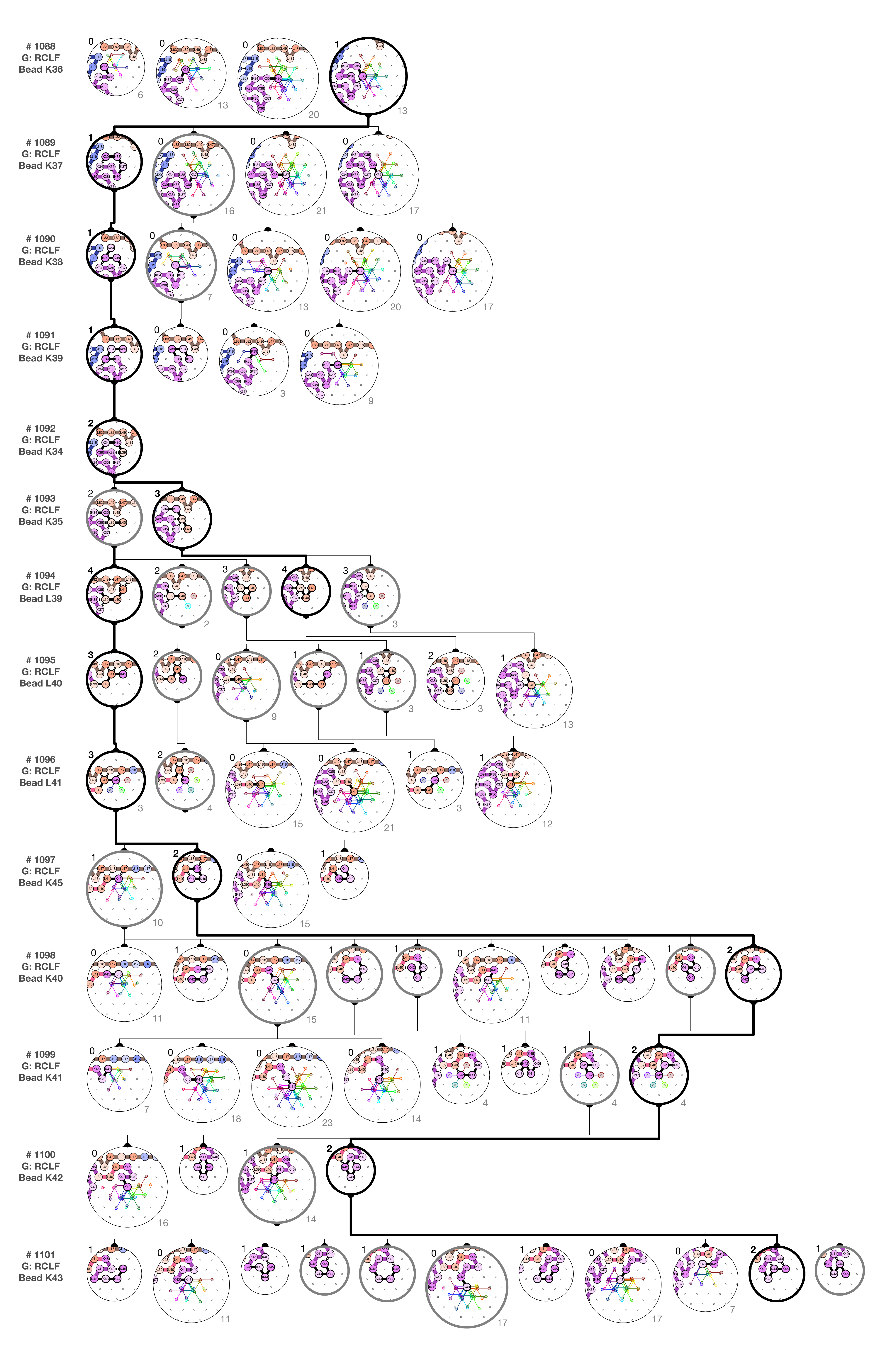}
\caption{Excerpt from the proof-tree certificate for the folding of  $\protect\ModuleG$  into $\protect\GReadZ$ when bouncing on a spike encoding a $\word0$.}
\label{fig:prooftree} 
\end{figure}

A \emph{proof-tree} is a compact representation of the enumeration of all the possible paths the molecule explores as it folds. Figure~\ref{fig:prooftree} presents the proof-tree for the folding of $\ModuleG$ when bouncing on a bump encoding a $\word0$ in $\GReadZ$. For the sake of readability, several paths are drawn in the same ball when they share the same beginning up to their last bond with the environment; then, as a sanity check, the grey number at the bottom left of the ball indicates how many paths are drawn in this ball. The black number in the top right corner of each ball indicates how many bonds are made by the paths with the environment. The ball(s) with the maximum number of bonds is(are) highlighted in black and go to the next round, together with the balls that place the first bead at the same position. 

These proof-trees are automatically generated as the molecule folds. Each environment (surrouding + the three beads currently folding) is given a number (written \#xxxx). When an already studied environment is encountered, the proof-tree is stopped, and the next (already encountered) environment number is written, allowing easy navigation in the proof --- note that Figure~\ref{fig:prooftree} is an excerpt from a larger proof-tree and does not show its beginning nor its end, this is why the navigation tag cannot be observed in this figure.

The complete proof certificates may be found on the website:

\begin{center}
\href{https://www.irif.fr/~nschaban/oritatami/prooftrees/}{\ttfamily https://www.irif.fr/$\sim$nschaban/oritatami/prooftrees/}
\end{center}

%% file: proof-trees.tex


The following tables refer to the proof-trees on the website:

\begin{center}
\href{https://www.irif.fr/~nschaban/oritatami/prooftrees/}{\ttfamily https://www.irif.fr/$\sim$nschaban/oritatami/prooftrees/}
\end{center}

proving the correctness of the folding of our design in every possible surroundings.

\newcommand{\ChibiFig}[1]{\includegraphics[scale=.3]{Chibi/#1}}
\newcommand{\refProofTree}[4]{%
	\mbox{\scriptsize\textsf{%
		\##1-#2 %
	}}}
\newcommand{\stackRefs}[1]{\begin{tabular}[t]{@{}c@{}}	#1	\end{tabular}}


\begin{center}
\begin{tabular}{ccccc}
\toprule
\multicolumn{5}{c}{\textsf{\bfseries ZIG-UP}}
\\
\midrule
\ModuleA
&	\ChibiFig{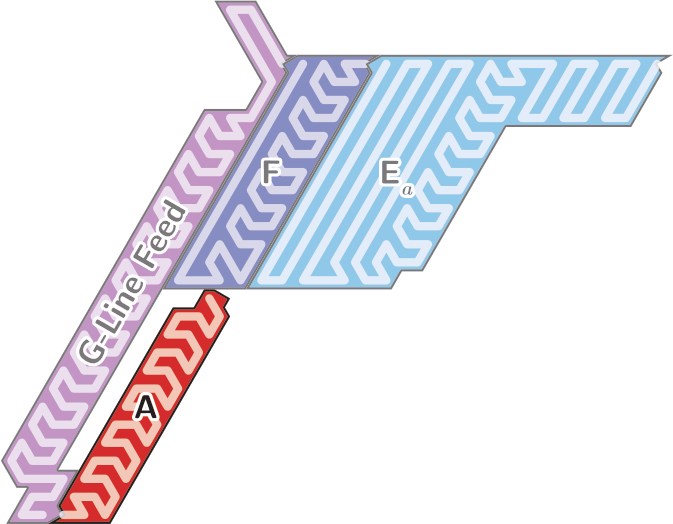}
&	\ChibiFig{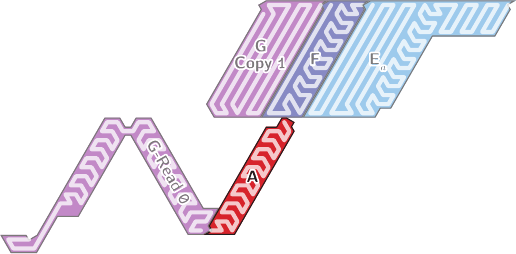}
&	\ChibiFig{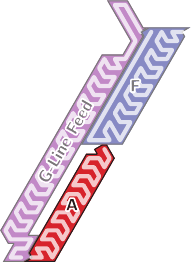}
&	\ChibiFig{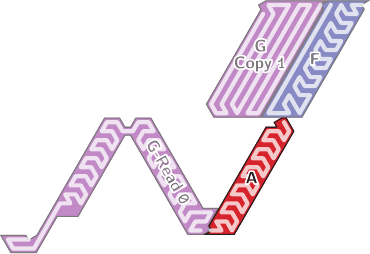}
\\
&	\refProofTree{0}{98}{}{}
&	\refProofTree{1286}{1312}{}{}
&	\refProofTree{}{}{}{}
&	\refProofTree{4995}{4997}{}{}
\\
\midrule
\ModuleB
&	\ChibiFig{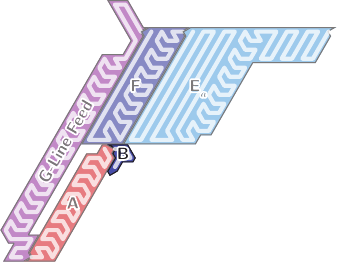}
&	\ChibiFig{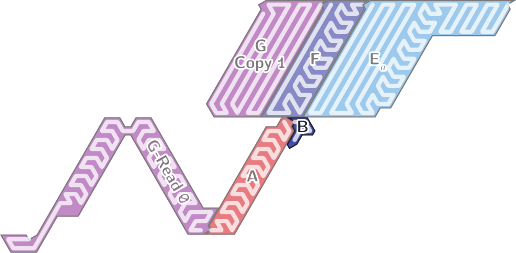}
&	\ChibiFig{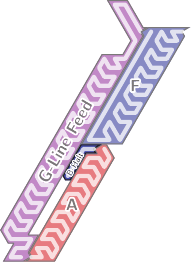}
&	\ChibiFig{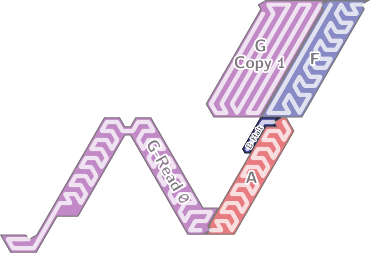}
\\
&	\refProofTree{99}{103}{}{}
&	\refProofTree{1313}{}{}{}
&	\refProofTree{}{}{}{}
&	\refProofTree{4998}{5000}{}{}
\\
\midrule
\ModuleC
&	\ChibiFig{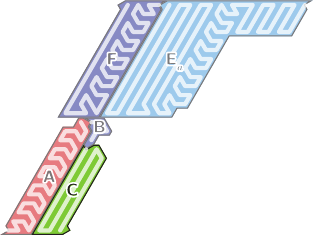}
\\
&	\stackRefs{	\refProofTree{104}{159}{}{}\\	\refProofTree{1314}{1315}{}{}	}
\\
\midrule
\ModuleDx{x}
&	\ChibiFig{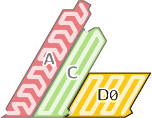}
&	\ChibiFig{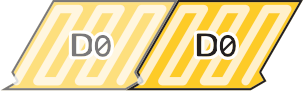}
\\
&	\refProofTree{160}{339}{}{}
&	\refProofTree{340}{384}{}{}
\\
\midrule
\ModuleE
&	\ChibiFig{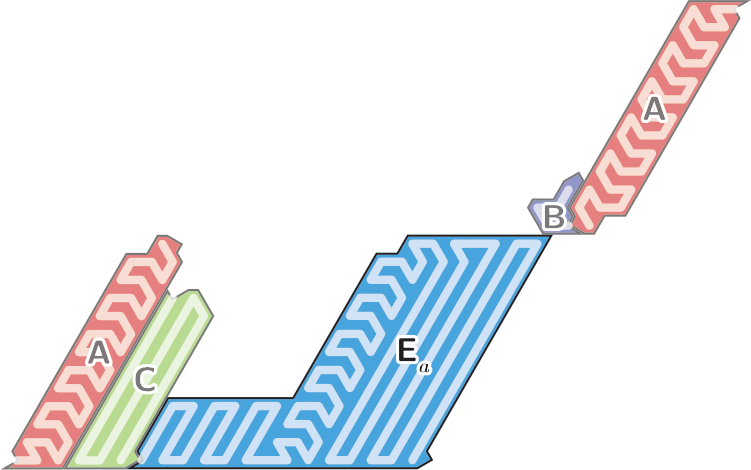}
&	\ChibiFig{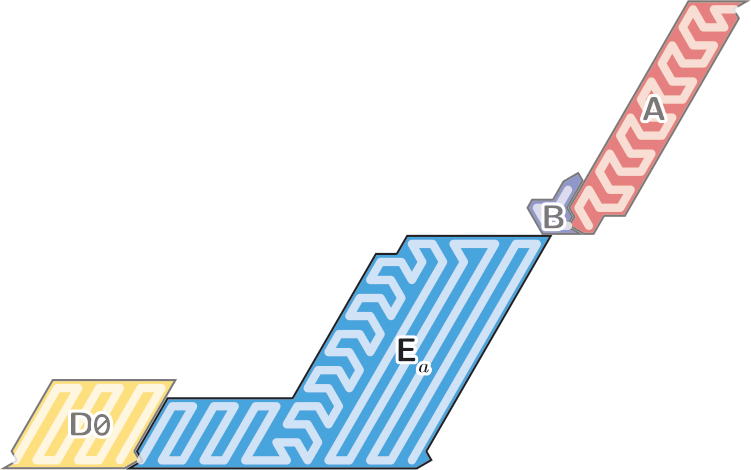}
\\
&	\refProofTree{1316}{1392}{}{}
&	\refProofTree{385}{749}{}{}
\\
\midrule
\ModuleF
&	\ChibiFig{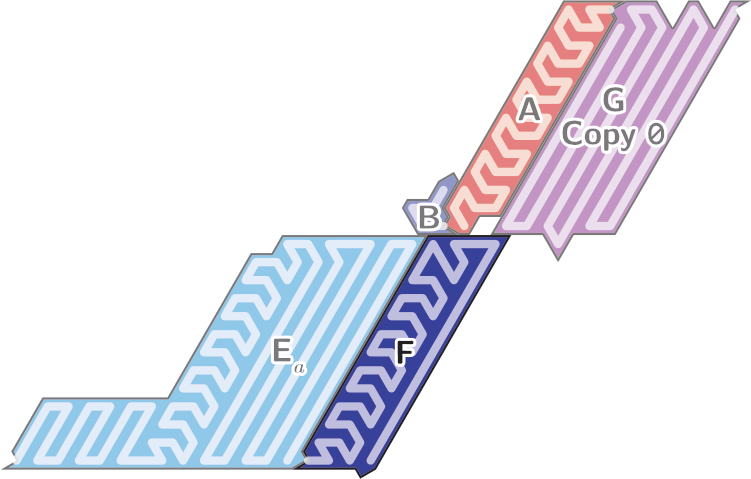}
&	\ChibiFig{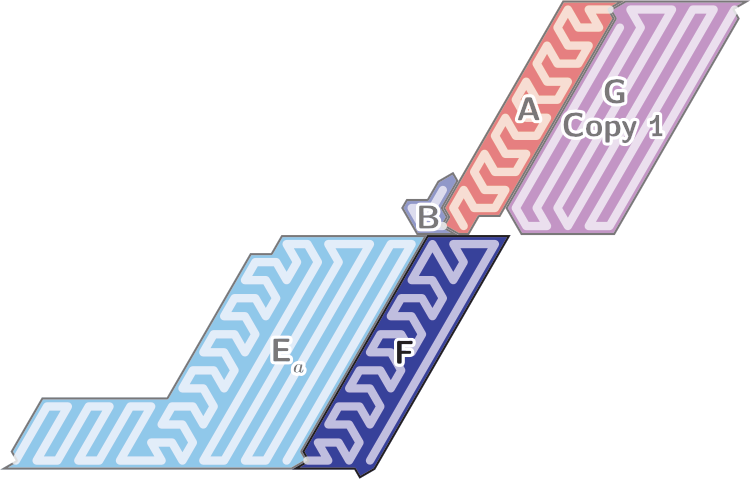}
\\
&	\refProofTree{750}{856}{}{}
&	\refProofTree{1393}{1401}{}{}
\\
\midrule
\ModuleG
&	\ChibiFig{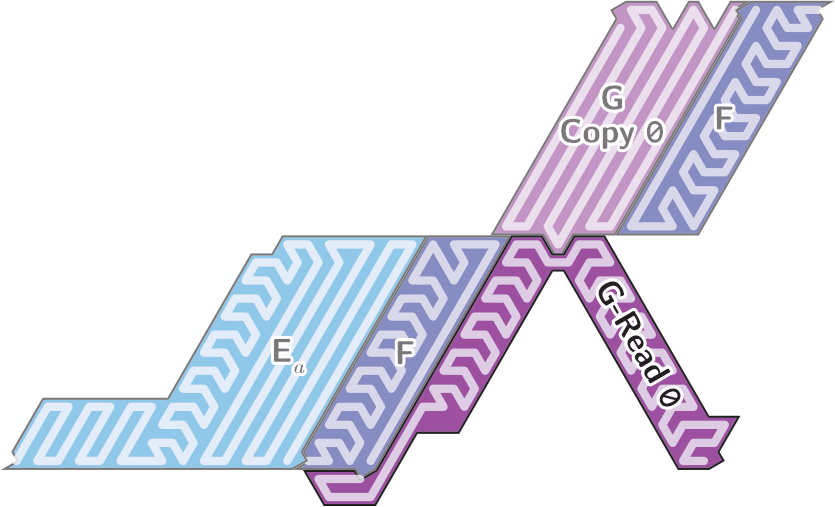}
&	\multicolumn{3}{c}{\ChibiFig{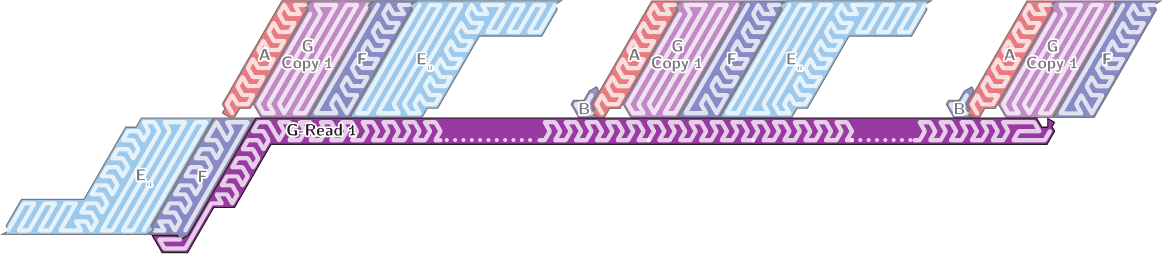}}
\\
&	\refProofTree{857}{1285}{}{}
&	\multicolumn{3}{c}{\refProofTree{1402}{1853}{}{}}
\\
\bottomrule
\end{tabular}


\begin{tabular}{cccccc}
\toprule
\multicolumn{6}{c}{\textsf{\bfseries ZIG-DOWN}}
\\
\midrule
\ModuleA
&	\ChibiFig{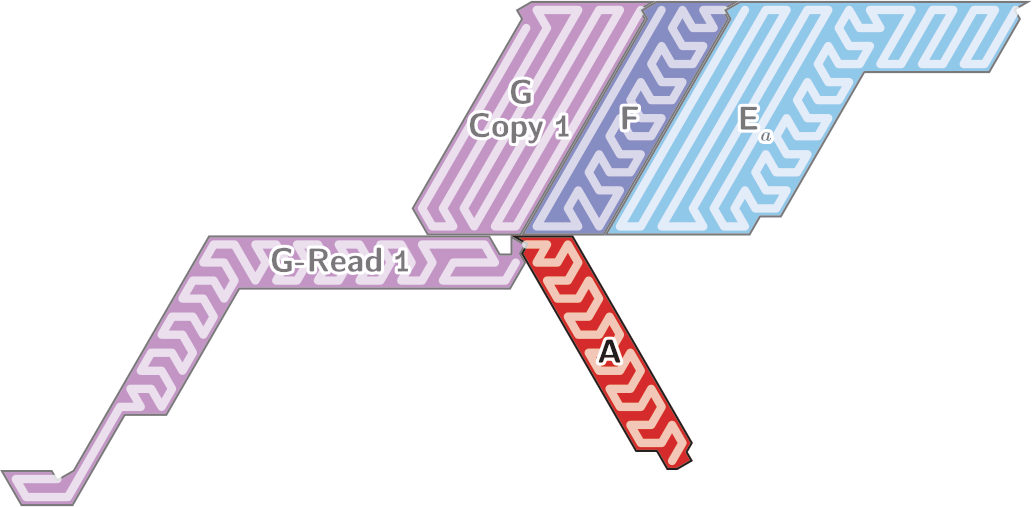}
&	\ChibiFig{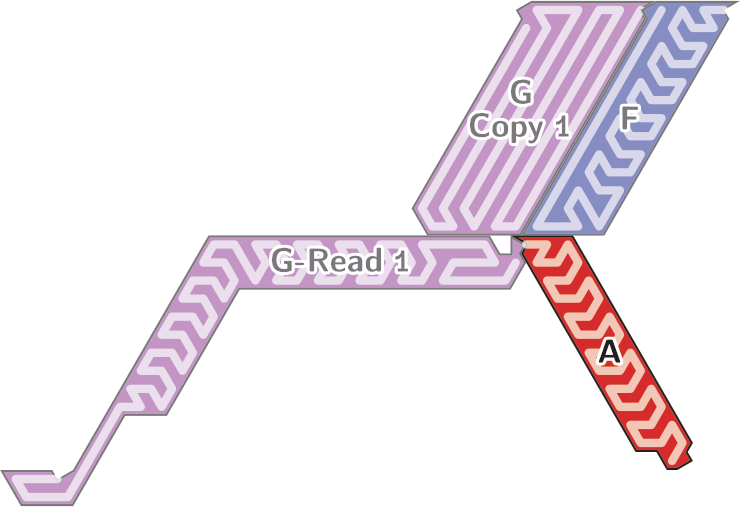}
&	\ChibiFig{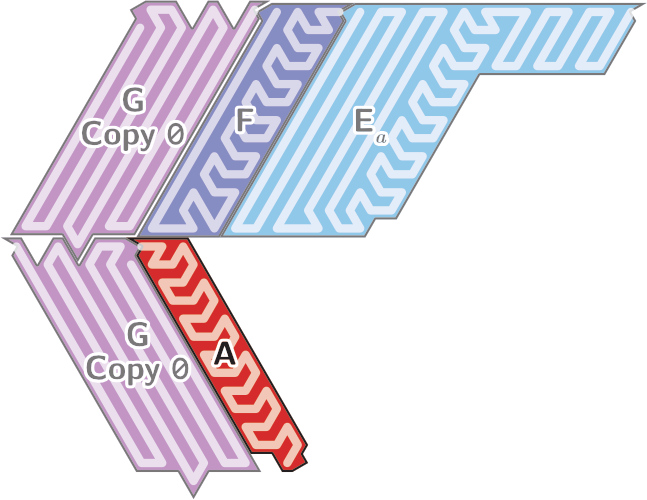}
&	\ChibiFig{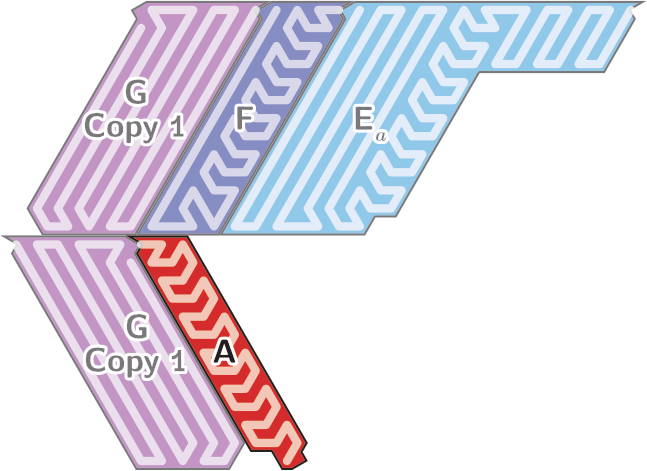}
&	\ChibiFig{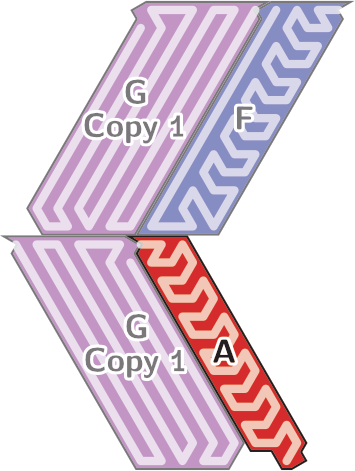}
\\
&	\refProofTree{1854}{1874}{}{}
&	\refProofTree{4745}{4752}{}{}
&	\refProofTree{2382}{2578}{}{}
&	\refProofTree{2745}{2755}{}{}
&	\refProofTree{2790}{2797}{}{}
\\
\midrule
\ModuleB
&	\ChibiFig{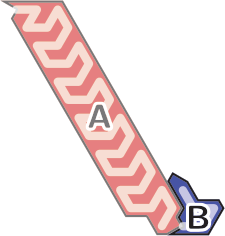}
&	\ChibiFig{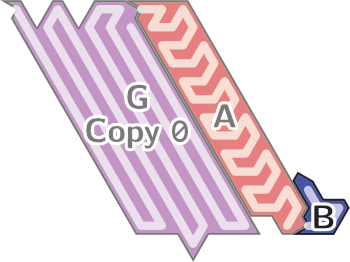}
&	\ChibiFig{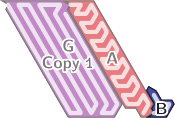}
\\
&	\refProofTree{1875}{1878}{}{}
&	\refProofTree{2579}{2580}{}{}
&	\refProofTree{2756}{}{}{}
\\
\midrule
\ModuleC
&	\ChibiFig{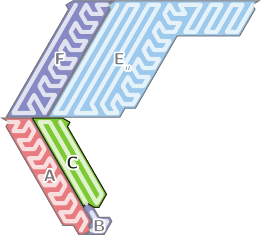}
&	\ChibiFig{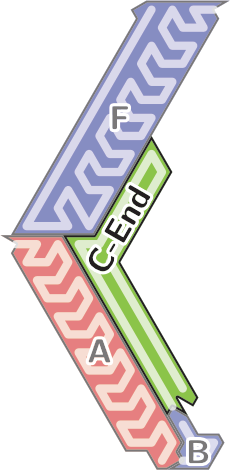}
\\
&	\stackRefs{	\refProofTree{1879}{1889}{}{}	\\	\refProofTree{2757}{2758}{}{}	}
&	\stackRefs{	\refProofTree{2798}{2838}{}{}	\\	\refProofTree{4701}{4702}{}{}	}
\\
\midrule
\ModuleDx{x}
&	\ChibiFig{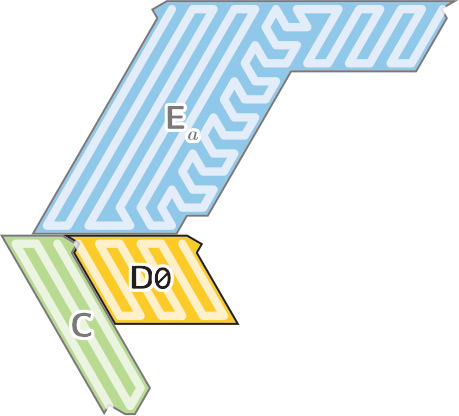}
&	\ChibiFig{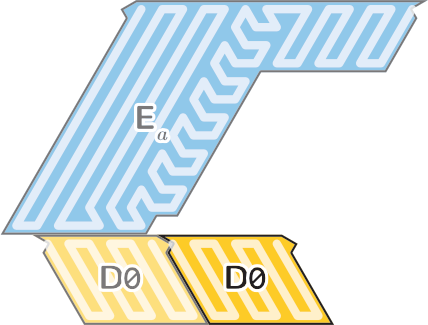}
&	\ChibiFig{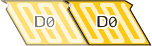}
\\
&	\stackRefs{	\refProofTree{1890}{1913}{}{}	\\	\refProofTree{2581}{2599}{}{}	\\	\refProofTree{2600}{2602}{}{}	}
&	\refProofTree{1914}{1932}{}{}
&	same as previous ones
\\
\midrule
\ModuleE
&	\ChibiFig{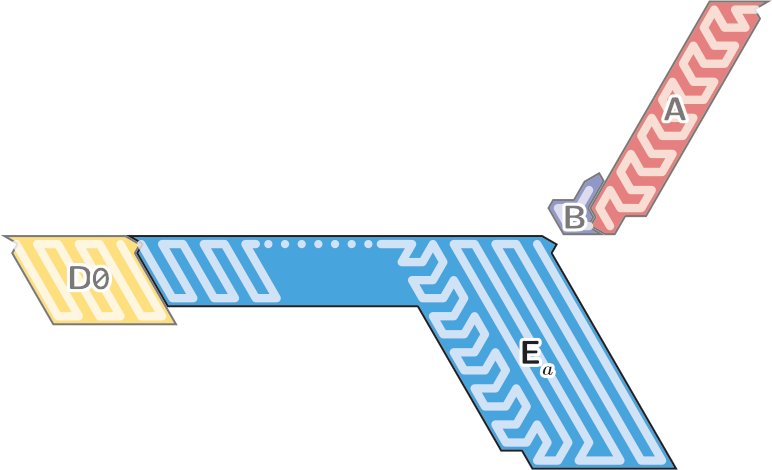}
&	\ChibiFig{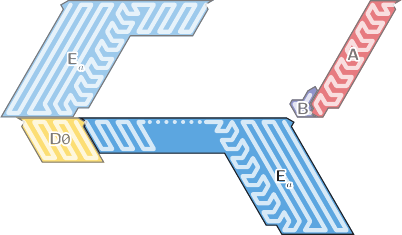}
&	\ChibiFig{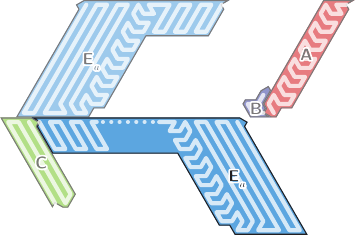}
\\
&	\refProofTree{1933}{2011}{}{}
&	\refProofTree{2603}{2632}{}{}
&	\refProofTree{2759}{2789}{}{}
\\
\midrule
\ModuleF
&	\ChibiFig{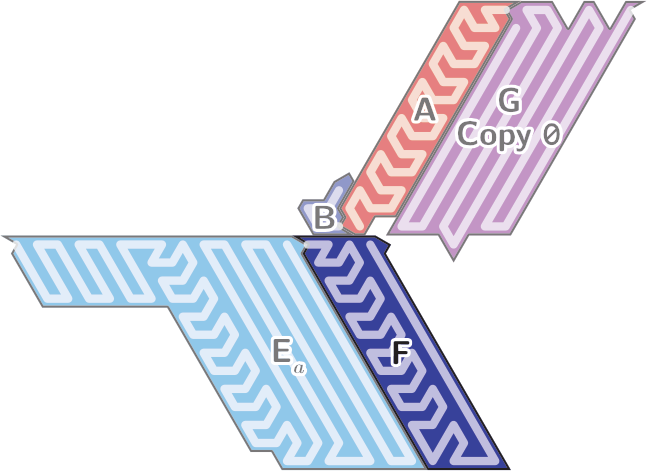}
&	\ChibiFig{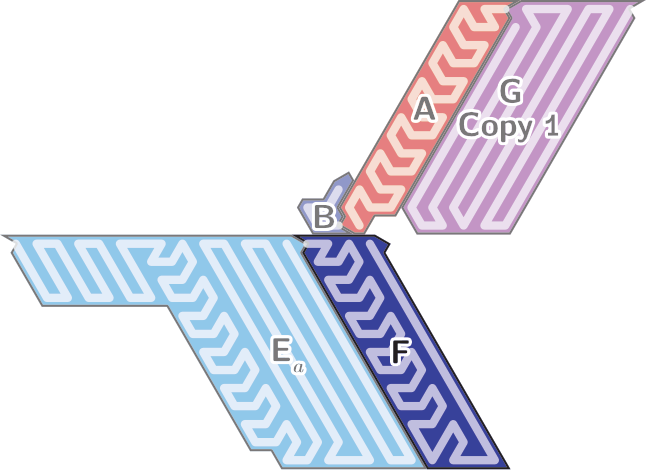}
\\
&	\refProofTree{2012}{2041}{}{}
&	\refProofTree{2633}{2643}{}{}
\\
\midrule
\ModuleG
&	\ChibiFig{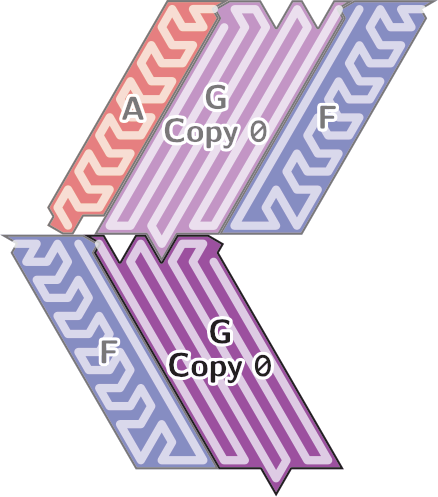}
&	\ChibiFig{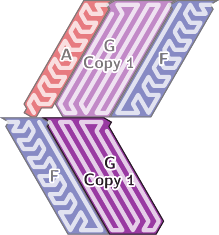}
\\
&	\refProofTree{2042}{2381}{}{}
&	\refProofTree{2644}{2744}{}{}
\\
\bottomrule
\end{tabular}


\begin{tabular}{ccc}
\toprule
\multicolumn{3}{c}{\textsf{\bfseries WRITE}}
\\
\midrule
\ModuleDx{x}
&	\ChibiFig{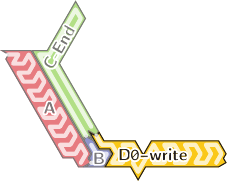}
&	\ChibiFig{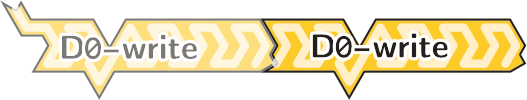}
\\
&	\refProofTree{2839}{2999}{}{}
&	\refProofTree{4753}{4786}{}{}
\\
\midrule
\ModuleE
&	\ChibiFig{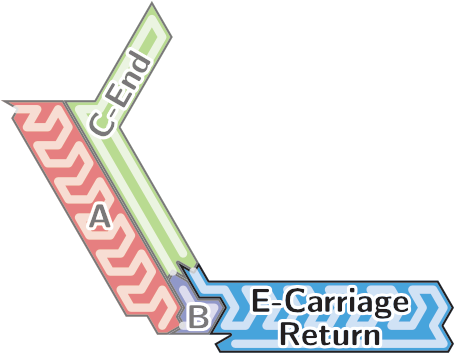}
&	\ChibiFig{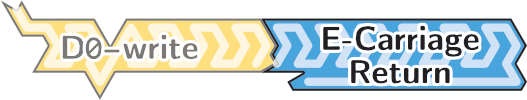}
\\
&	\refProofTree{4703}{4733}{}{}
&	\stackRefs{	\refProofTree{3000}{3749}{}{}	\\		\refProofTree{4787}{4945}{}{}	}
\\
\midrule
\ModuleF
&	\ChibiFig{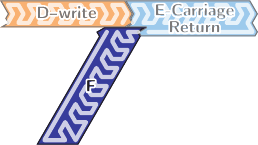}
&	\ChibiFig{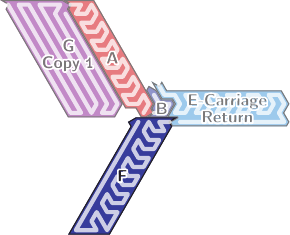}
\\
&		\stackRefs{	\refProofTree{3750}{3781}{}{}	\\		\refProofTree{4946}{4959}{}{}	}
&	\refProofTree{4734}{4744}{}{}
\\
\bottomrule
\end{tabular}


\begin{tabular}{cccc}
\toprule
\multicolumn{4}{c}{\textsf{\bfseries ZAG-WRITE}}
\\
\midrule
\ModuleA
&	\ChibiFig{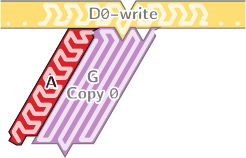}
&	\ChibiFig{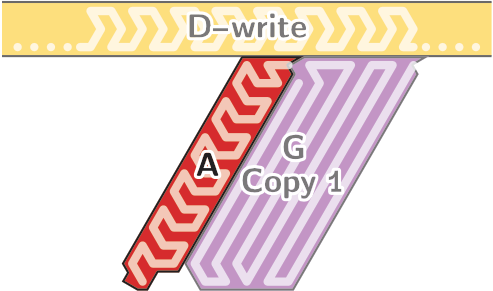}
\\
&	\refProofTree{4994}{}{}{}
&	\stackRefs{	\refProofTree{3806}{3816}{}{},	\refProofTree{4059}{4069}{}{},	\refProofTree{4215}{4225}{}{}	\\	\refProofTree{4310}{4320}{}{}		}
\\
\midrule
\ModuleC
&	\ChibiFig{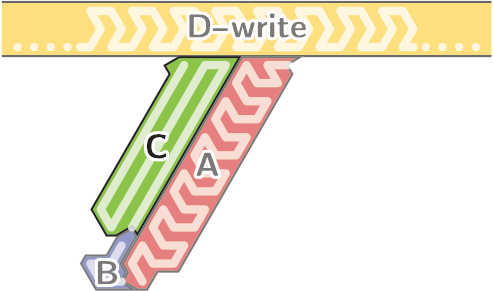}
\\
&	\stackRefs{	\refProofTree{3817}{3827}{}{},	\refProofTree{4070}{4080}{}{},	\refProofTree{4226}{4236}{}{}	\\	\refProofTree{4321}{4331}{}{},	\refProofTree{4459}{4460}{}{},	\refProofTree{4475}{4476}{}{}\\	\refProofTree{4550}{4551}{}{},	\refProofTree{4605}{4606}{}{}	}
\\
\midrule
\ModuleDx{x}
&	\ChibiFig{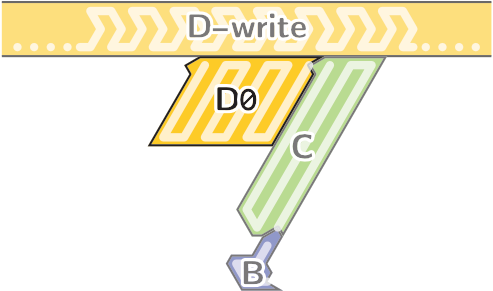}
&	\ChibiFig{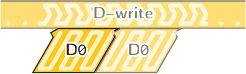}
\\
&	\stackRefs{	\refProofTree{3828}{3851}{}{} \\	\refProofTree{4237}{4263}{}{}	\\	\refProofTree{4332}{4355}{}{}	}
&	\stackRefs{		\refProofTree{3939}{3941}{}{},	\refProofTree{4434}{4439}{}{},	\refProofTree{4525}{4527}{}{}	\\	\refProofTree{4571}{4573}{}{},	\refProofTree{4587}{4589}{}{}	,	\refProofTree{4595}{4597}{}{}	\\	\refProofTree{4618}{4620}{}{}	}
\\
\midrule
\ModuleE
&	\ChibiFig{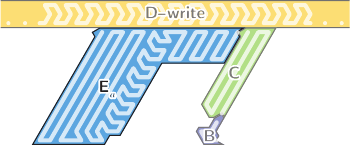}
&	\ChibiFig{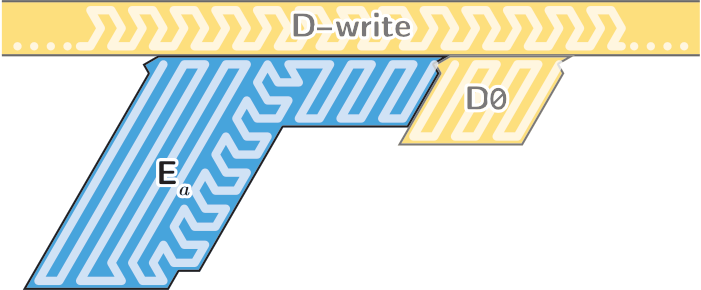}
\\
&	\stackRefs{	\refProofTree{4081}{4176}{}{},	\refProofTree{4461}{4474}{}{}	\\	\refProofTree{4552}{4570}{}{},	\refProofTree{4607}{4617}{}{}	}
&	\stackRefs{	\refProofTree{3852}{3924}{}{},	\refProofTree{3942}{4020}{}{},	\refProofTree{4264}{4271}{}{}\\	\refProofTree{4356}{4428}{}{},	\refProofTree{4440}{4458}{}{},	\refProofTree{4504}{4519}{}{}\\	\refProofTree{4528}{4549}{}{},	\refProofTree{4574}{4581}{}{},	\refProofTree{4590}{4594}{}{}\\	\refProofTree{4598}{4604}{}{},	\refProofTree{4621}{4644}{}{}	}
\\
\midrule
\ModuleF
&	\ChibiFig{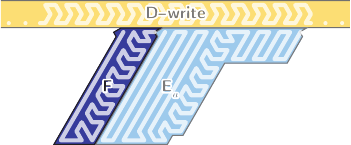}
&	\ChibiFig{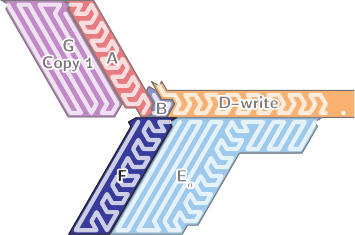}
&	\ChibiFig{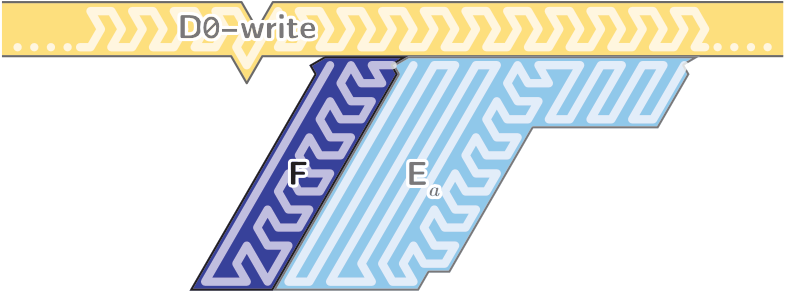}
\\
&	\stackRefs{	\refProofTree{3925}{3938}{}{},	\refProofTree{4021}{4034}{}{},	\refProofTree{4177}{4190}{}{}	\\	\refProofTree{4272}{4285}{}{},	\refProofTree{4429}{4433}{}{},	\refProofTree{4520}{4524}{}{}	\\	\refProofTree{4582}{4586}{}{}	}
&	\refProofTree{4645}{4653}{}{}
&	\refProofTree{4960}{4967}{}{}
\\
\midrule
\ModuleG
&	\ChibiFig{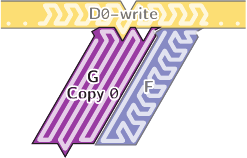}
&	\ChibiFig{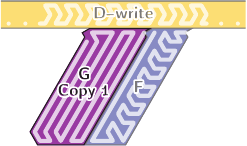}
\\
&	\refProofTree{4968}{4993}{}{}
&	\stackRefs{	\refProofTree{3782}{3805}{}{},	\refProofTree{4035}{4058}{}{},	\refProofTree{4191}{4214}{}{}	\\\refProofTree{4286}{4309}{}{}	}
\\
\bottomrule
\end{tabular}


\begin{tabular}{cccc}
\toprule
\multicolumn{4}{c}{\textsf{\bfseries ZAG}}
\\
\midrule
\ModuleA
&	\ChibiFig{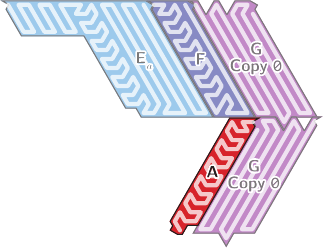}
&	\ChibiFig{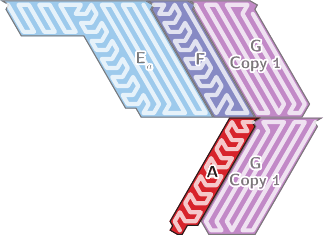}
\\
&	see Zig-Down 
&	see Zig-Down 
\\
\midrule
\ModuleB
&	\ChibiFig{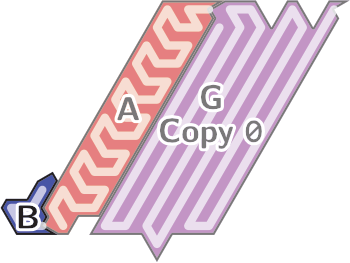}
&	\ChibiFig{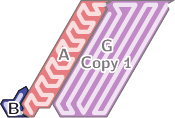}
\\
&	see Zig-Down 
&	see Zig-Down 
\\
\midrule
\ModuleC
&	\ChibiFig{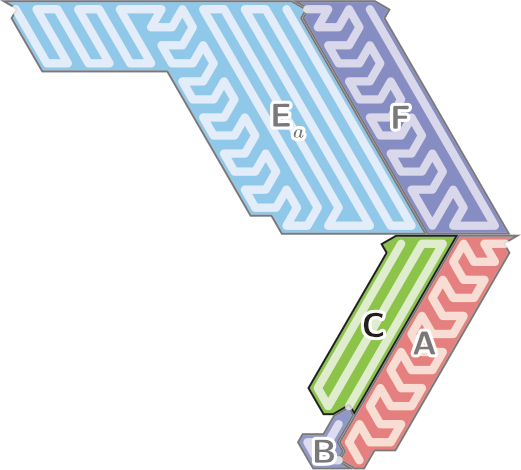}
\\
&	see Zig-Down 
\\
\midrule
\ModuleDx{x}
&	\ChibiFig{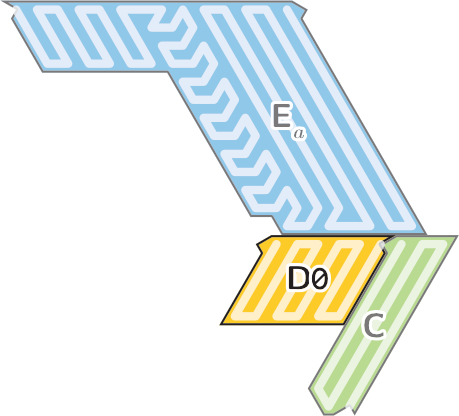}
&	\ChibiFig{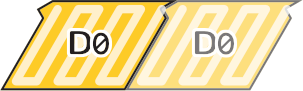}
&	\ChibiFig{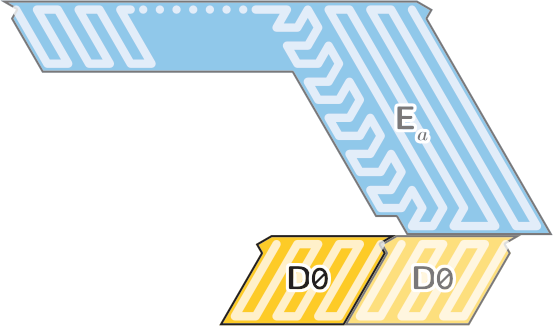}
\\
&	see Zig-Down 
&	see Zig-Down 
&	see Zig-Down 
\\
\midrule
\ModuleE
&	\ChibiFig{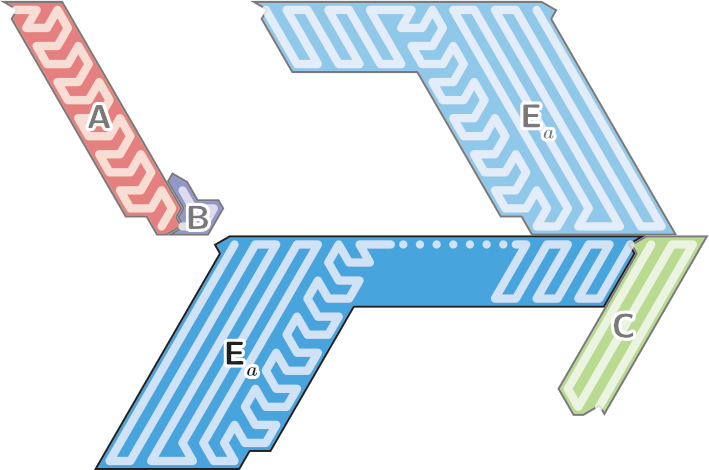}
&	\ChibiFig{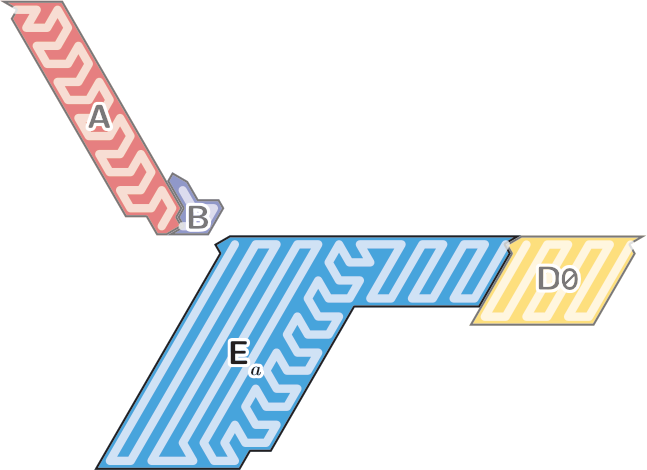}
&	\ChibiFig{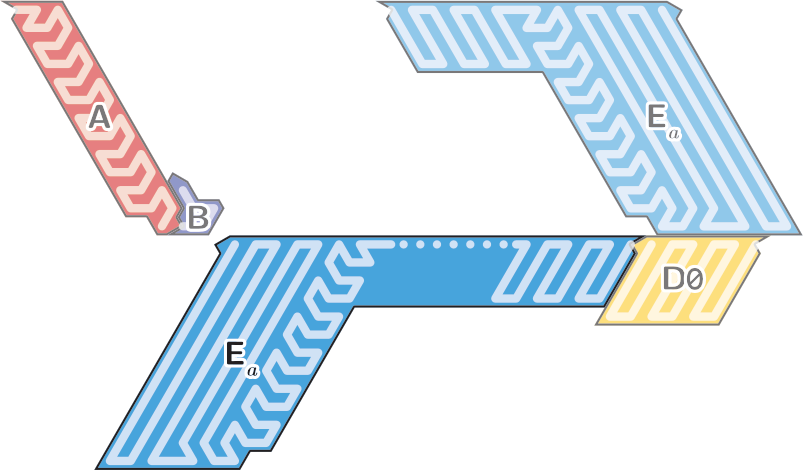}
\\
&	see Zig-Down 
&	see Zig-Down 
&	see Zig-Down 
\\
\midrule
\ModuleF
&	\ChibiFig{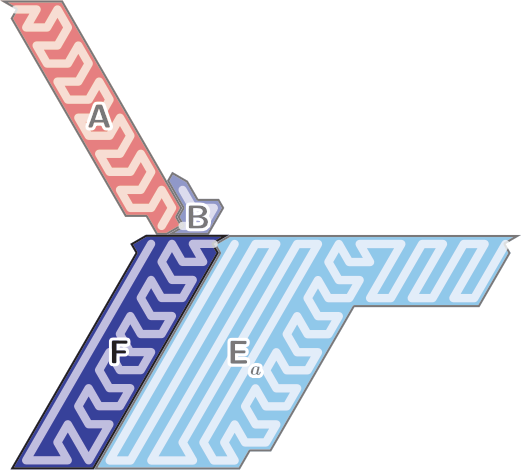}
&	\ChibiFig{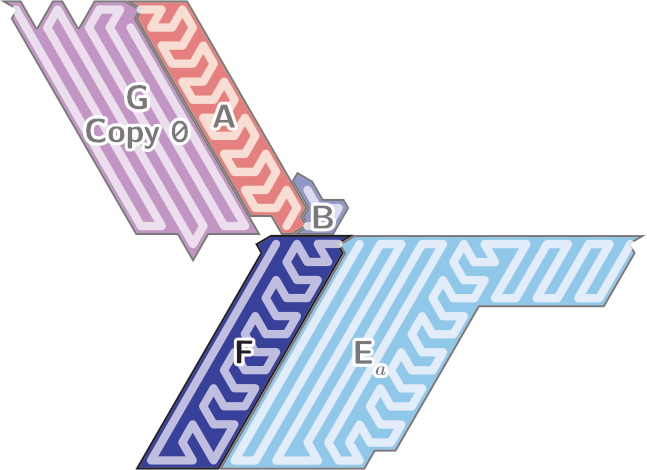}
&	\ChibiFig{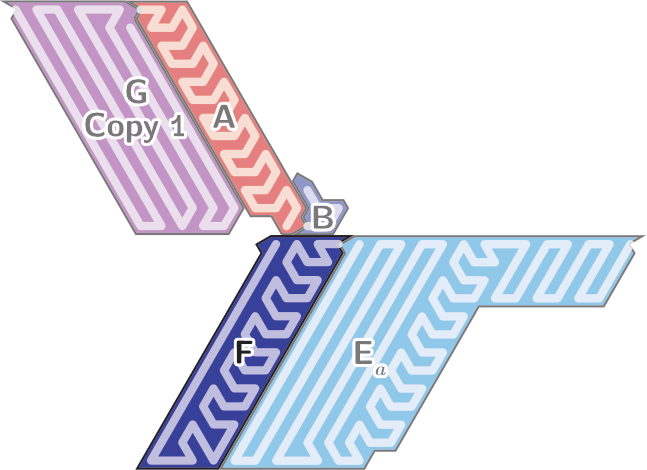}
\\
&	see Zig-Down 
&	see Zig-Down 
\\
\midrule
\ModuleG
&	\ChibiFig{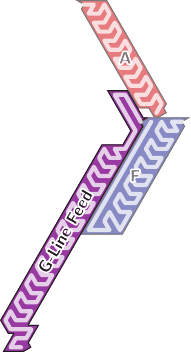}
&	\ChibiFig{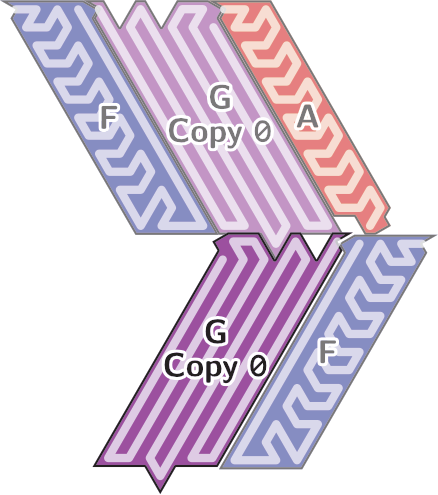}
&	\ChibiFig{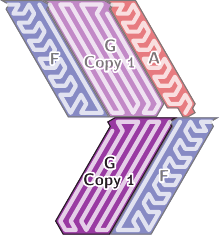}
\\
&	\refProofTree{4654}{4700}{}{}
&	see Zig-Down 
&	see Zig-Down 
\\
\bottomrule
\end{tabular}

\end{center}

%% file: rule.tex

\section{The complete attraction rule}
\label{sec:rule}

\begin{sidewaysfigure}
\includegraphics[width=\textwidth, height=.9\textheight, keepaspectratio]{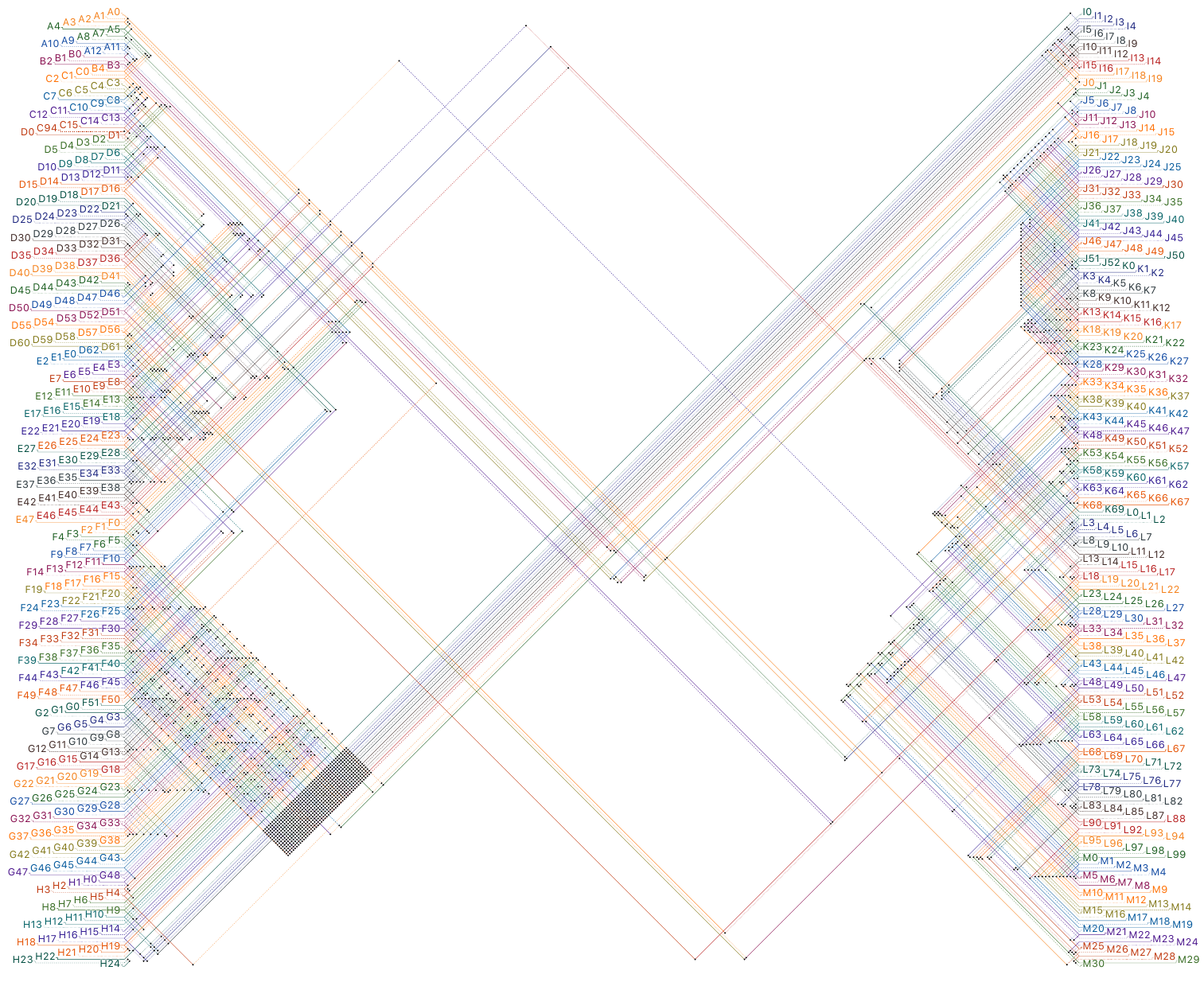}
\caption{\textsf{The rule $\protect\heart$ of the SCTS Oritatami system}: in this diagram, we have $b\protect\heart b'$ iff there is a bullet $\bullet$ at the intersection of one the two lines coming from $b$ and from $b'$; for instance, we have \texttt{A0$\protect\heart$A2} but not \texttt{A0$\protect\heart$A5}.}
\label{fig:rule:Turing}
\end{sidewaysfigure}

We first gives the rule in text. Fig.~\ref{fig:rule:Turing} displays it as a matrix.

\begin{multicols}{9}
\newcommand{\petit}[1]{\mbox{\ttfamily\scriptsize #1}}
\newcommand{\ruleTuringskip}{-.4em}
\noindent
\\
\noindent\hspace*{5mm}$\ModuleA$\\[.5em]
\petit{A00 $\heart$ A02 }\\[\ruleTuringskip]
\petit{A00 $\heart$ E02 }\\[\ruleTuringskip]
\petit{A00 $\heart$ E08 }\\[\ruleTuringskip]
\petit{A00 $\heart$ E14 }\\[\ruleTuringskip]
\petit{A00 $\heart$ E20 }\\[\ruleTuringskip]
\petit{A00 $\heart$ E26 }\\[\ruleTuringskip]
\petit{A00 $\heart$ E32 }\\[\ruleTuringskip]
\petit{A00 $\heart$ E38 }\\[\ruleTuringskip]
\petit{A00 $\heart$ E44 }\\[\ruleTuringskip]
\petit{A00 $\heart$ J18 }\\[\ruleTuringskip]
\petit{A00 $\heart$ L17 }\\[\ruleTuringskip]
\petit{A01 $\heart$ A03 }\\[\ruleTuringskip]
\petit{A01 $\heart$ A04 }\\[\ruleTuringskip]
\petit{A01 $\heart$ E03 }\\[\ruleTuringskip]
\petit{A01 $\heart$ E09 }\\[\ruleTuringskip]
\petit{A01 $\heart$ E15 }\\[\ruleTuringskip]
\petit{A01 $\heart$ E21 }\\[\ruleTuringskip]
\petit{A01 $\heart$ E27 }\\[\ruleTuringskip]
\petit{A01 $\heart$ E33 }\\[\ruleTuringskip]
\petit{A01 $\heart$ E39 }\\[\ruleTuringskip]
\petit{A01 $\heart$ E45 }\\[\ruleTuringskip]
\petit{A01 $\heart$ J18 }\\[\ruleTuringskip]
\petit{A02 $\heart$ A00}\\[\ruleTuringskip]
\petit{A02 $\heart$ A07 }\\[\ruleTuringskip]
\petit{A02 $\heart$ M26 }\\[\ruleTuringskip]
\petit{A03 $\heart$ A01}\\[\ruleTuringskip]
\petit{A04 $\heart$ A01}\\[\ruleTuringskip]
\petit{A04 $\heart$ C02 }\\[\ruleTuringskip]
\petit{A05 $\heart$ A10 }\\[\ruleTuringskip]
\petit{A05 $\heart$ C00 }\\[\ruleTuringskip]
\petit{A05 $\heart$ C01 }\\[\ruleTuringskip]
\petit{A05 $\heart$ C02 }\\[\ruleTuringskip]
\petit{A07 $\heart$ A02}\\[\ruleTuringskip]
\petit{A07 $\heart$ A08 }\\[\ruleTuringskip]
\petit{A07 $\heart$ L02 }\\[\ruleTuringskip]
\petit{A07 $\heart$ L03 }\\[\ruleTuringskip]
\petit{A08 $\heart$ A07}\\[\ruleTuringskip]
\petit{A08 $\heart$ L03 }\\[\ruleTuringskip]
\petit{A09 $\heart$ A11 }\\[\ruleTuringskip]
\petit{A09 $\heart$ A12 }\\[\ruleTuringskip]
\petit{A10 $\heart$ A05}\\[\ruleTuringskip]
\petit{A10 $\heart$ B04 }\\[\ruleTuringskip]
\petit{A11 $\heart$ A09}\\[\ruleTuringskip]
\petit{A11 $\heart$ B00 }\\[\ruleTuringskip]
\petit{A11 $\heart$ B03 }\\[\ruleTuringskip]
\petit{A11 $\heart$ B04 }\\[\ruleTuringskip]
\petit{A12 $\heart$ A09}\\[\ruleTuringskip]
\petit{A12 $\heart$ J17 }\\[\ruleTuringskip]
\\
\hspace*{5mm}$\ModuleB$\\[.5em]
\petit{B00 $\heart$ A11}\\[\ruleTuringskip]
\petit{B00 $\heart$ B03 }\\[\ruleTuringskip]
\petit{B00 $\heart$ J17 }\\[\ruleTuringskip]
\petit{B01 $\heart$ B03 }\\[\ruleTuringskip]
\petit{B01 $\heart$ D05 }\\[\ruleTuringskip]
\petit{B01 $\heart$ F05 }\\[\ruleTuringskip]
\petit{B01 $\heart$ J00 }\\[\ruleTuringskip]
\petit{B01 $\heart$ J01 }\\[\ruleTuringskip]
\petit{B01 $\heart$ J18 }\\[\ruleTuringskip]
\petit{B02 $\heart$ D03 }\\[\ruleTuringskip]
\petit{B02 $\heart$ D04 }\\[\ruleTuringskip]
\petit{B02 $\heart$ F03 }\\[\ruleTuringskip]
\petit{B02 $\heart$ F04 }\\[\ruleTuringskip]
\petit{B02 $\heart$ I17 }\\[\ruleTuringskip]
\petit{B02 $\heart$ I19 }\\[\ruleTuringskip]
\petit{B03 $\heart$ A11}\\[\ruleTuringskip]
\petit{B03 $\heart$ B00}\\[\ruleTuringskip]
\petit{B03 $\heart$ B01}\\[\ruleTuringskip]
\petit{B03 $\heart$ D02 }\\[\ruleTuringskip]
\petit{B03 $\heart$ F02 }\\[\ruleTuringskip]
\petit{B04 $\heart$ A10}\\[\ruleTuringskip]
\petit{B04 $\heart$ A11}\\[\ruleTuringskip]
\petit{B04 $\heart$ C08 }\\[\ruleTuringskip]
\petit{B04 $\heart$ D00 }\\[\ruleTuringskip]
\petit{B04 $\heart$ D02 }\\[\ruleTuringskip]
\petit{B04 $\heart$ F02 }\\[\ruleTuringskip]
\petit{B04 $\heart$ F48 }\\[\ruleTuringskip]
\\
\hspace*{5mm}$\ModuleC$\\[.5em]
\petit{C00 $\heart$ A05}\\[\ruleTuringskip]
\petit{C00 $\heart$ C05 }\\[\ruleTuringskip]
\petit{C00 $\heart$ C07 }\\[\ruleTuringskip]
\petit{C00 $\heart$ C08 }\\[\ruleTuringskip]
\petit{C00 $\heart$ C10 }\\[\ruleTuringskip]
\petit{C00 $\heart$ C13 }\\[\ruleTuringskip]
\petit{C01 $\heart$ A05}\\[\ruleTuringskip]
\petit{C01 $\heart$ C10 }\\[\ruleTuringskip]
\petit{C01 $\heart$ C13 }\\[\ruleTuringskip]
\petit{C02 $\heart$ A04}\\[\ruleTuringskip]
\petit{C02 $\heart$ A05}\\[\ruleTuringskip]
\petit{C02 $\heart$ C05 }\\[\ruleTuringskip]
\petit{C02 $\heart$ C08 }\\[\ruleTuringskip]
\petit{C03 $\heart$ C10 }\\[\ruleTuringskip]
\petit{C03 $\heart$ C11 }\\[\ruleTuringskip]
\petit{C03 $\heart$ C13 }\\[\ruleTuringskip]
\petit{C03 $\heart$ C14 }\\[\ruleTuringskip]
\petit{C03 $\heart$ C15 }\\[\ruleTuringskip]
\petit{C03 $\heart$ J08 }\\[\ruleTuringskip]
\petit{C03 $\heart$ J12 }\\[\ruleTuringskip]
\petit{C04 $\heart$ C08 }\\[\ruleTuringskip]
\petit{C04 $\heart$ J08 }\\[\ruleTuringskip]
\petit{C04 $\heart$ J11 }\\[\ruleTuringskip]
\petit{C05 $\heart$ C00}\\[\ruleTuringskip]
\petit{C05 $\heart$ C02}\\[\ruleTuringskip]
\petit{C05 $\heart$ C07 }\\[\ruleTuringskip]
\petit{C05 $\heart$ J07 }\\[\ruleTuringskip]
\petit{C06 $\heart$ C11 }\\[\ruleTuringskip]
\petit{C06 $\heart$ C14 }\\[\ruleTuringskip]
\petit{C07 $\heart$ C00}\\[\ruleTuringskip]
\petit{C07 $\heart$ C05}\\[\ruleTuringskip]
\petit{C08 $\heart$ B04}\\[\ruleTuringskip]
\petit{C08 $\heart$ C00}\\[\ruleTuringskip]
\petit{C08 $\heart$ C02}\\[\ruleTuringskip]
\petit{C08 $\heart$ C04}\\[\ruleTuringskip]
\petit{C08 $\heart$ C11 }\\[\ruleTuringskip]
\petit{C08 $\heart$ C14 }\\[\ruleTuringskip]
\petit{C09 $\heart$ D03 }\\[\ruleTuringskip]
\petit{C09 $\heart$ D15 }\\[\ruleTuringskip]
\petit{C09 $\heart$ D58 }\\[\ruleTuringskip]
\petit{C09 $\heart$ D59 }\\[\ruleTuringskip]
\petit{C09 $\heart$ E05 }\\[\ruleTuringskip]
\petit{C09 $\heart$ E11 }\\[\ruleTuringskip]
\petit{C09 $\heart$ F03 }\\[\ruleTuringskip]
\petit{C09 $\heart$ F09 }\\[\ruleTuringskip]
\petit{C10 $\heart$ C00}\\[\ruleTuringskip]
\petit{C10 $\heart$ C01}\\[\ruleTuringskip]
\petit{C10 $\heart$ C03}\\[\ruleTuringskip]
\petit{C10 $\heart$ D02 }\\[\ruleTuringskip]
\petit{C10 $\heart$ D08 }\\[\ruleTuringskip]
\petit{C10 $\heart$ D57 }\\[\ruleTuringskip]
\petit{C10 $\heart$ D58 }\\[\ruleTuringskip]
\petit{C10 $\heart$ F02 }\\[\ruleTuringskip]
\petit{C10 $\heart$ F08 }\\[\ruleTuringskip]
\petit{C11 $\heart$ C03}\\[\ruleTuringskip]
\petit{C11 $\heart$ C06}\\[\ruleTuringskip]
\petit{C11 $\heart$ C08}\\[\ruleTuringskip]
\petit{C11 $\heart$ D07 }\\[\ruleTuringskip]
\petit{C11 $\heart$ D13 }\\[\ruleTuringskip]
\petit{C11 $\heart$ D56 }\\[\ruleTuringskip]
\petit{C11 $\heart$ D57 }\\[\ruleTuringskip]
\petit{C11 $\heart$ D62 }\\[\ruleTuringskip]
\petit{C11 $\heart$ E02 }\\[\ruleTuringskip]
\petit{C11 $\heart$ E08 }\\[\ruleTuringskip]
\petit{C11 $\heart$ F01 }\\[\ruleTuringskip]
\petit{C11 $\heart$ F07 }\\[\ruleTuringskip]
\petit{C11 $\heart$ L17 }\\[\ruleTuringskip]
\petit{C12 $\heart$ D12 }\\[\ruleTuringskip]
\petit{C12 $\heart$ D13 }\\[\ruleTuringskip]
\petit{C12 $\heart$ D55 }\\[\ruleTuringskip]
\petit{C12 $\heart$ D56 }\\[\ruleTuringskip]
\petit{C12 $\heart$ D61 }\\[\ruleTuringskip]
\petit{C12 $\heart$ D62 }\\[\ruleTuringskip]
\petit{C12 $\heart$ E02 }\\[\ruleTuringskip]
\petit{C12 $\heart$ E08 }\\[\ruleTuringskip]
\petit{C12 $\heart$ F00 }\\[\ruleTuringskip]
\petit{C12 $\heart$ F06 }\\[\ruleTuringskip]
\petit{C12 $\heart$ L17 }\\[\ruleTuringskip]
\petit{C13 $\heart$ C00}\\[\ruleTuringskip]
\petit{C13 $\heart$ C01}\\[\ruleTuringskip]
\petit{C13 $\heart$ C03}\\[\ruleTuringskip]
\petit{C13 $\heart$ D11 }\\[\ruleTuringskip]
\petit{C13 $\heart$ D55 }\\[\ruleTuringskip]
\petit{C13 $\heart$ D60 }\\[\ruleTuringskip]
\petit{C13 $\heart$ D61 }\\[\ruleTuringskip]
\petit{C13 $\heart$ F05 }\\[\ruleTuringskip]
\petit{C13 $\heart$ F11 }\\[\ruleTuringskip]
\petit{C14 $\heart$ C03}\\[\ruleTuringskip]
\petit{C14 $\heart$ C06}\\[\ruleTuringskip]
\petit{C14 $\heart$ C08}\\[\ruleTuringskip]
\petit{C14 $\heart$ D04 }\\[\ruleTuringskip]
\petit{C14 $\heart$ D10 }\\[\ruleTuringskip]
\petit{C14 $\heart$ D16 }\\[\ruleTuringskip]
\petit{C14 $\heart$ D59 }\\[\ruleTuringskip]
\petit{C14 $\heart$ D60 }\\[\ruleTuringskip]
\petit{C14 $\heart$ E05 }\\[\ruleTuringskip]
\petit{C14 $\heart$ E11 }\\[\ruleTuringskip]
\petit{C14 $\heart$ F04 }\\[\ruleTuringskip]
\petit{C14 $\heart$ F10 }\\[\ruleTuringskip]
\petit{C15 $\heart$ C03}\\[\ruleTuringskip]
\petit{C15 $\heart$ D01 }\\[\ruleTuringskip]
%
\\
\hspace*{5mm}$\ModuleDx{}$\\[.5em]
\petit{D00 $\heart$ B04}\\[\ruleTuringskip]
\petit{D00 $\heart$ D02 }\\[\ruleTuringskip]
\petit{D00 $\heart$ D11 }\\[\ruleTuringskip]
\petit{D00 $\heart$ D45 }\\[\ruleTuringskip]
\petit{D01 $\heart$ C15}\\[\ruleTuringskip]
\petit{D01 $\heart$ D45 }\\[\ruleTuringskip]
\petit{D02 $\heart$ B03}\\[\ruleTuringskip]
\petit{D02 $\heart$ B04}\\[\ruleTuringskip]
\petit{D02 $\heart$ C10}\\[\ruleTuringskip]
\petit{D02 $\heart$ D00}\\[\ruleTuringskip]
\petit{D03 $\heart$ B02}\\[\ruleTuringskip]
\petit{D03 $\heart$ C09}\\[\ruleTuringskip]
\petit{D03 $\heart$ D08 }\\[\ruleTuringskip]
\petit{D03 $\heart$ D48 }\\[\ruleTuringskip]
\petit{D04 $\heart$ B02}\\[\ruleTuringskip]
\petit{D04 $\heart$ C14}\\[\ruleTuringskip]
\petit{D04 $\heart$ D48 }\\[\ruleTuringskip]
\petit{D05 $\heart$ B01}\\[\ruleTuringskip]
\petit{D06 $\heart$ E22 }\\[\ruleTuringskip]
\petit{D07 $\heart$ C11}\\[\ruleTuringskip]
\petit{D07 $\heart$ E00 }\\[\ruleTuringskip]
\petit{D07 $\heart$ E01 }\\[\ruleTuringskip]
\petit{D07 $\heart$ E02 }\\[\ruleTuringskip]
\petit{D07 $\heart$ E22 }\\[\ruleTuringskip]
\petit{D08 $\heart$ C10}\\[\ruleTuringskip]
\petit{D08 $\heart$ D03}\\[\ruleTuringskip]
\petit{D08 $\heart$ D16 }\\[\ruleTuringskip]
\petit{D09 $\heart$ E19 }\\[\ruleTuringskip]
\petit{D10 $\heart$ C14}\\[\ruleTuringskip]
\petit{D10 $\heart$ D14 }\\[\ruleTuringskip]
\petit{D10 $\heart$ E19 }\\[\ruleTuringskip]
\petit{D11 $\heart$ C13}\\[\ruleTuringskip]
\petit{D11 $\heart$ D00}\\[\ruleTuringskip]
\petit{D11 $\heart$ D13 }\\[\ruleTuringskip]
\petit{D12 $\heart$ C12}\\[\ruleTuringskip]
\petit{D12 $\heart$ E16 }\\[\ruleTuringskip]
\petit{D13 $\heart$ C11}\\[\ruleTuringskip]
\petit{D13 $\heart$ C12}\\[\ruleTuringskip]
\petit{D13 $\heart$ D11}\\[\ruleTuringskip]
\petit{D13 $\heart$ E16 }\\[\ruleTuringskip]
\petit{D14 $\heart$ D10}\\[\ruleTuringskip]
\petit{D15 $\heart$ C09}\\[\ruleTuringskip]
\petit{D15 $\heart$ E13 }\\[\ruleTuringskip]
\petit{D16 $\heart$ C14}\\[\ruleTuringskip]
\petit{D16 $\heart$ D08}\\[\ruleTuringskip]
\petit{D16 $\heart$ E13 }\\[\ruleTuringskip]
\petit{D17 $\heart$ D22 }\\[\ruleTuringskip]
\petit{D18 $\heart$ D27 }\\[\ruleTuringskip]
\petit{D18 $\heart$ D38 }\\[\ruleTuringskip]
\petit{D18 $\heart$ E18 }\\[\ruleTuringskip]
\petit{D18 $\heart$ E42 }\\[\ruleTuringskip]
\petit{D18 $\heart$ F03 }\\[\ruleTuringskip]
\petit{D18 $\heart$ F27 }\\[\ruleTuringskip]
\petit{D19 $\heart$ D26 }\\[\ruleTuringskip]
\petit{D19 $\heart$ D27 }\\[\ruleTuringskip]
\petit{D19 $\heart$ D37 }\\[\ruleTuringskip]
\petit{D19 $\heart$ D38 }\\[\ruleTuringskip]
\petit{D19 $\heart$ E23 }\\[\ruleTuringskip]
\petit{D19 $\heart$ E47 }\\[\ruleTuringskip]
\petit{D20 $\heart$ F02 }\\[\ruleTuringskip]
\petit{D20 $\heart$ F26 }\\[\ruleTuringskip]
\petit{D21 $\heart$ F00 }\\[\ruleTuringskip]
\petit{D21 $\heart$ F01 }\\[\ruleTuringskip]
\petit{D21 $\heart$ F24 }\\[\ruleTuringskip]
\petit{D21 $\heart$ F25 }\\[\ruleTuringskip]
\petit{D22 $\heart$ D17}\\[\ruleTuringskip]
\petit{D22 $\heart$ D23 }\\[\ruleTuringskip]
\petit{D22 $\heart$ D24 }\\[\ruleTuringskip]
\petit{D22 $\heart$ D34 }\\[\ruleTuringskip]
\petit{D22 $\heart$ D35 }\\[\ruleTuringskip]
\petit{D23 $\heart$ D22}\\[\ruleTuringskip]
\petit{D23 $\heart$ D45 }\\[\ruleTuringskip]
\petit{D23 $\heart$ D51 }\\[\ruleTuringskip]
\petit{D24 $\heart$ D22}\\[\ruleTuringskip]
\petit{D24 $\heart$ D45 }\\[\ruleTuringskip]
\petit{D25 $\heart$ D53 }\\[\ruleTuringskip]
\petit{D26 $\heart$ D19}\\[\ruleTuringskip]
\petit{D26 $\heart$ D48 }\\[\ruleTuringskip]
\petit{D26 $\heart$ D54 }\\[\ruleTuringskip]
\petit{D27 $\heart$ D18}\\[\ruleTuringskip]
\petit{D27 $\heart$ D19}\\[\ruleTuringskip]
\petit{D27 $\heart$ D48 }\\[\ruleTuringskip]
\petit{D28 $\heart$ E46 }\\[\ruleTuringskip]
\petit{D28 $\heart$ E47 }\\[\ruleTuringskip]
\petit{D29 $\heart$ E22 }\\[\ruleTuringskip]
\petit{D29 $\heart$ E46 }\\[\ruleTuringskip]
\petit{D30 $\heart$ D32 }\\[\ruleTuringskip]
\petit{D30 $\heart$ D33 }\\[\ruleTuringskip]
\petit{D30 $\heart$ E22 }\\[\ruleTuringskip]
\petit{D31 $\heart$ E43 }\\[\ruleTuringskip]
\petit{D31 $\heart$ E45 }\\[\ruleTuringskip]
\petit{D32 $\heart$ D30}\\[\ruleTuringskip]
\petit{D32 $\heart$ E08 }\\[\ruleTuringskip]
\petit{D32 $\heart$ E19 }\\[\ruleTuringskip]
\petit{D32 $\heart$ E42 }\\[\ruleTuringskip]
\petit{D32 $\heart$ E43 }\\[\ruleTuringskip]
\petit{D33 $\heart$ D30}\\[\ruleTuringskip]
\petit{D33 $\heart$ E07 }\\[\ruleTuringskip]
\petit{D33 $\heart$ E08 }\\[\ruleTuringskip]
\petit{D33 $\heart$ E19 }\\[\ruleTuringskip]
\petit{D33 $\heart$ E42 }\\[\ruleTuringskip]
\petit{D34 $\heart$ D22}\\[\ruleTuringskip]
\petit{D34 $\heart$ D46 }\\[\ruleTuringskip]
\petit{D34 $\heart$ D50 }\\[\ruleTuringskip]
\petit{D35 $\heart$ D22}\\[\ruleTuringskip]
\petit{D35 $\heart$ D50 }\\[\ruleTuringskip]
\petit{D36 $\heart$ D48 }\\[\ruleTuringskip]
\petit{D37 $\heart$ D19}\\[\ruleTuringskip]
\petit{D37 $\heart$ D49 }\\[\ruleTuringskip]
\petit{D37 $\heart$ D53 }\\[\ruleTuringskip]
\petit{D38 $\heart$ D18}\\[\ruleTuringskip]
\petit{D38 $\heart$ D19}\\[\ruleTuringskip]
\petit{D38 $\heart$ D53 }\\[\ruleTuringskip]
\petit{D39 $\heart$ E22 }\\[\ruleTuringskip]
\petit{D39 $\heart$ E23 }\\[\ruleTuringskip]
\petit{D40 $\heart$ E22 }\\[\ruleTuringskip]
\petit{D40 $\heart$ E46 }\\[\ruleTuringskip]
\petit{D41 $\heart$ D43 }\\[\ruleTuringskip]
\petit{D41 $\heart$ D44 }\\[\ruleTuringskip]
\petit{D41 $\heart$ E46 }\\[\ruleTuringskip]
\petit{D42 $\heart$ E19 }\\[\ruleTuringskip]
\petit{D42 $\heart$ E21 }\\[\ruleTuringskip]
\petit{D43 $\heart$ D41}\\[\ruleTuringskip]
\petit{D43 $\heart$ E18 }\\[\ruleTuringskip]
\petit{D43 $\heart$ E19 }\\[\ruleTuringskip]
\petit{D43 $\heart$ E32 }\\[\ruleTuringskip]
\petit{D43 $\heart$ E43 }\\[\ruleTuringskip]
\petit{D44 $\heart$ D41}\\[\ruleTuringskip]
\petit{D44 $\heart$ E18 }\\[\ruleTuringskip]
\petit{D44 $\heart$ E31 }\\[\ruleTuringskip]
\petit{D44 $\heart$ E32 }\\[\ruleTuringskip]
\petit{D44 $\heart$ E43 }\\[\ruleTuringskip]
\petit{D45 $\heart$ D00}\\[\ruleTuringskip]
\petit{D45 $\heart$ D01}\\[\ruleTuringskip]
\petit{D45 $\heart$ D23}\\[\ruleTuringskip]
\petit{D45 $\heart$ D24}\\[\ruleTuringskip]
\petit{D45 $\heart$ E18 }\\[\ruleTuringskip]
\petit{D46 $\heart$ D34}\\[\ruleTuringskip]
\petit{D47 $\heart$ E12 }\\[\ruleTuringskip]
\petit{D48 $\heart$ D03}\\[\ruleTuringskip]
\petit{D48 $\heart$ D04}\\[\ruleTuringskip]
\petit{D48 $\heart$ D26}\\[\ruleTuringskip]
\petit{D48 $\heart$ D27}\\[\ruleTuringskip]
\petit{D48 $\heart$ D36}\\[\ruleTuringskip]
\petit{D49 $\heart$ D37}\\[\ruleTuringskip]
\petit{D50 $\heart$ D34}\\[\ruleTuringskip]
\petit{D50 $\heart$ D35}\\[\ruleTuringskip]
\petit{D50 $\heart$ E42 }\\[\ruleTuringskip]
\petit{D51 $\heart$ D23}\\[\ruleTuringskip]
\petit{D52 $\heart$ E36 }\\[\ruleTuringskip]
\petit{D53 $\heart$ D25}\\[\ruleTuringskip]
\petit{D53 $\heart$ D37}\\[\ruleTuringskip]
\petit{D53 $\heart$ D38}\\[\ruleTuringskip]
\petit{D54 $\heart$ D26}\\[\ruleTuringskip]
\petit{D55 $\heart$ C12}\\[\ruleTuringskip]
\petit{D55 $\heart$ C13}\\[\ruleTuringskip]
\petit{D55 $\heart$ D58 }\\[\ruleTuringskip]
\petit{D55 $\heart$ D59 }\\[\ruleTuringskip]
\petit{D55 $\heart$ E15 }\\[\ruleTuringskip]
\petit{D55 $\heart$ E16 }\\[\ruleTuringskip]
\petit{D55 $\heart$ E39 }\\[\ruleTuringskip]
\petit{D55 $\heart$ E40 }\\[\ruleTuringskip]
\petit{D56 $\heart$ C11}\\[\ruleTuringskip]
\petit{D56 $\heart$ C12}\\[\ruleTuringskip]
\petit{D56 $\heart$ D58 }\\[\ruleTuringskip]
\petit{D56 $\heart$ E14 }\\[\ruleTuringskip]
\petit{D56 $\heart$ E15 }\\[\ruleTuringskip]
\petit{D56 $\heart$ E38 }\\[\ruleTuringskip]
\petit{D56 $\heart$ E39 }\\[\ruleTuringskip]
\petit{D56 $\heart$ L18 }\\[\ruleTuringskip]
\petit{D57 $\heart$ C10}\\[\ruleTuringskip]
\petit{D57 $\heart$ C11}\\[\ruleTuringskip]
\petit{D57 $\heart$ E13 }\\[\ruleTuringskip]
\petit{D57 $\heart$ E14 }\\[\ruleTuringskip]
\petit{D57 $\heart$ E37 }\\[\ruleTuringskip]
\petit{D57 $\heart$ E38 }\\[\ruleTuringskip]
\petit{D57 $\heart$ L31 }\\[\ruleTuringskip]
\petit{D57 $\heart$ L64 }\\[\ruleTuringskip]
\petit{D58 $\heart$ C09}\\[\ruleTuringskip]
\petit{D58 $\heart$ C10}\\[\ruleTuringskip]
\petit{D58 $\heart$ D55}\\[\ruleTuringskip]
\petit{D58 $\heart$ D56}\\[\ruleTuringskip]
\petit{D58 $\heart$ D62 }\\[\ruleTuringskip]
\petit{D58 $\heart$ E12 }\\[\ruleTuringskip]
\petit{D58 $\heart$ E13 }\\[\ruleTuringskip]
\petit{D58 $\heart$ E36 }\\[\ruleTuringskip]
\petit{D58 $\heart$ E37 }\\[\ruleTuringskip]
\petit{D58 $\heart$ L31 }\\[\ruleTuringskip]
\petit{D59 $\heart$ C09}\\[\ruleTuringskip]
\petit{D59 $\heart$ C14}\\[\ruleTuringskip]
\petit{D59 $\heart$ D55}\\[\ruleTuringskip]
\petit{D59 $\heart$ D61 }\\[\ruleTuringskip]
\petit{D59 $\heart$ D62 }\\[\ruleTuringskip]
\petit{D59 $\heart$ E11 }\\[\ruleTuringskip]
\petit{D59 $\heart$ E12 }\\[\ruleTuringskip]
\petit{D59 $\heart$ E35 }\\[\ruleTuringskip]
\petit{D59 $\heart$ E36 }\\[\ruleTuringskip]
\petit{D60 $\heart$ C13}\\[\ruleTuringskip]
\petit{D60 $\heart$ C14}\\[\ruleTuringskip]
\petit{D60 $\heart$ E05 }\\[\ruleTuringskip]
\petit{D60 $\heart$ E22 }\\[\ruleTuringskip]
\petit{D60 $\heart$ E23 }\\[\ruleTuringskip]
\petit{D60 $\heart$ E29 }\\[\ruleTuringskip]
\petit{D60 $\heart$ E46 }\\[\ruleTuringskip]
\petit{D60 $\heart$ E47 }\\[\ruleTuringskip]
\petit{D61 $\heart$ C12}\\[\ruleTuringskip]
\petit{D61 $\heart$ C13}\\[\ruleTuringskip]
\petit{D61 $\heart$ D59}\\[\ruleTuringskip]
\petit{D61 $\heart$ E21 }\\[\ruleTuringskip]
\petit{D61 $\heart$ E22 }\\[\ruleTuringskip]
\petit{D61 $\heart$ E45 }\\[\ruleTuringskip]
\petit{D61 $\heart$ E46 }\\[\ruleTuringskip]
\petit{D62 $\heart$ C11}\\[\ruleTuringskip]
\petit{D62 $\heart$ C12}\\[\ruleTuringskip]
\petit{D62 $\heart$ D58}\\[\ruleTuringskip]
\petit{D62 $\heart$ D59}\\[\ruleTuringskip]
\petit{D62 $\heart$ E20 }\\[\ruleTuringskip]
\petit{D62 $\heart$ E21 }\\[\ruleTuringskip]
\petit{D62 $\heart$ E44 }\\[\ruleTuringskip]
\petit{D62 $\heart$ E45 }\\[\ruleTuringskip]
\petit{E00 $\heart$ D07}\\[\ruleTuringskip]
\petit{E00 $\heart$ E05 }\\[\ruleTuringskip]
\petit{E00 $\heart$ E23 }\\[\ruleTuringskip]
\petit{E00 $\heart$ E46 }\\[\ruleTuringskip]
\petit{E01 $\heart$ D07}\\[\ruleTuringskip]
\petit{E01 $\heart$ E22 }\\[\ruleTuringskip]
\petit{E01 $\heart$ E46 }\\[\ruleTuringskip]
\petit{E02 $\heart$ A00}\\[\ruleTuringskip]
\petit{E02 $\heart$ C11}\\[\ruleTuringskip]
\petit{E02 $\heart$ C12}\\[\ruleTuringskip]
\petit{E02 $\heart$ D07}\\[\ruleTuringskip]
\petit{E02 $\heart$ E09 }\\[\ruleTuringskip]
\petit{E02 $\heart$ E45 }\\[\ruleTuringskip]
\petit{E03 $\heart$ A01}\\[\ruleTuringskip]
\petit{E03 $\heart$ E08 }\\[\ruleTuringskip]
\petit{E03 $\heart$ E20 }\\[\ruleTuringskip]
\petit{E04 $\heart$ E19 }\\[\ruleTuringskip]
\petit{E04 $\heart$ E42 }\\[\ruleTuringskip]
\petit{E05 $\heart$ C09}\\[\ruleTuringskip]
\petit{E05 $\heart$ C14}\\[\ruleTuringskip]
\petit{E05 $\heart$ D60}\\[\ruleTuringskip]
\petit{E05 $\heart$ E00}\\[\ruleTuringskip]
\petit{E05 $\heart$ E42 }\\[\ruleTuringskip]
\petit{E06 $\heart$ E11 }\\[\ruleTuringskip]
\petit{E06 $\heart$ E17 }\\[\ruleTuringskip]
\petit{E07 $\heart$ D33}\\[\ruleTuringskip]
\petit{E07 $\heart$ E16 }\\[\ruleTuringskip]
\petit{E07 $\heart$ E39 }\\[\ruleTuringskip]
\petit{E07 $\heart$ E40 }\\[\ruleTuringskip]
\petit{E08 $\heart$ A00}\\[\ruleTuringskip]
\petit{E08 $\heart$ C11}\\[\ruleTuringskip]
\petit{E08 $\heart$ C12}\\[\ruleTuringskip]
\petit{E08 $\heart$ D32}\\[\ruleTuringskip]
\petit{E08 $\heart$ D33}\\[\ruleTuringskip]
\petit{E08 $\heart$ E03}\\[\ruleTuringskip]
\petit{E08 $\heart$ E15 }\\[\ruleTuringskip]
\petit{E08 $\heart$ E38 }\\[\ruleTuringskip]
\petit{E08 $\heart$ E39 }\\[\ruleTuringskip]
\petit{E09 $\heart$ A01}\\[\ruleTuringskip]
\petit{E09 $\heart$ E02}\\[\ruleTuringskip]
\petit{E09 $\heart$ E14 }\\[\ruleTuringskip]
\petit{E09 $\heart$ L17 }\\[\ruleTuringskip]
\petit{E10 $\heart$ E13 }\\[\ruleTuringskip]
\petit{E10 $\heart$ E14 }\\[\ruleTuringskip]
\petit{E10 $\heart$ E36 }\\[\ruleTuringskip]
\petit{E11 $\heart$ C09}\\[\ruleTuringskip]
\petit{E11 $\heart$ C14}\\[\ruleTuringskip]
\petit{E11 $\heart$ D59}\\[\ruleTuringskip]
\petit{E11 $\heart$ E06}\\[\ruleTuringskip]
\petit{E11 $\heart$ E36 }\\[\ruleTuringskip]
\petit{E12 $\heart$ D47}\\[\ruleTuringskip]
\petit{E12 $\heart$ D58}\\[\ruleTuringskip]
\petit{E12 $\heart$ D59}\\[\ruleTuringskip]
\petit{E12 $\heart$ E17 }\\[\ruleTuringskip]
\petit{E12 $\heart$ E34 }\\[\ruleTuringskip]
\petit{E12 $\heart$ E35 }\\[\ruleTuringskip]
\petit{E13 $\heart$ D15}\\[\ruleTuringskip]
\petit{E13 $\heart$ D16}\\[\ruleTuringskip]
\petit{E13 $\heart$ D57}\\[\ruleTuringskip]
\petit{E13 $\heart$ D58}\\[\ruleTuringskip]
\petit{E13 $\heart$ E10}\\[\ruleTuringskip]
\petit{E13 $\heart$ F26 }\\[\ruleTuringskip]
\petit{E13 $\heart$ F27 }\\[\ruleTuringskip]
\petit{E14 $\heart$ A00}\\[\ruleTuringskip]
\petit{E14 $\heart$ D56}\\[\ruleTuringskip]
\petit{E14 $\heart$ D57}\\[\ruleTuringskip]
\petit{E14 $\heart$ E09}\\[\ruleTuringskip]
\petit{E14 $\heart$ E10}\\[\ruleTuringskip]
\petit{E14 $\heart$ E21 }\\[\ruleTuringskip]
\petit{E14 $\heart$ E32 }\\[\ruleTuringskip]
\petit{E15 $\heart$ A01}\\[\ruleTuringskip]
\petit{E15 $\heart$ D55}\\[\ruleTuringskip]
\petit{E15 $\heart$ D56}\\[\ruleTuringskip]
\petit{E15 $\heart$ E08}\\[\ruleTuringskip]
\petit{E15 $\heart$ E20 }\\[\ruleTuringskip]
\petit{E15 $\heart$ E31 }\\[\ruleTuringskip]
\petit{E15 $\heart$ E32 }\\[\ruleTuringskip]
\petit{E16 $\heart$ D12}\\[\ruleTuringskip]
\petit{E16 $\heart$ D13}\\[\ruleTuringskip]
\petit{E16 $\heart$ D55}\\[\ruleTuringskip]
\petit{E16 $\heart$ E07}\\[\ruleTuringskip]
\petit{E16 $\heart$ E31 }\\[\ruleTuringskip]
\petit{E16 $\heart$ F24 }\\[\ruleTuringskip]
\petit{E16 $\heart$ F35 }\\[\ruleTuringskip]
\petit{E17 $\heart$ E06}\\[\ruleTuringskip]
\petit{E17 $\heart$ E12}\\[\ruleTuringskip]
\petit{E18 $\heart$ D18}\\[\ruleTuringskip]
\petit{E18 $\heart$ D43}\\[\ruleTuringskip]
\petit{E18 $\heart$ D44}\\[\ruleTuringskip]
\petit{E18 $\heart$ D45}\\[\ruleTuringskip]
\petit{E18 $\heart$ E23 }\\[\ruleTuringskip]
\petit{E18 $\heart$ E28 }\\[\ruleTuringskip]
\petit{E18 $\heart$ E29 }\\[\ruleTuringskip]
\petit{E19 $\heart$ D09}\\[\ruleTuringskip]
\petit{E19 $\heart$ D10}\\[\ruleTuringskip]
\petit{E19 $\heart$ D32}\\[\ruleTuringskip]
\petit{E19 $\heart$ D33}\\[\ruleTuringskip]
\petit{E19 $\heart$ D42}\\[\ruleTuringskip]
\petit{E19 $\heart$ D43}\\[\ruleTuringskip]
\petit{E19 $\heart$ E04}\\[\ruleTuringskip]
\petit{E19 $\heart$ F32 }\\[\ruleTuringskip]
\petit{E19 $\heart$ F33 }\\[\ruleTuringskip]
\petit{E20 $\heart$ A00}\\[\ruleTuringskip]
\petit{E20 $\heart$ D62}\\[\ruleTuringskip]
\petit{E20 $\heart$ E03}\\[\ruleTuringskip]
\petit{E20 $\heart$ E15}\\[\ruleTuringskip]
\petit{E20 $\heart$ L18 }\\[\ruleTuringskip]
\petit{E21 $\heart$ A01}\\[\ruleTuringskip]
\petit{E21 $\heart$ D42}\\[\ruleTuringskip]
\petit{E21 $\heart$ D61}\\[\ruleTuringskip]
\petit{E21 $\heart$ D62}\\[\ruleTuringskip]
\petit{E21 $\heart$ E14}\\[\ruleTuringskip]
\petit{E21 $\heart$ E26 }\\[\ruleTuringskip]
\petit{E22 $\heart$ D06}\\[\ruleTuringskip]
\petit{E22 $\heart$ D07}\\[\ruleTuringskip]
\petit{E22 $\heart$ D29}\\[\ruleTuringskip]
\petit{E22 $\heart$ D30}\\[\ruleTuringskip]
\petit{E22 $\heart$ D39}\\[\ruleTuringskip]
\petit{E22 $\heart$ D40}\\[\ruleTuringskip]
\petit{E22 $\heart$ D60}\\[\ruleTuringskip]
\petit{E22 $\heart$ D61}\\[\ruleTuringskip]
\petit{E22 $\heart$ E01}\\[\ruleTuringskip]
\petit{E22 $\heart$ E24 }\\[\ruleTuringskip]
\petit{E22 $\heart$ E25 }\\[\ruleTuringskip]
\petit{E22 $\heart$ F29 }\\[\ruleTuringskip]
\petit{E22 $\heart$ F30 }\\[\ruleTuringskip]
\petit{E23 $\heart$ D19}\\[\ruleTuringskip]
\petit{E23 $\heart$ D39}\\[\ruleTuringskip]
\petit{E23 $\heart$ D60}\\[\ruleTuringskip]
\petit{E23 $\heart$ E00}\\[\ruleTuringskip]
\petit{E23 $\heart$ E18}\\[\ruleTuringskip]
\petit{E24 $\heart$ E22}\\[\ruleTuringskip]
\petit{E24 $\heart$ E29 }\\[\ruleTuringskip]
\petit{E24 $\heart$ E47 }\\[\ruleTuringskip]
\petit{E25 $\heart$ E22}\\[\ruleTuringskip]
\petit{E25 $\heart$ E46 }\\[\ruleTuringskip]
\petit{E26 $\heart$ A00}\\[\ruleTuringskip]
\petit{E26 $\heart$ E21}\\[\ruleTuringskip]
\petit{E26 $\heart$ E33 }\\[\ruleTuringskip]
\petit{E27 $\heart$ A01}\\[\ruleTuringskip]
\petit{E27 $\heart$ E32 }\\[\ruleTuringskip]
\petit{E27 $\heart$ E44 }\\[\ruleTuringskip]
\petit{E28 $\heart$ E18}\\[\ruleTuringskip]
\petit{E28 $\heart$ E43 }\\[\ruleTuringskip]
\petit{E29 $\heart$ D60}\\[\ruleTuringskip]
\petit{E29 $\heart$ E18}\\[\ruleTuringskip]
\petit{E29 $\heart$ E24}\\[\ruleTuringskip]
\petit{E30 $\heart$ E35 }\\[\ruleTuringskip]
\petit{E30 $\heart$ E41 }\\[\ruleTuringskip]
\petit{E31 $\heart$ D44}\\[\ruleTuringskip]
\petit{E31 $\heart$ E15}\\[\ruleTuringskip]
\petit{E31 $\heart$ E16}\\[\ruleTuringskip]
\petit{E31 $\heart$ E40 }\\[\ruleTuringskip]
\petit{E32 $\heart$ A00}\\[\ruleTuringskip]
\petit{E32 $\heart$ D43}\\[\ruleTuringskip]
\petit{E32 $\heart$ D44}\\[\ruleTuringskip]
\petit{E32 $\heart$ E14}\\[\ruleTuringskip]
\petit{E32 $\heart$ E15}\\[\ruleTuringskip]
\petit{E32 $\heart$ E27}\\[\ruleTuringskip]
\petit{E32 $\heart$ E39 }\\[\ruleTuringskip]
\petit{E33 $\heart$ A01}\\[\ruleTuringskip]
\petit{E33 $\heart$ E26}\\[\ruleTuringskip]
\petit{E33 $\heart$ E38 }\\[\ruleTuringskip]
\petit{E33 $\heart$ L17 }\\[\ruleTuringskip]
\petit{E34 $\heart$ E12}\\[\ruleTuringskip]
\petit{E34 $\heart$ E37 }\\[\ruleTuringskip]
\petit{E34 $\heart$ E38 }\\[\ruleTuringskip]
\petit{E35 $\heart$ D59}\\[\ruleTuringskip]
\petit{E35 $\heart$ E12}\\[\ruleTuringskip]
\petit{E35 $\heart$ E30}\\[\ruleTuringskip]
\petit{E36 $\heart$ D52}\\[\ruleTuringskip]
\petit{E36 $\heart$ D58}\\[\ruleTuringskip]
\petit{E36 $\heart$ D59}\\[\ruleTuringskip]
\petit{E36 $\heart$ E10}\\[\ruleTuringskip]
\petit{E36 $\heart$ E11}\\[\ruleTuringskip]
\petit{E36 $\heart$ E41 }\\[\ruleTuringskip]
\petit{E37 $\heart$ D57}\\[\ruleTuringskip]
\petit{E37 $\heart$ D58}\\[\ruleTuringskip]
\petit{E37 $\heart$ E34}\\[\ruleTuringskip]
\petit{E37 $\heart$ F02 }\\[\ruleTuringskip]
\petit{E37 $\heart$ F03 }\\[\ruleTuringskip]
\petit{E38 $\heart$ A00}\\[\ruleTuringskip]
\petit{E38 $\heart$ D56}\\[\ruleTuringskip]
\petit{E38 $\heart$ D57}\\[\ruleTuringskip]
\petit{E38 $\heart$ E08}\\[\ruleTuringskip]
\petit{E38 $\heart$ E33}\\[\ruleTuringskip]
\petit{E38 $\heart$ E34}\\[\ruleTuringskip]
\petit{E38 $\heart$ E45 }\\[\ruleTuringskip]
\petit{E39 $\heart$ A01}\\[\ruleTuringskip]
\petit{E39 $\heart$ D55}\\[\ruleTuringskip]
\petit{E39 $\heart$ D56}\\[\ruleTuringskip]
\petit{E39 $\heart$ E07}\\[\ruleTuringskip]
\petit{E39 $\heart$ E08}\\[\ruleTuringskip]
\petit{E39 $\heart$ E32}\\[\ruleTuringskip]
\petit{E39 $\heart$ E44 }\\[\ruleTuringskip]
\petit{E40 $\heart$ D55}\\[\ruleTuringskip]
\petit{E40 $\heart$ E07}\\[\ruleTuringskip]
\petit{E40 $\heart$ E31}\\[\ruleTuringskip]
\petit{E40 $\heart$ F00 }\\[\ruleTuringskip]
\petit{E40 $\heart$ F11 }\\[\ruleTuringskip]
\petit{E41 $\heart$ E30}\\[\ruleTuringskip]
\petit{E41 $\heart$ E36}\\[\ruleTuringskip]
\petit{E42 $\heart$ D18}\\[\ruleTuringskip]
\petit{E42 $\heart$ D32}\\[\ruleTuringskip]
\petit{E42 $\heart$ D33}\\[\ruleTuringskip]
\petit{E42 $\heart$ D50}\\[\ruleTuringskip]
\petit{E42 $\heart$ E04}\\[\ruleTuringskip]
\petit{E42 $\heart$ E05}\\[\ruleTuringskip]
\petit{E42 $\heart$ E47 }\\[\ruleTuringskip]
\petit{E43 $\heart$ D31}\\[\ruleTuringskip]
\petit{E43 $\heart$ D32}\\[\ruleTuringskip]
\petit{E43 $\heart$ D43}\\[\ruleTuringskip]
\petit{E43 $\heart$ D44}\\[\ruleTuringskip]
\petit{E43 $\heart$ E28}\\[\ruleTuringskip]
\petit{E43 $\heart$ F08 }\\[\ruleTuringskip]
\petit{E43 $\heart$ F09 }\\[\ruleTuringskip]
\petit{E44 $\heart$ A00}\\[\ruleTuringskip]
\petit{E44 $\heart$ D62}\\[\ruleTuringskip]
\petit{E44 $\heart$ E27}\\[\ruleTuringskip]
\petit{E44 $\heart$ E39}\\[\ruleTuringskip]
\petit{E44 $\heart$ L18 }\\[\ruleTuringskip]
\petit{E45 $\heart$ A01}\\[\ruleTuringskip]
\petit{E45 $\heart$ D31}\\[\ruleTuringskip]
\petit{E45 $\heart$ D61}\\[\ruleTuringskip]
\petit{E45 $\heart$ D62}\\[\ruleTuringskip]
\petit{E45 $\heart$ E02}\\[\ruleTuringskip]
\petit{E45 $\heart$ E38}\\[\ruleTuringskip]
\petit{E46 $\heart$ D28}\\[\ruleTuringskip]
\petit{E46 $\heart$ D29}\\[\ruleTuringskip]
\petit{E46 $\heart$ D40}\\[\ruleTuringskip]
\petit{E46 $\heart$ D41}\\[\ruleTuringskip]
\petit{E46 $\heart$ D60}\\[\ruleTuringskip]
\petit{E46 $\heart$ D61}\\[\ruleTuringskip]
\petit{E46 $\heart$ E00}\\[\ruleTuringskip]
\petit{E46 $\heart$ E01}\\[\ruleTuringskip]
\petit{E46 $\heart$ E25}\\[\ruleTuringskip]
\petit{E46 $\heart$ F05 }\\[\ruleTuringskip]
\petit{E46 $\heart$ F06 }\\[\ruleTuringskip]
\petit{E47 $\heart$ D19}\\[\ruleTuringskip]
\petit{E47 $\heart$ D28}\\[\ruleTuringskip]
\petit{E47 $\heart$ D60}\\[\ruleTuringskip]
\petit{E47 $\heart$ E24}\\[\ruleTuringskip]
\petit{E47 $\heart$ E42}\\[\ruleTuringskip]
\\
\hspace*{5mm}$\ModuleE$\\[.5em]
\petit{F00 $\heart$ C12}\\[\ruleTuringskip]
\petit{F00 $\heart$ D21}\\[\ruleTuringskip]
\petit{F00 $\heart$ E40}\\[\ruleTuringskip]
\petit{F00 $\heart$ F05 }\\[\ruleTuringskip]
\petit{F00 $\heart$ F23 }\\[\ruleTuringskip]
\petit{F00 $\heart$ F46 }\\[\ruleTuringskip]
\petit{F00 $\heart$ F51 }\\[\ruleTuringskip]
\petit{F00 $\heart$ G00 }\\[\ruleTuringskip]
\petit{F00 $\heart$ G04 }\\[\ruleTuringskip]
\petit{F00 $\heart$ G08 }\\[\ruleTuringskip]
\petit{F00 $\heart$ G12 }\\[\ruleTuringskip]
\petit{F00 $\heart$ G16 }\\[\ruleTuringskip]
\petit{F00 $\heart$ G20 }\\[\ruleTuringskip]
\petit{F00 $\heart$ G23 }\\[\ruleTuringskip]
\petit{F00 $\heart$ G24 }\\[\ruleTuringskip]
\petit{F00 $\heart$ G28 }\\[\ruleTuringskip]
\petit{F00 $\heart$ G32 }\\[\ruleTuringskip]
\petit{F00 $\heart$ G36 }\\[\ruleTuringskip]
\petit{F00 $\heart$ G40 }\\[\ruleTuringskip]
\petit{F00 $\heart$ G44 }\\[\ruleTuringskip]
\petit{F00 $\heart$ H03 }\\[\ruleTuringskip]
\petit{F00 $\heart$ H04 }\\[\ruleTuringskip]
\petit{F00 $\heart$ H07 }\\[\ruleTuringskip]
\petit{F00 $\heart$ H11 }\\[\ruleTuringskip]
\petit{F00 $\heart$ H15 }\\[\ruleTuringskip]
\petit{F00 $\heart$ H17 }\\[\ruleTuringskip]
\petit{F00 $\heart$ I00 }\\[\ruleTuringskip]
\petit{F00 $\heart$ I01 }\\[\ruleTuringskip]
\petit{F00 $\heart$ I02 }\\[\ruleTuringskip]
\petit{F00 $\heart$ I03 }\\[\ruleTuringskip]
\petit{F00 $\heart$ I04 }\\[\ruleTuringskip]
\petit{F00 $\heart$ I05 }\\[\ruleTuringskip]
\petit{F00 $\heart$ I06 }\\[\ruleTuringskip]
\petit{F00 $\heart$ I07 }\\[\ruleTuringskip]
\petit{F00 $\heart$ I08 }\\[\ruleTuringskip]
\petit{F00 $\heart$ I09 }\\[\ruleTuringskip]
\petit{F00 $\heart$ I10 }\\[\ruleTuringskip]
\petit{F00 $\heart$ I11 }\\[\ruleTuringskip]
\petit{F00 $\heart$ I12 }\\[\ruleTuringskip]
\petit{F00 $\heart$ I13 }\\[\ruleTuringskip]
\petit{F00 $\heart$ I14 }\\[\ruleTuringskip]
\petit{F00 $\heart$ J00 }\\[\ruleTuringskip]
\petit{F00 $\heart$ J01 }\\[\ruleTuringskip]
\petit{F01 $\heart$ C11}\\[\ruleTuringskip]
\petit{F01 $\heart$ D21}\\[\ruleTuringskip]
\petit{F01 $\heart$ F22 }\\[\ruleTuringskip]
\petit{F01 $\heart$ F46 }\\[\ruleTuringskip]
\petit{F01 $\heart$ G22 }\\[\ruleTuringskip]
\petit{F01 $\heart$ I00 }\\[\ruleTuringskip]
\petit{F01 $\heart$ I01 }\\[\ruleTuringskip]
\petit{F01 $\heart$ I02 }\\[\ruleTuringskip]
\petit{F01 $\heart$ I03 }\\[\ruleTuringskip]
\petit{F01 $\heart$ I04 }\\[\ruleTuringskip]
\petit{F01 $\heart$ I05 }\\[\ruleTuringskip]
\petit{F01 $\heart$ I06 }\\[\ruleTuringskip]
\petit{F01 $\heart$ I07 }\\[\ruleTuringskip]
\petit{F01 $\heart$ I08 }\\[\ruleTuringskip]
\petit{F01 $\heart$ I09 }\\[\ruleTuringskip]
\petit{F01 $\heart$ I10 }\\[\ruleTuringskip]
\petit{F01 $\heart$ I11 }\\[\ruleTuringskip]
\petit{F01 $\heart$ I12 }\\[\ruleTuringskip]
\petit{F01 $\heart$ I13 }\\[\ruleTuringskip]
\petit{F01 $\heart$ I14 }\\[\ruleTuringskip]
\petit{F02 $\heart$ B03}\\[\ruleTuringskip]
\petit{F02 $\heart$ B04}\\[\ruleTuringskip]
\petit{F02 $\heart$ C10}\\[\ruleTuringskip]
\petit{F02 $\heart$ D20}\\[\ruleTuringskip]
\petit{F02 $\heart$ E37}\\[\ruleTuringskip]
\petit{F02 $\heart$ F09 }\\[\ruleTuringskip]
\petit{F02 $\heart$ F45 }\\[\ruleTuringskip]
\petit{F02 $\heart$ G21 }\\[\ruleTuringskip]
\petit{F02 $\heart$ I00 }\\[\ruleTuringskip]
\petit{F02 $\heart$ I01 }\\[\ruleTuringskip]
\petit{F02 $\heart$ I02 }\\[\ruleTuringskip]
\petit{F02 $\heart$ I03 }\\[\ruleTuringskip]
\petit{F02 $\heart$ I04 }\\[\ruleTuringskip]
\petit{F02 $\heart$ I05 }\\[\ruleTuringskip]
\petit{F02 $\heart$ I06 }\\[\ruleTuringskip]
\petit{F02 $\heart$ I07 }\\[\ruleTuringskip]
\petit{F02 $\heart$ I08 }\\[\ruleTuringskip]
\petit{F02 $\heart$ I09 }\\[\ruleTuringskip]
\petit{F02 $\heart$ I10 }\\[\ruleTuringskip]
\petit{F02 $\heart$ I11 }\\[\ruleTuringskip]
\petit{F02 $\heart$ I12 }\\[\ruleTuringskip]
\petit{F02 $\heart$ I13 }\\[\ruleTuringskip]
\petit{F02 $\heart$ I14 }\\[\ruleTuringskip]
\petit{F03 $\heart$ B02}\\[\ruleTuringskip]
\petit{F03 $\heart$ C09}\\[\ruleTuringskip]
\petit{F03 $\heart$ D18}\\[\ruleTuringskip]
\petit{F03 $\heart$ E37}\\[\ruleTuringskip]
\petit{F03 $\heart$ F08 }\\[\ruleTuringskip]
\petit{F03 $\heart$ F20 }\\[\ruleTuringskip]
\petit{F03 $\heart$ F43 }\\[\ruleTuringskip]
\petit{F03 $\heart$ F44 }\\[\ruleTuringskip]
\petit{F03 $\heart$ G20 }\\[\ruleTuringskip]
\petit{F03 $\heart$ I00 }\\[\ruleTuringskip]
\petit{F03 $\heart$ I01 }\\[\ruleTuringskip]
\petit{F03 $\heart$ I02 }\\[\ruleTuringskip]
\petit{F03 $\heart$ I03 }\\[\ruleTuringskip]
\petit{F03 $\heart$ I04 }\\[\ruleTuringskip]
\petit{F03 $\heart$ I05 }\\[\ruleTuringskip]
\petit{F03 $\heart$ I06 }\\[\ruleTuringskip]
\petit{F03 $\heart$ I07 }\\[\ruleTuringskip]
\petit{F03 $\heart$ I08 }\\[\ruleTuringskip]
\petit{F03 $\heart$ I09 }\\[\ruleTuringskip]
\petit{F03 $\heart$ I10 }\\[\ruleTuringskip]
\petit{F03 $\heart$ I11 }\\[\ruleTuringskip]
\petit{F03 $\heart$ I12 }\\[\ruleTuringskip]
\petit{F03 $\heart$ I13 }\\[\ruleTuringskip]
\petit{F03 $\heart$ I14 }\\[\ruleTuringskip]
\petit{F04 $\heart$ B02}\\[\ruleTuringskip]
\petit{F04 $\heart$ C14}\\[\ruleTuringskip]
\petit{F04 $\heart$ F19 }\\[\ruleTuringskip]
\petit{F04 $\heart$ F42 }\\[\ruleTuringskip]
\petit{F04 $\heart$ G19 }\\[\ruleTuringskip]
\petit{F04 $\heart$ I00 }\\[\ruleTuringskip]
\petit{F04 $\heart$ I01 }\\[\ruleTuringskip]
\petit{F04 $\heart$ I02 }\\[\ruleTuringskip]
\petit{F04 $\heart$ I03 }\\[\ruleTuringskip]
\petit{F04 $\heart$ I04 }\\[\ruleTuringskip]
\petit{F04 $\heart$ I05 }\\[\ruleTuringskip]
\petit{F04 $\heart$ I06 }\\[\ruleTuringskip]
\petit{F04 $\heart$ I07 }\\[\ruleTuringskip]
\petit{F04 $\heart$ I08 }\\[\ruleTuringskip]
\petit{F04 $\heart$ I09 }\\[\ruleTuringskip]
\petit{F04 $\heart$ I10 }\\[\ruleTuringskip]
\petit{F04 $\heart$ I11 }\\[\ruleTuringskip]
\petit{F04 $\heart$ I12 }\\[\ruleTuringskip]
\petit{F04 $\heart$ I13 }\\[\ruleTuringskip]
\petit{F04 $\heart$ I14 }\\[\ruleTuringskip]
\petit{F05 $\heart$ B01}\\[\ruleTuringskip]
\petit{F05 $\heart$ C13}\\[\ruleTuringskip]
\petit{F05 $\heart$ E46}\\[\ruleTuringskip]
\petit{F05 $\heart$ F00}\\[\ruleTuringskip]
\petit{F05 $\heart$ F42 }\\[\ruleTuringskip]
\petit{F05 $\heart$ G02 }\\[\ruleTuringskip]
\petit{F05 $\heart$ G06 }\\[\ruleTuringskip]
\petit{F05 $\heart$ G10 }\\[\ruleTuringskip]
\petit{F05 $\heart$ G14 }\\[\ruleTuringskip]
\petit{F05 $\heart$ G18 }\\[\ruleTuringskip]
\petit{F05 $\heart$ G22 }\\[\ruleTuringskip]
\petit{F05 $\heart$ G26 }\\[\ruleTuringskip]
\petit{F05 $\heart$ G30 }\\[\ruleTuringskip]
\petit{F05 $\heart$ G34 }\\[\ruleTuringskip]
\petit{F05 $\heart$ G38 }\\[\ruleTuringskip]
\petit{F05 $\heart$ G42 }\\[\ruleTuringskip]
\petit{F05 $\heart$ G46 }\\[\ruleTuringskip]
\petit{F05 $\heart$ H02 }\\[\ruleTuringskip]
\petit{F05 $\heart$ H06 }\\[\ruleTuringskip]
\petit{F05 $\heart$ H10 }\\[\ruleTuringskip]
\petit{F05 $\heart$ H14 }\\[\ruleTuringskip]
\petit{F05 $\heart$ I00 }\\[\ruleTuringskip]
\petit{F05 $\heart$ I01 }\\[\ruleTuringskip]
\petit{F05 $\heart$ I02 }\\[\ruleTuringskip]
\petit{F05 $\heart$ I03 }\\[\ruleTuringskip]
\petit{F05 $\heart$ I04 }\\[\ruleTuringskip]
\petit{F05 $\heart$ I05 }\\[\ruleTuringskip]
\petit{F05 $\heart$ I06 }\\[\ruleTuringskip]
\petit{F05 $\heart$ I07 }\\[\ruleTuringskip]
\petit{F05 $\heart$ I08 }\\[\ruleTuringskip]
\petit{F05 $\heart$ I09 }\\[\ruleTuringskip]
\petit{F05 $\heart$ I10 }\\[\ruleTuringskip]
\petit{F05 $\heart$ I11 }\\[\ruleTuringskip]
\petit{F05 $\heart$ I12 }\\[\ruleTuringskip]
\petit{F05 $\heart$ I13 }\\[\ruleTuringskip]
\petit{F05 $\heart$ I14 }\\[\ruleTuringskip]
\petit{F05 $\heart$ J00 }\\[\ruleTuringskip]
\petit{F06 $\heart$ C12}\\[\ruleTuringskip]
\petit{F06 $\heart$ E46}\\[\ruleTuringskip]
\petit{F06 $\heart$ F11 }\\[\ruleTuringskip]
\petit{F06 $\heart$ F17 }\\[\ruleTuringskip]
\petit{F06 $\heart$ F50 }\\[\ruleTuringskip]
\petit{F06 $\heart$ F51 }\\[\ruleTuringskip]
\petit{F06 $\heart$ G01 }\\[\ruleTuringskip]
\petit{F06 $\heart$ G02 }\\[\ruleTuringskip]
\petit{F06 $\heart$ G06 }\\[\ruleTuringskip]
\petit{F06 $\heart$ G10 }\\[\ruleTuringskip]
\petit{F06 $\heart$ G13 }\\[\ruleTuringskip]
\petit{F06 $\heart$ G14 }\\[\ruleTuringskip]
\petit{F06 $\heart$ G17 }\\[\ruleTuringskip]
\petit{F06 $\heart$ G18 }\\[\ruleTuringskip]
\petit{F06 $\heart$ G22 }\\[\ruleTuringskip]
\petit{F06 $\heart$ G25 }\\[\ruleTuringskip]
\petit{F06 $\heart$ G26 }\\[\ruleTuringskip]
\petit{F06 $\heart$ G30 }\\[\ruleTuringskip]
\petit{F06 $\heart$ G34 }\\[\ruleTuringskip]
\petit{F06 $\heart$ G37 }\\[\ruleTuringskip]
\petit{F06 $\heart$ G38 }\\[\ruleTuringskip]
\petit{F06 $\heart$ G42 }\\[\ruleTuringskip]
\petit{F06 $\heart$ G46 }\\[\ruleTuringskip]
\petit{F06 $\heart$ H01 }\\[\ruleTuringskip]
\petit{F06 $\heart$ H05 }\\[\ruleTuringskip]
\petit{F06 $\heart$ H09 }\\[\ruleTuringskip]
\petit{F06 $\heart$ H13 }\\[\ruleTuringskip]
\petit{F06 $\heart$ H19 }\\[\ruleTuringskip]
\petit{F06 $\heart$ I00 }\\[\ruleTuringskip]
\petit{F06 $\heart$ I01 }\\[\ruleTuringskip]
\petit{F06 $\heart$ I02 }\\[\ruleTuringskip]
\petit{F06 $\heart$ I03 }\\[\ruleTuringskip]
\petit{F06 $\heart$ I04 }\\[\ruleTuringskip]
\petit{F06 $\heart$ I05 }\\[\ruleTuringskip]
\petit{F06 $\heart$ I06 }\\[\ruleTuringskip]
\petit{F06 $\heart$ I07 }\\[\ruleTuringskip]
\petit{F06 $\heart$ I08 }\\[\ruleTuringskip]
\petit{F06 $\heart$ I09 }\\[\ruleTuringskip]
\petit{F06 $\heart$ I10 }\\[\ruleTuringskip]
\petit{F06 $\heart$ I11 }\\[\ruleTuringskip]
\petit{F06 $\heart$ I12 }\\[\ruleTuringskip]
\petit{F06 $\heart$ I13 }\\[\ruleTuringskip]
\petit{F06 $\heart$ I14 }\\[\ruleTuringskip]
\petit{F07 $\heart$ C11}\\[\ruleTuringskip]
\petit{F07 $\heart$ F16 }\\[\ruleTuringskip]
\petit{F07 $\heart$ F39 }\\[\ruleTuringskip]
\petit{F07 $\heart$ F51 }\\[\ruleTuringskip]
\petit{F07 $\heart$ G16 }\\[\ruleTuringskip]
\petit{F07 $\heart$ I00 }\\[\ruleTuringskip]
\petit{F07 $\heart$ I01 }\\[\ruleTuringskip]
\petit{F07 $\heart$ I02 }\\[\ruleTuringskip]
\petit{F07 $\heart$ I03 }\\[\ruleTuringskip]
\petit{F07 $\heart$ I04 }\\[\ruleTuringskip]
\petit{F07 $\heart$ I05 }\\[\ruleTuringskip]
\petit{F07 $\heart$ I06 }\\[\ruleTuringskip]
\petit{F07 $\heart$ I07 }\\[\ruleTuringskip]
\petit{F07 $\heart$ I08 }\\[\ruleTuringskip]
\petit{F07 $\heart$ I09 }\\[\ruleTuringskip]
\petit{F07 $\heart$ I10 }\\[\ruleTuringskip]
\petit{F07 $\heart$ I11 }\\[\ruleTuringskip]
\petit{F07 $\heart$ I12 }\\[\ruleTuringskip]
\petit{F07 $\heart$ I13 }\\[\ruleTuringskip]
\petit{F07 $\heart$ I14 }\\[\ruleTuringskip]
\petit{F08 $\heart$ C10}\\[\ruleTuringskip]
\petit{F08 $\heart$ E43}\\[\ruleTuringskip]
\petit{F08 $\heart$ F03}\\[\ruleTuringskip]
\petit{F08 $\heart$ F15 }\\[\ruleTuringskip]
\petit{F08 $\heart$ F39 }\\[\ruleTuringskip]
\petit{F08 $\heart$ G15 }\\[\ruleTuringskip]
\petit{F08 $\heart$ I00 }\\[\ruleTuringskip]
\petit{F08 $\heart$ I01 }\\[\ruleTuringskip]
\petit{F08 $\heart$ I02 }\\[\ruleTuringskip]
\petit{F08 $\heart$ I03 }\\[\ruleTuringskip]
\petit{F08 $\heart$ I04 }\\[\ruleTuringskip]
\petit{F08 $\heart$ I05 }\\[\ruleTuringskip]
\petit{F08 $\heart$ I06 }\\[\ruleTuringskip]
\petit{F08 $\heart$ I07 }\\[\ruleTuringskip]
\petit{F08 $\heart$ I08 }\\[\ruleTuringskip]
\petit{F08 $\heart$ I09 }\\[\ruleTuringskip]
\petit{F08 $\heart$ I10 }\\[\ruleTuringskip]
\petit{F08 $\heart$ I11 }\\[\ruleTuringskip]
\petit{F08 $\heart$ I12 }\\[\ruleTuringskip]
\petit{F08 $\heart$ I13 }\\[\ruleTuringskip]
\petit{F08 $\heart$ I14 }\\[\ruleTuringskip]
\petit{F09 $\heart$ C09}\\[\ruleTuringskip]
\petit{F09 $\heart$ E43}\\[\ruleTuringskip]
\petit{F09 $\heart$ F02}\\[\ruleTuringskip]
\petit{F09 $\heart$ F14 }\\[\ruleTuringskip]
\petit{F09 $\heart$ G14 }\\[\ruleTuringskip]
\petit{F09 $\heart$ I00 }\\[\ruleTuringskip]
\petit{F09 $\heart$ I01 }\\[\ruleTuringskip]
\petit{F09 $\heart$ I02 }\\[\ruleTuringskip]
\petit{F09 $\heart$ I03 }\\[\ruleTuringskip]
\petit{F09 $\heart$ I04 }\\[\ruleTuringskip]
\petit{F09 $\heart$ I05 }\\[\ruleTuringskip]
\petit{F09 $\heart$ I06 }\\[\ruleTuringskip]
\petit{F09 $\heart$ I07 }\\[\ruleTuringskip]
\petit{F09 $\heart$ I08 }\\[\ruleTuringskip]
\petit{F09 $\heart$ I09 }\\[\ruleTuringskip]
\petit{F09 $\heart$ I10 }\\[\ruleTuringskip]
\petit{F09 $\heart$ I11 }\\[\ruleTuringskip]
\petit{F09 $\heart$ I12 }\\[\ruleTuringskip]
\petit{F09 $\heart$ I13 }\\[\ruleTuringskip]
\petit{F09 $\heart$ I14 }\\[\ruleTuringskip]
\petit{F10 $\heart$ C14}\\[\ruleTuringskip]
\petit{F10 $\heart$ F13 }\\[\ruleTuringskip]
\petit{F10 $\heart$ F14 }\\[\ruleTuringskip]
\petit{F10 $\heart$ F36 }\\[\ruleTuringskip]
\petit{F10 $\heart$ G13 }\\[\ruleTuringskip]
\petit{F10 $\heart$ I00 }\\[\ruleTuringskip]
\petit{F10 $\heart$ I01 }\\[\ruleTuringskip]
\petit{F10 $\heart$ I02 }\\[\ruleTuringskip]
\petit{F10 $\heart$ I03 }\\[\ruleTuringskip]
\petit{F10 $\heart$ I04 }\\[\ruleTuringskip]
\petit{F10 $\heart$ I05 }\\[\ruleTuringskip]
\petit{F10 $\heart$ I06 }\\[\ruleTuringskip]
\petit{F10 $\heart$ I07 }\\[\ruleTuringskip]
\petit{F10 $\heart$ I08 }\\[\ruleTuringskip]
\petit{F10 $\heart$ I09 }\\[\ruleTuringskip]
\petit{F10 $\heart$ I10 }\\[\ruleTuringskip]
\petit{F10 $\heart$ I11 }\\[\ruleTuringskip]
\petit{F10 $\heart$ I12 }\\[\ruleTuringskip]
\petit{F10 $\heart$ I13 }\\[\ruleTuringskip]
\petit{F10 $\heart$ I14 }\\[\ruleTuringskip]
\petit{F11 $\heart$ C13}\\[\ruleTuringskip]
\petit{F11 $\heart$ E40}\\[\ruleTuringskip]
\petit{F11 $\heart$ F06}\\[\ruleTuringskip]
\petit{F11 $\heart$ F36 }\\[\ruleTuringskip]
\petit{F11 $\heart$ G00 }\\[\ruleTuringskip]
\petit{F11 $\heart$ G04 }\\[\ruleTuringskip]
\petit{F11 $\heart$ G08 }\\[\ruleTuringskip]
\petit{F11 $\heart$ G12 }\\[\ruleTuringskip]
\petit{F11 $\heart$ G16 }\\[\ruleTuringskip]
\petit{F11 $\heart$ G20 }\\[\ruleTuringskip]
\petit{F11 $\heart$ G24 }\\[\ruleTuringskip]
\petit{F11 $\heart$ G28 }\\[\ruleTuringskip]
\petit{F11 $\heart$ G32 }\\[\ruleTuringskip]
\petit{F11 $\heart$ G36 }\\[\ruleTuringskip]
\petit{F11 $\heart$ G40 }\\[\ruleTuringskip]
\petit{F11 $\heart$ G44 }\\[\ruleTuringskip]
\petit{F11 $\heart$ H00 }\\[\ruleTuringskip]
\petit{F11 $\heart$ H04 }\\[\ruleTuringskip]
\petit{F11 $\heart$ H08 }\\[\ruleTuringskip]
\petit{F11 $\heart$ H12 }\\[\ruleTuringskip]
\petit{F11 $\heart$ H16 }\\[\ruleTuringskip]
\petit{F11 $\heart$ H18 }\\[\ruleTuringskip]
\petit{F11 $\heart$ I00 }\\[\ruleTuringskip]
\petit{F11 $\heart$ I01 }\\[\ruleTuringskip]
\petit{F11 $\heart$ I02 }\\[\ruleTuringskip]
\petit{F11 $\heart$ I03 }\\[\ruleTuringskip]
\petit{F11 $\heart$ I04 }\\[\ruleTuringskip]
\petit{F11 $\heart$ I05 }\\[\ruleTuringskip]
\petit{F11 $\heart$ I06 }\\[\ruleTuringskip]
\petit{F11 $\heart$ I07 }\\[\ruleTuringskip]
\petit{F11 $\heart$ I08 }\\[\ruleTuringskip]
\petit{F11 $\heart$ I09 }\\[\ruleTuringskip]
\petit{F11 $\heart$ I10 }\\[\ruleTuringskip]
\petit{F11 $\heart$ I11 }\\[\ruleTuringskip]
\petit{F11 $\heart$ I12 }\\[\ruleTuringskip]
\petit{F11 $\heart$ I13 }\\[\ruleTuringskip]
\petit{F11 $\heart$ I14 }\\[\ruleTuringskip]
\petit{F12 $\heart$ F17 }\\[\ruleTuringskip]
\petit{F12 $\heart$ F34 }\\[\ruleTuringskip]
\petit{F12 $\heart$ F35 }\\[\ruleTuringskip]
\petit{F12 $\heart$ F51 }\\[\ruleTuringskip]
\petit{F12 $\heart$ G00 }\\[\ruleTuringskip]
\petit{F12 $\heart$ G04 }\\[\ruleTuringskip]
\petit{F12 $\heart$ G08 }\\[\ruleTuringskip]
\petit{F12 $\heart$ G12 }\\[\ruleTuringskip]
\petit{F12 $\heart$ G16 }\\[\ruleTuringskip]
\petit{F12 $\heart$ G20 }\\[\ruleTuringskip]
\petit{F12 $\heart$ G24 }\\[\ruleTuringskip]
\petit{F12 $\heart$ G28 }\\[\ruleTuringskip]
\petit{F12 $\heart$ G32 }\\[\ruleTuringskip]
\petit{F12 $\heart$ G34 }\\[\ruleTuringskip]
\petit{F12 $\heart$ G35 }\\[\ruleTuringskip]
\petit{F12 $\heart$ G36 }\\[\ruleTuringskip]
\petit{F12 $\heart$ G40 }\\[\ruleTuringskip]
\petit{F12 $\heart$ G44 }\\[\ruleTuringskip]
\petit{F12 $\heart$ H03 }\\[\ruleTuringskip]
\petit{F12 $\heart$ H04 }\\[\ruleTuringskip]
\petit{F12 $\heart$ H07 }\\[\ruleTuringskip]
\petit{F12 $\heart$ H11 }\\[\ruleTuringskip]
\petit{F12 $\heart$ H15 }\\[\ruleTuringskip]
\petit{F12 $\heart$ H17 }\\[\ruleTuringskip]
\petit{F12 $\heart$ I00 }\\[\ruleTuringskip]
\petit{F12 $\heart$ I01 }\\[\ruleTuringskip]
\petit{F12 $\heart$ I02 }\\[\ruleTuringskip]
\petit{F12 $\heart$ I03 }\\[\ruleTuringskip]
\petit{F12 $\heart$ I04 }\\[\ruleTuringskip]
\petit{F12 $\heart$ I05 }\\[\ruleTuringskip]
\petit{F12 $\heart$ I06 }\\[\ruleTuringskip]
\petit{F12 $\heart$ I07 }\\[\ruleTuringskip]
\petit{F12 $\heart$ I08 }\\[\ruleTuringskip]
\petit{F12 $\heart$ I09 }\\[\ruleTuringskip]
\petit{F12 $\heart$ I10 }\\[\ruleTuringskip]
\petit{F12 $\heart$ I11 }\\[\ruleTuringskip]
\petit{F12 $\heart$ I12 }\\[\ruleTuringskip]
\petit{F12 $\heart$ I13 }\\[\ruleTuringskip]
\petit{F12 $\heart$ I14 }\\[\ruleTuringskip]
\petit{F13 $\heart$ F10}\\[\ruleTuringskip]
\petit{F13 $\heart$ G33 }\\[\ruleTuringskip]
\petit{F13 $\heart$ I00 }\\[\ruleTuringskip]
\petit{F13 $\heart$ I01 }\\[\ruleTuringskip]
\petit{F13 $\heart$ I02 }\\[\ruleTuringskip]
\petit{F13 $\heart$ I03 }\\[\ruleTuringskip]
\petit{F13 $\heart$ I04 }\\[\ruleTuringskip]
\petit{F13 $\heart$ I05 }\\[\ruleTuringskip]
\petit{F13 $\heart$ I06 }\\[\ruleTuringskip]
\petit{F13 $\heart$ I07 }\\[\ruleTuringskip]
\petit{F13 $\heart$ I08 }\\[\ruleTuringskip]
\petit{F13 $\heart$ I09 }\\[\ruleTuringskip]
\petit{F13 $\heart$ I10 }\\[\ruleTuringskip]
\petit{F13 $\heart$ I11 }\\[\ruleTuringskip]
\petit{F13 $\heart$ I12 }\\[\ruleTuringskip]
\petit{F13 $\heart$ I13 }\\[\ruleTuringskip]
\petit{F13 $\heart$ I14 }\\[\ruleTuringskip]
\petit{F14 $\heart$ F09}\\[\ruleTuringskip]
\petit{F14 $\heart$ F10}\\[\ruleTuringskip]
\petit{F14 $\heart$ F21 }\\[\ruleTuringskip]
\petit{F14 $\heart$ G32 }\\[\ruleTuringskip]
\petit{F14 $\heart$ I00 }\\[\ruleTuringskip]
\petit{F14 $\heart$ I01 }\\[\ruleTuringskip]
\petit{F14 $\heart$ I02 }\\[\ruleTuringskip]
\petit{F14 $\heart$ I03 }\\[\ruleTuringskip]
\petit{F14 $\heart$ I04 }\\[\ruleTuringskip]
\petit{F14 $\heart$ I05 }\\[\ruleTuringskip]
\petit{F14 $\heart$ I06 }\\[\ruleTuringskip]
\petit{F14 $\heart$ I07 }\\[\ruleTuringskip]
\petit{F14 $\heart$ I08 }\\[\ruleTuringskip]
\petit{F14 $\heart$ I09 }\\[\ruleTuringskip]
\petit{F14 $\heart$ I10 }\\[\ruleTuringskip]
\petit{F14 $\heart$ I11 }\\[\ruleTuringskip]
\petit{F14 $\heart$ I12 }\\[\ruleTuringskip]
\petit{F14 $\heart$ I13 }\\[\ruleTuringskip]
\petit{F14 $\heart$ I14 }\\[\ruleTuringskip]
\petit{F15 $\heart$ F08}\\[\ruleTuringskip]
\petit{F15 $\heart$ F20 }\\[\ruleTuringskip]
\petit{F15 $\heart$ F31 }\\[\ruleTuringskip]
\petit{F15 $\heart$ F32 }\\[\ruleTuringskip]
\petit{F15 $\heart$ G31 }\\[\ruleTuringskip]
\petit{F15 $\heart$ I00 }\\[\ruleTuringskip]
\petit{F15 $\heart$ I01 }\\[\ruleTuringskip]
\petit{F15 $\heart$ I02 }\\[\ruleTuringskip]
\petit{F15 $\heart$ I03 }\\[\ruleTuringskip]
\petit{F15 $\heart$ I04 }\\[\ruleTuringskip]
\petit{F15 $\heart$ I05 }\\[\ruleTuringskip]
\petit{F15 $\heart$ I06 }\\[\ruleTuringskip]
\petit{F15 $\heart$ I07 }\\[\ruleTuringskip]
\petit{F15 $\heart$ I08 }\\[\ruleTuringskip]
\petit{F15 $\heart$ I09 }\\[\ruleTuringskip]
\petit{F15 $\heart$ I10 }\\[\ruleTuringskip]
\petit{F15 $\heart$ I11 }\\[\ruleTuringskip]
\petit{F15 $\heart$ I12 }\\[\ruleTuringskip]
\petit{F15 $\heart$ I13 }\\[\ruleTuringskip]
\petit{F15 $\heart$ I14 }\\[\ruleTuringskip]
\petit{F16 $\heart$ F07}\\[\ruleTuringskip]
\petit{F16 $\heart$ G30 }\\[\ruleTuringskip]
\petit{F16 $\heart$ I00 }\\[\ruleTuringskip]
\petit{F16 $\heart$ I01 }\\[\ruleTuringskip]
\petit{F16 $\heart$ I02 }\\[\ruleTuringskip]
\petit{F16 $\heart$ I03 }\\[\ruleTuringskip]
\petit{F16 $\heart$ I04 }\\[\ruleTuringskip]
\petit{F16 $\heart$ I05 }\\[\ruleTuringskip]
\petit{F16 $\heart$ I06 }\\[\ruleTuringskip]
\petit{F16 $\heart$ I07 }\\[\ruleTuringskip]
\petit{F16 $\heart$ I08 }\\[\ruleTuringskip]
\petit{F16 $\heart$ I09 }\\[\ruleTuringskip]
\petit{F16 $\heart$ I10 }\\[\ruleTuringskip]
\petit{F16 $\heart$ I11 }\\[\ruleTuringskip]
\petit{F16 $\heart$ I12 }\\[\ruleTuringskip]
\petit{F16 $\heart$ I13 }\\[\ruleTuringskip]
\petit{F16 $\heart$ I14 }\\[\ruleTuringskip]
\petit{F17 $\heart$ F06}\\[\ruleTuringskip]
\petit{F17 $\heart$ F12}\\[\ruleTuringskip]
\petit{F17 $\heart$ F51 }\\[\ruleTuringskip]
\petit{F17 $\heart$ G02 }\\[\ruleTuringskip]
\petit{F17 $\heart$ G06 }\\[\ruleTuringskip]
\petit{F17 $\heart$ G10 }\\[\ruleTuringskip]
\petit{F17 $\heart$ G14 }\\[\ruleTuringskip]
\petit{F17 $\heart$ G18 }\\[\ruleTuringskip]
\petit{F17 $\heart$ G22 }\\[\ruleTuringskip]
\petit{F17 $\heart$ G26 }\\[\ruleTuringskip]
\petit{F17 $\heart$ G29 }\\[\ruleTuringskip]
\petit{F17 $\heart$ G30 }\\[\ruleTuringskip]
\petit{F17 $\heart$ G34 }\\[\ruleTuringskip]
\petit{F17 $\heart$ G38 }\\[\ruleTuringskip]
\petit{F17 $\heart$ G42 }\\[\ruleTuringskip]
\petit{F17 $\heart$ G46 }\\[\ruleTuringskip]
\petit{F17 $\heart$ H02 }\\[\ruleTuringskip]
\petit{F17 $\heart$ H06 }\\[\ruleTuringskip]
\petit{F17 $\heart$ H10 }\\[\ruleTuringskip]
\petit{F17 $\heart$ H14 }\\[\ruleTuringskip]
\petit{F17 $\heart$ I00 }\\[\ruleTuringskip]
\petit{F17 $\heart$ I01 }\\[\ruleTuringskip]
\petit{F17 $\heart$ I02 }\\[\ruleTuringskip]
\petit{F17 $\heart$ I03 }\\[\ruleTuringskip]
\petit{F17 $\heart$ I04 }\\[\ruleTuringskip]
\petit{F17 $\heart$ I05 }\\[\ruleTuringskip]
\petit{F17 $\heart$ I06 }\\[\ruleTuringskip]
\petit{F17 $\heart$ I07 }\\[\ruleTuringskip]
\petit{F17 $\heart$ I08 }\\[\ruleTuringskip]
\petit{F17 $\heart$ I09 }\\[\ruleTuringskip]
\petit{F17 $\heart$ I10 }\\[\ruleTuringskip]
\petit{F17 $\heart$ I11 }\\[\ruleTuringskip]
\petit{F17 $\heart$ I12 }\\[\ruleTuringskip]
\petit{F17 $\heart$ I13 }\\[\ruleTuringskip]
\petit{F17 $\heart$ I14 }\\[\ruleTuringskip]
\petit{F18 $\heart$ F23 }\\[\ruleTuringskip]
\petit{F18 $\heart$ F28 }\\[\ruleTuringskip]
\petit{F18 $\heart$ F29 }\\[\ruleTuringskip]
\petit{F18 $\heart$ F51 }\\[\ruleTuringskip]
\petit{F18 $\heart$ G01 }\\[\ruleTuringskip]
\petit{F18 $\heart$ G02 }\\[\ruleTuringskip]
\petit{F18 $\heart$ G06 }\\[\ruleTuringskip]
\petit{F18 $\heart$ G10 }\\[\ruleTuringskip]
\petit{F18 $\heart$ G13 }\\[\ruleTuringskip]
\petit{F18 $\heart$ G14 }\\[\ruleTuringskip]
\petit{F18 $\heart$ G18 }\\[\ruleTuringskip]
\petit{F18 $\heart$ G22 }\\[\ruleTuringskip]
\petit{F18 $\heart$ G25 }\\[\ruleTuringskip]
\petit{F18 $\heart$ G26 }\\[\ruleTuringskip]
\petit{F18 $\heart$ G28 }\\[\ruleTuringskip]
\petit{F18 $\heart$ G30 }\\[\ruleTuringskip]
\petit{F18 $\heart$ G34 }\\[\ruleTuringskip]
\petit{F18 $\heart$ G37 }\\[\ruleTuringskip]
\petit{F18 $\heart$ G38 }\\[\ruleTuringskip]
\petit{F18 $\heart$ G42 }\\[\ruleTuringskip]
\petit{F18 $\heart$ G46 }\\[\ruleTuringskip]
\petit{F18 $\heart$ H01 }\\[\ruleTuringskip]
\petit{F18 $\heart$ H05 }\\[\ruleTuringskip]
\petit{F18 $\heart$ H09 }\\[\ruleTuringskip]
\petit{F18 $\heart$ H13 }\\[\ruleTuringskip]
\petit{F18 $\heart$ H19 }\\[\ruleTuringskip]
\petit{F18 $\heart$ I00 }\\[\ruleTuringskip]
\petit{F18 $\heart$ I01 }\\[\ruleTuringskip]
\petit{F18 $\heart$ I02 }\\[\ruleTuringskip]
\petit{F18 $\heart$ I03 }\\[\ruleTuringskip]
\petit{F18 $\heart$ I04 }\\[\ruleTuringskip]
\petit{F18 $\heart$ I05 }\\[\ruleTuringskip]
\petit{F18 $\heart$ I06 }\\[\ruleTuringskip]
\petit{F18 $\heart$ I07 }\\[\ruleTuringskip]
\petit{F18 $\heart$ I08 }\\[\ruleTuringskip]
\petit{F18 $\heart$ I09 }\\[\ruleTuringskip]
\petit{F18 $\heart$ I10 }\\[\ruleTuringskip]
\petit{F18 $\heart$ I11 }\\[\ruleTuringskip]
\petit{F18 $\heart$ I12 }\\[\ruleTuringskip]
\petit{F18 $\heart$ I13 }\\[\ruleTuringskip]
\petit{F18 $\heart$ I14 }\\[\ruleTuringskip]
\petit{F19 $\heart$ F04}\\[\ruleTuringskip]
\petit{F19 $\heart$ F27 }\\[\ruleTuringskip]
\petit{F19 $\heart$ G27 }\\[\ruleTuringskip]
\petit{F19 $\heart$ I00 }\\[\ruleTuringskip]
\petit{F19 $\heart$ I01 }\\[\ruleTuringskip]
\petit{F19 $\heart$ I02 }\\[\ruleTuringskip]
\petit{F19 $\heart$ I03 }\\[\ruleTuringskip]
\petit{F19 $\heart$ I04 }\\[\ruleTuringskip]
\petit{F19 $\heart$ I05 }\\[\ruleTuringskip]
\petit{F19 $\heart$ I06 }\\[\ruleTuringskip]
\petit{F19 $\heart$ I07 }\\[\ruleTuringskip]
\petit{F19 $\heart$ I08 }\\[\ruleTuringskip]
\petit{F19 $\heart$ I09 }\\[\ruleTuringskip]
\petit{F19 $\heart$ I10 }\\[\ruleTuringskip]
\petit{F19 $\heart$ I11 }\\[\ruleTuringskip]
\petit{F19 $\heart$ I12 }\\[\ruleTuringskip]
\petit{F19 $\heart$ I13 }\\[\ruleTuringskip]
\petit{F19 $\heart$ I14 }\\[\ruleTuringskip]
\petit{F20 $\heart$ F03}\\[\ruleTuringskip]
\petit{F20 $\heart$ F15}\\[\ruleTuringskip]
\petit{F20 $\heart$ F27 }\\[\ruleTuringskip]
\petit{F20 $\heart$ G26 }\\[\ruleTuringskip]
\petit{F20 $\heart$ I00 }\\[\ruleTuringskip]
\petit{F20 $\heart$ I01 }\\[\ruleTuringskip]
\petit{F20 $\heart$ I02 }\\[\ruleTuringskip]
\petit{F20 $\heart$ I03 }\\[\ruleTuringskip]
\petit{F20 $\heart$ I04 }\\[\ruleTuringskip]
\petit{F20 $\heart$ I05 }\\[\ruleTuringskip]
\petit{F20 $\heart$ I06 }\\[\ruleTuringskip]
\petit{F20 $\heart$ I07 }\\[\ruleTuringskip]
\petit{F20 $\heart$ I08 }\\[\ruleTuringskip]
\petit{F20 $\heart$ I09 }\\[\ruleTuringskip]
\petit{F20 $\heart$ I10 }\\[\ruleTuringskip]
\petit{F20 $\heart$ I11 }\\[\ruleTuringskip]
\petit{F20 $\heart$ I12 }\\[\ruleTuringskip]
\petit{F20 $\heart$ I13 }\\[\ruleTuringskip]
\petit{F20 $\heart$ I14 }\\[\ruleTuringskip]
\petit{F21 $\heart$ F14}\\[\ruleTuringskip]
\petit{F21 $\heart$ F26 }\\[\ruleTuringskip]
\petit{F21 $\heart$ G25 }\\[\ruleTuringskip]
\petit{F21 $\heart$ I00 }\\[\ruleTuringskip]
\petit{F21 $\heart$ I01 }\\[\ruleTuringskip]
\petit{F21 $\heart$ I02 }\\[\ruleTuringskip]
\petit{F21 $\heart$ I03 }\\[\ruleTuringskip]
\petit{F21 $\heart$ I04 }\\[\ruleTuringskip]
\petit{F21 $\heart$ I05 }\\[\ruleTuringskip]
\petit{F21 $\heart$ I06 }\\[\ruleTuringskip]
\petit{F21 $\heart$ I07 }\\[\ruleTuringskip]
\petit{F21 $\heart$ I08 }\\[\ruleTuringskip]
\petit{F21 $\heart$ I09 }\\[\ruleTuringskip]
\petit{F21 $\heart$ I10 }\\[\ruleTuringskip]
\petit{F21 $\heart$ I11 }\\[\ruleTuringskip]
\petit{F21 $\heart$ I12 }\\[\ruleTuringskip]
\petit{F21 $\heart$ I13 }\\[\ruleTuringskip]
\petit{F21 $\heart$ I14 }\\[\ruleTuringskip]
\petit{F22 $\heart$ F01}\\[\ruleTuringskip]
\petit{F22 $\heart$ F24 }\\[\ruleTuringskip]
\petit{F22 $\heart$ F25 }\\[\ruleTuringskip]
\petit{F22 $\heart$ F49 }\\[\ruleTuringskip]
\petit{F22 $\heart$ G24 }\\[\ruleTuringskip]
\petit{F22 $\heart$ I00 }\\[\ruleTuringskip]
\petit{F22 $\heart$ I01 }\\[\ruleTuringskip]
\petit{F22 $\heart$ I02 }\\[\ruleTuringskip]
\petit{F22 $\heart$ I03 }\\[\ruleTuringskip]
\petit{F22 $\heart$ I04 }\\[\ruleTuringskip]
\petit{F22 $\heart$ I05 }\\[\ruleTuringskip]
\petit{F22 $\heart$ I06 }\\[\ruleTuringskip]
\petit{F22 $\heart$ I07 }\\[\ruleTuringskip]
\petit{F22 $\heart$ I08 }\\[\ruleTuringskip]
\petit{F22 $\heart$ I09 }\\[\ruleTuringskip]
\petit{F22 $\heart$ I10 }\\[\ruleTuringskip]
\petit{F22 $\heart$ I11 }\\[\ruleTuringskip]
\petit{F22 $\heart$ I12 }\\[\ruleTuringskip]
\petit{F22 $\heart$ I13 }\\[\ruleTuringskip]
\petit{F22 $\heart$ I14 }\\[\ruleTuringskip]
\petit{F23 $\heart$ F00}\\[\ruleTuringskip]
\petit{F23 $\heart$ F18}\\[\ruleTuringskip]
\petit{F23 $\heart$ F49 }\\[\ruleTuringskip]
\petit{F23 $\heart$ G00 }\\[\ruleTuringskip]
\petit{F23 $\heart$ G04 }\\[\ruleTuringskip]
\petit{F23 $\heart$ G08 }\\[\ruleTuringskip]
\petit{F23 $\heart$ G12 }\\[\ruleTuringskip]
\petit{F23 $\heart$ G16 }\\[\ruleTuringskip]
\petit{F23 $\heart$ G20 }\\[\ruleTuringskip]
\petit{F23 $\heart$ G24 }\\[\ruleTuringskip]
\petit{F23 $\heart$ G28 }\\[\ruleTuringskip]
\petit{F23 $\heart$ G32 }\\[\ruleTuringskip]
\petit{F23 $\heart$ G36 }\\[\ruleTuringskip]
\petit{F23 $\heart$ G40 }\\[\ruleTuringskip]
\petit{F23 $\heart$ G44 }\\[\ruleTuringskip]
\petit{F23 $\heart$ H00 }\\[\ruleTuringskip]
\petit{F23 $\heart$ H04 }\\[\ruleTuringskip]
\petit{F23 $\heart$ H08 }\\[\ruleTuringskip]
\petit{F23 $\heart$ H12 }\\[\ruleTuringskip]
\petit{F23 $\heart$ H16 }\\[\ruleTuringskip]
\petit{F23 $\heart$ H18 }\\[\ruleTuringskip]
\petit{F23 $\heart$ I00 }\\[\ruleTuringskip]
\petit{F23 $\heart$ I01 }\\[\ruleTuringskip]
\petit{F23 $\heart$ I02 }\\[\ruleTuringskip]
\petit{F23 $\heart$ I03 }\\[\ruleTuringskip]
\petit{F23 $\heart$ I04 }\\[\ruleTuringskip]
\petit{F23 $\heart$ I05 }\\[\ruleTuringskip]
\petit{F23 $\heart$ I06 }\\[\ruleTuringskip]
\petit{F23 $\heart$ I07 }\\[\ruleTuringskip]
\petit{F23 $\heart$ I08 }\\[\ruleTuringskip]
\petit{F23 $\heart$ I09 }\\[\ruleTuringskip]
\petit{F23 $\heart$ I10 }\\[\ruleTuringskip]
\petit{F23 $\heart$ I11 }\\[\ruleTuringskip]
\petit{F23 $\heart$ I12 }\\[\ruleTuringskip]
\petit{F23 $\heart$ I13 }\\[\ruleTuringskip]
\petit{F23 $\heart$ I14 }\\[\ruleTuringskip]
\petit{F24 $\heart$ D21}\\[\ruleTuringskip]
\petit{F24 $\heart$ E16}\\[\ruleTuringskip]
\petit{F24 $\heart$ F22}\\[\ruleTuringskip]
\petit{F24 $\heart$ F29 }\\[\ruleTuringskip]
\petit{F24 $\heart$ F47 }\\[\ruleTuringskip]
\petit{F24 $\heart$ F51 }\\[\ruleTuringskip]
\petit{F24 $\heart$ G00 }\\[\ruleTuringskip]
\petit{F24 $\heart$ G04 }\\[\ruleTuringskip]
\petit{F24 $\heart$ G08 }\\[\ruleTuringskip]
\petit{F24 $\heart$ G12 }\\[\ruleTuringskip]
\petit{F24 $\heart$ G16 }\\[\ruleTuringskip]
\petit{F24 $\heart$ G20 }\\[\ruleTuringskip]
\petit{F24 $\heart$ G24 }\\[\ruleTuringskip]
\petit{F24 $\heart$ G28 }\\[\ruleTuringskip]
\petit{F24 $\heart$ G32 }\\[\ruleTuringskip]
\petit{F24 $\heart$ G36 }\\[\ruleTuringskip]
\petit{F24 $\heart$ G40 }\\[\ruleTuringskip]
\petit{F24 $\heart$ G44 }\\[\ruleTuringskip]
\petit{F24 $\heart$ G47 }\\[\ruleTuringskip]
\petit{F24 $\heart$ H03 }\\[\ruleTuringskip]
\petit{F24 $\heart$ H04 }\\[\ruleTuringskip]
\petit{F24 $\heart$ H07 }\\[\ruleTuringskip]
\petit{F24 $\heart$ H11 }\\[\ruleTuringskip]
\petit{F24 $\heart$ H15 }\\[\ruleTuringskip]
\petit{F24 $\heart$ H17 }\\[\ruleTuringskip]
\petit{F24 $\heart$ I00 }\\[\ruleTuringskip]
\petit{F24 $\heart$ I01 }\\[\ruleTuringskip]
\petit{F24 $\heart$ I02 }\\[\ruleTuringskip]
\petit{F24 $\heart$ I03 }\\[\ruleTuringskip]
\petit{F24 $\heart$ I04 }\\[\ruleTuringskip]
\petit{F24 $\heart$ I05 }\\[\ruleTuringskip]
\petit{F24 $\heart$ I06 }\\[\ruleTuringskip]
\petit{F24 $\heart$ I07 }\\[\ruleTuringskip]
\petit{F24 $\heart$ I08 }\\[\ruleTuringskip]
\petit{F24 $\heart$ I09 }\\[\ruleTuringskip]
\petit{F24 $\heart$ I10 }\\[\ruleTuringskip]
\petit{F24 $\heart$ I11 }\\[\ruleTuringskip]
\petit{F24 $\heart$ I12 }\\[\ruleTuringskip]
\petit{F24 $\heart$ I13 }\\[\ruleTuringskip]
\petit{F24 $\heart$ I14 }\\[\ruleTuringskip]
\petit{F24 $\heart$ J00 }\\[\ruleTuringskip]
\petit{F24 $\heart$ J01 }\\[\ruleTuringskip]
\petit{F25 $\heart$ D21}\\[\ruleTuringskip]
\petit{F25 $\heart$ F22}\\[\ruleTuringskip]
\petit{F25 $\heart$ F46 }\\[\ruleTuringskip]
\petit{F25 $\heart$ G46 }\\[\ruleTuringskip]
\petit{F25 $\heart$ I00 }\\[\ruleTuringskip]
\petit{F25 $\heart$ I01 }\\[\ruleTuringskip]
\petit{F25 $\heart$ I02 }\\[\ruleTuringskip]
\petit{F25 $\heart$ I03 }\\[\ruleTuringskip]
\petit{F25 $\heart$ I04 }\\[\ruleTuringskip]
\petit{F25 $\heart$ I05 }\\[\ruleTuringskip]
\petit{F25 $\heart$ I06 }\\[\ruleTuringskip]
\petit{F25 $\heart$ I07 }\\[\ruleTuringskip]
\petit{F25 $\heart$ I08 }\\[\ruleTuringskip]
\petit{F25 $\heart$ I09 }\\[\ruleTuringskip]
\petit{F25 $\heart$ I10 }\\[\ruleTuringskip]
\petit{F25 $\heart$ I11 }\\[\ruleTuringskip]
\petit{F25 $\heart$ I12 }\\[\ruleTuringskip]
\petit{F25 $\heart$ I13 }\\[\ruleTuringskip]
\petit{F25 $\heart$ I14 }\\[\ruleTuringskip]
\petit{F26 $\heart$ D20}\\[\ruleTuringskip]
\petit{F26 $\heart$ E13}\\[\ruleTuringskip]
\petit{F26 $\heart$ F21}\\[\ruleTuringskip]
\petit{F26 $\heart$ F33 }\\[\ruleTuringskip]
\petit{F26 $\heart$ G45 }\\[\ruleTuringskip]
\petit{F26 $\heart$ I00 }\\[\ruleTuringskip]
\petit{F26 $\heart$ I01 }\\[\ruleTuringskip]
\petit{F26 $\heart$ I02 }\\[\ruleTuringskip]
\petit{F26 $\heart$ I03 }\\[\ruleTuringskip]
\petit{F26 $\heart$ I04 }\\[\ruleTuringskip]
\petit{F26 $\heart$ I05 }\\[\ruleTuringskip]
\petit{F26 $\heart$ I06 }\\[\ruleTuringskip]
\petit{F26 $\heart$ I07 }\\[\ruleTuringskip]
\petit{F26 $\heart$ I08 }\\[\ruleTuringskip]
\petit{F26 $\heart$ I09 }\\[\ruleTuringskip]
\petit{F26 $\heart$ I10 }\\[\ruleTuringskip]
\petit{F26 $\heart$ I11 }\\[\ruleTuringskip]
\petit{F26 $\heart$ I12 }\\[\ruleTuringskip]
\petit{F26 $\heart$ I13 }\\[\ruleTuringskip]
\petit{F26 $\heart$ I14 }\\[\ruleTuringskip]
\petit{F27 $\heart$ D18}\\[\ruleTuringskip]
\petit{F27 $\heart$ E13}\\[\ruleTuringskip]
\petit{F27 $\heart$ F19}\\[\ruleTuringskip]
\petit{F27 $\heart$ F20}\\[\ruleTuringskip]
\petit{F27 $\heart$ F32 }\\[\ruleTuringskip]
\petit{F27 $\heart$ F44 }\\[\ruleTuringskip]
\petit{F27 $\heart$ G44 }\\[\ruleTuringskip]
\petit{F27 $\heart$ I00 }\\[\ruleTuringskip]
\petit{F27 $\heart$ I01 }\\[\ruleTuringskip]
\petit{F27 $\heart$ I02 }\\[\ruleTuringskip]
\petit{F27 $\heart$ I03 }\\[\ruleTuringskip]
\petit{F27 $\heart$ I04 }\\[\ruleTuringskip]
\petit{F27 $\heart$ I05 }\\[\ruleTuringskip]
\petit{F27 $\heart$ I06 }\\[\ruleTuringskip]
\petit{F27 $\heart$ I07 }\\[\ruleTuringskip]
\petit{F27 $\heart$ I08 }\\[\ruleTuringskip]
\petit{F27 $\heart$ I09 }\\[\ruleTuringskip]
\petit{F27 $\heart$ I10 }\\[\ruleTuringskip]
\petit{F27 $\heart$ I11 }\\[\ruleTuringskip]
\petit{F27 $\heart$ I12 }\\[\ruleTuringskip]
\petit{F27 $\heart$ I13 }\\[\ruleTuringskip]
\petit{F27 $\heart$ I14 }\\[\ruleTuringskip]
\petit{F28 $\heart$ F18}\\[\ruleTuringskip]
\petit{F28 $\heart$ F43 }\\[\ruleTuringskip]
\petit{F28 $\heart$ G43 }\\[\ruleTuringskip]
\petit{F28 $\heart$ I00 }\\[\ruleTuringskip]
\petit{F28 $\heart$ I01 }\\[\ruleTuringskip]
\petit{F28 $\heart$ I02 }\\[\ruleTuringskip]
\petit{F28 $\heart$ I03 }\\[\ruleTuringskip]
\petit{F28 $\heart$ I04 }\\[\ruleTuringskip]
\petit{F28 $\heart$ I05 }\\[\ruleTuringskip]
\petit{F28 $\heart$ I06 }\\[\ruleTuringskip]
\petit{F28 $\heart$ I07 }\\[\ruleTuringskip]
\petit{F28 $\heart$ I08 }\\[\ruleTuringskip]
\petit{F28 $\heart$ I09 }\\[\ruleTuringskip]
\petit{F28 $\heart$ I10 }\\[\ruleTuringskip]
\petit{F28 $\heart$ I11 }\\[\ruleTuringskip]
\petit{F28 $\heart$ I12 }\\[\ruleTuringskip]
\petit{F28 $\heart$ I13 }\\[\ruleTuringskip]
\petit{F28 $\heart$ I14 }\\[\ruleTuringskip]
\petit{F29 $\heart$ E22}\\[\ruleTuringskip]
\petit{F29 $\heart$ F18}\\[\ruleTuringskip]
\petit{F29 $\heart$ F24}\\[\ruleTuringskip]
\petit{F29 $\heart$ G02 }\\[\ruleTuringskip]
\petit{F29 $\heart$ G06 }\\[\ruleTuringskip]
\petit{F29 $\heart$ G10 }\\[\ruleTuringskip]
\petit{F29 $\heart$ G14 }\\[\ruleTuringskip]
\petit{F29 $\heart$ G18 }\\[\ruleTuringskip]
\petit{F29 $\heart$ G22 }\\[\ruleTuringskip]
\petit{F29 $\heart$ G26 }\\[\ruleTuringskip]
\petit{F29 $\heart$ G30 }\\[\ruleTuringskip]
\petit{F29 $\heart$ G34 }\\[\ruleTuringskip]
\petit{F29 $\heart$ G38 }\\[\ruleTuringskip]
\petit{F29 $\heart$ G42 }\\[\ruleTuringskip]
\petit{F29 $\heart$ G46 }\\[\ruleTuringskip]
\petit{F29 $\heart$ H02 }\\[\ruleTuringskip]
\petit{F29 $\heart$ H06 }\\[\ruleTuringskip]
\petit{F29 $\heart$ H10 }\\[\ruleTuringskip]
\petit{F29 $\heart$ H14 }\\[\ruleTuringskip]
\petit{F29 $\heart$ I00 }\\[\ruleTuringskip]
\petit{F29 $\heart$ I01 }\\[\ruleTuringskip]
\petit{F29 $\heart$ I02 }\\[\ruleTuringskip]
\petit{F29 $\heart$ I03 }\\[\ruleTuringskip]
\petit{F29 $\heart$ I04 }\\[\ruleTuringskip]
\petit{F29 $\heart$ I05 }\\[\ruleTuringskip]
\petit{F29 $\heart$ I06 }\\[\ruleTuringskip]
\petit{F29 $\heart$ I07 }\\[\ruleTuringskip]
\petit{F29 $\heart$ I08 }\\[\ruleTuringskip]
\petit{F29 $\heart$ I09 }\\[\ruleTuringskip]
\petit{F29 $\heart$ I10 }\\[\ruleTuringskip]
\petit{F29 $\heart$ I11 }\\[\ruleTuringskip]
\petit{F29 $\heart$ I12 }\\[\ruleTuringskip]
\petit{F29 $\heart$ I13 }\\[\ruleTuringskip]
\petit{F29 $\heart$ I14 }\\[\ruleTuringskip]
\petit{F29 $\heart$ J00 }\\[\ruleTuringskip]
\petit{F30 $\heart$ E22}\\[\ruleTuringskip]
\petit{F30 $\heart$ F35 }\\[\ruleTuringskip]
\petit{F30 $\heart$ F41 }\\[\ruleTuringskip]
\petit{F30 $\heart$ F51 }\\[\ruleTuringskip]
\petit{F30 $\heart$ G01 }\\[\ruleTuringskip]
\petit{F30 $\heart$ G02 }\\[\ruleTuringskip]
\petit{F30 $\heart$ G06 }\\[\ruleTuringskip]
\petit{F30 $\heart$ G10 }\\[\ruleTuringskip]
\petit{F30 $\heart$ G13 }\\[\ruleTuringskip]
\petit{F30 $\heart$ G14 }\\[\ruleTuringskip]
\petit{F30 $\heart$ G18 }\\[\ruleTuringskip]
\petit{F30 $\heart$ G22 }\\[\ruleTuringskip]
\petit{F30 $\heart$ G25 }\\[\ruleTuringskip]
\petit{F30 $\heart$ G26 }\\[\ruleTuringskip]
\petit{F30 $\heart$ G30 }\\[\ruleTuringskip]
\petit{F30 $\heart$ G34 }\\[\ruleTuringskip]
\petit{F30 $\heart$ G37 }\\[\ruleTuringskip]
\petit{F30 $\heart$ G38 }\\[\ruleTuringskip]
\petit{F30 $\heart$ G41 }\\[\ruleTuringskip]
\petit{F30 $\heart$ G42 }\\[\ruleTuringskip]
\petit{F30 $\heart$ G46 }\\[\ruleTuringskip]
\petit{F30 $\heart$ H01 }\\[\ruleTuringskip]
\petit{F30 $\heart$ H05 }\\[\ruleTuringskip]
\petit{F30 $\heart$ H09 }\\[\ruleTuringskip]
\petit{F30 $\heart$ H13 }\\[\ruleTuringskip]
\petit{F30 $\heart$ H19 }\\[\ruleTuringskip]
\petit{F30 $\heart$ I00 }\\[\ruleTuringskip]
\petit{F30 $\heart$ I01 }\\[\ruleTuringskip]
\petit{F30 $\heart$ I02 }\\[\ruleTuringskip]
\petit{F30 $\heart$ I03 }\\[\ruleTuringskip]
\petit{F30 $\heart$ I04 }\\[\ruleTuringskip]
\petit{F30 $\heart$ I05 }\\[\ruleTuringskip]
\petit{F30 $\heart$ I06 }\\[\ruleTuringskip]
\petit{F30 $\heart$ I07 }\\[\ruleTuringskip]
\petit{F30 $\heart$ I08 }\\[\ruleTuringskip]
\petit{F30 $\heart$ I09 }\\[\ruleTuringskip]
\petit{F30 $\heart$ I10 }\\[\ruleTuringskip]
\petit{F30 $\heart$ I11 }\\[\ruleTuringskip]
\petit{F30 $\heart$ I12 }\\[\ruleTuringskip]
\petit{F30 $\heart$ I13 }\\[\ruleTuringskip]
\petit{F30 $\heart$ I14 }\\[\ruleTuringskip]
\petit{F31 $\heart$ F15}\\[\ruleTuringskip]
\petit{F31 $\heart$ F40 }\\[\ruleTuringskip]
\petit{F31 $\heart$ F51 }\\[\ruleTuringskip]
\petit{F31 $\heart$ G40 }\\[\ruleTuringskip]
\petit{F31 $\heart$ I00 }\\[\ruleTuringskip]
\petit{F31 $\heart$ I01 }\\[\ruleTuringskip]
\petit{F31 $\heart$ I02 }\\[\ruleTuringskip]
\petit{F31 $\heart$ I03 }\\[\ruleTuringskip]
\petit{F31 $\heart$ I04 }\\[\ruleTuringskip]
\petit{F31 $\heart$ I05 }\\[\ruleTuringskip]
\petit{F31 $\heart$ I06 }\\[\ruleTuringskip]
\petit{F31 $\heart$ I07 }\\[\ruleTuringskip]
\petit{F31 $\heart$ I08 }\\[\ruleTuringskip]
\petit{F31 $\heart$ I09 }\\[\ruleTuringskip]
\petit{F31 $\heart$ I10 }\\[\ruleTuringskip]
\petit{F31 $\heart$ I11 }\\[\ruleTuringskip]
\petit{F31 $\heart$ I12 }\\[\ruleTuringskip]
\petit{F31 $\heart$ I13 }\\[\ruleTuringskip]
\petit{F31 $\heart$ I14 }\\[\ruleTuringskip]
\petit{F32 $\heart$ E19}\\[\ruleTuringskip]
\petit{F32 $\heart$ F15}\\[\ruleTuringskip]
\petit{F32 $\heart$ F27}\\[\ruleTuringskip]
\petit{F32 $\heart$ F39 }\\[\ruleTuringskip]
\petit{F32 $\heart$ G39 }\\[\ruleTuringskip]
\petit{F32 $\heart$ I00 }\\[\ruleTuringskip]
\petit{F32 $\heart$ I01 }\\[\ruleTuringskip]
\petit{F32 $\heart$ I02 }\\[\ruleTuringskip]
\petit{F32 $\heart$ I03 }\\[\ruleTuringskip]
\petit{F32 $\heart$ I04 }\\[\ruleTuringskip]
\petit{F32 $\heart$ I05 }\\[\ruleTuringskip]
\petit{F32 $\heart$ I06 }\\[\ruleTuringskip]
\petit{F32 $\heart$ I07 }\\[\ruleTuringskip]
\petit{F32 $\heart$ I08 }\\[\ruleTuringskip]
\petit{F32 $\heart$ I09 }\\[\ruleTuringskip]
\petit{F32 $\heart$ I10 }\\[\ruleTuringskip]
\petit{F32 $\heart$ I11 }\\[\ruleTuringskip]
\petit{F32 $\heart$ I12 }\\[\ruleTuringskip]
\petit{F32 $\heart$ I13 }\\[\ruleTuringskip]
\petit{F32 $\heart$ I14 }\\[\ruleTuringskip]
\petit{F33 $\heart$ E19}\\[\ruleTuringskip]
\petit{F33 $\heart$ F26}\\[\ruleTuringskip]
\petit{F33 $\heart$ F38 }\\[\ruleTuringskip]
\petit{F33 $\heart$ G38 }\\[\ruleTuringskip]
\petit{F33 $\heart$ I00 }\\[\ruleTuringskip]
\petit{F33 $\heart$ I01 }\\[\ruleTuringskip]
\petit{F33 $\heart$ I02 }\\[\ruleTuringskip]
\petit{F33 $\heart$ I03 }\\[\ruleTuringskip]
\petit{F33 $\heart$ I04 }\\[\ruleTuringskip]
\petit{F33 $\heart$ I05 }\\[\ruleTuringskip]
\petit{F33 $\heart$ I06 }\\[\ruleTuringskip]
\petit{F33 $\heart$ I07 }\\[\ruleTuringskip]
\petit{F33 $\heart$ I08 }\\[\ruleTuringskip]
\petit{F33 $\heart$ I09 }\\[\ruleTuringskip]
\petit{F33 $\heart$ I10 }\\[\ruleTuringskip]
\petit{F33 $\heart$ I11 }\\[\ruleTuringskip]
\petit{F33 $\heart$ I12 }\\[\ruleTuringskip]
\petit{F33 $\heart$ I13 }\\[\ruleTuringskip]
\petit{F33 $\heart$ I14 }\\[\ruleTuringskip]
\petit{F34 $\heart$ F12}\\[\ruleTuringskip]
\petit{F34 $\heart$ F37 }\\[\ruleTuringskip]
\petit{F34 $\heart$ F38 }\\[\ruleTuringskip]
\petit{F34 $\heart$ G37 }\\[\ruleTuringskip]
\petit{F34 $\heart$ I00 }\\[\ruleTuringskip]
\petit{F34 $\heart$ I01 }\\[\ruleTuringskip]
\petit{F34 $\heart$ I02 }\\[\ruleTuringskip]
\petit{F34 $\heart$ I03 }\\[\ruleTuringskip]
\petit{F34 $\heart$ I04 }\\[\ruleTuringskip]
\petit{F34 $\heart$ I05 }\\[\ruleTuringskip]
\petit{F34 $\heart$ I06 }\\[\ruleTuringskip]
\petit{F34 $\heart$ I07 }\\[\ruleTuringskip]
\petit{F34 $\heart$ I08 }\\[\ruleTuringskip]
\petit{F34 $\heart$ I09 }\\[\ruleTuringskip]
\petit{F34 $\heart$ I10 }\\[\ruleTuringskip]
\petit{F34 $\heart$ I11 }\\[\ruleTuringskip]
\petit{F34 $\heart$ I12 }\\[\ruleTuringskip]
\petit{F34 $\heart$ I13 }\\[\ruleTuringskip]
\petit{F34 $\heart$ I14 }\\[\ruleTuringskip]
\petit{F35 $\heart$ E16}\\[\ruleTuringskip]
\petit{F35 $\heart$ F12}\\[\ruleTuringskip]
\petit{F35 $\heart$ F30}\\[\ruleTuringskip]
\petit{F35 $\heart$ G00 }\\[\ruleTuringskip]
\petit{F35 $\heart$ G04 }\\[\ruleTuringskip]
\petit{F35 $\heart$ G08 }\\[\ruleTuringskip]
\petit{F35 $\heart$ G12 }\\[\ruleTuringskip]
\petit{F35 $\heart$ G16 }\\[\ruleTuringskip]
\petit{F35 $\heart$ G20 }\\[\ruleTuringskip]
\petit{F35 $\heart$ G24 }\\[\ruleTuringskip]
\petit{F35 $\heart$ G28 }\\[\ruleTuringskip]
\petit{F35 $\heart$ G32 }\\[\ruleTuringskip]
\petit{F35 $\heart$ G36 }\\[\ruleTuringskip]
\petit{F35 $\heart$ G40 }\\[\ruleTuringskip]
\petit{F35 $\heart$ G44 }\\[\ruleTuringskip]
\petit{F35 $\heart$ H00 }\\[\ruleTuringskip]
\petit{F35 $\heart$ H04 }\\[\ruleTuringskip]
\petit{F35 $\heart$ H08 }\\[\ruleTuringskip]
\petit{F35 $\heart$ H12 }\\[\ruleTuringskip]
\petit{F35 $\heart$ H16 }\\[\ruleTuringskip]
\petit{F35 $\heart$ H18 }\\[\ruleTuringskip]
\petit{F35 $\heart$ I00 }\\[\ruleTuringskip]
\petit{F35 $\heart$ I01 }\\[\ruleTuringskip]
\petit{F35 $\heart$ I02 }\\[\ruleTuringskip]
\petit{F35 $\heart$ I03 }\\[\ruleTuringskip]
\petit{F35 $\heart$ I04 }\\[\ruleTuringskip]
\petit{F35 $\heart$ I05 }\\[\ruleTuringskip]
\petit{F35 $\heart$ I06 }\\[\ruleTuringskip]
\petit{F35 $\heart$ I07 }\\[\ruleTuringskip]
\petit{F35 $\heart$ I08 }\\[\ruleTuringskip]
\petit{F35 $\heart$ I09 }\\[\ruleTuringskip]
\petit{F35 $\heart$ I10 }\\[\ruleTuringskip]
\petit{F35 $\heart$ I11 }\\[\ruleTuringskip]
\petit{F35 $\heart$ I12 }\\[\ruleTuringskip]
\petit{F35 $\heart$ I13 }\\[\ruleTuringskip]
\petit{F35 $\heart$ I14 }\\[\ruleTuringskip]
\petit{F36 $\heart$ F10}\\[\ruleTuringskip]
\petit{F36 $\heart$ F11}\\[\ruleTuringskip]
\petit{F36 $\heart$ F41 }\\[\ruleTuringskip]
\petit{F36 $\heart$ F51 }\\[\ruleTuringskip]
\petit{F36 $\heart$ G00 }\\[\ruleTuringskip]
\petit{F36 $\heart$ G04 }\\[\ruleTuringskip]
\petit{F36 $\heart$ G08 }\\[\ruleTuringskip]
\petit{F36 $\heart$ G10 }\\[\ruleTuringskip]
\petit{F36 $\heart$ G11 }\\[\ruleTuringskip]
\petit{F36 $\heart$ G12 }\\[\ruleTuringskip]
\petit{F36 $\heart$ G16 }\\[\ruleTuringskip]
\petit{F36 $\heart$ G20 }\\[\ruleTuringskip]
\petit{F36 $\heart$ G24 }\\[\ruleTuringskip]
\petit{F36 $\heart$ G28 }\\[\ruleTuringskip]
\petit{F36 $\heart$ G32 }\\[\ruleTuringskip]
\petit{F36 $\heart$ G36 }\\[\ruleTuringskip]
\petit{F36 $\heart$ G40 }\\[\ruleTuringskip]
\petit{F36 $\heart$ G44 }\\[\ruleTuringskip]
\petit{F36 $\heart$ H03 }\\[\ruleTuringskip]
\petit{F36 $\heart$ H04 }\\[\ruleTuringskip]
\petit{F36 $\heart$ H07 }\\[\ruleTuringskip]
\petit{F36 $\heart$ H11 }\\[\ruleTuringskip]
\petit{F36 $\heart$ H15 }\\[\ruleTuringskip]
\petit{F36 $\heart$ H17 }\\[\ruleTuringskip]
\petit{F36 $\heart$ I00 }\\[\ruleTuringskip]
\petit{F36 $\heart$ I01 }\\[\ruleTuringskip]
\petit{F36 $\heart$ I02 }\\[\ruleTuringskip]
\petit{F36 $\heart$ I03 }\\[\ruleTuringskip]
\petit{F36 $\heart$ I04 }\\[\ruleTuringskip]
\petit{F36 $\heart$ I05 }\\[\ruleTuringskip]
\petit{F36 $\heart$ I06 }\\[\ruleTuringskip]
\petit{F36 $\heart$ I07 }\\[\ruleTuringskip]
\petit{F36 $\heart$ I08 }\\[\ruleTuringskip]
\petit{F36 $\heart$ I09 }\\[\ruleTuringskip]
\petit{F36 $\heart$ I10 }\\[\ruleTuringskip]
\petit{F36 $\heart$ I11 }\\[\ruleTuringskip]
\petit{F36 $\heart$ I12 }\\[\ruleTuringskip]
\petit{F36 $\heart$ I13 }\\[\ruleTuringskip]
\petit{F36 $\heart$ I14 }\\[\ruleTuringskip]
\petit{F37 $\heart$ F34}\\[\ruleTuringskip]
\petit{F37 $\heart$ G09 }\\[\ruleTuringskip]
\petit{F37 $\heart$ I00 }\\[\ruleTuringskip]
\petit{F37 $\heart$ I01 }\\[\ruleTuringskip]
\petit{F37 $\heart$ I02 }\\[\ruleTuringskip]
\petit{F37 $\heart$ I03 }\\[\ruleTuringskip]
\petit{F37 $\heart$ I04 }\\[\ruleTuringskip]
\petit{F37 $\heart$ I05 }\\[\ruleTuringskip]
\petit{F37 $\heart$ I06 }\\[\ruleTuringskip]
\petit{F37 $\heart$ I07 }\\[\ruleTuringskip]
\petit{F37 $\heart$ I08 }\\[\ruleTuringskip]
\petit{F37 $\heart$ I09 }\\[\ruleTuringskip]
\petit{F37 $\heart$ I10 }\\[\ruleTuringskip]
\petit{F37 $\heart$ I11 }\\[\ruleTuringskip]
\petit{F37 $\heart$ I12 }\\[\ruleTuringskip]
\petit{F37 $\heart$ I13 }\\[\ruleTuringskip]
\petit{F37 $\heart$ I14 }\\[\ruleTuringskip]
\petit{F38 $\heart$ F33}\\[\ruleTuringskip]
\petit{F38 $\heart$ F34}\\[\ruleTuringskip]
\petit{F38 $\heart$ F45 }\\[\ruleTuringskip]
\petit{F38 $\heart$ G08 }\\[\ruleTuringskip]
\petit{F38 $\heart$ I00 }\\[\ruleTuringskip]
\petit{F38 $\heart$ I01 }\\[\ruleTuringskip]
\petit{F38 $\heart$ I02 }\\[\ruleTuringskip]
\petit{F38 $\heart$ I03 }\\[\ruleTuringskip]
\petit{F38 $\heart$ I04 }\\[\ruleTuringskip]
\petit{F38 $\heart$ I05 }\\[\ruleTuringskip]
\petit{F38 $\heart$ I06 }\\[\ruleTuringskip]
\petit{F38 $\heart$ I07 }\\[\ruleTuringskip]
\petit{F38 $\heart$ I08 }\\[\ruleTuringskip]
\petit{F38 $\heart$ I09 }\\[\ruleTuringskip]
\petit{F38 $\heart$ I10 }\\[\ruleTuringskip]
\petit{F38 $\heart$ I11 }\\[\ruleTuringskip]
\petit{F38 $\heart$ I12 }\\[\ruleTuringskip]
\petit{F38 $\heart$ I13 }\\[\ruleTuringskip]
\petit{F38 $\heart$ I14 }\\[\ruleTuringskip]
\petit{F39 $\heart$ F07}\\[\ruleTuringskip]
\petit{F39 $\heart$ F08}\\[\ruleTuringskip]
\petit{F39 $\heart$ F32}\\[\ruleTuringskip]
\petit{F39 $\heart$ F44 }\\[\ruleTuringskip]
\petit{F39 $\heart$ G07 }\\[\ruleTuringskip]
\petit{F39 $\heart$ I00 }\\[\ruleTuringskip]
\petit{F39 $\heart$ I01 }\\[\ruleTuringskip]
\petit{F39 $\heart$ I02 }\\[\ruleTuringskip]
\petit{F39 $\heart$ I03 }\\[\ruleTuringskip]
\petit{F39 $\heart$ I04 }\\[\ruleTuringskip]
\petit{F39 $\heart$ I05 }\\[\ruleTuringskip]
\petit{F39 $\heart$ I06 }\\[\ruleTuringskip]
\petit{F39 $\heart$ I07 }\\[\ruleTuringskip]
\petit{F39 $\heart$ I08 }\\[\ruleTuringskip]
\petit{F39 $\heart$ I09 }\\[\ruleTuringskip]
\petit{F39 $\heart$ I10 }\\[\ruleTuringskip]
\petit{F39 $\heart$ I11 }\\[\ruleTuringskip]
\petit{F39 $\heart$ I12 }\\[\ruleTuringskip]
\petit{F39 $\heart$ I13 }\\[\ruleTuringskip]
\petit{F39 $\heart$ I14 }\\[\ruleTuringskip]
\petit{F40 $\heart$ F31}\\[\ruleTuringskip]
\petit{F40 $\heart$ G06 }\\[\ruleTuringskip]
\petit{F40 $\heart$ I00 }\\[\ruleTuringskip]
\petit{F40 $\heart$ I01 }\\[\ruleTuringskip]
\petit{F40 $\heart$ I02 }\\[\ruleTuringskip]
\petit{F40 $\heart$ I03 }\\[\ruleTuringskip]
\petit{F40 $\heart$ I04 }\\[\ruleTuringskip]
\petit{F40 $\heart$ I05 }\\[\ruleTuringskip]
\petit{F40 $\heart$ I06 }\\[\ruleTuringskip]
\petit{F40 $\heart$ I07 }\\[\ruleTuringskip]
\petit{F40 $\heart$ I08 }\\[\ruleTuringskip]
\petit{F40 $\heart$ I09 }\\[\ruleTuringskip]
\petit{F40 $\heart$ I10 }\\[\ruleTuringskip]
\petit{F40 $\heart$ I11 }\\[\ruleTuringskip]
\petit{F40 $\heart$ I12 }\\[\ruleTuringskip]
\petit{F40 $\heart$ I13 }\\[\ruleTuringskip]
\petit{F40 $\heart$ I14 }\\[\ruleTuringskip]
\petit{F41 $\heart$ F30}\\[\ruleTuringskip]
\petit{F41 $\heart$ F36}\\[\ruleTuringskip]
\petit{F41 $\heart$ F51 }\\[\ruleTuringskip]
\petit{F41 $\heart$ G02 }\\[\ruleTuringskip]
\petit{F41 $\heart$ G05 }\\[\ruleTuringskip]
\petit{F41 $\heart$ G06 }\\[\ruleTuringskip]
\petit{F41 $\heart$ G10 }\\[\ruleTuringskip]
\petit{F41 $\heart$ G14 }\\[\ruleTuringskip]
\petit{F41 $\heart$ G18 }\\[\ruleTuringskip]
\petit{F41 $\heart$ G22 }\\[\ruleTuringskip]
\petit{F41 $\heart$ G26 }\\[\ruleTuringskip]
\petit{F41 $\heart$ G30 }\\[\ruleTuringskip]
\petit{F41 $\heart$ G34 }\\[\ruleTuringskip]
\petit{F41 $\heart$ G38 }\\[\ruleTuringskip]
\petit{F41 $\heart$ G42 }\\[\ruleTuringskip]
\petit{F41 $\heart$ G46 }\\[\ruleTuringskip]
\petit{F41 $\heart$ H02 }\\[\ruleTuringskip]
\petit{F41 $\heart$ H06 }\\[\ruleTuringskip]
\petit{F41 $\heart$ H10 }\\[\ruleTuringskip]
\petit{F41 $\heart$ H14 }\\[\ruleTuringskip]
\petit{F41 $\heart$ I00 }\\[\ruleTuringskip]
\petit{F41 $\heart$ I01 }\\[\ruleTuringskip]
\petit{F41 $\heart$ I02 }\\[\ruleTuringskip]
\petit{F41 $\heart$ I03 }\\[\ruleTuringskip]
\petit{F41 $\heart$ I04 }\\[\ruleTuringskip]
\petit{F41 $\heart$ I05 }\\[\ruleTuringskip]
\petit{F41 $\heart$ I06 }\\[\ruleTuringskip]
\petit{F41 $\heart$ I07 }\\[\ruleTuringskip]
\petit{F41 $\heart$ I08 }\\[\ruleTuringskip]
\petit{F41 $\heart$ I09 }\\[\ruleTuringskip]
\petit{F41 $\heart$ I10 }\\[\ruleTuringskip]
\petit{F41 $\heart$ I11 }\\[\ruleTuringskip]
\petit{F41 $\heart$ I12 }\\[\ruleTuringskip]
\petit{F41 $\heart$ I13 }\\[\ruleTuringskip]
\petit{F41 $\heart$ I14 }\\[\ruleTuringskip]
\petit{F42 $\heart$ F04}\\[\ruleTuringskip]
\petit{F42 $\heart$ F05}\\[\ruleTuringskip]
\petit{F42 $\heart$ F47 }\\[\ruleTuringskip]
\petit{F42 $\heart$ F51 }\\[\ruleTuringskip]
\petit{F42 $\heart$ G01 }\\[\ruleTuringskip]
\petit{F42 $\heart$ G02 }\\[\ruleTuringskip]
\petit{F42 $\heart$ G04 }\\[\ruleTuringskip]
\petit{F42 $\heart$ G06 }\\[\ruleTuringskip]
\petit{F42 $\heart$ G10 }\\[\ruleTuringskip]
\petit{F42 $\heart$ G13 }\\[\ruleTuringskip]
\petit{F42 $\heart$ G14 }\\[\ruleTuringskip]
\petit{F42 $\heart$ G18 }\\[\ruleTuringskip]
\petit{F42 $\heart$ G22 }\\[\ruleTuringskip]
\petit{F42 $\heart$ G25 }\\[\ruleTuringskip]
\petit{F42 $\heart$ G26 }\\[\ruleTuringskip]
\petit{F42 $\heart$ G30 }\\[\ruleTuringskip]
\petit{F42 $\heart$ G34 }\\[\ruleTuringskip]
\petit{F42 $\heart$ G37 }\\[\ruleTuringskip]
\petit{F42 $\heart$ G38 }\\[\ruleTuringskip]
\petit{F42 $\heart$ G42 }\\[\ruleTuringskip]
\petit{F42 $\heart$ G46 }\\[\ruleTuringskip]
\petit{F42 $\heart$ H01 }\\[\ruleTuringskip]
\petit{F42 $\heart$ H05 }\\[\ruleTuringskip]
\petit{F42 $\heart$ H09 }\\[\ruleTuringskip]
\petit{F42 $\heart$ H13 }\\[\ruleTuringskip]
\petit{F42 $\heart$ H19 }\\[\ruleTuringskip]
\petit{F42 $\heart$ I00 }\\[\ruleTuringskip]
\petit{F42 $\heart$ I01 }\\[\ruleTuringskip]
\petit{F42 $\heart$ I02 }\\[\ruleTuringskip]
\petit{F42 $\heart$ I03 }\\[\ruleTuringskip]
\petit{F42 $\heart$ I04 }\\[\ruleTuringskip]
\petit{F42 $\heart$ I05 }\\[\ruleTuringskip]
\petit{F42 $\heart$ I06 }\\[\ruleTuringskip]
\petit{F42 $\heart$ I07 }\\[\ruleTuringskip]
\petit{F42 $\heart$ I08 }\\[\ruleTuringskip]
\petit{F42 $\heart$ I09 }\\[\ruleTuringskip]
\petit{F42 $\heart$ I10 }\\[\ruleTuringskip]
\petit{F42 $\heart$ I11 }\\[\ruleTuringskip]
\petit{F42 $\heart$ I12 }\\[\ruleTuringskip]
\petit{F42 $\heart$ I13 }\\[\ruleTuringskip]
\petit{F42 $\heart$ I14 }\\[\ruleTuringskip]
\petit{F43 $\heart$ F03}\\[\ruleTuringskip]
\petit{F43 $\heart$ F28}\\[\ruleTuringskip]
\petit{F43 $\heart$ G03 }\\[\ruleTuringskip]
\petit{F43 $\heart$ I00 }\\[\ruleTuringskip]
\petit{F43 $\heart$ I01 }\\[\ruleTuringskip]
\petit{F43 $\heart$ I02 }\\[\ruleTuringskip]
\petit{F43 $\heart$ I03 }\\[\ruleTuringskip]
\petit{F43 $\heart$ I04 }\\[\ruleTuringskip]
\petit{F43 $\heart$ I05 }\\[\ruleTuringskip]
\petit{F43 $\heart$ I06 }\\[\ruleTuringskip]
\petit{F43 $\heart$ I07 }\\[\ruleTuringskip]
\petit{F43 $\heart$ I08 }\\[\ruleTuringskip]
\petit{F43 $\heart$ I09 }\\[\ruleTuringskip]
\petit{F43 $\heart$ I10 }\\[\ruleTuringskip]
\petit{F43 $\heart$ I11 }\\[\ruleTuringskip]
\petit{F43 $\heart$ I12 }\\[\ruleTuringskip]
\petit{F43 $\heart$ I13 }\\[\ruleTuringskip]
\petit{F43 $\heart$ I14 }\\[\ruleTuringskip]
\petit{F44 $\heart$ F03}\\[\ruleTuringskip]
\petit{F44 $\heart$ F27}\\[\ruleTuringskip]
\petit{F44 $\heart$ F39}\\[\ruleTuringskip]
\petit{F44 $\heart$ G02 }\\[\ruleTuringskip]
\petit{F44 $\heart$ I00 }\\[\ruleTuringskip]
\petit{F44 $\heart$ I01 }\\[\ruleTuringskip]
\petit{F44 $\heart$ I02 }\\[\ruleTuringskip]
\petit{F44 $\heart$ I03 }\\[\ruleTuringskip]
\petit{F44 $\heart$ I04 }\\[\ruleTuringskip]
\petit{F44 $\heart$ I05 }\\[\ruleTuringskip]
\petit{F44 $\heart$ I06 }\\[\ruleTuringskip]
\petit{F44 $\heart$ I07 }\\[\ruleTuringskip]
\petit{F44 $\heart$ I08 }\\[\ruleTuringskip]
\petit{F44 $\heart$ I09 }\\[\ruleTuringskip]
\petit{F44 $\heart$ I10 }\\[\ruleTuringskip]
\petit{F44 $\heart$ I11 }\\[\ruleTuringskip]
\petit{F44 $\heart$ I12 }\\[\ruleTuringskip]
\petit{F44 $\heart$ I13 }\\[\ruleTuringskip]
\petit{F44 $\heart$ I14 }\\[\ruleTuringskip]
\petit{F45 $\heart$ F02}\\[\ruleTuringskip]
\petit{F45 $\heart$ F38}\\[\ruleTuringskip]
\petit{F45 $\heart$ G01 }\\[\ruleTuringskip]
\petit{F45 $\heart$ I00 }\\[\ruleTuringskip]
\petit{F45 $\heart$ I01 }\\[\ruleTuringskip]
\petit{F45 $\heart$ I02 }\\[\ruleTuringskip]
\petit{F45 $\heart$ I03 }\\[\ruleTuringskip]
\petit{F45 $\heart$ I04 }\\[\ruleTuringskip]
\petit{F45 $\heart$ I05 }\\[\ruleTuringskip]
\petit{F45 $\heart$ I06 }\\[\ruleTuringskip]
\petit{F45 $\heart$ I07 }\\[\ruleTuringskip]
\petit{F45 $\heart$ I08 }\\[\ruleTuringskip]
\petit{F45 $\heart$ I09 }\\[\ruleTuringskip]
\petit{F45 $\heart$ I10 }\\[\ruleTuringskip]
\petit{F45 $\heart$ I11 }\\[\ruleTuringskip]
\petit{F45 $\heart$ I12 }\\[\ruleTuringskip]
\petit{F45 $\heart$ I13 }\\[\ruleTuringskip]
\petit{F45 $\heart$ I14 }\\[\ruleTuringskip]
\petit{F46 $\heart$ F00}\\[\ruleTuringskip]
\petit{F46 $\heart$ F01}\\[\ruleTuringskip]
\petit{F46 $\heart$ F25}\\[\ruleTuringskip]
\petit{F46 $\heart$ G00 }\\[\ruleTuringskip]
\petit{F46 $\heart$ I00 }\\[\ruleTuringskip]
\petit{F46 $\heart$ I01 }\\[\ruleTuringskip]
\petit{F46 $\heart$ I02 }\\[\ruleTuringskip]
\petit{F46 $\heart$ I03 }\\[\ruleTuringskip]
\petit{F46 $\heart$ I04 }\\[\ruleTuringskip]
\petit{F46 $\heart$ I05 }\\[\ruleTuringskip]
\petit{F46 $\heart$ I06 }\\[\ruleTuringskip]
\petit{F46 $\heart$ I07 }\\[\ruleTuringskip]
\petit{F46 $\heart$ I08 }\\[\ruleTuringskip]
\petit{F46 $\heart$ I09 }\\[\ruleTuringskip]
\petit{F46 $\heart$ I10 }\\[\ruleTuringskip]
\petit{F46 $\heart$ I11 }\\[\ruleTuringskip]
\petit{F46 $\heart$ I12 }\\[\ruleTuringskip]
\petit{F46 $\heart$ I13 }\\[\ruleTuringskip]
\petit{F46 $\heart$ I14 }\\[\ruleTuringskip]
\petit{F47 $\heart$ F24}\\[\ruleTuringskip]
\petit{F47 $\heart$ F42}\\[\ruleTuringskip]
\petit{F47 $\heart$ G00 }\\[\ruleTuringskip]
\petit{F47 $\heart$ G04 }\\[\ruleTuringskip]
\petit{F47 $\heart$ G08 }\\[\ruleTuringskip]
\petit{F47 $\heart$ G12 }\\[\ruleTuringskip]
\petit{F47 $\heart$ G16 }\\[\ruleTuringskip]
\petit{F47 $\heart$ G20 }\\[\ruleTuringskip]
\petit{F47 $\heart$ G24 }\\[\ruleTuringskip]
\petit{F47 $\heart$ G28 }\\[\ruleTuringskip]
\petit{F47 $\heart$ G32 }\\[\ruleTuringskip]
\petit{F47 $\heart$ G36 }\\[\ruleTuringskip]
\petit{F47 $\heart$ G40 }\\[\ruleTuringskip]
\petit{F47 $\heart$ G44 }\\[\ruleTuringskip]
\petit{F47 $\heart$ H00 }\\[\ruleTuringskip]
\petit{F47 $\heart$ H04 }\\[\ruleTuringskip]
\petit{F47 $\heart$ H08 }\\[\ruleTuringskip]
\petit{F47 $\heart$ H12 }\\[\ruleTuringskip]
\petit{F47 $\heart$ H16 }\\[\ruleTuringskip]
\petit{F47 $\heart$ H18 }\\[\ruleTuringskip]
\petit{F47 $\heart$ I00 }\\[\ruleTuringskip]
\petit{F47 $\heart$ I01 }\\[\ruleTuringskip]
\petit{F47 $\heart$ I02 }\\[\ruleTuringskip]
\petit{F47 $\heart$ I03 }\\[\ruleTuringskip]
\petit{F47 $\heart$ I04 }\\[\ruleTuringskip]
\petit{F47 $\heart$ I05 }\\[\ruleTuringskip]
\petit{F47 $\heart$ I06 }\\[\ruleTuringskip]
\petit{F47 $\heart$ I07 }\\[\ruleTuringskip]
\petit{F47 $\heart$ I08 }\\[\ruleTuringskip]
\petit{F47 $\heart$ I09 }\\[\ruleTuringskip]
\petit{F47 $\heart$ I10 }\\[\ruleTuringskip]
\petit{F47 $\heart$ I11 }\\[\ruleTuringskip]
\petit{F47 $\heart$ I12 }\\[\ruleTuringskip]
\petit{F47 $\heart$ I13 }\\[\ruleTuringskip]
\petit{F47 $\heart$ I14 }\\[\ruleTuringskip]
\petit{F48 $\heart$ B04}\\[\ruleTuringskip]
\petit{F49 $\heart$ F22}\\[\ruleTuringskip]
\petit{F49 $\heart$ F23}\\[\ruleTuringskip]
\petit{F50 $\heart$ F06}\\[\ruleTuringskip]
\petit{F51 $\heart$ F00}\\[\ruleTuringskip]
\petit{F51 $\heart$ F06}\\[\ruleTuringskip]
\petit{F51 $\heart$ F07}\\[\ruleTuringskip]
\petit{F51 $\heart$ F12}\\[\ruleTuringskip]
\petit{F51 $\heart$ F17}\\[\ruleTuringskip]
\petit{F51 $\heart$ F18}\\[\ruleTuringskip]
\petit{F51 $\heart$ F24}\\[\ruleTuringskip]
\petit{F51 $\heart$ F30}\\[\ruleTuringskip]
\petit{F51 $\heart$ F31}\\[\ruleTuringskip]
\petit{F51 $\heart$ F36}\\[\ruleTuringskip]
\petit{F51 $\heart$ F41}\\[\ruleTuringskip]
\petit{F51 $\heart$ F42}\\[\ruleTuringskip]
\petit{F51 $\heart$ G07 }\\[\ruleTuringskip]
\petit{F51 $\heart$ G19 }\\[\ruleTuringskip]
\petit{F51 $\heart$ G31 }\\[\ruleTuringskip]
\petit{F51 $\heart$ G43 }\\[\ruleTuringskip]
\petit{G00 $\heart$ F00}\\[\ruleTuringskip]
\petit{G00 $\heart$ F11}\\[\ruleTuringskip]
\petit{G00 $\heart$ F12}\\[\ruleTuringskip]
\petit{G00 $\heart$ F23}\\[\ruleTuringskip]
\petit{G00 $\heart$ F24}\\[\ruleTuringskip]
\petit{G00 $\heart$ F35}\\[\ruleTuringskip]
\petit{G00 $\heart$ F36}\\[\ruleTuringskip]
\petit{G00 $\heart$ F46}\\[\ruleTuringskip]
\petit{G00 $\heart$ F47}\\[\ruleTuringskip]
\petit{G00 $\heart$ G23 }\\[\ruleTuringskip]
\petit{G00 $\heart$ G46 }\\[\ruleTuringskip]
\petit{G00 $\heart$ G48 }\\[\ruleTuringskip]
\petit{G01 $\heart$ F06}\\[\ruleTuringskip]
\petit{G01 $\heart$ F18}\\[\ruleTuringskip]
\petit{G01 $\heart$ F30}\\[\ruleTuringskip]
\petit{G01 $\heart$ F42}\\[\ruleTuringskip]
\petit{G01 $\heart$ F45}\\[\ruleTuringskip]
\petit{G01 $\heart$ G46 }\\[\ruleTuringskip]
\petit{G02 $\heart$ F05}\\[\ruleTuringskip]
\petit{G02 $\heart$ F06}\\[\ruleTuringskip]
\petit{G02 $\heart$ F17}\\[\ruleTuringskip]
\petit{G02 $\heart$ F18}\\[\ruleTuringskip]
\petit{G02 $\heart$ F29}\\[\ruleTuringskip]
\petit{G02 $\heart$ F30}\\[\ruleTuringskip]
\petit{G02 $\heart$ F41}\\[\ruleTuringskip]
\petit{G02 $\heart$ F42}\\[\ruleTuringskip]
\petit{G02 $\heart$ F44}\\[\ruleTuringskip]
\petit{G02 $\heart$ G21 }\\[\ruleTuringskip]
\petit{G02 $\heart$ G44 }\\[\ruleTuringskip]
\petit{G03 $\heart$ F43}\\[\ruleTuringskip]
\petit{G04 $\heart$ F00}\\[\ruleTuringskip]
\petit{G04 $\heart$ F11}\\[\ruleTuringskip]
\petit{G04 $\heart$ F12}\\[\ruleTuringskip]
\petit{G04 $\heart$ F23}\\[\ruleTuringskip]
\petit{G04 $\heart$ F24}\\[\ruleTuringskip]
\petit{G04 $\heart$ F35}\\[\ruleTuringskip]
\petit{G04 $\heart$ F36}\\[\ruleTuringskip]
\petit{G04 $\heart$ F42}\\[\ruleTuringskip]
\petit{G04 $\heart$ F47}\\[\ruleTuringskip]
\petit{G04 $\heart$ G19 }\\[\ruleTuringskip]
\petit{G04 $\heart$ G42 }\\[\ruleTuringskip]
\petit{G05 $\heart$ F41}\\[\ruleTuringskip]
\petit{G06 $\heart$ F05}\\[\ruleTuringskip]
\petit{G06 $\heart$ F06}\\[\ruleTuringskip]
\petit{G06 $\heart$ F17}\\[\ruleTuringskip]
\petit{G06 $\heart$ F18}\\[\ruleTuringskip]
\petit{G06 $\heart$ F29}\\[\ruleTuringskip]
\petit{G06 $\heart$ F30}\\[\ruleTuringskip]
\petit{G06 $\heart$ F40}\\[\ruleTuringskip]
\petit{G06 $\heart$ F41}\\[\ruleTuringskip]
\petit{G06 $\heart$ F42}\\[\ruleTuringskip]
\petit{G06 $\heart$ G17 }\\[\ruleTuringskip]
\petit{G06 $\heart$ G40 }\\[\ruleTuringskip]
\petit{G07 $\heart$ F39}\\[\ruleTuringskip]
\petit{G07 $\heart$ F51}\\[\ruleTuringskip]
\petit{G08 $\heart$ F00}\\[\ruleTuringskip]
\petit{G08 $\heart$ F11}\\[\ruleTuringskip]
\petit{G08 $\heart$ F12}\\[\ruleTuringskip]
\petit{G08 $\heart$ F23}\\[\ruleTuringskip]
\petit{G08 $\heart$ F24}\\[\ruleTuringskip]
\petit{G08 $\heart$ F35}\\[\ruleTuringskip]
\petit{G08 $\heart$ F36}\\[\ruleTuringskip]
\petit{G08 $\heart$ F38}\\[\ruleTuringskip]
\petit{G08 $\heart$ F47}\\[\ruleTuringskip]
\petit{G08 $\heart$ G15 }\\[\ruleTuringskip]
\petit{G08 $\heart$ G38 }\\[\ruleTuringskip]
\petit{G09 $\heart$ F37}\\[\ruleTuringskip]
\petit{G09 $\heart$ G14 }\\[\ruleTuringskip]
\petit{G10 $\heart$ F05}\\[\ruleTuringskip]
\petit{G10 $\heart$ F06}\\[\ruleTuringskip]
\petit{G10 $\heart$ F17}\\[\ruleTuringskip]
\petit{G10 $\heart$ F18}\\[\ruleTuringskip]
\petit{G10 $\heart$ F29}\\[\ruleTuringskip]
\petit{G10 $\heart$ F30}\\[\ruleTuringskip]
\petit{G10 $\heart$ F36}\\[\ruleTuringskip]
\petit{G10 $\heart$ F41}\\[\ruleTuringskip]
\petit{G10 $\heart$ F42}\\[\ruleTuringskip]
\petit{G10 $\heart$ G13 }\\[\ruleTuringskip]
\petit{G10 $\heart$ G36 }\\[\ruleTuringskip]
\petit{G11 $\heart$ F36}\\[\ruleTuringskip]
\petit{G11 $\heart$ G13 }\\[\ruleTuringskip]
\petit{G11 $\heart$ G36 }\\[\ruleTuringskip]
\petit{G12 $\heart$ F00}\\[\ruleTuringskip]
\petit{G12 $\heart$ F11}\\[\ruleTuringskip]
\petit{G12 $\heart$ F12}\\[\ruleTuringskip]
\petit{G12 $\heart$ F23}\\[\ruleTuringskip]
\petit{G12 $\heart$ F24}\\[\ruleTuringskip]
\petit{G12 $\heart$ F35}\\[\ruleTuringskip]
\petit{G12 $\heart$ F36}\\[\ruleTuringskip]
\petit{G12 $\heart$ F47}\\[\ruleTuringskip]
\petit{G12 $\heart$ G34 }\\[\ruleTuringskip]
\petit{G12 $\heart$ G35 }\\[\ruleTuringskip]
\petit{G13 $\heart$ F06}\\[\ruleTuringskip]
\petit{G13 $\heart$ F10}\\[\ruleTuringskip]
\petit{G13 $\heart$ F18}\\[\ruleTuringskip]
\petit{G13 $\heart$ F30}\\[\ruleTuringskip]
\petit{G13 $\heart$ F42}\\[\ruleTuringskip]
\petit{G13 $\heart$ G10}\\[\ruleTuringskip]
\petit{G13 $\heart$ G11}\\[\ruleTuringskip]
\petit{G14 $\heart$ F05}\\[\ruleTuringskip]
\petit{G14 $\heart$ F06}\\[\ruleTuringskip]
\petit{G14 $\heart$ F09}\\[\ruleTuringskip]
\petit{G14 $\heart$ F17}\\[\ruleTuringskip]
\petit{G14 $\heart$ F18}\\[\ruleTuringskip]
\petit{G14 $\heart$ F29}\\[\ruleTuringskip]
\petit{G14 $\heart$ F30}\\[\ruleTuringskip]
\petit{G14 $\heart$ F41}\\[\ruleTuringskip]
\petit{G14 $\heart$ F42}\\[\ruleTuringskip]
\petit{G14 $\heart$ G09}\\[\ruleTuringskip]
\petit{G14 $\heart$ G32 }\\[\ruleTuringskip]
\petit{G15 $\heart$ F08}\\[\ruleTuringskip]
\petit{G15 $\heart$ G08}\\[\ruleTuringskip]
\petit{G16 $\heart$ F00}\\[\ruleTuringskip]
\petit{G16 $\heart$ F07}\\[\ruleTuringskip]
\petit{G16 $\heart$ F11}\\[\ruleTuringskip]
\petit{G16 $\heart$ F12}\\[\ruleTuringskip]
\petit{G16 $\heart$ F23}\\[\ruleTuringskip]
\petit{G16 $\heart$ F24}\\[\ruleTuringskip]
\petit{G16 $\heart$ F35}\\[\ruleTuringskip]
\petit{G16 $\heart$ F36}\\[\ruleTuringskip]
\petit{G16 $\heart$ F47}\\[\ruleTuringskip]
\petit{G16 $\heart$ G30 }\\[\ruleTuringskip]
\petit{G17 $\heart$ F06}\\[\ruleTuringskip]
\petit{G17 $\heart$ G06}\\[\ruleTuringskip]
\petit{G18 $\heart$ F05}\\[\ruleTuringskip]
\petit{G18 $\heart$ F06}\\[\ruleTuringskip]
\petit{G18 $\heart$ F17}\\[\ruleTuringskip]
\petit{G18 $\heart$ F18}\\[\ruleTuringskip]
\petit{G18 $\heart$ F29}\\[\ruleTuringskip]
\petit{G18 $\heart$ F30}\\[\ruleTuringskip]
\petit{G18 $\heart$ F41}\\[\ruleTuringskip]
\petit{G18 $\heart$ F42}\\[\ruleTuringskip]
\petit{G18 $\heart$ G28 }\\[\ruleTuringskip]
\petit{G19 $\heart$ F04}\\[\ruleTuringskip]
\petit{G19 $\heart$ F51}\\[\ruleTuringskip]
\petit{G19 $\heart$ G04}\\[\ruleTuringskip]
\petit{G20 $\heart$ F00}\\[\ruleTuringskip]
\petit{G20 $\heart$ F03}\\[\ruleTuringskip]
\petit{G20 $\heart$ F11}\\[\ruleTuringskip]
\petit{G20 $\heart$ F12}\\[\ruleTuringskip]
\petit{G20 $\heart$ F23}\\[\ruleTuringskip]
\petit{G20 $\heart$ F24}\\[\ruleTuringskip]
\petit{G20 $\heart$ F35}\\[\ruleTuringskip]
\petit{G20 $\heart$ F36}\\[\ruleTuringskip]
\petit{G20 $\heart$ F47}\\[\ruleTuringskip]
\petit{G20 $\heart$ G26 }\\[\ruleTuringskip]
\petit{G21 $\heart$ F02}\\[\ruleTuringskip]
\petit{G21 $\heart$ G02}\\[\ruleTuringskip]
\petit{G21 $\heart$ H02 }\\[\ruleTuringskip]
\petit{G22 $\heart$ F01}\\[\ruleTuringskip]
\petit{G22 $\heart$ F05}\\[\ruleTuringskip]
\petit{G22 $\heart$ F06}\\[\ruleTuringskip]
\petit{G22 $\heart$ F17}\\[\ruleTuringskip]
\petit{G22 $\heart$ F18}\\[\ruleTuringskip]
\petit{G22 $\heart$ F29}\\[\ruleTuringskip]
\petit{G22 $\heart$ F30}\\[\ruleTuringskip]
\petit{G22 $\heart$ F41}\\[\ruleTuringskip]
\petit{G22 $\heart$ F42}\\[\ruleTuringskip]
\petit{G22 $\heart$ G24 }\\[\ruleTuringskip]
\petit{G22 $\heart$ G25 }\\[\ruleTuringskip]
\petit{G23 $\heart$ F00}\\[\ruleTuringskip]
\petit{G23 $\heart$ G00}\\[\ruleTuringskip]
\petit{G24 $\heart$ F00}\\[\ruleTuringskip]
\petit{G24 $\heart$ F11}\\[\ruleTuringskip]
\petit{G24 $\heart$ F12}\\[\ruleTuringskip]
\petit{G24 $\heart$ F22}\\[\ruleTuringskip]
\petit{G24 $\heart$ F23}\\[\ruleTuringskip]
\petit{G24 $\heart$ F24}\\[\ruleTuringskip]
\petit{G24 $\heart$ F35}\\[\ruleTuringskip]
\petit{G24 $\heart$ F36}\\[\ruleTuringskip]
\petit{G24 $\heart$ F47}\\[\ruleTuringskip]
\petit{G24 $\heart$ G22}\\[\ruleTuringskip]
\petit{G24 $\heart$ G47 }\\[\ruleTuringskip]
\petit{G24 $\heart$ G48 }\\[\ruleTuringskip]
\petit{G25 $\heart$ F06}\\[\ruleTuringskip]
\petit{G25 $\heart$ F18}\\[\ruleTuringskip]
\petit{G25 $\heart$ F21}\\[\ruleTuringskip]
\petit{G25 $\heart$ F30}\\[\ruleTuringskip]
\petit{G25 $\heart$ F42}\\[\ruleTuringskip]
\petit{G25 $\heart$ G22}\\[\ruleTuringskip]
\petit{G26 $\heart$ F05}\\[\ruleTuringskip]
\petit{G26 $\heart$ F06}\\[\ruleTuringskip]
\petit{G26 $\heart$ F17}\\[\ruleTuringskip]
\petit{G26 $\heart$ F18}\\[\ruleTuringskip]
\petit{G26 $\heart$ F20}\\[\ruleTuringskip]
\petit{G26 $\heart$ F29}\\[\ruleTuringskip]
\petit{G26 $\heart$ F30}\\[\ruleTuringskip]
\petit{G26 $\heart$ F41}\\[\ruleTuringskip]
\petit{G26 $\heart$ F42}\\[\ruleTuringskip]
\petit{G26 $\heart$ G20}\\[\ruleTuringskip]
\petit{G26 $\heart$ G45 }\\[\ruleTuringskip]
\petit{G27 $\heart$ F19}\\[\ruleTuringskip]
\petit{G28 $\heart$ F00}\\[\ruleTuringskip]
\petit{G28 $\heart$ F11}\\[\ruleTuringskip]
\petit{G28 $\heart$ F12}\\[\ruleTuringskip]
\petit{G28 $\heart$ F18}\\[\ruleTuringskip]
\petit{G28 $\heart$ F23}\\[\ruleTuringskip]
\petit{G28 $\heart$ F24}\\[\ruleTuringskip]
\petit{G28 $\heart$ F35}\\[\ruleTuringskip]
\petit{G28 $\heart$ F36}\\[\ruleTuringskip]
\petit{G28 $\heart$ F47}\\[\ruleTuringskip]
\petit{G28 $\heart$ G18}\\[\ruleTuringskip]
\petit{G28 $\heart$ G43 }\\[\ruleTuringskip]
\petit{G29 $\heart$ F17}\\[\ruleTuringskip]
\petit{G30 $\heart$ F05}\\[\ruleTuringskip]
\petit{G30 $\heart$ F06}\\[\ruleTuringskip]
\petit{G30 $\heart$ F16}\\[\ruleTuringskip]
\petit{G30 $\heart$ F17}\\[\ruleTuringskip]
\petit{G30 $\heart$ F18}\\[\ruleTuringskip]
\petit{G30 $\heart$ F29}\\[\ruleTuringskip]
\petit{G30 $\heart$ F30}\\[\ruleTuringskip]
\petit{G30 $\heart$ F41}\\[\ruleTuringskip]
\petit{G30 $\heart$ F42}\\[\ruleTuringskip]
\petit{G30 $\heart$ G16}\\[\ruleTuringskip]
\petit{G30 $\heart$ G41 }\\[\ruleTuringskip]
\petit{G31 $\heart$ F15}\\[\ruleTuringskip]
\petit{G31 $\heart$ F51}\\[\ruleTuringskip]
\petit{G32 $\heart$ F00}\\[\ruleTuringskip]
\petit{G32 $\heart$ F11}\\[\ruleTuringskip]
\petit{G32 $\heart$ F12}\\[\ruleTuringskip]
\petit{G32 $\heart$ F14}\\[\ruleTuringskip]
\petit{G32 $\heart$ F23}\\[\ruleTuringskip]
\petit{G32 $\heart$ F24}\\[\ruleTuringskip]
\petit{G32 $\heart$ F35}\\[\ruleTuringskip]
\petit{G32 $\heart$ F36}\\[\ruleTuringskip]
\petit{G32 $\heart$ F47}\\[\ruleTuringskip]
\petit{G32 $\heart$ G14}\\[\ruleTuringskip]
\petit{G32 $\heart$ G39 }\\[\ruleTuringskip]
\petit{G33 $\heart$ F13}\\[\ruleTuringskip]
\petit{G33 $\heart$ G38 }\\[\ruleTuringskip]
\petit{G34 $\heart$ F05}\\[\ruleTuringskip]
\petit{G34 $\heart$ F06}\\[\ruleTuringskip]
\petit{G34 $\heart$ F12}\\[\ruleTuringskip]
\petit{G34 $\heart$ F17}\\[\ruleTuringskip]
\petit{G34 $\heart$ F18}\\[\ruleTuringskip]
\petit{G34 $\heart$ F29}\\[\ruleTuringskip]
\petit{G34 $\heart$ F30}\\[\ruleTuringskip]
\petit{G34 $\heart$ F41}\\[\ruleTuringskip]
\petit{G34 $\heart$ F42}\\[\ruleTuringskip]
\petit{G34 $\heart$ G12}\\[\ruleTuringskip]
\petit{G34 $\heart$ G37 }\\[\ruleTuringskip]
\petit{G35 $\heart$ F12}\\[\ruleTuringskip]
\petit{G35 $\heart$ G12}\\[\ruleTuringskip]
\petit{G35 $\heart$ G37 }\\[\ruleTuringskip]
\petit{G36 $\heart$ F00}\\[\ruleTuringskip]
\petit{G36 $\heart$ F11}\\[\ruleTuringskip]
\petit{G36 $\heart$ F12}\\[\ruleTuringskip]
\petit{G36 $\heart$ F23}\\[\ruleTuringskip]
\petit{G36 $\heart$ F24}\\[\ruleTuringskip]
\petit{G36 $\heart$ F35}\\[\ruleTuringskip]
\petit{G36 $\heart$ F36}\\[\ruleTuringskip]
\petit{G36 $\heart$ F47}\\[\ruleTuringskip]
\petit{G36 $\heart$ G10}\\[\ruleTuringskip]
\petit{G36 $\heart$ G11}\\[\ruleTuringskip]
\petit{G37 $\heart$ F06}\\[\ruleTuringskip]
\petit{G37 $\heart$ F18}\\[\ruleTuringskip]
\petit{G37 $\heart$ F30}\\[\ruleTuringskip]
\petit{G37 $\heart$ F34}\\[\ruleTuringskip]
\petit{G37 $\heart$ F42}\\[\ruleTuringskip]
\petit{G37 $\heart$ G34}\\[\ruleTuringskip]
\petit{G37 $\heart$ G35}\\[\ruleTuringskip]
\petit{G38 $\heart$ F05}\\[\ruleTuringskip]
\petit{G38 $\heart$ F06}\\[\ruleTuringskip]
\petit{G38 $\heart$ F17}\\[\ruleTuringskip]
\petit{G38 $\heart$ F18}\\[\ruleTuringskip]
\petit{G38 $\heart$ F29}\\[\ruleTuringskip]
\petit{G38 $\heart$ F30}\\[\ruleTuringskip]
\petit{G38 $\heart$ F33}\\[\ruleTuringskip]
\petit{G38 $\heart$ F41}\\[\ruleTuringskip]
\petit{G38 $\heart$ F42}\\[\ruleTuringskip]
\petit{G38 $\heart$ G08}\\[\ruleTuringskip]
\petit{G38 $\heart$ G33}\\[\ruleTuringskip]
\petit{G39 $\heart$ F32}\\[\ruleTuringskip]
\petit{G39 $\heart$ G32}\\[\ruleTuringskip]
\petit{G40 $\heart$ F00}\\[\ruleTuringskip]
\petit{G40 $\heart$ F11}\\[\ruleTuringskip]
\petit{G40 $\heart$ F12}\\[\ruleTuringskip]
\petit{G40 $\heart$ F23}\\[\ruleTuringskip]
\petit{G40 $\heart$ F24}\\[\ruleTuringskip]
\petit{G40 $\heart$ F31}\\[\ruleTuringskip]
\petit{G40 $\heart$ F35}\\[\ruleTuringskip]
\petit{G40 $\heart$ F36}\\[\ruleTuringskip]
\petit{G40 $\heart$ F47}\\[\ruleTuringskip]
\petit{G40 $\heart$ G06}\\[\ruleTuringskip]
\petit{G41 $\heart$ F30}\\[\ruleTuringskip]
\petit{G41 $\heart$ G30}\\[\ruleTuringskip]
\petit{G42 $\heart$ F05}\\[\ruleTuringskip]
\petit{G42 $\heart$ F06}\\[\ruleTuringskip]
\petit{G42 $\heart$ F17}\\[\ruleTuringskip]
\petit{G42 $\heart$ F18}\\[\ruleTuringskip]
\petit{G42 $\heart$ F29}\\[\ruleTuringskip]
\petit{G42 $\heart$ F30}\\[\ruleTuringskip]
\petit{G42 $\heart$ F41}\\[\ruleTuringskip]
\petit{G42 $\heart$ F42}\\[\ruleTuringskip]
\petit{G42 $\heart$ G04}\\[\ruleTuringskip]
\petit{G43 $\heart$ F28}\\[\ruleTuringskip]
\petit{G43 $\heart$ F51}\\[\ruleTuringskip]
\petit{G43 $\heart$ G28}\\[\ruleTuringskip]
\petit{G44 $\heart$ F00}\\[\ruleTuringskip]
\petit{G44 $\heart$ F11}\\[\ruleTuringskip]
\petit{G44 $\heart$ F12}\\[\ruleTuringskip]
\petit{G44 $\heart$ F23}\\[\ruleTuringskip]
\petit{G44 $\heart$ F24}\\[\ruleTuringskip]
\petit{G44 $\heart$ F27}\\[\ruleTuringskip]
\petit{G44 $\heart$ F35}\\[\ruleTuringskip]
\petit{G44 $\heart$ F36}\\[\ruleTuringskip]
\petit{G44 $\heart$ F47}\\[\ruleTuringskip]
\petit{G44 $\heart$ G02}\\[\ruleTuringskip]
\petit{G45 $\heart$ F26}\\[\ruleTuringskip]
\petit{G45 $\heart$ G26}\\[\ruleTuringskip]
\petit{G45 $\heart$ H02 }\\[\ruleTuringskip]
\petit{G46 $\heart$ F05}\\[\ruleTuringskip]
\petit{G46 $\heart$ F06}\\[\ruleTuringskip]
\petit{G46 $\heart$ F17}\\[\ruleTuringskip]
\petit{G46 $\heart$ F18}\\[\ruleTuringskip]
\petit{G46 $\heart$ F25}\\[\ruleTuringskip]
\petit{G46 $\heart$ F29}\\[\ruleTuringskip]
\petit{G46 $\heart$ F30}\\[\ruleTuringskip]
\petit{G46 $\heart$ F41}\\[\ruleTuringskip]
\petit{G46 $\heart$ F42}\\[\ruleTuringskip]
\petit{G46 $\heart$ G00}\\[\ruleTuringskip]
\petit{G46 $\heart$ G01}\\[\ruleTuringskip]
\petit{G47 $\heart$ F24}\\[\ruleTuringskip]
\petit{G47 $\heart$ G24}\\[\ruleTuringskip]
\petit{G48 $\heart$ G00}\\[\ruleTuringskip]
\petit{G48 $\heart$ G24}\\[\ruleTuringskip]
\petit{H00 $\heart$ F11}\\[\ruleTuringskip]
\petit{H00 $\heart$ F23}\\[\ruleTuringskip]
\petit{H00 $\heart$ F35}\\[\ruleTuringskip]
\petit{H00 $\heart$ F47}\\[\ruleTuringskip]
\petit{H00 $\heart$ H02 }\\[\ruleTuringskip]
\petit{H01 $\heart$ F06}\\[\ruleTuringskip]
\petit{H01 $\heart$ F18}\\[\ruleTuringskip]
\petit{H01 $\heart$ F30}\\[\ruleTuringskip]
\petit{H01 $\heart$ F42}\\[\ruleTuringskip]
\petit{H01 $\heart$ H03 }\\[\ruleTuringskip]
\petit{H01 $\heart$ H04 }\\[\ruleTuringskip]
\petit{H02 $\heart$ F05}\\[\ruleTuringskip]
\petit{H02 $\heart$ F17}\\[\ruleTuringskip]
\petit{H02 $\heart$ F29}\\[\ruleTuringskip]
\petit{H02 $\heart$ F41}\\[\ruleTuringskip]
\petit{H02 $\heart$ G21}\\[\ruleTuringskip]
\petit{H02 $\heart$ G45}\\[\ruleTuringskip]
\petit{H02 $\heart$ H00}\\[\ruleTuringskip]
\petit{H02 $\heart$ H07 }\\[\ruleTuringskip]
\petit{H03 $\heart$ F00}\\[\ruleTuringskip]
\petit{H03 $\heart$ F12}\\[\ruleTuringskip]
\petit{H03 $\heart$ F24}\\[\ruleTuringskip]
\petit{H03 $\heart$ F36}\\[\ruleTuringskip]
\petit{H03 $\heart$ H01}\\[\ruleTuringskip]
\petit{H04 $\heart$ F00}\\[\ruleTuringskip]
\petit{H04 $\heart$ F11}\\[\ruleTuringskip]
\petit{H04 $\heart$ F12}\\[\ruleTuringskip]
\petit{H04 $\heart$ F23}\\[\ruleTuringskip]
\petit{H04 $\heart$ F24}\\[\ruleTuringskip]
\petit{H04 $\heart$ F35}\\[\ruleTuringskip]
\petit{H04 $\heart$ F36}\\[\ruleTuringskip]
\petit{H04 $\heart$ F47}\\[\ruleTuringskip]
\petit{H04 $\heart$ H01}\\[\ruleTuringskip]
\petit{H04 $\heart$ I18 }\\[\ruleTuringskip]
\petit{H05 $\heart$ F06}\\[\ruleTuringskip]
\petit{H05 $\heart$ F18}\\[\ruleTuringskip]
\petit{H05 $\heart$ F30}\\[\ruleTuringskip]
\petit{H05 $\heart$ F42}\\[\ruleTuringskip]
\petit{H05 $\heart$ H10 }\\[\ruleTuringskip]
\petit{H05 $\heart$ I01 }\\[\ruleTuringskip]
\petit{H05 $\heart$ I03 }\\[\ruleTuringskip]
\petit{H05 $\heart$ I05 }\\[\ruleTuringskip]
\petit{H06 $\heart$ F05}\\[\ruleTuringskip]
\petit{H06 $\heart$ F17}\\[\ruleTuringskip]
\petit{H06 $\heart$ F29}\\[\ruleTuringskip]
\petit{H06 $\heart$ F41}\\[\ruleTuringskip]
\petit{H07 $\heart$ F00}\\[\ruleTuringskip]
\petit{H07 $\heart$ F12}\\[\ruleTuringskip]
\petit{H07 $\heart$ F24}\\[\ruleTuringskip]
\petit{H07 $\heart$ F36}\\[\ruleTuringskip]
\petit{H07 $\heart$ H02}\\[\ruleTuringskip]
\petit{H07 $\heart$ H14 }\\[\ruleTuringskip]
\petit{H08 $\heart$ F11}\\[\ruleTuringskip]
\petit{H08 $\heart$ F23}\\[\ruleTuringskip]
\petit{H08 $\heart$ F35}\\[\ruleTuringskip]
\petit{H08 $\heart$ F47}\\[\ruleTuringskip]
\petit{H08 $\heart$ H13 }\\[\ruleTuringskip]
\petit{H08 $\heart$ H19 }\\[\ruleTuringskip]
\petit{H09 $\heart$ F06}\\[\ruleTuringskip]
\petit{H09 $\heart$ F18}\\[\ruleTuringskip]
\petit{H09 $\heart$ F30}\\[\ruleTuringskip]
\petit{H09 $\heart$ F42}\\[\ruleTuringskip]
\petit{H10 $\heart$ F05}\\[\ruleTuringskip]
\petit{H10 $\heart$ F17}\\[\ruleTuringskip]
\petit{H10 $\heart$ F29}\\[\ruleTuringskip]
\petit{H10 $\heart$ F41}\\[\ruleTuringskip]
\petit{H10 $\heart$ H05}\\[\ruleTuringskip]
\petit{H10 $\heart$ I00 }\\[\ruleTuringskip]
\petit{H10 $\heart$ I02 }\\[\ruleTuringskip]
\petit{H10 $\heart$ I04 }\\[\ruleTuringskip]
\petit{H11 $\heart$ F00}\\[\ruleTuringskip]
\petit{H11 $\heart$ F12}\\[\ruleTuringskip]
\petit{H11 $\heart$ F24}\\[\ruleTuringskip]
\petit{H11 $\heart$ F36}\\[\ruleTuringskip]
\petit{H11 $\heart$ H16 }\\[\ruleTuringskip]
\petit{H11 $\heart$ I01 }\\[\ruleTuringskip]
\petit{H11 $\heart$ I03 }\\[\ruleTuringskip]
\petit{H11 $\heart$ I05 }\\[\ruleTuringskip]
\petit{H12 $\heart$ F11}\\[\ruleTuringskip]
\petit{H12 $\heart$ F23}\\[\ruleTuringskip]
\petit{H12 $\heart$ F35}\\[\ruleTuringskip]
\petit{H12 $\heart$ F47}\\[\ruleTuringskip]
\petit{H13 $\heart$ F06}\\[\ruleTuringskip]
\petit{H13 $\heart$ F18}\\[\ruleTuringskip]
\petit{H13 $\heart$ F30}\\[\ruleTuringskip]
\petit{H13 $\heart$ F42}\\[\ruleTuringskip]
\petit{H13 $\heart$ H08}\\[\ruleTuringskip]
\petit{H14 $\heart$ F05}\\[\ruleTuringskip]
\petit{H14 $\heart$ F17}\\[\ruleTuringskip]
\petit{H14 $\heart$ F29}\\[\ruleTuringskip]
\petit{H14 $\heart$ F41}\\[\ruleTuringskip]
\petit{H14 $\heart$ H07}\\[\ruleTuringskip]
\petit{H15 $\heart$ F00}\\[\ruleTuringskip]
\petit{H15 $\heart$ F12}\\[\ruleTuringskip]
\petit{H15 $\heart$ F24}\\[\ruleTuringskip]
\petit{H15 $\heart$ F36}\\[\ruleTuringskip]
\petit{H16 $\heart$ F11}\\[\ruleTuringskip]
\petit{H16 $\heart$ F23}\\[\ruleTuringskip]
\petit{H16 $\heart$ F35}\\[\ruleTuringskip]
\petit{H16 $\heart$ F47}\\[\ruleTuringskip]
\petit{H16 $\heart$ H11}\\[\ruleTuringskip]
\petit{H16 $\heart$ I00 }\\[\ruleTuringskip]
\petit{H16 $\heart$ I02 }\\[\ruleTuringskip]
\petit{H16 $\heart$ I04 }\\[\ruleTuringskip]
\petit{H17 $\heart$ F00}\\[\ruleTuringskip]
\petit{H17 $\heart$ F12}\\[\ruleTuringskip]
\petit{H17 $\heart$ F24}\\[\ruleTuringskip]
\petit{H17 $\heart$ F36}\\[\ruleTuringskip]
\petit{H17 $\heart$ H22 }\\[\ruleTuringskip]
\petit{H17 $\heart$ I03 }\\[\ruleTuringskip]
\petit{H18 $\heart$ F11}\\[\ruleTuringskip]
\petit{H18 $\heart$ F23}\\[\ruleTuringskip]
\petit{H18 $\heart$ F35}\\[\ruleTuringskip]
\petit{H18 $\heart$ F47}\\[\ruleTuringskip]
\petit{H18 $\heart$ H20 }\\[\ruleTuringskip]
\petit{H18 $\heart$ H21 }\\[\ruleTuringskip]
\petit{H19 $\heart$ F06}\\[\ruleTuringskip]
\petit{H19 $\heart$ F18}\\[\ruleTuringskip]
\petit{H19 $\heart$ F30}\\[\ruleTuringskip]
\petit{H19 $\heart$ F42}\\[\ruleTuringskip]
\petit{H19 $\heart$ H08}\\[\ruleTuringskip]
\petit{H20 $\heart$ H18}\\[\ruleTuringskip]
\petit{H21 $\heart$ H18}\\[\ruleTuringskip]
\petit{H21 $\heart$ H23 }\\[\ruleTuringskip]
\petit{H21 $\heart$ H24 }\\[\ruleTuringskip]
\petit{H22 $\heart$ H17}\\[\ruleTuringskip]
\petit{H22 $\heart$ I02 }\\[\ruleTuringskip]
\petit{H23 $\heart$ H21}\\[\ruleTuringskip]
\petit{H24 $\heart$ H21}\\[\ruleTuringskip]
\petit{I00 $\heart$ F00}\\[\ruleTuringskip]
\petit{I00 $\heart$ F01}\\[\ruleTuringskip]
\petit{I00 $\heart$ F02}\\[\ruleTuringskip]
\petit{I00 $\heart$ F03}\\[\ruleTuringskip]
\petit{I00 $\heart$ F04}\\[\ruleTuringskip]
\petit{I00 $\heart$ F05}\\[\ruleTuringskip]
\petit{I00 $\heart$ F06}\\[\ruleTuringskip]
\petit{I00 $\heart$ F07}\\[\ruleTuringskip]
\petit{I00 $\heart$ F08}\\[\ruleTuringskip]
\petit{I00 $\heart$ F09}\\[\ruleTuringskip]
\petit{I00 $\heart$ F10}\\[\ruleTuringskip]
\petit{I00 $\heart$ F11}\\[\ruleTuringskip]
\petit{I00 $\heart$ F12}\\[\ruleTuringskip]
\petit{I00 $\heart$ F13}\\[\ruleTuringskip]
\petit{I00 $\heart$ F14}\\[\ruleTuringskip]
\petit{I00 $\heart$ F15}\\[\ruleTuringskip]
\petit{I00 $\heart$ F16}\\[\ruleTuringskip]
\petit{I00 $\heart$ F17}\\[\ruleTuringskip]
\petit{I00 $\heart$ F18}\\[\ruleTuringskip]
\petit{I00 $\heart$ F19}\\[\ruleTuringskip]
\petit{I00 $\heart$ F20}\\[\ruleTuringskip]
\petit{I00 $\heart$ F21}\\[\ruleTuringskip]
\petit{I00 $\heart$ F22}\\[\ruleTuringskip]
\petit{I00 $\heart$ F23}\\[\ruleTuringskip]
\petit{I00 $\heart$ F24}\\[\ruleTuringskip]
\petit{I00 $\heart$ F25}\\[\ruleTuringskip]
\petit{I00 $\heart$ F26}\\[\ruleTuringskip]
\petit{I00 $\heart$ F27}\\[\ruleTuringskip]
\petit{I00 $\heart$ F28}\\[\ruleTuringskip]
\petit{I00 $\heart$ F29}\\[\ruleTuringskip]
\petit{I00 $\heart$ F30}\\[\ruleTuringskip]
\petit{I00 $\heart$ F31}\\[\ruleTuringskip]
\petit{I00 $\heart$ F32}\\[\ruleTuringskip]
\petit{I00 $\heart$ F33}\\[\ruleTuringskip]
\petit{I00 $\heart$ F34}\\[\ruleTuringskip]
\petit{I00 $\heart$ F35}\\[\ruleTuringskip]
\petit{I00 $\heart$ F36}\\[\ruleTuringskip]
\petit{I00 $\heart$ F37}\\[\ruleTuringskip]
\petit{I00 $\heart$ F38}\\[\ruleTuringskip]
\petit{I00 $\heart$ F39}\\[\ruleTuringskip]
\petit{I00 $\heart$ F40}\\[\ruleTuringskip]
\petit{I00 $\heart$ F41}\\[\ruleTuringskip]
\petit{I00 $\heart$ F42}\\[\ruleTuringskip]
\petit{I00 $\heart$ F43}\\[\ruleTuringskip]
\petit{I00 $\heart$ F44}\\[\ruleTuringskip]
\petit{I00 $\heart$ F45}\\[\ruleTuringskip]
\petit{I00 $\heart$ F46}\\[\ruleTuringskip]
\petit{I00 $\heart$ F47}\\[\ruleTuringskip]
\petit{I00 $\heart$ H10}\\[\ruleTuringskip]
\petit{I00 $\heart$ H16}\\[\ruleTuringskip]
\petit{I00 $\heart$ I07 }\\[\ruleTuringskip]
\petit{I00 $\heart$ I08 }\\[\ruleTuringskip]
\petit{I01 $\heart$ F00}\\[\ruleTuringskip]
\petit{I01 $\heart$ F01}\\[\ruleTuringskip]
\petit{I01 $\heart$ F02}\\[\ruleTuringskip]
\petit{I01 $\heart$ F03}\\[\ruleTuringskip]
\petit{I01 $\heart$ F04}\\[\ruleTuringskip]
\petit{I01 $\heart$ F05}\\[\ruleTuringskip]
\petit{I01 $\heart$ F06}\\[\ruleTuringskip]
\petit{I01 $\heart$ F07}\\[\ruleTuringskip]
\petit{I01 $\heart$ F08}\\[\ruleTuringskip]
\petit{I01 $\heart$ F09}\\[\ruleTuringskip]
\petit{I01 $\heart$ F10}\\[\ruleTuringskip]
\petit{I01 $\heart$ F11}\\[\ruleTuringskip]
\petit{I01 $\heart$ F12}\\[\ruleTuringskip]
\petit{I01 $\heart$ F13}\\[\ruleTuringskip]
\petit{I01 $\heart$ F14}\\[\ruleTuringskip]
\petit{I01 $\heart$ F15}\\[\ruleTuringskip]
\petit{I01 $\heart$ F16}\\[\ruleTuringskip]
\petit{I01 $\heart$ F17}\\[\ruleTuringskip]
\petit{I01 $\heart$ F18}\\[\ruleTuringskip]
\petit{I01 $\heart$ F19}\\[\ruleTuringskip]
\petit{I01 $\heart$ F20}\\[\ruleTuringskip]
\petit{I01 $\heart$ F21}\\[\ruleTuringskip]
\petit{I01 $\heart$ F22}\\[\ruleTuringskip]
\petit{I01 $\heart$ F23}\\[\ruleTuringskip]
\petit{I01 $\heart$ F24}\\[\ruleTuringskip]
\petit{I01 $\heart$ F25}\\[\ruleTuringskip]
\petit{I01 $\heart$ F26}\\[\ruleTuringskip]
\petit{I01 $\heart$ F27}\\[\ruleTuringskip]
\petit{I01 $\heart$ F28}\\[\ruleTuringskip]
\petit{I01 $\heart$ F29}\\[\ruleTuringskip]
\petit{I01 $\heart$ F30}\\[\ruleTuringskip]
\petit{I01 $\heart$ F31}\\[\ruleTuringskip]
\petit{I01 $\heart$ F32}\\[\ruleTuringskip]
\petit{I01 $\heart$ F33}\\[\ruleTuringskip]
\petit{I01 $\heart$ F34}\\[\ruleTuringskip]
\petit{I01 $\heart$ F35}\\[\ruleTuringskip]
\petit{I01 $\heart$ F36}\\[\ruleTuringskip]
\petit{I01 $\heart$ F37}\\[\ruleTuringskip]
\petit{I01 $\heart$ F38}\\[\ruleTuringskip]
\petit{I01 $\heart$ F39}\\[\ruleTuringskip]
\petit{I01 $\heart$ F40}\\[\ruleTuringskip]
\petit{I01 $\heart$ F41}\\[\ruleTuringskip]
\petit{I01 $\heart$ F42}\\[\ruleTuringskip]
\petit{I01 $\heart$ F43}\\[\ruleTuringskip]
\petit{I01 $\heart$ F44}\\[\ruleTuringskip]
\petit{I01 $\heart$ F45}\\[\ruleTuringskip]
\petit{I01 $\heart$ F46}\\[\ruleTuringskip]
\petit{I01 $\heart$ F47}\\[\ruleTuringskip]
\petit{I01 $\heart$ H05}\\[\ruleTuringskip]
\petit{I01 $\heart$ H11}\\[\ruleTuringskip]
\petit{I01 $\heart$ I07 }\\[\ruleTuringskip]
\petit{I01 $\heart$ I13 }\\[\ruleTuringskip]
\petit{I01 $\heart$ I15 }\\[\ruleTuringskip]
\petit{I01 $\heart$ I16 }\\[\ruleTuringskip]
\petit{I01 $\heart$ I19 }\\[\ruleTuringskip]
\petit{I01 $\heart$ J00 }\\[\ruleTuringskip]
\petit{I01 $\heart$ J02 }\\[\ruleTuringskip]
\petit{I02 $\heart$ F00}\\[\ruleTuringskip]
\petit{I02 $\heart$ F01}\\[\ruleTuringskip]
\petit{I02 $\heart$ F02}\\[\ruleTuringskip]
\petit{I02 $\heart$ F03}\\[\ruleTuringskip]
\petit{I02 $\heart$ F04}\\[\ruleTuringskip]
\petit{I02 $\heart$ F05}\\[\ruleTuringskip]
\petit{I02 $\heart$ F06}\\[\ruleTuringskip]
\petit{I02 $\heart$ F07}\\[\ruleTuringskip]
\petit{I02 $\heart$ F08}\\[\ruleTuringskip]
\petit{I02 $\heart$ F09}\\[\ruleTuringskip]
\petit{I02 $\heart$ F10}\\[\ruleTuringskip]
\petit{I02 $\heart$ F11}\\[\ruleTuringskip]
\petit{I02 $\heart$ F12}\\[\ruleTuringskip]
\petit{I02 $\heart$ F13}\\[\ruleTuringskip]
\petit{I02 $\heart$ F14}\\[\ruleTuringskip]
\petit{I02 $\heart$ F15}\\[\ruleTuringskip]
\petit{I02 $\heart$ F16}\\[\ruleTuringskip]
\petit{I02 $\heart$ F17}\\[\ruleTuringskip]
\petit{I02 $\heart$ F18}\\[\ruleTuringskip]
\petit{I02 $\heart$ F19}\\[\ruleTuringskip]
\petit{I02 $\heart$ F20}\\[\ruleTuringskip]
\petit{I02 $\heart$ F21}\\[\ruleTuringskip]
\petit{I02 $\heart$ F22}\\[\ruleTuringskip]
\petit{I02 $\heart$ F23}\\[\ruleTuringskip]
\petit{I02 $\heart$ F24}\\[\ruleTuringskip]
\petit{I02 $\heart$ F25}\\[\ruleTuringskip]
\petit{I02 $\heart$ F26}\\[\ruleTuringskip]
\petit{I02 $\heart$ F27}\\[\ruleTuringskip]
\petit{I02 $\heart$ F28}\\[\ruleTuringskip]
\petit{I02 $\heart$ F29}\\[\ruleTuringskip]
\petit{I02 $\heart$ F30}\\[\ruleTuringskip]
\petit{I02 $\heart$ F31}\\[\ruleTuringskip]
\petit{I02 $\heart$ F32}\\[\ruleTuringskip]
\petit{I02 $\heart$ F33}\\[\ruleTuringskip]
\petit{I02 $\heart$ F34}\\[\ruleTuringskip]
\petit{I02 $\heart$ F35}\\[\ruleTuringskip]
\petit{I02 $\heart$ F36}\\[\ruleTuringskip]
\petit{I02 $\heart$ F37}\\[\ruleTuringskip]
\petit{I02 $\heart$ F38}\\[\ruleTuringskip]
\petit{I02 $\heart$ F39}\\[\ruleTuringskip]
\petit{I02 $\heart$ F40}\\[\ruleTuringskip]
\petit{I02 $\heart$ F41}\\[\ruleTuringskip]
\petit{I02 $\heart$ F42}\\[\ruleTuringskip]
\petit{I02 $\heart$ F43}\\[\ruleTuringskip]
\petit{I02 $\heart$ F44}\\[\ruleTuringskip]
\petit{I02 $\heart$ F45}\\[\ruleTuringskip]
\petit{I02 $\heart$ F46}\\[\ruleTuringskip]
\petit{I02 $\heart$ F47}\\[\ruleTuringskip]
\petit{I02 $\heart$ H10}\\[\ruleTuringskip]
\petit{I02 $\heart$ H16}\\[\ruleTuringskip]
\petit{I02 $\heart$ H22}\\[\ruleTuringskip]
\petit{I02 $\heart$ I12 }\\[\ruleTuringskip]
\petit{I02 $\heart$ I13 }\\[\ruleTuringskip]
\petit{I02 $\heart$ I18 }\\[\ruleTuringskip]
\petit{I03 $\heart$ F00}\\[\ruleTuringskip]
\petit{I03 $\heart$ F01}\\[\ruleTuringskip]
\petit{I03 $\heart$ F02}\\[\ruleTuringskip]
\petit{I03 $\heart$ F03}\\[\ruleTuringskip]
\petit{I03 $\heart$ F04}\\[\ruleTuringskip]
\petit{I03 $\heart$ F05}\\[\ruleTuringskip]
\petit{I03 $\heart$ F06}\\[\ruleTuringskip]
\petit{I03 $\heart$ F07}\\[\ruleTuringskip]
\petit{I03 $\heart$ F08}\\[\ruleTuringskip]
\petit{I03 $\heart$ F09}\\[\ruleTuringskip]
\petit{I03 $\heart$ F10}\\[\ruleTuringskip]
\petit{I03 $\heart$ F11}\\[\ruleTuringskip]
\petit{I03 $\heart$ F12}\\[\ruleTuringskip]
\petit{I03 $\heart$ F13}\\[\ruleTuringskip]
\petit{I03 $\heart$ F14}\\[\ruleTuringskip]
\petit{I03 $\heart$ F15}\\[\ruleTuringskip]
\petit{I03 $\heart$ F16}\\[\ruleTuringskip]
\petit{I03 $\heart$ F17}\\[\ruleTuringskip]
\petit{I03 $\heart$ F18}\\[\ruleTuringskip]
\petit{I03 $\heart$ F19}\\[\ruleTuringskip]
\petit{I03 $\heart$ F20}\\[\ruleTuringskip]
\petit{I03 $\heart$ F21}\\[\ruleTuringskip]
\petit{I03 $\heart$ F22}\\[\ruleTuringskip]
\petit{I03 $\heart$ F23}\\[\ruleTuringskip]
\petit{I03 $\heart$ F24}\\[\ruleTuringskip]
\petit{I03 $\heart$ F25}\\[\ruleTuringskip]
\petit{I03 $\heart$ F26}\\[\ruleTuringskip]
\petit{I03 $\heart$ F27}\\[\ruleTuringskip]
\petit{I03 $\heart$ F28}\\[\ruleTuringskip]
\petit{I03 $\heart$ F29}\\[\ruleTuringskip]
\petit{I03 $\heart$ F30}\\[\ruleTuringskip]
\petit{I03 $\heart$ F31}\\[\ruleTuringskip]
\petit{I03 $\heart$ F32}\\[\ruleTuringskip]
\petit{I03 $\heart$ F33}\\[\ruleTuringskip]
\petit{I03 $\heart$ F34}\\[\ruleTuringskip]
\petit{I03 $\heart$ F35}\\[\ruleTuringskip]
\petit{I03 $\heart$ F36}\\[\ruleTuringskip]
\petit{I03 $\heart$ F37}\\[\ruleTuringskip]
\petit{I03 $\heart$ F38}\\[\ruleTuringskip]
\petit{I03 $\heart$ F39}\\[\ruleTuringskip]
\petit{I03 $\heart$ F40}\\[\ruleTuringskip]
\petit{I03 $\heart$ F41}\\[\ruleTuringskip]
\petit{I03 $\heart$ F42}\\[\ruleTuringskip]
\petit{I03 $\heart$ F43}\\[\ruleTuringskip]
\petit{I03 $\heart$ F44}\\[\ruleTuringskip]
\petit{I03 $\heart$ F45}\\[\ruleTuringskip]
\petit{I03 $\heart$ F46}\\[\ruleTuringskip]
\petit{I03 $\heart$ F47}\\[\ruleTuringskip]
\petit{I03 $\heart$ H05}\\[\ruleTuringskip]
\petit{I03 $\heart$ H11}\\[\ruleTuringskip]
\petit{I03 $\heart$ H17}\\[\ruleTuringskip]
\petit{I03 $\heart$ I07 }\\[\ruleTuringskip]
\petit{I03 $\heart$ I08 }\\[\ruleTuringskip]
\petit{I04 $\heart$ F00}\\[\ruleTuringskip]
\petit{I04 $\heart$ F01}\\[\ruleTuringskip]
\petit{I04 $\heart$ F02}\\[\ruleTuringskip]
\petit{I04 $\heart$ F03}\\[\ruleTuringskip]
\petit{I04 $\heart$ F04}\\[\ruleTuringskip]
\petit{I04 $\heart$ F05}\\[\ruleTuringskip]
\petit{I04 $\heart$ F06}\\[\ruleTuringskip]
\petit{I04 $\heart$ F07}\\[\ruleTuringskip]
\petit{I04 $\heart$ F08}\\[\ruleTuringskip]
\petit{I04 $\heart$ F09}\\[\ruleTuringskip]
\petit{I04 $\heart$ F10}\\[\ruleTuringskip]
\petit{I04 $\heart$ F11}\\[\ruleTuringskip]
\petit{I04 $\heart$ F12}\\[\ruleTuringskip]
\petit{I04 $\heart$ F13}\\[\ruleTuringskip]
\petit{I04 $\heart$ F14}\\[\ruleTuringskip]
\petit{I04 $\heart$ F15}\\[\ruleTuringskip]
\petit{I04 $\heart$ F16}\\[\ruleTuringskip]
\petit{I04 $\heart$ F17}\\[\ruleTuringskip]
\petit{I04 $\heart$ F18}\\[\ruleTuringskip]
\petit{I04 $\heart$ F19}\\[\ruleTuringskip]
\petit{I04 $\heart$ F20}\\[\ruleTuringskip]
\petit{I04 $\heart$ F21}\\[\ruleTuringskip]
\petit{I04 $\heart$ F22}\\[\ruleTuringskip]
\petit{I04 $\heart$ F23}\\[\ruleTuringskip]
\petit{I04 $\heart$ F24}\\[\ruleTuringskip]
\petit{I04 $\heart$ F25}\\[\ruleTuringskip]
\petit{I04 $\heart$ F26}\\[\ruleTuringskip]
\petit{I04 $\heart$ F27}\\[\ruleTuringskip]
\petit{I04 $\heart$ F28}\\[\ruleTuringskip]
\petit{I04 $\heart$ F29}\\[\ruleTuringskip]
\petit{I04 $\heart$ F30}\\[\ruleTuringskip]
\petit{I04 $\heart$ F31}\\[\ruleTuringskip]
\petit{I04 $\heart$ F32}\\[\ruleTuringskip]
\petit{I04 $\heart$ F33}\\[\ruleTuringskip]
\petit{I04 $\heart$ F34}\\[\ruleTuringskip]
\petit{I04 $\heart$ F35}\\[\ruleTuringskip]
\petit{I04 $\heart$ F36}\\[\ruleTuringskip]
\petit{I04 $\heart$ F37}\\[\ruleTuringskip]
\petit{I04 $\heart$ F38}\\[\ruleTuringskip]
\petit{I04 $\heart$ F39}\\[\ruleTuringskip]
\petit{I04 $\heart$ F40}\\[\ruleTuringskip]
\petit{I04 $\heart$ F41}\\[\ruleTuringskip]
\petit{I04 $\heart$ F42}\\[\ruleTuringskip]
\petit{I04 $\heart$ F43}\\[\ruleTuringskip]
\petit{I04 $\heart$ F44}\\[\ruleTuringskip]
\petit{I04 $\heart$ F45}\\[\ruleTuringskip]
\petit{I04 $\heart$ F46}\\[\ruleTuringskip]
\petit{I04 $\heart$ F47}\\[\ruleTuringskip]
\petit{I04 $\heart$ H10}\\[\ruleTuringskip]
\petit{I04 $\heart$ H16}\\[\ruleTuringskip]
\petit{I05 $\heart$ F00}\\[\ruleTuringskip]
\petit{I05 $\heart$ F01}\\[\ruleTuringskip]
\petit{I05 $\heart$ F02}\\[\ruleTuringskip]
\petit{I05 $\heart$ F03}\\[\ruleTuringskip]
\petit{I05 $\heart$ F04}\\[\ruleTuringskip]
\petit{I05 $\heart$ F05}\\[\ruleTuringskip]
\petit{I05 $\heart$ F06}\\[\ruleTuringskip]
\petit{I05 $\heart$ F07}\\[\ruleTuringskip]
\petit{I05 $\heart$ F08}\\[\ruleTuringskip]
\petit{I05 $\heart$ F09}\\[\ruleTuringskip]
\petit{I05 $\heart$ F10}\\[\ruleTuringskip]
\petit{I05 $\heart$ F11}\\[\ruleTuringskip]
\petit{I05 $\heart$ F12}\\[\ruleTuringskip]
\petit{I05 $\heart$ F13}\\[\ruleTuringskip]
\petit{I05 $\heart$ F14}\\[\ruleTuringskip]
\petit{I05 $\heart$ F15}\\[\ruleTuringskip]
\petit{I05 $\heart$ F16}\\[\ruleTuringskip]
\petit{I05 $\heart$ F17}\\[\ruleTuringskip]
\petit{I05 $\heart$ F18}\\[\ruleTuringskip]
\petit{I05 $\heart$ F19}\\[\ruleTuringskip]
\petit{I05 $\heart$ F20}\\[\ruleTuringskip]
\petit{I05 $\heart$ F21}\\[\ruleTuringskip]
\petit{I05 $\heart$ F22}\\[\ruleTuringskip]
\petit{I05 $\heart$ F23}\\[\ruleTuringskip]
\petit{I05 $\heart$ F24}\\[\ruleTuringskip]
\petit{I05 $\heart$ F25}\\[\ruleTuringskip]
\petit{I05 $\heart$ F26}\\[\ruleTuringskip]
\petit{I05 $\heart$ F27}\\[\ruleTuringskip]
\petit{I05 $\heart$ F28}\\[\ruleTuringskip]
\petit{I05 $\heart$ F29}\\[\ruleTuringskip]
\petit{I05 $\heart$ F30}\\[\ruleTuringskip]
\petit{I05 $\heart$ F31}\\[\ruleTuringskip]
\petit{I05 $\heart$ F32}\\[\ruleTuringskip]
\petit{I05 $\heart$ F33}\\[\ruleTuringskip]
\petit{I05 $\heart$ F34}\\[\ruleTuringskip]
\petit{I05 $\heart$ F35}\\[\ruleTuringskip]
\petit{I05 $\heart$ F36}\\[\ruleTuringskip]
\petit{I05 $\heart$ F37}\\[\ruleTuringskip]
\petit{I05 $\heart$ F38}\\[\ruleTuringskip]
\petit{I05 $\heart$ F39}\\[\ruleTuringskip]
\petit{I05 $\heart$ F40}\\[\ruleTuringskip]
\petit{I05 $\heart$ F41}\\[\ruleTuringskip]
\petit{I05 $\heart$ F42}\\[\ruleTuringskip]
\petit{I05 $\heart$ F43}\\[\ruleTuringskip]
\petit{I05 $\heart$ F44}\\[\ruleTuringskip]
\petit{I05 $\heart$ F45}\\[\ruleTuringskip]
\petit{I05 $\heart$ F46}\\[\ruleTuringskip]
\petit{I05 $\heart$ F47}\\[\ruleTuringskip]
\petit{I05 $\heart$ H05}\\[\ruleTuringskip]
\petit{I05 $\heart$ H11}\\[\ruleTuringskip]
\petit{I06 $\heart$ F00}\\[\ruleTuringskip]
\petit{I06 $\heart$ F01}\\[\ruleTuringskip]
\petit{I06 $\heart$ F02}\\[\ruleTuringskip]
\petit{I06 $\heart$ F03}\\[\ruleTuringskip]
\petit{I06 $\heart$ F04}\\[\ruleTuringskip]
\petit{I06 $\heart$ F05}\\[\ruleTuringskip]
\petit{I06 $\heart$ F06}\\[\ruleTuringskip]
\petit{I06 $\heart$ F07}\\[\ruleTuringskip]
\petit{I06 $\heart$ F08}\\[\ruleTuringskip]
\petit{I06 $\heart$ F09}\\[\ruleTuringskip]
\petit{I06 $\heart$ F10}\\[\ruleTuringskip]
\petit{I06 $\heart$ F11}\\[\ruleTuringskip]
\petit{I06 $\heart$ F12}\\[\ruleTuringskip]
\petit{I06 $\heart$ F13}\\[\ruleTuringskip]
\petit{I06 $\heart$ F14}\\[\ruleTuringskip]
\petit{I06 $\heart$ F15}\\[\ruleTuringskip]
\petit{I06 $\heart$ F16}\\[\ruleTuringskip]
\petit{I06 $\heart$ F17}\\[\ruleTuringskip]
\petit{I06 $\heart$ F18}\\[\ruleTuringskip]
\petit{I06 $\heart$ F19}\\[\ruleTuringskip]
\petit{I06 $\heart$ F20}\\[\ruleTuringskip]
\petit{I06 $\heart$ F21}\\[\ruleTuringskip]
\petit{I06 $\heart$ F22}\\[\ruleTuringskip]
\petit{I06 $\heart$ F23}\\[\ruleTuringskip]
\petit{I06 $\heart$ F24}\\[\ruleTuringskip]
\petit{I06 $\heart$ F25}\\[\ruleTuringskip]
\petit{I06 $\heart$ F26}\\[\ruleTuringskip]
\petit{I06 $\heart$ F27}\\[\ruleTuringskip]
\petit{I06 $\heart$ F28}\\[\ruleTuringskip]
\petit{I06 $\heart$ F29}\\[\ruleTuringskip]
\petit{I06 $\heart$ F30}\\[\ruleTuringskip]
\petit{I06 $\heart$ F31}\\[\ruleTuringskip]
\petit{I06 $\heart$ F32}\\[\ruleTuringskip]
\petit{I06 $\heart$ F33}\\[\ruleTuringskip]
\petit{I06 $\heart$ F34}\\[\ruleTuringskip]
\petit{I06 $\heart$ F35}\\[\ruleTuringskip]
\petit{I06 $\heart$ F36}\\[\ruleTuringskip]
\petit{I06 $\heart$ F37}\\[\ruleTuringskip]
\petit{I06 $\heart$ F38}\\[\ruleTuringskip]
\petit{I06 $\heart$ F39}\\[\ruleTuringskip]
\petit{I06 $\heart$ F40}\\[\ruleTuringskip]
\petit{I06 $\heart$ F41}\\[\ruleTuringskip]
\petit{I06 $\heart$ F42}\\[\ruleTuringskip]
\petit{I06 $\heart$ F43}\\[\ruleTuringskip]
\petit{I06 $\heart$ F44}\\[\ruleTuringskip]
\petit{I06 $\heart$ F45}\\[\ruleTuringskip]
\petit{I06 $\heart$ F46}\\[\ruleTuringskip]
\petit{I06 $\heart$ F47}\\[\ruleTuringskip]
\petit{I06 $\heart$ I11 }\\[\ruleTuringskip]
\petit{I06 $\heart$ I15 }\\[\ruleTuringskip]
\petit{I07 $\heart$ F00}\\[\ruleTuringskip]
\petit{I07 $\heart$ F01}\\[\ruleTuringskip]
\petit{I07 $\heart$ F02}\\[\ruleTuringskip]
\petit{I07 $\heart$ F03}\\[\ruleTuringskip]
\petit{I07 $\heart$ F04}\\[\ruleTuringskip]
\petit{I07 $\heart$ F05}\\[\ruleTuringskip]
\petit{I07 $\heart$ F06}\\[\ruleTuringskip]
\petit{I07 $\heart$ F07}\\[\ruleTuringskip]
\petit{I07 $\heart$ F08}\\[\ruleTuringskip]
\petit{I07 $\heart$ F09}\\[\ruleTuringskip]
\petit{I07 $\heart$ F10}\\[\ruleTuringskip]
\petit{I07 $\heart$ F11}\\[\ruleTuringskip]
\petit{I07 $\heart$ F12}\\[\ruleTuringskip]
\petit{I07 $\heart$ F13}\\[\ruleTuringskip]
\petit{I07 $\heart$ F14}\\[\ruleTuringskip]
\petit{I07 $\heart$ F15}\\[\ruleTuringskip]
\petit{I07 $\heart$ F16}\\[\ruleTuringskip]
\petit{I07 $\heart$ F17}\\[\ruleTuringskip]
\petit{I07 $\heart$ F18}\\[\ruleTuringskip]
\petit{I07 $\heart$ F19}\\[\ruleTuringskip]
\petit{I07 $\heart$ F20}\\[\ruleTuringskip]
\petit{I07 $\heart$ F21}\\[\ruleTuringskip]
\petit{I07 $\heart$ F22}\\[\ruleTuringskip]
\petit{I07 $\heart$ F23}\\[\ruleTuringskip]
\petit{I07 $\heart$ F24}\\[\ruleTuringskip]
\petit{I07 $\heart$ F25}\\[\ruleTuringskip]
\petit{I07 $\heart$ F26}\\[\ruleTuringskip]
\petit{I07 $\heart$ F27}\\[\ruleTuringskip]
\petit{I07 $\heart$ F28}\\[\ruleTuringskip]
\petit{I07 $\heart$ F29}\\[\ruleTuringskip]
\petit{I07 $\heart$ F30}\\[\ruleTuringskip]
\petit{I07 $\heart$ F31}\\[\ruleTuringskip]
\petit{I07 $\heart$ F32}\\[\ruleTuringskip]
\petit{I07 $\heart$ F33}\\[\ruleTuringskip]
\petit{I07 $\heart$ F34}\\[\ruleTuringskip]
\petit{I07 $\heart$ F35}\\[\ruleTuringskip]
\petit{I07 $\heart$ F36}\\[\ruleTuringskip]
\petit{I07 $\heart$ F37}\\[\ruleTuringskip]
\petit{I07 $\heart$ F38}\\[\ruleTuringskip]
\petit{I07 $\heart$ F39}\\[\ruleTuringskip]
\petit{I07 $\heart$ F40}\\[\ruleTuringskip]
\petit{I07 $\heart$ F41}\\[\ruleTuringskip]
\petit{I07 $\heart$ F42}\\[\ruleTuringskip]
\petit{I07 $\heart$ F43}\\[\ruleTuringskip]
\petit{I07 $\heart$ F44}\\[\ruleTuringskip]
\petit{I07 $\heart$ F45}\\[\ruleTuringskip]
\petit{I07 $\heart$ F46}\\[\ruleTuringskip]
\petit{I07 $\heart$ F47}\\[\ruleTuringskip]
\petit{I07 $\heart$ I00}\\[\ruleTuringskip]
\petit{I07 $\heart$ I01}\\[\ruleTuringskip]
\petit{I07 $\heart$ I03}\\[\ruleTuringskip]
\petit{I07 $\heart$ I10 }\\[\ruleTuringskip]
\petit{I07 $\heart$ I11 }\\[\ruleTuringskip]
\petit{I07 $\heart$ I16 }\\[\ruleTuringskip]
\petit{I07 $\heart$ I19 }\\[\ruleTuringskip]
\petit{I08 $\heart$ F00}\\[\ruleTuringskip]
\petit{I08 $\heart$ F01}\\[\ruleTuringskip]
\petit{I08 $\heart$ F02}\\[\ruleTuringskip]
\petit{I08 $\heart$ F03}\\[\ruleTuringskip]
\petit{I08 $\heart$ F04}\\[\ruleTuringskip]
\petit{I08 $\heart$ F05}\\[\ruleTuringskip]
\petit{I08 $\heart$ F06}\\[\ruleTuringskip]
\petit{I08 $\heart$ F07}\\[\ruleTuringskip]
\petit{I08 $\heart$ F08}\\[\ruleTuringskip]
\petit{I08 $\heart$ F09}\\[\ruleTuringskip]
\petit{I08 $\heart$ F10}\\[\ruleTuringskip]
\petit{I08 $\heart$ F11}\\[\ruleTuringskip]
\petit{I08 $\heart$ F12}\\[\ruleTuringskip]
\petit{I08 $\heart$ F13}\\[\ruleTuringskip]
\petit{I08 $\heart$ F14}\\[\ruleTuringskip]
\petit{I08 $\heart$ F15}\\[\ruleTuringskip]
\petit{I08 $\heart$ F16}\\[\ruleTuringskip]
\petit{I08 $\heart$ F17}\\[\ruleTuringskip]
\petit{I08 $\heart$ F18}\\[\ruleTuringskip]
\petit{I08 $\heart$ F19}\\[\ruleTuringskip]
\petit{I08 $\heart$ F20}\\[\ruleTuringskip]
\petit{I08 $\heart$ F21}\\[\ruleTuringskip]
\petit{I08 $\heart$ F22}\\[\ruleTuringskip]
\petit{I08 $\heart$ F23}\\[\ruleTuringskip]
\petit{I08 $\heart$ F24}\\[\ruleTuringskip]
\petit{I08 $\heart$ F25}\\[\ruleTuringskip]
\petit{I08 $\heart$ F26}\\[\ruleTuringskip]
\petit{I08 $\heart$ F27}\\[\ruleTuringskip]
\petit{I08 $\heart$ F28}\\[\ruleTuringskip]
\petit{I08 $\heart$ F29}\\[\ruleTuringskip]
\petit{I08 $\heart$ F30}\\[\ruleTuringskip]
\petit{I08 $\heart$ F31}\\[\ruleTuringskip]
\petit{I08 $\heart$ F32}\\[\ruleTuringskip]
\petit{I08 $\heart$ F33}\\[\ruleTuringskip]
\petit{I08 $\heart$ F34}\\[\ruleTuringskip]
\petit{I08 $\heart$ F35}\\[\ruleTuringskip]
\petit{I08 $\heart$ F36}\\[\ruleTuringskip]
\petit{I08 $\heart$ F37}\\[\ruleTuringskip]
\petit{I08 $\heart$ F38}\\[\ruleTuringskip]
\petit{I08 $\heart$ F39}\\[\ruleTuringskip]
\petit{I08 $\heart$ F40}\\[\ruleTuringskip]
\petit{I08 $\heart$ F41}\\[\ruleTuringskip]
\petit{I08 $\heart$ F42}\\[\ruleTuringskip]
\petit{I08 $\heart$ F43}\\[\ruleTuringskip]
\petit{I08 $\heart$ F44}\\[\ruleTuringskip]
\petit{I08 $\heart$ F45}\\[\ruleTuringskip]
\petit{I08 $\heart$ F46}\\[\ruleTuringskip]
\petit{I08 $\heart$ F47}\\[\ruleTuringskip]
\petit{I08 $\heart$ I00}\\[\ruleTuringskip]
\petit{I08 $\heart$ I03}\\[\ruleTuringskip]
\petit{I09 $\heart$ F00}\\[\ruleTuringskip]
\petit{I09 $\heart$ F01}\\[\ruleTuringskip]
\petit{I09 $\heart$ F02}\\[\ruleTuringskip]
\petit{I09 $\heart$ F03}\\[\ruleTuringskip]
\petit{I09 $\heart$ F04}\\[\ruleTuringskip]
\petit{I09 $\heart$ F05}\\[\ruleTuringskip]
\petit{I09 $\heart$ F06}\\[\ruleTuringskip]
\petit{I09 $\heart$ F07}\\[\ruleTuringskip]
\petit{I09 $\heart$ F08}\\[\ruleTuringskip]
\petit{I09 $\heart$ F09}\\[\ruleTuringskip]
\petit{I09 $\heart$ F10}\\[\ruleTuringskip]
\petit{I09 $\heart$ F11}\\[\ruleTuringskip]
\petit{I09 $\heart$ F12}\\[\ruleTuringskip]
\petit{I09 $\heart$ F13}\\[\ruleTuringskip]
\petit{I09 $\heart$ F14}\\[\ruleTuringskip]
\petit{I09 $\heart$ F15}\\[\ruleTuringskip]
\petit{I09 $\heart$ F16}\\[\ruleTuringskip]
\petit{I09 $\heart$ F17}\\[\ruleTuringskip]
\petit{I09 $\heart$ F18}\\[\ruleTuringskip]
\petit{I09 $\heart$ F19}\\[\ruleTuringskip]
\petit{I09 $\heart$ F20}\\[\ruleTuringskip]
\petit{I09 $\heart$ F21}\\[\ruleTuringskip]
\petit{I09 $\heart$ F22}\\[\ruleTuringskip]
\petit{I09 $\heart$ F23}\\[\ruleTuringskip]
\petit{I09 $\heart$ F24}\\[\ruleTuringskip]
\petit{I09 $\heart$ F25}\\[\ruleTuringskip]
\petit{I09 $\heart$ F26}\\[\ruleTuringskip]
\petit{I09 $\heart$ F27}\\[\ruleTuringskip]
\petit{I09 $\heart$ F28}\\[\ruleTuringskip]
\petit{I09 $\heart$ F29}\\[\ruleTuringskip]
\petit{I09 $\heart$ F30}\\[\ruleTuringskip]
\petit{I09 $\heart$ F31}\\[\ruleTuringskip]
\petit{I09 $\heart$ F32}\\[\ruleTuringskip]
\petit{I09 $\heart$ F33}\\[\ruleTuringskip]
\petit{I09 $\heart$ F34}\\[\ruleTuringskip]
\petit{I09 $\heart$ F35}\\[\ruleTuringskip]
\petit{I09 $\heart$ F36}\\[\ruleTuringskip]
\petit{I09 $\heart$ F37}\\[\ruleTuringskip]
\petit{I09 $\heart$ F38}\\[\ruleTuringskip]
\petit{I09 $\heart$ F39}\\[\ruleTuringskip]
\petit{I09 $\heart$ F40}\\[\ruleTuringskip]
\petit{I09 $\heart$ F41}\\[\ruleTuringskip]
\petit{I09 $\heart$ F42}\\[\ruleTuringskip]
\petit{I09 $\heart$ F43}\\[\ruleTuringskip]
\petit{I09 $\heart$ F44}\\[\ruleTuringskip]
\petit{I09 $\heart$ F45}\\[\ruleTuringskip]
\petit{I09 $\heart$ F46}\\[\ruleTuringskip]
\petit{I09 $\heart$ F47}\\[\ruleTuringskip]
\petit{I09 $\heart$ I13 }\\[\ruleTuringskip]
\petit{I09 $\heart$ I14 }\\[\ruleTuringskip]
\petit{I10 $\heart$ F00}\\[\ruleTuringskip]
\petit{I10 $\heart$ F01}\\[\ruleTuringskip]
\petit{I10 $\heart$ F02}\\[\ruleTuringskip]
\petit{I10 $\heart$ F03}\\[\ruleTuringskip]
\petit{I10 $\heart$ F04}\\[\ruleTuringskip]
\petit{I10 $\heart$ F05}\\[\ruleTuringskip]
\petit{I10 $\heart$ F06}\\[\ruleTuringskip]
\petit{I10 $\heart$ F07}\\[\ruleTuringskip]
\petit{I10 $\heart$ F08}\\[\ruleTuringskip]
\petit{I10 $\heart$ F09}\\[\ruleTuringskip]
\petit{I10 $\heart$ F10}\\[\ruleTuringskip]
\petit{I10 $\heart$ F11}\\[\ruleTuringskip]
\petit{I10 $\heart$ F12}\\[\ruleTuringskip]
\petit{I10 $\heart$ F13}\\[\ruleTuringskip]
\petit{I10 $\heart$ F14}\\[\ruleTuringskip]
\petit{I10 $\heart$ F15}\\[\ruleTuringskip]
\petit{I10 $\heart$ F16}\\[\ruleTuringskip]
\petit{I10 $\heart$ F17}\\[\ruleTuringskip]
\petit{I10 $\heart$ F18}\\[\ruleTuringskip]
\petit{I10 $\heart$ F19}\\[\ruleTuringskip]
\petit{I10 $\heart$ F20}\\[\ruleTuringskip]
\petit{I10 $\heart$ F21}\\[\ruleTuringskip]
\petit{I10 $\heart$ F22}\\[\ruleTuringskip]
\petit{I10 $\heart$ F23}\\[\ruleTuringskip]
\petit{I10 $\heart$ F24}\\[\ruleTuringskip]
\petit{I10 $\heart$ F25}\\[\ruleTuringskip]
\petit{I10 $\heart$ F26}\\[\ruleTuringskip]
\petit{I10 $\heart$ F27}\\[\ruleTuringskip]
\petit{I10 $\heart$ F28}\\[\ruleTuringskip]
\petit{I10 $\heart$ F29}\\[\ruleTuringskip]
\petit{I10 $\heart$ F30}\\[\ruleTuringskip]
\petit{I10 $\heart$ F31}\\[\ruleTuringskip]
\petit{I10 $\heart$ F32}\\[\ruleTuringskip]
\petit{I10 $\heart$ F33}\\[\ruleTuringskip]
\petit{I10 $\heart$ F34}\\[\ruleTuringskip]
\petit{I10 $\heart$ F35}\\[\ruleTuringskip]
\petit{I10 $\heart$ F36}\\[\ruleTuringskip]
\petit{I10 $\heart$ F37}\\[\ruleTuringskip]
\petit{I10 $\heart$ F38}\\[\ruleTuringskip]
\petit{I10 $\heart$ F39}\\[\ruleTuringskip]
\petit{I10 $\heart$ F40}\\[\ruleTuringskip]
\petit{I10 $\heart$ F41}\\[\ruleTuringskip]
\petit{I10 $\heart$ F42}\\[\ruleTuringskip]
\petit{I10 $\heart$ F43}\\[\ruleTuringskip]
\petit{I10 $\heart$ F44}\\[\ruleTuringskip]
\petit{I10 $\heart$ F45}\\[\ruleTuringskip]
\petit{I10 $\heart$ F46}\\[\ruleTuringskip]
\petit{I10 $\heart$ F47}\\[\ruleTuringskip]
\petit{I10 $\heart$ I07}\\[\ruleTuringskip]
\petit{I10 $\heart$ I13 }\\[\ruleTuringskip]
\petit{I10 $\heart$ I16 }\\[\ruleTuringskip]
\petit{I10 $\heart$ I18 }\\[\ruleTuringskip]
\petit{I11 $\heart$ F00}\\[\ruleTuringskip]
\petit{I11 $\heart$ F01}\\[\ruleTuringskip]
\petit{I11 $\heart$ F02}\\[\ruleTuringskip]
\petit{I11 $\heart$ F03}\\[\ruleTuringskip]
\petit{I11 $\heart$ F04}\\[\ruleTuringskip]
\petit{I11 $\heart$ F05}\\[\ruleTuringskip]
\petit{I11 $\heart$ F06}\\[\ruleTuringskip]
\petit{I11 $\heart$ F07}\\[\ruleTuringskip]
\petit{I11 $\heart$ F08}\\[\ruleTuringskip]
\petit{I11 $\heart$ F09}\\[\ruleTuringskip]
\petit{I11 $\heart$ F10}\\[\ruleTuringskip]
\petit{I11 $\heart$ F11}\\[\ruleTuringskip]
\petit{I11 $\heart$ F12}\\[\ruleTuringskip]
\petit{I11 $\heart$ F13}\\[\ruleTuringskip]
\petit{I11 $\heart$ F14}\\[\ruleTuringskip]
\petit{I11 $\heart$ F15}\\[\ruleTuringskip]
\petit{I11 $\heart$ F16}\\[\ruleTuringskip]
\petit{I11 $\heart$ F17}\\[\ruleTuringskip]
\petit{I11 $\heart$ F18}\\[\ruleTuringskip]
\petit{I11 $\heart$ F19}\\[\ruleTuringskip]
\petit{I11 $\heart$ F20}\\[\ruleTuringskip]
\petit{I11 $\heart$ F21}\\[\ruleTuringskip]
\petit{I11 $\heart$ F22}\\[\ruleTuringskip]
\petit{I11 $\heart$ F23}\\[\ruleTuringskip]
\petit{I11 $\heart$ F24}\\[\ruleTuringskip]
\petit{I11 $\heart$ F25}\\[\ruleTuringskip]
\petit{I11 $\heart$ F26}\\[\ruleTuringskip]
\petit{I11 $\heart$ F27}\\[\ruleTuringskip]
\petit{I11 $\heart$ F28}\\[\ruleTuringskip]
\petit{I11 $\heart$ F29}\\[\ruleTuringskip]
\petit{I11 $\heart$ F30}\\[\ruleTuringskip]
\petit{I11 $\heart$ F31}\\[\ruleTuringskip]
\petit{I11 $\heart$ F32}\\[\ruleTuringskip]
\petit{I11 $\heart$ F33}\\[\ruleTuringskip]
\petit{I11 $\heart$ F34}\\[\ruleTuringskip]
\petit{I11 $\heart$ F35}\\[\ruleTuringskip]
\petit{I11 $\heart$ F36}\\[\ruleTuringskip]
\petit{I11 $\heart$ F37}\\[\ruleTuringskip]
\petit{I11 $\heart$ F38}\\[\ruleTuringskip]
\petit{I11 $\heart$ F39}\\[\ruleTuringskip]
\petit{I11 $\heart$ F40}\\[\ruleTuringskip]
\petit{I11 $\heart$ F41}\\[\ruleTuringskip]
\petit{I11 $\heart$ F42}\\[\ruleTuringskip]
\petit{I11 $\heart$ F43}\\[\ruleTuringskip]
\petit{I11 $\heart$ F44}\\[\ruleTuringskip]
\petit{I11 $\heart$ F45}\\[\ruleTuringskip]
\petit{I11 $\heart$ F46}\\[\ruleTuringskip]
\petit{I11 $\heart$ F47}\\[\ruleTuringskip]
\petit{I11 $\heart$ I06}\\[\ruleTuringskip]
\petit{I11 $\heart$ I07}\\[\ruleTuringskip]
\petit{I12 $\heart$ F00}\\[\ruleTuringskip]
\petit{I12 $\heart$ F01}\\[\ruleTuringskip]
\petit{I12 $\heart$ F02}\\[\ruleTuringskip]
\petit{I12 $\heart$ F03}\\[\ruleTuringskip]
\petit{I12 $\heart$ F04}\\[\ruleTuringskip]
\petit{I12 $\heart$ F05}\\[\ruleTuringskip]
\petit{I12 $\heart$ F06}\\[\ruleTuringskip]
\petit{I12 $\heart$ F07}\\[\ruleTuringskip]
\petit{I12 $\heart$ F08}\\[\ruleTuringskip]
\petit{I12 $\heart$ F09}\\[\ruleTuringskip]
\petit{I12 $\heart$ F10}\\[\ruleTuringskip]
\petit{I12 $\heart$ F11}\\[\ruleTuringskip]
\petit{I12 $\heart$ F12}\\[\ruleTuringskip]
\petit{I12 $\heart$ F13}\\[\ruleTuringskip]
\petit{I12 $\heart$ F14}\\[\ruleTuringskip]
\petit{I12 $\heart$ F15}\\[\ruleTuringskip]
\petit{I12 $\heart$ F16}\\[\ruleTuringskip]
\petit{I12 $\heart$ F17}\\[\ruleTuringskip]
\petit{I12 $\heart$ F18}\\[\ruleTuringskip]
\petit{I12 $\heart$ F19}\\[\ruleTuringskip]
\petit{I12 $\heart$ F20}\\[\ruleTuringskip]
\petit{I12 $\heart$ F21}\\[\ruleTuringskip]
\petit{I12 $\heart$ F22}\\[\ruleTuringskip]
\petit{I12 $\heart$ F23}\\[\ruleTuringskip]
\petit{I12 $\heart$ F24}\\[\ruleTuringskip]
\petit{I12 $\heart$ F25}\\[\ruleTuringskip]
\petit{I12 $\heart$ F26}\\[\ruleTuringskip]
\petit{I12 $\heart$ F27}\\[\ruleTuringskip]
\petit{I12 $\heart$ F28}\\[\ruleTuringskip]
\petit{I12 $\heart$ F29}\\[\ruleTuringskip]
\petit{I12 $\heart$ F30}\\[\ruleTuringskip]
\petit{I12 $\heart$ F31}\\[\ruleTuringskip]
\petit{I12 $\heart$ F32}\\[\ruleTuringskip]
\petit{I12 $\heart$ F33}\\[\ruleTuringskip]
\petit{I12 $\heart$ F34}\\[\ruleTuringskip]
\petit{I12 $\heart$ F35}\\[\ruleTuringskip]
\petit{I12 $\heart$ F36}\\[\ruleTuringskip]
\petit{I12 $\heart$ F37}\\[\ruleTuringskip]
\petit{I12 $\heart$ F38}\\[\ruleTuringskip]
\petit{I12 $\heart$ F39}\\[\ruleTuringskip]
\petit{I12 $\heart$ F40}\\[\ruleTuringskip]
\petit{I12 $\heart$ F41}\\[\ruleTuringskip]
\petit{I12 $\heart$ F42}\\[\ruleTuringskip]
\petit{I12 $\heart$ F43}\\[\ruleTuringskip]
\petit{I12 $\heart$ F44}\\[\ruleTuringskip]
\petit{I12 $\heart$ F45}\\[\ruleTuringskip]
\petit{I12 $\heart$ F46}\\[\ruleTuringskip]
\petit{I12 $\heart$ F47}\\[\ruleTuringskip]
\petit{I12 $\heart$ I02}\\[\ruleTuringskip]
\petit{I13 $\heart$ F00}\\[\ruleTuringskip]
\petit{I13 $\heart$ F01}\\[\ruleTuringskip]
\petit{I13 $\heart$ F02}\\[\ruleTuringskip]
\petit{I13 $\heart$ F03}\\[\ruleTuringskip]
\petit{I13 $\heart$ F04}\\[\ruleTuringskip]
\petit{I13 $\heart$ F05}\\[\ruleTuringskip]
\petit{I13 $\heart$ F06}\\[\ruleTuringskip]
\petit{I13 $\heart$ F07}\\[\ruleTuringskip]
\petit{I13 $\heart$ F08}\\[\ruleTuringskip]
\petit{I13 $\heart$ F09}\\[\ruleTuringskip]
\petit{I13 $\heart$ F10}\\[\ruleTuringskip]
\petit{I13 $\heart$ F11}\\[\ruleTuringskip]
\petit{I13 $\heart$ F12}\\[\ruleTuringskip]
\petit{I13 $\heart$ F13}\\[\ruleTuringskip]
\petit{I13 $\heart$ F14}\\[\ruleTuringskip]
\petit{I13 $\heart$ F15}\\[\ruleTuringskip]
\petit{I13 $\heart$ F16}\\[\ruleTuringskip]
\petit{I13 $\heart$ F17}\\[\ruleTuringskip]
\petit{I13 $\heart$ F18}\\[\ruleTuringskip]
\petit{I13 $\heart$ F19}\\[\ruleTuringskip]
\petit{I13 $\heart$ F20}\\[\ruleTuringskip]
\petit{I13 $\heart$ F21}\\[\ruleTuringskip]
\petit{I13 $\heart$ F22}\\[\ruleTuringskip]
\petit{I13 $\heart$ F23}\\[\ruleTuringskip]
\petit{I13 $\heart$ F24}\\[\ruleTuringskip]
\petit{I13 $\heart$ F25}\\[\ruleTuringskip]
\petit{I13 $\heart$ F26}\\[\ruleTuringskip]
\petit{I13 $\heart$ F27}\\[\ruleTuringskip]
\petit{I13 $\heart$ F28}\\[\ruleTuringskip]
\petit{I13 $\heart$ F29}\\[\ruleTuringskip]
\petit{I13 $\heart$ F30}\\[\ruleTuringskip]
\petit{I13 $\heart$ F31}\\[\ruleTuringskip]
\petit{I13 $\heart$ F32}\\[\ruleTuringskip]
\petit{I13 $\heart$ F33}\\[\ruleTuringskip]
\petit{I13 $\heart$ F34}\\[\ruleTuringskip]
\petit{I13 $\heart$ F35}\\[\ruleTuringskip]
\petit{I13 $\heart$ F36}\\[\ruleTuringskip]
\petit{I13 $\heart$ F37}\\[\ruleTuringskip]
\petit{I13 $\heart$ F38}\\[\ruleTuringskip]
\petit{I13 $\heart$ F39}\\[\ruleTuringskip]
\petit{I13 $\heart$ F40}\\[\ruleTuringskip]
\petit{I13 $\heart$ F41}\\[\ruleTuringskip]
\petit{I13 $\heart$ F42}\\[\ruleTuringskip]
\petit{I13 $\heart$ F43}\\[\ruleTuringskip]
\petit{I13 $\heart$ F44}\\[\ruleTuringskip]
\petit{I13 $\heart$ F45}\\[\ruleTuringskip]
\petit{I13 $\heart$ F46}\\[\ruleTuringskip]
\petit{I13 $\heart$ F47}\\[\ruleTuringskip]
\petit{I13 $\heart$ I01}\\[\ruleTuringskip]
\petit{I13 $\heart$ I02}\\[\ruleTuringskip]
\petit{I13 $\heart$ I09}\\[\ruleTuringskip]
\petit{I13 $\heart$ I10}\\[\ruleTuringskip]
\petit{I13 $\heart$ I17 }\\[\ruleTuringskip]
\petit{I14 $\heart$ F00}\\[\ruleTuringskip]
\petit{I14 $\heart$ F01}\\[\ruleTuringskip]
\petit{I14 $\heart$ F02}\\[\ruleTuringskip]
\petit{I14 $\heart$ F03}\\[\ruleTuringskip]
\petit{I14 $\heart$ F04}\\[\ruleTuringskip]
\petit{I14 $\heart$ F05}\\[\ruleTuringskip]
\petit{I14 $\heart$ F06}\\[\ruleTuringskip]
\petit{I14 $\heart$ F07}\\[\ruleTuringskip]
\petit{I14 $\heart$ F08}\\[\ruleTuringskip]
\petit{I14 $\heart$ F09}\\[\ruleTuringskip]
\petit{I14 $\heart$ F10}\\[\ruleTuringskip]
\petit{I14 $\heart$ F11}\\[\ruleTuringskip]
\petit{I14 $\heart$ F12}\\[\ruleTuringskip]
\petit{I14 $\heart$ F13}\\[\ruleTuringskip]
\petit{I14 $\heart$ F14}\\[\ruleTuringskip]
\petit{I14 $\heart$ F15}\\[\ruleTuringskip]
\petit{I14 $\heart$ F16}\\[\ruleTuringskip]
\petit{I14 $\heart$ F17}\\[\ruleTuringskip]
\petit{I14 $\heart$ F18}\\[\ruleTuringskip]
\petit{I14 $\heart$ F19}\\[\ruleTuringskip]
\petit{I14 $\heart$ F20}\\[\ruleTuringskip]
\petit{I14 $\heart$ F21}\\[\ruleTuringskip]
\petit{I14 $\heart$ F22}\\[\ruleTuringskip]
\petit{I14 $\heart$ F23}\\[\ruleTuringskip]
\petit{I14 $\heart$ F24}\\[\ruleTuringskip]
\petit{I14 $\heart$ F25}\\[\ruleTuringskip]
\petit{I14 $\heart$ F26}\\[\ruleTuringskip]
\petit{I14 $\heart$ F27}\\[\ruleTuringskip]
\petit{I14 $\heart$ F28}\\[\ruleTuringskip]
\petit{I14 $\heart$ F29}\\[\ruleTuringskip]
\petit{I14 $\heart$ F30}\\[\ruleTuringskip]
\petit{I14 $\heart$ F31}\\[\ruleTuringskip]
\petit{I14 $\heart$ F32}\\[\ruleTuringskip]
\petit{I14 $\heart$ F33}\\[\ruleTuringskip]
\petit{I14 $\heart$ F34}\\[\ruleTuringskip]
\petit{I14 $\heart$ F35}\\[\ruleTuringskip]
\petit{I14 $\heart$ F36}\\[\ruleTuringskip]
\petit{I14 $\heart$ F37}\\[\ruleTuringskip]
\petit{I14 $\heart$ F38}\\[\ruleTuringskip]
\petit{I14 $\heart$ F39}\\[\ruleTuringskip]
\petit{I14 $\heart$ F40}\\[\ruleTuringskip]
\petit{I14 $\heart$ F41}\\[\ruleTuringskip]
\petit{I14 $\heart$ F42}\\[\ruleTuringskip]
\petit{I14 $\heart$ F43}\\[\ruleTuringskip]
\petit{I14 $\heart$ F44}\\[\ruleTuringskip]
\petit{I14 $\heart$ F45}\\[\ruleTuringskip]
\petit{I14 $\heart$ F46}\\[\ruleTuringskip]
\petit{I14 $\heart$ F47}\\[\ruleTuringskip]
\petit{I14 $\heart$ I09}\\[\ruleTuringskip]
\petit{I15 $\heart$ I01}\\[\ruleTuringskip]
\petit{I15 $\heart$ I06}\\[\ruleTuringskip]
\petit{I15 $\heart$ I15}\\[\ruleTuringskip]
\petit{I15 $\heart$ I15 }\\[\ruleTuringskip]
\petit{I16 $\heart$ I01}\\[\ruleTuringskip]
\petit{I16 $\heart$ I07}\\[\ruleTuringskip]
\petit{I16 $\heart$ I10}\\[\ruleTuringskip]
\petit{I17 $\heart$ B02}\\[\ruleTuringskip]
\petit{I17 $\heart$ I13}\\[\ruleTuringskip]
\petit{I17 $\heart$ I19 }\\[\ruleTuringskip]
\petit{I18 $\heart$ H04}\\[\ruleTuringskip]
\petit{I18 $\heart$ I02}\\[\ruleTuringskip]
\petit{I18 $\heart$ I10}\\[\ruleTuringskip]
\petit{I19 $\heart$ B02}\\[\ruleTuringskip]
\petit{I19 $\heart$ I01}\\[\ruleTuringskip]
\petit{I19 $\heart$ I07}\\[\ruleTuringskip]
\petit{I19 $\heart$ I17}\\[\ruleTuringskip]
\petit{I19 $\heart$ I19}\\[\ruleTuringskip]
\petit{I19 $\heart$ I19 }\\[\ruleTuringskip]
\petit{I19 $\heart$ J12 }\\[\ruleTuringskip]
\\
\hspace*{5mm}$\ModuleF$\\[.5em]
\petit{J00 $\heart$ B01}\\[\ruleTuringskip]
\petit{J00 $\heart$ F00}\\[\ruleTuringskip]
\petit{J00 $\heart$ F05}\\[\ruleTuringskip]
\petit{J00 $\heart$ F24}\\[\ruleTuringskip]
\petit{J00 $\heart$ F29}\\[\ruleTuringskip]
\petit{J00 $\heart$ I01}\\[\ruleTuringskip]
\petit{J00 $\heart$ J02 }\\[\ruleTuringskip]
\petit{J00 $\heart$ L02 }\\[\ruleTuringskip]
\petit{J01 $\heart$ B01}\\[\ruleTuringskip]
\petit{J01 $\heart$ F00}\\[\ruleTuringskip]
\petit{J01 $\heart$ F24}\\[\ruleTuringskip]
\petit{J01 $\heart$ J03 }\\[\ruleTuringskip]
\petit{J01 $\heart$ J04 }\\[\ruleTuringskip]
\petit{J01 $\heart$ L01 }\\[\ruleTuringskip]
\petit{J02 $\heart$ I01}\\[\ruleTuringskip]
\petit{J02 $\heart$ J00}\\[\ruleTuringskip]
\petit{J02 $\heart$ J07 }\\[\ruleTuringskip]
\petit{J03 $\heart$ J01}\\[\ruleTuringskip]
\petit{J04 $\heart$ J01}\\[\ruleTuringskip]
\petit{J04 $\heart$ J48 }\\[\ruleTuringskip]
\petit{J04 $\heart$ L00 }\\[\ruleTuringskip]
\petit{J05 $\heart$ J10 }\\[\ruleTuringskip]
\petit{J05 $\heart$ J14 }\\[\ruleTuringskip]
\petit{J05 $\heart$ J21 }\\[\ruleTuringskip]
\petit{J05 $\heart$ J23 }\\[\ruleTuringskip]
\petit{J05 $\heart$ J25 }\\[\ruleTuringskip]
\petit{J05 $\heart$ J27 }\\[\ruleTuringskip]
\petit{J05 $\heart$ J29 }\\[\ruleTuringskip]
\petit{J05 $\heart$ J31 }\\[\ruleTuringskip]
\petit{J05 $\heart$ J33 }\\[\ruleTuringskip]
\petit{J05 $\heart$ J35 }\\[\ruleTuringskip]
\petit{J05 $\heart$ J37 }\\[\ruleTuringskip]
\petit{J05 $\heart$ J39 }\\[\ruleTuringskip]
\petit{J05 $\heart$ J41 }\\[\ruleTuringskip]
\petit{J05 $\heart$ J43 }\\[\ruleTuringskip]
\petit{J05 $\heart$ J45 }\\[\ruleTuringskip]
\petit{J05 $\heart$ J47 }\\[\ruleTuringskip]
\petit{J05 $\heart$ J48 }\\[\ruleTuringskip]
\petit{J05 $\heart$ J49 }\\[\ruleTuringskip]
\petit{J05 $\heart$ J51 }\\[\ruleTuringskip]
\petit{J06 $\heart$ J11 }\\[\ruleTuringskip]
\petit{J07 $\heart$ C05}\\[\ruleTuringskip]
\petit{J07 $\heart$ J02}\\[\ruleTuringskip]
\petit{J07 $\heart$ J08 }\\[\ruleTuringskip]
\petit{J08 $\heart$ C03}\\[\ruleTuringskip]
\petit{J08 $\heart$ C04}\\[\ruleTuringskip]
\petit{J08 $\heart$ J07}\\[\ruleTuringskip]
\petit{J10 $\heart$ J05}\\[\ruleTuringskip]
\petit{J10 $\heart$ J22 }\\[\ruleTuringskip]
\petit{J10 $\heart$ J24 }\\[\ruleTuringskip]
\petit{J10 $\heart$ J26 }\\[\ruleTuringskip]
\petit{J10 $\heart$ J28 }\\[\ruleTuringskip]
\petit{J10 $\heart$ J30 }\\[\ruleTuringskip]
\petit{J10 $\heart$ J32 }\\[\ruleTuringskip]
\petit{J10 $\heart$ J34 }\\[\ruleTuringskip]
\petit{J10 $\heart$ J36 }\\[\ruleTuringskip]
\petit{J10 $\heart$ J38 }\\[\ruleTuringskip]
\petit{J10 $\heart$ J40 }\\[\ruleTuringskip]
\petit{J10 $\heart$ J42 }\\[\ruleTuringskip]
\petit{J10 $\heart$ J44 }\\[\ruleTuringskip]
\petit{J10 $\heart$ J46 }\\[\ruleTuringskip]
\petit{J10 $\heart$ J48 }\\[\ruleTuringskip]
\petit{J10 $\heart$ J49 }\\[\ruleTuringskip]
\petit{J10 $\heart$ J50 }\\[\ruleTuringskip]
\petit{J10 $\heart$ J52 }\\[\ruleTuringskip]
\petit{J11 $\heart$ C04}\\[\ruleTuringskip]
\petit{J11 $\heart$ J06}\\[\ruleTuringskip]
\petit{J11 $\heart$ J13 }\\[\ruleTuringskip]
\petit{J12 $\heart$ C03}\\[\ruleTuringskip]
\petit{J12 $\heart$ I19}\\[\ruleTuringskip]
\petit{J12 $\heart$ J16 }\\[\ruleTuringskip]
\petit{J13 $\heart$ J11}\\[\ruleTuringskip]
\petit{J14 $\heart$ J05}\\[\ruleTuringskip]
\petit{J14 $\heart$ J20 }\\[\ruleTuringskip]
\petit{J15 $\heart$ J17 }\\[\ruleTuringskip]
\petit{J15 $\heart$ J19 }\\[\ruleTuringskip]
\petit{J16 $\heart$ J12}\\[\ruleTuringskip]
\petit{J17 $\heart$ A12}\\[\ruleTuringskip]
\petit{J17 $\heart$ B00}\\[\ruleTuringskip]
\petit{J17 $\heart$ J15}\\[\ruleTuringskip]
\petit{J17 $\heart$ J19 }\\[\ruleTuringskip]
\petit{J18 $\heart$ A00}\\[\ruleTuringskip]
\petit{J18 $\heart$ A01}\\[\ruleTuringskip]
\petit{J18 $\heart$ B01}\\[\ruleTuringskip]
\petit{J18 $\heart$ L16 }\\[\ruleTuringskip]
\petit{J18 $\heart$ L17 }\\[\ruleTuringskip]
\petit{J19 $\heart$ J15}\\[\ruleTuringskip]
\petit{J19 $\heart$ J17}\\[\ruleTuringskip]
\petit{J19 $\heart$ L15 }\\[\ruleTuringskip]
\petit{J20 $\heart$ J14}\\[\ruleTuringskip]
\petit{J20 $\heart$ L14 }\\[\ruleTuringskip]
\petit{J21 $\heart$ J05}\\[\ruleTuringskip]
\petit{J21 $\heart$ L16 }\\[\ruleTuringskip]
\petit{J22 $\heart$ J10}\\[\ruleTuringskip]
\petit{J22 $\heart$ L12 }\\[\ruleTuringskip]
\petit{J23 $\heart$ J05}\\[\ruleTuringskip]
\petit{J23 $\heart$ L11 }\\[\ruleTuringskip]
\petit{J24 $\heart$ J10}\\[\ruleTuringskip]
\petit{J24 $\heart$ K03 }\\[\ruleTuringskip]
\petit{J24 $\heart$ K04 }\\[\ruleTuringskip]
\petit{J25 $\heart$ J05}\\[\ruleTuringskip]
\petit{J25 $\heart$ K00 }\\[\ruleTuringskip]
\petit{J25 $\heart$ K05 }\\[\ruleTuringskip]
\petit{J26 $\heart$ J10}\\[\ruleTuringskip]
\petit{J26 $\heart$ K01 }\\[\ruleTuringskip]
\petit{J26 $\heart$ K06 }\\[\ruleTuringskip]
\petit{J27 $\heart$ J05}\\[\ruleTuringskip]
\petit{J27 $\heart$ K02 }\\[\ruleTuringskip]
\petit{J27 $\heart$ K07 }\\[\ruleTuringskip]
\petit{J28 $\heart$ J10}\\[\ruleTuringskip]
\petit{J28 $\heart$ K08 }\\[\ruleTuringskip]
\petit{J28 $\heart$ L07 }\\[\ruleTuringskip]
\petit{J29 $\heart$ J05}\\[\ruleTuringskip]
\petit{J29 $\heart$ K09 }\\[\ruleTuringskip]
\petit{J29 $\heart$ L08 }\\[\ruleTuringskip]
\petit{J30 $\heart$ J10}\\[\ruleTuringskip]
\petit{J30 $\heart$ K10 }\\[\ruleTuringskip]
\petit{J30 $\heart$ L09 }\\[\ruleTuringskip]
\petit{J31 $\heart$ J05}\\[\ruleTuringskip]
\petit{J31 $\heart$ K11 }\\[\ruleTuringskip]
\petit{J31 $\heart$ L10 }\\[\ruleTuringskip]
\petit{J32 $\heart$ J10}\\[\ruleTuringskip]
\petit{J32 $\heart$ K12 }\\[\ruleTuringskip]
\petit{J33 $\heart$ J05}\\[\ruleTuringskip]
\petit{J33 $\heart$ K13 }\\[\ruleTuringskip]
\petit{J34 $\heart$ J10}\\[\ruleTuringskip]
\petit{J34 $\heart$ K14 }\\[\ruleTuringskip]
\petit{J35 $\heart$ J05}\\[\ruleTuringskip]
\petit{J35 $\heart$ K15 }\\[\ruleTuringskip]
\petit{J36 $\heart$ J10}\\[\ruleTuringskip]
\petit{J36 $\heart$ K16 }\\[\ruleTuringskip]
\petit{J37 $\heart$ J05}\\[\ruleTuringskip]
\petit{J37 $\heart$ K17 }\\[\ruleTuringskip]
\petit{J38 $\heart$ J10}\\[\ruleTuringskip]
\petit{J38 $\heart$ K18 }\\[\ruleTuringskip]
\petit{J39 $\heart$ J05}\\[\ruleTuringskip]
\petit{J39 $\heart$ K00 }\\[\ruleTuringskip]
\petit{J39 $\heart$ K19 }\\[\ruleTuringskip]
\petit{J40 $\heart$ J10}\\[\ruleTuringskip]
\petit{J40 $\heart$ K01 }\\[\ruleTuringskip]
\petit{J40 $\heart$ K20 }\\[\ruleTuringskip]
\petit{J41 $\heart$ J05}\\[\ruleTuringskip]
\petit{J41 $\heart$ K02 }\\[\ruleTuringskip]
\petit{J41 $\heart$ K21 }\\[\ruleTuringskip]
\petit{J42 $\heart$ J10}\\[\ruleTuringskip]
\petit{J42 $\heart$ K22 }\\[\ruleTuringskip]
\petit{J43 $\heart$ J05}\\[\ruleTuringskip]
\petit{J43 $\heart$ K23 }\\[\ruleTuringskip]
\petit{J44 $\heart$ J10}\\[\ruleTuringskip]
\petit{J44 $\heart$ K24 }\\[\ruleTuringskip]
\petit{J45 $\heart$ J05}\\[\ruleTuringskip]
\petit{J45 $\heart$ K25 }\\[\ruleTuringskip]
\petit{J46 $\heart$ J10}\\[\ruleTuringskip]
\petit{J46 $\heart$ K26 }\\[\ruleTuringskip]
\petit{J47 $\heart$ J05}\\[\ruleTuringskip]
\petit{J47 $\heart$ K27 }\\[\ruleTuringskip]
\petit{J48 $\heart$ J04}\\[\ruleTuringskip]
\petit{J48 $\heart$ J05}\\[\ruleTuringskip]
\petit{J48 $\heart$ J10}\\[\ruleTuringskip]
\petit{J48 $\heart$ K03 }\\[\ruleTuringskip]
\petit{J48 $\heart$ L03 }\\[\ruleTuringskip]
\petit{J48 $\heart$ L06 }\\[\ruleTuringskip]
\petit{J48 $\heart$ L08 }\\[\ruleTuringskip]
\petit{J49 $\heart$ J05}\\[\ruleTuringskip]
\petit{J49 $\heart$ J10}\\[\ruleTuringskip]
\petit{J49 $\heart$ K00 }\\[\ruleTuringskip]
\petit{J49 $\heart$ L00 }\\[\ruleTuringskip]
\petit{J49 $\heart$ L04 }\\[\ruleTuringskip]
\petit{J50 $\heart$ J10}\\[\ruleTuringskip]
\petit{J50 $\heart$ K01 }\\[\ruleTuringskip]
\petit{J50 $\heart$ L05 }\\[\ruleTuringskip]
\petit{J51 $\heart$ J05}\\[\ruleTuringskip]
\petit{J51 $\heart$ K02 }\\[\ruleTuringskip]
\petit{J51 $\heart$ L02 }\\[\ruleTuringskip]
\petit{J51 $\heart$ L06 }\\[\ruleTuringskip]
\petit{J52 $\heart$ J10}\\[\ruleTuringskip]
\petit{J52 $\heart$ K01 }\\[\ruleTuringskip]
\\
\hspace*{5mm}$\ModuleG$\\[.5em]
\petit{K00 $\heart$ J25}\\[\ruleTuringskip]
\petit{K00 $\heart$ J39}\\[\ruleTuringskip]
\petit{K00 $\heart$ J49}\\[\ruleTuringskip]
\petit{K00 $\heart$ K29 }\\[\ruleTuringskip]
\petit{K00 $\heart$ K31 }\\[\ruleTuringskip]
\petit{K00 $\heart$ K33 }\\[\ruleTuringskip]
\petit{K01 $\heart$ J26}\\[\ruleTuringskip]
\petit{K01 $\heart$ J40}\\[\ruleTuringskip]
\petit{K01 $\heart$ J50}\\[\ruleTuringskip]
\petit{K01 $\heart$ J52}\\[\ruleTuringskip]
\petit{K01 $\heart$ K28 }\\[\ruleTuringskip]
\petit{K01 $\heart$ K30 }\\[\ruleTuringskip]
\petit{K01 $\heart$ K32 }\\[\ruleTuringskip]
\petit{K02 $\heart$ J27}\\[\ruleTuringskip]
\petit{K02 $\heart$ J41}\\[\ruleTuringskip]
\petit{K02 $\heart$ J51}\\[\ruleTuringskip]
\petit{K02 $\heart$ K29 }\\[\ruleTuringskip]
\petit{K02 $\heart$ K31 }\\[\ruleTuringskip]
\petit{K02 $\heart$ K33 }\\[\ruleTuringskip]
\petit{K03 $\heart$ J24}\\[\ruleTuringskip]
\petit{K03 $\heart$ J48}\\[\ruleTuringskip]
\petit{K03 $\heart$ K28 }\\[\ruleTuringskip]
\petit{K03 $\heart$ K30 }\\[\ruleTuringskip]
\petit{K03 $\heart$ K32 }\\[\ruleTuringskip]
\petit{K04 $\heart$ J24}\\[\ruleTuringskip]
\petit{K04 $\heart$ K32 }\\[\ruleTuringskip]
\petit{K05 $\heart$ J25}\\[\ruleTuringskip]
\petit{K05 $\heart$ K06 }\\[\ruleTuringskip]
\petit{K05 $\heart$ K12 }\\[\ruleTuringskip]
\petit{K05 $\heart$ K18 }\\[\ruleTuringskip]
\petit{K05 $\heart$ K24 }\\[\ruleTuringskip]
\petit{K05 $\heart$ K31 }\\[\ruleTuringskip]
\petit{K06 $\heart$ J26}\\[\ruleTuringskip]
\petit{K06 $\heart$ K05}\\[\ruleTuringskip]
\petit{K06 $\heart$ K11 }\\[\ruleTuringskip]
\petit{K06 $\heart$ K17 }\\[\ruleTuringskip]
\petit{K06 $\heart$ K23 }\\[\ruleTuringskip]
\petit{K06 $\heart$ K30 }\\[\ruleTuringskip]
\petit{K06 $\heart$ L11 }\\[\ruleTuringskip]
\petit{K07 $\heart$ J27}\\[\ruleTuringskip]
\petit{K07 $\heart$ K29 }\\[\ruleTuringskip]
\petit{K08 $\heart$ J28}\\[\ruleTuringskip]
\petit{K08 $\heart$ K09 }\\[\ruleTuringskip]
\petit{K08 $\heart$ K15 }\\[\ruleTuringskip]
\petit{K08 $\heart$ K21 }\\[\ruleTuringskip]
\petit{K08 $\heart$ K27 }\\[\ruleTuringskip]
\petit{K08 $\heart$ K28 }\\[\ruleTuringskip]
\petit{K09 $\heart$ J29}\\[\ruleTuringskip]
\petit{K09 $\heart$ K08}\\[\ruleTuringskip]
\petit{K09 $\heart$ K14 }\\[\ruleTuringskip]
\petit{K09 $\heart$ K20 }\\[\ruleTuringskip]
\petit{K09 $\heart$ K26 }\\[\ruleTuringskip]
\petit{K09 $\heart$ K33 }\\[\ruleTuringskip]
\petit{K09 $\heart$ L14 }\\[\ruleTuringskip]
\petit{K10 $\heart$ J30}\\[\ruleTuringskip]
\petit{K10 $\heart$ K32 }\\[\ruleTuringskip]
\petit{K11 $\heart$ J31}\\[\ruleTuringskip]
\petit{K11 $\heart$ K06}\\[\ruleTuringskip]
\petit{K11 $\heart$ K12 }\\[\ruleTuringskip]
\petit{K11 $\heart$ K18 }\\[\ruleTuringskip]
\petit{K11 $\heart$ K24 }\\[\ruleTuringskip]
\petit{K11 $\heart$ K31 }\\[\ruleTuringskip]
\petit{K12 $\heart$ J32}\\[\ruleTuringskip]
\petit{K12 $\heart$ K05}\\[\ruleTuringskip]
\petit{K12 $\heart$ K11}\\[\ruleTuringskip]
\petit{K12 $\heart$ K17 }\\[\ruleTuringskip]
\petit{K12 $\heart$ K23 }\\[\ruleTuringskip]
\petit{K12 $\heart$ K30 }\\[\ruleTuringskip]
\petit{K12 $\heart$ L11 }\\[\ruleTuringskip]
\petit{K13 $\heart$ J33}\\[\ruleTuringskip]
\petit{K13 $\heart$ K29 }\\[\ruleTuringskip]
\petit{K14 $\heart$ J34}\\[\ruleTuringskip]
\petit{K14 $\heart$ K09}\\[\ruleTuringskip]
\petit{K14 $\heart$ K15 }\\[\ruleTuringskip]
\petit{K14 $\heart$ K21 }\\[\ruleTuringskip]
\petit{K14 $\heart$ K27 }\\[\ruleTuringskip]
\petit{K14 $\heart$ K28 }\\[\ruleTuringskip]
\petit{K14 $\heart$ L08 }\\[\ruleTuringskip]
\petit{K15 $\heart$ J35}\\[\ruleTuringskip]
\petit{K15 $\heart$ K08}\\[\ruleTuringskip]
\petit{K15 $\heart$ K14}\\[\ruleTuringskip]
\petit{K15 $\heart$ K20 }\\[\ruleTuringskip]
\petit{K15 $\heart$ K26 }\\[\ruleTuringskip]
\petit{K15 $\heart$ K33 }\\[\ruleTuringskip]
\petit{K15 $\heart$ L14 }\\[\ruleTuringskip]
\petit{K16 $\heart$ J36}\\[\ruleTuringskip]
\petit{K16 $\heart$ K32 }\\[\ruleTuringskip]
\petit{K17 $\heart$ J37}\\[\ruleTuringskip]
\petit{K17 $\heart$ K06}\\[\ruleTuringskip]
\petit{K17 $\heart$ K12}\\[\ruleTuringskip]
\petit{K17 $\heart$ K18 }\\[\ruleTuringskip]
\petit{K17 $\heart$ K24 }\\[\ruleTuringskip]
\petit{K17 $\heart$ K31 }\\[\ruleTuringskip]
\petit{K18 $\heart$ J38}\\[\ruleTuringskip]
\petit{K18 $\heart$ K05}\\[\ruleTuringskip]
\petit{K18 $\heart$ K11}\\[\ruleTuringskip]
\petit{K18 $\heart$ K17}\\[\ruleTuringskip]
\petit{K18 $\heart$ K23 }\\[\ruleTuringskip]
\petit{K18 $\heart$ K30 }\\[\ruleTuringskip]
\petit{K18 $\heart$ L11 }\\[\ruleTuringskip]
\petit{K19 $\heart$ J39}\\[\ruleTuringskip]
\petit{K19 $\heart$ K29 }\\[\ruleTuringskip]
\petit{K20 $\heart$ J40}\\[\ruleTuringskip]
\petit{K20 $\heart$ K09}\\[\ruleTuringskip]
\petit{K20 $\heart$ K15}\\[\ruleTuringskip]
\petit{K20 $\heart$ K21 }\\[\ruleTuringskip]
\petit{K20 $\heart$ K27 }\\[\ruleTuringskip]
\petit{K20 $\heart$ K28 }\\[\ruleTuringskip]
\petit{K21 $\heart$ J41}\\[\ruleTuringskip]
\petit{K21 $\heart$ K08}\\[\ruleTuringskip]
\petit{K21 $\heart$ K14}\\[\ruleTuringskip]
\petit{K21 $\heart$ K20}\\[\ruleTuringskip]
\petit{K21 $\heart$ K26 }\\[\ruleTuringskip]
\petit{K21 $\heart$ K33 }\\[\ruleTuringskip]
\petit{K21 $\heart$ L14 }\\[\ruleTuringskip]
\petit{K22 $\heart$ J42}\\[\ruleTuringskip]
\petit{K22 $\heart$ K32 }\\[\ruleTuringskip]
\petit{K23 $\heart$ J43}\\[\ruleTuringskip]
\petit{K23 $\heart$ K06}\\[\ruleTuringskip]
\petit{K23 $\heart$ K12}\\[\ruleTuringskip]
\petit{K23 $\heart$ K18}\\[\ruleTuringskip]
\petit{K23 $\heart$ K24 }\\[\ruleTuringskip]
\petit{K23 $\heart$ K31 }\\[\ruleTuringskip]
\petit{K24 $\heart$ J44}\\[\ruleTuringskip]
\petit{K24 $\heart$ K05}\\[\ruleTuringskip]
\petit{K24 $\heart$ K11}\\[\ruleTuringskip]
\petit{K24 $\heart$ K17}\\[\ruleTuringskip]
\petit{K24 $\heart$ K23}\\[\ruleTuringskip]
\petit{K24 $\heart$ K30 }\\[\ruleTuringskip]
\petit{K24 $\heart$ L11 }\\[\ruleTuringskip]
\petit{K25 $\heart$ J45}\\[\ruleTuringskip]
\petit{K25 $\heart$ K29 }\\[\ruleTuringskip]
\petit{K26 $\heart$ J46}\\[\ruleTuringskip]
\petit{K26 $\heart$ K09}\\[\ruleTuringskip]
\petit{K26 $\heart$ K15}\\[\ruleTuringskip]
\petit{K26 $\heart$ K21}\\[\ruleTuringskip]
\petit{K26 $\heart$ K27 }\\[\ruleTuringskip]
\petit{K26 $\heart$ K28 }\\[\ruleTuringskip]
\petit{K27 $\heart$ J47}\\[\ruleTuringskip]
\petit{K27 $\heart$ K08}\\[\ruleTuringskip]
\petit{K27 $\heart$ K14}\\[\ruleTuringskip]
\petit{K27 $\heart$ K20}\\[\ruleTuringskip]
\petit{K27 $\heart$ K26}\\[\ruleTuringskip]
\petit{K27 $\heart$ K33 }\\[\ruleTuringskip]
\petit{K27 $\heart$ L14 }\\[\ruleTuringskip]
\petit{K28 $\heart$ K01}\\[\ruleTuringskip]
\petit{K28 $\heart$ K03}\\[\ruleTuringskip]
\petit{K28 $\heart$ K08}\\[\ruleTuringskip]
\petit{K28 $\heart$ K14}\\[\ruleTuringskip]
\petit{K28 $\heart$ K20}\\[\ruleTuringskip]
\petit{K28 $\heart$ K26}\\[\ruleTuringskip]
\petit{K28 $\heart$ K33 }\\[\ruleTuringskip]
\petit{K28 $\heart$ K38 }\\[\ruleTuringskip]
\petit{K28 $\heart$ K44 }\\[\ruleTuringskip]
\petit{K28 $\heart$ L07 }\\[\ruleTuringskip]
\petit{K28 $\heart$ L25 }\\[\ruleTuringskip]
\petit{K28 $\heart$ L40 }\\[\ruleTuringskip]
\petit{K28 $\heart$ L41 }\\[\ruleTuringskip]
\petit{K29 $\heart$ K00}\\[\ruleTuringskip]
\petit{K29 $\heart$ K02}\\[\ruleTuringskip]
\petit{K29 $\heart$ K07}\\[\ruleTuringskip]
\petit{K29 $\heart$ K13}\\[\ruleTuringskip]
\petit{K29 $\heart$ K19}\\[\ruleTuringskip]
\petit{K29 $\heart$ K25}\\[\ruleTuringskip]
\petit{K29 $\heart$ K37 }\\[\ruleTuringskip]
\petit{K29 $\heart$ K43 }\\[\ruleTuringskip]
\petit{K29 $\heart$ L39 }\\[\ruleTuringskip]
\petit{K29 $\heart$ L40 }\\[\ruleTuringskip]
\petit{K29 $\heart$ L41 }\\[\ruleTuringskip]
\petit{K30 $\heart$ K01}\\[\ruleTuringskip]
\petit{K30 $\heart$ K03}\\[\ruleTuringskip]
\petit{K30 $\heart$ K06}\\[\ruleTuringskip]
\petit{K30 $\heart$ K12}\\[\ruleTuringskip]
\petit{K30 $\heart$ K18}\\[\ruleTuringskip]
\petit{K30 $\heart$ K24}\\[\ruleTuringskip]
\petit{K30 $\heart$ K31 }\\[\ruleTuringskip]
\petit{K30 $\heart$ K36 }\\[\ruleTuringskip]
\petit{K30 $\heart$ K42 }\\[\ruleTuringskip]
\petit{K30 $\heart$ L24 }\\[\ruleTuringskip]
\petit{K30 $\heart$ L39 }\\[\ruleTuringskip]
\petit{K30 $\heart$ L40 }\\[\ruleTuringskip]
\petit{K31 $\heart$ K00}\\[\ruleTuringskip]
\petit{K31 $\heart$ K02}\\[\ruleTuringskip]
\petit{K31 $\heart$ K05}\\[\ruleTuringskip]
\petit{K31 $\heart$ K11}\\[\ruleTuringskip]
\petit{K31 $\heart$ K17}\\[\ruleTuringskip]
\petit{K31 $\heart$ K23}\\[\ruleTuringskip]
\petit{K31 $\heart$ K30}\\[\ruleTuringskip]
\petit{K31 $\heart$ K35 }\\[\ruleTuringskip]
\petit{K31 $\heart$ K41 }\\[\ruleTuringskip]
\petit{K31 $\heart$ L10 }\\[\ruleTuringskip]
\petit{K31 $\heart$ L28 }\\[\ruleTuringskip]
\petit{K31 $\heart$ L39 }\\[\ruleTuringskip]
\petit{K32 $\heart$ K01}\\[\ruleTuringskip]
\petit{K32 $\heart$ K03}\\[\ruleTuringskip]
\petit{K32 $\heart$ K04}\\[\ruleTuringskip]
\petit{K32 $\heart$ K10}\\[\ruleTuringskip]
\petit{K32 $\heart$ K16}\\[\ruleTuringskip]
\petit{K32 $\heart$ K22}\\[\ruleTuringskip]
\petit{K32 $\heart$ K34 }\\[\ruleTuringskip]
\petit{K32 $\heart$ K40 }\\[\ruleTuringskip]
\petit{K32 $\heart$ L09 }\\[\ruleTuringskip]
\petit{K32 $\heart$ L38 }\\[\ruleTuringskip]
\petit{K33 $\heart$ K00}\\[\ruleTuringskip]
\petit{K33 $\heart$ K02}\\[\ruleTuringskip]
\petit{K33 $\heart$ K09}\\[\ruleTuringskip]
\petit{K33 $\heart$ K15}\\[\ruleTuringskip]
\petit{K33 $\heart$ K21}\\[\ruleTuringskip]
\petit{K33 $\heart$ K27}\\[\ruleTuringskip]
\petit{K33 $\heart$ K28}\\[\ruleTuringskip]
\petit{K33 $\heart$ K39 }\\[\ruleTuringskip]
\petit{K33 $\heart$ K45 }\\[\ruleTuringskip]
\petit{K33 $\heart$ L08 }\\[\ruleTuringskip]
\petit{K33 $\heart$ L21 }\\[\ruleTuringskip]
\petit{K33 $\heart$ L41 }\\[\ruleTuringskip]
\petit{K34 $\heart$ K32}\\[\ruleTuringskip]
\petit{K34 $\heart$ K39 }\\[\ruleTuringskip]
\petit{K34 $\heart$ K45 }\\[\ruleTuringskip]
\petit{K34 $\heart$ K50 }\\[\ruleTuringskip]
\petit{K35 $\heart$ K31}\\[\ruleTuringskip]
\petit{K35 $\heart$ K45 }\\[\ruleTuringskip]
\petit{K35 $\heart$ K49 }\\[\ruleTuringskip]
\petit{K35 $\heart$ K50 }\\[\ruleTuringskip]
\petit{K35 $\heart$ L40 }\\[\ruleTuringskip]
\petit{K35 $\heart$ L41 }\\[\ruleTuringskip]
\petit{K35 $\heart$ L56 }\\[\ruleTuringskip]
\petit{K35 $\heart$ L57 }\\[\ruleTuringskip]
\petit{K36 $\heart$ K30}\\[\ruleTuringskip]
\petit{K36 $\heart$ K37 }\\[\ruleTuringskip]
\petit{K36 $\heart$ L35 }\\[\ruleTuringskip]
\petit{K37 $\heart$ K29}\\[\ruleTuringskip]
\petit{K37 $\heart$ K36}\\[\ruleTuringskip]
\petit{K37 $\heart$ K47 }\\[\ruleTuringskip]
\petit{K37 $\heart$ L39 }\\[\ruleTuringskip]
\petit{K37 $\heart$ L40 }\\[\ruleTuringskip]
\petit{K38 $\heart$ K28}\\[\ruleTuringskip]
\petit{K38 $\heart$ K46 }\\[\ruleTuringskip]
\petit{K38 $\heart$ K47 }\\[\ruleTuringskip]
\petit{K38 $\heart$ L39 }\\[\ruleTuringskip]
\petit{K39 $\heart$ K33}\\[\ruleTuringskip]
\petit{K39 $\heart$ K34}\\[\ruleTuringskip]
\petit{K39 $\heart$ L38 }\\[\ruleTuringskip]
\petit{K40 $\heart$ K32}\\[\ruleTuringskip]
\petit{K40 $\heart$ K45 }\\[\ruleTuringskip]
\petit{K40 $\heart$ K46 }\\[\ruleTuringskip]
\petit{K40 $\heart$ K50 }\\[\ruleTuringskip]
\petit{K41 $\heart$ K31}\\[\ruleTuringskip]
\petit{K41 $\heart$ K44 }\\[\ruleTuringskip]
\petit{K41 $\heart$ K49 }\\[\ruleTuringskip]
\petit{K41 $\heart$ K50 }\\[\ruleTuringskip]
\petit{K41 $\heart$ K51 }\\[\ruleTuringskip]
\petit{K41 $\heart$ L24 }\\[\ruleTuringskip]
\petit{K41 $\heart$ L41 }\\[\ruleTuringskip]
\petit{K41 $\heart$ L55 }\\[\ruleTuringskip]
\petit{K42 $\heart$ K30}\\[\ruleTuringskip]
\petit{K42 $\heart$ K43 }\\[\ruleTuringskip]
\petit{K42 $\heart$ L40 }\\[\ruleTuringskip]
\petit{K42 $\heart$ L44 }\\[\ruleTuringskip]
\petit{K42 $\heart$ L46 }\\[\ruleTuringskip]
\petit{K42 $\heart$ L48 }\\[\ruleTuringskip]
\petit{K43 $\heart$ K29}\\[\ruleTuringskip]
\petit{K43 $\heart$ K42}\\[\ruleTuringskip]
\petit{K43 $\heart$ K47 }\\[\ruleTuringskip]
\petit{K43 $\heart$ L42 }\\[\ruleTuringskip]
\petit{K43 $\heart$ L43 }\\[\ruleTuringskip]
\petit{K43 $\heart$ L45 }\\[\ruleTuringskip]
\petit{K43 $\heart$ L47 }\\[\ruleTuringskip]
\petit{K44 $\heart$ K28}\\[\ruleTuringskip]
\petit{K44 $\heart$ K41}\\[\ruleTuringskip]
\petit{K44 $\heart$ K46 }\\[\ruleTuringskip]
\petit{K44 $\heart$ K47 }\\[\ruleTuringskip]
\petit{K45 $\heart$ K33}\\[\ruleTuringskip]
\petit{K45 $\heart$ K34}\\[\ruleTuringskip]
\petit{K45 $\heart$ K35}\\[\ruleTuringskip]
\petit{K45 $\heart$ K40}\\[\ruleTuringskip]
\petit{K45 $\heart$ L17 }\\[\ruleTuringskip]
\petit{K45 $\heart$ L82 }\\[\ruleTuringskip]
\petit{K46 $\heart$ K38}\\[\ruleTuringskip]
\petit{K46 $\heart$ K40}\\[\ruleTuringskip]
\petit{K46 $\heart$ K44}\\[\ruleTuringskip]
\petit{K46 $\heart$ K47 }\\[\ruleTuringskip]
\petit{K46 $\heart$ K56 }\\[\ruleTuringskip]
\petit{K46 $\heart$ L41 }\\[\ruleTuringskip]
\petit{K47 $\heart$ K37}\\[\ruleTuringskip]
\petit{K47 $\heart$ K38}\\[\ruleTuringskip]
\petit{K47 $\heart$ K43}\\[\ruleTuringskip]
\petit{K47 $\heart$ K44}\\[\ruleTuringskip]
\petit{K47 $\heart$ K46}\\[\ruleTuringskip]
\petit{K47 $\heart$ K55 }\\[\ruleTuringskip]
\petit{K47 $\heart$ K57 }\\[\ruleTuringskip]
\petit{K47 $\heart$ L40 }\\[\ruleTuringskip]
\petit{K47 $\heart$ L41 }\\[\ruleTuringskip]
\petit{K47 $\heart$ L75 }\\[\ruleTuringskip]
\petit{K48 $\heart$ K51 }\\[\ruleTuringskip]
\petit{K48 $\heart$ K54 }\\[\ruleTuringskip]
\petit{K48 $\heart$ K56 }\\[\ruleTuringskip]
\petit{K48 $\heart$ L72 }\\[\ruleTuringskip]
\petit{K48 $\heart$ L73 }\\[\ruleTuringskip]
\petit{K48 $\heart$ L74 }\\[\ruleTuringskip]
\petit{K49 $\heart$ K35}\\[\ruleTuringskip]
\petit{K49 $\heart$ K41}\\[\ruleTuringskip]
\petit{K49 $\heart$ K50 }\\[\ruleTuringskip]
\petit{K49 $\heart$ K53 }\\[\ruleTuringskip]
\petit{K49 $\heart$ L58 }\\[\ruleTuringskip]
\petit{K49 $\heart$ L60 }\\[\ruleTuringskip]
\petit{K49 $\heart$ L62 }\\[\ruleTuringskip]
\petit{K49 $\heart$ L64 }\\[\ruleTuringskip]
\petit{K50 $\heart$ K34}\\[\ruleTuringskip]
\petit{K50 $\heart$ K35}\\[\ruleTuringskip]
\petit{K50 $\heart$ K40}\\[\ruleTuringskip]
\petit{K50 $\heart$ K41}\\[\ruleTuringskip]
\petit{K50 $\heart$ K49}\\[\ruleTuringskip]
\petit{K50 $\heart$ K52 }\\[\ruleTuringskip]
\petit{K50 $\heart$ L56 }\\[\ruleTuringskip]
\petit{K50 $\heart$ L57 }\\[\ruleTuringskip]
\petit{K50 $\heart$ L59 }\\[\ruleTuringskip]
\petit{K50 $\heart$ L61 }\\[\ruleTuringskip]
\petit{K50 $\heart$ L63 }\\[\ruleTuringskip]
\petit{K51 $\heart$ K41}\\[\ruleTuringskip]
\petit{K51 $\heart$ K48}\\[\ruleTuringskip]
\petit{K51 $\heart$ K57 }\\[\ruleTuringskip]
\petit{K52 $\heart$ K50}\\[\ruleTuringskip]
\petit{K52 $\heart$ K57 }\\[\ruleTuringskip]
\petit{K52 $\heart$ K59 }\\[\ruleTuringskip]
\petit{K52 $\heart$ K63 }\\[\ruleTuringskip]
\petit{K52 $\heart$ K69 }\\[\ruleTuringskip]
\petit{K52 $\heart$ L92 }\\[\ruleTuringskip]
\petit{K52 $\heart$ L98 }\\[\ruleTuringskip]
\petit{K52 $\heart$ M04 }\\[\ruleTuringskip]
\petit{K52 $\heart$ M10 }\\[\ruleTuringskip]
\petit{K52 $\heart$ M16 }\\[\ruleTuringskip]
\petit{K53 $\heart$ K49}\\[\ruleTuringskip]
\petit{K53 $\heart$ K56 }\\[\ruleTuringskip]
\petit{K53 $\heart$ K58 }\\[\ruleTuringskip]
\petit{K53 $\heart$ K62 }\\[\ruleTuringskip]
\petit{K53 $\heart$ K63 }\\[\ruleTuringskip]
\petit{K53 $\heart$ K68 }\\[\ruleTuringskip]
\petit{K53 $\heart$ K69 }\\[\ruleTuringskip]
\petit{K53 $\heart$ L55 }\\[\ruleTuringskip]
\petit{K53 $\heart$ L90 }\\[\ruleTuringskip]
\petit{K53 $\heart$ L91 }\\[\ruleTuringskip]
\petit{K53 $\heart$ L92 }\\[\ruleTuringskip]
\petit{K53 $\heart$ L97 }\\[\ruleTuringskip]
\petit{K53 $\heart$ L98 }\\[\ruleTuringskip]
\petit{K53 $\heart$ M03 }\\[\ruleTuringskip]
\petit{K53 $\heart$ M04 }\\[\ruleTuringskip]
\petit{K53 $\heart$ M09 }\\[\ruleTuringskip]
\petit{K53 $\heart$ M10 }\\[\ruleTuringskip]
\petit{K53 $\heart$ M15 }\\[\ruleTuringskip]
\petit{K53 $\heart$ M16 }\\[\ruleTuringskip]
\petit{K54 $\heart$ K48}\\[\ruleTuringskip]
\petit{K54 $\heart$ K55 }\\[\ruleTuringskip]
\petit{K54 $\heart$ L75 }\\[\ruleTuringskip]
\petit{K54 $\heart$ L78 }\\[\ruleTuringskip]
\petit{K54 $\heart$ L80 }\\[\ruleTuringskip]
\petit{K55 $\heart$ K47}\\[\ruleTuringskip]
\petit{K55 $\heart$ K54}\\[\ruleTuringskip]
\petit{K55 $\heart$ K60 }\\[\ruleTuringskip]
\petit{K55 $\heart$ K66 }\\[\ruleTuringskip]
\petit{K55 $\heart$ L74 }\\[\ruleTuringskip]
\petit{K55 $\heart$ L76 }\\[\ruleTuringskip]
\petit{K55 $\heart$ L77 }\\[\ruleTuringskip]
\petit{K55 $\heart$ L79 }\\[\ruleTuringskip]
\petit{K55 $\heart$ L95 }\\[\ruleTuringskip]
\petit{K55 $\heart$ M01 }\\[\ruleTuringskip]
\petit{K55 $\heart$ M07 }\\[\ruleTuringskip]
\petit{K55 $\heart$ M13 }\\[\ruleTuringskip]
\petit{K55 $\heart$ M19 }\\[\ruleTuringskip]
\petit{K55 $\heart$ M22 }\\[\ruleTuringskip]
\petit{K56 $\heart$ K46}\\[\ruleTuringskip]
\petit{K56 $\heart$ K48}\\[\ruleTuringskip]
\petit{K56 $\heart$ K53}\\[\ruleTuringskip]
\petit{K56 $\heart$ K59 }\\[\ruleTuringskip]
\petit{K56 $\heart$ K60 }\\[\ruleTuringskip]
\petit{K56 $\heart$ K65 }\\[\ruleTuringskip]
\petit{K56 $\heart$ K66 }\\[\ruleTuringskip]
\petit{K56 $\heart$ L94 }\\[\ruleTuringskip]
\petit{K56 $\heart$ L95 }\\[\ruleTuringskip]
\petit{K56 $\heart$ M00 }\\[\ruleTuringskip]
\petit{K56 $\heart$ M01 }\\[\ruleTuringskip]
\petit{K56 $\heart$ M06 }\\[\ruleTuringskip]
\petit{K56 $\heart$ M07 }\\[\ruleTuringskip]
\petit{K56 $\heart$ M12 }\\[\ruleTuringskip]
\petit{K56 $\heart$ M13 }\\[\ruleTuringskip]
\petit{K56 $\heart$ M18 }\\[\ruleTuringskip]
\petit{K56 $\heart$ M19 }\\[\ruleTuringskip]
\petit{K56 $\heart$ M21 }\\[\ruleTuringskip]
\petit{K56 $\heart$ M22 }\\[\ruleTuringskip]
\petit{K57 $\heart$ K47}\\[\ruleTuringskip]
\petit{K57 $\heart$ K51}\\[\ruleTuringskip]
\petit{K57 $\heart$ K52}\\[\ruleTuringskip]
\petit{K57 $\heart$ M05 }\\[\ruleTuringskip]
\petit{K58 $\heart$ K53}\\[\ruleTuringskip]
\petit{K58 $\heart$ K61 }\\[\ruleTuringskip]
\petit{K59 $\heart$ K52}\\[\ruleTuringskip]
\petit{K59 $\heart$ K56}\\[\ruleTuringskip]
\petit{K59 $\heart$ K60 }\\[\ruleTuringskip]
\petit{K60 $\heart$ K55}\\[\ruleTuringskip]
\petit{K60 $\heart$ K56}\\[\ruleTuringskip]
\petit{K60 $\heart$ K59}\\[\ruleTuringskip]
\petit{K60 $\heart$ K69 }\\[\ruleTuringskip]
\petit{K60 $\heart$ L75 }\\[\ruleTuringskip]
\petit{K61 $\heart$ K58}\\[\ruleTuringskip]
\petit{K61 $\heart$ K68 }\\[\ruleTuringskip]
\petit{K62 $\heart$ K53}\\[\ruleTuringskip]
\petit{K62 $\heart$ K63 }\\[\ruleTuringskip]
\petit{K62 $\heart$ L93 }\\[\ruleTuringskip]
\petit{K62 $\heart$ L95 }\\[\ruleTuringskip]
\petit{K62 $\heart$ L97 }\\[\ruleTuringskip]
\petit{K62 $\heart$ L99 }\\[\ruleTuringskip]
\petit{K62 $\heart$ M01 }\\[\ruleTuringskip]
\petit{K62 $\heart$ M03 }\\[\ruleTuringskip]
\petit{K63 $\heart$ K52}\\[\ruleTuringskip]
\petit{K63 $\heart$ K53}\\[\ruleTuringskip]
\petit{K63 $\heart$ K62}\\[\ruleTuringskip]
\petit{K63 $\heart$ L91 }\\[\ruleTuringskip]
\petit{K63 $\heart$ L92 }\\[\ruleTuringskip]
\petit{K63 $\heart$ L94 }\\[\ruleTuringskip]
\petit{K63 $\heart$ L96 }\\[\ruleTuringskip]
\petit{K63 $\heart$ L98 }\\[\ruleTuringskip]
\petit{K63 $\heart$ M00 }\\[\ruleTuringskip]
\petit{K63 $\heart$ M02 }\\[\ruleTuringskip]
\petit{K63 $\heart$ M04 }\\[\ruleTuringskip]
\petit{K64 $\heart$ K69 }\\[\ruleTuringskip]
\petit{K65 $\heart$ K56}\\[\ruleTuringskip]
\petit{K65 $\heart$ K68 }\\[\ruleTuringskip]
\petit{K66 $\heart$ K55}\\[\ruleTuringskip]
\petit{K66 $\heart$ K56}\\[\ruleTuringskip]
\petit{K66 $\heart$ K67 }\\[\ruleTuringskip]
\petit{K67 $\heart$ K66}\\[\ruleTuringskip]
\petit{K67 $\heart$ M22 }\\[\ruleTuringskip]
\petit{K68 $\heart$ K53}\\[\ruleTuringskip]
\petit{K68 $\heart$ K61}\\[\ruleTuringskip]
\petit{K68 $\heart$ K65}\\[\ruleTuringskip]
\petit{K68 $\heart$ M21 }\\[\ruleTuringskip]
\petit{K69 $\heart$ K52}\\[\ruleTuringskip]
\petit{K69 $\heart$ K53}\\[\ruleTuringskip]
\petit{K69 $\heart$ K60}\\[\ruleTuringskip]
\petit{K69 $\heart$ K64}\\[\ruleTuringskip]
\petit{L00 $\heart$ J04}\\[\ruleTuringskip]
\petit{L00 $\heart$ J49}\\[\ruleTuringskip]
\petit{L00 $\heart$ L04 }\\[\ruleTuringskip]
\petit{L00 $\heart$ L05 }\\[\ruleTuringskip]
\petit{L00 $\heart$ L31 }\\[\ruleTuringskip]
\petit{L01 $\heart$ J01}\\[\ruleTuringskip]
\petit{L01 $\heart$ L30 }\\[\ruleTuringskip]
\petit{L02 $\heart$ A07}\\[\ruleTuringskip]
\petit{L02 $\heart$ J00}\\[\ruleTuringskip]
\petit{L02 $\heart$ J51}\\[\ruleTuringskip]
\petit{L02 $\heart$ L05 }\\[\ruleTuringskip]
\petit{L02 $\heart$ L06 }\\[\ruleTuringskip]
\petit{L02 $\heart$ L29 }\\[\ruleTuringskip]
\petit{L03 $\heart$ A07}\\[\ruleTuringskip]
\petit{L03 $\heart$ A08}\\[\ruleTuringskip]
\petit{L03 $\heart$ J48}\\[\ruleTuringskip]
\petit{L03 $\heart$ L28 }\\[\ruleTuringskip]
\petit{L04 $\heart$ J49}\\[\ruleTuringskip]
\petit{L04 $\heart$ L00}\\[\ruleTuringskip]
\petit{L04 $\heart$ L27 }\\[\ruleTuringskip]
\petit{L05 $\heart$ J50}\\[\ruleTuringskip]
\petit{L05 $\heart$ L00}\\[\ruleTuringskip]
\petit{L05 $\heart$ L02}\\[\ruleTuringskip]
\petit{L06 $\heart$ J48}\\[\ruleTuringskip]
\petit{L06 $\heart$ J51}\\[\ruleTuringskip]
\petit{L06 $\heart$ L02}\\[\ruleTuringskip]
\petit{L06 $\heart$ L25 }\\[\ruleTuringskip]
\petit{L07 $\heart$ J28}\\[\ruleTuringskip]
\petit{L07 $\heart$ K28}\\[\ruleTuringskip]
\petit{L07 $\heart$ L09 }\\[\ruleTuringskip]
\petit{L07 $\heart$ L10 }\\[\ruleTuringskip]
\petit{L08 $\heart$ J29}\\[\ruleTuringskip]
\petit{L08 $\heart$ J48}\\[\ruleTuringskip]
\petit{L08 $\heart$ K14}\\[\ruleTuringskip]
\petit{L08 $\heart$ K33}\\[\ruleTuringskip]
\petit{L09 $\heart$ J30}\\[\ruleTuringskip]
\petit{L09 $\heart$ K32}\\[\ruleTuringskip]
\petit{L09 $\heart$ L07}\\[\ruleTuringskip]
\petit{L10 $\heart$ J31}\\[\ruleTuringskip]
\petit{L10 $\heart$ K31}\\[\ruleTuringskip]
\petit{L10 $\heart$ L07}\\[\ruleTuringskip]
\petit{L11 $\heart$ J23}\\[\ruleTuringskip]
\petit{L11 $\heart$ K06}\\[\ruleTuringskip]
\petit{L11 $\heart$ K12}\\[\ruleTuringskip]
\petit{L11 $\heart$ K18}\\[\ruleTuringskip]
\petit{L11 $\heart$ K24}\\[\ruleTuringskip]
\petit{L12 $\heart$ J22}\\[\ruleTuringskip]
\petit{L12 $\heart$ L17 }\\[\ruleTuringskip]
\petit{L12 $\heart$ L23 }\\[\ruleTuringskip]
\petit{L13 $\heart$ L22 }\\[\ruleTuringskip]
\petit{L14 $\heart$ J20}\\[\ruleTuringskip]
\petit{L14 $\heart$ K09}\\[\ruleTuringskip]
\petit{L14 $\heart$ K15}\\[\ruleTuringskip]
\petit{L14 $\heart$ K21}\\[\ruleTuringskip]
\petit{L14 $\heart$ K27}\\[\ruleTuringskip]
\petit{L15 $\heart$ J19}\\[\ruleTuringskip]
\petit{L15 $\heart$ L20 }\\[\ruleTuringskip]
\petit{L16 $\heart$ J18}\\[\ruleTuringskip]
\petit{L16 $\heart$ J21}\\[\ruleTuringskip]
\petit{L16 $\heart$ L19 }\\[\ruleTuringskip]
\petit{L16 $\heart$ L20 }\\[\ruleTuringskip]
\petit{L17 $\heart$ A00}\\[\ruleTuringskip]
\petit{L17 $\heart$ C11}\\[\ruleTuringskip]
\petit{L17 $\heart$ C12}\\[\ruleTuringskip]
\petit{L17 $\heart$ E09}\\[\ruleTuringskip]
\petit{L17 $\heart$ E33}\\[\ruleTuringskip]
\petit{L17 $\heart$ J18}\\[\ruleTuringskip]
\petit{L17 $\heart$ K45}\\[\ruleTuringskip]
\petit{L17 $\heart$ L12}\\[\ruleTuringskip]
\petit{L17 $\heart$ L65 }\\[\ruleTuringskip]
\petit{L18 $\heart$ D56}\\[\ruleTuringskip]
\petit{L18 $\heart$ E20}\\[\ruleTuringskip]
\petit{L18 $\heart$ E44}\\[\ruleTuringskip]
\petit{L18 $\heart$ L23 }\\[\ruleTuringskip]
\petit{L18 $\heart$ L41 }\\[\ruleTuringskip]
\petit{L18 $\heart$ L47 }\\[\ruleTuringskip]
\petit{L18 $\heart$ L64 }\\[\ruleTuringskip]
\petit{L18 $\heart$ L65 }\\[\ruleTuringskip]
\petit{L18 $\heart$ M22 }\\[\ruleTuringskip]
\petit{L19 $\heart$ L16}\\[\ruleTuringskip]
\petit{L19 $\heart$ L46 }\\[\ruleTuringskip]
\petit{L20 $\heart$ L15}\\[\ruleTuringskip]
\petit{L20 $\heart$ L16}\\[\ruleTuringskip]
\petit{L20 $\heart$ L45 }\\[\ruleTuringskip]
\petit{L21 $\heart$ K33}\\[\ruleTuringskip]
\petit{L21 $\heart$ L44 }\\[\ruleTuringskip]
\petit{L22 $\heart$ L13}\\[\ruleTuringskip]
\petit{L22 $\heart$ L43 }\\[\ruleTuringskip]
\petit{L23 $\heart$ L12}\\[\ruleTuringskip]
\petit{L23 $\heart$ L18}\\[\ruleTuringskip]
\petit{L23 $\heart$ L42 }\\[\ruleTuringskip]
\petit{L24 $\heart$ K30}\\[\ruleTuringskip]
\petit{L24 $\heart$ K41}\\[\ruleTuringskip]
\petit{L25 $\heart$ K28}\\[\ruleTuringskip]
\petit{L25 $\heart$ L06}\\[\ruleTuringskip]
\petit{L25 $\heart$ L37 }\\[\ruleTuringskip]
\petit{L26 $\heart$ L31 }\\[\ruleTuringskip]
\petit{L26 $\heart$ L36 }\\[\ruleTuringskip]
\petit{L27 $\heart$ L04}\\[\ruleTuringskip]
\petit{L27 $\heart$ L31 }\\[\ruleTuringskip]
\petit{L27 $\heart$ L35 }\\[\ruleTuringskip]
\petit{L28 $\heart$ K31}\\[\ruleTuringskip]
\petit{L28 $\heart$ L03}\\[\ruleTuringskip]
\petit{L28 $\heart$ L33 }\\[\ruleTuringskip]
\petit{L28 $\heart$ L34 }\\[\ruleTuringskip]
\petit{L29 $\heart$ L02}\\[\ruleTuringskip]
\petit{L29 $\heart$ L33 }\\[\ruleTuringskip]
\petit{L29 $\heart$ L34 }\\[\ruleTuringskip]
\petit{L30 $\heart$ L01}\\[\ruleTuringskip]
\petit{L30 $\heart$ L32 }\\[\ruleTuringskip]
\petit{L30 $\heart$ L33 }\\[\ruleTuringskip]
\petit{L31 $\heart$ D57}\\[\ruleTuringskip]
\petit{L31 $\heart$ D58}\\[\ruleTuringskip]
\petit{L31 $\heart$ L00}\\[\ruleTuringskip]
\petit{L31 $\heart$ L26}\\[\ruleTuringskip]
\petit{L31 $\heart$ L27}\\[\ruleTuringskip]
\petit{L31 $\heart$ L48 }\\[\ruleTuringskip]
\petit{L31 $\heart$ L49 }\\[\ruleTuringskip]
\petit{L32 $\heart$ L30}\\[\ruleTuringskip]
\petit{L32 $\heart$ L37 }\\[\ruleTuringskip]
\petit{L32 $\heart$ L63 }\\[\ruleTuringskip]
\petit{L32 $\heart$ L83 }\\[\ruleTuringskip]
\petit{L33 $\heart$ L28}\\[\ruleTuringskip]
\petit{L33 $\heart$ L29}\\[\ruleTuringskip]
\petit{L33 $\heart$ L30}\\[\ruleTuringskip]
\petit{L33 $\heart$ L62 }\\[\ruleTuringskip]
\petit{L33 $\heart$ L63 }\\[\ruleTuringskip]
\petit{L34 $\heart$ L28}\\[\ruleTuringskip]
\petit{L34 $\heart$ L29}\\[\ruleTuringskip]
\petit{L35 $\heart$ K36}\\[\ruleTuringskip]
\petit{L35 $\heart$ L27}\\[\ruleTuringskip]
\petit{L35 $\heart$ L60 }\\[\ruleTuringskip]
\petit{L36 $\heart$ L26}\\[\ruleTuringskip]
\petit{L36 $\heart$ L59 }\\[\ruleTuringskip]
\petit{L36 $\heart$ L60 }\\[\ruleTuringskip]
\petit{L37 $\heart$ L25}\\[\ruleTuringskip]
\petit{L37 $\heart$ L32}\\[\ruleTuringskip]
\petit{L38 $\heart$ K32}\\[\ruleTuringskip]
\petit{L38 $\heart$ K39}\\[\ruleTuringskip]
\petit{L38 $\heart$ L57 }\\[\ruleTuringskip]
\petit{L39 $\heart$ K29}\\[\ruleTuringskip]
\petit{L39 $\heart$ K30}\\[\ruleTuringskip]
\petit{L39 $\heart$ K31}\\[\ruleTuringskip]
\petit{L39 $\heart$ K37}\\[\ruleTuringskip]
\petit{L39 $\heart$ K38}\\[\ruleTuringskip]
\petit{L39 $\heart$ L41 }\\[\ruleTuringskip]
\petit{L40 $\heart$ K28}\\[\ruleTuringskip]
\petit{L40 $\heart$ K29}\\[\ruleTuringskip]
\petit{L40 $\heart$ K30}\\[\ruleTuringskip]
\petit{L40 $\heart$ K35}\\[\ruleTuringskip]
\petit{L40 $\heart$ K37}\\[\ruleTuringskip]
\petit{L40 $\heart$ K42}\\[\ruleTuringskip]
\petit{L40 $\heart$ K47}\\[\ruleTuringskip]
\petit{L41 $\heart$ K28}\\[\ruleTuringskip]
\petit{L41 $\heart$ K29}\\[\ruleTuringskip]
\petit{L41 $\heart$ K33}\\[\ruleTuringskip]
\petit{L41 $\heart$ K35}\\[\ruleTuringskip]
\petit{L41 $\heart$ K41}\\[\ruleTuringskip]
\petit{L41 $\heart$ K46}\\[\ruleTuringskip]
\petit{L41 $\heart$ K47}\\[\ruleTuringskip]
\petit{L41 $\heart$ L18}\\[\ruleTuringskip]
\petit{L41 $\heart$ L39}\\[\ruleTuringskip]
\petit{L41 $\heart$ L47 }\\[\ruleTuringskip]
\petit{L42 $\heart$ K43}\\[\ruleTuringskip]
\petit{L42 $\heart$ L23}\\[\ruleTuringskip]
\petit{L42 $\heart$ L55 }\\[\ruleTuringskip]
\petit{L43 $\heart$ K43}\\[\ruleTuringskip]
\petit{L43 $\heart$ L22}\\[\ruleTuringskip]
\petit{L43 $\heart$ L53 }\\[\ruleTuringskip]
\petit{L44 $\heart$ K42}\\[\ruleTuringskip]
\petit{L44 $\heart$ L21}\\[\ruleTuringskip]
\petit{L44 $\heart$ L52 }\\[\ruleTuringskip]
\petit{L44 $\heart$ L53 }\\[\ruleTuringskip]
\petit{L45 $\heart$ K43}\\[\ruleTuringskip]
\petit{L45 $\heart$ L20}\\[\ruleTuringskip]
\petit{L46 $\heart$ K42}\\[\ruleTuringskip]
\petit{L46 $\heart$ L19}\\[\ruleTuringskip]
\petit{L46 $\heart$ L50 }\\[\ruleTuringskip]
\petit{L46 $\heart$ L51 }\\[\ruleTuringskip]
\petit{L47 $\heart$ K43}\\[\ruleTuringskip]
\petit{L47 $\heart$ L18}\\[\ruleTuringskip]
\petit{L47 $\heart$ L41}\\[\ruleTuringskip]
\petit{L47 $\heart$ L49 }\\[\ruleTuringskip]
\petit{L47 $\heart$ L50 }\\[\ruleTuringskip]
\petit{L48 $\heart$ K42}\\[\ruleTuringskip]
\petit{L48 $\heart$ L31}\\[\ruleTuringskip]
\petit{L48 $\heart$ L64 }\\[\ruleTuringskip]
\petit{L48 $\heart$ L82 }\\[\ruleTuringskip]
\petit{L48 $\heart$ M21 }\\[\ruleTuringskip]
\petit{L48 $\heart$ M22 }\\[\ruleTuringskip]
\petit{L49 $\heart$ L31}\\[\ruleTuringskip]
\petit{L49 $\heart$ L47}\\[\ruleTuringskip]
\petit{L49 $\heart$ L81 }\\[\ruleTuringskip]
\petit{L49 $\heart$ L82 }\\[\ruleTuringskip]
\petit{L50 $\heart$ L46}\\[\ruleTuringskip]
\petit{L50 $\heart$ L47}\\[\ruleTuringskip]
\petit{L50 $\heart$ L80 }\\[\ruleTuringskip]
\petit{L51 $\heart$ L46}\\[\ruleTuringskip]
\petit{L51 $\heart$ L79 }\\[\ruleTuringskip]
\petit{L52 $\heart$ L44}\\[\ruleTuringskip]
\petit{L52 $\heart$ L78 }\\[\ruleTuringskip]
\petit{L53 $\heart$ L43}\\[\ruleTuringskip]
\petit{L53 $\heart$ L44}\\[\ruleTuringskip]
\petit{L53 $\heart$ L77 }\\[\ruleTuringskip]
\petit{L54 $\heart$ L76 }\\[\ruleTuringskip]
\petit{L55 $\heart$ K41}\\[\ruleTuringskip]
\petit{L55 $\heart$ K53}\\[\ruleTuringskip]
\petit{L55 $\heart$ L42}\\[\ruleTuringskip]
\petit{L56 $\heart$ K35}\\[\ruleTuringskip]
\petit{L56 $\heart$ K50}\\[\ruleTuringskip]
\petit{L56 $\heart$ L72 }\\[\ruleTuringskip]
\petit{L56 $\heart$ L73 }\\[\ruleTuringskip]
\petit{L57 $\heart$ K35}\\[\ruleTuringskip]
\petit{L57 $\heart$ K50}\\[\ruleTuringskip]
\petit{L57 $\heart$ L38}\\[\ruleTuringskip]
\petit{L57 $\heart$ L71 }\\[\ruleTuringskip]
\petit{L58 $\heart$ K49}\\[\ruleTuringskip]
\petit{L58 $\heart$ L70 }\\[\ruleTuringskip]
\petit{L59 $\heart$ K50}\\[\ruleTuringskip]
\petit{L59 $\heart$ L36}\\[\ruleTuringskip]
\petit{L59 $\heart$ L69 }\\[\ruleTuringskip]
\petit{L60 $\heart$ K49}\\[\ruleTuringskip]
\petit{L60 $\heart$ L35}\\[\ruleTuringskip]
\petit{L60 $\heart$ L36}\\[\ruleTuringskip]
\petit{L60 $\heart$ L68 }\\[\ruleTuringskip]
\petit{L61 $\heart$ K50}\\[\ruleTuringskip]
\petit{L61 $\heart$ L67 }\\[\ruleTuringskip]
\petit{L62 $\heart$ K49}\\[\ruleTuringskip]
\petit{L62 $\heart$ L33}\\[\ruleTuringskip]
\petit{L62 $\heart$ L66 }\\[\ruleTuringskip]
\petit{L63 $\heart$ K50}\\[\ruleTuringskip]
\petit{L63 $\heart$ L32}\\[\ruleTuringskip]
\petit{L63 $\heart$ L33}\\[\ruleTuringskip]
\petit{L64 $\heart$ D57}\\[\ruleTuringskip]
\petit{L64 $\heart$ K49}\\[\ruleTuringskip]
\petit{L64 $\heart$ L18}\\[\ruleTuringskip]
\petit{L64 $\heart$ L48}\\[\ruleTuringskip]
\petit{L65 $\heart$ L17}\\[\ruleTuringskip]
\petit{L65 $\heart$ L18}\\[\ruleTuringskip]
\petit{L65 $\heart$ M30 }\\[\ruleTuringskip]
\petit{L66 $\heart$ L62}\\[\ruleTuringskip]
\petit{L66 $\heart$ M30 }\\[\ruleTuringskip]
\petit{L67 $\heart$ L61}\\[\ruleTuringskip]
\petit{L67 $\heart$ M29 }\\[\ruleTuringskip]
\petit{L68 $\heart$ L60}\\[\ruleTuringskip]
\petit{L68 $\heart$ M28 }\\[\ruleTuringskip]
\petit{L69 $\heart$ L59}\\[\ruleTuringskip]
\petit{L69 $\heart$ M27 }\\[\ruleTuringskip]
\petit{L69 $\heart$ M28 }\\[\ruleTuringskip]
\petit{L70 $\heart$ L58}\\[\ruleTuringskip]
\petit{L70 $\heart$ M27 }\\[\ruleTuringskip]
\petit{L71 $\heart$ L57}\\[\ruleTuringskip]
\petit{L71 $\heart$ M25 }\\[\ruleTuringskip]
\petit{L72 $\heart$ K48}\\[\ruleTuringskip]
\petit{L72 $\heart$ L56}\\[\ruleTuringskip]
\petit{L72 $\heart$ M24 }\\[\ruleTuringskip]
\petit{L72 $\heart$ M25 }\\[\ruleTuringskip]
\petit{L73 $\heart$ K48}\\[\ruleTuringskip]
\petit{L73 $\heart$ L56}\\[\ruleTuringskip]
\petit{L74 $\heart$ K48}\\[\ruleTuringskip]
\petit{L74 $\heart$ K55}\\[\ruleTuringskip]
\petit{L74 $\heart$ L82 }\\[\ruleTuringskip]
\petit{L75 $\heart$ K47}\\[\ruleTuringskip]
\petit{L75 $\heart$ K54}\\[\ruleTuringskip]
\petit{L75 $\heart$ K60}\\[\ruleTuringskip]
\petit{L75 $\heart$ L81 }\\[\ruleTuringskip]
\petit{L76 $\heart$ K55}\\[\ruleTuringskip]
\petit{L76 $\heart$ L54}\\[\ruleTuringskip]
\petit{L76 $\heart$ L90 }\\[\ruleTuringskip]
\petit{L77 $\heart$ K55}\\[\ruleTuringskip]
\petit{L77 $\heart$ L53}\\[\ruleTuringskip]
\petit{L77 $\heart$ L88 }\\[\ruleTuringskip]
\petit{L78 $\heart$ K54}\\[\ruleTuringskip]
\petit{L78 $\heart$ L52}\\[\ruleTuringskip]
\petit{L78 $\heart$ L87 }\\[\ruleTuringskip]
\petit{L78 $\heart$ L88 }\\[\ruleTuringskip]
\petit{L79 $\heart$ K55}\\[\ruleTuringskip]
\petit{L79 $\heart$ L51}\\[\ruleTuringskip]
\petit{L80 $\heart$ K54}\\[\ruleTuringskip]
\petit{L80 $\heart$ L50}\\[\ruleTuringskip]
\petit{L80 $\heart$ L85 }\\[\ruleTuringskip]
\petit{L81 $\heart$ L49}\\[\ruleTuringskip]
\petit{L81 $\heart$ L75}\\[\ruleTuringskip]
\petit{L81 $\heart$ L83 }\\[\ruleTuringskip]
\petit{L81 $\heart$ L84 }\\[\ruleTuringskip]
\petit{L81 $\heart$ L85 }\\[\ruleTuringskip]
\petit{L82 $\heart$ K45}\\[\ruleTuringskip]
\petit{L82 $\heart$ L48}\\[\ruleTuringskip]
\petit{L82 $\heart$ L49}\\[\ruleTuringskip]
\petit{L82 $\heart$ L74}\\[\ruleTuringskip]
\petit{L82 $\heart$ L84 }\\[\ruleTuringskip]
\petit{L82 $\heart$ M20 }\\[\ruleTuringskip]
\petit{L82 $\heart$ M21 }\\[\ruleTuringskip]
\petit{L83 $\heart$ L32}\\[\ruleTuringskip]
\petit{L83 $\heart$ L81}\\[\ruleTuringskip]
\petit{L83 $\heart$ M20 }\\[\ruleTuringskip]
\petit{L84 $\heart$ L81}\\[\ruleTuringskip]
\petit{L84 $\heart$ L82}\\[\ruleTuringskip]
\petit{L85 $\heart$ L80}\\[\ruleTuringskip]
\petit{L85 $\heart$ L81}\\[\ruleTuringskip]
\petit{L87 $\heart$ L78}\\[\ruleTuringskip]
\petit{L88 $\heart$ L77}\\[\ruleTuringskip]
\petit{L88 $\heart$ L78}\\[\ruleTuringskip]
\petit{L90 $\heart$ K53}\\[\ruleTuringskip]
\petit{L90 $\heart$ L76}\\[\ruleTuringskip]
\petit{L91 $\heart$ K53}\\[\ruleTuringskip]
\petit{L91 $\heart$ K63}\\[\ruleTuringskip]
\petit{L91 $\heart$ M19 }\\[\ruleTuringskip]
\petit{L92 $\heart$ K52}\\[\ruleTuringskip]
\petit{L92 $\heart$ K53}\\[\ruleTuringskip]
\petit{L92 $\heart$ K63}\\[\ruleTuringskip]
\petit{L92 $\heart$ M17 }\\[\ruleTuringskip]
\petit{L93 $\heart$ K62}\\[\ruleTuringskip]
\petit{L93 $\heart$ M16 }\\[\ruleTuringskip]
\petit{L94 $\heart$ K56}\\[\ruleTuringskip]
\petit{L94 $\heart$ K63}\\[\ruleTuringskip]
\petit{L94 $\heart$ M15 }\\[\ruleTuringskip]
\petit{L95 $\heart$ K55}\\[\ruleTuringskip]
\petit{L95 $\heart$ K56}\\[\ruleTuringskip]
\petit{L95 $\heart$ K62}\\[\ruleTuringskip]
\petit{L95 $\heart$ M14 }\\[\ruleTuringskip]
\petit{L96 $\heart$ K63}\\[\ruleTuringskip]
\petit{L96 $\heart$ M13 }\\[\ruleTuringskip]
\petit{L97 $\heart$ K53}\\[\ruleTuringskip]
\petit{L97 $\heart$ K62}\\[\ruleTuringskip]
\petit{L97 $\heart$ M12 }\\[\ruleTuringskip]
\petit{L98 $\heart$ K52}\\[\ruleTuringskip]
\petit{L98 $\heart$ K53}\\[\ruleTuringskip]
\petit{L98 $\heart$ K63}\\[\ruleTuringskip]
\petit{L98 $\heart$ M11 }\\[\ruleTuringskip]
\petit{L99 $\heart$ K62}\\[\ruleTuringskip]
\petit{L99 $\heart$ M10 }\\[\ruleTuringskip]
\petit{M00 $\heart$ K56}\\[\ruleTuringskip]
\petit{M00 $\heart$ K63}\\[\ruleTuringskip]
\petit{M00 $\heart$ M09 }\\[\ruleTuringskip]
\petit{M01 $\heart$ K55}\\[\ruleTuringskip]
\petit{M01 $\heart$ K56}\\[\ruleTuringskip]
\petit{M01 $\heart$ K62}\\[\ruleTuringskip]
\petit{M01 $\heart$ M08 }\\[\ruleTuringskip]
\petit{M02 $\heart$ K63}\\[\ruleTuringskip]
\petit{M02 $\heart$ M07 }\\[\ruleTuringskip]
\petit{M03 $\heart$ K53}\\[\ruleTuringskip]
\petit{M03 $\heart$ K62}\\[\ruleTuringskip]
\petit{M03 $\heart$ M06 }\\[\ruleTuringskip]
\petit{M04 $\heart$ K52}\\[\ruleTuringskip]
\petit{M04 $\heart$ K53}\\[\ruleTuringskip]
\petit{M04 $\heart$ K63}\\[\ruleTuringskip]
\petit{M04 $\heart$ M06 }\\[\ruleTuringskip]
\petit{M05 $\heart$ K57}\\[\ruleTuringskip]
\petit{M06 $\heart$ K56}\\[\ruleTuringskip]
\petit{M06 $\heart$ M03}\\[\ruleTuringskip]
\petit{M06 $\heart$ M04}\\[\ruleTuringskip]
\petit{M07 $\heart$ K55}\\[\ruleTuringskip]
\petit{M07 $\heart$ K56}\\[\ruleTuringskip]
\petit{M07 $\heart$ M02}\\[\ruleTuringskip]
\petit{M08 $\heart$ M01}\\[\ruleTuringskip]
\petit{M09 $\heart$ K53}\\[\ruleTuringskip]
\petit{M09 $\heart$ M00}\\[\ruleTuringskip]
\petit{M10 $\heart$ K52}\\[\ruleTuringskip]
\petit{M10 $\heart$ K53}\\[\ruleTuringskip]
\petit{M10 $\heart$ L99}\\[\ruleTuringskip]
\petit{M11 $\heart$ L98}\\[\ruleTuringskip]
\petit{M12 $\heart$ K56}\\[\ruleTuringskip]
\petit{M12 $\heart$ L97}\\[\ruleTuringskip]
\petit{M13 $\heart$ K55}\\[\ruleTuringskip]
\petit{M13 $\heart$ K56}\\[\ruleTuringskip]
\petit{M13 $\heart$ L96}\\[\ruleTuringskip]
\petit{M14 $\heart$ L95}\\[\ruleTuringskip]
\petit{M15 $\heart$ K53}\\[\ruleTuringskip]
\petit{M15 $\heart$ L94}\\[\ruleTuringskip]
\petit{M16 $\heart$ K52}\\[\ruleTuringskip]
\petit{M16 $\heart$ K53}\\[\ruleTuringskip]
\petit{M16 $\heart$ L93}\\[\ruleTuringskip]
\petit{M17 $\heart$ L92}\\[\ruleTuringskip]
\petit{M18 $\heart$ K56}\\[\ruleTuringskip]
\petit{M19 $\heart$ K55}\\[\ruleTuringskip]
\petit{M19 $\heart$ K56}\\[\ruleTuringskip]
\petit{M19 $\heart$ L91}\\[\ruleTuringskip]
\petit{M20 $\heart$ L82}\\[\ruleTuringskip]
\petit{M20 $\heart$ L83}\\[\ruleTuringskip]
\petit{M20 $\heart$ M25 }\\[\ruleTuringskip]
\petit{M21 $\heart$ K56}\\[\ruleTuringskip]
\petit{M21 $\heart$ K68}\\[\ruleTuringskip]
\petit{M21 $\heart$ L48}\\[\ruleTuringskip]
\petit{M21 $\heart$ L82}\\[\ruleTuringskip]
\petit{M21 $\heart$ M24 }\\[\ruleTuringskip]
\petit{M22 $\heart$ K55}\\[\ruleTuringskip]
\petit{M22 $\heart$ K56}\\[\ruleTuringskip]
\petit{M22 $\heart$ K67}\\[\ruleTuringskip]
\petit{M22 $\heart$ L18}\\[\ruleTuringskip]
\petit{M22 $\heart$ L48}\\[\ruleTuringskip]
\petit{M23 $\heart$ M28 }\\[\ruleTuringskip]
\petit{M23 $\heart$ M29 }\\[\ruleTuringskip]
\petit{M24 $\heart$ L72}\\[\ruleTuringskip]
\petit{M24 $\heart$ M21}\\[\ruleTuringskip]
\petit{M24 $\heart$ M28 }\\[\ruleTuringskip]
\petit{M25 $\heart$ L71}\\[\ruleTuringskip]
\petit{M25 $\heart$ L72}\\[\ruleTuringskip]
\petit{M25 $\heart$ M20}\\[\ruleTuringskip]
\petit{M25 $\heart$ M27 }\\[\ruleTuringskip]
\petit{M25 $\heart$ M28 }\\[\ruleTuringskip]
\petit{M26 $\heart$ A02}\\[\ruleTuringskip]
\petit{M27 $\heart$ L69}\\[\ruleTuringskip]
\petit{M27 $\heart$ L70}\\[\ruleTuringskip]
\petit{M27 $\heart$ M25}\\[\ruleTuringskip]
\petit{M27 $\heart$ M30 }\\[\ruleTuringskip]
\petit{M28 $\heart$ L68}\\[\ruleTuringskip]
\petit{M28 $\heart$ L69}\\[\ruleTuringskip]
\petit{M28 $\heart$ M23}\\[\ruleTuringskip]
\petit{M28 $\heart$ M24}\\[\ruleTuringskip]
\petit{M28 $\heart$ M25}\\[\ruleTuringskip]
\petit{M28 $\heart$ M30 }\\[\ruleTuringskip]
\petit{M29 $\heart$ L67}\\[\ruleTuringskip]
\petit{M29 $\heart$ M23}\\[\ruleTuringskip]
\petit{M30 $\heart$ L65}\\[\ruleTuringskip]
\petit{M30 $\heart$ L66}\\[\ruleTuringskip]
\petit{M30 $\heart$ M27}\\[\ruleTuringskip]
\petit{M30 $\heart$ M28}\\[\ruleTuringskip]
\end{multicols}

\afterpage\clearpage

%% file: _main.bbl
\begin{thebibliography}{10}

\bibitem{cody1}
J.~Boyle, G.~Robillard, and S.~Kim.
\newblock Sequential folding of transfer {RNA}. a nuclear magnetic resonance
  study of successively longer t{RNA} fragments with a common 5' end.
\newblock {\em J. Mol. Biol.}, 139:601--625, 1980.

\bibitem{Cook2004}
M.~Cook.
\newblock Universality in elementary cellular automata.
\newblock {\em Complex Systems}, 15:1--40, 2004.

\bibitem{shape2018DNA}
E.D. Demaine, J.~Hendricks, M.~Olsen, M.J. Patitz, T.~Rogers~N. Schabanel,
  S.~Seki, and H.~Thomas.
\newblock Know when to fold 'em: Self-assembly of shapes by folding in
  oritatami.
\newblock In {\em DNA}, 2018.
\newblock To be published.

\bibitem{cody2}
K.~L. Frieda and S.~M. Block.
\newblock Direct observation of cotranscriptional folding in an adenine
  riboswitch.
\newblock {\em Science}, 338(6105):397--400, 2012.

\bibitem{GeMeScSe2016}
C.~Geary, P.-\'E. Meunier, N.~Schabanel, and S.~Seki.
\newblock Programming biomolecules that fold greedily during transcription.
\newblock In {\em MFCS 2016}, volume LIPIcs 58, pages 43:1--43:14, 2016.

\bibitem{GearyRothemundAndersen2014}
C.~Geary, P.~W.~K. Rothemund, and E.~S. Andersen.
\newblock A single-stranded architecture for cotranscriptional folding of {RNA}
  nanostructures.
\newblock {\em Science}, 345:799--804, 2014.

\bibitem{gacs1997reliable}
P.~Gács.
\newblock Reliable cellular automata with self-organization.
\newblock In {\em FOCS}, pages 90--99, 1997.

\bibitem{HanKim2017}
Y-S. Han and H.~Kim.
\newblock Ruleset optimization on isomorphic oritatami systems.
\newblock In {\em Proc.~DNA23}, LNCS 10467, pages 33--45. Springer, 2017.

\bibitem{HaKiOtSe2016}
Y.-S. Han, H.~Kim, M.~Ota, and S.~Seki.
\newblock Nondeterministic seedless oritatami systems and hardness of testing
  their equivalence.
\newblock In {\em DNA}, volume LNCS 9818, pages 19--34, 2016.

\bibitem{DBLP:conf/icalp/KariKMPS15}
L.~Kari, S.~Kopecki, P.{-}{\'{E}}. Meunier, M.~J. Patitz, and S.~Seki.
\newblock Binary pattern tile set synthesis is np-hard.
\newblock In {\em ICALP}, pages 1022--1034, 2015.

\bibitem{MaShUb2018}
Yusei Masuda, Shinnosuke Seki, and Yuki Ubukata.
\newblock Towards the algorithmic molecular self-assembly of fractals by
  cotranscriptional folding.
\newblock In {\em CIAA 2018}, LNCS, 2018.
\newblock in press.

\bibitem{NearyPhD}
Turlough Neary.
\newblock {\em Small universal Turing machines}.
\newblock PhD thesis, NUI, Maynooth, 2008.

\bibitem{Ollinger2002ICALP}
Nicolas Ollinger.
\newblock The quest for small universal cellular automata.
\newblock In {\em ICALP}, volume LNCS 2380, pages 318--329, 2002.

\bibitem{OtaSeki2017}
Makoto Ota and Shinnosuke Seki.
\newblock Ruleset design problems for oritatami systems.
\newblock {\em Theor. Comput. Sci.}, 671:26--35, 2017.

\bibitem{WoodsNeary2006}
Damien Woods and Turlough Neary.
\newblock On the time complexity of 2-tag systems and small universal {T}uring
  machines.
\newblock In {\em FOCS}, pages 439--448, 2006.

\end{thebibliography}
